\documentclass{junopaper}
\pdfoutput=1            
\usepackage[utf8]{inputenc}
\usepackage{color}
\usepackage{graphicx}
\usepackage{subfigure}
\usepackage[space]{grffile}
\usepackage[T1]{fontenc}
\usepackage{lmodern}
\usepackage{xspace}
\usepackage{rotating}
\usepackage{lipsum}
\usepackage{amsmath}
\usepackage{multirow}
\usepackage{longtable,booktabs}
\usepackage[section]{placeins}
\usepackage[affil-it]{authblk}

\usepackage{fancyhdr} 

\fancypagestyle{empty}{
  \fancyhf{}

}
\fancypagestyle{plain}{
  \fancyhf{}
  \fancyfoot[C]{\bfseries \thepage}

}
\fancypagestyle{simple}{%
  \fancyhf{}
  \fancyhead[RO,LE]{\textsf{\footnotesize Conceptual Design Report}}
  \fancyhead[LO,RE]{\textsf{\footnotesize JUNO Collaboration}}
  \fancyfoot[C]{\bfseries \thepage}
  
}

\fancypagestyle{references}{%
  \fancyhf{}
  \fancyhead[RO,LE]{\textsf{\footnotesize \bfseries \thepage}}
  \fancyhead[LO,RE]{\textsf{\footnotesize References}}
  \fancyfoot[CO,CE]{\textsf{\footnotesize \bfseries \@title}}
}

\usepackage[subfigure]{tocloft}
\tocloftpagestyle{plain}

\usepackage[detect-all=true,group-digits=true,group-separator={,},group-minimum-digits=4]{siunitx}
\DeclareSIUnit \kton {\kilo\tonne}
\DeclareSIUnit \kt {\kilo\tonne}
\DeclareSIUnit \Mt {\mega\tonne}
\DeclareSIUnit \eV {\electronvolt}
\DeclareSIUnit \keV {\kilo\electronvolt}
\DeclareSIUnit \MeV {\mega\electronvolt}
\DeclareSIUnit \GeV {\giga\electronvolt}
\DeclareSIUnit \km {\kilo\meter}
\DeclareSIUnit \kW {\kilo\watt}
\DeclareSIUnit \MW {\mega\watt}
\DeclareSIUnit \MHz {\mega\hertz}
\DeclareSIUnit \mrad {\milli\radian}
\DeclareSIUnit \year {year}
\DeclareSIUnit \POT {POT}
\DeclareSIUnit \sig {$\sigma$}
\DeclareSIUnit\parsec{pc}
\DeclareSIUnit\lightyear{ly}
\DeclareSIUnit\foot{ft}
\DeclareSIUnit\feet{ft}
\DeclareSIUnit\ft{ft}

\usepackage[pdftex,bookmarks]{hyperref}
\hypersetup{
    colorlinks=false,
    linktoc=section,
    linkcolor=black,
    citecolor=blue,
    urlcolor=blue,
    filecolor=black,
    pdfpagemode=UseOutlines,
}

\usepackage{geometry}
\begin{document}
\input{maintex/frontmatter}
\cleardoublepage
\renewcommand{\thepage}{\arabic{page}}
\setcounter{page}{1}
\pagestyle{simple}

\cleardoublepage

\chapter{Introduction}
\label{chap:intro}

\section{Neutrino Physics: Status and Prospect}
\subsection{Fundamentals of Massive Neutrinos}
\label{subsec:mixing}

In the Standard Model \cite{Weinberg} of particle physics, the weak
charged-current interactions of leptons and quarks can be written as
\begin{eqnarray}
{\cal L}^{}_{\rm cc} = &-&\frac{g}{\sqrt{2}} \left[ \overline{\left(e
\hspace{0.3cm} \mu \hspace{0.3cm} \tau \right)^{}_{\rm L}} \
\gamma^\mu \ U \left(\begin{matrix} \nu^{}_1 \cr \nu^{}_2 \cr
\nu^{}_3 \cr \end{matrix}\right)^{}_{\rm L} W^-_\mu \right. \nonumber \\ &+&
\left.\overline{\left(u \hspace{0.3cm} c \hspace{0.3cm} t \right)^{}_{\rm
L}} \ \gamma^\mu \ V \left(\begin{matrix} d \cr s \cr b \cr
\end{matrix}\right)^{}_{\rm L} W^+_\mu \right] + {\rm h.c.} \; ,
\end{eqnarray}
where all the fermion fields are the mass eigenstates, $U$ is the
$3\times 3$ Maki-Nakagawa-Sakata-Pontecorvo (MNSP) matrix
\cite{MNS}, and $V$ denotes the $3\times 3$
Cabibbo-Kobayashi-Maskawa (CKM) matrix \cite{CKM}.

Given the basis in which the flavor eigenstates of the three charged
leptons are identified with their mass eigenstates, the flavor
eigenstates of the three active neutrinos and $n$ sterile neutrinos read
as
\begin{eqnarray}
\left(\begin{matrix} \nu^{}_e \cr \nu^{}_\mu \cr \nu^{}_\tau \cr
\vdots \cr
\end{matrix}
\right) = \left(\begin{matrix} U^{}_{e1} & U^{}_{e2} & U^{}_{e3} &
\cdots \cr U^{}_{\mu 1} & U^{}_{\mu 2} & U^{}_{\mu 3} & \cdots \cr
U^{}_{\tau 1} & U^{}_{\tau 2} & U^{}_{\tau 3} & \cdots \cr \vdots &
\vdots & \vdots & \ddots \cr \end{matrix} \right)
\left(\begin{matrix} \nu^{}_1 \cr \nu^{}_2 \cr \nu^{}_3 \cr \vdots
\cr
\end{matrix}
\right) \; .
\end{eqnarray}
where $\nu^{}_i$ is a neutrino mass eigenstate with the physical
mass $m^{}_i$ (for $i=1,2, \cdots, 3+n$). Equation (1.1) tells us that a
$\nu^{}_\alpha$ neutrino can be produced from the $W^+ +
\ell^-_\alpha \to \nu^{}_\alpha$ interaction, and a $\nu^{}_\beta$
neutrino can be detected through the $\nu^{}_\beta \to W^+ +
\ell^-_\beta$ interaction (for $\alpha, \beta = e, \mu, \tau$). So
oscillation may happen if the $\nu^{}_i$ beam with energy $E \gg
m^{}_i$ travels a proper distance $L$. In vacuum, the oscillation
probability of the $\bar\nu^{}_e \to \bar\nu^{}_e$ transition turns
out to be
\begin{eqnarray}
P(\overline{\nu}^{}_e \to \overline{\nu}^{}_e) = 1 -
\frac{4}{\displaystyle \left(\sum_i |U^{}_{e i}|^2 \right)^2}
\sum^{}_{i<j} \left(|U^{}_{e i}|^2 |U^{}_{e j}|^2 \sin^2
\frac{\Delta m^{2}_{ij} L}{4 E} \right) \; ,
\end{eqnarray}
with $\Delta m^2_{ij} \equiv m^2_i - m^2_j$ being the mass-squared
difference. Note that the denominator on the right-hand side of Equation
(1.3) is not equal to one if there are heavy sterile antineutrinos
which mix with the active antineutrinos but do not take part in the
flavor oscillations. Note also that the terrestrial matter effects
on $P(\overline{\nu}^{}_e \to \overline{\nu}^{}_e)$ are negligibly
small, because the typical value of $E$ is only a few MeV and that
of $L$ is usually less than several hundred km for a realistic reactor-based
$\overline{\nu}^{}_e \to \overline{\nu}^{}_e$ oscillation
experiment.

If the $3\times 3$ MNSP matrix $U$ is exactly unitary, it can be
parameterized in terms of three flavor mixing angles and three
CP-violating phases in the following standard way \cite{PDG}:
\begin{eqnarray}
U & \hspace{-0.2cm} = \hspace{-0.2cm} & \left( \begin{matrix} 1 & 0
& 0 \cr 0 & c^{}_{23} & s^{}_{23} \cr 0 & -s^{}_{23} & c^{}_{23} \cr
\end{matrix} \right) \left( \begin{matrix} c^{}_{13} & 0 & s^{}_{13}
e^{-{\rm i}\delta} \cr 0 & 1 & 0 \cr -s^{}_{13} e^{{\rm i}\delta} &
0 & c^{}_{13} \cr
\end{matrix} \right)
\left( \begin{matrix} c^{}_{12} & s^{}_{12} & 0 \cr -s^{}_{12} &
c^{}_{12} & 0 \cr 0 & 0 & 1 \cr \end{matrix} \right) P^{}_\nu
\nonumber \\
& \hspace{-0.2cm} = \hspace{-0.2cm} & \left( \begin{matrix}
c^{}_{12} c^{}_{13} & s^{}_{12} c^{}_{13} & s^{}_{13} e^{-{\rm i}
\delta} \cr -s^{}_{12} c^{}_{23} - c^{}_{12} s^{}_{13} s^{}_{23}
e^{{\rm i} \delta} & c^{}_{12} c^{}_{23} - s^{}_{12} s^{}_{13}
s^{}_{23} e^{{\rm i} \delta} & c^{}_{13} s^{}_{23} \cr s^{}_{12}
s^{}_{23} - c^{}_{12} s^{}_{13} c^{}_{23} e^{{\rm i} \delta} &
-c^{}_{12} s^{}_{23} - s^{}_{12} s^{}_{13} c^{}_{23} e^{{\rm i}
\delta} & c^{}_{13} c^{}_{23} \cr
\end{matrix} \right) P^{}_\nu \; ,
\end{eqnarray}
where $c^{}_{ij} \equiv \cos\theta^{}_{ij}$ and $s^{}_{ij} \equiv
\sin\theta^{}_{ij}$ (for $ij = 12, 13, 23$) are defined, and
$P^{}_\nu = {\rm Diag}\{e^{{\rm i}\rho}, e^{{\rm i}\sigma}, 1\}$
denotes the diagonal Majorana phase matrix which has nothing to do
with neutrino oscillations. In this case,
\begin{eqnarray}
P(\overline{\nu}^{}_e \to \overline{\nu}^{}_e) &=& 1 - \sin^2
2\theta^{}_{12} c^4_{13} \sin^2 \frac{\Delta m^{2}_{21} L}{4 E}\nonumber \\ &-&
\sin^2 2\theta^{}_{13} \left[ c^2_{12} \sin^2 \frac{\Delta
m^{2}_{31} L}{4 E} + s^2_{12} \sin^2 \frac{\Delta m^{2}_{32} L}{4 E}
\right] \; ,
\end{eqnarray}
in which $\Delta m^2_{32} = \Delta m^2_{31} - \Delta m^2_{21}$. The
oscillation terms driven by $\Delta m^2_{21}$ and $\Delta m^2_{31}
\simeq \Delta m^2_{32}$ can therefore be used to determine
$\theta^{}_{12}$ and $\theta^{}_{13}$, respectively.

\subsection{Open Issues of Massive Neutrinos}
\label{subsec:openissue}

\begin{table}[t]
\vspace{-0.25cm}
\begin{center}
\caption{The best-fit values, together with the 1$\sigma$
and 3$\sigma$ intervals, for the six three-flavor neutrino
oscillation parameters from a global analysis of current
experimental data \cite{GF1}.} \vspace{0.5cm}
\begin{tabular}{c|c|c|c}
\hline \hline
Parameter & Best fit & 1$\sigma$ range & 3$\sigma$ range \\
\hline \multicolumn{4}{c}{Normal neutrino mass hierarchy $(m^{}_1 <
m^{}_2 < m^{}_3$)} \\ \hline
$\Delta m^2_{21}/10^{-5} ~{\rm eV}^2$
& $7.54$  & 7.32 --- 7.80 & 6.99 --- 8.18 \\
$\Delta m^2_{31}/10^{-3} ~ {\rm eV}^2$~ & $2.47$
& 2.41 --- 2.53 & 2.27 --- 2.65 \\
$\sin^2\theta_{12}/10^{-1}$
& $3.08$ & 2.91 --- 3.25  & 2.59 --- 3.59 \\
$\sin^2\theta_{13}/10^{-2}$
& $2.34$ & 2.15 --- 2.54 & 1.76 --- 2.95 \\
$\sin^2\theta_{23}/10^{-1}$
& $4.37$  & 4.14 --- 4.70 & 3.74 --- 6.26 \\
$\delta/180^\circ$ &  $1.39$ & 1.12 --- 1.77 & 0.00 --- 2.00 \\ \hline
\multicolumn{4}{c}{Inverted neutrino mass hierarchy $(m^{}_3 < m^{}_1
< m^{}_2$)} \\ \hline
$\Delta m^2_{21}/10^{-5} ~{\rm eV}^2$
& $7.54$  & 7.32 --- 7.80 & 6.99 --- 8.18 \\
$\Delta m^2_{31}/10^{-3} ~ {\rm eV}^2$~ & $2.34$
& 2.28 --- 2.40 & 2.15 --- 2.52 \\
$\sin^2\theta_{12}/10^{-1}$
& $3.08$ & 2.91 --- 3.25  & 2.59 --- 3.59 \\
$\sin^2\theta_{13}/10^{-2}$
& $2.40$ & 2.18 --- 2.59  & 1.78 --- 2.98 \\
$\sin^2\theta_{23}/10^{-1}$ & $4.55$  & 4.24 --- 5.94
 & 3.80 --- 6.41 \\
$\delta/180^\circ$ &  $1.31$ & 0.98 --- 1.60 & 0.00 --- 2.00 \\ \hline\hline
\end{tabular}
\end{center}
\end{table}
Although we have known quite a lot about massive neutrinos,
where the current status of neutrino oscillation measurements can be
summarized in Table~2.1, we have many open questions about their
fundamental
properties and their unique roles in the Universe
\cite{XZ}. In the following we concentrate on some intrinsic flavor issues
of massive neutrinos which may related to future neutrino
experiments:

(a) The nature of neutrinos and their mass spectrum;

{\it Question (1): Dirac or Majorana nature?}

{\it Question (2): Normal or inverted mass hierarchy?}

{\it Question (3): The absolute mass scale?}

(b) Lepton flavor mixing pattern and CP violation;

{\it Question (4): The octant of $\theta^{}_{23}$?}

{\it Question (5): The Dirac CP-violating phase $\delta$?}

{\it Question (6): The Majorana CP-violating phases $\rho$ and
$\sigma$?}

(c) Extra neutrino species and unitarity tests;

{\it Question (7): Extra light or heavy sterile neutrinos?}

{\it Question (8): Direct and indirect non-unitary effects?}

\section{JUNO Experiment}
\label{sec:juno}

\begin{figure}[htb!]
\centering
\includegraphics[width=0.7\textwidth]{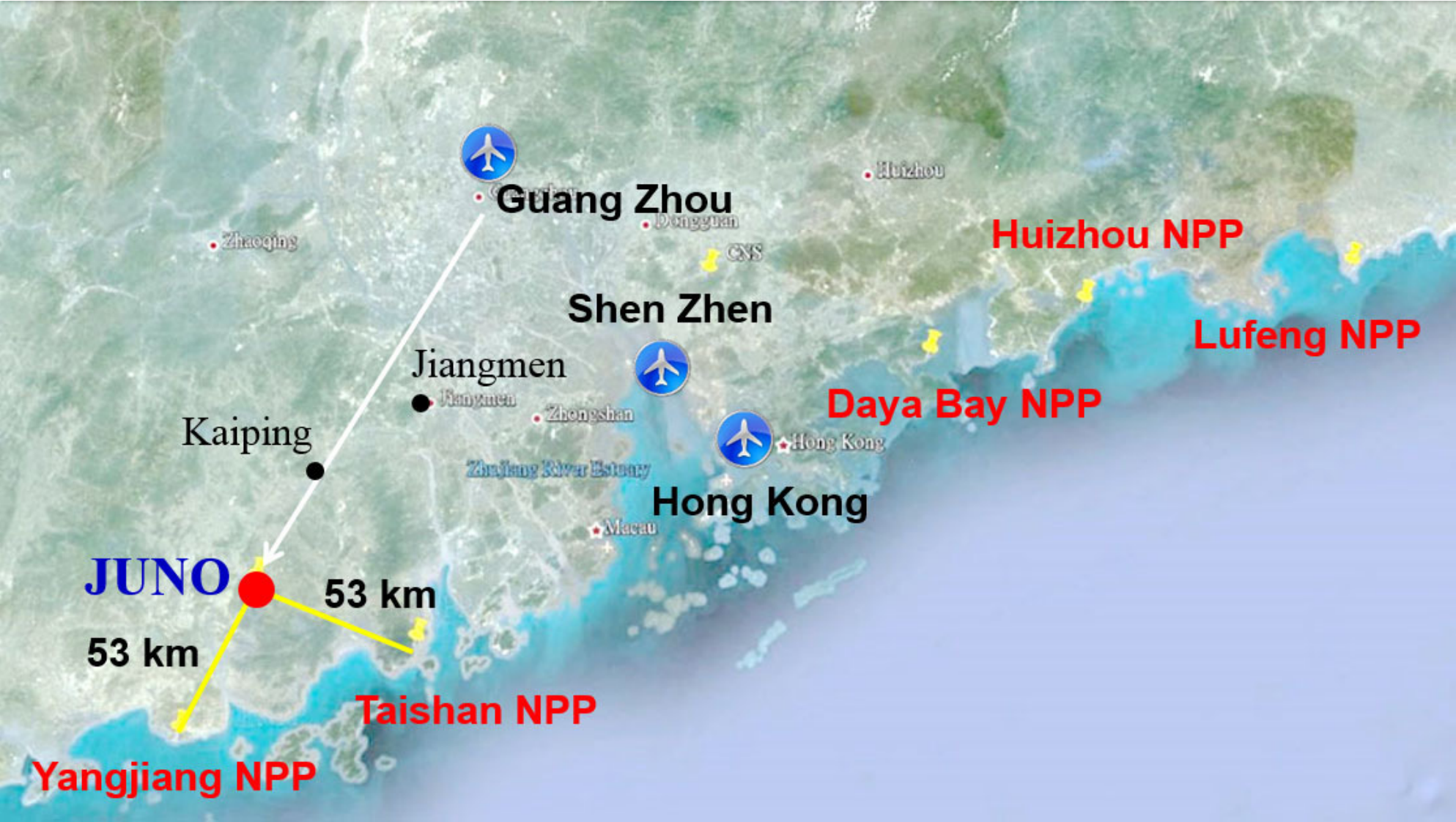}
\caption{Location of the JUNO site. The distances to the nearby
Yangjiang NPP and Taishan NPP are both 53 km. Daya Bay NPP is 215 km
away. Huizhou and Lufeng NPPs have not been approved yet. Three
metropolises, Hong Kong, Shenzhen, and Guangzhou, are also shown.
\label{fig:intro:location} }
\end{figure}

The Jiangmen Underground Neutrino Observatory (JUNO) is a multi-purpose neutrino experiment.
It was proposed in 2008 for neutrino mass hierarchy (MH) determination by detecting reactor
antineutrinos from nuclear power plants (NPPs)~\cite{zhanl2008,
yfwang2008,caoj2009}. The site location is optimized to have the
best sensitivity for mass hierarchy determination, which is at 53~km
from both the Yangjiang and Taishan NPPs. The neutrino detector is a
liquid scintillator (LS) detector with a 20~kton fiducial mass,
deployed in a laboratory 700 meters underground. The JUNO experiment
is located in Jinji town, Kaiping city, Jiangmen city, Guangdong
province which is shown in Fig.~\ref{fig:intro:location}. The
thermal power and baselines are listed in Table~\ref{tab:intro:NPP}.
\begin{table}[htb]
\centering
\begin{tabular}{|c|c|c|c|c|c|c|}\hline\hline
Cores & YJ-C1 & YJ-C2 & YJ-C3 & YJ-C4 & YJ-C5  & YJ-C6 \\
\hline Power (GW) & 2.9 & 2.9 & 2.9 & 2.9 & 2.9 & 2.9 \\ \hline
Baseline(km) & 52.75 & 52.84 & 52.42 & 52.51 & 52.12 & 52.21 \\
\hline\hline
Cores & TS-C1 & TS-C2 & TS-C3 & TS-C4 & DYB  & HZ \\
\hline Power (GW) & 4.6 & 4.6 & 4.6 & 4.6 & 17.4 & 17.4 \\ \hline
Baseline(km) & 52.76 & 52.63 & 52.32 & 52.20 & 215 & 265 \\
\hline
\end{tabular}
\caption{Summary of the thermal power and baseline to the JUNO
detector for the Yangjiang (YJ) and Taishan (TS) reactor cores, as
well as the remote reactors of Daya Bay (DYB) and Huizhou
(HZ).\label{tab:intro:NPP}}
\end{table}

\begin{figure}[htb!]
\centering
\includegraphics[width=0.7\textwidth]{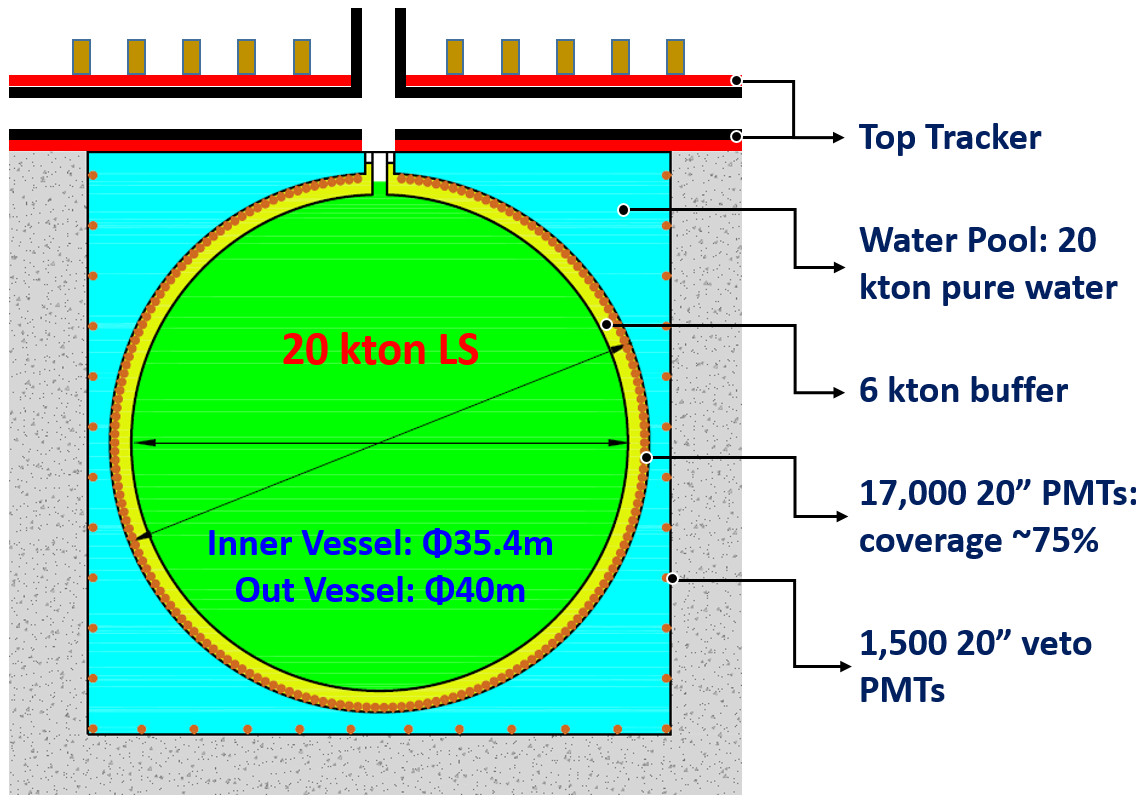}
\caption{A schematic view of the JUNO detector.
\label{fig:intro:det} }
\end{figure}

JUNO consists of a central detector, a water Cherenkov
detector and a muon tracker (shown in Fig.~\ref{fig:intro:det}). The
central detector is a LS detector of 20~kton target mass and
$3\%/\sqrt{E{\rm (MeV)}}$ energy resolution. The central detector is
submerged in a water pool to be shielded from natural
radioactivities from the surrounding rock and air. The water pool is
equipped with PMTs to detect the Cherenkov light from muons. On top
of the water pool, there is another muon detector to accurately
measure the muon track.

It is crucial to achieve a $3\%/\sqrt{E{\rm (MeV)}}$ energy resolution for the determination of the MH.
A Monte Carlo simulation has been developed based on
the Daya Bay Monte Carlo. The photoelectron yield has been tuned
according to the real data of Daya Bay. The required energy resolution can
be reached with the following improvements from Daya Bay \cite{DYB}:
\begin{itemize}
\item The PMT photocathode covergage $\geq 75$\%.
\item The PMT photocathode quantum efficiency $\geq 35$\%.
\item The attenuation length of the liquid scintillator $\geq 20$~m at 430~nm,
which corresponds to an absorption length of 60~m with a Rayleigh
scattering length of 30~m.
\end{itemize}
For the real experimental environments, there are many other factors beyond the photoelectron
statistics
that can alter the energy resolution, including the dark noise from the PMTs and electronics,
the detector non-uniformity and vertex resolution, and the PMT charge resolution. A generic parametrization
for the detector energy resolution is defined as
\begin{equation}
\frac{\sigma_{E}}{E} = \sqrt{\left(\frac{a}{\sqrt{E}}\right)^2+b^2+\left(\frac{c}{E}\right)^2}\,\;,\label{eq:mh:abcterms}
\end{equation}
where the visible energy $E$ is in MeV.

Based on the numerical simulation as shown in the next section, a requirement for the resolution of
${a}/{\sqrt{E}}$ better than $3\%$ is approximately equivalent to the following requirement,
\begin{equation}
\sqrt{\left({a}\right)^2+\left({1.6\times b}\right)^2+\left(\frac{c}{1.6}\right)^2}\leq 3\%\;.\label{eq:mh:abc}
\end{equation}

\section{Physics Potentials}
\label{sec:potentials}

\subsection{Mass Hierarchy}
\label{subsubsec:MH}

\begin{figure}
\begin{center}
\begin{tabular}{c}
\includegraphics*[width=0.6\textwidth]{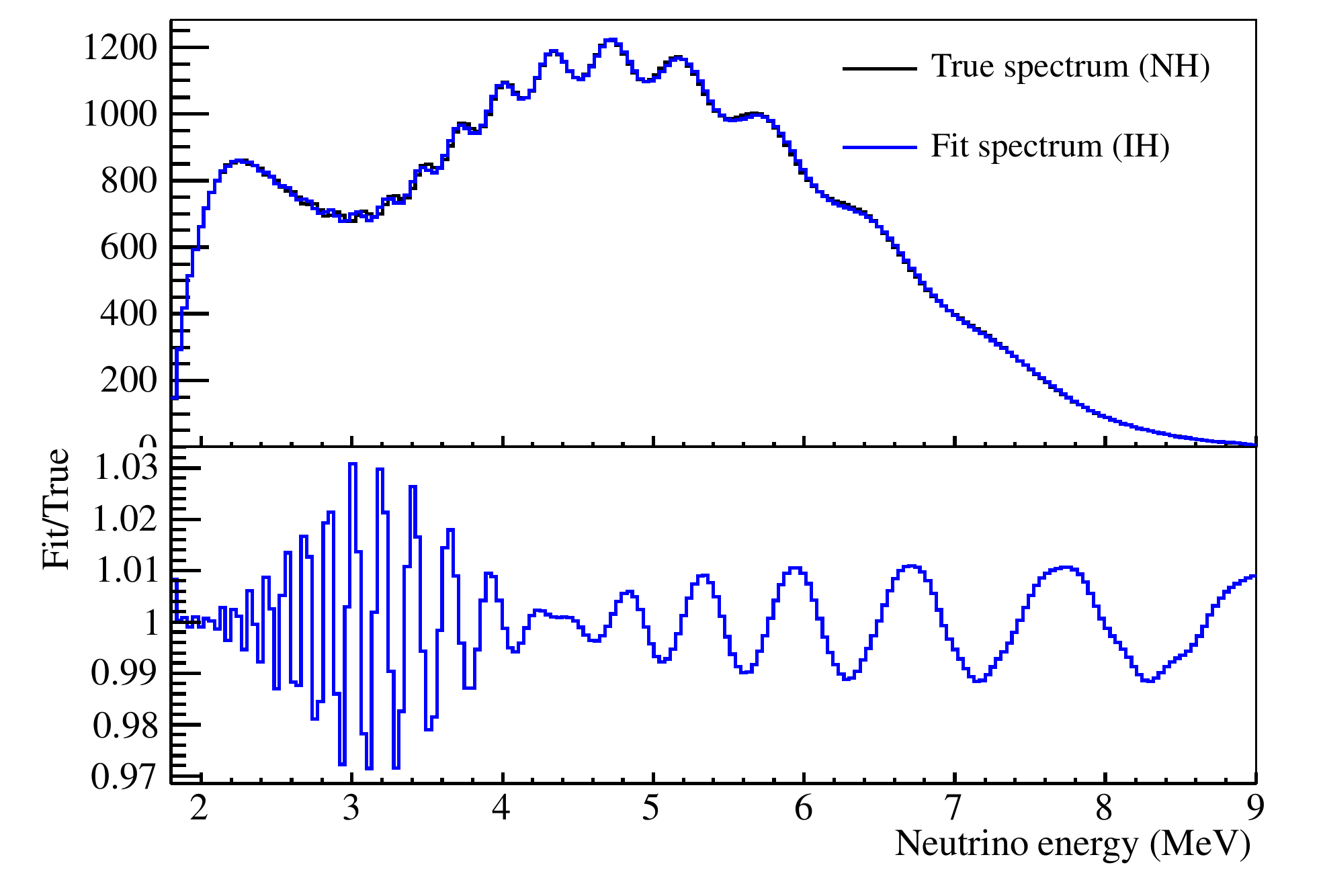}
\end{tabular}
\end{center}
\caption{Neutrino energy spectra (upper panel) and their ratio (lower panel) with the true (NH) and fit (IH) MHs.}
\label{fig:Espec}
\end{figure}
The neutrino mass hierarchy (MH) answers the question whether the
third generation ($\nu_3$ mass eigenstate) is heavier or lighter
than the first two generations ($\nu_1$ and $\nu_2$). The normal
mass hierarchy (NH) refers to $m_3
> m_1$ and the inverted mass hierarchy (IH) refers to $m_3 < m_1$.
JUNO is designed to resolve the neutrino MH using precision spectral
measurements of reactor antineutrino oscillations, where the general
principle is shown in Fig.~\ref{fig:Espec}.

In the JUNO simulation, we assume a 20~kton LS detector and
a total thermal power of two reactor complexes of 36~GW. We assume
nominal running time of six years, 300 effective days per year,
$80\%$ detection efficiency and a detector energy resolution
$3\%/\sqrt{E{\rm (MeV)}}$ as a benchmark. The simulation details can
be found in Ref.~\cite{JUNO}. In Fig.~\ref{fig:Espec},
the expected neutrino energy spectrum of the true normal MH, and best-fit
neutrino spectrum of the wrong inverted MH are shown in the upper panel, and
the ratio of two spectra is illustrated in the lower panel. One can observe that
the distinct feature of two MHs lies in the fine structures of the neutrino spectrum.

To quantify the sensitivity of MH determination, we define the following quantity as the MH discriminator,
\begin{equation}
\Delta \chi^2_{\text{MH}}=|\chi^2_{\rm min}(\rm NH)-\chi^2_{\rm
min}(\rm IH)|, \label{eq:mh:chisquare}
\end{equation}
where the minimization process is implemented for the oscillation parameters and systematics.

\begin{figure}
\begin{center}
\begin{tabular}{cc}
\includegraphics*[bb=20 20 290 232, width=0.43\textwidth]{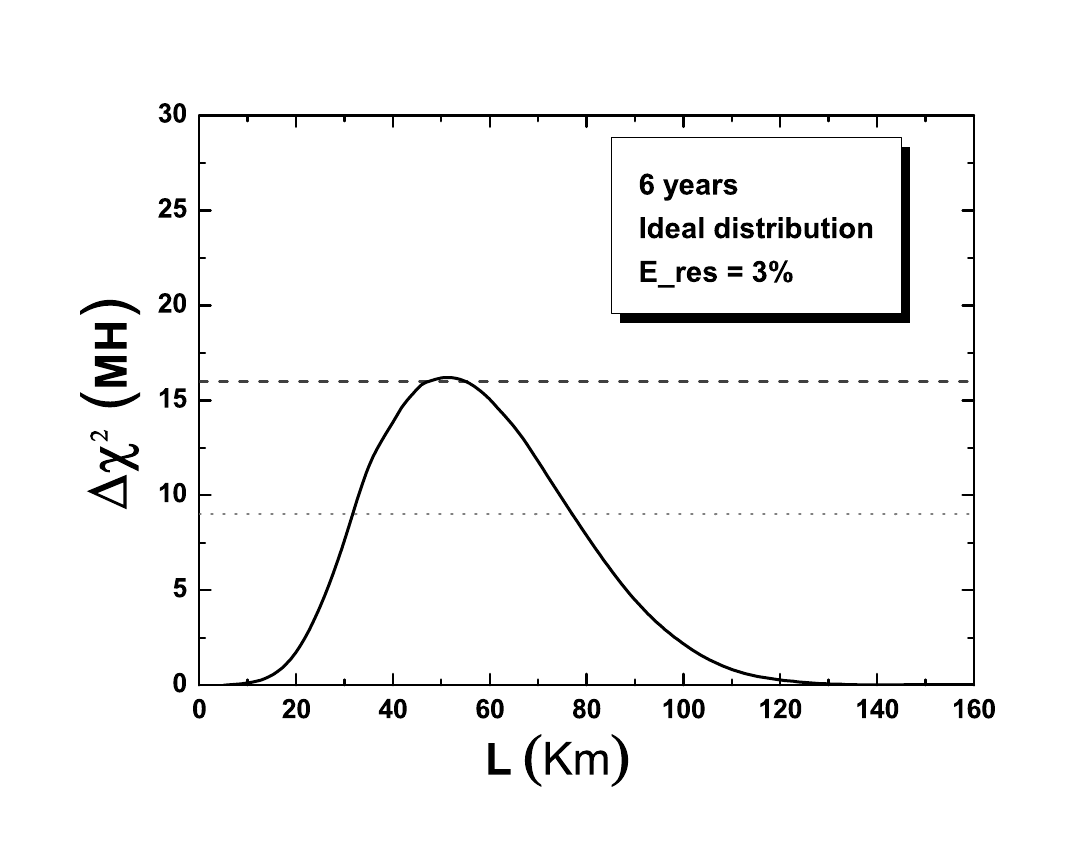}
&
\includegraphics*[bb=20 16 284 214, width=0.42\textwidth]{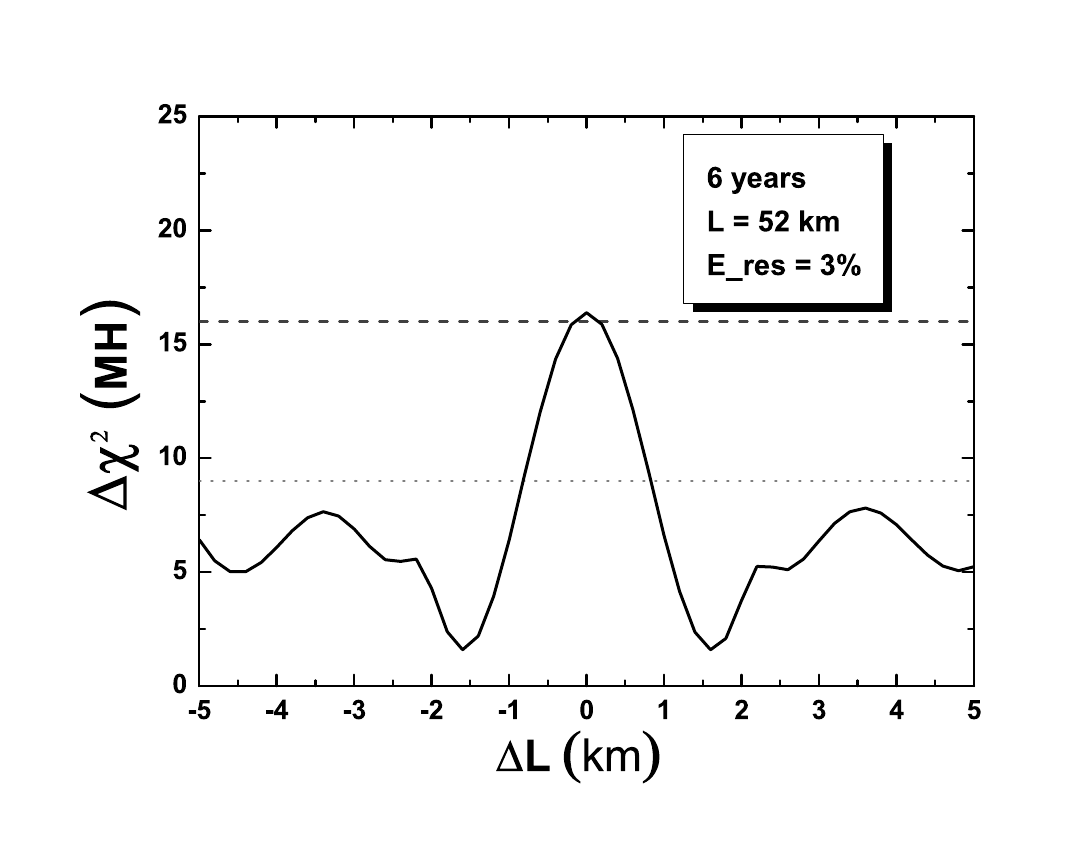}
\end{tabular}
\end{center}
\caption{The MH discrimination ability as the function of the
baseline (left panel) and as the function of the baseline difference of two
reactors (right panel).} \label{fig:mh:baseline}
\end{figure}

\begin{figure}
\begin{center}
\begin{tabular}{c}
\includegraphics*[bb=26 22 292 222, width=0.5\textwidth]{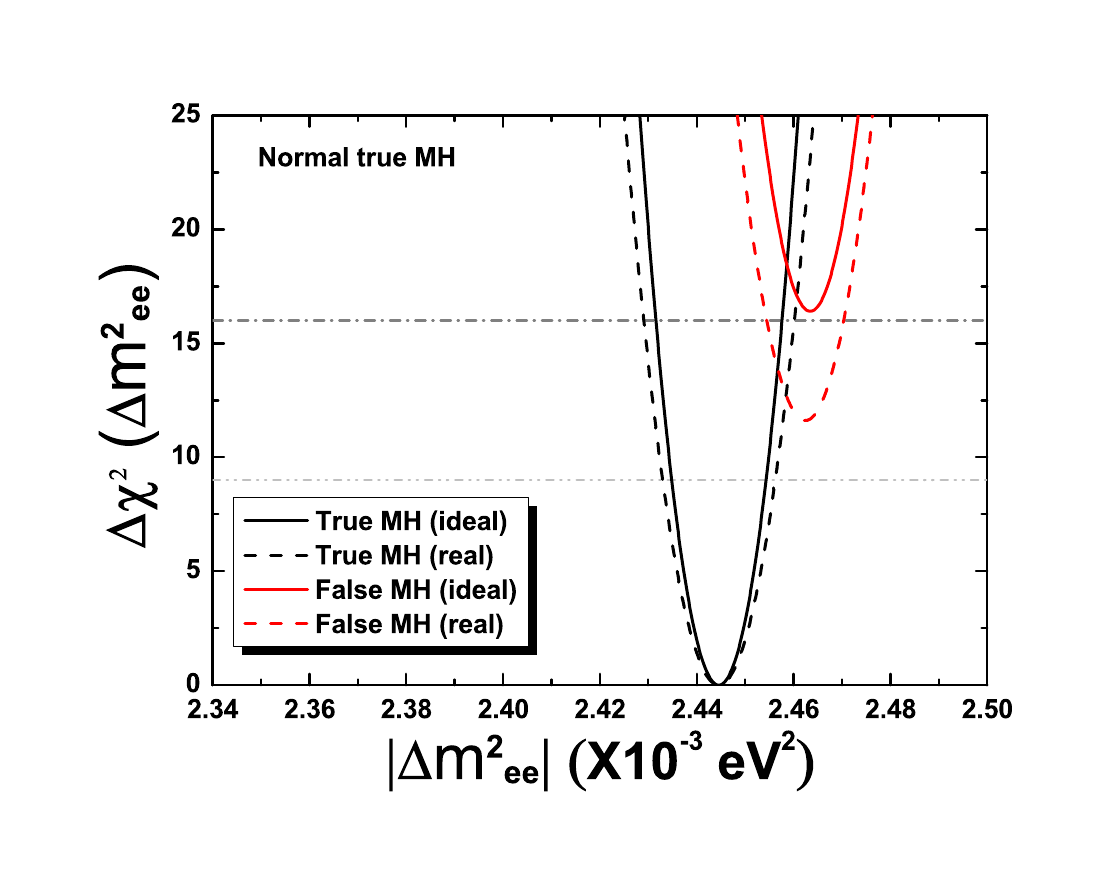}
\end{tabular}
\end{center}
\caption{The comparison of the MH sensitivity for the ideal and
actual distributions of the reactor cores and baselines. The real
distribution gives a degradation of $\Delta
\chi^2_{\text{MH}}\simeq5$.} \label{fig:mh:ideal}
\end{figure}

The discriminator defined in Eq.~(\ref{eq:mh:chisquare}) can be used
to obtain the optimal baseline, which is shown in the left panel of
Fig.~\ref{fig:mh:baseline}. An optimal sensitivity of $\Delta
\chi^2_{\text{MH}}\simeq16$ can be obtained for the ideal case with
identical baseline at around 50~km, where the oscillation effect of $\delta m^2_{21}$ is
maximal.
The impact of the baseline
difference due to multiple reactor cores is shown in the right panel
of Fig.~\ref{fig:mh:baseline}, by keeping the baseline of one
reactor unchanged and varying that of the other. A rapid oscillatory
behavior is observed, which demonstrates the importance of baseline
differences for the reactor cores. The worst case is at $\Delta L
\sim 1.7$~km, where the $|\Delta m^2_{ee}|$ related oscillation is
canceled between the two reactors. Taking into account the reactor power
and baseline distribution of the real experimental site of JUNO, we
show the reduction of the MH sensitivity in Fig.~\ref{fig:mh:ideal},
which gives a degradation of $\Delta \chi^2_{\text{MH}}\simeq5$.

\begin{figure}
\begin{center}
\begin{tabular}{c}
\includegraphics*[bb=25 20 295 228, width=0.5\textwidth]{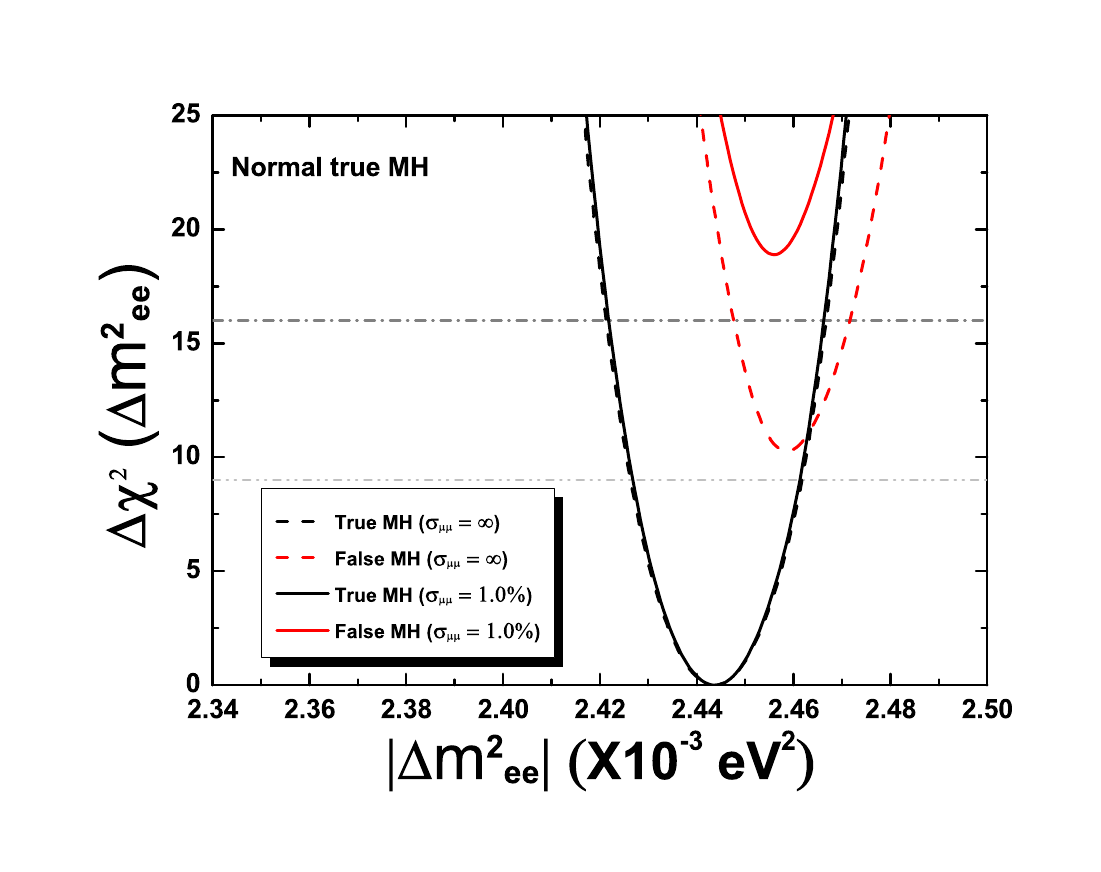}
\end{tabular}
\end{center}
\caption{the reactor-only (dashed) and combined (solid)
distributions of the $\Delta\chi^2$ function, where a $1\%$ relative
error of $\Delta m^2_{\mu\mu}$ is assumed and the CP-violating phase
($\delta$) is assigned to be $90^\circ/270^\circ$ ($\cos\delta=0$)
for illustration. The black and red lines are for the true (normal)
and false (inverted) neutrino MH, respectively.
\label{fig:mh:chi2eemumu}}
\end{figure}

There are also reactor and detector related uncertainties that
affect the MH sensitivity. Rate uncertainties are
negligible, because most of the MH sensitivity is derived from the spectral
information. On the other hand, the energy-related uncertainties are
important, including the reactor spectrum error, the detector
bin-to-bin error, and the energy non-linearity error. By considering
realistic spectral uncertainties and taking into account the
self-calibration of oscillation patterns of reactor antineutrino
oscillations, we obtain the nominal MH sensitivity of JUNO as shown
with the dashed lines in Fig.~\ref{fig:mh:chi2eemumu}. In addition,
due to the difference between $|\Delta m^2_{ee}|$ and $|\Delta
m^2_{\mu\mu}|$, precise measurements of the two mass-squared difference
can provide additional sensitivity to MH, besides the sensitivity
from the interference effects. In Fig.~\ref{fig:mh:chi2eemumu}, we
show with the solid lines the improvement by adding a $|\Delta
m^2_{\mu\mu}|$ measurement of $1\%$ precision, where an increase of
$\Delta \chi^2_{\text{MH}}\simeq8$ is achieved for the MH
sensitivity.

\subsection{Precision Measurement of Mixing Parameters}
\label{subsec:prec}

JUNO is a precision experiment in terms of huge statistics (${\cal
O}(100\rm k)$ inverse beta decay (IBD) events), optimal baseline, unprecedented energy
resolution ($3\%/\sqrt{E}$) and accurate energy response (better
than $1\%$). Therefore, besides the determination of the neutrino
mass hierarchy (MH) \cite{JUNO}, the ${\cal O}(100\rm k)$ IBD events
allow JUNO to probe the fundamental properties of neutrino
oscillations and access four additional parameters $\theta_{12}$,
$\theta_{13}$, $\Delta m^2_{21}$, and $|\Delta m^2_{ee}|$.

\begin{table}[!htb]\footnotesize
\begin{center}
\begin{tabular}[c]{l|l|l|l|l|l} \hline\hline
  & $\Delta m^2_{21}$ & $|\Delta m^2_{31}|$ & $\sin^2\theta_{12}$ &
  $\sin^2\theta_{13}$ & $\sin^2\theta_{23}$ \\ \hline
 Dominant Exps. & KamLAND  & MINOS & SNO & Daya Bay & SK/T2K  \\  \hline
 Individual 1$\sigma$ & 2.7\% \cite{KLloe} & 4.1\% \cite{MINOS} & 6.7\% \cite{SNO}
 & 6\% \cite{DBloe} & 14\% \cite{SKth23,T2Kth23} \\  \hline
 Global 1$\sigma$ & 2.6\% & 2.7\% & 4.1\% & 5\% & 11\%  \\
\hline\hline
\end{tabular}
\caption{\label{tab:prec:current} Current precision for the five
known oscillation parameters. The dominant experiments and their
corresponding 1$\sigma$ accuracy and global precision from global
fitting groups \cite{GF1} are shown in the first, second and third
row, respectively.}
\end{center}
\end{table}
\begin{table}[!htb]
\begin{center}
\begin{tabular}[c]{l|l|l|l|l|l} \hline\hline
  & Nominal &  + B2B (1\%)  & + BG &
  + EL (1\%) &  + NL (1\%) \\ \hline
 $\sin^2\theta_{12}$ & 0.54\%  & 0.60\% & 0.62\% & 0.64\% & 0.67\%  \\  \hline
 $\Delta m^2_{21}$ & 0.24\% & 0.27\%  & 0.29\% & 0.44\% & 0.59\% \\  \hline
 $|\Delta m^2_{ee}|$ & 0.27\% & 0.31\% & 0.31\% & 0.35\% & 0.44\%  \\
\hline\hline
\end{tabular}
\caption{\label{tab:prec:syst} Precision of $\sin^2\theta_{12}$,
$\Delta m^2_{21}$ and $|\Delta m^2_{ee}|$ from the nominal setup to
those with more systematic uncertainties. The systematics are added
one by one from the left cell to right cell.}
\end{center}
\end{table}
Current precision for the five known oscillation parameters is
summarized in Table~\ref{tab:prec:current}, where both the results
from individual experiments and from global analyses \cite{GF1} are
presented. We notice that most of the oscillation parameters have
been measured to better than $10\%$.

Among all the four parameters that are accessible in JUNO, the $\theta_{13}$
measurement from JUNO is less accurate than that of Daya Bay because the
designed baseline is much larger than the optimized one ($\sim2\,\rm
km$) and only one single detector is considered in the design
concept. The ultimate $\sin^22\theta_{13}$ sensitivity of Daya Bay will be about $3\%$, and would lead this precision level in
the foreseeable future. Therefore, we shall consider the precision
measurements of $\theta_{12}$, $\Delta m^2_{21}$ and $|\Delta
m^2_{ee}|$\footnote{There will be two degenerate solutions for
$|\Delta m^2_{ee}|$ in case of undetermined MH. We consider
the correct MH in the following studies.}.

With the nominal setup as in the study of the MH
measurement \cite{JUNO}, we estimate the precision of the three relevant
parameters, $\sin^2\theta_{12}$, $\Delta
m^2_{21}$ and $\Delta m^2_{ee}$, which can achieve the level of $0.54\%$, $0.24\%$
and $0.27\%$, respectively.

Moreover, the effects of important systematic errors, such as
the bin-to-bin (B2B) energy uncorrelated uncertainty,
the energy linear scale (EL) uncertainty and the energy non-linear
(NL) uncertainty, and the influence of background
(BG) are presented. As a benchmark, a $1\%$ precision for all the
systematic errors considered is assumed. The background level and
uncertainties are the same as in the previous chapter for the MH
determination. In Table~\ref{tab:prec:syst}, we show the precision
of $\sin^2\theta_{12}$, $\Delta m^2_{21}$ and $|\Delta m^2_{ee}|$
from the nominal setup to those with more systematic uncertainties.
We can see that the energy-related uncertainties are more
important because most of the sensitivity is derived from the spectrum
distortion due to neutrino oscillations.

In summary, we can achieve the precision of $0.5\% - 0.7\%$ for
the three oscillation parameters $\sin^2\theta_{12}$, $\Delta m^2_{21}$
and $|\Delta m^2_{ee}|$. Precision tests of the unitarity
of the lepton mixing matrix and mass sum rule are possible with
the unprecedented precision of these measurements.

\subsection{Supernova Neutrinos}
\label{subsec:sn}

\begin{figure}
\centering
\includegraphics[width=1\textwidth]{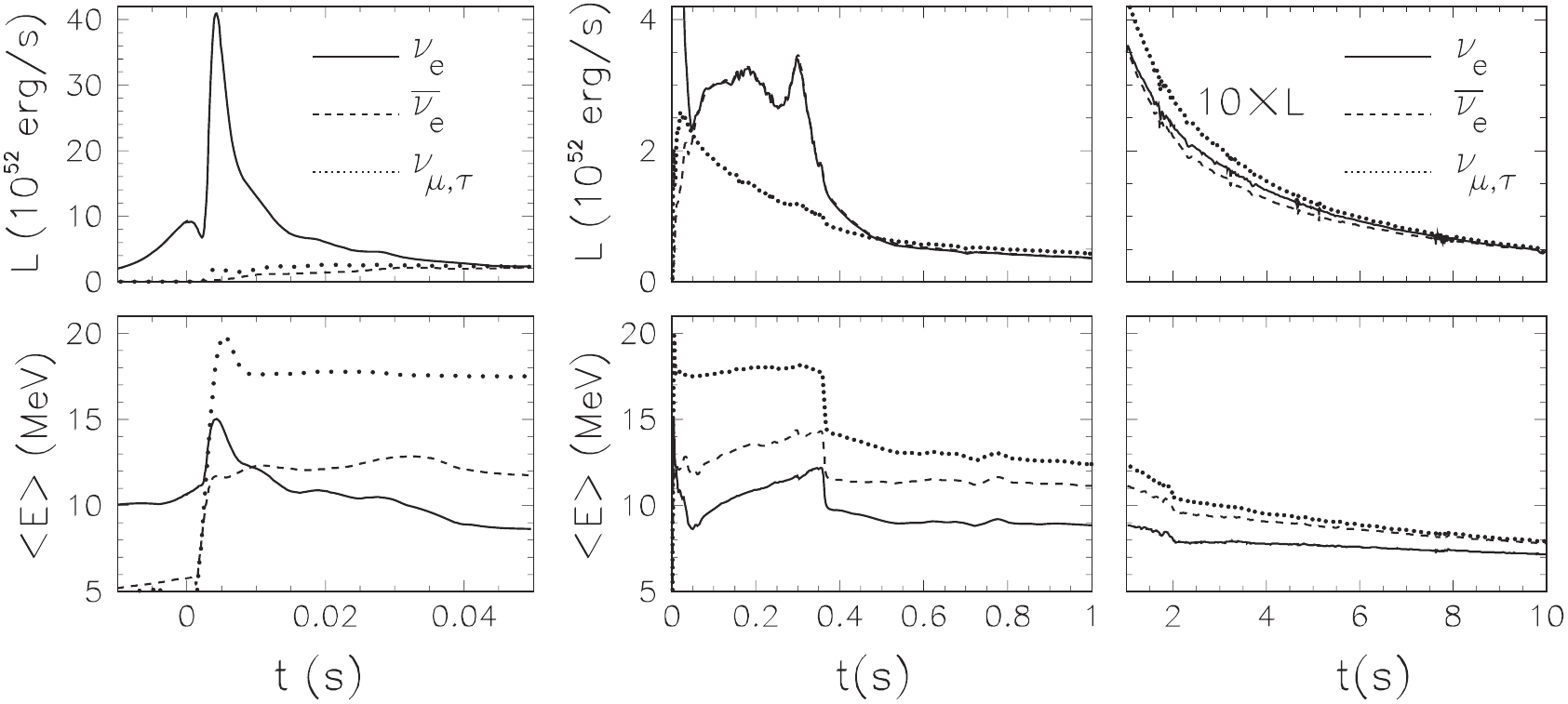}
\caption{Three phases of neutrino emission from a core-collapse SN,
from left to right: (1)~Infall, bounce and initial shock-wave
propagation, including prompt $\nu_e$ burst. (2)~Accretion phase
with significant flavor differences of fluxes and spectra and time
variations of the signal. (3)~Cooling of the newly formed neutron
star, only small flavor differences between fluxes and spectra.
\label{fig:sn:SNburst}}
\end{figure}

Measuring the neutrino burst from the next nearby supernova (SN) is
a premier goal of low-energy neutrino physics and astrophysics.
According to current understanding and numerical simulation, one
expects a neutrino signal with three characteristic phases as shown
in Fig.~\ref{fig:sn:SNburst}. In a high-statistics observation one
should consider these essentially as three different experiments,
each holding different and characteristic lessons for particle and
astrophysics.
\begin{itemize}
\item[\em 1.]{\em Infall, Bounce and Shock Propagation}.---Few tens of ms
    after bounce. Prompt $\nu_e$ burst, emission of $\bar\nu_e$ at first
    suppressed and emission of other flavors begins. 

\item[\em 2.]{\em Accretion Phase (Shock Stagnation)}.---Few tens to few
    hundreds of ms, depending on progenitor properties and other
    parameters. Neutrino emission is powered by accretion flow.
    Luminosity in $\nu_e$ and $\bar\nu_e$ perhaps as much as a factor of
    two larger than each of the $\nu_x$ fluxes.

\item[\em 3.]{\em Neutron-star cooling}.---Lasts until 10--20~s, powered by
cooling and deleptonization of the inner core on a diffusion time scale.
No strong asymmetries. Details depend on final neutron-star mass and
nuclear equation of state.
\end{itemize}

In order to estimate the expected neutrino signals at JUNO, we
assume a fiducial mass of 20~kton LS. For a typical galactic SN at 10~kpc,
we take the time-integrated neutrino
spectra as $f_\nu(E_\nu) \propto E^\alpha_\nu
\exp[-(1+\alpha)E_\nu/\langle E_\nu \rangle]$ with a nominal index
$\alpha = 3$ and $\langle E_\nu \rangle$ being the average neutrino
energy. Furthermore, a total neutrino energy of $E_{\rm tot} = 3 \times
10^{53}~{\rm erg}$ is assumed to be equally distributed among neutrinos
and antineutrinos of three flavors. As average neutrino energies are
both flavor- and time-dependent, we calculate the event rates for
three representative values $\langle E_\nu \rangle = 12~{\rm MeV}$,
$14~{\rm MeV}$ and $16~{\rm MeV}$, and in each case the average
energy is assumed to be equal for all flavors. The total numbers of
neutrino events in JUNO are summarized in Table~\ref{table:events}.
Some comments on the detection channels are presented as follows.

(1) The inverse beta decay (IBD) is the dominant channel for
supernova neutrino detection at both scintillator and
water-Cherenkov detectors. In the IBD reaction
\begin{equation}
\overline{\nu}_e + p \to e^+ + n \; , \label{eq: IBD}
\end{equation}
the neutrino energy threshold is $E^{\rm th}_\nu \approx 1.8~{\rm MeV}$.
The deposition of
positron kinetic energy and the annihilation of the positron
give rise to a prompt signal. In addition, the
neutron is captured on free protons with a lifetime of about $200~{\rm \mu s}$,
producing a $2.2~{\rm MeV}$ $\gamma$. Hence the coincidence of
prompt and delayed signals increases greatly the power of background rejection.
\begin{table}[!t]
\centering
\begin{tabular}{ccccccccc}
\hline
\multicolumn{1}{c}{\multirow {2}{*}{Channel}} & \multicolumn{1}{c}{} & \multicolumn{1}{c}{\multirow {2}{*}{Type}} & \multicolumn{1}{c}{} & \multicolumn{5}{c}{Events for different $\langle E_\nu \rangle$ values} \\
\cline{5-9} \multicolumn{1}{c}{} & \multicolumn{1}{c}{} & \multicolumn{1}{c}{} & \multicolumn{1}{c}{} & \multicolumn{1}{c}{$12~{\rm MeV}$} & \multicolumn{1}{c}{} & \multicolumn{1}{c}{$14~{\rm MeV}$} & \multicolumn{1}{c}{} & \multicolumn{1}{c}{$16~{\rm MeV}$} \\
\hline
\multicolumn{1}{l}{$\overline{\nu}_e + p \to e^+ + n$} & \multicolumn{1}{c}{} & \multicolumn{1}{c}{CC} & \multicolumn{1}{c}{} & \multicolumn{1}{c}{$4.3\times 10^3$} & \multicolumn{1}{c}{} & \multicolumn{1}{c}{$5.0\times 10^3$} & \multicolumn{1}{c}{} & \multicolumn{1}{c}{$5.7\times 10^3$} \\
\multicolumn{1}{l}{$\nu + p \to \nu + p$} & \multicolumn{1}{c}{} & \multicolumn{1}{c}{NC} & \multicolumn{1}{c}{} & \multicolumn{1}{c}{$6.0\times 10^2$} & \multicolumn{1}{c}{} & \multicolumn{1}{c}{$1.2\times 10^3$} & \multicolumn{1}{c}{} & \multicolumn{1}{c}{$2.0\times 10^3$} \\
\multicolumn{1}{l}{$\nu + e \to \nu + e$} & \multicolumn{1}{c}{} & \multicolumn{1}{c}{NC} & \multicolumn{1}{c}{} & \multicolumn{1}{c}{$3.6\times 10^2$} & \multicolumn{1}{c}{} & \multicolumn{1}{c}{$3.6\times 10^2$} & \multicolumn{1}{c}{} & \multicolumn{1}{c}{$3.6\times 10^2$} \\
\multicolumn{1}{l}{$\nu +~^{12}{\rm C} \to \nu +~^{12}{\rm C}^*$} & \multicolumn{1}{c}{} & \multicolumn{1}{c}{NC} & \multicolumn{1}{c}{} & \multicolumn{1}{c}{$1.7\times 10^2$} & \multicolumn{1}{c}{} & \multicolumn{1}{c}{$3.2\times 10^2$} & \multicolumn{1}{c}{} & \multicolumn{1}{c}{$5.2\times 10^2$} \\
\multicolumn{1}{l}{$\nu_e +~^{12}{\rm C} \to e^- +~^{12}{\rm N}$} & \multicolumn{1}{c}{} & \multicolumn{1}{c}{CC} & \multicolumn{1}{c}{} & \multicolumn{1}{c}{$4.7\times 10^1$} & \multicolumn{1}{c}{} & \multicolumn{1}{c}{$9.4\times 10^1$} & \multicolumn{1}{c}{} & \multicolumn{1}{c}{$1.6\times 10^2$} \\
\multicolumn{1}{l}{$\overline{\nu}_e +~^{12}{\rm C} \to e^+ +~^{12}{\rm B}$} & \multicolumn{1}{c}{} & \multicolumn{1}{c}{CC} & \multicolumn{1}{c}{} & \multicolumn{1}{c}{$6.0\times 10^1$} & \multicolumn{1}{c}{} & \multicolumn{1}{c}{$1.1\times 10^2$} & \multicolumn{1}{c}{} & \multicolumn{1}{c}{$1.6\times 10^2$} \\
\hline
\end{tabular}
\caption{Numbers of neutrino events in JUNO for a SN at a galactic distance
of 10 kpc.}
    \label{table:events}
\end{table}

(2) As an advantage of the LS detector, the
charged-current interaction on $^{12}{\rm C}$ takes place for both
$\nu_e$ and $\overline{\nu}_e$ via
\begin{eqnarray}
&& \nu_e +~^{12}{\rm C} \to e^- +~^{12}{\rm B} \; , \label{eq: CCnue}\\
&& \overline{\nu}_e +~^{12}{\rm C} \to e^+ +~^{12}{\rm N} \; .
\label{eq: CCnueb}
\end{eqnarray}
The energy threshold for $\nu_e$ is $17.34~{\rm MeV}$, while that
for $\overline{\nu}_e$ is $14.39~{\rm MeV}$. The subsequent beta
decays of $^{12}{\rm B}$ and $^{12}{\rm N}$ with a $20.2~{\rm ms}$
and $11~{\rm ms}$ half-life, respectively, lead to a prompt-delayed
coincident signal. Hence the charged-current reactions in
Eqs.~(\ref{eq: CCnue}) and (\ref{eq: CCnueb}) provide a possibility
to detect separately $\nu_e$ and $\overline{\nu}_e$.

(3) Neutral-current interaction on $^{12}{\rm C}$ is of crucial
importance to probe neutrinos of non-electron flavors, i.e.,
\begin{equation}
\nu +~^{12}{\rm C} \to \nu +~^{12}{\rm C}^* \; , \label{eq: NCnux}
\end{equation}
where $\nu$ stands for neutrinos or antineutrinos of all three
flavors. A $15.11~{\rm MeV}$ $\gamma$ from the de-excitation of
$^{12}{\rm C}^*$ to its ground state is a clear signal of SN
neutrinos. Since neutrinos of non-electron flavors $\nu_x$ have
higher average energies, the neutral-current interaction is more
sensitive to $\nu_x$, providing a possibility to pin down the flavor
content of supernova neutrinos. However,
it is impossible to reconstruct the neutrino energy in this channel.

(4) Elastic scattering of neutrinos on electrons will carry the
directional information of incident neutrinos, and thus can be used
to locate the SN. This is extremely important if a SN is hidden by
other stars and the optical signal is obscured by galactic dust. The
elastic scattering
\begin{equation}
\nu + e^- \to \nu + e^- \label{eq:ESe}
\end{equation}
is most sensitive to $\nu_e$ because of the larger cross section.
However, it is difficult to determine the direction of the scattered
electron in the scintillator detector. At this point,
large water-Cherenkov detectors, such as Super-Kamiokande,
are complementary to scintillator detectors.

(5) Elastic scattering of neutrinos on protons has been proposed as
a promising channel to measure supernova neutrinos of non-electron
flavors~\cite{Beacom:2002hs,Dasgupta:2011wg}:
\begin{equation}
\nu + p \to \nu + p \; . \label{eq: ESp}
\end{equation}
Although the total cross section is about four times smaller than
that of IBD reaction, the contributions from all the neutrinos and
antineutrinos of three flavors will compensate for the reduction of
cross section. As shown in
Refs.~\cite{Beacom:2002hs,Dasgupta:2011wg}, it is possible to
reconstruct the energy spectrum of $\nu_x$ at a large scintillator
detector, which is very important to establish flavor conversions or
spectral splits of SN neutrinos.
For a realistic measurement of $\nu_x$ spectrum, a low-energy
threshold and a satisfactory reconstruction of proton recoil energy
are required. Taking into account the quenching effect, the total number of events is about
$2240$ above a threshold $0.2~{\rm MeV}$.

In summary, we show the time-integrated neutrino event spectra of SN neutrinos
with respect to the visible energy $E^{}_{\rm d}$ in the JUNO detector for a SN at 10 kpc,
where no neutrino flavor conversions are assumed for illustration
and the average neutrino energies are $\langle E^{}_{\nu_e}\rangle = 12~{\rm MeV}$,
$\langle E^{}_{\overline{\nu}_e}\rangle = 14~{\rm MeV}$ and $\langle E^{}_{\nu_x}\rangle = 16~{\rm MeV}$.
The main reaction channels are shown together with the threshold of neutrino energies:
(1) IBD (black and solid curve), $E_{\rm d} = E^{}_\nu - 0.8~{\rm MeV}$;
(2) Elastic $\nu$-$p$ scattering (red and dashed curve), $E_{\rm d}$ stands for the recoil energy of proton;
(3) Elastic $\nu$-$e$ scattering (blue and double-dotted-dashed curve), $E_{\rm d}$ denotes the recoil energy of electron;
(4) Neutral-current reaction ${^{12}{\rm C}}(\nu, \nu^\prime){^{12}{\rm C}^*}$ (orange and dotted curve), $E_{\rm d} \approx 15.1~{\rm MeV}$;
(5) Charged-current reaction ${^{12}{\rm C}}(\nu_e, e^-){^{12}{\rm N}}$ (green and dotted-dashed curve), $E_{\rm d} = E_\nu - 17.3~{\rm MeV}$;
(6) Charged-current reaction ${^{12}{\rm C}}(\overline{\nu}_e, e^+){^{12}{\rm B}}$ (magenta and double-dotted curve), $E_{\rm d} = E_\nu - 13.9~{\rm MeV}$.

\begin{figure}[!t]
    \centering
    \includegraphics[width=0.7\textwidth]{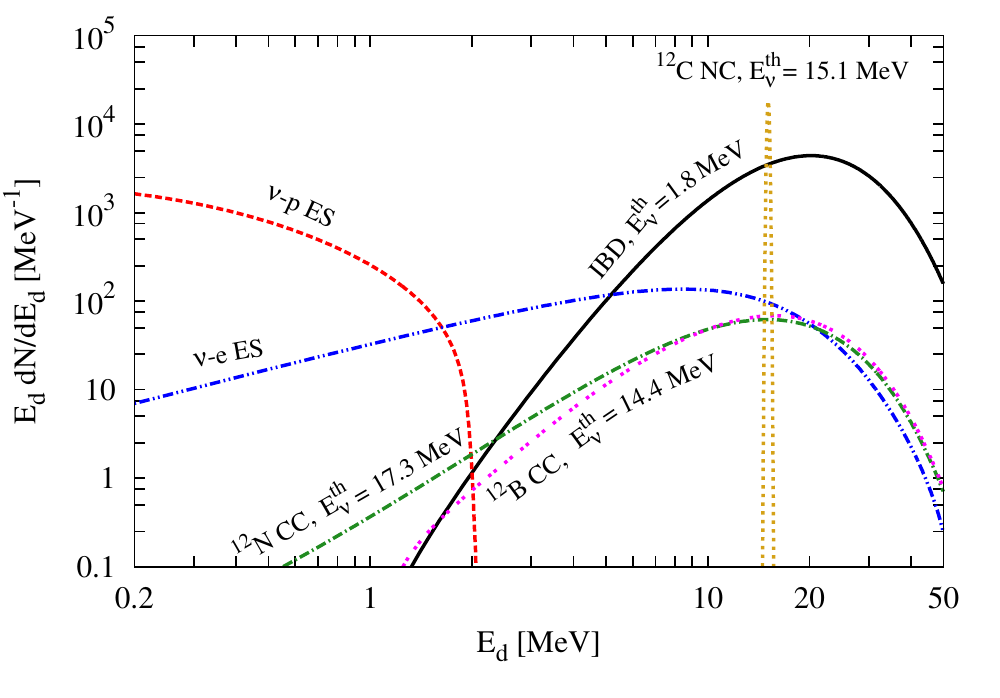}
    \vspace{-0.4cm}
    \caption{The neutrino event spectra with respect to the visible energy $E^{}_{\rm d}$ in the JUNO detector for a SN at 10 kpc, where no neutrino flavor conversions are assumed for illustration and the average neutrino energies are $\langle E^{}_{\nu_e}\rangle = 12~{\rm MeV}$, $\langle E^{}_{\overline{\nu}_e}\rangle = 14~{\rm MeV}$ and $\langle E^{}_{\nu_x}\rangle = 16~{\rm MeV}$. (See the text for details)}
\label{fig:spectra}
\end{figure}

\subsection{Diffuse Supernova Neutrino Background}
\label{subsec:DSNB}

The integrated neutrino flux from all past core-collapse events in
the visible universe forms the diffuse supernova neutrino background
(DSNB), holding information on the cosmic star-formation rate, the
average core-collapse neutrino spectrum, and the rate of failed
supernovae. The Super-Kamiokande water Cherenkov detector has
provided the first limits \cite{Malek:2002ns,Bays:2011si} and
eventually may achieve a measurement at the rate of a few events per
year, depending on the implementation of its gadolinium upgrade.

The JUNO detector has the potential to achieve a measurement comparable to Super-Kamiokande,
benefiting from the excellent intrinsic capabilities of
liquid scintillator detectors for antineutrino tagging and
background rejection.
Sources of correlated background events must be taken into account
when defining the energy window, fiducial volume and pulse shape selection
criteria for the DSNB detection.
Reactor and atmospheric neutrino IBD signals define an observational window reaching from 11 to $\sim$30\,MeV.
Taking into account the JUNO detector simulation of the expected signal and background rates,
a $3\sigma$ signal is conceivable after 10~years of running for typical assumptions of DSNB parameters.
A non-detection would strongly improve current limits and exclude a significant
range of the DSNB parameter space.

\begin{table}[!htb]
\begin{center}
\begin{tabular}{|c|cc|cc|}
\hline
Syst. uncertainty BG & \multicolumn{2}{c|}{5\,\%}& \multicolumn{2}{c|}{20\,\%}\\
\hline
$\mathrm{\langle E_{\bar\nu_e}\rangle}$ & rate only & spectral fit & rate only & spectral fit \\
\hline
12\,MeV & $2.3\,\sigma$ & $2.5\,\sigma$ & $2.0\,\sigma$ & $2.3\,\sigma$\\
15\,MeV & $3.5\,\sigma$ & $3.7\,\sigma$ & $3.2\,\sigma$ & $3.3\,\sigma$\\
18\,MeV & $4.6\,\sigma$ & $4.8\,\sigma$ & $4.1\,\sigma$ & $4.3\,\sigma$\\
21\,MeV & $5.5\,\sigma$ & $5.8\,\sigma$ & $4.9\,\sigma$ & $5.1\,\sigma$\\
\hline
\end{tabular}
\caption{The expected detection significance after 10 years of data taking for different DSNB models
with $\langle E_{\bar\nu_e} \rangle$ ranging from 12\,MeV to 21\,MeV ($\Phi=\Phi_0$). Results are given based on either a rate-only or spectral fit analysis and assuming 5\% or 20\% for background uncertainty.}
\label{tab:snd:det_sig}
\end{center}
\end{table}
\begin{figure}
\centering
\includegraphics[width=0.66\textwidth]{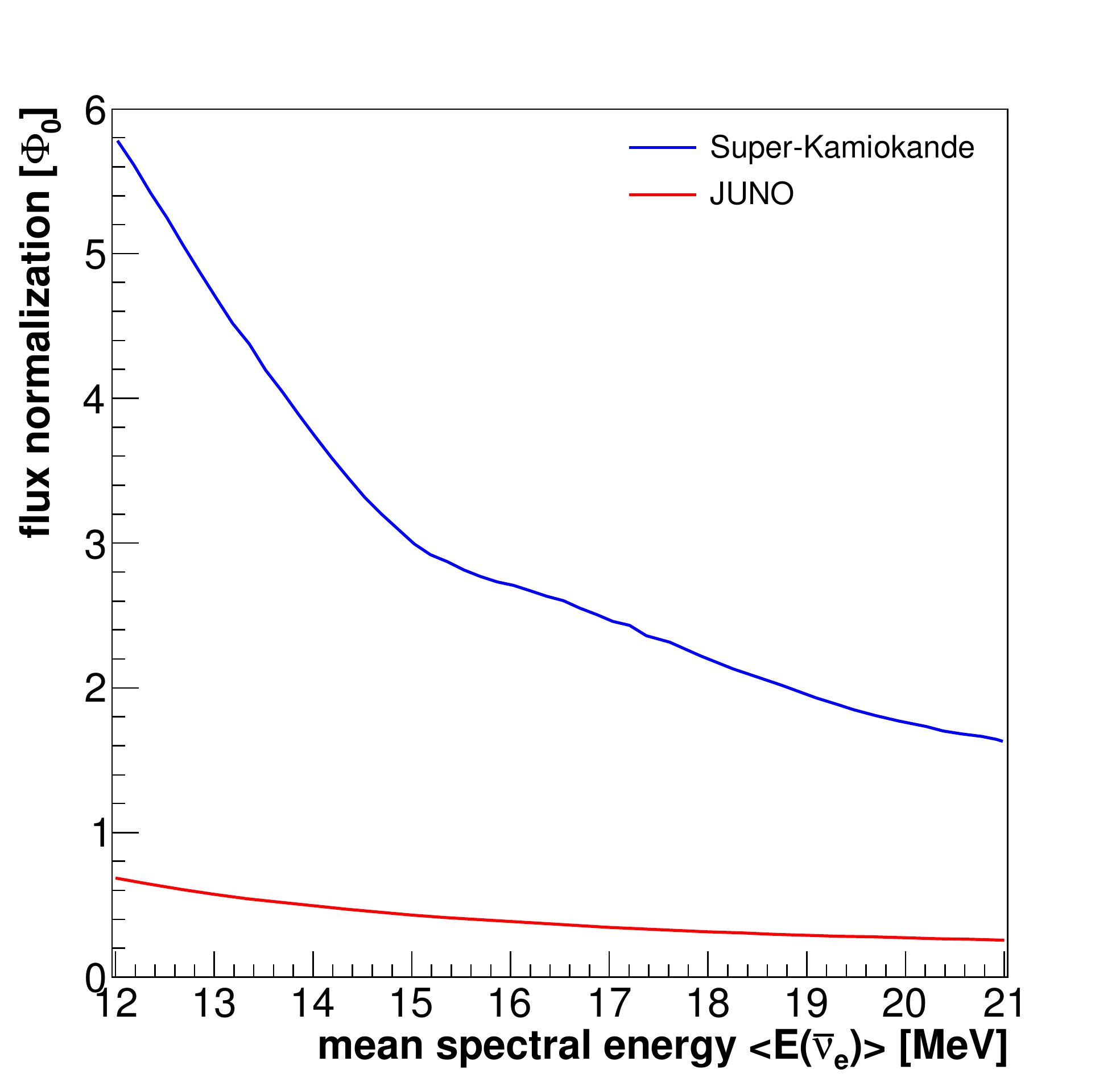}
\caption{Exclusion contour at 90\%~C.L.\ as a function of the mean energy of the SN spectrum $\langle E_{\bar\nu_e}\rangle$ and
the flux amplitude (SN rate times total emitted energy per SN).
We assume 5\% background uncertainty and no DSNB signal detection ($N_ {\rm det}=\langle N_{\rm bg} \rangle$) after~10\,yrs.}
\label{fig:snd:exclusion}
\end{figure}

Following the method described in Ref.~\cite{Rolke:2004mj}, we show in
Tab.~\ref{tab:snd:det_sig} the expected detection significance after
10 years of data taking for different DSNB models
with $\langle E_{\bar\nu_e} \rangle$ ranging from 12\,MeV to 21\,MeV.
Results are given based on either a rate-only or spectral fit
analysis and assuming 5\% or 20\% for the background uncertainty.
For each DSNB model it was assumed that the number of
detected events equals the sum of the expected signal and background
events. After 10\,years the DSNB can be detected with $>3\,\sigma$
significance if $\mathrm{\langle E_{\bar\nu_e} \rangle}\ge 15\,$MeV.

If there is no positive detection of the DSNB, the current limit can be significantly improved.
Assuming that the detected event spectrum equals the background expectation in the
overall normalization and shape, the upper limit on the DSNB flux above 17.3\,MeV would be
$\sim 0.2\,{\rm cm}^{-2}{\rm s}^{-1}$ (90\%~C.L.) after 10 years for $\langle E_{\bar\nu_e}\rangle=18$\,MeV.
This limit is almost an order of magnitude better than the current value from Super-Kamiokande~\cite{Bays:2011si}.
In Fig.~\ref{fig:snd:exclusion} we show the corresponding exclusion contour at 90\%~C.L. as a function of the mean energy
with $\langle E_{\bar\nu_e} \rangle$ ranging from 12\,MeV to 21\,MeV.

\subsection{Solar Neutrinos}
\label{subsec:solar}

\begin{figure}
\begin{center}
\begin{tabular}{c}
\centerline{\includegraphics[width=0.6\linewidth]{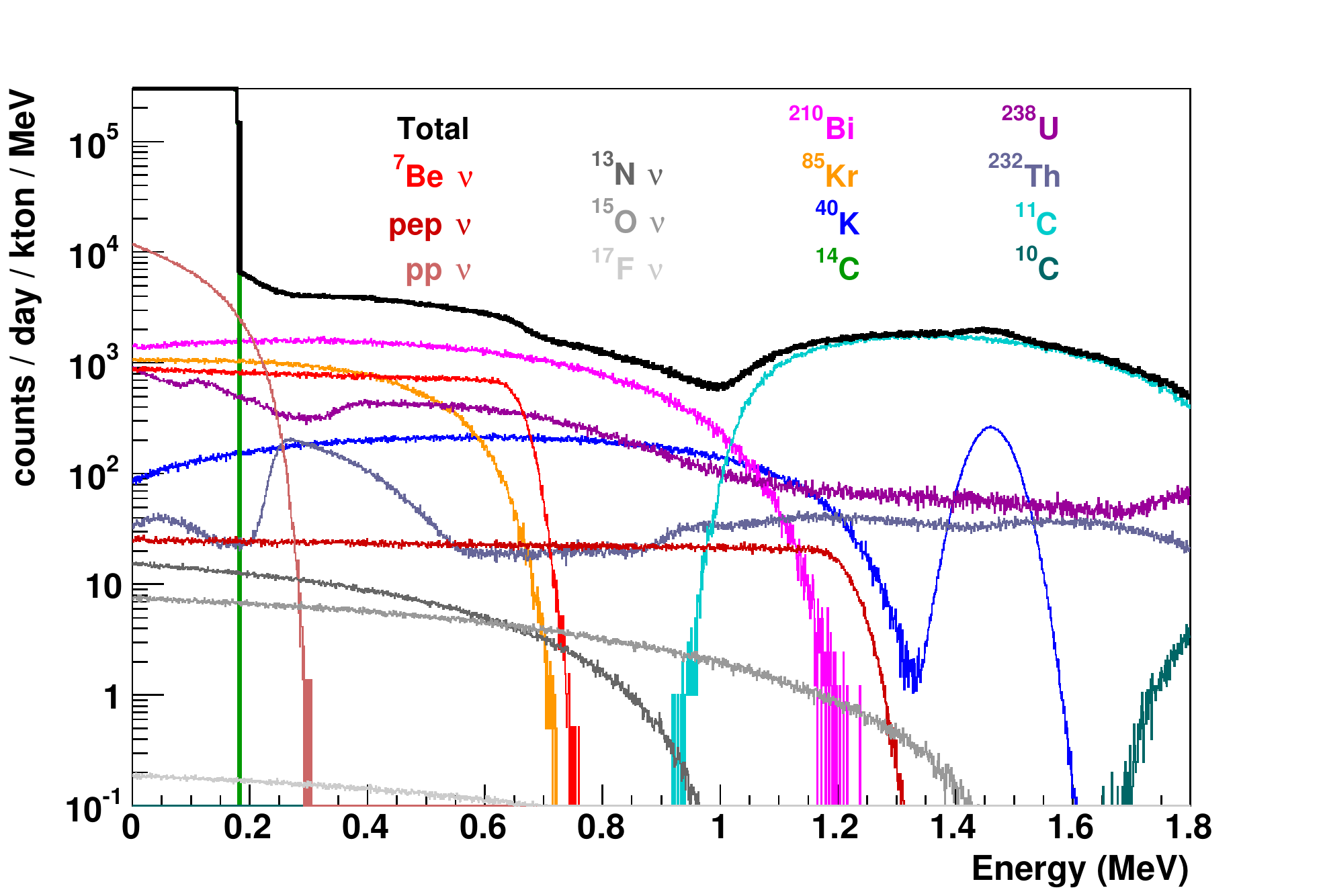}}
\\
\centerline{\includegraphics[width=0.6\linewidth]{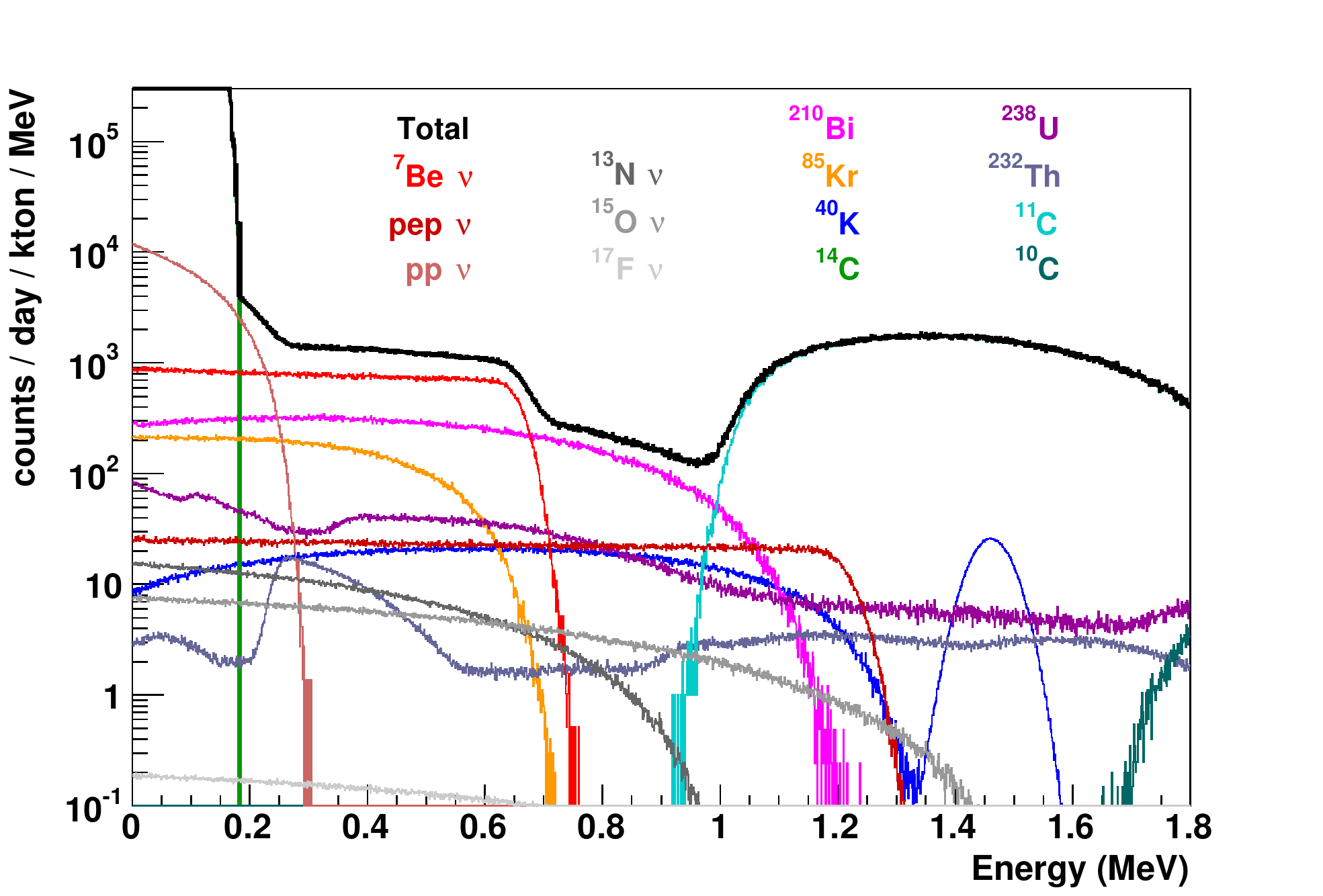}}
\end{tabular}
\end{center}
\caption{(a)On the top, the expected singles spectra at JUNO with the radio purity assumption
$10^{-16} {\rm g/g}$ for $^{238}{\rm U}$ and $^{232}{\rm Th}$, $10^{-17} {\rm g/g}$ for $^{40}{\rm K}$
and $^{14}{\rm C}$. (b) On the bottom, assuming a background level one order of magnitude better for
$^{238}{\rm U}$, $^{232}{\rm Th}$ and for $^{40}{\rm K}$, $^{14}{\rm C}$.}
\label{fig:solar:simul2}
\end{figure}
More than forty years of experiments and phenomenological analyses brought to a crystal
clear solution of the Solar neutrino problem in terms of oscillating massive neutrinos
interacting with matter, with only the LMA region surviving in the
mixing parameter space.
The rather coherent picture emerging from all of these solar neutrino experiments is in
general agreement with the values of the mixing parameters extracted from
KamLAND data~\cite{KL-recent}.
A global three flavor analysis, including all the solar neutrino experiments and KamLAND data and
assuming CPT invariance, gives the following values for the mixing angles and the differences
of the squared mass eigenvalues~\cite{Borexino-fase-I}:
$
{\rm tan}^2 \, \theta_{12} = 0.457^{+0.038}_{-0.025} \, ;
{\rm sin}^2 \, \theta_{13} = 0.023^{+0.014}_{-0.018} \, ;
\Delta m_{21}^2 = 7.50^{+0.18}_{-0.21} \, \times \, 10^{-5} \, {\rm eV}^2\, .
$

The measured values of the neutrino fluxes produced in different
phases of the $pp$ chain and of the CNO cycle are consistent with the Standard Solar Model (SSM).
But they are not accurate enough to solve the metallicity
problem~\cite{Villante-Serenelli-2014}, discriminating between the
different (high Z versus low Z) versions of the SSM.
Because most of the experimental results fall somehow in the middle between the high and
the low Z predictions and the uncertainties are still too high.
In order to operate this discrimination, a future experimental challenge would be an even more
accurate determination of the $^8{\rm B}$ and $^7{\rm Be}$ fluxes,
combined with the measurement of CNO neutrinos.

We present in Fig.~\ref{fig:solar:simul2}(a) the expected singles spectra at JUNO with the
radio purity assumption of $10^{-16} {\rm g/g}$ for $^{238}{\rm U}$ and $^{232}{\rm Th}$, $10^{-17} {\rm g/g}$ for $^{40}{\rm K}$ and $^{14}{\rm C}$, and Fig.~\ref{fig:solar:simul2}(b) the expected singles spectra at JUNO with an ideal radio purity assumption of $10^{-17} {\rm g/g}$ for $^{238}{\rm U}$ and $^{232}{\rm Th}$, $10^{-18} {\rm g/g}$ for $^{40}{\rm K}$ and $^{14}{\rm C}$. The $^7{\rm Be}$ spectrum clearly stands out all backgrounds. More studies on the capability sensitivity deserve further simulations.

\subsection{Atmospheric Neutrinos}
\label{subsubsec:atm}

\begin{figure}
\begin{center}
\begin{tabular}{c}
\centerline{\includegraphics[width=0.6\linewidth]{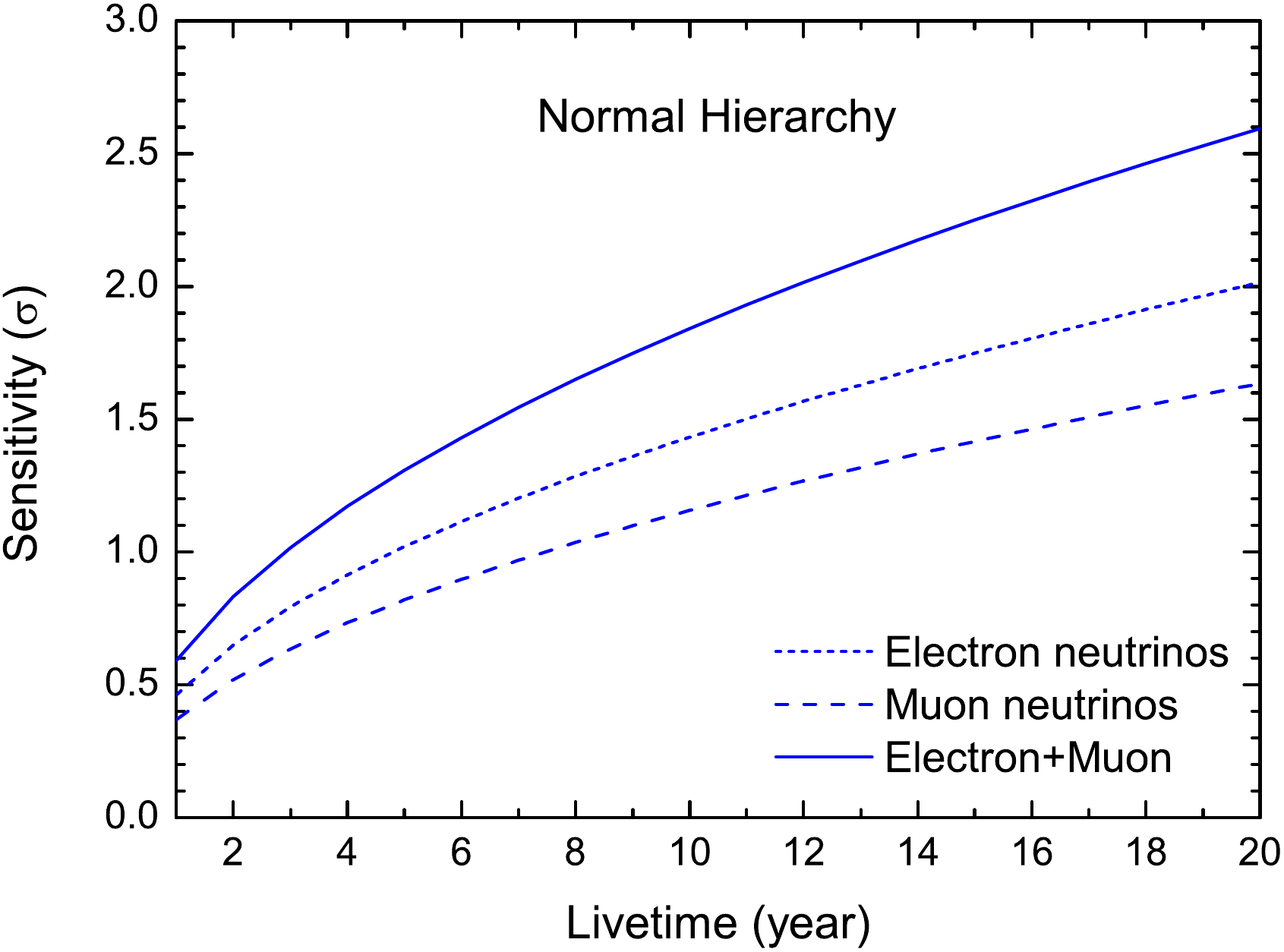}}
\\
\centerline{\includegraphics[width=0.6\linewidth]{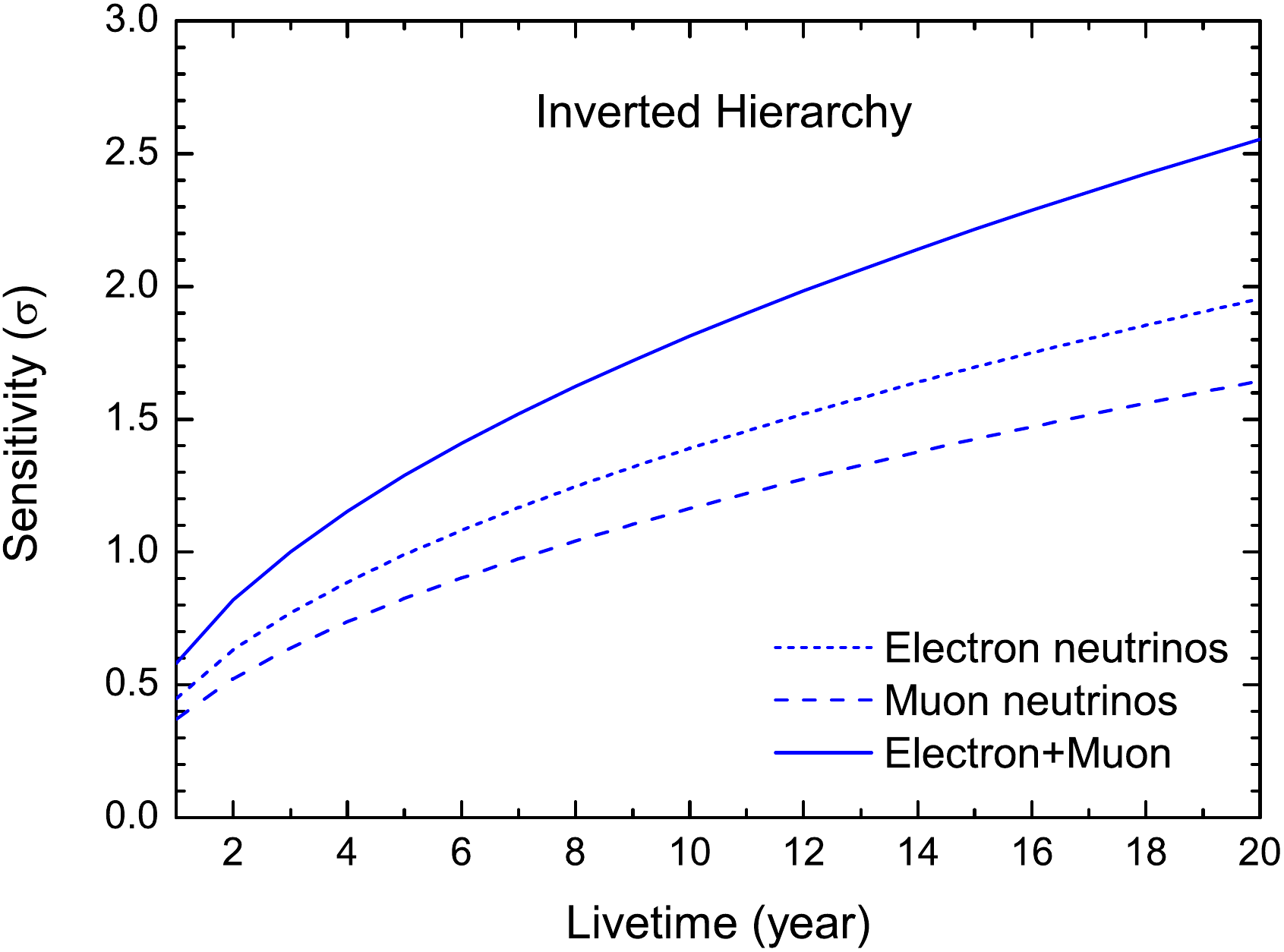}}
\end{tabular}
\end{center}
\caption{The MH sensitivities of atmospheric neutrinos as a function of livetime for the true NH (upper panel) and IH (lower panel) cases.}
\label{fig:atm:MHyears}
\end{figure}
Atmospheric neutrinos are one of the most important neutrino sources for
neutrino oscillation studies. Atmospheric neutrinos have broad
ranges in the baseline length (15~km $\sim$ 13000~km) and energy (0.1
~GeV $\sim$ 10~TeV). When atmospheric neutrinos pass through the
Earth, the matter effect will affect the
oscillations of the neutrinos. The
Mikheyev-Smirnov-Wolfenstein resonance enhancement may occur in the
normal mass hierarchy (NH) for neutrinos and inverted mass hierarchy
(IH) for antineutrinos. Therefore, JUNO has the capability to measure
the neutrino MH through detecting the upward atmospheric neutrinos.
This is complementary to the JUNO reactor antineutrino results.
JUNO's ability in MH determination is at the 1$\sigma$-2$\sigma$ level
for 10 years of data taking. An optimistic estimation of the MH sensitivity is obtained based
on several reasonable assumptions:
\begin{itemize}
\item[$\bullet$] An angular resolution of $10^\circ$ for the neutrino direction and
$\sigma_{E_{vis}} = 0.01 \sqrt{E_{\rm vis}/{\rm GeV}}$ for the visible energy of neutrino events are assumed.
\item[$\bullet$] The $\nu_e/\bar{\nu}_e$ CC events are identified and
reconstructed in the $e^\pm$ visible energy range with $E^e_{\rm vis} >1$ GeV
and $Y_{\rm vis} < 0.5$, where $Y_{\rm vis}$ is defined as the ratio of the hadron visible energy
in the visible energy of an atmospheric neutrino event.
\item[$\bullet$] The selected $\nu_e/\bar{\nu}_e$ events are divided into two samples in terms of the numbers
of Michael electrons with $N_e = 0$ ($\bar\nu_e$-like events) or $N_e \geq 1$ ($\nu_e$-like events).
\item[$\bullet$] The $\nu_\mu/\bar{\nu}_\mu$ CC events are selected with $L_{\mu} > 3 $ m,
where $L_{\mu}$ is the track length inside the LS region for the charge lepton $\mu^\pm$.
\item[$\bullet$] The selected $\nu_\mu/\bar{\nu}_\mu$ events are divided into four samples. First, these events are
grouped as partially contained (PC) and fully contained (FC) events depending on whether $\mu^\pm$ can escape from the LS region or not.
Second, FC events with $N_e \geq 2$ or $\mu^{-}$ capture on $^{12}$C or $Y_{\rm vis}>0.5$ are
defined as FC $\nu_\mu$-like events, and the residual FC events are defined as FC $\bar\nu_\mu$-like events.
Finally, PC events with $N_e \geq 1$ or $Y_{\rm vis}>0.5$ are classified as PC $\nu_\mu$-like events, and all the other PC events are PC $\bar\nu_\mu$-like events.
\end{itemize}
The MH sensitivity of the separate contribution of muon and electron neutrino events and their combinations
in the cases of both normal and inverted MHs are presented in Fig.~\ref{fig:atm:MHyears}.
We can observe from the figure that
the $\nu_e/\bar{\nu}_e$ events have better sensitivity than $\nu_\mu/\bar{\nu}_\mu$ events because
more $\nu_e/\bar{\nu}_e$ events with higher energies can deposit their whole visible energy in
the LS region. The combined sensitivity can reach 1.8$\sigma$ (2.6$\sigma$) after 10 (20) years of running.
Finally, we want to mention that
atmospheric neutrinos can also be used to search for CP violation effects and
precisely measure the atmospheric mixing angle $\theta_{23}$.

\subsection{Geo-Neutrinos}
\label{subsubsec:geo}

For half a century we have established with considerable precision the Earth's surface
heat flow $46 \pm 3 $~TW ($10^{12}$~watts), however we are vigorously debating what fraction of
this power comes from primordial versus radioactive sources. This debate touches on the composition
of the Earth, the question of chemical layering in the mantle, the nature of mantle convection,
the energy needed to drive Plate Tectonics, and the power source of the geodynamo, the magnetosphere
that shields the Earth for the harmful cosmic ray flux.

Over the last decade particle physicists have detected the Earth's geoneutrino flux, the
planetary emission of electron anti-neutrinos that are derived from naturally occurring,
radioactive beta-decay events inside the Earth \cite{bib3,bib4}. Matter, including the Earth,
is mostly transparent to these elusive messengers that reveal the sources of heat inside the
Earth. By detecting a few events per years we are now measuring the geoneutrino flux from Thorium
and Uranium inside the planet, which in turn allows us to determine the amount of radiogenic
power driving the Earth's engine.

Predicting the geoneutrino signal at JUNO demands that we accumulate the
basic geological, geochemical and geophysical data for the regional area
surrounding the detector.  Experience tells us that in the continents the closest 500~km to the
detector contributes half of the signal and it is this region that needs to be critically
evaluated \cite{bib5}.  This goal demands that the physical (density and structure) and
chemical (abundance and distribution of Th and U) nature of the continent must be specified
for the region. The main tasks include surveys and descriptions of the geology, seismology,
heat flow, and geochemistry of the regional lithosphere.

The dominate background for the geo-neutrino observation is the reactor antineutrinos.
In addition, other backgrounds include the cosmic-muons spallation
products ($^{9}$Li-$^{8}$He isotopes, fast neutrons), accidental coincidences of non-correlated
events, and backgrounds induced by radioactive contaminants of scintillator and
detector materials (i.e., $^{13}$C ($\alpha$, n)$^{16}$O).
In Tab.~\ref{tab:geo:Nev}, we summarize the predicted geo-neutrino signal
and backgrounds considered in the sensitivity study.
 \begin{table}
\label{tab:exp}
\vspace{0.4cm}
\begin{center}
\begin{tabular}{ll}
\hline \hline
Source &
Events/year
\\ \hline
Geoneutrinos 	& $408\pm 60$   	\\
U chain      	& $311\pm 55$    	\\
Th chain 	& $92\pm 37$   		\\
Reactors 	& $16100\pm 900$ 	\\
Fast neutrons	& $3.65\pm 3.65$ 	\\
$^{9}$Li - $^{8}$He 		& $657\pm 130$ 		\\
$^{13}$C$(\alpha,n)^{16}$O & $18.2\pm 9.1$  \\
Accidental coincidences & $401\pm 4$ 	\\
\hline \hline
\end{tabular}
\end{center}
\caption{Signal and backgrounds considered in the geoneutrino sensitivity study:
the number of expected events for all components contributing to the IBD spectrum in the 0.7 - 12 MeV energy region
of the prompt signal. We have assumed 80\% antineutrino detection efficiency and 17.2\,m radial cut (18.35\,kton of liquid scintillator).}
\label{tab:geo:Nev}
\end{table}

Precision of the reconstruction of geoneutrino signal is shown in
Tab.~\ref{tab:geo:Fit}, for the running time in 1, 3, 5, and 10 years.
Different columns refer to the measurement of geo-neutrino signal with fixed Th/U ratio, and
U and Th signals fit as free and independent components.
The given numbers are the position and root mean square (RMS) of the Gaussian fit to the distribution of the
ratios between the number of reconstructed and generated events.
It can be seen that while RMS is decreasing with longer data acquisition time, there are
some systematic effects which do not depend on the acquired statistics.
With 1, 3, 5, and 10 years of data, the statistical error amounts to 17\%, 10\%, 8\%, and 6\% respectively with
the fixed chondritic Th/U ratio.
\begin{table}
\footnotesize
\label{tab:exp}
\vspace{0.4cm}
\begin{center}
\begin{tabular}{cccc}
\hline \hline
Number of years & U + Th (fixed chondritic Th/U ratio) & U (free) & Th (free) \\
\hline
1	& $0.96\pm 0.17$   & $1.01\pm 0.32$	& $0.79\pm 0.66$	\\
3     & $0.96\pm 0.10$   & $1.03\pm 0.19$	& $0.80\pm 0.37$	\\
5 	& $0.96\pm 0.08$   & $1.03\pm 0.15$	& $0.80\pm 0.30$	\\
10 	& $0.96\pm 0.06$   & $1.03\pm 0.11$	& $0.80\pm 0.21$	\\
\hline \hline
\end{tabular}
\end{center}
\caption{Precision of the reconstruction of geoneutrino signal. See the text for details.}
\label{tab:geo:Fit}
\end{table}


\subsection{Nucleon Decay}
\label{subsubsec:nucleon}

Being a large underground LS detector, JUNO is
in an excellent position in search for nucleon decays. In
particular, in the SUSY favored decay channel $p \to K^+ + \bar\nu$,
JUNO will be competitive and complementary to other experiments
using water Cherenkov and liquid argon detectors ~\cite{kearns:isoup}.

The protons in the JUNO detector are provided by both the hydrogen
nuclei and the carbon nuclei. Using the Daya Bay liquid scintillator
as a reference, the H to C molar ratio is 1.639. For a 20~kt
fiducial volume detector, the number of protons from hydrogen (free
protons) is $1.45\times10^{33}$ and the number of protons from
carbon (bound protons) is $5.30\times10^{33}$.

If the proton is from hydrogen, it decays at rest. The kinetic energy of the $K^+$ is fixed by kinematics to
be 105~MeV, which gives a prompt signal in the liquid scintillator.
The $K^{+}$ has a lifetime of 12.4 nanoseconds and would quickly decay
into five different channels,
$K^{+} \to \mu^+ \nu_{\mu}$ and $K^{+} \to \pi^+  \pi^0$ are two most probably decay modes.
In either case, there is a short-delayed ($\sim$12~ns) signal from the $K^{+}$ daughters.

If the $K^+$ decays into $\mu^+ \nu_{\mu}$, the delayed signal comes
from the $\mu^+$, which has a fixed kinetic energy of 152~MeV from
the kinematics. The $\mu^+$ itself decays 2.2~$\mu s$ later into
$e^+ \nu_e \bar\nu_{\mu}$, which gives a third long-delayed signal
with well known (Michel electron) energy spectrum. If the $K^+$
decays into $\pi^+ \pi^0$, the $\pi^+$ deposits its kinetic energy
(108~MeV) and the $\pi^0$ instantaneously decays ($\tau =
8.4\times10^{-17} s$) into primarily two gamma rays (98.80\%) with
the sum of deposited energy equal to the total energy of $\pi^0$
(246~MeV). The delayed signal includes all of the aforementioned
deposited energy. Then, the $\pi^+$ decays ($\tau=26$ ns) primarily
into $\mu^+ \nu_{\mu}$ (99.99\%).  The $\mu^+$ itself has very low
kinetic energy (4.1 MeV), but it decays 2.2~$\mu s$ later into $e^+
\nu_e \bar\nu_{\mu}$, which gives the third long-delayed decay
positron signal.

\begin{figure}
\centering
\includegraphics[width=0.7\textwidth]{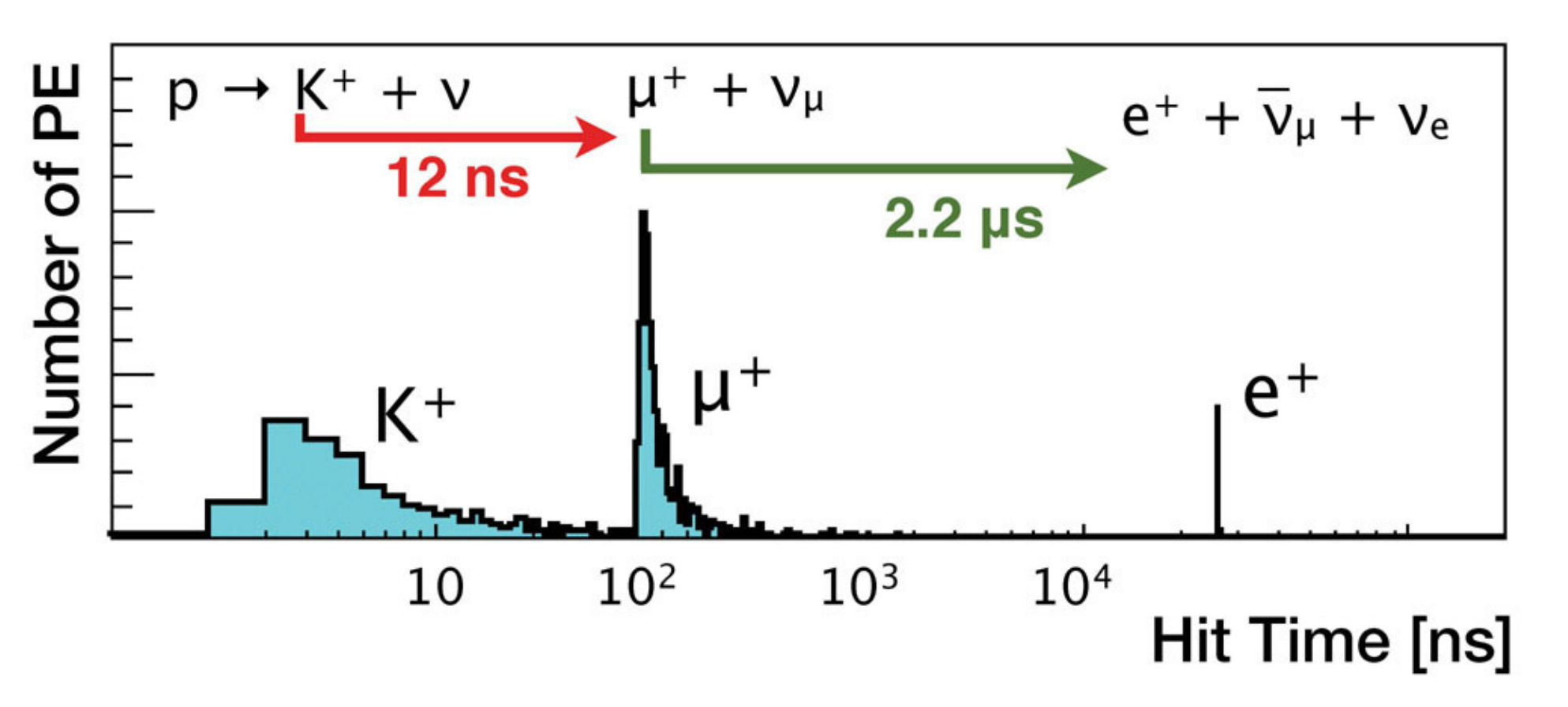}
\caption{The simulated hit time distribution of photoelectrons (PEs)
from a $K^{+} \to \mu^+ \nu_{\mu}$ event at JUNO.}
\label{fig:nd:pdk_event}
\end{figure}

The simulated hit time distribution of a $K^{+} \to \mu^+
\nu_{\mu}$ event is shown in Fig.~\ref{fig:nd:pdk_event}, which
displays a clear three-fold coincidence:
\begin{itemize}
    \item A prompt signal from $K^+$ and a delayed signal from its decay daughters with a time coincidence of 12 ns.
    \item Both the prompt and delayed signals have well-defined energy.
    \item One and only one decay positron with a time coincidence of 2.2 $\mu$s from the prompt signals.
\end{itemize}
The time coincidence and the well-defined energy provide a powerful
tool to reject background, which is crucial in the proton decay
search.

The sensitivity to proton lifetime, can be calculated as $\tau (p
\to K^+ + \bar\nu) = N_p T R \epsilon / S$, where $N_p$
($6.75\times10^{33}$) is the total number of protons, $T$ is the
measuring time where we assumed 10 years, $R$ (84.5\%) is the $K^+$
decay branching ratio included in the analysis, and $\epsilon$
(65\%) is the total signal efficiency. $S$ is the upper limit of
number of signal events at certain confidence interval, which
depends on the number of observed events as well as the expected
number of background events. The expected background is 0.5 events
in 10 years. If no event is observed, the 90\% C.L upper limit is $S
= 1.94$. The corresponding sensitivity to proton lifetime is $\tau >
1.9\times10^{34}$ yr. This represents a factor of three improvement
over the current best limit from Super-Kamiokande, and starts to
approach the region of interest predicted by various GUTs models. In
a real experiment, the sensitivity may decrease if background
fluctuates high. In the case that one event is observed (30\%
probability), the 90\% C.L upper limit is $S = 3.86$. The
corresponding sensitivity to proton lifetime is $\tau >
9.6\times10^{33}$ yrs. If two events are observed (7.6\%
probability), the sensitivity is further reduced to $\tau >
6.8\times10^{33}$ yrs.
\begin{figure}
\centering
\includegraphics[width=0.7\textwidth]{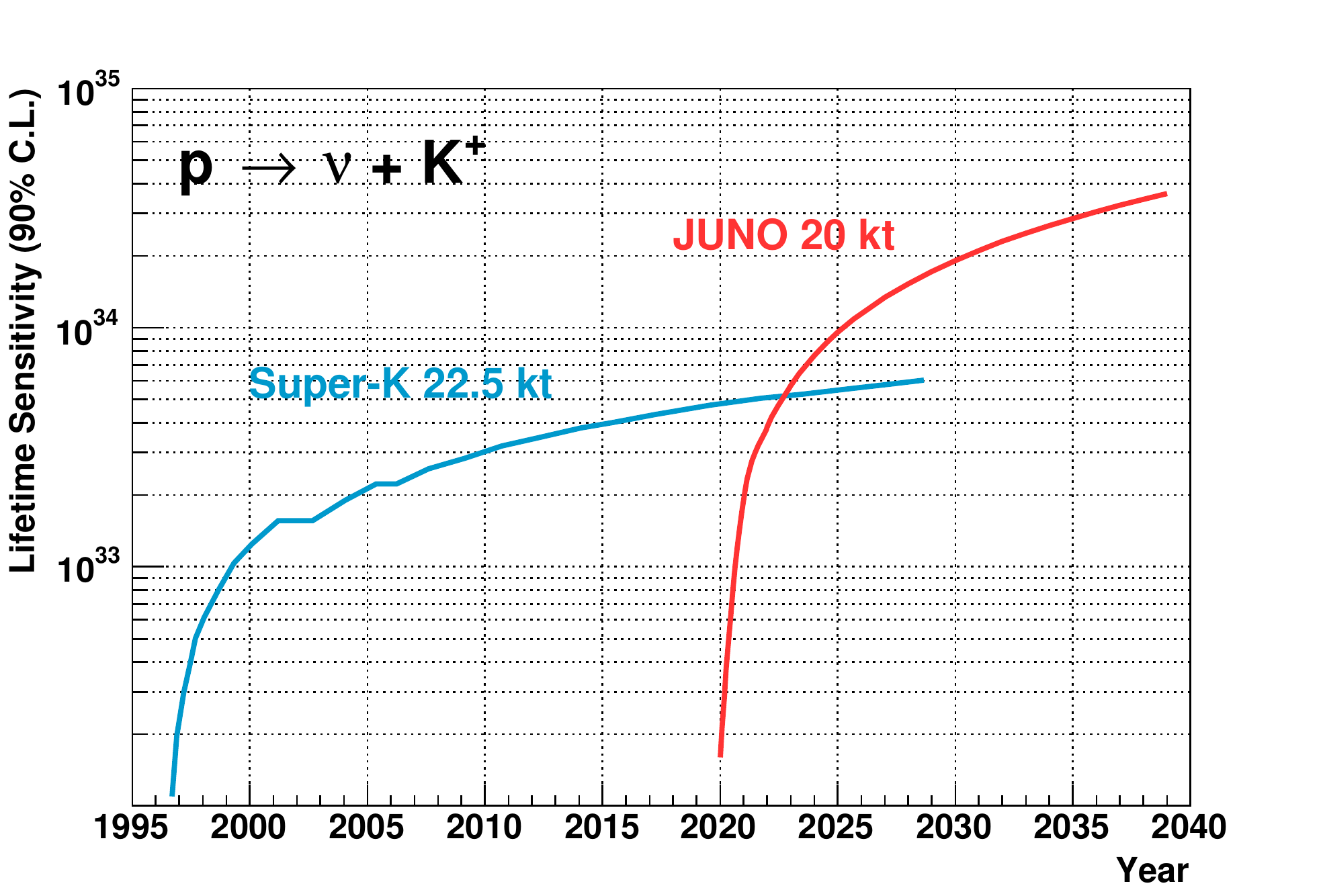}
\caption{The 90\% C.L. sensitivity to the proton lifetime in the decay
mode $p \to K^+ + \overline\nu$ at JUNO as a function of time. In
comparison, Super-Kamiokande's sensitivity is also projected \cite{kearns:isoup}.} \label{fig:nd:pdk_sens}
\end{figure}
In Fig.~\ref{fig:nd:pdk_sens} we plot the 90\% C.L. sensitivity
to the proton lifetime in the decay mode $p \to K^+ + \overline\nu$ at JUNO
as a function of the running time. Due to the high efficiency in measuring this
mode, JUNO's sensitivity will surpass Super-Kamiokande's after
3 yrs of data taking.

\subsection{Light Sterile Neutrinos}
\label{subsubsec:sterile}

Motivated from the anomalies of LSND \cite{Aguilar:2001ty}, MiniBooNE \cite{Aguilar-Arevalo:2013pmq},
the reactor antineutrino anomaly \cite{Mention:2011rk}, and Gallium anomaly \cite{Giunti:2010zu},
the light sterile neutrino \cite{Giunti:2013aea,Kopp:2013vaa} is regarded as one of the most promising possibilities for new physics
beyond the three neutrino oscillation paradigm. Therefore, future experimental oscillation searches at short baselines are required
to test the light sterile neutrino hypothesis \cite{Lasserre:2014ita}.

Several possible methods of sterile neutrino studies are considered at JUNO. The first one is the use of existing reactor antineutrinos,
which can test the active-sterile mixing with the mass-squared difference ranging from $10^{-5}$ to $10^{-2}$ eV$^{2}$. This parameter space
is irrelevant to the short baseline oscillation, but could be tested as the sub-leading effect of solar neutrino oscillations.

The direct test of short baseline oscillations at JUNO requires additional neutrino sources placed near or inside the detector.
Using 50 kCi $^{144}$Ce-$^{144}$Pr as the antineutrino source at the detector center, JUNO can reach the $10^{-2}$ level of active-sterile mixing
at 95\% C.L. after 1.5 yrs of data taking. On the other hand, a cyclotron-driven $^8$Li source can be employed as the decay-at-rest (DAR) neutrino source near
the detector. The high DAR flux coupled with the large size of the JUNO detector
allows us to make an extremely sensitive search for antineutrino disappearance in the region of short baseline oscillation anomalies.
With a 60 MeV/amu cyclotron accelerator, JUNO can reach the $10^{-3}$ level of active-sterile mixing at 5$\sigma$ C.L.
after 5 yrs of data taking. IsoDAR$@$JUNO will be able to map out definitively the oscillation wave.
There is no other planned experiment that matches this sensitivity.

\subsection{Indirect Dark Matter Search}
\label{subsubsec:dark}

The existence of non-baryonic dark matter (DM) in the Universe has been
well established by astronomical observations. DM can also be detected indirectly by looking for the neutrino
signature from DM annihilation or decays in the Galactic halo, the
Sun or the Earth. In particular, the search for the DM-induced
neutrino signature from the Sun has given quite tight constraints on
the spin-dependent (SD) DM-proton scattering cross section $\sigma^{\rm
SD}_{\chi p}$~\cite{Aartsen:2012kia}.
\begin{figure}
\centering
\includegraphics[width=0.7\textwidth]{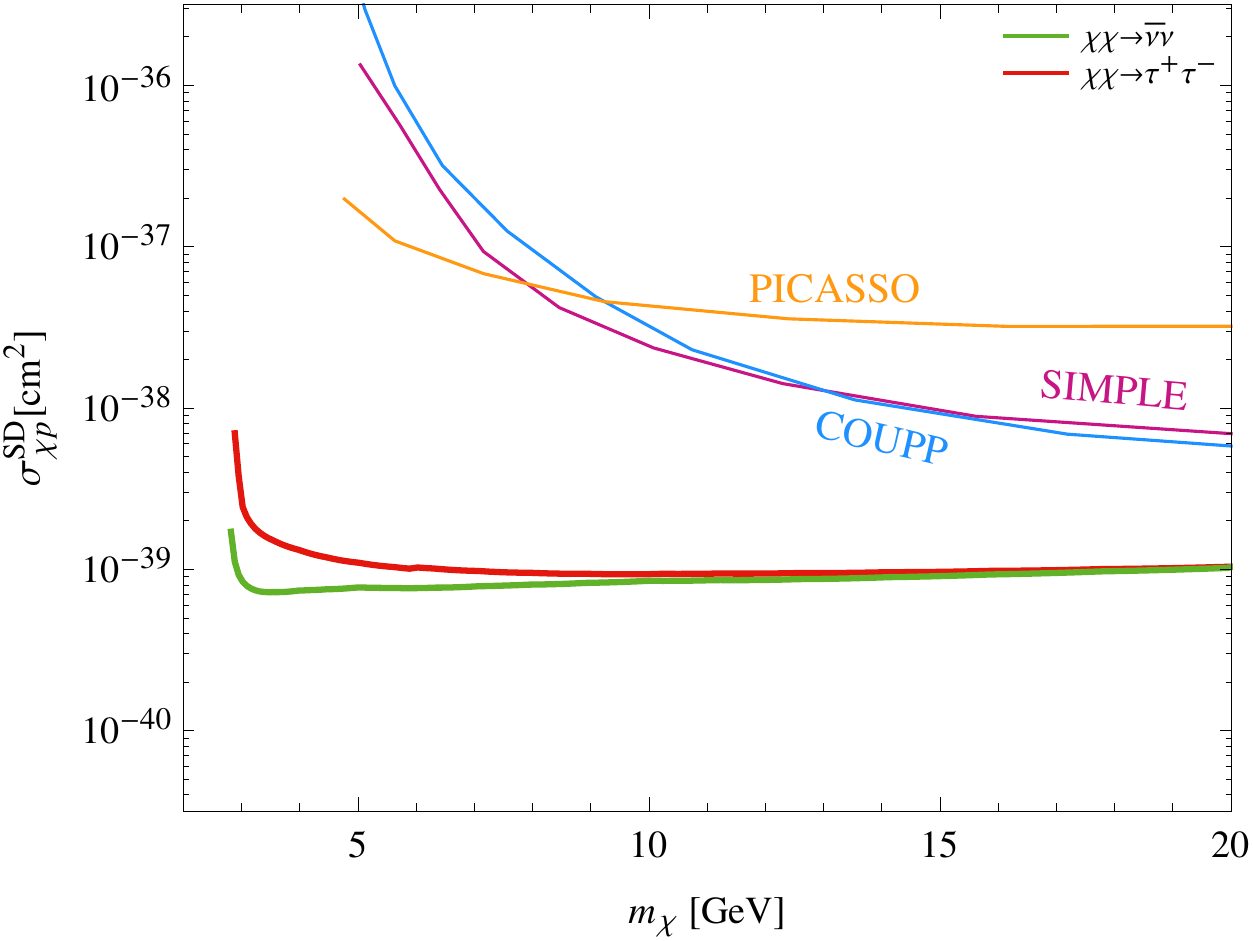}\\
\caption{ The JUNO $2\sigma$ sensitivity to the spin-dependent
cross section $\sigma^{\rm SD}_{\chi p}$ after 5 yrs of data taking. The
constraints from the direct detection experiments are also
shown for comparison. } \label{fig:SD}
\end{figure}

In general, DM inside the Sun can annihilate into leptons, quarks
and gauge bosons. The neutrino flux results from the decays of such
final-state particles.
Here we consider two annihilation modes $\chi\chi\to \tau^+\tau^-$ and $\chi\chi\to \nu\bar{\nu}$  as a
benchmark. The sensitivity calculations for $\sigma^{\rm SD}_{\chi p}$ are given in Fig.~\ref{fig:SD},
where one can see that the JUNO sensitivity is much better than current direct
detection constraints set by COUPP~\cite{Behnke:2012ys},
SIMPLE~\cite{Felizardo:2011uw} and PICASSO~\cite{Archambault:2012pm}
experiments. Notice that the sensitivity for $m_{\chi}< 3$ GeV
becomes poor due to the DM evaporation from the Sun.

\subsection{Other Exotic Searches}
\label{subsubsec:exotic}

The standard three-neutrino mixing paradigm can describe most of the phenomena in
solar, atmospheric, reactor, and accelerator neutrino oscillations \cite{PDG}.
However, other new physics mechanisms can operate at a sub-leading level,
and may appear at the stage of precision measurements of neutrino oscillations.
Therefore, other exotic searches can be performed at JUNO using reactor neutrinos,
solar neutrinos, atmospheric neutrinos and supernova neutrinos.

With the reactor neutrino oscillation, JUNO can:
\begin{itemize}
    \item test the nonstandard neutrino interactions at the source and detector \cite{ohlsson:14a,Li:2014mlo};
    \item test the Lorentz and CPT invariance violation \cite{Li:2014rya};
    \item test the mass-varying properties of neutrino propagation \cite{Schwetz:2005fy}.
\end{itemize}
Meanwhile, using solar and atmospheric neutrino oscillations, JUNO can
\begin{itemize}
    \item test the nonstandard neutrino interactions during the propagation \cite{Ohlsson:2012kf,Bolanos:2008km};
    \item test the long range forces \cite{Joshipura:2003jh};
    \item test the super light sterile neutrinos \cite{deHolanda:2010am};
    \item test the anomalous magnetic moment \cite{Arpesella:2008mt,Giunti:2014ixa}.
\end{itemize}
Therefore, with precision measurements JUNO will be a powerful playground to test different exotic phenomena
of new physics beyond the Standard Model.

\clearpage

\cleardoublepage
\setcounter{secnumdepth}{3}
\chapter{Central Detector}
\label{ch:Central Detector}
\section{Introduction and requirement of the central detector}
The central detector of JUNO aims to measure the neutrino energy spectrum using
$\sim$20~kt liquid scintillator (LS) and $\sim$17,000 PMTs. An inner sphere with a diameter
of around $\sim$35.4~m is designed to contain the huge volume of LS, and an outer
structure with a diameter of around $\sim$40~m is needed to support the inner sphere
as well as PMTs. In order to get the $3\%/\sqrt{E}$ energy resolution, the
central detector is required to maximize the collection of optical signals from
LS meanwhile minimize the background from a variety of radioactive sources.
Since it is not possible to repair during the operation, this detector
must have a long life time and a high reliability.

Since the LS and PMTs are separately documented in Chapter 4 and Chapter 6, this chapter is mainly focused on the challenges:

1) Construction of the large structure of the sphere and its support;

2) Test and installation of the $\sim$17,000~PMTs;

3) Filling of LS and long term operation of the detector.

Throughout the design effort, the strength and stability of the
structure is the drive. In addition, the working conditions in the
underground hall, safety issues during the construction, cost, and the overall
time schedule should be taken into account. The ideal design of the JUNO
central detector is a stainless-steel tank plus an acrylic sphere, where the
stainless-steel tank is used to separate the shielding liquid (mineral oil or
LAB, to be decided) from the pure water in the water pool, and the acrylic
sphere is to hold the $\sim$20~kt LS, as indicated in Fig.~\ref{fig:intro:det}. Due to the limitation
of space, it is very difficult to simultaneously construct the two spheres and
there is also a high risk and significant schedule delay if the two spheres were
constructed in series. As a result, alternative options could be:

1) the stainless-steel tank is replaced by an open space truss or other steel supporting structures;

2) the acrylic sphere is replaced by an off-site fabricated balloon.

Currently, the baseline is the first one, i.e. the so called acrylic sphere
plus stainless-steel support option, which will be detailed in this document.
The second option of the balloon plus stainless-steel sphere is a backup, and
will be described also in this chapter.
%

The main requirements for the JUNO central detector are the following:

1) The detector should meet the physics requirements of the JUNO experiment,
with $\sim$20~kt high purity LS and $\sim$17000 high Q.E. PMTs to reach an energy
resolution of $3\%/\sqrt{E}$;

2) The detector should minimize the radioactive background from different
sources, including environment, structure, materials and the pollution from
the construction process;

3) The detector structure should be reliable and its design should meet the
standards in fields of large vessel, civil architecture and engineering. Leakage is not allowed between different spheres. The structure should be safe up to a seismic intensity of Richter scale 5.5, and should not be
sensitive to temperature variation. The structure must have no single point of failure;

4) Materials used for the detector should have a long-term compatibility (about $\sim$30~years) with the liquid scintillator and pure water;

5) The central detector should provide a proper interface to other systems, i.e. VETO, calibration, electronics and so on;

6) The total construction duration should be reasonably short and not longer than 18~months;

7) The lifetime of the detector should be longer than 20~years, and during this period, no significant repair is needed;

8) The cost of the central detector should also be at a reasonable level.

\section{Baseline option: Acrylic Sphere with Stainless-Steel Support}

\subsection{Introduction and the design requirements}
\begin{figure}[!htbp]
\begin{center}
\includegraphics[width=15cm]{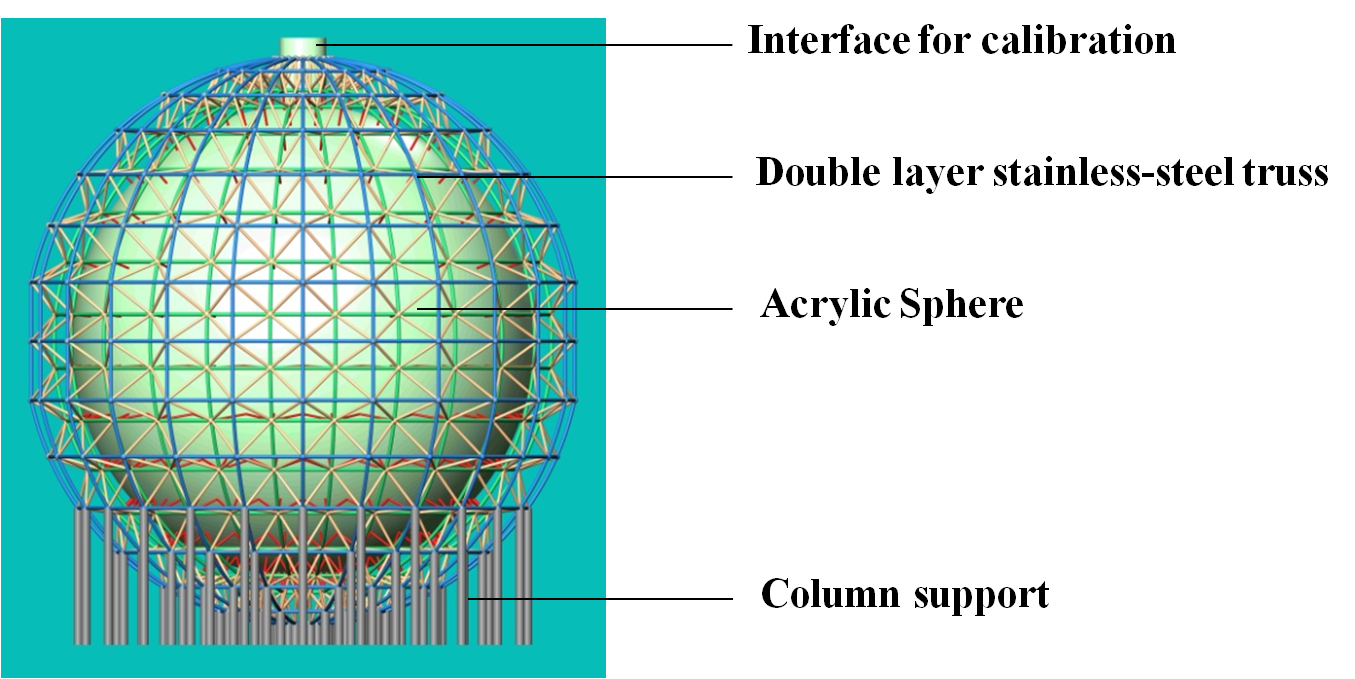}
\caption[Logo in CentralDetector]{Schematic view of acrylic sphere plus stainless-steel double layer truss}
\label{fig:cd3-2}
\end{center}
\end{figure}

In the baseline design the inner sphere is made of acrylic with a thickness of
$\sim$12~cm and an inner diameter of 35.4~m. Surrounding the acrylic sphere there is a
stainless-steel structure which supports the acrylic sphere and also the PMTs.
Currently the stainless steel support structure has three options:  the double
layer truss, the single layer truss and the equator supporting method. For the
first two options, the truss can also be used to fix the PMTs, but for the
equator supporting method, an extra frame is needed for PMT installation. The
inner diameter of the stainless-steel truss is ~40~m in the present design,
supported by a number of columns which are built on the base of the water pool.
For the acrylic sphere, a chimney with an inner diameter of $\sim$1~m will extend up
from the top of the sphere, serving as the interface for the calibration system.
The relative pressure between the LS and the water has a great effect on the sphere.  The
chimney will be a few meters higher than the water level in order to provide
flexibility in setting the LS relative height and the resulting stress in the
sphere. Due to the height of the chimney above the sphere, optical isolation is
needed to block the background induced light from being detected by the
PMTs. Between the acrylic sphere and the truss, there are about $\sim$17000~
inward-facing PMTs to collect optical signals produced by the LS. There is
an opaque layer behind the PMTs to separate the central detector from outside
veto detector. As an example, Fig.~\ref{fig:cd3-2} shows the acrylic sphere and the double
layer stainless-steel truss.

Many factors need to be considered for the baseline design, such as:

$\bullet$ Stability and load status of the structure at different stages, such as during the construction, after the construction and
installation of PMTs, during LS and water filling, and long-term operation of the
detector.

$\bullet$ The acrylic sphere  under different conditions such as earthquake, temperature variation and so on.

$\bullet$ The maximum stress on the sphere for long term operation < 5~MPa and for shorter durations < 10~MPa.

$\bullet$ Structure of the stainless-steel support should be reliable, and should meet the specifications of related fields.

$\bullet$ The joint of the acrylic and stainless steel should be reliable, and
the impact induced by local failures should be minimal.

\subsection{Stainless-steel Supporting Structure}
The acrylic sphere is the most critical part of the whole central detector and
is supported by the stainless-steel structure on its outer sphere. The design
of the joint of the acrylic and stainless steel structure is critical and the
ball head connection for flexibility and a rubber layer for buffering are
adopted in the design. Currently, there are three design options under
consideration for the stainless steel structure: double layer truss, single
layer truss and equator support. The stainless steel is chosen to be the 316
type and the main parameters are the following: density of 8.0~$g/cm^2$, elastic modulus of
200~GPa, poisson ratio of 0.33 and the yield stress of 240~MPa.

\subsubsection{Double Layer Stainless-Steel Truss}
\begin{figure}[!htbp]
\begin{center}
\includegraphics[width=12cm]{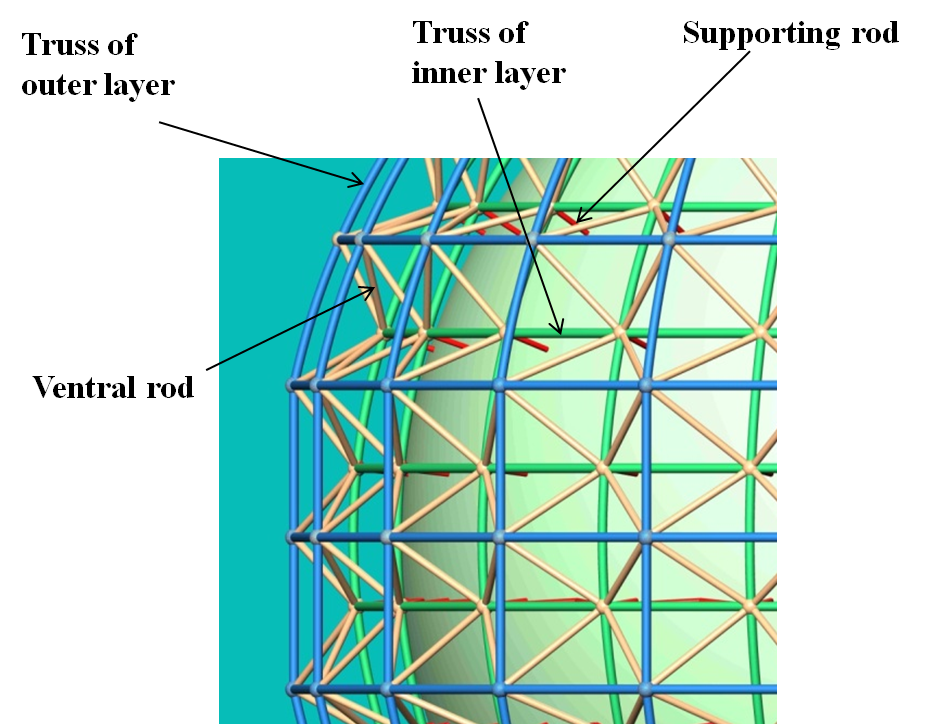}
\caption[Logo in CentralDetector]{Double layer stainless-steel truss}
\label{fig:cd3-3}
\end{center}
\end{figure}

The space truss has been a popular choice in the civil construction domain given its
properties of light-weight, small-size truss member, ease of handling and
transportation, high rigidity, short construction time, low cost and good
seismic resistance,etc. For the central detector, the  truss is selected to be
the outer structure which will not only support the acrylic sphere but also the
PMTs. For the design of the truss, load-carrying capacity and stability were
considered under different conditions. The truss will be the square pyramid
space grid which is a multiple statically indeterminate structure. This type of
truss can withstand the load coming from different directions, and has better
load-carrying capability than the plane truss. The truss members have good
regularity and high stiffness, and they are connected to each other by
bolt-sphere joints which transmit only axial compression or tension, hence
there are no moments or torsional resistance. Since the truss can be built up
from simple, prefabricated units of standard size and shape which will be
mass-produced industrially, the units can be assembled on-site easily and
rapidly which greatly reduces the time for construction.  With the present
design, the diameter of the inner layer truss is $\sim$38.5~m, while that of the
outer layer is $\sim$42.5~m at the two poles and $\sim$40.5~m at the equator. This
shrinkage from poles to equator can significantly reduce the size of
experimental hall hence the cost and time for civil construction. There are
hundreds of supporting rods between the acrylic sphere and the inner layer of
the truss. These rods will be connected to the truss at one end and inserted
into the acrylic sphere at the other end. The acrylic sphere is supported by
those rods directly. To reduce the axial load on each rod and hence the stress
in acrylic, some other rods are added between the sphere and the outer layer of
the truss. The truss itself will installed on the bottom of the water pool by
supporting columns at the lower hemisphere. Fig.~\ref{fig:cd3-3} shows the truss structure
and some of its details.

According to JGJ7-2010, which is the technical specification for space frame
structures, the truss members will be designed as the compression members and
the slenderness ratio should be less than 150. Referring to the preliminary
finite element analysis, the size of the chord members is selected to be
$\phi$273~mm $\times$ 8~mm, and the ventral members are $\phi$219~mm $\times$ 8~mm. The
supporting rods (brace members) between acrylic and truss are $\phi$102~mm $\times$
12~mm, and the columns for supporting truss are designed to be $\phi$400~mm $\times$
20~mm. The final size will be determined in engineering design after further stress analysis.

\subsubsection{Single Layer Stainless-Steel Truss}
\begin{figure}[!htbp]
\centering
\subfigure[Single layer stainless-steel truss with spiral support]{
\label{fig:subfig:a} 
\includegraphics[width=7.5cm]{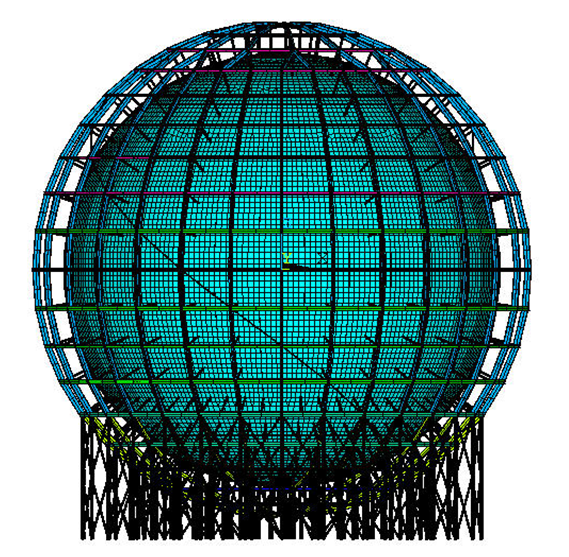}}
\subfigure[Reduction of the number of truss members in pole region]{
\label{fig:subfig:b} 
\includegraphics[width=7.5cm]{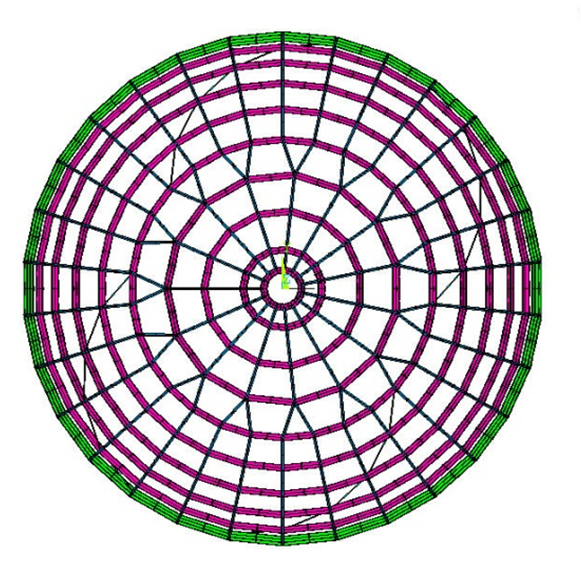}}
\caption{Single layer steel truss}
\label{fig:subfig:siglelayer} 
\end{figure}

As shown in Fig.~\ref{fig:subfig:siglelayer} (a), this single layer truss is made of I-shaped unistrut
in both longitudinal and latitudinal directions. Similar to the double layer
truss, the supporting rods are also used to connect the sphere to the truss.
The truss itself is supported on the base of the water pool by a number of
columns. To improve the stability of the single layer truss and to avoid the
possible torsion, a ring of spiral bracings are added in the truss grids to
prevent any occurrence of torsional vibration shape. In addition, due to space
limitation in the pole region of the truss, the square shaped structure is
replaced by a triangle shaped one, so the number of truss members is reduced,
as sketched in Fig.~\ref{fig:subfig:siglelayer}(b). This optimization gives more space for PMT
installation and keeps the grid size of the truss in a reasonable range.

Compared to the double layer truss, the single layer truss can save a
significant amount of space and hence the civil construction cost. Another
advantage of this option is that PMT installation is much easier since there is
no interference caused by the truss members.

\subsubsection{Equator Supporting Method}
\begin{figure}[!htbp]
\centering
\subfigure[Overall view of acrylic sphere supported at equator]{
\label{fig:subfig:a} 
\includegraphics[width=7.5cm,height=6.5cm]{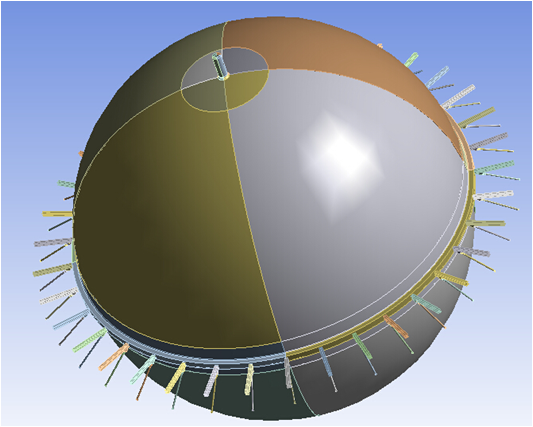}}
\subfigure[Detailed view of the support at equator by steel ring]{
\label{fig:subfig:b} 
\includegraphics[width=7.5cm,height=7.5cm]{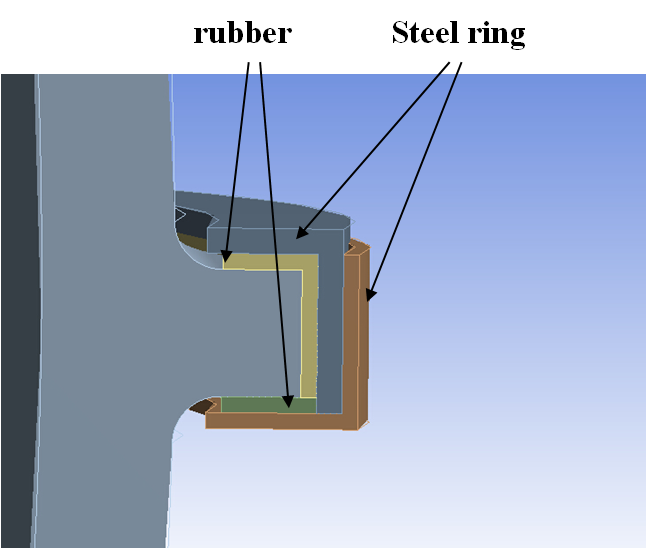}}
\caption{Acrylic sphere supported at equator}
\label{fig:subfig:equator} 
\end{figure}

This supporting structure is indicated in Fig.~\ref{fig:subfig:equator}. The acrylic sphere is
supported at the equator area with stainless-steel rings which connect to the
wall or bottom of the water pool by rods. A radial extension growing from the
sphere at the equator functions as supporting points where the stainless-steel
rings are clamped to it. A rubber layer is used as buffer between the acrylic
and steel ring to avoid any stress concentration, and this buffer layer is also
very useful to improve the stability of the whole structure, especially under
seismic conditions. The rings are separated into several parts along the
circumference, each part anchored to the water pool by an H-shaped unistrut and a tilted rod.

\begin{figure}[!htbp]
\centering
\subfigure[]{
\label{fig:subfig:a} 
\includegraphics[width=7.5cm,height=7cm]{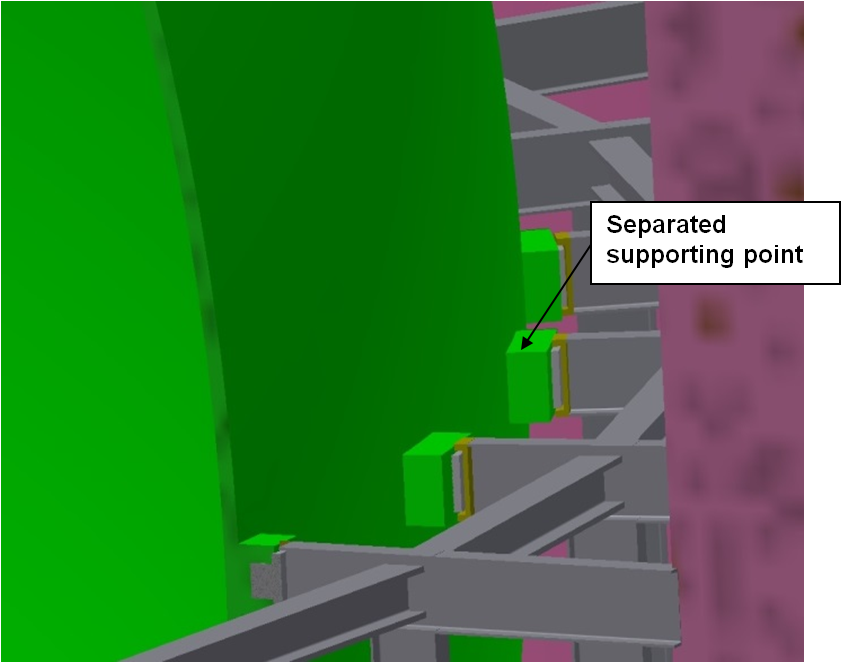}}
\subfigure[]{
\label{fig:subfig:b} 
\includegraphics[width=7.5cm,height=7cm]{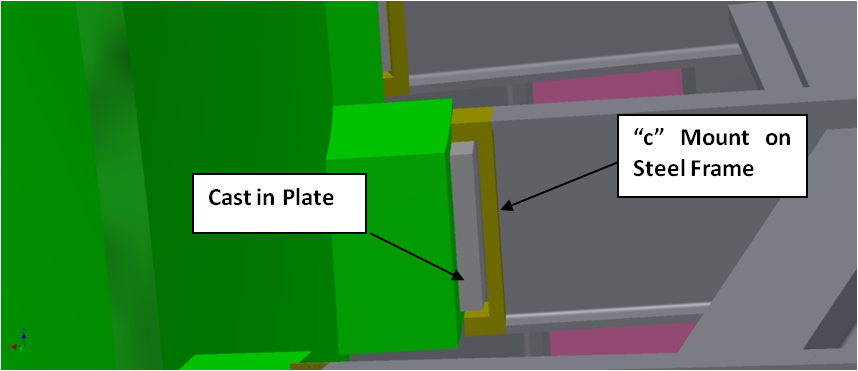}}
\caption{Detailed view of the support at equator by separated points}
\label{fig:subfig:equator-design} 
\end{figure}

To minimize the light blocking by the steel ring,  another method is using a
number of separated supporting points which are distributed at the equator. In
this design there are 32~blocks cast into the sheets that are assembled at the
equator.  Into each of these blocks will be cast a steel plate.  The sphere is
supported at each of these steel plates so there is no direct bearing on the
acrylic. Fig.~\ref{fig:subfig:equator-design} shows details of this concept. The 32 support points on the
sphere will be supported by an external steel structure that is made of columns
braced against the concrete wall. A beam will project forward from the columns
and through the layer of PMTs which will surround the sphere. At the end of this
beam a plate is welded to the profile. The height and
depth of the "C" will be much larger than the plate that is cast into the
acrylic sphere. The resulting gap between the plate and the "C" accommodates
the assembly tolerances. The sphere will be built up from the bottom using
temporary supports until the equator ring with the 32 support points is
completed. The external steel structure will be in place and the "C" plate
will be surrounding the plate cast into the sphere. At this point shims will be
placed around the "C" which will lock the cast plate into place. Once shims
have been installed on all 32 support points the bottom temporary supports can
be removed and the entire load of the sphere will be transferred to the "C"
plates and external steel structure. The top half of the sphere can then be
assembled.

\subsection{The Acrylic Sphere}
\begin{figure}[htp]
\begin{center}
\includegraphics[width=10cm,height=11cm]{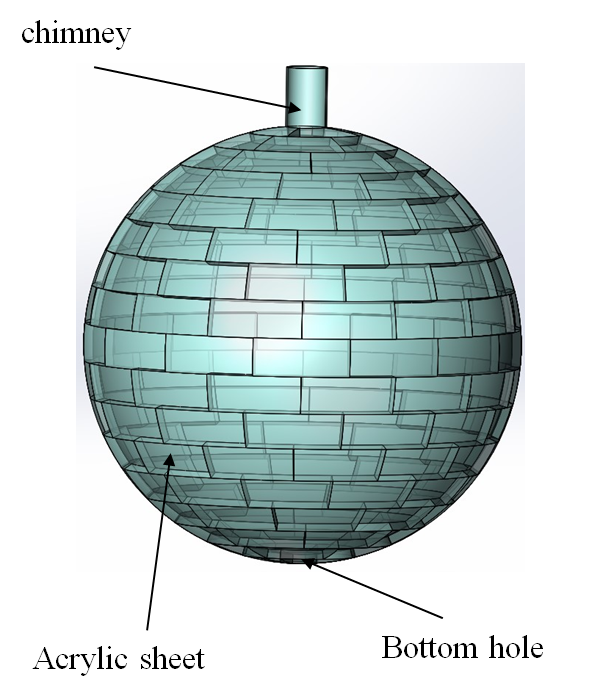}
\caption{Schematic view of the acrylic sphere}
\label{fig:cd3-7}
\end{center}
\end{figure}

The inner structure of the central detector is a transparent acrylic sphere
with $\sim$35.4~m inner diameter and 120~mm in shell thickness. This acrylic sphere
will be assembled and bonded with bulk polymerization by a number of acrylic
sheets, as shown in Fig.~\ref{fig:cd3-7}. Considering the production capacity and
transportation limit, the acrylic shell will be divided into more than 170
sheets, each is about 3~m~$\times$~8~m in dimension. The final number and
size of the sheet may be modified after further consideration and discussion
with manufacturers. To reduce the stress on the sphere at the supporting
points, some appended acrylic pieces will be bonded on top of the
sheets. The supporting structure of the sphere will be connected to these
appended acrylics. A chimney of about $\sim$1~m diameter is designed on top of the
acrylic sphere, which will be used as the filling port and interface to the
calibration system. An outlet may be designed on the bottom of the sphere for
cleaning of the sphere and LS recycling during detector running.

\subsection{Joint of the Acrylic and Truss}
\begin{figure}[!htbp]
\begin{center}
\includegraphics[width=14cm]{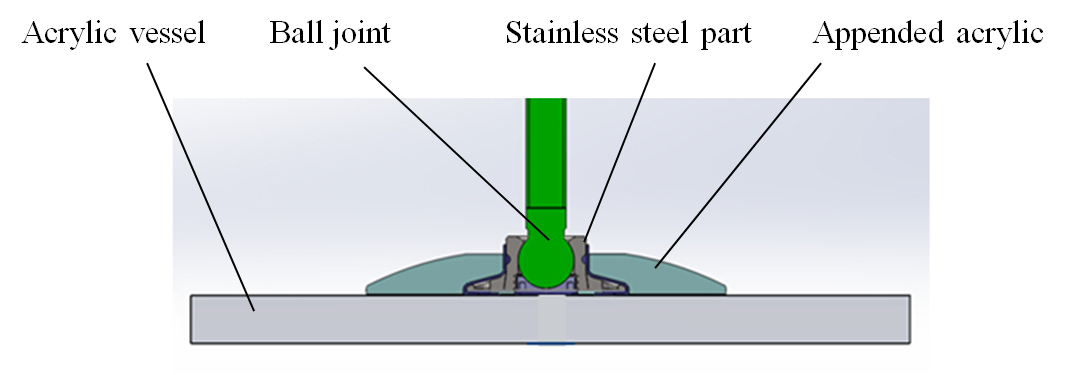}
\caption[Logo in CentralDetector]{Supporting joint of the acrylic sphere and steel truss}
\label{fig:cd3-8}
\end{center}
\end{figure}

In the baseline design for the truss the concept of the joint between the acrylic
sphere and the steel truss comes from the idea of the connecting structure used
in the glass curtain wall. For each acrylic sheet, there will be one or two
stainless steel disks which are embedded in it as a connecting structure, as
shown in Fig.~\ref{fig:cd3-8}. At the location of each steel disk, there is an appended
acrylic piece with $\sim$100~mm thickness on top of it and bonded to the sphere, and
a rubber layer is placed between the acrylic and steel for buffering. The steel
disk is designed to have a ball-type head, and a supporting rod will connect it
to the truss. The load on the acrylic sphere will be transferred to the truss
through these rods. The dimensions of the appended acrylic and the steel disk
may be changed according to a further analysis of the stress.

\begin{figure}[!htbp]
\begin{center}
\includegraphics[width=14cm]{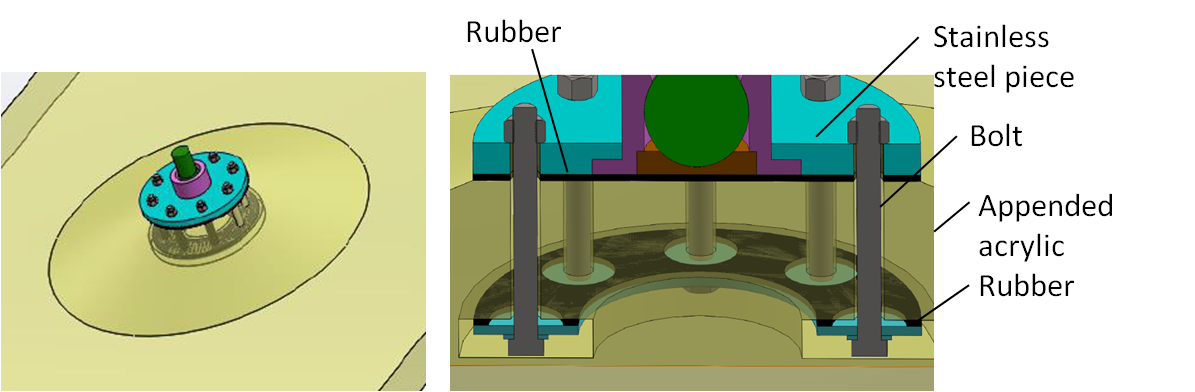}
\caption[Logo in CentralDetector]{Structure of the joint with bolt}
\label{fig:cd3-9}
\end{center}
\end{figure}

In addition, there is another type of joint structure under design, as shown in
Fig.~\ref{fig:cd3-9}. Instead of using the steel disk which is heavy and of large size, a
steel ring is embedded into the acrylic, and connected to a steel plate on top
of the appended acrylic by bolts. Rubber is placed in between the different
components, and the bolt is surrounded by plastic bushing to avoid direct contact with the acrylic.
This new structure is easier for construction and also can reduce the amount of stainless steel and hence the
radioactive background.

Currently the two design options are both under optimization, with a main goal of
reducing the stress level to less than 5~MPa. Several prototype tests have been
finished for the first option, and test for the second option will be done
in the near future.

\subsection{Finite Element Analysis and Testing}
\subsubsection{Global FEA}
We have finished a global FEA for the central detector under the following conditions.

1) sphere is empty after the completion of acrylic sphere and steel truss: the load is just the self-weight of the structure.

2) All PMTs are installed on the truss: the weight of total $\sim$17000 PMTs is about $\sim$1700 KN, which will load on the truss.

3) Sphere is filled with liquid and running for long term: the linear liquid
pressure will be loaded on the sphere, and the total buoyance of the sphere is
about $\sim$3100~t; the PMT's buoyancy is about $\sim$1000~t, which will be distributed on the joint nodes of the truss.

During the analysis a load factor of 1.35 was used for the dead load. The FEA
results include stress, deflection and stability for the three loading
conditions. The effect caused by seismic load, temperature or relative liquid
level difference has also been analyzed. Following are the main conclusions:

(1) Stress and deflection

\begin{figure}
\centering
\subfigure[condition 1]{
\label{fig:subfig:a} 
\includegraphics[width=7.5cm]{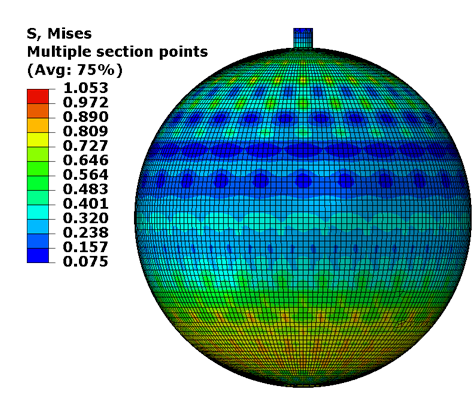}}
\subfigure[condition 2]{
\label{fig:subfig:b} 
\includegraphics[width=7.5cm]{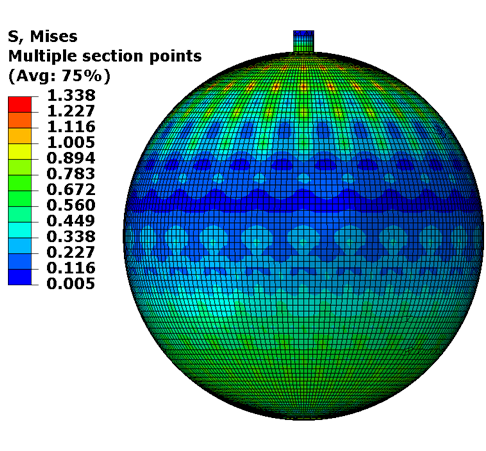}}
\subfigure[condition 3]{
\label{fig:subfig:c} 
\includegraphics[width=7.5cm]{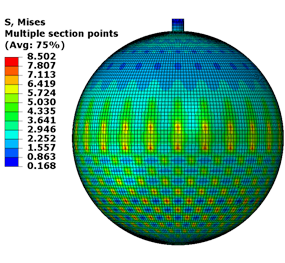}}
\caption{Stress distribution on the acrylic sphere}
\label{fig:cd3-10} 
\end{figure}

\begin{figure}
\centering
\subfigure[condition 1]{
\label{fig:subfig:a} 
\includegraphics[width=7.5cm]{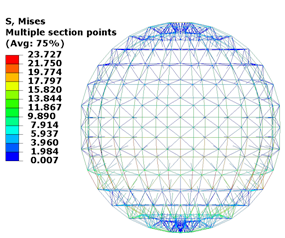}}
\subfigure[condition 2]{
\label{fig:subfig:b} 
\includegraphics[width=7.5cm]{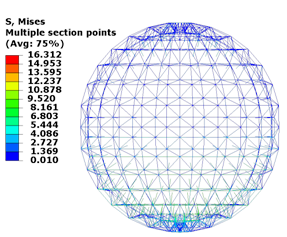}}
\subfigure[condition 3]{
\label{fig:subfig:c} 
\includegraphics[width=7.5cm]{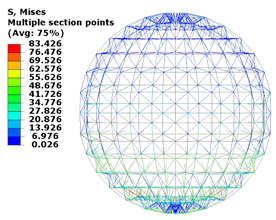}}
\caption{Stress distribution of the steel truss}
\label{fig:cd3-11} 
\end{figure}

\begin{figure}
\centering
\subfigure[condition 1]{
\label{fig:subfig:a} 
\includegraphics[width=7.5cm]{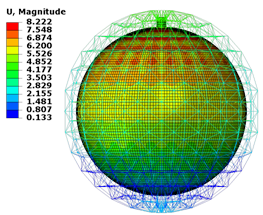}}
\subfigure[condtion 2]{
\label{fig:subfig:b} 
\includegraphics[width=7.5cm]{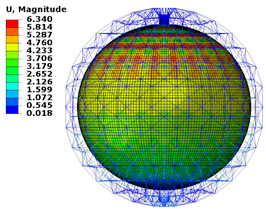}}
\subfigure[condition 3]{
\label{fig:subfig:c} 
\includegraphics[width=7.5cm]{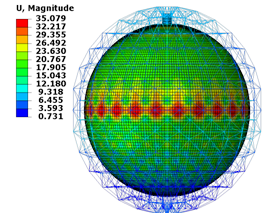}}
\caption{Deflection of the central detector}
\label{fig:cd3-12} 
\end{figure}

From Table~\ref{table1}, Fig.~\ref{fig:cd3-10} to Fig.~\ref{fig:cd3-12}, we can see that the maximum
stress on the acrylic sphere is $\sim$8.5~MPa in condition (3). If the load factor
of 1.35 is removed, the stress is $\sim$6.3~MPa. The maximum displacement is $\sim$35~mm,
occurring at the bottom of the structure, which is only $\sim$1/1000 of the sphere
span and meets the technical specification for space frame structures
(JGJ7-2010).

\begin{table}[!htbp]
\centering
\caption{Maximum stress and deformation for each loading condition\label{table1}}
\newcommand{\tabincell}[2]{\begin{tabular}{@{}#1@{}}#2\end{tabular}}
\begin{tabular}{p{3cm}<{\centering}|p{3cm}<{\centering}|p{3cm}<{\centering}|p{3cm}<{\centering}}
\hline
	Loading condition &		Max. stress on sphere(MPa) & 	Max. stress on truss(MPa) & 	Max. general deflection (mm) \\
\hline
\tabincell{c}{Condition 1}  & 1.1 	&	 23.7 &  8.2	\\
\hline
\tabincell{c}{Condition 2}    & 1.3 	&	 16.3 &  6.3	\\
\hline
\tabincell{c}{Condition 3}       & 8.5 	&	 83.4 &  35.1	\\
\hline
\end{tabular}
\end{table}

(2) Analysis of the seismic load

\begin{figure}
\centering
\subfigure[Stress on the acrylic sphere under seismic condition]{
\label{fig:subfig:a} 
\includegraphics[width=7.5cm]{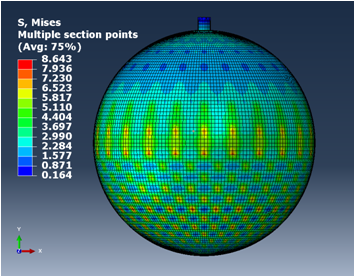}}
\subfigure[Stress on the steel truss under seismic condition]{
\label{fig:subfig:b} 
\includegraphics[width=7.5cm]{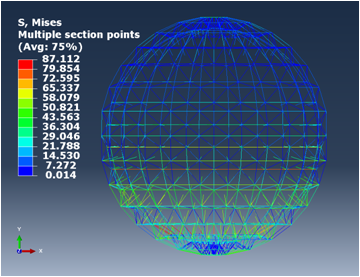}}
\caption{Stress on the acrylic sphere and steel truss under seismic condition}
\label{fig:cd3-13} 
\end{figure}

Based on the local geological condition, a seismic fortification
intensity at Richter scale 5.5 was considered, and a seismic load of 0.1~g was taken in our
analysis. Adding 0.1~g seismic load into the finite element model, the detector
structure was analyzed for condition 3 and the result is shown in Fig.~\ref{fig:cd3-13}.
We can see that the stress on the acrylic sphere is increased by only $\sim$1.6\%
and on the steel truss by only $\sim$4.4\%. The small change shows that the structure
of the detector is safe under the seismic load.

(3)Stability analysis

\begin{figure}
\centering
\subfigure[Instability mode]{
\label{fig:subfig:a} 
\includegraphics[width=7.5cm,height=6cm]{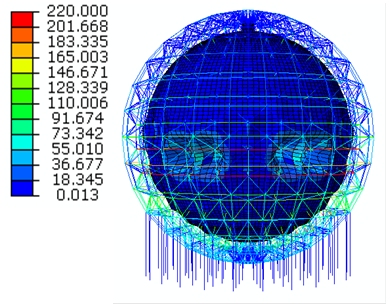}}
\subfigure[Deflection-load coefficient curve]{
\label{fig:subfig:b} 
\includegraphics[width=7.5cm,height=6cm]{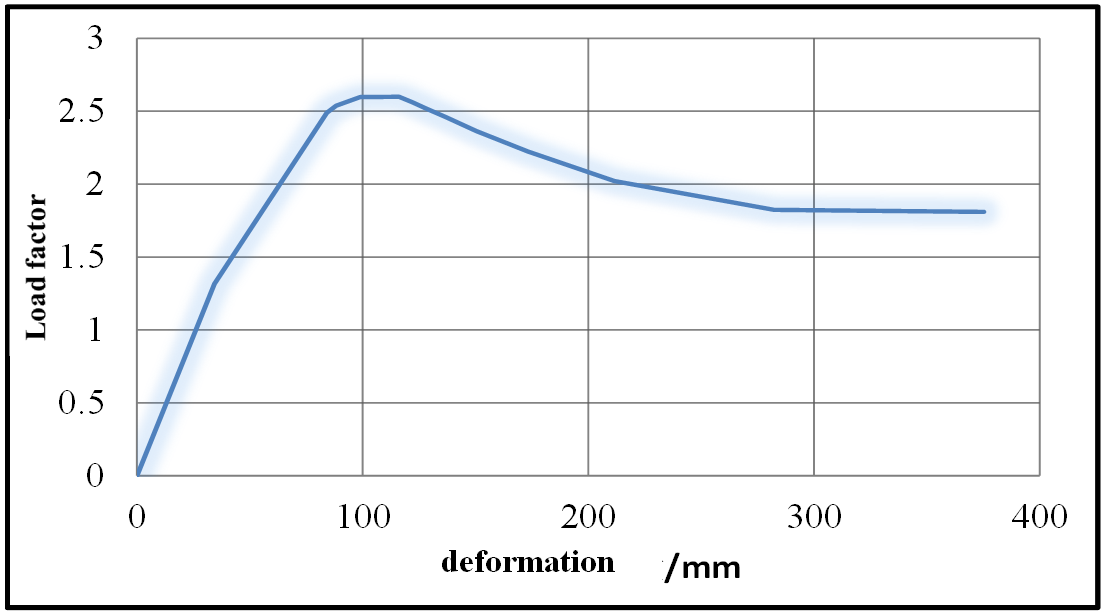}}
\caption{Stability analysis of the central detector}
\label{fig:cd3-14} 
\end{figure}

The safety factor for the buckling of the acrylic sphere is increased by the
support from the steel truss. The double non-linearity of the material and the
geometry were taken into account for the global stability analysis, and the
first elastic buckling mode of the structure was taken as the initial
imperfection in calculation. Fig.~\ref{fig:cd3-14}(a) shows the results of the instability
mode. Fig.~\ref{fig:cd3-14}(b) shows the deflection-load coefficient curve and we
can see from it the stability factor is 2.61, which meets the specification of JGJ7-2010.

(4) Effect of liquid level difference

Liquid level of the LS in the acrylic sphere will affect the stress status of
the sphere. We did the analysis for several cases of the relative liquid level.
A difference of -2~m, -1~m, 0~m, 1~m, 2~m and 3~m between the LS and water( LS
level - water level) were taken into account. The results are shown in Table
\ref{table2}.  We can see that the variation of the relative liquid level impacts stress
significantly. If the LS level in the sphere is higher than the level of water
outside, the stress will be reduced significantly. Raising the LS level may be
considered in our final design.

\begin{table}[!htbp]
\centering
\caption{Impact analysis of liquid level\label{table2}}
\newcommand{\tabincell}[2]{\begin{tabular}{@{}#1@{}}#2\end{tabular}}
\begin{tabular}{p{4cm}<{\centering}|p{4cm}<{\centering}|p{4cm}<{\centering}}
\hline
	 Liquid level difference H(m)&		Maximum stress on sphere (MPa) & 	Stress on supporting rod (MPa)  \\
\hline
\tabincell{c} {-2} 	&	 10.2 &  67.7	\\
\hline
\tabincell{c} {-1} 	&	 9.4 &  61.6	\\
\hline
\tabincell{c} { 0} 	&	 8.4 &  57.6	\\
\hline
\tabincell{c} { 1} 	&	 7.4 &  53.7	\\
\hline
\tabincell{c} { 2} 	&	 6.4 &  49.7	\\
\hline
\tabincell{c} { 3} 	&	 5.6 &  46.5	\\
\hline
\end{tabular}
\end{table}

(5) Analysis of temperature impact

The detector will be constructed in a hall $\sim$728~m below the ground and the
yearly temperature variation is rather small. Since the thermal expansion
coefficients of the stainless steel and acrylic is different, the impact of
temperature on the stress still needs to be considered. The impact of
increasing or decreasing temperature by 10$^{\circ}$C has been analyzed. Table
\ref{table3} shows the results. For the stainless steel truss, the stress is always less
than 100~MPa and the safety level will not be affected by temperature change.
For the acrylic sphere, temperature elevation is helpful while decreasing of
temperature will lead to a larger stress.So avoiding temperature decrease is
necessary during detector running period.

\begin{table}[!htbp]
\centering
\caption{Temperature impact on stress\label{table3}}
\newcommand{\tabincell}[2]{\begin{tabular}{@{}#1@{}}#2\end{tabular}}
\begin{tabular}{p{5cm}<{\centering}|p{3.5cm}<{\centering}|p{3.5cm}<{\centering}}
\hline
	Condition&		Maximum stress on sphere (MPa) & 	Maximum stress on truss (MPa)  \\
\hline
\tabincell{c} {Room temperature} 	&	 8.5 &  83.4	\\
\hline
\tabincell{c} {Room temperature +10$^{\circ}$C} 	&	 8.0 &  77.5	\\
\hline
\tabincell{c} {Room temperature -10$^{\circ}$C} 	&	 10.1 &  95.0	\\
\hline
\end{tabular}
\end{table}

(6) Failure analysis

The central detector is required to have a life-time of at least 20 years, and
any crucial failure is not allowed during this period. For the acrylic sphere
and steel truss, the key component is the support points on the sphere. The
effect of one or more failure (for example, if a rod is broken) was
evaluated. Four kinds of possible failures were analyzed, which are:

$\bullet$ condition 1: failure appeared at the maximum stress point;

$\bullet$ condition 2: failure appeared at all points on the same latitudinal layer where the maximum stress is located;

$\bullet$ condition 3: failure appeared at the points located in one
longitudinal band in the lower hemisphere;

$\bullet$ condition 4. failure appeared at several points randomly distributed.

\begin{figure}[!htbp]
\centering
\includegraphics[width=14cm]{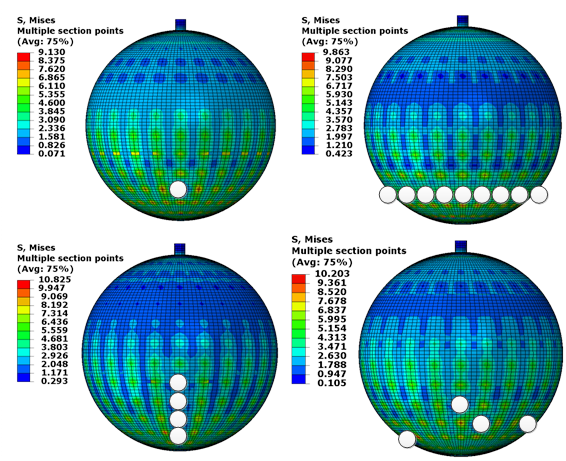}
\caption{Failure analysis of the central detector for (a) condition 1,
(b)condition 2, (c)condition 3 and (d)condition 4.}
\label{fig:cd3-15}
\end{figure}

The analysis results of those cases are shown in Fig.~\ref{fig:cd3-15}.

In the four failure cases, the maximum stress on the acrylic sphere is 10.8~Mpa
(8~Mpa in dead load), an increase of 27.3\% after failure. In fact,
the possibility is very small for failures appearing at the same latitudinal or
longitudinal band. One point failure is the most probable case and the stress
variation is 7\% in this case. The above analysis shows that the risk to the
central detector is controllable.

\subsubsection{Local FEA of the Supporting Point of the Acrylic Sphere}

The grid size of the FEA model affects the accuracy of the calculation,
especially for a singular point and the area around it. The above analysis
shows that the maximum stress of the acrylic sphere appears at the joint of
the stainless steel supporting rod and the acrylic sphere.
Since this node is an intersection of the beam element and the shell element,
it's a typical singular point. For FEA, a local fine model around the singular
point is often employed in order to improve the accuracy. A validation test is
necessary to check the analysis.

\begin{figure}[!htbp]
\centering
\includegraphics{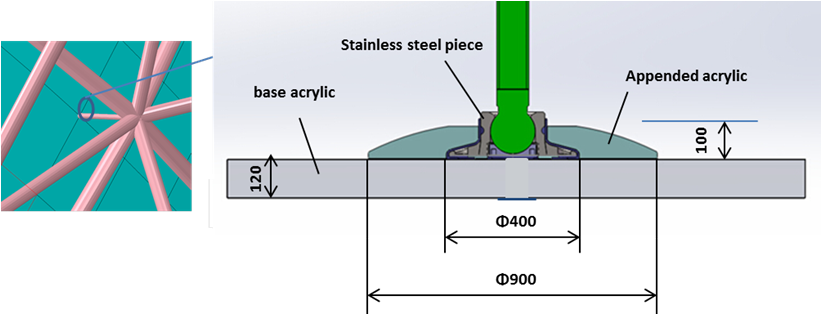}
\caption{the structure of the connecting joint}
\label{fig:cd3-16}
\end{figure}

Fig.~\ref{fig:cd3-16} shows the structure of the connecting point between the acrylic
sphere and stainless steel truss. The structure at this joint involves the base
acrylic shell, the appended acrylic piece and the stainless steel part.

\begin{figure}
\centering
\subfigure[stress distribution on the base acrylic. (Max: 4.2~MPa)]{
\label{fig:subfig:a} 
\includegraphics[width=7cm,height=6cm]{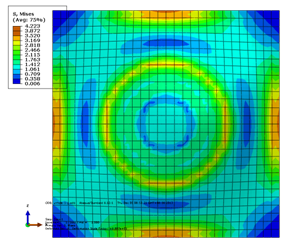}}
\subfigure[stress distribution on the appended acrylic. (Max: 7.2~MPa)]{
\label{fig:subfig:b} 
\includegraphics[width=7cm,height=6cm]{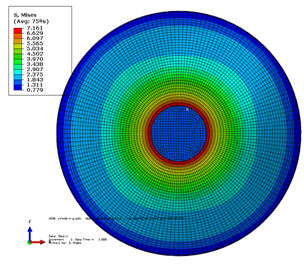}}
\subfigure[stress distribution on the steel disk. (Max: 68.4~MPa)]{
\label{fig:subfig:c} 
\includegraphics[width=7cm,height=6cm]{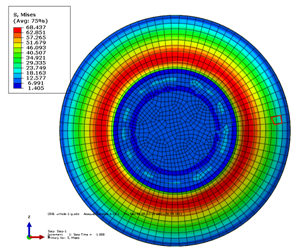}}
\subfigure[deformation distribution on the whole prototype. (Max: 2.6~mm)]{
\label{fig:subfig:c} 
\includegraphics[width=7cm,height=6cm]{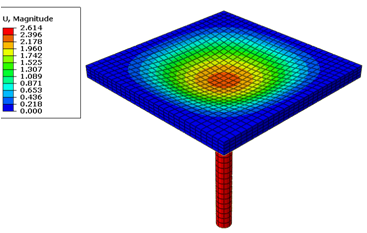}}
\caption{Analysis results for fixed constraint}
\label{fig:cd3-17} 
\end{figure}

\begin{figure}
\centering
\subfigure[stress distribution on the base acrylic. (Max: 11.6~MPa)]{
\label{fig:subfig:a} 
\includegraphics[width=7.5cm,height=6cm]{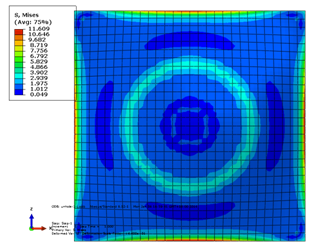}}
\subfigure[stress distribution on the appended acrylic. (Max: 7.9~MPa)]{
\label{fig:subfig:b} 
\includegraphics[width=7.5cm,height=6cm]{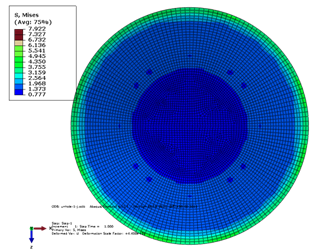}}
\subfigure[stress distribution on the steel disk. (Max: 69.8~MPa)]{
\label{fig:subfig:c} 
\includegraphics[width=7.5cm,height=6cm]{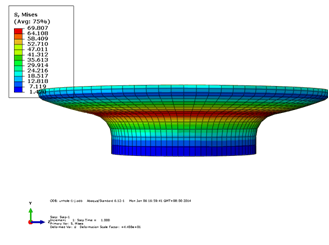}}
\subfigure[deformation distribution on the whole prototype. (Max: 4.1~mm)]{
\label{fig:subfig:c} 
\includegraphics[width=7.5cm,height=6cm]{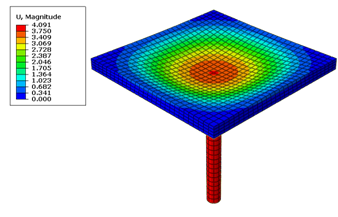}}
\caption{Analysis results for simple constraint}
\label{fig:cd3-18} 
\end{figure}

For the local finite model with 3D solid elements, the base acrylic shell is
2.6~m $\times$ 2.6~m with a thickness of 120~mm, the appended acrylic is 100~mm in
thickness and 900~mm in outer diameter, and the diameter of the stainless-steel
disk is about 400~mm. The axial load of 14~t in the stainless-steel rod was
taken into account for local FEA. Two different boundary conditions around the
base acrylic were applied for analysis respectively, one being a fixed
constraint and the other a simple constraint. Fig.~\ref{fig:cd3-17} shows the analysis
results of the fixed constraint and Fig.~\ref{fig:cd3-18} the results of the simple
constraint.

The results show that there is no large stress difference for the appended
acrylic and stainless steel disk under the two boundary conditions for the base
acrylic. The maximum stress in the simply supported model is 63.6\% higher than
that for the fixed constraint. The main reason for this is the stress concentration
at boundary line due to the applied constraint. Removing these singular points,
the stress difference for the two conditions is going to be small. For the
central detector, the real loading status is between the fixed and simple
constraint, so the maximum stress on the acrylic should be less than 10~Mpa.

\subsubsection{Prototype and Testing of the Supporting Structure of the Acrylic Sphere}
\begin{figure}
\centering
\subfigure[the joint prototype]{
\label{fig:subfig:a} 
\includegraphics[width=7.5cm]{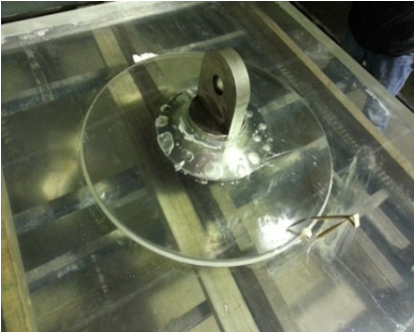}}
\subfigure[Installation of the prototype]{
\label{fig:subfig:b} 
\includegraphics[width=7.5cm]{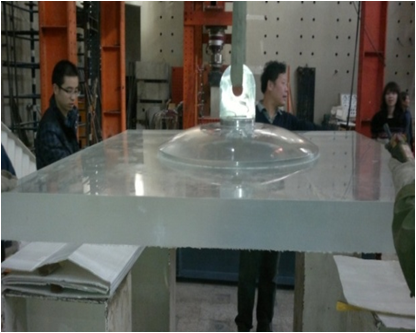}}
\subfigure[Preparation of the prototype test]{
\label{fig:subfig:c} 
\includegraphics[width=7.5cm]{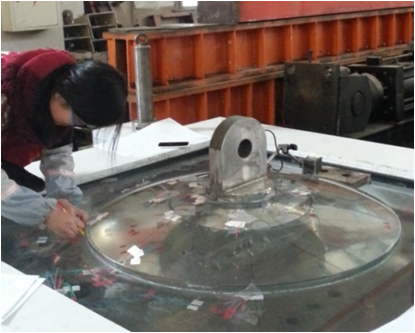}}
\subfigure[load-carrying test of the prototype]{
\label{fig:subfig:d} 
\includegraphics[width=7.5cm]{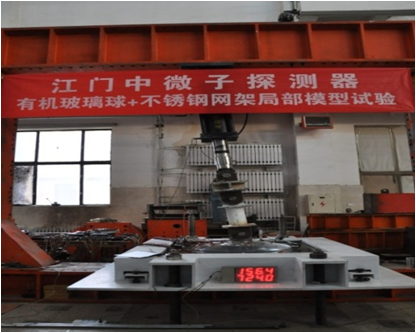}}
\caption{the prototype and test}
\label{fig:cd3-19} 
\end{figure}

A prototype of the supporting structure was constructed to study the technique
of fabricating the acrylic with steel built-in. Tests of the prototype were
performed to understand the load-carrying capacity of the structure and to
measure the stress and deformation to check FEA results. The prototype and the
test can provide a reference and useful parameters for further design.
Fig.~\ref{fig:cd3-19} shows the prototype and some pictures taken during the test.

\begin{figure}
\begin{minipage}[t]{0.5\textwidth}
\centering
\includegraphics[width=7cm,height=7cm]{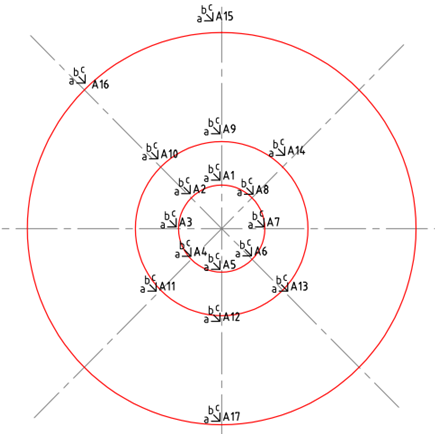}
\caption{layout of the measuring points}
\label{fig:cd3-20}
\end{minipage}
\begin{minipage}[t]{0.5\textwidth}\centering\includegraphics[width=7cm,height=7cm]{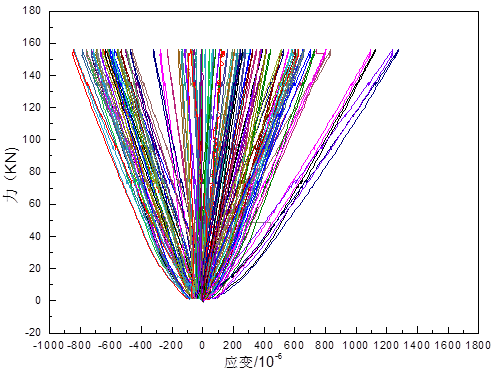}
\caption{load-strain curve of the measurement points}
\label{fig:cd3-21}
\end{minipage}
\end{figure}

The stress and deformation on the measured points in this test are shown in
Fig.~\ref{fig:cd3-20}. The measured points are distributed in the edge of the base
acrylic, the border between the base acrylic and the appended acrylic, and the
opening area of the appended acrylic. Fig.~\ref{fig:cd3-21} is the load-strain curve of
each measured point, which shows that all points have good elastic
characteristics. All the measured strain data were analyzed to get the stress
status of the prototype.

\begin{figure}[!htbp]
\centering
\includegraphics{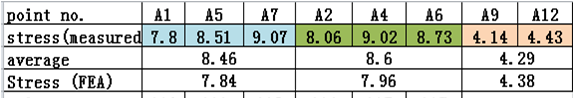}
\caption{Comparison of stress between measurement and FEA}
\label{fig:cd3-22}
\end{figure}

Under the axial load of 14~t, the maximum stress of the measured points is
8.6~Mpa which is located on the top of the appended acrylic. The measured
stress was compared with the result from FEA as shown in Fig.~\ref{fig:cd3-22}. The
deviation is less than 8\%, which demonstrates that the FEA is reasonable and
can simulate the real stress on the joint by using the local refinement method.

\begin{figure}
\centering
\subfigure[the third prototype of joint]{
\label{fig:subfig:a} 
\includegraphics[width=7cm,height=7cm]{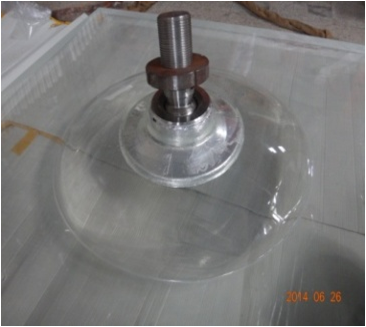}}
\subfigure[Test of the third prototype]{
\label{fig:subfig:b} 
\includegraphics[width=7cm,height=7cm]{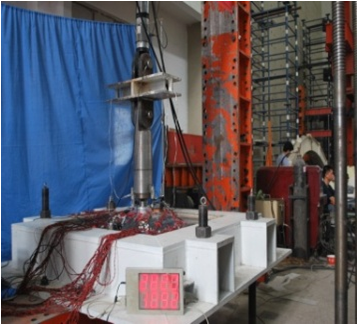}}
\subfigure[The crack pattern of the third prototype]{
\label{fig:subfig:c} 
\includegraphics[width=7cm,height=6cm]{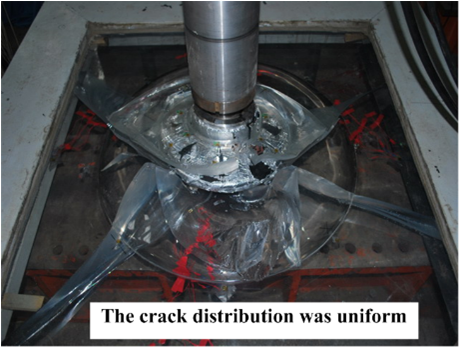}}
\caption{Third prototype test}
\label{fig:cd3-23} 
\end{figure}

The above prototypes and tests have verified that the supporting structure of
the sphere is safe under 14~t load as well as the reliability of FEA. Since
defects existed in fabrication of those prototypes(such as polymerization not
being completely finished due to the too low working temperature, the ball
joint not working as expected due to being accidentally glued to the appended
acrylic), a third prototype has been designed and tested with the known
problems fixed. In addition, to test the maximum load-carrying capacity of the
joint structure, an orthogonal load was applied to the prototype. The new
prototype and test are shown in Fig.~\ref{fig:cd3-23}.

Given the improvements in the new prototype, the maximum load that the
prototype can bear is up to 51 tons, which is significantly larger than the 28
tons obtained from the previous prototypes. Since the maximum design load is 14
tons, a 3.6 safety factor is reached by the third prototype. Additionally, the
measurements confirm the FEA results, and the maximum stress is about 8.4~MPa,
as shown in Table~\ref{table4}:

\begin{table}[!htbp]
\centering
\caption{Comparison of FEA and measurement for the third prototype\label{table4}}
\newcommand{\tabincell}[2]{\begin{tabular}{@{}#1@{}}#2\end{tabular}}
\begin{tabular}{p{2.5cm}<{\centering}|p{3.5cm}<{\centering}|p{2cm}<{\centering}|p{2cm}<{\centering}|p{2cm}<{\centering}}
\hline
		Load & 	Properties for comparison & 	Measured point & 	data & 	FEA\\
\hline
\tabincell{c}140KN & Stress/MPa 	&	 A7-3 &  8.467 &8.392	\\
\hline
\tabincell{c}{ } &{  }	&	 B7-3 &  3.479  &  2.963	\\
\hline
\tabincell{c}{ } &{ } 	&	 C1-3 &  3.262  &  3.093	\\
\hline
\tabincell{c}{ }& Displacement /mm&	 W4-3 &  0.512 & 0.630	\\
\hline
\end{tabular}
\end{table}


Since we require that the maximum stress on the acrylic should not be larger
than 5~MPa, optimization of the supporting structure is still needed to further
reduce the stress and get a larger safety factor.

\subsubsection{Optimization of the Detector Structure}
From the global and local FEA of the detector structure as well as the
prototype test, we can see that the maximum stress in the acrylic shell (not
including the appended acrylic) is less than 5~MPa and meet our requirement,
but stress in the joints (on the appended acrylic) is still larger than 8~MPa
and further optimization is needed. The optimization and modification can be
done by:

$\bullet$ reducing the axial load on the connecting rod;

$\bullet$ raising the LS liquid level in the sphere;

$\bullet$ improving the uniformity of the whole structure to get a better loading status;

\begin{figure}
\centering
\subfigure[Inner layer truss before sparsifying the truss members]{
\label{fig:subfig:a} 
\includegraphics[width=6cm,height=6cm]{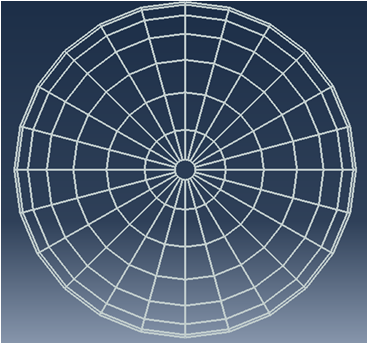}}
\subfigure[Inner layer truss after sparsifying]{
\label{fig:subfig:b} 
\includegraphics[width=7.4cm,height=6cm]{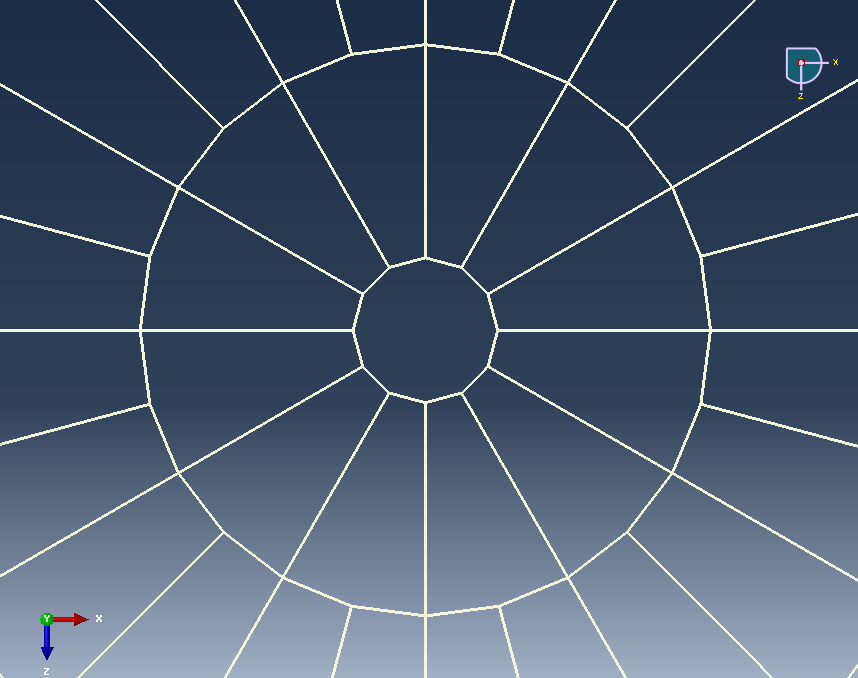}}
\caption{Inner layer truss after sparsifying the truss members}
\label{fig:cd3-24} 
\end{figure}

$\bullet$ reducing the number of connecting rods in the top and bottom area of the sphere to simplify the construction (see Fig.~\ref{fig:cd3-24}).

The results after those optimizations are shown in Table~\ref{table5}.

\begin{table}[!htbp]
\centering
\caption{Results from FEA after optimization\label{table5}}
\newcommand{\tabincell}[2]{\begin{tabular}{@{}#1@{}}#2\end{tabular}}
\begin{tabular}{p{3cm}<{\centering}|p{2.7cm}<{\centering}|p{2.7cm}<{\centering}|p{1.3cm}<{\centering}|p{1.3cm}<{\centering}|p{2.5cm}<{\centering}}
\hline
	Conditions &		Max-stress on sphere(Singular point excluded) /MPa& 	Max-stress on sphere(Singular point included) /MPa & 	Max-stress on truss /MPa & 	Max-axial load /KN & 	Max-displacement /mm\\
\hline
\tabincell{c}{Optimization 1 }& 9.6 &6.4	&	74.7 &  164.7  &  47.4	\\
\hline
\tabincell{c} {Optimization 2 }& 6.9 &5.2&	 115.1 &  135.9  &  27.3\\
\hline
\tabincell{c} {Optimization 3 }& 6.5 &4.9&	 91.3 &  126.1  &  28.4\\
\hline
\tabincell{c}  { No 1.35 factor}& 4.8& 3.6 	&	 67.6 &  93.4 & 21.0	\\
\hline
\end{tabular}
\end{table}

Note for Table~\ref{table5}:

$\bullet$ In all conditions, the liquid level difference is $\sim$3~m;

$\bullet$ Before optimization: the thickness of the upper acrylic hemisphere is
8~cm and the lower hemisphere is 12~cm. The number of supporting rods is 503
and no support comes from the outer layer of the truss to the upper hemisphere.

$\bullet$ Optimization step 1: the thickness of the sphere is changed to 10~cm.
The number of the supporting rods is increased to 646 and symmetrically placed
on the sphere;

$\bullet$ Optimization step 2: the supporting rods at both poles are sparsified, and the total number is 610.

From Table~\ref{table5}, we can see the stress in the acrylic is 3.6~MPa if the support
point is excluded, and the maximum axial load in the supporting rods is only
93.4~KN. Since the stress is proportional to the load, it means that the stress
should be also reduced a lot, and the FEA modeling to confirm this is currently
underway.

\subsection{Study of the Acrylic Properties}
Acrylic is the main material used for the central detector construction. Since
the detector will involve LS, LAB and high-purity water during long-time operation,
the acrylic materials in the liquids should be studied. At present, a number of tests have been finished or are in progress.

(1) Normal mechanical properties test

Referring to the standards of ASTM-D638, ASTM-D695 and ASTM-D790, the tensile
strength, compressive strength and flexural strength of the acrylic had been
tested and the results are 68.2~MPa, 108.3~MPa and 88.0~MPa, respectively.
Since the thickness of the acrylic sphere is 120~mm, it needs 2 or 3 layers of
thinner sheet to be bonded together. In order to know the strength after
bonding, we also did the test for the double-layer material. The test result
is 66.8~MPa for the tensile strength and 101.7~MPa for the flexural strength.

(2) Aging test in LS

\begin{figure}
\centering
\subfigure[Tensile strength]{
\label{fig:subfig:a} 
\includegraphics[width=7cm,height=7cm]{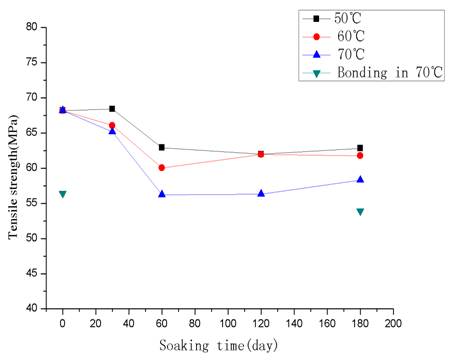}}
\subfigure[Compressive strength]{
\label{fig:subfig:b} 
\includegraphics[width=7cm,height=7cm]{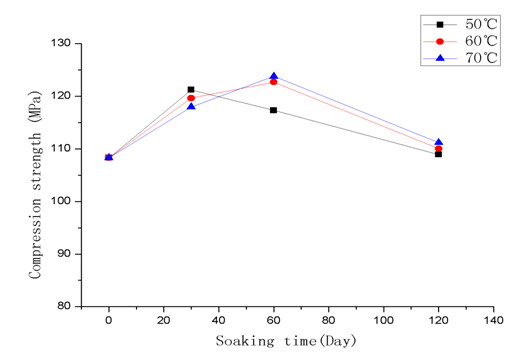}}
\subfigure[Flextual strength]{
\label{fig:subfig:c} 
\includegraphics[width=7cm,height=7cm]{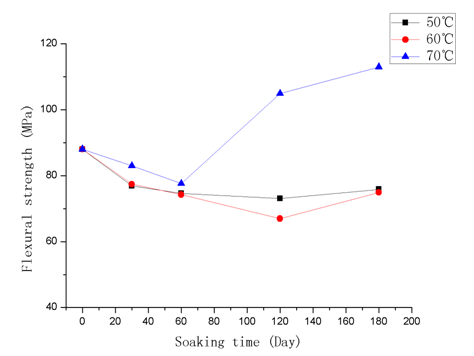}}
\caption{Performance of the acrylic after aging}
\label{fig:cd3-25} 
\end{figure}

The acrylic samples were submerged into LS at temperatures of 50$^{\circ}$C,
60$^{\circ}$C, and 70$^{\circ}$C. The testing time lasted 30, 60, 120, and 180
days for each temperature point. After that the mechanical properties of the
aged acrylic was measured to compare with the fresh one. Results are shown
in Fig.~\ref{fig:cd3-25}. Testing shows that the material has different aging behavior for
different mechanical properties. Here, we mainly focus on the tensile strength
since it is the weakest strength compared to others. It is known that 60
days at 70 degrees corresponds to 16 years at 20 degrees, so we can predict
that the tensile strength will drop by 18\% after 16 years of running. To
validate this conclusion, further tests are in progress.

(3) Test of the old acrylic pieces

To study the influence of the surrounding environment such as ultraviolet
radiation, sunlight, temperature and humidity, measurements have been made for
old acrylic pieces after 14 and 17 years in outdoor. Results are shown in Table~\ref{table6}.

\begin{table}[!htbp]
\centering
\caption{Strength of the acrylic after outdoor exposure\label{table6}}
\newcommand{\tabincell}[2]{\begin{tabular}{@{}#1@{}}#2\end{tabular}}
\begin{tabular}{p{4cm}<{\centering}|p{4cm}<{\centering}|p{4cm}<{\centering}}
\hline
	 Explosure time&		14 years & 	17 years  \\
\hline
\tabincell{c} {Elastic modulus /MPa} 	&	803 & 942	\\
\hline
\tabincell{c} {Poisson ration} 	&	 0.405&  0.405	\\
\hline
\tabincell{c} { Tensile strength} 	&	 56.6 &  53.57	\\
\hline
\end{tabular}
\end{table}

The tensile strength of the normal acrylics is about 68~MPa. From the table,
we can see that it drops by 17\% and 21\% respectively, for 14 years and 17 years in outer door.

(4) Test for the new bonding techniques

\begin{figure}[!htbp]
\centering
\includegraphics[width=12cm]{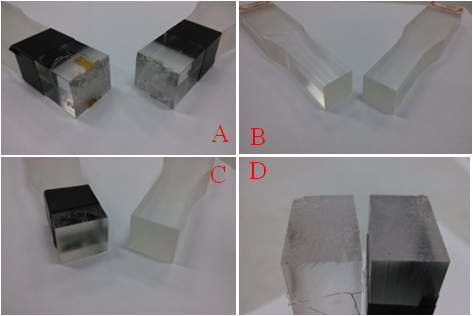}
\caption{cross section of  the broken acrylic}
\label{fig:cd3-26}
\end{figure}

The construction of the acrylic sphere will require a large amount of bonding.
Since the normal bonding technique needs a long time for curing, and therefore a
very long construction time, the manufacturers are studying two new bonding
techniques(ultraviolet irradiation and fast bonding) to reduce the curing time.
The tensile strength with those new techniques has been tested and compared
with the normal ones. The strength for ultraviolet irradiation is 46.5~MPa and
for fast bonding is 55.3~MPa. The latter corresponds to 80\% of the full
strength of the acrylic. Details are shown in Table~\ref{table7}. Fig.~\ref{fig:cd3-26} shows the
cross-section where the break happens; it can be seen that the cross-section of normal bonding
(A) and fast bonding (B) are similar, while for the other two
(C and D) it has a very flat surface, indicating that the bonding syrup is not
completely adhering with the original acrylic, hence giving a lower strength.

\begin{table}[!htbp]
\centering
\caption{Comparison of the different bonding techniques\label{table7}}
\newcommand{\tabincell}[2]{\begin{tabular}{@{}#1@{}}#2\end{tabular}}
\begin{tabular}{p{4cm}<{\centering}|p{2.5cm}<{\centering}|p{3cm}<{\centering}|p{2.5cm}<{\centering}}
\hline
	 Bonding techniques&		time /hours & 	Tensile strength /MPa & 	Ratio  \\
\hline
\tabincell{c} {Without bonding} 	&{ }& 68.2 & 100\%	\\
\hline
\tabincell{c} {Normal bonding} 	&	 12& 56.4&  82.7\%	\\
\hline
\tabincell{c} {Ultraviolet irradiation \\ without annealing} 	&3 &	 37.2 &  54.5\%	\\
\hline
\tabincell{c} {Ultraviolet irradiation \\ with annealing} 	&7 &	 46.5 &  68.2\%	\\
\hline
\tabincell{c} {Fast bonding} 	&4 &	 55.3 &  81.1\%	\\
\hline
\end{tabular}
\end{table}

(5) Crazing test

Users and manufacturers of plastics have been aware of a phenomenon that glassy
plastics will develop micro-cracks, which start at the surface of the material
and grow perpendicularly to the direction of stress. Studies have shown that
there are at least five factors which may cause glassy plastics to craze, as listed in the following:

$\bullet$ Applied or residual tensile stress;

$\bullet$ Stress generated on the surface of the acrylic due to differential temperature;

$\bullet$ Stress generated on the surface of the acrylic due to absorption and desorption of moisture;

$\bullet$ Weathering;

$\bullet$ Contact with organic solvents.

Crazes are initiated and propagated only when the tensile stress on the
material surface exceeds some critical value.
For the central detector of JUNO, the acrylic sphere will be immersed in LS and
water under stress loading. The problems of crazing become important concerning
the reliability and longevity of the structural parts manufactured from
acrylic.

We did the following tests on acrylic samples in laboratory. Samples with 4.8~MPa
stress applied were immersed in the LS and the liquid mixture of 80\% LS+20\% C$_{9}$H$_{12}$, at the room temperature
and 60$^{\circ}$C. The results are summarized as the following:

1) No crazing appeared by submerging in LS for 90 days at the room temperature;

2) No crazing appeared by submerging in the mixture of 80\% LS+20\% C$_{9}$H$_{12}$ for 90
days at the room temperature;

3) By submerging samples in LS at 60$^{\circ}$C, two of the
samples showed initial crazing after 30 days, and the other two had no crazing.
After 90 days, the initial crazing had grown along the direction of thickness. More tests and studies on crazing will be made in future.

(6) Test of light transmittance

The test results indicate that the light transmittance of the acrylic with
4~cm-8~cm thickness is over 96\%, and the thickness has no significant effect
on it. The measured attenuation length in air is $\sim$110~cm at 410~nm and $\sim$278~cm
at 450~nm.

(7) Creep test

\begin{figure}[!htbp]
\centering
\includegraphics{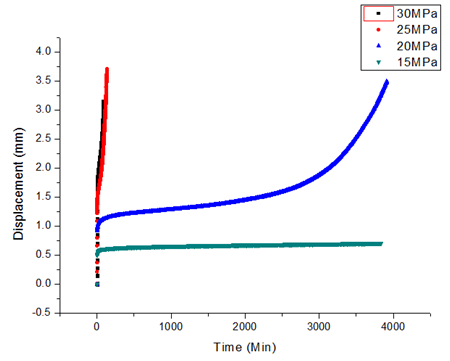}
\caption{Creep curve of the acrylic under different stress }
\label{fig:cd3-27}
\end{figure}

The creep test has been partially finished up to now for the acrylic submerged in LS
at 20 degrees under a stress of 30~MPa, 25~MPa, 20~MPa, and 15~MPa. The
results are shown in Fig.~\ref{fig:cd3-27}. This measured curve is consistent with the
theoretical prediction, and the time before breaking increases dramatically at lower
stress level. To understand the allowable maximum stress, more data are needed.
Currently the creep tests are ongoing, including the test in pure water.

\section{Backup option: Balloon in a steel tank}
\subsection{Introduction}
\begin{figure}[[!htbp]]
\centering
\includegraphics[width=14cm]{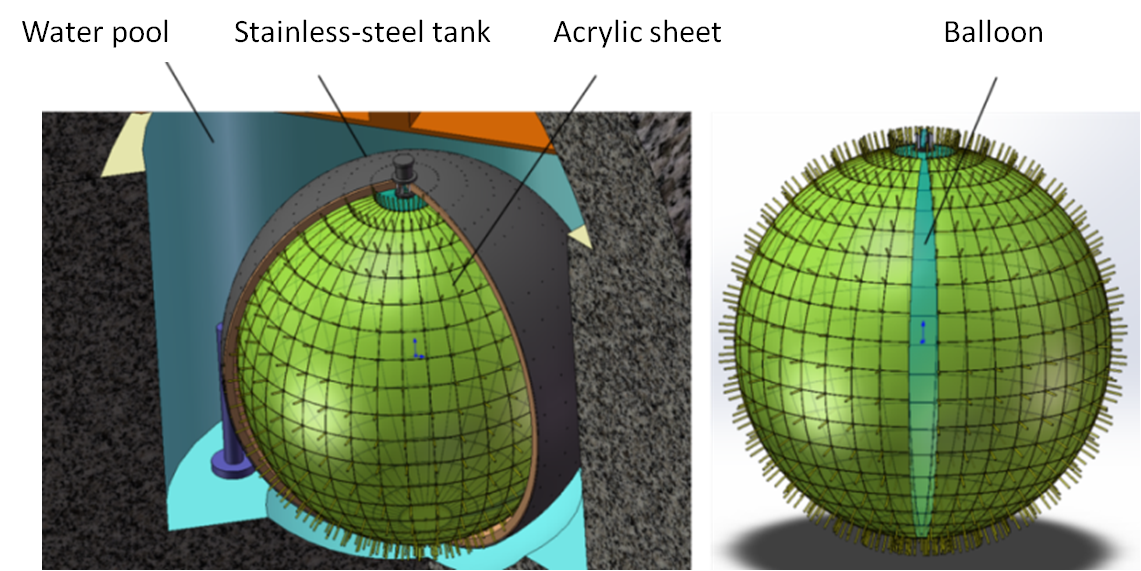}
\caption{Schematic view of stainless-steel tank and the balloon}
\label{fig:cd3-28}
\end{figure}

The backup option of the central detector is a balloon with a stainless-steel tank.
The tank, located in the center of the experimental hall, is about $\sim$40~m in diameter.
A balloon of $\sim$35.4~m in diameter is installed in the steel tank.
This balloon, welded from transparent films, is filled with $\sim$20~kt of LS as a
target. The buffer liquid between the tank and balloon is LAB or MO. The
density of LS in the balloon is 0.3-0.5\% larger than the buffer liquid. The
soft balloon needs a reliable supporting structure to keep it spherical and
reduce the stress, so a thin acrylic layer surrounding the balloon is designed for
this purpose. Both the balloon and the acrylic structure have the same
diameter. The inner wall of the steel tank is covered by more than $\sim$17000
20~inch PMTs. Fig.~\ref{fig:cd3-28} shows the structural sketch of this option.

\subsection{Balloon}
The balloon is the most important structure in this option. The candidate
materials for the balloon include PA, ETFE, PET, PE/PA and FEP. There are two
flanges with a pipe on the top and bottom of the balloon, which will provide
the interface for filling and recycling. As the balloon is supported by the
acrylic layer, the stress of the balloon is very small, and we do not have to worry
about the balloon's shape. Therefore, the main study on the balloon is as
follows:

$\bullet$ the compatibility of the balloon material in the liquid;

$\bullet$ the radioactivity of the material;

$\bullet$ the transparency and aging properties of the material;

$\bullet$ the cleanliness of the balloon during its production and installation;

$\bullet$ sealing of the balloon.

The primary task is to understand the material properties. Present studies show that the yield strength of ETFE film is about 18~MPa. This film has
a good compatibility with all the liquids, good aging resistance, strength and impact
resistance. It is self-cleaning, but its light transmittance is poor.
For example, the transmittance of a 50 micron ETFE film is only 93\%. The test
result shows that PE/PA film is compatible with MO and the compatibility test with
liquid scintillator is ongoing. PET film has good compatibility and light
transmittance, but its flexibility is slightly worse. FEP film has good light
transmittance, but there is no wide application for this film so far.

In order to keep the cleanness and minimize the dust, the balloon must be made and
installed in a clean room. The cleanliness of the workshop in the factory
should hopefully be at the 1000 level. Constant temperature and humidity
are also needed during the production in the workshop. Providing clean gas to
remove radon is necessary during the balloon production. The balloon should be
carefully packaged, not only for safe transportation and ease of opening in the experimental hall, but also for avoiding contamination and damage.

Leak tests of the balloon should be done in the workshop and at the
experimental site. Assuming that the concentration of PPO penetrating into the
buffer liquid is less than 10~ppm in 10~years, the leakage of the balloon
should be about 5~$\times$~$10^{-2}$~cc/s at 3~mbar hydrostatic pressure. In the workshop, the
method of leak test can refer to that of the Borexino experiment based on SF6 gas
\cite{Borexino}.Sealed into another balloon, pumping the SF6 gas into
the inner balloon and getting it into every corner, we can measure and obtain the curve of concentration-time of
SF6 in the outer balloon and calculate the leakage. A vacuum leak test is also very useful.  After the transportation to the underground hall,
the balloon will be installed into the stainless-steel tank and the final leak test will be done onsite. The balloon
can be filled with purified gas mixed with a certain concentration of SF6. The balloon's
leakage is determined by testing the concentration of SF6 in the volume between tank and balloon. Further studies and  tests should be done
to get a detailed solution for the leak test.

\subsection{Acrylic Structure for the Balloon Support}
\begin{figure}[[!htbp]]
\centering
\includegraphics[width=8cm]{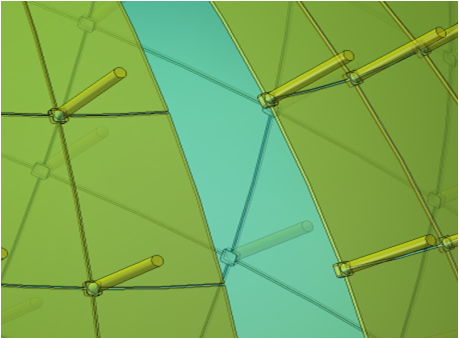}
\caption{The acrylic supporting structure for the balloon}
\label{fig:cd3-29}
\end{figure}

Acrylic here is designed as the supporting structure for the
balloon. It is a sphere with $\sim$35.4~m diameter, fixed on the steel tank by supporting legs, as shown in Fig.~\ref{fig:cd3-29}. The acrylic layer is made
up of many acrylic sheets and it is not necessary to seal it. In other words, it
will reduce the difficulty and working time for onsite installation.
The thickness of acrylic sheets in this option is 30~mm, which allows us to have a wide choice from commercial products.

\begin{figure}
\centering
\subfigure[the stress distribution]{
\label{fig:subfig:a} 
\includegraphics[width=7cm,height=7cm]{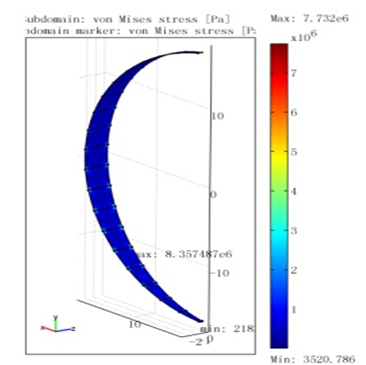}}
\subfigure[the deflection distribution]{
\label{fig:subfig:b} 
\includegraphics[width=7cm,height=7cm]{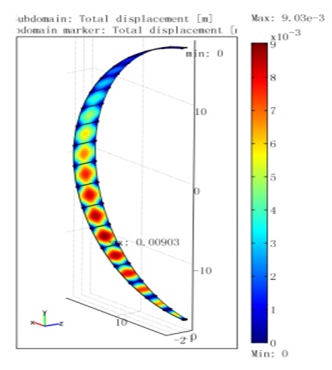}}
\caption{the stress and deformation of the acrylic support}
\label{fig:cd3-30} 
\end{figure}

The density of liquid inside and outside the balloon has a
small difference, expected to be 0.3-0.5\%. In stress analysis, we set a
0.5\% density difference as a starting condition, and the others conditions
being hydrostatic pressure towards the internal surface of the balloon and
external surface of the supporting structure. In this preliminary analysis, we used
tetrahedral finite elements to analyze a 9$^{\circ}$ partial spherical layer
with 25~mm thickness, and acrylic supporting legs were set as fixed ends. The
results showed that the stress of the main part is below 2~MPa and the total
displacement is about 9~mm, as shown in Fig.~\ref{fig:cd3-30}. Since the element density
affects the local stress on the supporting points, further analysis with more
detailed supporting structure will be done.

\subsection{Stainless-steel Spherical Tank}
\begin{figure}[[!htbp]]
\centering
\includegraphics[width=8cm]{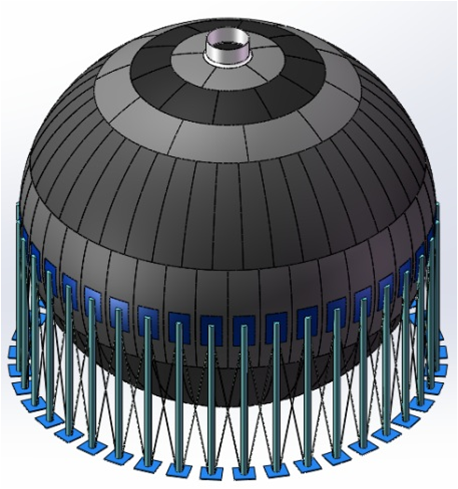}
\caption{A preliminary structure of the stainless-steel tank}
\label{fig:cd3-31}
\end{figure}

In the backup option, the outer structure is a spherical stainless-steel tank.
It will not only bear the weight of the liquid inside and support the
balloon and the acrylic structure, but also provide an installation
structure for the PMTs. The tank is not the same as a conventional pressure
tank; its design should consider the installation process and on-site
conditions as well as issues like cost and reliability. Based on the requirements
above, the preliminary structure of the tank is designed as shown in Fig.~\ref{fig:cd3-31}.

The tank is supported by landing legs, which are welded to the tank at its
equatorial belt. The tank has a chimney on the top, and its size should satisfy requirements such as the installation of a calibration
facility, the entrance for lifting the PMTs, ventilation and lighting of the
tank, arrangement of the cables, etc. The chimney reaches the liquid level, so it is
easy to seal. At the bottom of the tank, there is a hole to be used to
dispose the waste liquid from cleaning and as a manhole during the installation stage.

The tank is made of $\sim$170 welded steel plates, which are located in 9 belts of
the spherical tank. These plates can be produced in the factory and then be
assembled and welded after being transported to the site. Based on our
investigation, there are many professional companies who have experience in making large
spherical tanks. In addition, the largest spherical tank in China is up to 50~m
in diameter, while the largest stainless-steel spherical tank is only 17
meters in diameter. Although there is a certain degree of difficulty and
challenge for the $\phi$40~m stainless-steel tank in this option, there are many examples to follow.

\subsection{Installation Issues}
After the construction of the stainless-steel tank in the experimental hall, the PMTs and acrylic support sheets will be installed.
To reduce the time for installation, several PMTs will be pre-assembled as modules. The bottom of the PMT module is made of stainless steel plate and the PMTs can be mounted
on it. The ribs on the bottom plate are designed to increase its rigidity, as
each PMT suffers about $\sim$70~kg buoyancy. There has to be about 10~cm clearance
between the bottom plate of the PMT module and the inner wall of the
stainless-steel tank, which will provide space for cable arrangement.

During the installation, scaffolding will be installed in the stainless-steel
tank. After the tank is completed, the scaffolding can be used to wash the tank
from the top to the bottom. If the cable and PMT modules can be installed
separately, the pre-layout of the cables can be finished in the inner wall of
the tank before the PMT installation is started. A sling can also be hung along
the inner wall of the tank for lifting the PMT module. After the tank cleaning
and cable arrangement, the scaffolding can be removed. The pipe for filling the
liquid scintillator and LAB will be installed outside the tank, and
the pipe outlet, which is also the entrance into the steel tank, is at the
bottom of the tank. The PMT modules will be installed from bottom to top one by
one. The PMT modules can be moved by a manual chain which is hung by the sling.
The cable can be connected to each PMT after it is installed. The PMTs in each
module will be tested after the installation of each layer.

After the PMT modules are installed, acrylic rods will be installed at the back
plate of each module, and the acrylic sheet will be mounted on the rod  to support
the balloon. Once PMTs are tested, water can be filled into the tank. When
water reaches the top of that layer of PMTs, installation of the next layer
of PMTs can be started. After installing all PMTs and acrylic sheets, the water can be drained from the bottom of the tank. During this process, it
is necessary to guarantee that the air coming into the tank is clean. After all
water is removed, the balloon will be lifted into the tank from the top, and the
filling pipe can be connected to the bottom flange of the balloon through the
manhole. The top hole and bottom manhole will then be sealed to prevent
dust getting into the tank. The flange on the top of the balloon will be
sealed, and the gas for the leak test will go through the pipeline for filling
liquid scintillator into the balloon to replace the air in it. Once the balloon leak test completed,
the manhole at the bottom will be sealed by the flange.
After making sure that every part works well, liquid
scintillator will be filled into the balloon. During the process, the filling speed of LS, LAB and water outside the tank should be
controlled strictly to keep the level of different liquids the same.

\subsection{Conclusions}
Since the balloon can be made in the factory and then delivered to underground lab for
the final installation and leak check, the time needed onsite is relatively
short. The cost of fabricating a balloon is much lower compared to the acrylic
sphere so more than one balloon can be made. If one balloon has
trouble in the leak checking or any other damage, it can be replaced in a short
time. On the other hand, the balloon is a soft structure and not easily
damaged under a shock load (such as seismic load). The use of the acrylic layer
as support for the balloon reduces the stress and greatly improves the safety.
Since the acrylic sheets in this option do not need to be sealed, the
difficulty and time needed for on-site construction is reduced.  Because the
detector construction in this option involves the tank welding onsite,
installation of the acrylic sheets and the balloon in the tank, cleaning and
leak test of the balloon, the procedure will be more complex. Therefore this option is
considered as a backup.

\section{PMT assembly}
\subsection{PMT testing and validation}
The JUNO central detector will use 20~inch PMTs with highest possible quantum efficiency and large photo-cathode.
The main specifications of the PMT include:

$\bullet$ Response to single photoelectrons

$\bullet$ Quantum efficiency of photocathode

$\bullet$ Gain-HV function and gain nonlinearity

$\bullet$ Transit time spread, ratio of after pulse and time structure

$\bullet$ Dark noise rate

$\bullet$ Uniformity of detection efficiency

To precisely measure those parameters, a test bench integrated together with
software and hardware needs to be setup. Based on the requirements, a dark room
with shielding to earth's magnetic field is necessary. Efficient testing can be
implemented by using a light source with multiple optical fiber coupling.
Testing of PMTs consists of several steps: single sample test, test of a group of samples and batch test.
Details will be finalized later.

\subsection{High Voltage System of PMTs}
A high voltage system is needed to provide stable power to the PMTs. This
includes high voltage power supply, cable and water-proof structure. The three
options below are under consideration:

1) One to one option:  one HV channel supplies one PMT, and a divider is needed
to produce different voltages for each electrode. This option can have a control
to each PMT HV independently but a huge number of cables are needed.

2) Multi-core cable option: a group of HV channels are carried by a single
cable with multiple cores, via an exchange box close to the PMTs, the HV supplies a bunch of PMTs. In this option: the number of
cables are decreased, while each PMT is still controlled independently. However, a dry box is needed for HV exchange and distribution, and a
connector with water-proof design is needed.

3) Front-end HV module option: for this option, the low voltage is firstly
delivered to the front end near the detector, then the low voltage is
transformed to high voltage which is supplied to the PMTs. The advantage of this
option is to save cost and significantly reduce the number of cables.

The voltage of the PMTs is provided by a Cockroft Walton capacitor diode, which
converts the low voltage carried by a flat cable to high voltage and finally
supplies the whole power grid. Compared to the traditional HV power based on
dividers, this method of voltage multiplier has some clear advantages:

1) the expensive high voltage cable and connector is replaced by a cheap flex cable. This is an efficient way to save cost, reduce the number of cable and improve the overall quality.

2) the high efficiency of the voltage multiplier can reduce heat power
consumption significantly. In general this kind of device will consume power
less than 50~mW without light load, while under maximum light load, the power
consumption is less than 200~mW. As a comparison, the option using dividers
will consume 3.6~W in the form of heat loss.

3) the high voltage is very stable due to a deep negative feedback designed in the module, and complete load characters are provided.

4) the cost of the high voltage power supply with the multiplier itself is low,
about 3-4 times lower than the other options.

The main parameters of the JUNO 20'' PMT high voltage power supply are shown in Table~ref{table8}, Table~ref{table9} and Table~ref{table10} below:

\begin{table}[!htbp]
\centering
\caption{USB bus adapter\label{table8}}
\newcommand{\tabincell}[2]{\begin{tabular}{@{}#1@{}}#2\end{tabular}}
\begin{tabular}{p{7cm}<{\centering}|p{6cm}<{\centering}}
\hline
	Maximum number of channel&		127  \\
\hline
\tabincell{c} {Power supply} 	&	 computer USB port	\\
\hline
\tabincell{c} {BUS  voltage/ V} 	&	 LV - 5V,BV - 24	\\
\hline
\tabincell{c} {Interface} 	&	 RS-485	\\
\hline
\tabincell{c} {Communication line to computer} 	&	USB-2.0	\\
\hline
\tabincell{c} {Working temperature/degree} 	&	 (- 10) - (+40)	\\
\hline
\tabincell{c} {Working humidity/ \%} 	&	 0 - 80	\\
\hline
\tabincell{c} {size/ mm x mm x mm} 	&	70x50x22	\\
\hline
\tabincell{c} {Weight/ kg} 	&	 0.15	\\
\hline
\end{tabular}
\end{table}

\begin{table}[!htbp]
\centering
\caption{20'' PMT HV module\label{table9}}
\newcommand{\tabincell}[2]{\begin{tabular}{@{}#1@{}}#2\end{tabular}}
\begin{tabular}{p{7cm}<{\centering}|p{6cm}<{\centering}}
\hline
	PMT connection, grounding&		cathode  \\
\hline
\tabincell{c} {Range of HV for anode / V} 	&	1500-2500*	\\
\hline
\tabincell{c} {Operating voltage/ V} 	&	+2300*	\\
\hline
\tabincell{c} {Precision of anode HV/ V} 	&	~ 0.25	\\
\hline
\tabincell{c} {Error of HV output} 	&	 3\%	\\
\hline
\tabincell{c} {Stability of HV / \%} 	&	 0.05\%	\\
\hline
\tabincell{c} {Temperature coefficient/ ppm/degree} 	&	100	\\
\hline
\end{tabular}
\end{table}

\begin{table}[!htbp]
\centering
\caption{HV value of PMT dynode\label{table10}}
\newcommand{\tabincell}[2]{\begin{tabular}{@{}#1@{}}#2\end{tabular}}
\begin{tabular}{p{8cm}<{\centering}|p{5cm}<{\centering}}
\hline
	focus / V&		+300  \\
\hline
\tabincell{c} {D1/ V} 	&	+400	\\
\hline
\tabincell{c} {D2/ V} 	&	+1200	\\
\hline
\tabincell{c} {D3/ V} 	&	+1300	\\
\hline
\tabincell{c} {D4/ V} 	&	+2100\\
\hline
\tabincell{c} {Anode/ V} 	&	 +2300	\\
\hline
\tabincell{c} {Up limit of anode current/ mA} 	&	100	\\
\hline
\tabincell{c} {Cross talk/ mV} 	&	20	\\
\hline
\tabincell{c} {Nominal voltage/ V} 	&	+5,+24	\\
\hline
\tabincell{c} {Power consumption of single module /W} 	&	0.1	\\
\hline
\tabincell{c} {Communication protocol } 	&	RS-485	\\
\hline
\end{tabular}
\end{table}

\subsection{PMT related Structure Design}
PMT related structures include PMT voltage divider, implosion protection
structure, potting structure and mechanical supporting structure. JUNO will use
20'' PMTs, each one weighting about 10kg and the buoyancy in water of 70~kg.
The PMTs are mounted on the stainless-steel truss.

A preliminary design and test of the MCP-PMT divider has been carried out. The
method of positive HV powering has been adopted, to reduce the number of cables
and connectors, and the signal is readout by a splitter.

\begin{figure}[!htbp]
\centering
\includegraphics{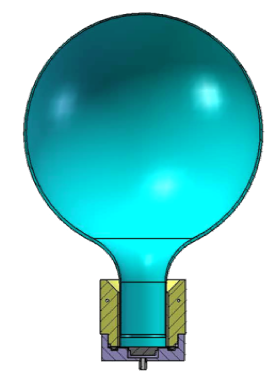}
\caption{conceptual view of PMT potting structure}
\label{fig:cd3-32}
\end{figure}

A potting structure is needed since the PMTs are working in water, and a
conceptual view is shown as Fig.~\ref{fig:cd3-32}. The potting will use a plastic cover and
the water-proof adhesive to seal the PMT base, electronics and cable from
contact with water. The waterproof design needs to consider long-term
compatibility with high purity water and 4~atm hydrostatic pressure. After
finishing the design and related tests, we will set up a batch test system to
test each of the PMTs before installation.

The following sections will discuss the implosion protection structure of the
PMTs. Currently there are three ideas under consideration to protect the PMTs
from implosion in 40~m depth of water:

A. the PMT is covered by an acrylic shell for the upper half sphere and
the lower part by another shell made from FRP, PE or ABS.
A few small holes are opened on the shell to allow air flow and liquid
circulation. When the PMT collapses, the amount of ingoing water is minimized
by those holes, hence the strength of the induced shock wave is suppressed;

B. an air chamber is placed around the lower half of the PMT. When
collapse happens, the air chamber will quickly expand due to a sudden drop of
the local water pressure, hence the amount of ingoing water is also reduced and the shock wave is weakened;

C. a thin layer of film is attached to the PMT surface, this film will prolong the PMT collapse time, and hence the intensity of inrushing water.
The film is transparent so no light will be blocked.

(1) Study of PMT implosion mechanism

\begin{figure}[!htbp]
\centering
\subfigure[]{
\label{fig:subfig:a} 
\includegraphics[width=6.3cm,height=7cm]{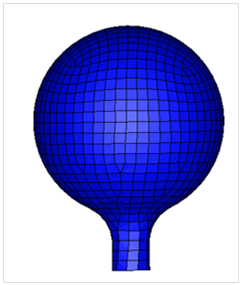}}
\subfigure[]{
\label{fig:subfig:b} 
\includegraphics[width=8.4cm,height=7cm]{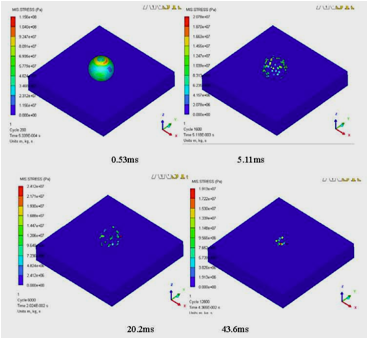}}
\caption{Simulation of PMT implosion}
\label{fig:cd3-33} 
\end{figure}

The shell of the PMT is made of fragile glass, so an implosion could happen
when damaged glass shell is immersed in water. As was experienced in Super-K, implosion of one PMT could lead to a
cascaded implosion and eventually all the PMTs would be destroyed. A simulation
of the PMT implosion is shown in Fig.~\ref{fig:cd3-33}. The PMT glass shell crashes in 5~ms
due to an outer pressure of 5 atm, the water then rushes inward suddenly. After
about 40~ms, due to the collision of inrushing water, the implosion happens,
and a shock wave is produced which propagates outwards.

\begin{figure}[!htbp]
\begin{minipage}[t]{0.5\textwidth}
\centering
\includegraphics[width=7.5cm,height=6cm]{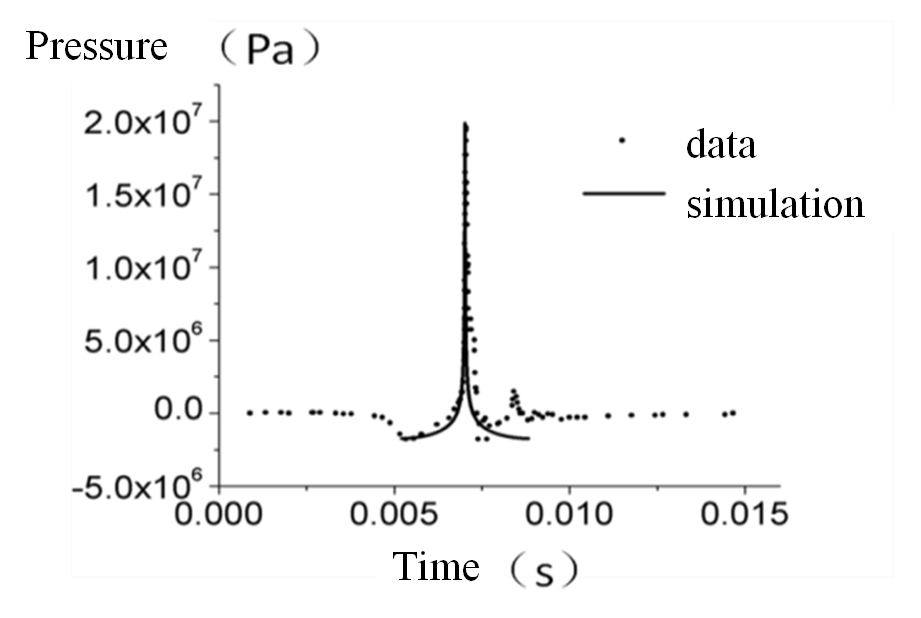}
\caption{Peak pressure of the shock wave}
\label{fig:cd3-34}
\end{minipage}
\begin{minipage}[t]{0.5\textwidth}\centering\includegraphics[width=7.5cm,height=6cm]{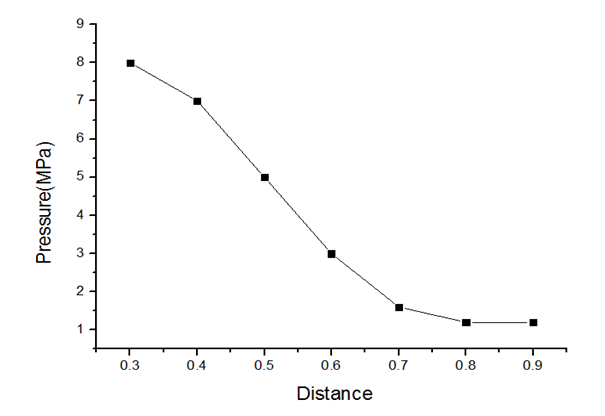}
\caption{strength of the shock wave as a function of distance}
\label{fig:cd3-35}
\end{minipage}
\end{figure}

\begin{figure}[!htbp]
\begin{minipage}[t]{0.5\textwidth}
\centering
\includegraphics[width=7.5cm,height=6cm]{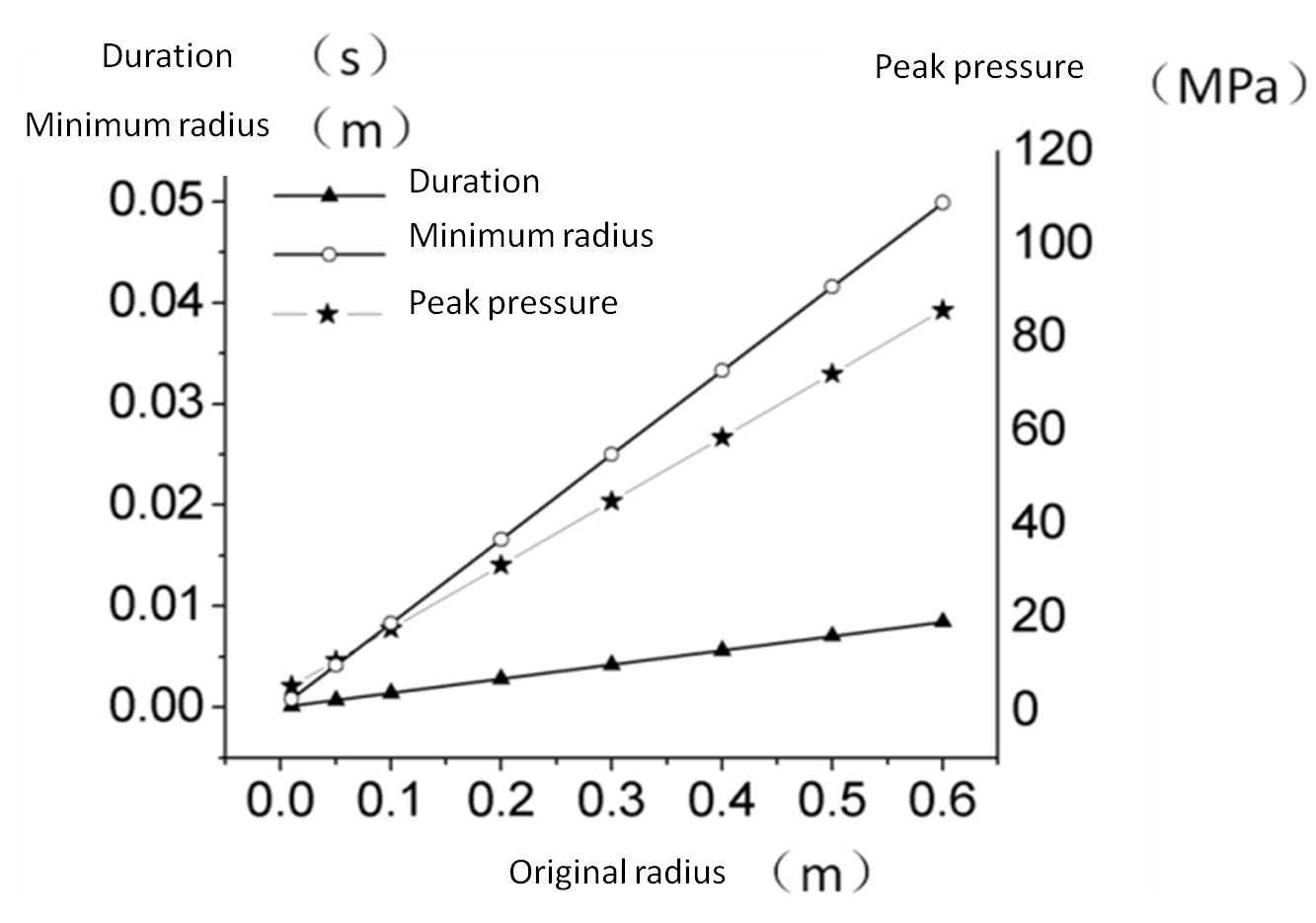}
\caption{strength of the shock wave as a function of PMT volume}
\label{fig:cd3-36}
\end{minipage}
\begin{minipage}[t]{0.5\textwidth}\centering\includegraphics[width=7.5cm,height=6cm]{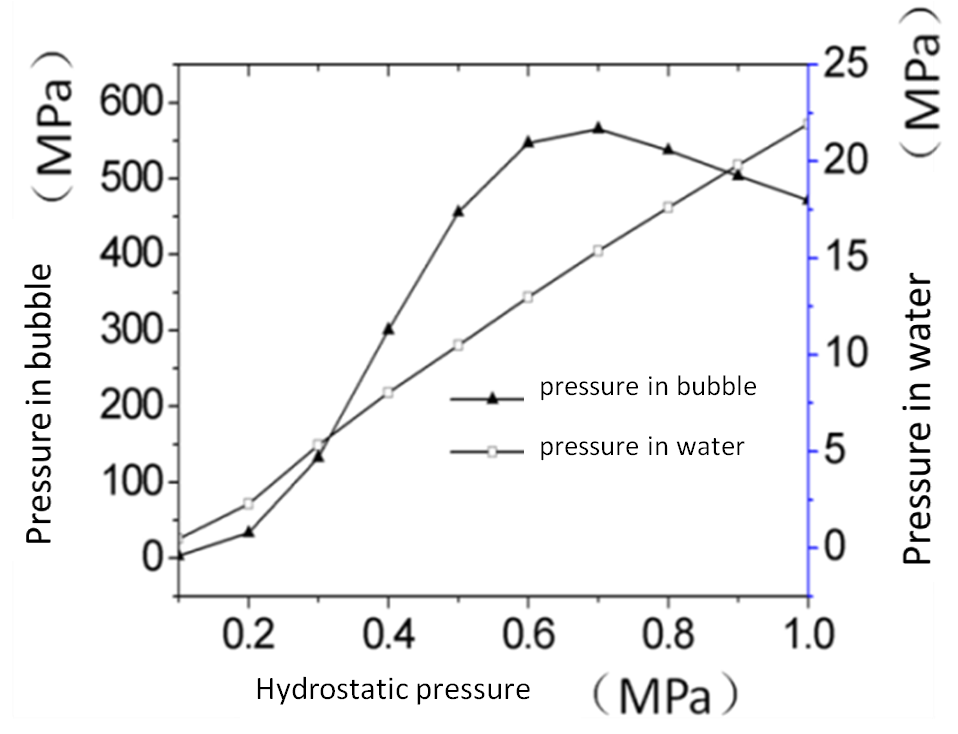}
\caption{strength of the shock wave as a function of pressure}
\label{fig:cd3-37}
\end{minipage}
\end{figure}

A preliminary study of the implosion behaviors such as shock wave strength as a
function of time, distance, PMT volume and water pressure has been
performed. As shown in Fig.~\ref{fig:cd3-34}, the strength from simulation is consistent
with the real test. The strength follows a 1/r relation to distance as seen from
Fig.~\ref{fig:cd3-35}, and is proportional to PMT volume and water pressure as shown in
Fig.~\ref{fig:cd3-36} and Fig.~\ref{fig:cd3-37}, respectively.

(2) PMT implosion test and simulation

\begin{figure}[!htbp]
\centering
\subfigure[Schematic view of the implosion test]{
\label{fig:subfig:a} 
\includegraphics[width=7.5cm,height=6cm]{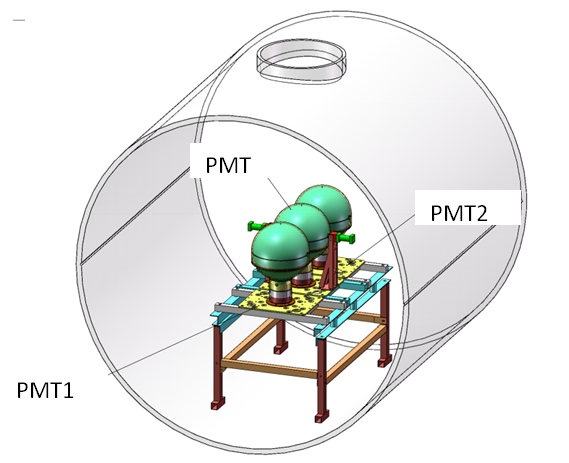}}
\subfigure[the pressure tank for implosion test]{
\label{fig:subfig:b} 
\includegraphics[width=7.5cm,height=6cm]{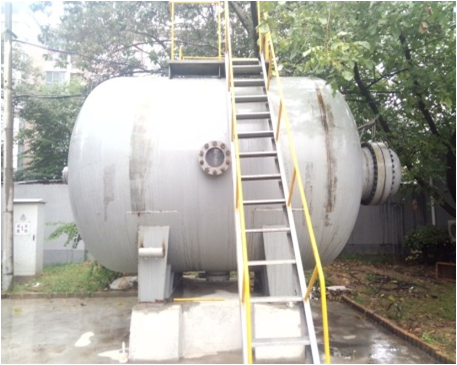}}
\caption{the setup for implosion test}
\label{fig:cd3-38} 
\end{figure}

The purpose of the PMT implosion test is to measure the shock wave parameters,
including production, propagation and relation to space and time. In addition,
the test can verify the function of the acrylic cover. Fig.~\ref{fig:cd3-38} shows the
schematic view of the implosion test and the pressurized tank.
The tests are currently underway.

\begin{figure}[!htbp]
\centering
\subfigure[the 2D model for simulation]{
\label{fig:subfig:a} 
\includegraphics[width=7cm,height=4cm]{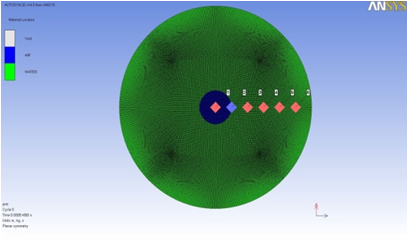}}
\subfigure[the pressure-time curve for point 1]{
\label{fig:subfig:b} 
\includegraphics[width=7cm,height=4cm]{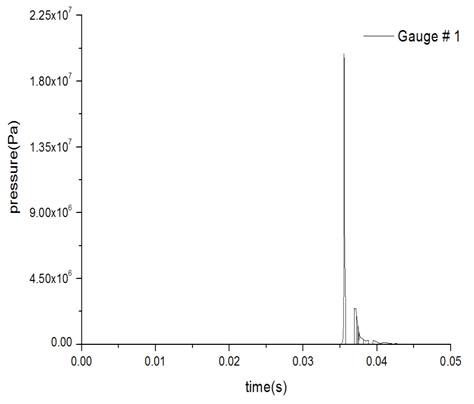}}
\subfigure[Pressure-time curve of different points]{
\label{fig:subfig:c} 
\includegraphics[width=7cm,height=4cm]{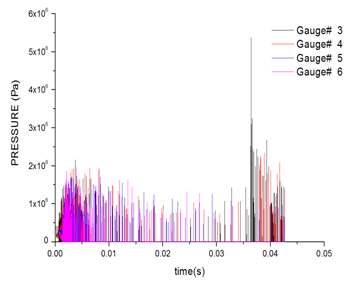}}
\subfigure[Peak pressure as a function of distance]{
\label{fig:subfig:d} 
\includegraphics[width=7cm,height=4cm]{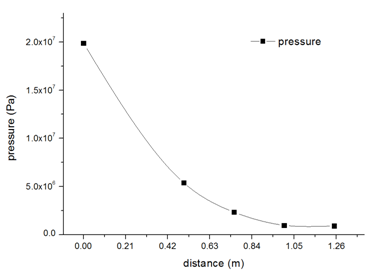}}
\caption{Simulation of the PMT implosion}
\label{fig:cd3-39} 
\end{figure}

To get the correct implosion model, simulation have been done following the
real boundary conditions in the test. Fig.~\ref{fig:cd3-39}(a) shows a 2D simulation of single PMT without protection enclosure, in which the diameter of the water
tank is 1.5~m, the diameter of the air bubble (PMT volume) is 0.5~m, the
hydrostatic pressure is 4 atms, and the gas pressure in PMT is about $10^{-5}$ Pa. The
boundary is set as a rigid border. There are 6 points in simulation which are
uniformly distributed from the center to the border, in which point 2 is
floating and movable to measure the size variation of the air bubble, and the
other 5 points are fixed to measure the dynamic pressure induced by the shock
wave. Fig.~\ref{fig:cd3-39}(b) shows that the peak pressure appears at t=35.2~ms. From
Fig.~\ref{fig:cd3-39}(c), multiple shock waves are emitted during the process of
compression and rebound of the air bubble, and the waves frequently happen at
beginning or the end of the compression. The air bubble is finally compressed
to a minimum size with the pressure reaching a maximum value of 20~MPa
inside of the bubble, then the strongest shock wave is initiated. Fig.~\ref{fig:cd3-39}(d)
shows the pressure for different points, where the pressure drops
along with the distance. At the border of the PMT (d=25cm), the strength of the
shock wave is about 11~MPa.

The data from the real test will validate the simulation described above. Tests of multiple PMT implosion, and simulation of the effect of the
acrylic cover are also in progress.

\section{Prototype}
\subsection{Motivations of the Prototype Detector}

Following the progress and schedule of the JUNO experiment, considering the requirements of each sub-system, a prototype detector was proposed to test key technical issues:

1) Test and study of the PMT candidates:

The JUNO experiment is designed to use large area, high quantum efficiency PMTs to realize the highest photo-cathode coverage. Candidate PMTs include Hamamatsu, Hainan Zhan Chuang (HZC), and a newly developed MCP-PMT. It is necessary to compare their performance in a real scintillator detector. From the prototype, we expect to extract the effective PMT parameters and compare them.

2) Test and study of the liquid scintillator:

Liquid scintillator is another key component of the JUNO detector, a lot of experiences show that the final LS parameters in a detector, such as light yield, attenuation length, stability (or aging) and background level, are different from what were measured in the lab. With the prototype, many of those parameters can be extracted and optimized.

3) Test and study of electronics:

The planned PMT waveform readout electronics will be very flexible for different physics requirements. The electronics system should be tested in a real LS detector.

4) Waveform data analysis algorithms and detector performance study:

With the prototype, algorithms of PMT waveform readout scheme will be developed and detector performances will be understood.

\subsection{Prototype Design}
\begin{figure}[!htbp]
\centering
\subfigure[]{
\label{fig:subfig:a} 
\includegraphics[width=6cm,height=5cm]{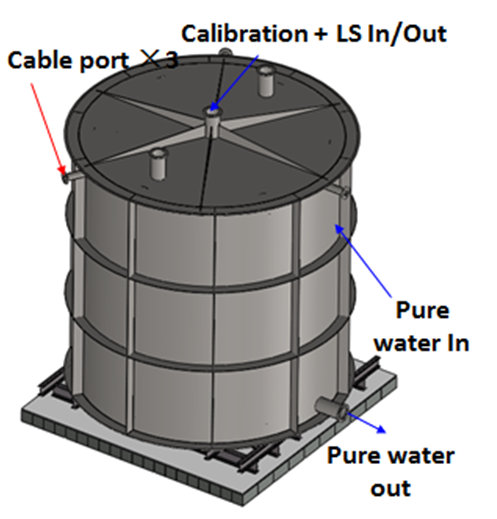}}
\subfigure[]{
\label{fig:subfig:b} 
\includegraphics[width=5cm,height=5cm]{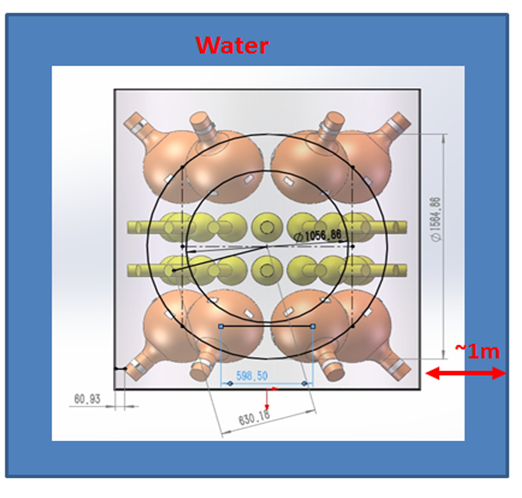}}
\caption{the prototype design}
\label{fig:cd3-40} 
\end{figure}

The prototype detector is shown in Fig.~\ref{fig:cd3-40}, which re-uses the Daya Bay prototype as the main container. An acrylic sphere locates at the center of the stainless-steel tank (SST) as a LS vessel, and is viewed by 51 PMTs dipped in the pure water. The expected photo-cathode coverage is ~55\%. The water tankand the PP/Lead layer is designed to have 1 m.w.e. shielding,  aiming to reduce the radioactivity coming from the outside of SST.
The expected trigger rate is less than 100Hz@0.7MeV, and the energy resolution is ~4\%@1MeV.

\begin{figure}[[!htbp]]
\centering
\includegraphics[width=6cm,height=6cm]{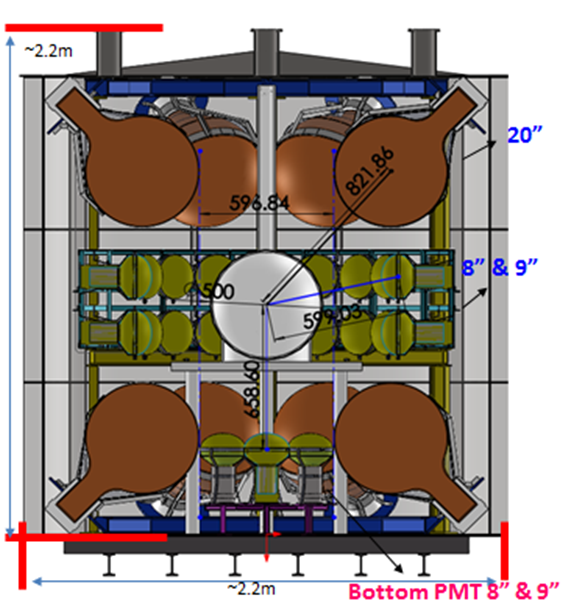}
\caption{The side view of the prototype detector}
\label{fig:cd3-41}
\end{figure}

As shown in Fig.~\ref{fig:cd3-41}, the diameter of the acrylic sphere is 50 cm with a thickness of ~1 cm, and a tube with a diameter of 5 cm and a length of 70 cm is located at the top of the acrylic sphere for filling and calibration. The 51 high quantum efficiency PMTs, in the diameter of 8'', 9'' and  20'', are uniformly arrayed in 4 layers facing to the center of the acrylic sphere. As follows is the detailed arrangement of the PMTs:

$\bullet$ Top layer: which includes 4 MCP-PMTs of 20'' and 2 Hamamatsu PMTs of 20'';

$\bullet$ Middle layers: which are divided into two layers and each layer including 8 MCP-PMTs of 8'', 4 HZC-PMTs of 9'' and 4 Hamamatsu PMTs of 8'';

$\bullet$ Bottom layers: which are divided into two rings, the inner ring includes  3 MCP-PMTs of 8'', 2 HZC-PMTs of 9''and 2 Hamamatsu PMTs of 8'', and the outer ring has the same arrangement as the top layer.

\begin{figure}[!htbp]
\centering
\subfigure[schematics of the shielding structure]{
\label{fig:subfig:a} 
\includegraphics[width=5.5cm,height=4cm]{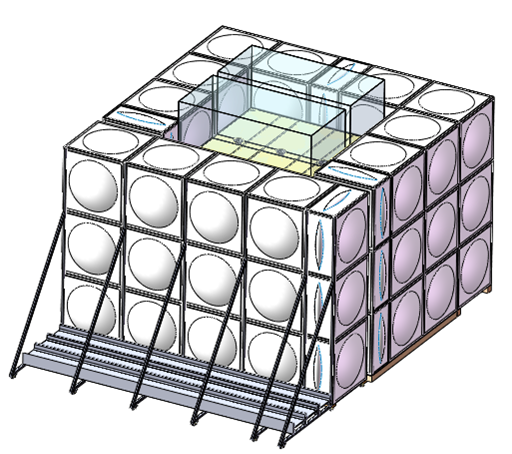}}
\subfigure[one side of the water tank opened]{
\label{fig:subfig:b} 
\includegraphics[width=5.5cm,height=4cm]{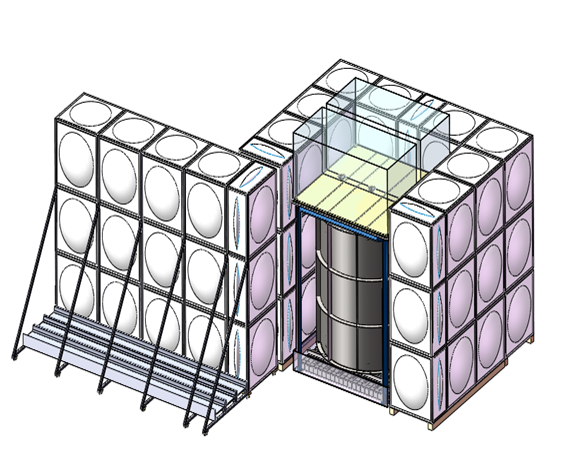}}
\subfigure[bottom and top shielding design with lead and pp plate]{
\label{fig:subfig:c} 
\includegraphics[width=5.5cm,height=4cm]{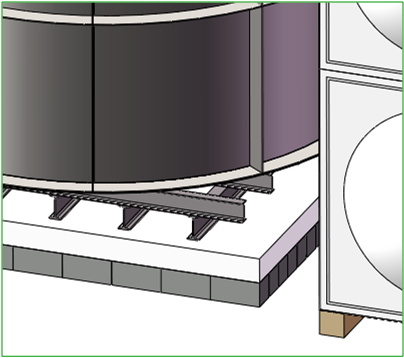}}
\caption{the shielding system}
\label{fig:cd3-42} 
\end{figure}

Fig.~\ref{fig:cd3-42} is the shielding system. The bottom and top of this system are covered by 10 cm thick lead plus 10 cm thick pp plate, while each of the other 4 sides is shielded by a customized water tank in 1m x 5m x 3m dimension constructed by standard stainless-steel elements which are widely used by water and conditioning system. One of the 4 sides is movable to allow the internal detector installation.

\subsection{Expected Performance}

\begin{figure}[[!htbp]]
\centering
\includegraphics[width=8cm]{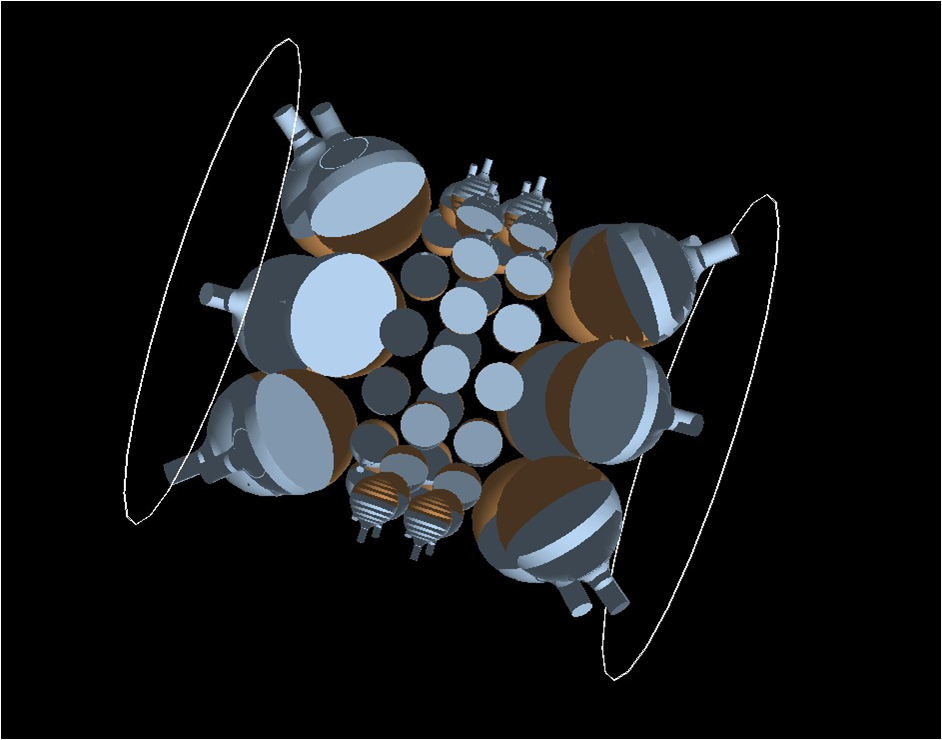}
\caption{Geant4 simulation geometry}
\label{fig:cd3-43}
\end{figure}

With preliminary Geant4 simulation (simulation geometry is shown in Fig.~\ref{fig:cd3-43}), we confirm that the detector will have a good measurement of electron and alpha for LS study. Gammas have a large energy leakage due to the limited detector dimension, we will need a spectral fitting to extract more parameters. Measurements of Muon and calibration will provide more opportunities for detailed study.
According to the prototype design, we expect to measure PMTs' parameters, liquid scintillator, electronics and background, to achieve the goals.

The prototype is now under preparation and will be operational soon.

\section{Construction of the central detector}
\subsection{Assembly of the Stainless-steel Truss}
The stainless-steel truss used in the central detector has a bolted connection,
which makes the on-site assembly easy and safe without any welding. The design
and the construction of the truss should follow the technical standards, design
specifications and procedures. In China, the space truss technique is very popular for building construction.
Many companies have capability and rich experience in design and production of steel trusses.

There are two main stages for the construction of the stainless-steel truss.
The first stage is the off-site preparation in the factory, and the second stage is assembly
onsite. All the truss members, bolts and joints will be finished for the
machining and production in factory before transportation, and the machining
precision and quality should be ensured at this stage. When all parts are
transported to the experimental hall, some important tests need to be done
onsite to make sure that all conditions meet the requirements for construction
before truss assembly. Theses tests include: checking the foundation of the
truss area; double checking the parts of the truss; checking the height mark,
axial and gradient of the embedded part, and so on. Since the complexity of
truss assembly depends on the accuracy and roundness of the first three
circles, the beginning preparation of axis positioning and height level is more
important for the whole assembly of the truss. To control and check the
coordinate position of each truss joint, some reference and checking points
should be installed at different areas of the hall. When all conditions for the construction
meets the requirements of design and standards, the truss assembly can be
started following the detailed construction procedure. Theodolites or other
survey instruments will be used during the truss assembly to control the
error.

For the central detector, the stainless-steel truss and acrylic sphere are not
independent and need overall consideration of the construction sequence and
procedure. It should be better to make the truss and acrylic sphere layer by
layer alternately. The detailed procedure needs further discussion with the
factory. In general, the whole procedure needs to ensure the safety and
reliability of the sphere and truss.

\subsection{Construction of the Acrylic Sphere}
As Fig.~\ref{fig:cd3-7} shows, the acrylic sphere is made up of many acrylic sheets which
are bonded together to form the sphere.  Satisfying the production capacity and
transportation limit, the size of each acrylic sheet should be as big as
possible to reduce the total bonding length onsite. For the preliminary
consideration and design of the sphere, there are a total of $\sim$170 acrylic sheets
and divided into 17 layers. Each sheet is less than 3~m $\times$ 8~m in size. The
total bonding length of the sphere is about 1.8~km.

Production and machining of the acrylic sheets will be finished in the factory
before construction onsite. In the factory, the first step is to produce a flat
sheet of acrylic with 120~mm thickness. The next step is to thermoform the flat
acrylic into a spherical sheet with 35.4~m inner diameter in a concave
spherical mold, and these thermo-formed sheets should be stored carefully to
prevent damage to the acrylic surface. After that are the steps of machining,
milling, polishing and annealing. Some measurements should be done for each
step to control the quality in the factory. One such step is the thickness
measurement of each flat sheet used as the thermoforming blank. The second
measurement is for the thickness of the sheet after thermoforming. In general,
the acrylic blank increases in thickness around its border and decreases in
thickness at the center after the thermoforming operation in the female mold.
The third measurement is the acrylic thickness after annealing. The curvature
of the acrylic blank should also be measured to ensure the assembly requirement
\cite{Handbook}. In the factory, the fixtures for positioning and assembly
should be machined.

After all the part production and fixture preparation is finished by the
manufacturer, they will be delivered to the underground to do the construction
of the acrylic sphere onsite. The acrylic sphere can be constructed with the
help of the truss. The acrylic sectors can be supported by the truss joints
during assembling, positioning and bonding layer by layer. The truss can also
be used as the scaffolding or platform for sphere construction. The acrylic
sphere will be built from the bottom to the top, layer by layer. The quality
control and check of dimension, positioning, roundness, bonding status and
annealing need to be ensured for each layer throughout the period of
construction. Inspections of the finished spherical sphere consist of careful
visual checks, dimensional measurements, and stress checking.

For the construction of the detector, two cranes will be needed in the hall.
One is to carry and lift the components of the detector, and the other is a
lift to carry people to access the working area in the pool. The working
platforms should be designed and installed inside and outside the acrylic
sphere for the processes of positioning, assembly, sanding and polishing.
Safety nets should be installed surrounding the sphere during construction to
prevent any dropping of objects which will damage the detector or people.
Special requirements may be needed for the ventilation system of the assembly
cavern. There are special rules and regulations governing working underground
to which the project must conform.

\subsection{PMT Installation}
When the acrylic sphere and stainless-steel truss are finished, about 16000
PMTs need to be installed on the truss facing inwards to receive signal from
the LS. These PMTs should be arranged as close to each other as possible to
provide more than 75\% optical coverage. The PMT support structure must provide
a stable and accurate platform for the mounting and positioning of the PMT
assemblies to view the active detector regions. The PMT support structure
should also provide a mechanism to protect adjacent PMTs from the risks of
cascaded implosion and permit the replacement of the PMT assemblies. To reduce
the installation time, PMTs will be installed by PMT module. Several PMTs are
assembled to be one module in the surface assembly building. The PMT modules
will then be delivered underground for on-site installation. At present, two
options for PMT installation are considered.

\begin{figure}[!htbp]
\begin{minipage}[t]{0.5\textwidth}
\centering
\includegraphics[width=8cm]{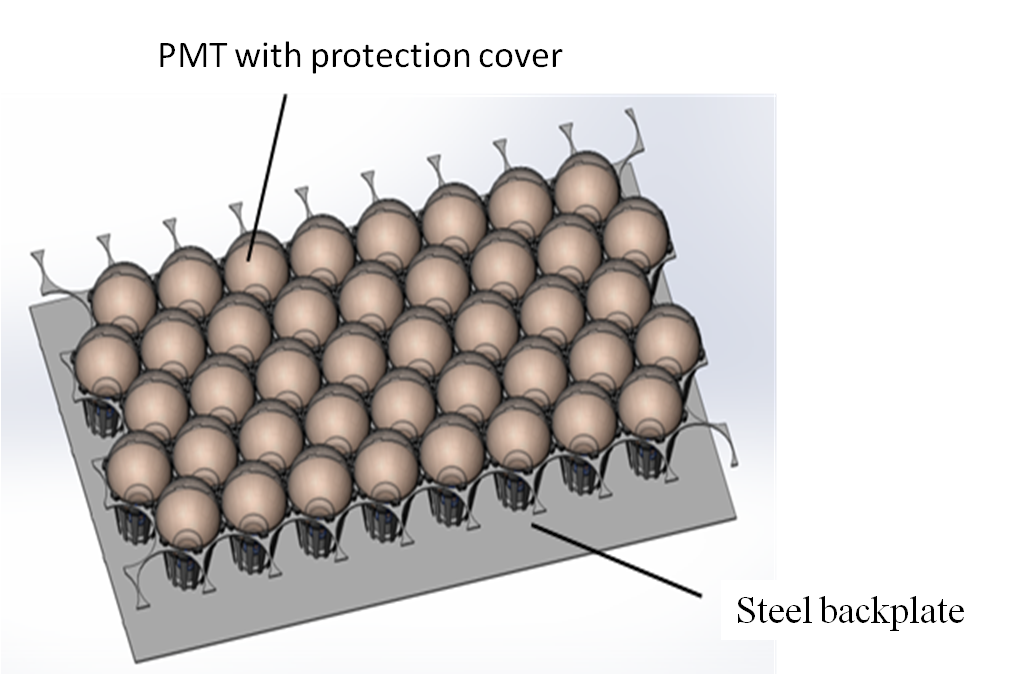}
\caption{PMT module with the stainless-steel backplate}
\label{fig:cd3-44}
\end{minipage}
\begin{minipage}[t]{0.5\textwidth}\centering\includegraphics[width=8cm]{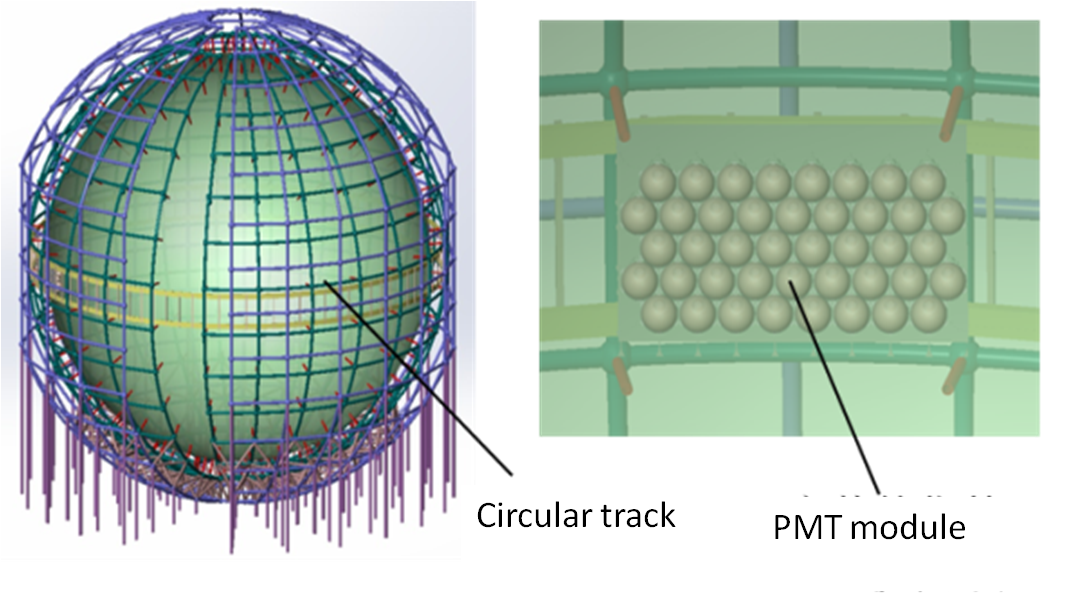}
\caption{Installation of the PMT module}
\label{fig:cd3-45}
\end{minipage}
\end{figure}

One option is to use stainless steel as the support panel of the PMT module,
and each PMT with its protecting cover will be mounted using a clamp on the
panel as Fig.~\ref{fig:cd3-44} shows. Some slots on the truss should be left for the
lifting and installation of the PMT module. The circular rails along the
latitude will be installed on the joint of the inner layer of the truss. There
are pulleys at the back of the stainless-steel panel to allow the PMT element
to slide to the expected location and finish the mounting. Fig.~\ref{fig:cd3-45} shows the
sketch of this option. The design of the module size and shape should consider
the location and number of the joints and supporting rods between acrylic
sphere and truss to get the maximum optical coverage for the central detector.

\begin{figure}[[!htbp]]
\centering
\includegraphics[width=10cm]{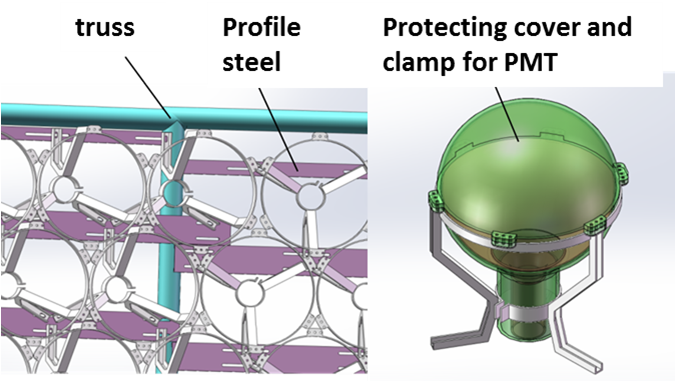}
\caption{Using profile steel for PMT installation}
\label{fig:cd3-46}
\end{figure}

The other option is to use profile stainless steel as the supporting structure.
PMTs will be mounted on the profile steel one by one. This profile steel will
be pre-installed on the inner layer of the truss. Fig.~\ref{fig:cd3-46} shows the sketch of
this option. This option allows many work groups to do the installation at
different working areas simultaneously.

\subsection{Filling system for the Liquid Scintillator}
Filling will start when the construction of central detector is finished. The
filling liquid involves LS and water for the main option and LAB needs to be
considered for the backup option. The veto system will be responsible for water
filling and the central detector system is responsible for LS and LAB filling.
The two systems should coordinate their filling speed to balance the pressure.
During filling, cleanness, background, temperature, liquid storage tank, pipe
design and process monitoring should be considered and controlled.

1) Controlling the Cleanness and low background

During filling, the detector and LS need to be kept clean, especially to
prevent the radioactive background and radon gas from pollution. The central
detector will be cleaned using purified water before filling. For the acrylic
sphere plus stainless-steel option, a special cleaning robot will be designed
and used for cleaning the acrylic sphere and the other parts such as the steel
truss. The PMTs will be washed first. The acrylic sphere can be washed with LS
at the end. For the backup option, the inner wall of the stainless-steel tank
and the outer surface of balloon will be water flushed and  the inner surface
of the balloon can be cleaned by filling and pumping water repeatedly. The
filling pipes and pump, which will come into contact with LS, need special care
and cleaning, and they should be compatible with LS. Before filling, the whole
filling system will be cleaned with LS several times and the used LS will be
disposed of. Spaces where the LS is exposed to air, such as inside the acrylic
or balloon and the LS storage tank, will be filled with nitrogen to prevent
radon pollution and contact with oxygen.

2) Pure water exchange or nitrogen exchange for filling

Pure water exchange means we need to firstly fill the acrylic sphere or balloon
with water and then fill LS into the vessel while draining the water out.
Because of its lower density, the LS will be float at the top of the water. This
process filters the LS and is helpful to remove the background. The other
method is exchanging with pure nitrogen, which means filling the acrylic sphere
or balloon with nitrogen first and then putting pure LS into the vessel to
replace the nitrogen.

3) Controlling the liquid temperature

The LS temperature will be high if it is produced and stored on the ground.
Since we hope the liquid for filling can be kept at 20$^{\circ}$C, the liquid
needs to be cooled before filling the detector. The place for cooling should be
selected on ground or underground according to the general deployment of the
civil construction, and the cooling method can be water cooling or
air-conditioning.

4) The preliminary plan for the LS pipe, storage tank, overflow tank and LS cycling system.

\begin{figure}[!htb]
\centering
\includegraphics[width=14cm]{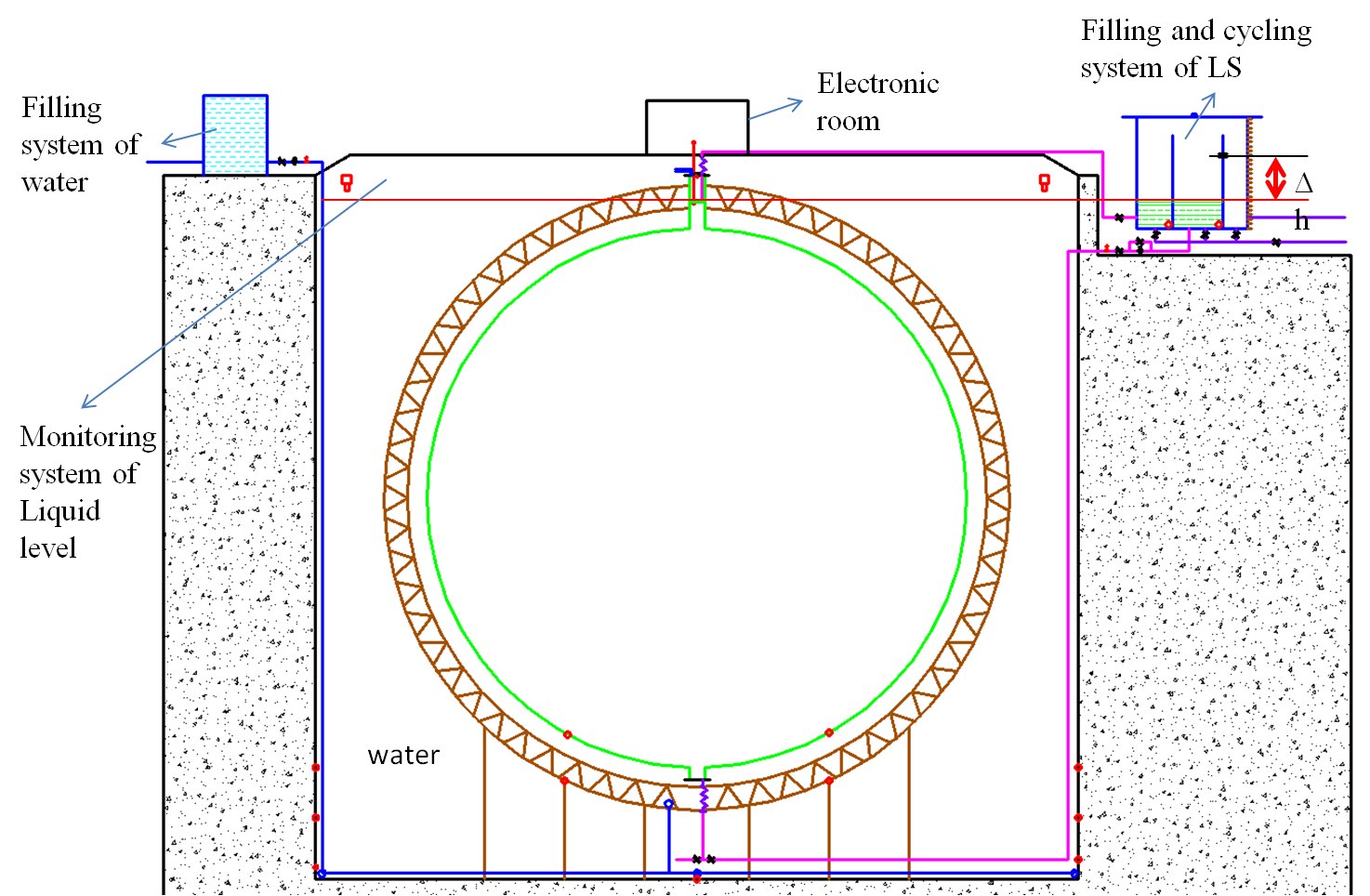}
\caption{Sketch of the LS filling system}
\label{fig:cd3-47}
\end{figure}

As shown in Fig.~\ref{fig:cd3-47}, the filling port of the central detector is designed at
the bottom of the sphere. The filling pipe comes from the LS storage tank which
sits in the storage room, and goes down along the water pool wall to connect to
the filling port. When the LS filling is finished, the pipes between the
storage tank and central detector make them communicating vessels and then the
storage tank can be used as the overflow tank. If the LS needs recycling during
running, we will remove the LS from top of the detector and transport it to the
LS recycling equipment, and then pump it back into LS storage tank. Finally the
LS will be filled back into the central detector through filling pipes from the storage tank.

5) Filling process control

Filling process control involves the speed control, monitoring of liquid
levels, monitoring of stress and deformation of the detector, monitoring of
stress, temperature, and so on. The design of filling speed and capacity of the
filling system should consider the requirements of total filling time, the
capacity of LS storage and production, and cost. The design of the relative
liquid level difference should consider the load stress on the detector.

\section{Reliability and Risk analysis}

$\bullet$ Structural reliability analysis

The baseline option takes large glass buildings and techniques used for glass
curtain walls as an important reference, given that those buildings and
techniques have been matured for long time in the field of architecture. In our
case, the material of glass is replaced by acrylic, and the crucial parts
include bulk-polymerization of the acrylic sheets, different liquid located
inside and outside of the acrylic and long term water pressure and buoyancy.
Reliability of the stainless-steel structure is less important than that of the
acrylic sphere in this option. We should consider:

1) The aging properties of acrylic under stress and submersion in organic or
inorganic liquid needs further tests, to make sure it will be stable during the
20-year running.

2) Acrylic is a brittle material, if there is a crack it will grow under
stress. Cracks can be repaired during the construction but will be fatal after
the liquid filling is finished.

3) The stress is generally larger in the joints of the acrylic and truss than
other areas, so needs to be controlled to keep it low enough.

4) As for the requirements of the SNO and Daya Bay, the stress on the acrylic
should be less than 5~MPa, and more work is needed to reach this goal.

$\bullet$ Construction reliability

Fabrication of a thick acrylic sheet in a spherical shape is not a mature technique, and on-site construction of the large acrylic sphere also needs more R$\&$D effort.

$\bullet$ The reliability of FEA

The analysis needs to be compared with test results, and the accuracy of
analysis should be verified by joint tests and small prototype tests. Maximum errors should be taken as the input parameters for the analysis.

$\bullet$ Construction risk

Construction of the truss should be under control since there is a mature specification. For construction of the acrylic sphere, since there will be a large mount of work done onsite, a standard procedure should be followed and strict quality control should be applied especially on the key parameters of the structure such as stress, deformation and stability.

$\bullet$ Local failure

Local failure means a failure of the joint  between the acrylic sphere and
truss, so the stress of the whole structure would need to be re-analysed.
According to the current analysis, if the failure happens on the truss, it will
not be too serious. If it happens on the acrylic, however it will result in
severe damage, especially when the acrylic sphere has been filled with liquid
scintillator. This kind of failure should be avoided at all costs.

$\bullet$ Long term compatibility and leakage problem

It has been proved that acrylic is compatible with liquid scintillator, and
bulk-polymerization will not lead to leakage problems.

$\bullet$ Long term performance

The lifetime of acrylic is up to 30 years and the acrylic sphere for the SNO
\cite{SNO} experiment has already been running for nearly 20 years. In our
case, however the aging effect of the acrylic still needs more study.

\section{Schedule}

The schedules of the central detector are the following:

$\bullet$ 2013

1) Conceptual design of the detector, including FEA and prototype test

2) Consideration of the central detector schedule, budget, manpower related issues.

$\bullet$ 2014

1) Review existing options, select one main option and one candidate option. Perform extensive study on the two options, solve the key design parameters by analysis and prototype.

2) Preliminary design and test of implosion-proof structure of PMT.

3) Start to study PMT potting and LS filling.

4) Start to study batch test for PMT.

$\bullet$ 2015

1) Optimize the detector design, conduct prototype design, fabrication and test. Start to consider on-site construction of the detector.

2) Further study the water-proof of PMT

3) Finish a preliminary design of the implosion-proof structure of PMT

4) Study the batch test system of PMT, including electronics performance test, water-proof and pressure test.

5) Determine the final central detector structure.

$\bullet$ 2016

1) Continue detector structure design and on-site construction study

2) Determination of the batch test design for PMT

3) Start to design the liquid filling system and monitoring system.

4) Determination of the high voltage option for PMT

5) Start to manufacture the implosion-proof structure for PMT

6) Finish the engineering design of central detector.

$\bullet$ 2017

1) Finish bidding for the central detector structure, determine the
construction company and start to produce the components.

2) Determine the high voltage option and water-proof design of PMT.

3) Start PMT batch test.

4) Determine liquid filling option, bidding for the system.

5) Design of nitrogen protection system for liquid scintillator.

$\bullet$ 2018

1) Start to construct the central detector.

2) Continue PMT batch test.

3) Start to install detector monitor and other necessary devices.

$\bullet$ 2019

1) Finish central detector structure construction.

2) Finish PMT installation.

3) Finish detector monitor installation.

4) Finish installation of liquid filling system, start to fill the liquid.

$\bullet$ 2020

1) Finish liquid filling.

2) Finish detector commissioning and start data taking.

\vbox{}

The total time for the central detector construction is estimated about 18
months, as follows:

$\bullet$ Construction of the stainless-steel truss: 3 months (January 2018 -
April 2018). This includes interface(column supports on the base of the water
pool), fixtures, hoisting, measurement, fixing, surveys etc.

$\bullet$ Construction of the acrylic sphere: 8 months (May 2018 - November 2018). This includes acrylic sheet hoisting, positioning, fixture installation, joint connection and adjustment, bulk-polymerization, sanding, stress monitoring, installation of the LS filling device and the interface of calibration.

$\bullet$ Survey of the acrylic sphere and leak test: about 1 month (December 2018)

$\bullet$ Cleaning of the acrylic sphere. The total area of the inner and outer
surface of the sphere is about 8000 $m^{2}$. A high cleanliness level is needed
for the inner surface and it is hard for humans to clean, so a robot cleaner is
under consideration. An outlet need to be reserved in the bottom of the sphere. It will be closed after cleaning.

$\bullet$ Installation of PMTs and monitors: 5.5 months (February 2019 - June
2019). A number of work groups are considered to work in parallel at a rate of
2 PMTs/hour and 8 working hours/day, so 144 PMTs can be installed in one day.
In total this requires about 120 working days, meaning 5.5 months, for the $\sim$17000 PMTs.

$\bullet$ Cleaning of the whole detector after installation. This is not
included in the timetable of the central detector since it can be done together with VETO.

\clearpage

\cleardoublepage
\chapter{Liquid Scintillator}
\label{ch:LiquidScintillator}

\section{Introduction}

The innermost part of the JUNO detector is formed by 20,000 tons of liquid scintillator (LS) contained inside an acrylic sphere of 35\,m diameter. The LS serves as target medium for the detection of neutrinos and antineutrinos. The primary reactions for reactor electron antineutrinos ($\bar\nu_e$) is the inverse beta decay on free protons, $\bar\nu_e+p\to n+e^+$, resulting in a prompt positron and a delayed signal from the neutron capture on hydrogen ($\tau\sim200\,\mu$s).

The scintillator itself is a specific organic material containing molecules featuring benzene rings that can be excited by ionizing particles. Compared to other materials, this process is very efficient: About $10^4$ photons in the near UV and blue are emitted per MeV of deposited energy. The time profile of the light emission is dictated by the decay constants of the excited states of the organic molecules but also depends on the type and energy of the incident particle.

The LS is composed of several materials: The solvent liquid, linear alkyl benzene (LAB), forms the bulk of the target material and is excited by the ionizing particles. This passes on the excitation to a two-component system of a fluor (PPO) and a wavelength-shifter (Bis-MSB), that are added at low concentration (a few g/l resp.~mg/l) and by subsequent Stokes' shifts increase the wavelength of the emitted photons to $\sim 430$\,nm. This shift is crucial as the long wavelength avoids spectral self-absorption by the solvent and allows the photons generated by a reaction at the very center of the target volume to reach the photosensors (PMTs) mounted on a scaffolding outside the target volume. At 430\,nm, the transparency of the compound liquid will be largely governed by the Rayleigh scattering of photons on the solvent molecules. It can be reduced by the presence of unwanted organic molecules featuring absorption bands in the wavelength region of interest.

The energy resolution of JUNO is specified with 3\,\% at 1\,MeV, corresponding to at least 1,100 photoelectrons (pe) per MeV of deposited energy. Compared to Borexino, this corresponds to more than twice the pe yield. To meet this demanding requirements, both the initial light yield and the transparency of the liquid have to be optimized simultaneously.

In spite of the large number of neutrinos crossing the detector and the huge target mass, the minuscule cross-section of weak interaction allows the detection of only a few tens of neutrino events per day. Low-background conditions are therefore absolutely crucial. From the point of view of the LS, this means that the concentration of radioactive impurities inside the liquid should result in an activity on the same level or below the rate of neutrino events. This is less critical in case of  the inverse beta decay channel for $\bar\nu_e$ detection. The reaction provides a fast coincidence signature that can be used to suppress single-event background.

As a consequence, light yield, fluorescence time profile, transparency and radiopurity are the key features of the LS. These quantities will be mainly defined by the original materials, but as well by the presence of organic and radioactive impurities, temperature and aging effects. For the next years, we foresee a broad spectrum of laboratory measurements to characterize different brands of LAB and wavelength shifters. Multiple experimental methods will be used, like GC-MS, LS-MS, UV-VIS, and ICP-MS, amongst others. A close contact with the producing companies has been established which will allow the optimization of the production quality within certain limits. Moreover, we will study a variety of purification techniques to optimize both optical performance and radiopurity of the LS. Finally, a large-scale test will be conducted in one of the Antineutrino Detectors (ADs) of the Daya-Bay experiment, which will allow a to study the effect of purification on a sample of about 20 tons of LS.

The final aim is to set up a scheme for the mass production of 20,000 tons of LS. This study will comprise all steps from the initial production, the transport on site, the local facilities for liquid-handling, plants for purification and instrumentation for quality control as well as safety and cleaning methods~\cite{An:2012eh}.

\section{Specifications}

The basic specifications and requirements for the liquid scintillator are:

\begin{itemize}

\item {\bf Total mass:} 20,000 tons.

\item {\bf Composition:}\\
Scintillator solvent: Linear alkyl benzene (LAB)\\
Scintillating fluor: PPO (2,5- diphenyloxazole) at 3 g/L\\
Wavelength shifter: Bis-MSB at 15~mg/L

\item {\bf Light yield:}\\
Minimum of 1,100 photoelectrons (pe) per MeV

\item {\bf Transparency:}\\
Attenuation length @ 430\,nm: > 22~m\\
Absorption coefficient of less than $3\times10^{-3}$ in spectral range 430$-$600\,nm

\item {\bf Radiopurity:}\\
$\bar\nu_e$ detection: Concentrations of $^{238}$U, $^{232}$Th $\leq10^{-15}$\,g/g,  $^{40}$K $\leq10^{-16}$\,g/g. \\*
$\nu_e$ detection: Concentrations of $^{238}$U, $^{232}$Th $\leq10^{-17}$\,g/g,  $^{40}$K $\leq10^{-18}$\,g/g.
\end{itemize}

\section{Study of Optical Properties}

\subsection{Light Yield}

The light yield and emission time profile of the LS are key aspects of JUNO. For a given solvent (LAB), they depend mainly on the  the concentrations of the fluor PPO and the wavelength-shifter Bis-MSB. The resulting light yield and fluorescence times of the composite LS will be measured as a function of these concentrations. While for both quantities higher concentrations will in general increase the performance, the optimum concentration of PPO and Bis-MSB will be determined taking into regard the detrimental effects of self-absorption on the attenuation length and the radioimpurities introduced into the scintillator.

The studies presented in the following section mainly concentrate on the characterization of the light yield as a function of fluor concentration and temperature. Moreover, the non-linearity of the light yield as a function of the energy of incident electrons will be studied as it has implications on the linearity in the energy response of positrons. Beyond the setups described here, first studies of the fluorescence properties in laboratory experiments have already been performed at the INFN institutes in Milano and Perugia and will be intensified in the future.

\subsubsection{Effect of Fluor Concentrations}
\label{sec:ls:ly-fluor}

The light yield (LY) of an LS sample depends greatly on the concentration of fluor(s) added to the solvent. At low concentrations, it will increase almost linearly with the fluor concentration. However, the increase will become less steep once a concentration of 1$-$2\,g/L is reached. To optimize the amount of collected light with regard to self-absorption effects which are expected to play a role for concentrations of several g/L and more, the LY of a variety of LAB samples featuring varying concentrations of PPO and Bis-MSB will be studied. This will be done by a setup based on a Compton coincidence detector, an established technique for a comparative measurement of the LY of LS samples.

\begin{figure}[htb]
\begin{center}
\includegraphics[width=5cm]{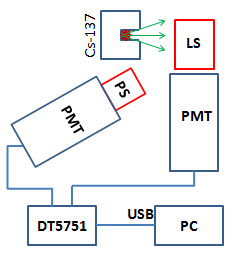}
\caption[Schematic of the setup for light-out measurements]{Schematic setup for the relative determination of the light out}
\label{fig:ls:ly}
\end{center}
\end{figure}

Figure~\ref{fig:ls:ly} shows the basic principle of the experimental setup: A PMT is attached to a vessel containing the LS sample that is excited by gamma-rays from a nearby source. Without a coincidence detector, the pulse height spectrum observed by the PMT will correspond to a Compton shoulder, which will require a relatively complicated fit to evaluate the light out. Adding a coincidence detector to record the scattered gamma-rays allows to define a fixed energy deposition in the LS sample by fixing the scattering geometry. In these circumstances, the pulse height spectrum can be fitted by a simple Gaussian, providing a more accurate result.

\begin{figure}[htb]
\begin{center}
\includegraphics[width=12cm]{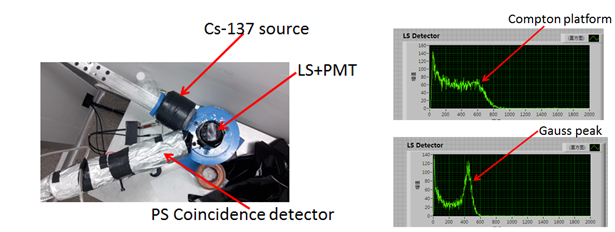}
\caption[Foto of the setup for LY measurements]{Setup for measuring the LS light out and preliminary result}
\label{fig:ls:ly2}
\end{center}
\end{figure}

The experimental setup shown in Fig.~\ref{fig:ls:ly2} has been designed to determine the LY with less than 2\,\% uncertainty. The panels on the right display the pulse-height spectra obtained in a sample measurement and demonstrate the impact of the coincidence measurement. Beyond fluor concentrations, this setup will be used in the long term to test the effects of purification techniques, LS temperature and aging.

\subsubsection{Temperature Dependence}

\begin{figure}[htb]
\begin{center}
\includegraphics[width=10cm]{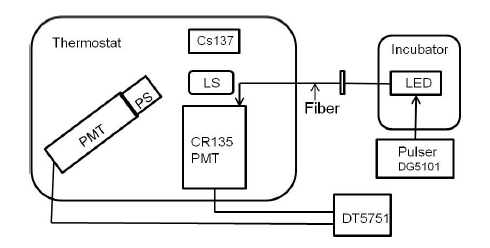}
\caption[Setup for measuring temperature-dependence of LY]{Experimental setup to determine the temperature dependence of the LY}
\label{fig:ls:ly3}
\end{center}
\end{figure}

An increase in scintillation LY has been observed when lowering the temperature of the LS. This will be studied in a somewhat modified setup shown in Fig.~\ref{fig:ls:ly3} that has been adjusted for the application of different temperature levels to the LS sample. The $^{137}$Cs $\gamma$-source, the LS vessel, the PMT and the coincidence detector are put into an enclosed thermostat, allowing to vary the temperature from -40~$^{\circ}$C to 30~$^{\circ}$C. The signals from the PMT and the coincidence detector are recorded by a CAEN DT5751 FADC unit. A temperature-resistant optical fiber is used to monitor the temperature response of the PMT.

\subsubsection{Energy Non-linearity of Electron Signals}

While the light output of LS is an almost linear function of the deposited energy, additional small non-linear terms will introduce an additional systematic uncertainty for the energy reconstruction in JUNO. Of special relevance is the non-linearity in case of positron signals that might affect the sensitivity of the mass hierarchy measurement. To take these effects correctly into account, it is important to study the energy response in small-scale laboratory setups. For practical reasons, these experiments will rely on electrons instead of positrons.

A corresponding experimental setup has been realized at IHEP, designed to measure the energy non-linearity of electron signals below 1\,MeV at a precision of better than 1\,\%. The measurement is performed by inducing low-energy recoil electrons by Compton scattering of $\gamma$-quanta in a small LS sample. As the incident energy $E_\gamma$ is known, a specific electron energy $E_e$ can be selected by fixing the scattering angle $\theta$ (see above). It holds
\begin{equation}
E_e=E_\gamma\bigg[1-\frac{1}{1+\alpha(1-\cos\theta)}\bigg],
\end{equation}
where $\alpha=\frac{E\gamma}{m_e}$, with $m_e$ the electron mass. The non-linearity of the energy response is obtained by determining the ratio of observed light output and computed electron energy as a function of $\theta$.

\begin{figure}[htb]
\begin{center}
\includegraphics[width=14cm]{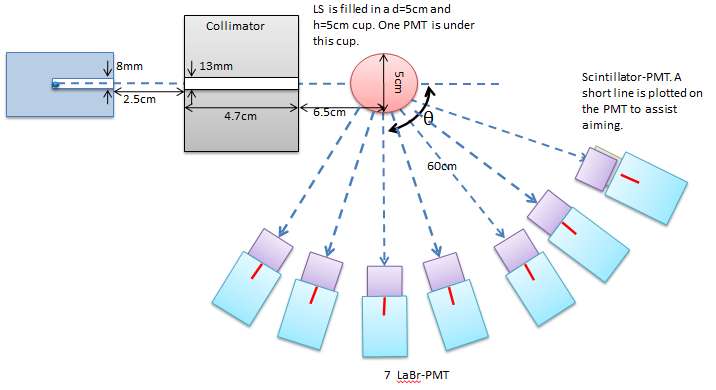}
\caption[Conceptual drawing of the electron non-linearity setup]{Conceptual drawing of the electron energy non-linearity setup}
\label{fig:ls:enl}
\end{center}
\end{figure}

\begin{figure}[htb]
\begin{center}
\includegraphics[width=12cm]{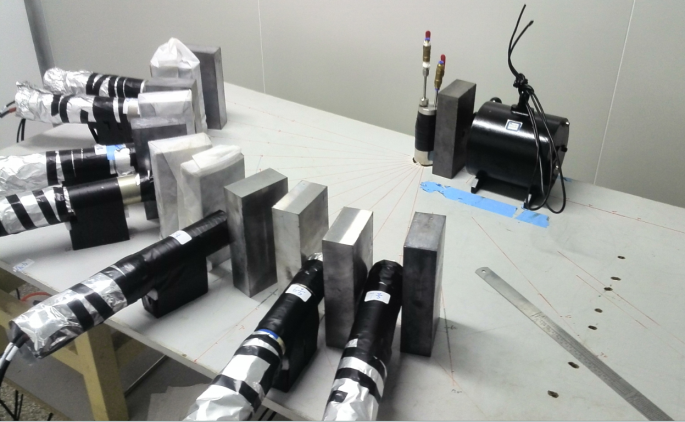}
\caption[Photograph of the electron non-linearity setup]{Photograph of the electron energy non-linearity setup}
\label{fig:ls:enl2}
\end{center}
\end{figure}

Figure~\ref{fig:ls:enl} shows a conceptual drawing of the experiment, while Fig.~\ref{fig:ls:enl2} displays the actual laboratory setup.The gamma source used in the setup is  $^{22}$Na with an activity of 0.3\,mCi. From $\beta^+$ decay to an excited state, it provides two $\gamma$-rays of 0.511\, MeV and one at 1.275 MeV. After passing through a lead collimator of 9\,mm diameter, the $\gamma$-rays scatter in the LS sample, which is held in a cylindrical silica cup of 5\,cm diameter and height. The light generated by the recoil electron is collected by a PMT (XP2020) attached to the LS cup. Seven coincidence detectors are placed at various angles ($20^{\circ}$, $30^{\circ}$, $50^{\circ}$, $60^{\circ}$, $80^{\circ}$, $100^{\circ}$, $110^{\circ}$) at 60\,cm distance from the LS cup.These coincidence detectors consist of an inorganic crystal scintillator (LaBr) and a PMT (XP2020). The signals acquired simultaneously by all PMTs pass a Fan-In-Fan-Out unit (CAEN N625) before being sent to a trigger board (CAENN405) and a FADC (CAEN N6742). Using an array of PMTs rather than a single PMT mounted to a rotatable arm is reducing the required acquisition time and avoids variations of the PMT response induced by the rotation of the dynode chain in magnetic stray fields.

Preliminary results are expected for late 2015. At the present state, the uncertainty on electron energy non-linearity is estimated to be smaller than 2\,\%. The results will be used as well in the data analysis of the Daya Bay experiment.

\subsection{Optical Transparency}

Compared with other neutrino experiments, the optical transparency of the LS in JUNO is absolutely crucial. By scaling up the target volume from several hundreds to 20,000 tons, the diameter of the target volume in JUNO will be 35.4~m. If light transmission in the liquid scintillator is too low, scintillation photons from a neutrino interaction in the center of the detector will be absorbed by the liquid itself before reaching the photomultipliers positioned at the verge of the volume.

The light transmission is described by the attenuation length which quantifies the distance over which the original light intensity is reduced to a fraction of $1/e$. This attenuation is due to two kind of processes: Light absorption on organic impurities, in which case the photon can either be fully absorbed or re-emitted, and light scattering off the solvent molecules. In the latter case, light is not lost but merely re-directed and may still be detected.
Several laboratory scale experiments will be conducted to characterize LS samples regarding their transparency: Measurements include the wavelength-dependent attenuation spectrum of LS in relatively short cells (10\,cm), precise measurements of the LS attenuation length over large samples (several meters) and independent measurements of the scattering length of LS. Moreover, the fraction of organic impurities of different samples will be determined and the long-term stability of LS samples tested.

\subsubsection{Characterization of Attenuation Spectrum}

The attenuation spectrum of a material corresponds to the fraction of incident radiation absorbed by the sample over a certain thickness as a function of the wavelength of the incident light. The measurement is performed by commercial UV-Vis spectrometers that usually are sensitive to a wavelength range from 190 to 900\,nm. These measurements require only a low amount of time and a relatively small amount of LS as typical sample cells contain only 30\,ml or less. Therefore, this method is very suitable for the real-time monitoring of solvent or LS samples. Periodic measurements can be used to study the long-term stability of LS samples.

\begin{figure}[htb]
\begin{center}
\includegraphics[width=12cm]{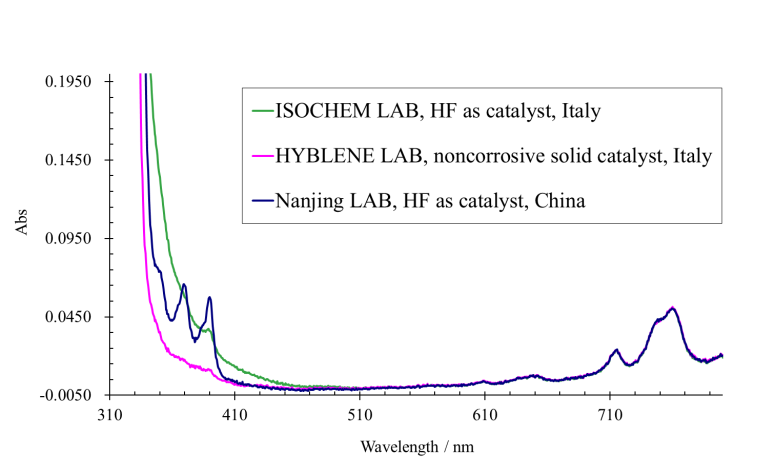}
\caption[UV-Vis attenuation spectra of LAB samples]{The UV-Vis attenuation spectra of LAB from different manufacturers}
\label{fig:ls:specsolvent}
\end{center}
\end{figure}

A first characterization of samples can be done based only on the solvent LAB (without adding the fluors). It is found that the resulting UV-Vis spectra vary for LAB samples from different providers. Fig.~\ref{fig:ls:specsolvent} shows the attenuation curves of LAB samples produced by several companies that apply different manufacturing technologies. Substantial divergences are present in the wavelength region below 500\,nm which is most important for the propagation of scintillation light.

\begin{figure}[htb]
\begin{center}
\includegraphics[width=12cm]{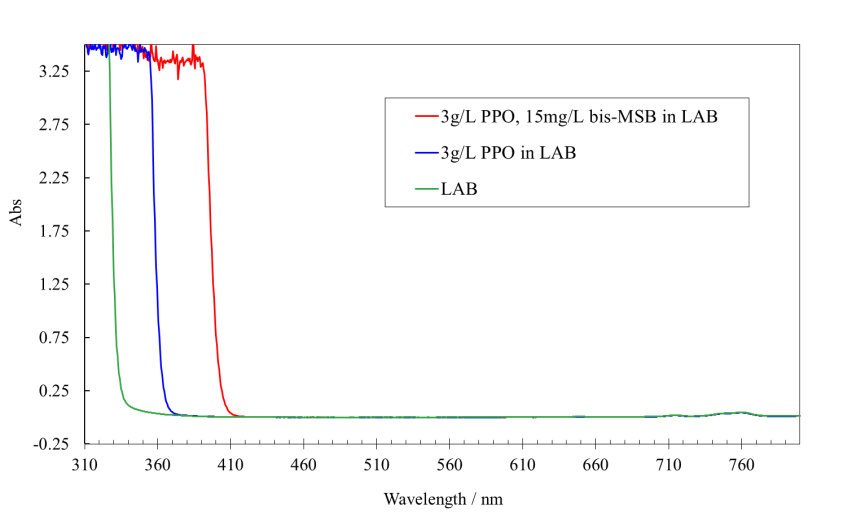}
\caption[UV-Vis attenuation spectrum: Impact of fluors]{The UV-Vis spectra of LAB and LAB-based liquid scintillators}
\label{fig:ls:specfluors}
\end{center}
\end{figure}

When loading the solvent with the fluors necessary to form the liquid scintillator, the attenuation spectrum changes. The shift of the absorption band to longer wavelengths is illustrated in Fig.~\ref{fig:ls:specfluors}.

\subsubsection{Long-range Measurement of Attenuation Length}

As stated above, the light intensity generated by particle interactions in the bulk of the LS volume will be attenuated by scattering and absorption before reaching the PMTs. The relation between the intensity $I(d)$ remaining after propagation by a distance $d$ in the LS and the initial light intensity $I_0$ can be expressed by
\begin{equation}
\label{eq:ls:att}
I(d)=I_0 \exp(-d/\ell_{\rm att}).
\end{equation}
The length scale $\ell_{\rm att}$ quantifying the distance after which $I(d)$ is reduced to $1/e$ of the initial value $I_0$ is the attenuation length.

\begin{figure}[htb]
\begin{center}
\includegraphics[width=12cm]{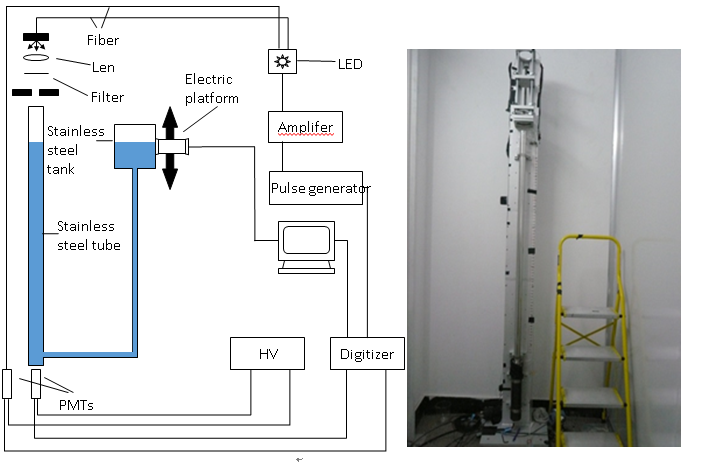}
\caption[Setup for attenuation length measurement]{Setup for a long-range measurement of the attenuation length}
\label{fig:ls:att}
\end{center}
\end{figure}

The accuracy that can be reached by UV-vis spectrometers on the attenuation length is limited if $\ell_{\rm att}\geq10$\,m. This is due to the very small decrease of the light intensity when passing through a sample cell of less than 10\,cm in length and thus  considerably shorter than the expected attenuation length of the LS ($\ell_{\rm att}>20$\,m). Considering the importance of exact knowledge of $\ell_{\rm att}$ for light propagation in the large LS volume, dedicated measurements are performed in the wavelength range around 430\,nm which corresponds to the emission band of Bis-MSB and at the same time represents the optimum intersection of scintillator transparency and PMT photosensitivity. Figure~\ref{fig:ls:att} shows the corresponding laboratory device at IHEP and a schematic sketch of the setup. A similar device with an even longer sample cell is currently developed at TU Munich (Germany). Measurements of this kind will be especially valuable to assess the effect of purification techniques on the optical transparency of the LS.

\subsubsection{Determination of Scattering Lengths}

The attenuation of the initial scintillation light can be caused by a variety of processes: Absorption by organic impurities, self-absorption by the fluors (primarily Bis-MSB), Mie scattering off dust or particulates in suspension in the LS, or Rayleigh scattering of the solvent molecules. In general, the relation between the corresponding optical length scales can be given as:
\begin{equation}
\frac{1}{\ell_{\rm att}}=\frac{1}{\ell_{\rm abs}}+\frac{1}{\ell_{\rm are}}+\frac{1}{\ell_{\rm scat}},
\end{equation}
where $\ell_{\rm abs}$ describes the absorption length without re-emission, $\ell_{\rm are}$ means absorption followed by re-emission, and $\ell_{\rm scat}$ is the Rayleigh scattering length. Only the light absorbed without re-emission will be fully lost for event reconstruction. All other processes merely divert a photon from its original path so that it can still be detected by a PMT.

For an ideal scintillator without impurities, $\ell_{\rm att}\approx \ell_{\rm Ray}$ as Rayleigh scattering is the only irreducible process. In comparison, a conceivable contribution from Mie scattering on dust particles is expected to be negligible because the LS will be virtually dust-free after purification. Absorption-reemission might play a role as the long-wavelength tail of the Bis-MSB absorption band reaches into its emission spectrum centered around 430\,nm and might thus be relevant for light propagation.

The physics requirements of JUNO necessitate to come at least close to this ideal state in order to maximize the light output of the target LS. On the other hand, light emitted in a neutrino interaction and scattered while traveling to the PMTs will be deflected from a straight line of sight, influencing both spatial reconstruction and pulse-shaping capabilities of the experiment.
Therefore, a precise measurement of the Rayleigh scattering length will allow to determine the ultimate transparency limit for an LS based on the given solvent (LAB in case of JUNO), and at the same time provide valuable input for MC simulations that study the expected detector performance. Moreover, the impact of different purification techniques on the composition of the scintillator can be monitored by studying the combined results of attenuation and scattering lengths measurements (see below).

\begin{figure}[htb]
\begin{center}
\includegraphics[width=12cm]{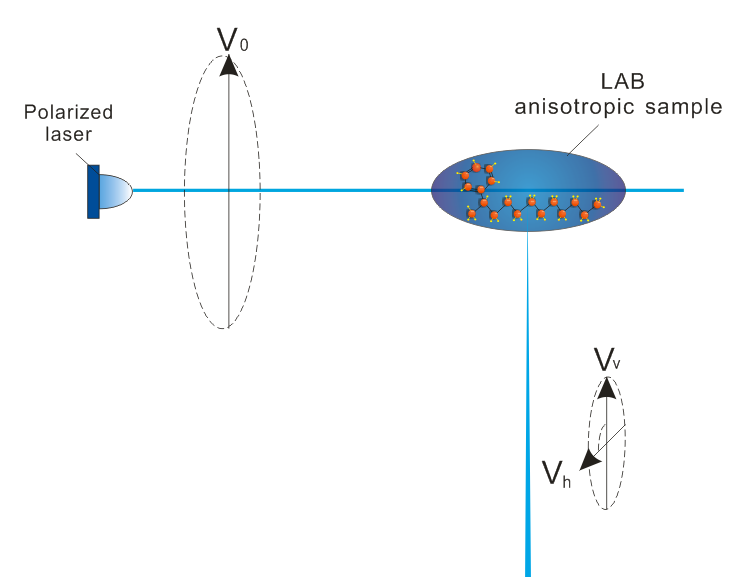}
\caption[Schematic setup for scattering length measurements]{Schematic diagram of Rayleigh scattering measuremens}
\label{fig:ls:scat1}
\end{center}
\end{figure}

The Rayleigh scattering length expected for a given material can be calculated based on the Einstein-Smoluchowski-Cabannes (ESC) formula,
\begin{equation}
\frac{1}{l_{ray}}=\frac{8\pi^3}{3\lambda^4}kT\bigg[\rho\bigg(\frac{\partial\epsilon}{\partial\rho}\bigg)_T\bigg]\beta_T\bigg(\frac{6+3\delta}{6-7\delta}\bigg)
\end{equation}
where $\lambda$ is the wavelength of scattered light, $\rho$ is the density and $\epsilon$ the average dielectric constant of the liquid, $k$ is the Boltzmann constant, $T$ is the temperature, $\beta_T$ is the isothermal compressibility and $\delta$ is the depolarization ratio. Based on the ESC formula, the Rayleigh scattering of LAB can be obtained by measuring the reflectivity,  isothermal compressibility, and the depolarization. Comparing to direct measurements of the scattered intensity (see below), this relative measurement can lower the systematic error on the depolarization ratio and directly derive the scattering length as a function of the incident wavelengths $\lambda$ of the scattered lights. Figure~\ref{fig:ls:scat1} shows a schematic drawing of the setup located at Wuhan University.

Complementarily to the described measurements, a direct measurement of the amount of scattered light will be performed at the University of Mainz (Germany). By investigating the dependence of the scattered intensity on the incident wavelength, polarization and scattering angle, contributions from different scattering processes as well as absorption/re-emission can be resolved. On the one hand, this will provide an important cross-check of the Rayleigh scattering length measurement. On the other hand, it will allow to scrutinize purified samples for residual dust particles and organic impurities and, more importantly, to assess the impact of Bis-MSB self-absorption on light propagation. Measurements of this kind have been already performed in Ref.~\cite{Wurm:2010ad}.

\subsubsection{Organic Impurities from GC-MS Measurements}
\label{sec:ls:gcms}

Gas chromatography combined with mass spectrometry (GC-MS) is a commonly used technique to analyze the composition of  complex organic mixtures. GC-MS analysis combines the high resolution of gas chromatography with the high sensitivity of mass spectrometry. Type and concentration of impurities in LAB as well as the primary LAB components can be resolved and identified by the GC-MS method. The results will provide useful information for improving the purification scheme.

\begin{figure}[htb]
\begin{center}
\includegraphics[width=14cm]{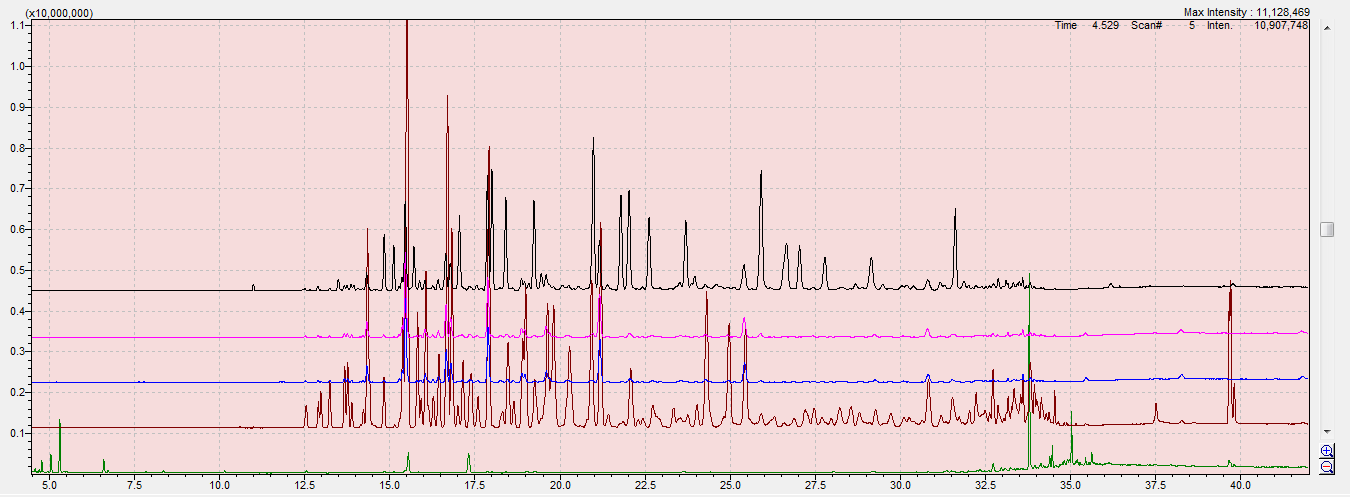}
\caption[GC-MS Chromatogram of LAB]{Total ion chromatogram of LAB and impurities in LAB from GC-MS measurements}
\label{fig:ls:gcms}
\end{center}
\end{figure}

We have analyzed a sample of commercially available LAB produced by the Nanjing LAB factory. In Fig.~\ref{fig:ls:gcms}, the black curve indicates the components of LAB. The remaining curves stem from the impurities. The result suggests that
\begin{enumerate}
\item There are many impurities in LAB, which can be divided into several categories: fluorene, naphathelene derivatives, biphenyl derivatives, diphenyl alkane, and small amounts of alcohol, ketone, and ester.
\item The retention times of impurities coincide with LAB components. The impurities in LAB must be separated and enriched before GC-MS analysis.
\item The resolution of the chromatogram is not satisfying. The analytical conditions of GC-MS must be optimized in order to perform precise qualitative and quantitative analyses.
\end{enumerate}
Therefore, both the LAB pre-treatment methods and instrumental conditions have to be optimized in order to establish the GC-MS method for studying type and level of impurities in LAB samples.

GC-MS is an important addition to the characterization of the LS as it provides data on the chemical composition complementary to the optical measurements. We expect that following the adaptation of the method we will be able to clearly identify the main organic impurities in different brands of LAB. This will allow us to cooperate with the LAB manufacturers in order to achieve an optimization of the chemical purity of the LS raw materials.

\subsection{Long-term Stability}

JUNO is supposed to be in operation for at least 10 years. This poses the question of the long-term stability of the composite liquid and especially of its scintillation and optical properties. A common method of investigation often applied in chemistry are aging tests relying on heating the samples. According to the Van't Hoff equation, the rate of possible deteriorating processes will increase by a factor 2 to 4 for every 10~C$^{\circ}$ over room temperature (25~$^{\circ}$C). Therefore, a 6-months aging test performed  at 40~$^{\circ}$C is equivalent to 1.5 - 4 years at 25~$^{\circ}$C. The temperature-treated samples will be characterized by their light out, attenuation length and absorption spectrum.

\begin{figure}[htb]
\begin{center}
\includegraphics[width=12cm]{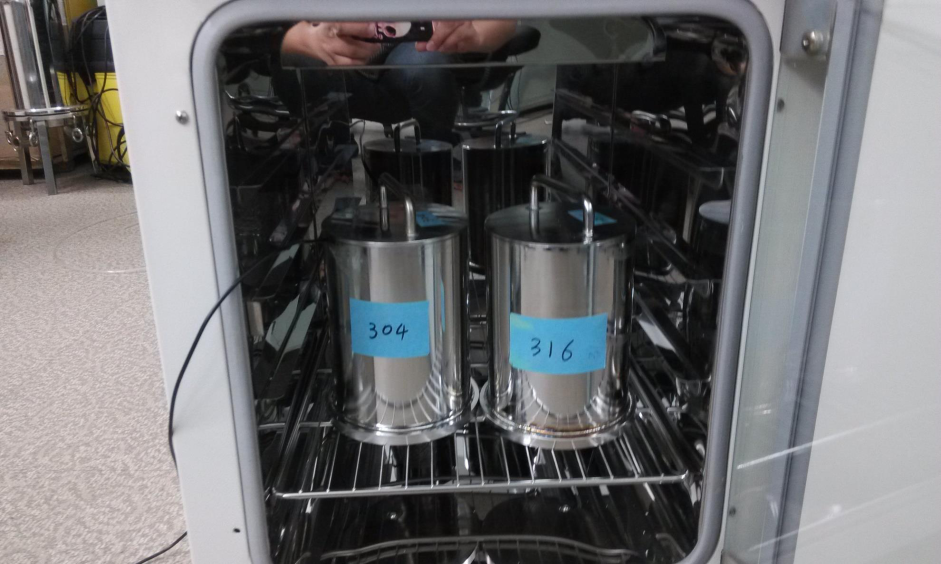}
\caption[Setup for LS aging tests]{Laboratory setup for LS aging tests}
\label{fig:ls:aging}
\end{center}
\end{figure}

During the tests, the LS samples will be filled into containers of stainless steel. This will allow the performance of a material compatibility test with the LS at the same time. Different types of stainless steel (SS316 or SS304) will be investigated.  Some ageing experiments are being conducted on LS. Figure~\ref{fig:ls:aging} shows two 15-liter stainless steel vessels.

\subsection{Selection of Raw Materials}

\subsubsection{Solvent}

LAB is a family of organic compounds with the formula C$_6$H$_5$C$_n$H$_{2n+1}$. When used as a detergent, $n$ lies typically between 10 and 16. For JUNO, a specific product selected for $n=10-13$ will be chosen as it shows better optical properties.

In China, the list of LAB manufacturers includes the Nanjing LAB factories of Jinling Petrochemical Company, the Fushun wash \& chemical factory of the Fushun Petrochemical Company, and the Jintung Petrochemical Corporation Ltd. The annual output of each factory is very large. Only several months will be needed to produce 20 ktons of LAB for JUNO.

In general, LAB production in China is based on a HF catalyst (HF-acid method). Outside China, an alternative technology based on a non-corrosive solid acid catalyst (Solid-acid method) is more common. The advantages of the HF acid method are the high conversion efficiency and the long-term stability in operation. The disadvantage is that it generates more by-products because the syntheisizing reaction is more violent than in case of the solid acid method. Therefore, the general expectation is that production by the solid-acid method should result in better optical properties of LAB.

\begin{figure}[htb]
\begin{center}
\includegraphics[width=12cm]{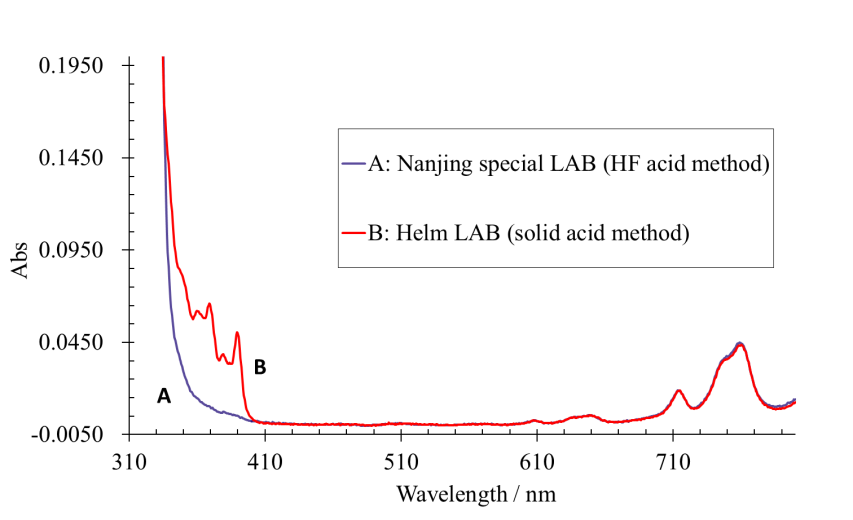}
\caption[UV-Vis attenuation spectra of LAB from different providers]{UV-Vis attenuation spectra of LAB produced by HF acid method and solid acid method}
\label{fig:ls:optpur1}
\end{center}
\end{figure}

Instead, our investigations based on the attenuation spectra of LAB samples suggest that the method of synthesis is not the critical factor for the resulting optical properties. Figure~\ref{fig:ls:optpur1} compiles the attenuation spectra of LAB samples produced both by the HF-acid method in the Nanjing LAB factory (specially ordered by optimizing production flow, not commercially available) and by the solid-acid method in an Egyptian plant (Helm AG). The sample from Nanjing shows much lower absorption for the spectral range below 410\,nm. As the spectral measurement is not sufficiently sensitive at the most interesting wavelength of 430\,nm, the sample was also tested in the long cell at IHEP. Again, the Nanjing sample showed a much longer attenuation length $\ell_{\rm att}\approx 20$\,m, while the LAB by Helm featured only $\ell_{\rm att}\approx12$\,m.

Based on these results, it seems that Chinese manufacturers using the HF-acid method should be able to provide LAB of sufficient quality for JUNO as long as the quality of the raw materials is assured and the production flow is optimized. For this, we foresee a cooperation with domestic LAB factories. In addition, we will further investigate samples of LAB from foreign companies.

\subsubsection{Fluor (PPO) and Wavelength Shifter (bis-MSB)}

As described above, a system of two fluors will be used in JUNO: LAB will non-radiatively pass on the excitations to the primary fluor PPO, which will in turn perform a mainly non-radiative transfer to the secondary wavelength shifter Bis-MSB. As a consequence, the effective light emission (including optical propagation effects) will be shifted to 430\,nm.

\begin{figure}[htb]
\begin{center}
\includegraphics[width=12cm]{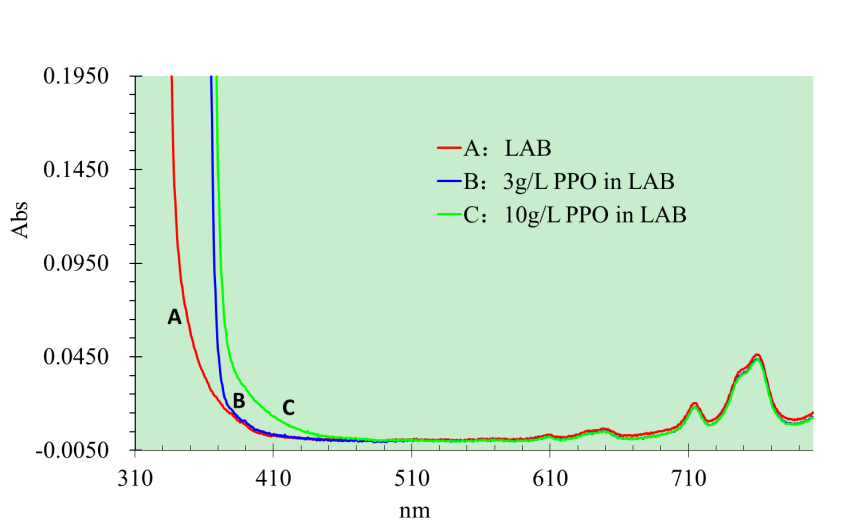}
\caption[UV-vis attenuation spectra of LAB doped with PPO]{UV-Vis attenuation spectra of LAB doped with PPO}
\label{fig:ls:fluor}
\end{center}
\end{figure}

Due to the self-absorption of the fluors, the absorption spectra of the complete LS will differ significantly from pure LAB. This might be further enhanced by optical impurities introduced along with the fluors. Figure~\ref{fig:ls:fluor} illustrates the influence of PPO on the absorption spectrum of LS. If PPO concentration is increased from the baseline value of 3\,g/L to 10\,g/L, an effect induced by the tails of the absorption band of PPO becomes visible that extends up to 460\,nm. This clearly demonstrates that the optimum PPO concentration must be balanced between light yield and optical transparency in the 430\,nm range, taking re-emission processes into account.

\begin{figure}[htb]
\begin{center}
\includegraphics[width=12cm]{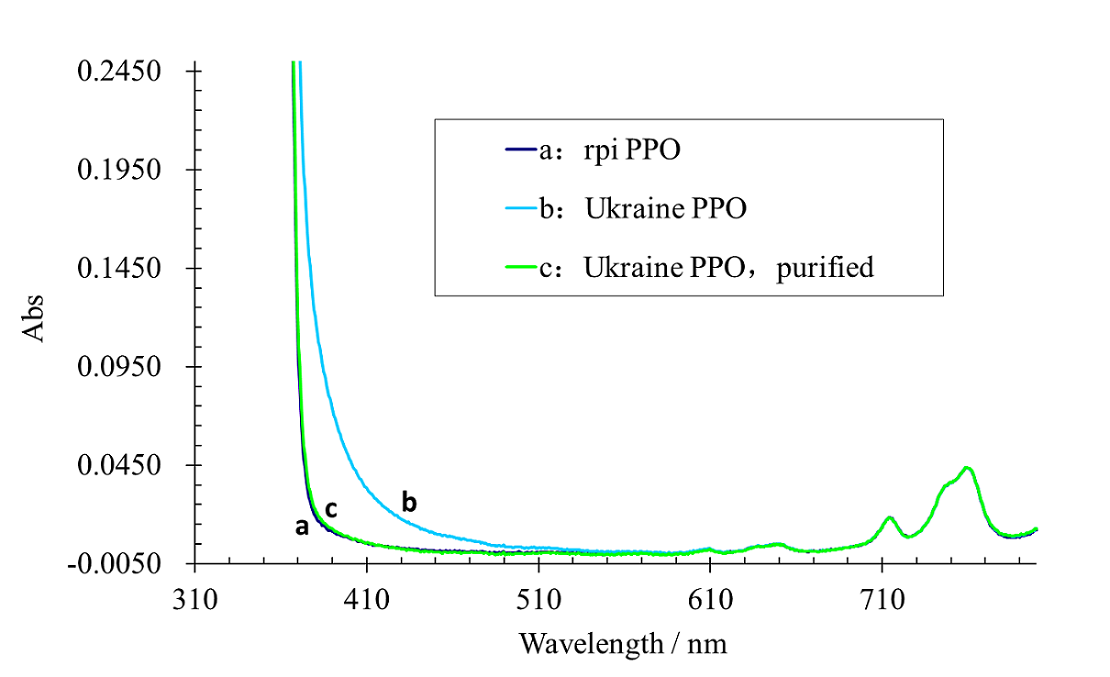}
\caption[UV-vis attenuation spectra with PPO from selected providers]{UV-vis attenuation spectra of LAB with PPO from selected providers}
\label{fig:ls:fluor2}
\end{center}
\end{figure}

Organic impurities introduced with PPO will as well have an impact on attenuation. Figure~\ref{fig:ls:fluor2} shows the attenuation spectra of LAB samples containing 10\,g/L PPO. The PPO was obtained from two different sources: one from rpi (Research Products International Corp), one from the Ukraine. For the latter, spectra for both the original and purified PPO (by Haiso pharmaceutical chemical Co., Ltd., Hubei, China) are shown separately. The positive impact of purification on transparency is clearly visible. The original quality of PPO from rpi seems to be comparable to purified PPO from Ukraine.

\begin{figure}[htb]
\begin{center}
\includegraphics[width=12cm]{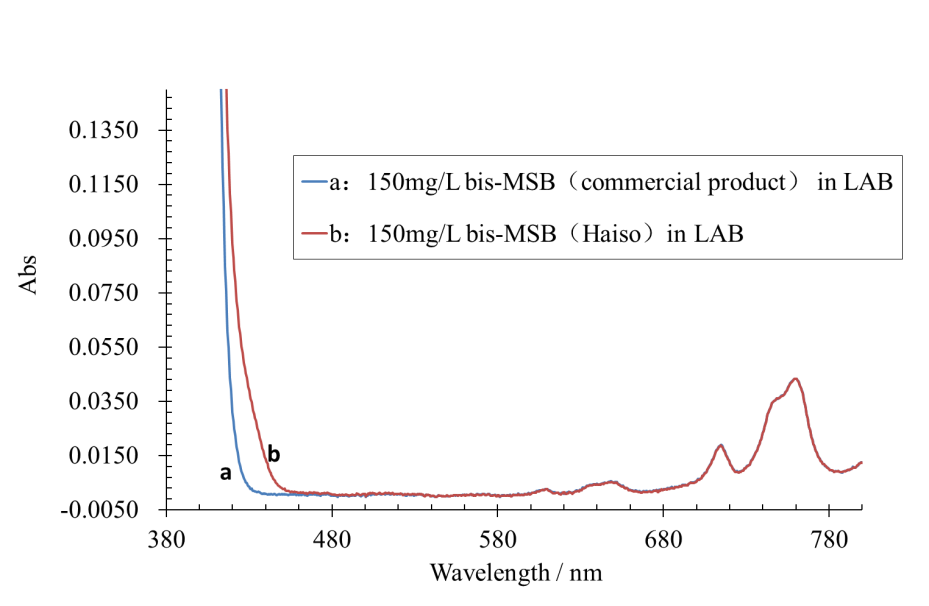}
\caption[UV-vis attenuation spectra of LAB with Bis-MSB]{UV-vis attenuation spectra of Bis-MSB in LAB solutions}
\label{fig:ls:fluor3}
\end{center}
\end{figure}

A similar or greater impact on the attenuation spectra is expected from the addition of the secondary fluor Bis-MSB. Like in the case of PPO, the chosen concentration has to be optimized and the impact of impurities studied. Two measured attenuation spectra of Bis-MSB in LAB are given in Fig.~\ref{fig:ls:fluor3}. While the same concentration of Bis-MSB was realized and the effective light yield of both solutions is identical, the absorption spectra differ.

\subsection{Purification Techniques}
\label{sec:ls:opt_purification}

\begin{figure}[htb]
\begin{center}
\includegraphics[width=12cm]{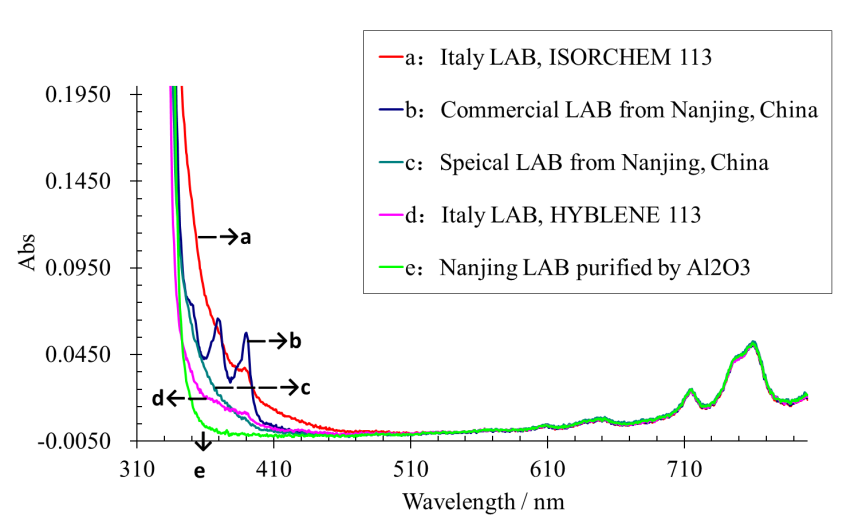}
\caption[UV-vis attenuation spectra before and after purification]{UV-vis attenuation spectra of LAB samples from different providers. In case of Nanjing LAB spectra taken before and after purification in an aluminum column are shown.}
\label{fig:ls:optpur2}
\end{center}
\end{figure}

Beyond the choice of suitable raw materials, optical properties can be further improved by dedicated purification methods. Several methods of purification have been investigated at IHEP, including distillation, column purification, water extraction and nitrogen stripping. Figure~\ref{fig:ls:optpur2} displays a compilation of attenuation spectra of LAB samples by different manufacturers. For comparison, the spectrum of Nanjing LAB after purification in an aluminum column is shown. It clearly shows the best performance in the wavelength region of interest. Similar studies are currently performed at INFN and German institutes~\cite{Ford:2011zza}.

\subsubsection{Fractional Distillation of LAB}

Distillation relies on the (partial) evaporation of the initial liquid to separate its constituents during vapor condensation. However, simple one-stage distillation usually does not meet the requirements for separating complex mixtures. This is usually achieved by multi-stage or fractional distillation: The technique relies on a heat exchange between upward-streaming vapor and downward-streaming condensate inside the distillation column. By this re-heating, the concentration of volatile components in the upper vapor increases, while less volatile components are enriched in the lower condensate. As the vapor keeps rising and the condensate keeps falling,  several equilibrium states between gas and liquid are established along the column. Therefore, distillation occurs in multiple stages. If the fractional column is sufficiently long, an efficient separation of volatile from solid components is achieved.

\begin{figure}[htb]
\begin{center}
\includegraphics[width=10.3cm]{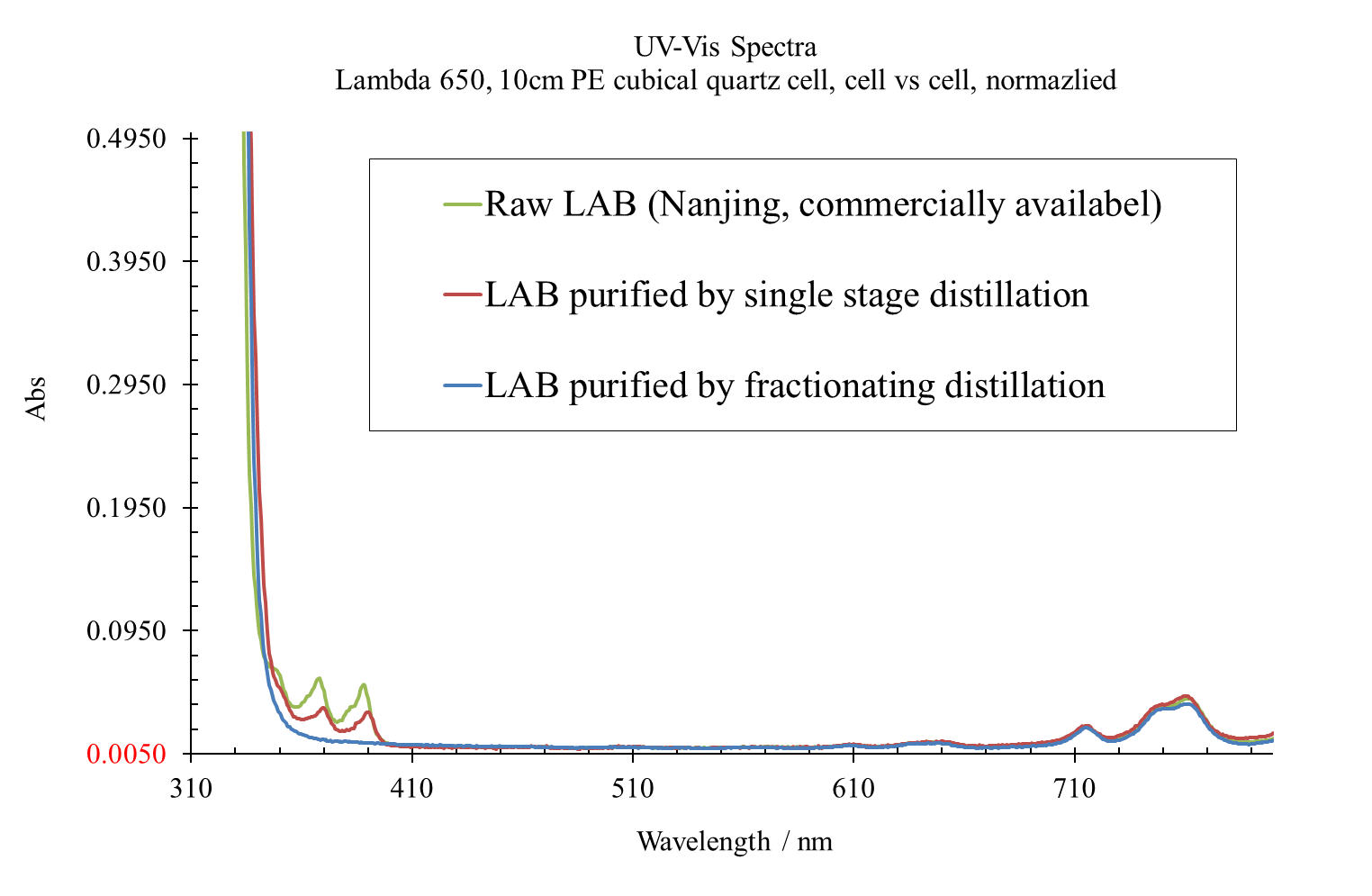}
\caption[UV-Vis spectra of LAB before and after distillation]{UV-Vis spectra of LAB before and after distillation}
\label{fig:ls:dist}
\end{center}
\end{figure}

In a first step, a setup including a single-stage distillation column was built up at IHEP. However, the resulting purification efficiency shown in Fig.~\ref{fig:ls:dist} proved insufficient. Therefore, the setup was upgraded to fractional distillation by the addition of a Vigreux column. This resulted in a substantial improvement due to the increased number of theoretical plates.

Compared to single-stage distillation, fractional distillation is much more efficient in the removal of the impurities causing absorption lines in the region from 330 to 410\,nm. However, in case of the IHEP setup, neither method provides a visible improvement in region around 430\,nm. Possible explanation comprise:
\begin{itemize}
\item The boiling points of the organic impurities influencing absorption at 430\,nm are too close to LAB to be removed by distillation.
\item New organic impurities are generated by distillation, e.g.~by the break-up of LAB molecules.
\item UV-vis spectroscopy is not sufficiently accurate to resolve an improvement.
\end{itemize}
While the unexpected absence of a beneficial affect of distillation in the 430-nm range requires further exploration, this method  clearly improves transparency in the wavelength region below 410~nm. However, First results from a similar test setup at INFN Perugia in fact show an increase of the attenuation length in the region of interest.

The Vigreux column proved to be simple and easy to operate for laboratory-scale operations. However, a system based on fractional columns will be necessary for the distillation of large amounts of LAB for JUNO.

\subsubsection{Purification of LAB by Aluminum Columns}

The impact of purification in an aluminum (Al$_2$O$_3$) column on the optical properties of LAB has been studied in detail at IHEP. The results show that column chromatography is very effective in removing optical impurities.

\begin{figure}[htb]
\begin{center}
\includegraphics[width=10.1cm]{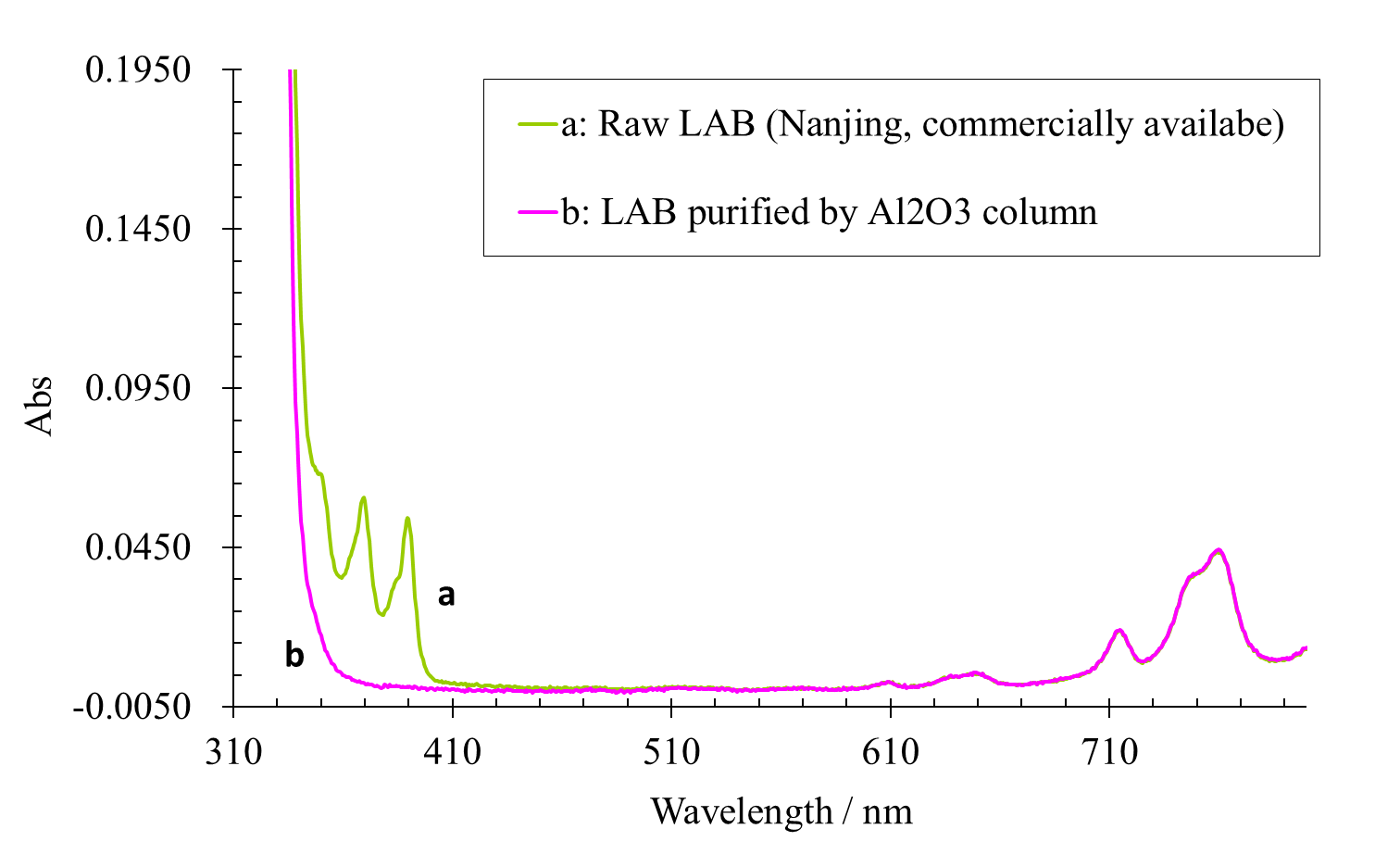}
\caption[UV-Vis spectra of LAB before and after column purification]{UV-Vis spectra of LAB before and after purification in an aluminum column}
\label{fig:ls:column}
\end{center}
\end{figure}

The attenuation spectra of Fig.~\ref{fig:ls:column} clearly demonstrate that column purification does not only remove the impurities featuring absorption in the region from 330 to 410\,nm, but also improves transparency in the critical range around 430~nm. These findings have been confirmed by long-range attenuation length measurements. Column purification increased the attenuation length of an LAB sample originally featuring $\ell_{\rm att}\approx 9$\,m to 20\,m.
Moreover, the most transparent raw product, Nanjing special LAB, featured an attenuation length of 25\,m after purification. Provided the manufacturer can produce this high-quality LAB in large quantities, aluminum column purification will be sufficient to obtain a LS meeting the experimental specifications.

Further laboratory studies are foreseen for the future:
\begin{enumerate}
\item Optimization of the Al$_2$O$_3$ column purification
\item Study of alternative column packings, such as a molecular sieve
\item Regeneration of the used Al$_2$O$_3$
\item Types and concentration of impurities in LAB which are removed by the  Al$_2$O$_3$ column.
\end{enumerate}

\subsubsection{Further Purification Techniques Applied to LS}

Distillation cannot completely remove the radioactive contaminants dissolved in LAB. Thus, water extraction and nitrogen stripping are applied after mixing of distilled LAB, PPO and Bis-MSB.
\medskip\\
{\bf Water extraction} utilizes the polarity of water molecules to extract polarized impurities and radioactive free-state metal ions from LAB. The method efficiently removes metallic radionuclides such as $^{238}$Th, $^{232}$U, $^{210}$Bi and $^{40}$K.
\medskip\\
{\bf Gas stripping} is used for purging the LS from radioactive gases, mainly $^{85}$Kr and $^{39}$Ar, oxygen (which will decrease the light yield due to photon quenching) and water (introduced by water extraction).

\subsubsection{Purification of Fluors}

The purification of the fluors has to be carefully considered for the JUNO experiment. In the Daya Bay experiment, purification of PPO was performed in cooperation with a domestic chemical company. A domestic producer has also been identified for Bis-MSB: The provided sample meets the requirements concerning light yield, while it will have to be purified to obtain the necessary optical transparency. Moreover, experience from Borexino suggests that the radiopurity of the fluors is an important issue that has to be studied by laboratory experiments.

\section{Radiopurity Studies}

An important prerequisite for the detection of low-energy neutrinos is the radiopurity of the detector and especially the target material, i.e.~the LS. The residual contamination aimed for is on the level of the neutrino event rate, so in the order of hundreds of counts per day within the target volume. In case of antineutrino detection, the inverse beta decay coincidence signal somewhat relaxes the requirement on radiopurity as the fast double signature allows to discriminate most of the single event backgrounds.

\begin{table}
\begin{center}
\begin{tabular}{ccc}
\hline
Isotope & Antineutrinos & Solar neutrinos\\
\hline
$^{232}$Th & $\leq 10^{-15}$\,g/g & $\leq 10^{-17}$\,g/g \\
$^{238}$U & $\leq 10^{-15}$\,g/g & $\leq 10^{-17}$\,g/g \\
$^{40}$K & $\leq 10^{-16}$\,g/g & $\leq 10^{-18}$\,g/g \\
\hline
\end{tabular}
\caption[Requirements for LS radiopurity]{Requirements for LS radiopurity concerning uranium, thorium and potassium, listed both for antineutrino and for solar neutrino detection.}
\label{tab:ls:radiopurity}
\end{center}
\end{table}

Radiopurity levels are usually specified by the concentration of $^{232}$Th, $^{238}$U and $^{40}$K in the LS. The basic requirements  for the JUNO LS are listed in Tab.~\ref{tab:ls:radiopurity}: The baseline scenario assumes a contamination on the level of $10^{-15}$ gram of U/Th and $10^{-16}$ gram of $^{40}$K per gram of LS, which will be sufficient for the detection of reactor antineutrinos. More stringent limits have to be met for the detection of solar neutrinos by elastic neutrino-electron scattering. Here, $10^{-17}$\,g/g resp.~$10^{-18}$\,g/g should be reached. These requirements are more than a factor 1000 resp.~10 above the current world-record held by the LS of the Borexino experiment. However, due to the scale of the project, achieving these radiopurity levels is a demanding task and will have to be studied and planned in considerate detail.
\medskip\\
While members of the natural $^{232}$Th and $^{238}$U decay chains are the most common contaminants, also other sources of radioactive impurities for the LS have to be taken into account. Moreover, the contamination may arise from different sources that have to be avoided or at least controlled:
\begin{itemize}
\item {\bf Dust particles} containing elements of the natural U/Th decay chains as well as radioactive potassium $^{40}$K.

\item {\bf Radon emanation}, especially the more long-lived $^{222}$Rn that is released from building materials (such as granite, brick sand, cement, gypsum, etc.) but also plastics, cables etc.

\item The {\bf radioactive noble gases $^{85}$Kr and $^{39}$Ar} can be introduced as residual contamination of the nitrogen used to create an inert atmosphere in the liquid handling system or by exposure to the ambient air

\item {\bf Surface depositions of $^{210}$Pb:} $^{222}$Rn will settle on surfaces exposed to air and decay to the long-lived isotope $^{210}$Pb ($\tau \sim 30$\,yrs), which in turn decays into $^{210}$Po and $^{210}$Bi.

\item {\bf $^{210}$Po} from surface contaminations, distilled water or other unknown sources.

\item {\bf Radioimpurities of the fluors}, mostly $^{40}$K.

\item {\bf Radioactive carbon $^{14}$C } intrinsic to the hydrocarbons of the LS.
\end{itemize}
While the majority of these studies is currently carried out at Chinese institutes, the European collaborators and especially the fraction that is involved in the Borexino experiment has a considerable experience in radio-purification. Several lab-scale studies are foreseen for the immediate future.

\subsection{Experimental Setups}

The aimed-for radiopurity levels of the JUNO specifications are too low to be measured by standard laboratory-scale experiments. Gamma spectrometers can only establish upper limits on the level of contamination. However, the effectiveness of purification techniques can be tested by artificially loading LS samples with greater amounts of radioactive elements and measuring the activity before and after applying a purification step. Moreover, the close relations with the Daya Bay experiment will allow a test of LS radiopurity by filling a large sample of LS ($\sim$20 tons) into one of the subvolumes of a Daya Bay Antineutrino Detector (AD). The AD will both offer the necessary capacity to hold a significant quantity of LS and a suitable low-background environment to test radiopurity to the level of the specificationsf~\cite{Ford:2011zza,Mark:2008gc,Mike:2008gc}.

\subsubsection{Low-background Gamma Spectrometer}

A standard laboratory technique to measure level and type of radioactive impurities inside a LS sample are gamma spectrometers. The device currently in use has been employed to measure LS samples before purification. Corresponding to the minimum sensitivity of the setup, only an upper limit can be set on the sample activity, which is smaller than 0.1\,Bq/kg. The corresponding concentration of $^{238}$U should be smaller than $8.1\cdot10^{-9}$\,g/g, so 6$-$7 orders of magnitude above the JUNO specification levels. The activity corresponding to the specified level of $10^{-15}$\,g/g is on the order of $10^{-8}$\,Bq/kg and thus clearly beyond the sensitivity of this and comparable setups.

\subsubsection{Assays with Loaded Scintillator Samples}

\begin{figure}[htb]
\begin{center}
\includegraphics[width=12cm]{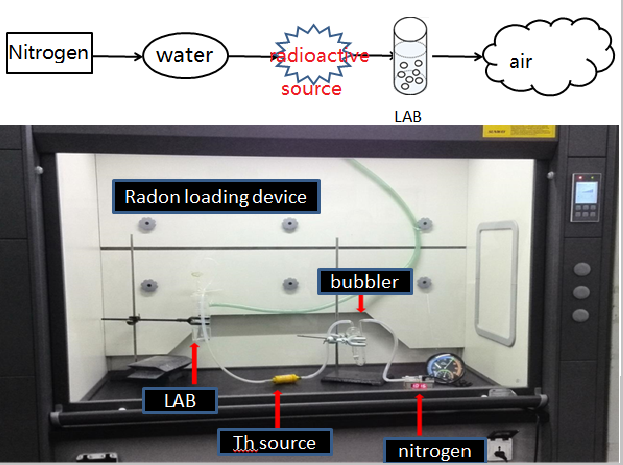}
\caption[Laboratory setup for radon-loading of LS samples]{Laboratory setup for radon-loading of LS samples}
\label{fig:ls:rnload}
\end{center}
\end{figure}

\begin{figure}[htb]
\begin{center}
\includegraphics[width=6cm]{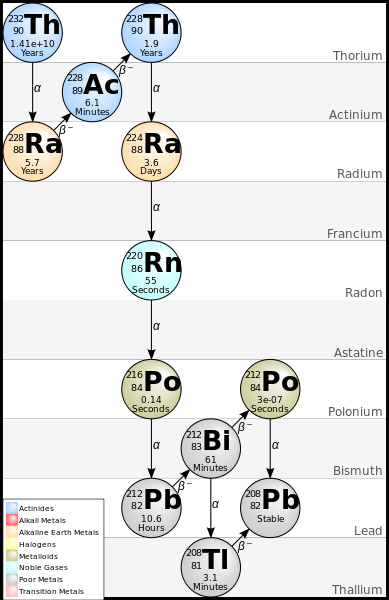}
\caption[Natural thorium decay chain]{The natural decay chain of $^{232}$Th}
\label{fig:ls:rnchain}
\end{center}
\end{figure}

\begin{figure}[htb]
\begin{center}
\includegraphics[width=12cm]{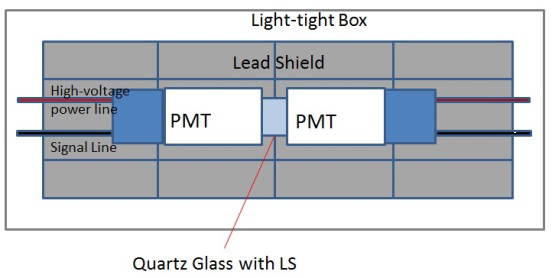}
\caption[Setup for measuring radio-purification efficiency]{Laboratory setup for measuring the efficiency of radio-purification}
\label{fig:ls:counter}
\end{center}
\end{figure}

\begin{figure}[htb]
\begin{center}
\includegraphics[width=14cm]{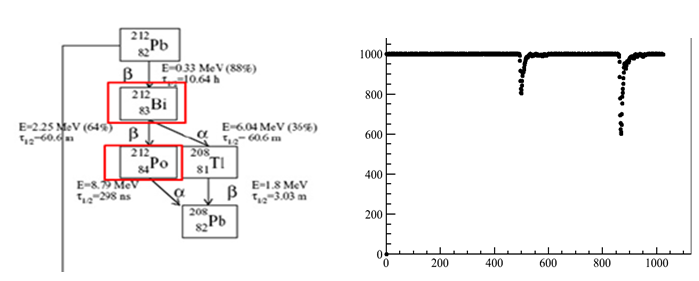}
\caption[Sample pulse of a fast $^{212}$Bi-$^{212}$Po coincidence]{Sampel pulse of a fast $^{212}$Bi-$^{212}$Po coincidence.}
\label{fig:ls:bipo}
\end{center}
\end{figure}

In order to investigate the effect of purification schemes on LS in laboratory experiments, the only solution is an artificial pollution of the samples with radioactivity, bringing the activity to a level accessible by small-scale counting experiments. This is achieved by loading the LS with suitable radioisotopes.

The experience gained in the purification campaigns carried out in Borexino suggests that the radioisotope most difficult to remove from the LS is the long-lived $^{210}$Pb. Therefore, it is crucial that purification tests will be carried out using$^{212}$Pb as a tracer. While easier to tag in lab experiments, its chemical properties correspond to $^{210}$Pb. However, direct deposition of lead in LS is difficult as metallic powders will not dissolve in the organic liquid. Instead, the method commonly chosen for loading is the exposition of an LS sample to radon which will in turn decay into lead.

The corresponding setup inside a glove box is shown in Fig.~\ref{fig:ls:rnload}: A solid $^{232}$Th source contained in a small box is used to produce the noble gas $^{220}$Rn ($T_{1/2}=56\,$sec)  by emanation (cf.~Fig.~\ref{fig:ls:rnchain}). This box is constantly ventilated by an air pump, creating a stream of radon-enriched air that is lead by a hose to a further container holding the LS sample. As a noble gas, radon is easily dissolved into the LS by purging the sample. It rapidly decays to $^{216}$Po and then $^{212}$Pb, which is $\beta$-unstable with a half-life of $T_{1/2}=10.6\,$h and is thus well-adjusted for performing purification tests. Its daughters $^{212}$Bi and $^{212}$Po decay in a fast coincidence that provides a further experimental tag.

The activity is measured by the experimental setup depicted in Fig.~\ref{fig:ls:counter}. In a light-tight box, a pair of 2'' PMTs is placed on both sides of an LS sample cell to detect the light produced by the $^{212}$Pb decays. Gamma rays from ambient radioactivity are attenuated by a shielding of low-activity lead bricks. Thus, the background rate for the $^{212}$Pb measurement is 0.5\,s$^{-1}$. If the fast coincidence of $^{212}$Bi and $^{212}$Po is used, the background rate is further reduced to only 2\,d$^{-1}$. A sample waveform of a Bi-Po event acquired in this setup is shown in Fig.~\ref{fig:ls:bipo}.

\subsubsection{Purification Test with a Daya Bay AD}

Once different purification schemes have been tested on the laboratory scale, the test in a medium-sized setup containing several tons of LS will be an important next step to assure that purification techniques are efficient down to the extremely low radiopurity levels of the JUNO specifications. The Antineutrino Detectors (ADs) of the Daya Bay Experiment are ideally suited to do such a study. The low-background conditions provided by the underground labs (the Daya Bay Near Experimental Halls are more than 100 meters underground) as well as the low-background material used in constructing the ADs themselves provide favorable conditions to measure the residual radioactivity of up to 20 tons of LS filled into one of the acrylic volumes of the AD in question.

Moreover, an AD test setup allows the practicability of the purification scheme on the scale of a system able to handle $\sim$100 liters of LS per hour. For this purpose, a circular system will be set up. After initial filling with an LAB-based LS replacing the original LS in the AD, purification techniques will be probed in loop mode by monitoring the change in radiopurity levels. This setup will provide important information for the design of a system able to handle the 20,000 tons of LS for the filling of the JUNO detector.

Beyond radiopurity, the filled AD will also provide information on the scintillation and optical properties of the LS. The large-scale setup allows to study the effect of purification on both the light yield and the light attenuation in the scintillator under realistic conditions.

\subsection{Purification Methods}
\label{sec:ls:radiopurity}

For the moment, the methods investigated for their purification efficiency concerning radioactive contaminants comprise aluminum columns, distillation, water extraction, and nitrogen stripping (cf.~Sec.~\ref{sec:ls:opt_purification}).

\subsubsection{Aluminum Columns}

Purification in aluminum Al$_2$O$_3$ columns proves to be very efficient in removing organic impurities from LAB, resulting in increased transparency. Beyond this, laboratory studies also show a decrease in radioactive impurities by adsorption on Al$_2$O$_3$.

The Al$_2$O$_3$ best suited for radio-purification was selected by gamma spectroscopy. Results of a comparative study are shown in Table~\ref{tab:ls:al2o3}. The Al$_2$O$_3$ featuring the lowest activity levels in $^{238}$U, $^{232}$Th, and $^{40}$K is produced by the Zibo Juteng chemical company ($\gamma$-type).

\begin{table}[!htb]
\begin{center}
\begin{tabular}{cccccc}
\hline
Nuclide & energy & XP type  & $\gamma$ type & (GY)Al$_2$O$_3$  & hamamasu low  \\
 &   & Al$_2$O$_3$ & Al$_2$O$_3$ & specific activity & background glass \\
 & keV & Bq/kg & Bq/kg & Bq/kg & Bq/kg \\
\hline
$^{232}$Th & 583.1 & 3.94 & 0.07 & 0.52 & 2.02  \\
$^{238}$U & 609.4 &3.93 & 0.16 & 0.60 & 5.46 \\
$^{40}$K & 1460.8 & 5.49 & 0.54 & 4.30 & 24.14 \\
\hline
\end{tabular}
\caption[Comparison of Al$_2$O$_3$ radioactivity levels]{Comparison of Al$_2$O$_3$ radioactivity levels.}
\label{tab:ls:al2o3}
\end{center}
\end{table}

The radiopurity assay was based on a  $^{220}$Rn-loaded LAB sample of 211\,ml volume. Measurements based on counting of $^{212}$Bi/Po coincidences were performed before and after column purification. A spin-type column was used, with a fill height of 20\,cm of Al$_2$O$_3$. Based on the  $^{212}$Bi/Po count rate, a radio-purification efficiency of 99.4\,\% could be determined.

For the future, an extension of these studies is foreseen at European institutes, especially INFN and JINR Dubna, which will also comprise adsorbants other than aluminum oxide.

\subsubsection{Distillation of LAB}

Distillation relies on the difference in boiling points and volatility of the LS components and possible impurities (see above). In particular, excellent efficiency is expected for the removal of radioactive metal ions because of the vast difference in boiling points compared to LS components. However, there are also large differences between the boiling points of the solvent LAB and the solutes PPO and Bis-MSB, which make the application of this technique very demanding once the three components are mixed for forming the LS. Thus, distillation is most suited for a purifications stage before mixing.

\begin{figure}[htb]
\begin{center}
\includegraphics[width=14cm]{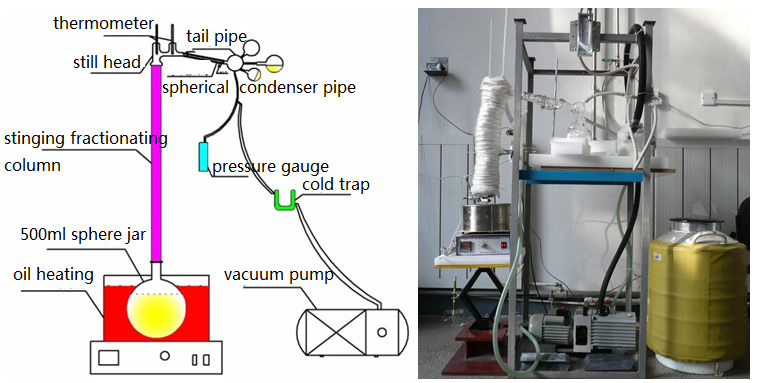}
\caption[Setup for distillation tests]{Laboratory setup for distillation tests}
\label{fig:ls:dist2}
\end{center}
\end{figure}

The effectiveness of distillation regarding radio-purification has been tested in the laboratory setup depicted in Fig.~\ref{fig:ls:dist2}. Radioactivity levels of a $^{220}$Rn-loaded LAB sample have been determined before and after distillation. Based on $^{212}$Bi/Po coincidence counting the efficiency has been determined to 98.4\,\%. Further tests will be performed after adding a reflux unit that should further improve the purification efficiency.

In the current setup, we find that after distillation with high temperatures or long heating times the LS samples feature a reduced attenuation length, change color and release a strong smell. A possible reason is the chemical break-up of organic impurities by the heating process. Possible solutions are an optimization of the distillation process or a reduction of heat-sensitive impurities in the LAB raw materials. It should be noted that similar laboratory-scale distillation tests at INFN Perugia did not show a degradation of the LS.

In the final large-scale setup for JUNO, vacuum distillation will be used for a reduction of the necessary heating temperature and the related energy consumption. To reduce the energy consumption on the experimental site,  the distillation plant could be realized directly at the LAB factory. The preliminary design of the installation comprises multistage plates and a vacuum
distillation tower for LAB with a processing capacity of up to 12,000 liters per hour, operated at a temperature  of $220\,^{\circ}$C. These preliminary working parameters will be refined by further studies carried out at INFN institutes.

\subsubsection{Water Extraction}

Purification by water extraction relies on the polarity of water molecules to separate polarized impurities, e.g. free-state ions of radioactive metals, from the non-polar LAB and fluor molecules. Mixing of water and LS is performed in a purification tower that contains a counter-current water extraction column. Pure water is filled at the top while LS enters at the bottom of extraction column. The mixing of the two liquids is supported by a porous filling. As soon as a state of full mixing has been reached, the LS leaves at the top of the column while the waste water is collected by the drainage system at the bottom.

The water extraction method is highly efficient for removing metal ions: Radium is removed at the level of 96.5\%. For lead and polonium, the purification efficiency of  82$-$87\,\% is somewhat lower as both can form chemical bonds with the organic molecules of the solvent that are fairly stable at room temperature.

\subsubsection{Nitrogen Stripping}

Purging the LS by nitrogen aims both at the removal of dissolved radioactive noble gases (argon, krypton, xenon, and radon) in the liquid and the removal of oxygen, which acts as a quencher lowering the light yield by oxidation of the fluor molecules. Beyond active purification by purging, ultrapure nitrogen will be used for establishing a nitrogen atmosphere or blanket inside the liquid handling system and in the final detector to avoid gas contamination of the LS.

\begin{figure}[htb]
\begin{center}
\includegraphics[width=11cm]{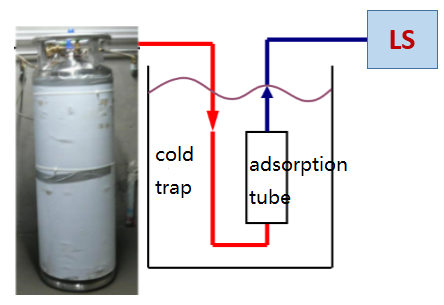}
\caption[Schematic drawing of ultrapure N2 production unit]{Schematic setup of the production unit for ultrapure nitrogen.}
\label{fig:ls:coldtrap}

\end{center}
\end{figure}

For laboratory purposes, industrial-grade high-purity nitrogen is mostly sufficient. However, the high solubility of noble gases in the LS makes it necessary to add a further purification step for the nitrogen before bringing it in contact with the liquid. For this, high-purity nitrogen of grade 6.0 (99.9999\,\% pure nitrogen) is passing through a cold trap at liquid argon temperatures. The resulting ultrapure nitrogen can reach radioactive background levels of 10$^{-17}$\,g/g, corresponding to 0.36\,ppm of $^{39}$Ar and 0.16\,ppt of $^{85}$Kr. Figure~\ref{fig:ls:coldtrap} shows a schematic drawing of a laboratory-scale setup for the production of ultrapure nitrogen based on an active carbon absorber submerged in a bath of liquid argon. The resulting radiopurity will be tested by GC-MS measurements (Agilent 7890A/5975, cf.~Sec.~\ref{sec:ls:gcms}).

%
%
%
%
%
%
%
%
%
%
%
%
%

\section{Mass Production of 20 kton LS}

{\bf Production. } The production of 20 kilotons of LS will be achieved in three or four cycles over a period of 1$-$2 years. Based on the current status of laboratory studies (see above), the solvent LAB will most likely be provided by the company from Nanjing. Criteria for quality control will be developed jointly by IHEP and the company. Similar arrangements will be made for the solutes PPO and Bis-MSB.
\medskip\\
{\bf Transport.} A professional logistics company will take care of the transport of the LAB from the production line at Nanjing to the JUNO experimental site. Suitable transport containers will be manufactured by a company specifically for this purpose, meeting specifications set by IHEP concerning radiopurity, air tightness and material compatibility with LAB.

\begin{figure}[htb]
\begin{center}
\includegraphics[width=15cm]{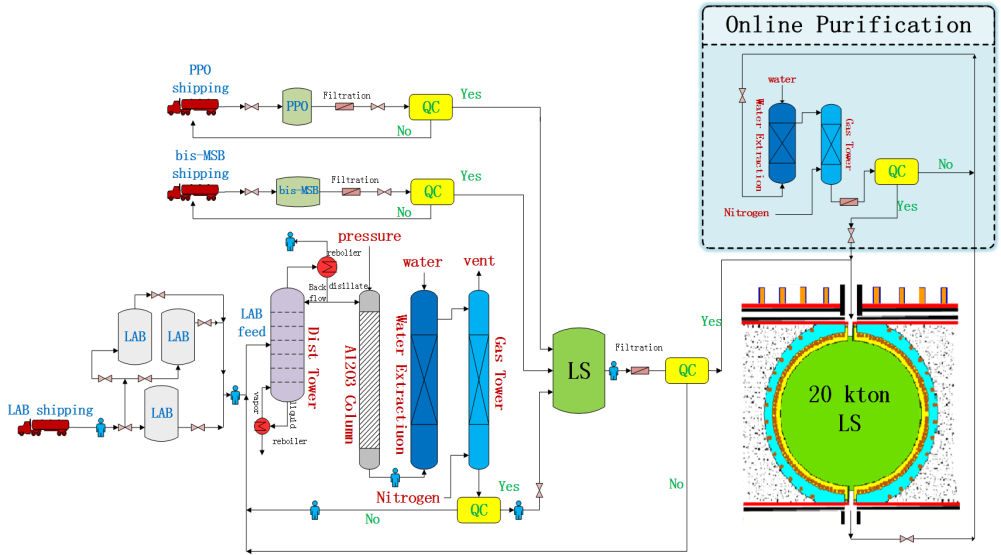}
\caption[Schematic drawing of LS purification plants]{Schematic drawing of the system of LS purification plants}
\label{fig:ls:filling}
\end{center}
\end{figure}
\bigskip
\noindent The on-site LS handling system is depicted in Fig.~\ref{fig:ls:filling}. The facilities include:
\medskip\\
{\bf Storage tanks.} Since the LAB will arrive in 3 to 4 batches of 5\,kt each, the foreseen storage area comprises about 5,000 square meters, featuring several tanks with a volume of 3 ktons and another set of smaller tanks on the scale of several hundred tons. This includes not only tanks for LAB, but also for white oil and temporal containers needed during unloading.  In addition, an area of 30\,m$^2$ is to be reserved for storage of the fluors.
\medskip\\
{\bf Surface purification plants.} Fig.~\ref{fig:ls:filling} depicts the basic scheme for the on-site system of purification plants: The current design mainly relies on an initial purification of the individual LS components (solvents and solutes) before mixing them to obtain the final LS. The major task of purifying LAB is to be performed on surface. The foreseen purification steps include distillation, adsorption in an aluminum oxide column, water extraction and nitrogen purging (cf.~sec.~\ref{sec:ls:radiopurity}).

Assuming a total mass of 20\,kt of LS and a period of 200 days reserved for continuous purification and detector filling, the required processing capacity of the purification plants is 100 tons per day. Based on the power consumption of similar (but smaller-scale) purification plants realized for the Borexino, KamLAND and SNO+ experiments, the corresponding power consumption is about 2\,MW.
In addition, a supply of ultrapure nitrogen on a rate of 200\,m$^3$/h as well as ultrapure water of 6 tons per hour has to be assured. The system will require a surface clean room of 200\,m$^2$ area and 10\,m height for its installation. In addition, rooms for the production plants of ultrapure water and nitrogen are needed (30\,m$^2$ each).
\medskip\\
{\bf LS mixing plants.} After purification of the individual components, LS mixing will be performed. A complication arises from the low solubility of Bis-MSB on LAB. To reduce the processing time and assure a constant amount of the fluor in the LS, batches of highly concentrated LAB-Bis-MSB solution (150\,mg/l) will be prepared beforehand and then watered down with pure LAB during mixing.

The mixing tanks will be made from organic glass to ensure both the material compatibility of the LS and to allow for a visual inspection of the mixing product. The mixing system will be housed in a room of 300\,m$^2$.
\medskip\\
{\bf Quality assurance and control.} Constant control of the LS leaving the purification and mixing plants is of uttermost importance for the success of JUNO. As the LS is produced in cycles, even a single low-quality batch passing the controls unnoticed and filled into the detector could potentially spoil the high-quality of all LS produced up to this point. Therefore, a careful program of QA/QC measures before and after purification and mixing has to be developed. Separate teams of operators and inspectors should be formed and be supervised by an on-site manager, following a strict protocol.

Two inspection rooms of a total area of 60\,m$^2$ will house systems for QC, including GC-MS for detection of organic impurities and small-scale experiments to monitor attenuation length and spectrum as well as light out of the LS.
\medskip\\
{\bf Underground installations.} Approved batches of LS will be sent by a pipeline to an underground storage area where it remains until detector filling. The pipe system should allow the flow of LS in both directions and feature high-pressure resistance to allow for a safe transport of the LS. Two storage tanks of 200 tons for LS and white oil are foreseen close to the detector cavern.

Beyond the surface plants for the initial purification of the LS, a secondary system is foreseen underground to perform loop-mode purification of the LS while the experiment is running. This online purification system will rely on water extraction and nitrogen stripping. The system will require two underground halls of 500\,m$^2$ area and 10\,m in height. The necessary supply of water, electricity and gas will be roughly equivalent to the requirements of the surface system.
\medskip\\
{\bf Cleanliness conditions.} Ensuring the cleanliness and purity of all materials in contact with the LS and its raw materials will be decisive for meeting the radiopurity requirements of JUNO. All pipes, pumps storage and processing vessels through which the LS passes from arrival on-site until filling of the detector will have to meet the requirements for chemical compatibility, cleanliness of all surfaces and air-tightness.

All surfaces have to be thoroughly cleaned and probably rinsed with LAB before being brought into contact with the LS. Removal of dust on surfaces is mandatory: At an average activity of 200 Bq/kg in U and Th, not more than 10\,g of dust are allowed to be dissolved in the total mass of 20\,kt of LS. What is more, piping and tank materials will be containing U and Th themselves and thus will potentially emanate radon that will diffuse into the liquid. This process can be mitigated by applying suitable surface liners stopping the diffusion.

The expected radon level in the underground air is about 100\,Bq/m$^3$. Contact of LS and process water to air has to be strictly avoided as the solubility of radon in LS is far greater than in water. To reach this goal, all the liquid handling system must be absolutely air-tight, and an atmosphere of ultrapure nitrogen will be established inside the system.
\medskip\\
All surface and underground facilities will be equipped with cameras, fire protection, monitors for oxygen and radon as well as alarm systems for poisonous gases.


\section{Risk Assessment}

\begin{table}
\begin{center}
\begin{tabular}{lcc}
\hline
Properties & LAB & Mineral Oil\\
\hline
Quantity & 20\,kt & 7\,kt \\
\hline
Appearance & odorless, & odorless,\\
	& colorless, & colorless,\\
	& transparent & transparent \\
Flammable & yes & yes \\
Toxic & no & no \\	
Corrosive & no & no \\
\hline
Chemical formula & C$_6$H$_5$C$_n$H$_{2n+1}$ & C$_n$H$_{2n+2}$ \\
 & ($n=10-13$) & \\
Density [g/cm$^3$] & 0.855$-$0.870 & 0.815$-$0.840\\
Freezing point & -50\,$^{\circ}$C & \\
Flash point & 135\,$^{\circ}$C & 145\,$^{\circ}$C\\
Vapor pressure (20$^\circ$C) & 0.1\,mmHg & 0.1\,mmHg \\
\hline
\end{tabular}
\caption[Table of material properties]{Table of material properties}
\label{tab:ls:matprop}
\end{center}
\end{table}

The JUNO detector will contain both LS and and mineral oil. Material properties are listed in Table~\ref{tab:ls:matprop}.
\medskip\\
{\bf LAB. } The main component of LS is linear alkyl benzene (LAB), of which 20 ktons are required. In industry, it is mainly used in the production of detergents. For this, LAB is sulfonated to form linear alkylbenzene sulfonate (LAS), which in turn serves as raw material for washing powders, liquid detergents, pesticide emulsifiers, lubricants and dispersants. It is non-poisonous and odorless. For use in JUNO, the only safety concern is its flammability.
\medskip\\
{\bf Mineral oil.} The buffer volume surrounding the acrylic sphere might be filled with 7,000 tons of mineral oil. The foreseen type is food grade mineral oil 10\#. It is a liquid by-product of the distillation of crude oil which is performed during the production of gasoline or other petroleum-based products. Again, the only safety concern is flammability. Mineral oil is widely used in biomedicine, cell cultures, veterinary uses, cosmetics, mechanical, electrical and industrial engineering, food preparation etc.
\medskip\\
Due to the flammability of both liquids, the main risk to be taken into account is fire safety. The necessary precautions have to be taken into account during transport, on-site storage, purification and mixing of the LS components as well as during detector operation. 

From an experimental point of view, the intactness of the LS poses additional constraints on all operations. Cleanliness and tightness of the liquid handling system have to be considered carefully.

Further risks will be induced by the installation of the LS purification plants and storage tanks in an underground cavern at a depth of 700\,m. This will result in complicated hoisting equipment, high-altitude operations, and a need for accurate positioning. Purification underground will pose additional risks because a possible spillage of nitrogen are more serious under these confined conditions. All operations of this kind will have to be carefully planned beforehand.

\section{Schedule}

\begin{itemize}
\item[2013] Realization of laboratory setups for LS properties.

First explorative measurements of LS properties.

Measurement of the depolarization ratio to determine the Rayleigh scattering length.

\item[2014] Realization of laboratory-scale setups for testing different purification method:\\
Low-pressure distillation, aluminum-oxide column, nitrogen stripping and water extraction.

First steps towards the optimization of the purification scheme:

Tests of optical purification: Measurements of light out, attenuation length and spectrum, and water content of the purified LAB.

Tests of radio-purification: Establish the technology for injecting $^{212}$Pb and $^{224}$Ra in LS samples and setup of the corresponding measurement device to study the effect of purification on radioactive contamination levels.

Start of joint research programs with companies providing LAB.

\item[2015] Design and realization of a purification system.

Final determination of purification parameters.

Start of measurements for electron energy non-linearity.

Design and setup of the quality control devices.

Start of joint research activities with providers of PPO and Bis-MSB.\\

Conception and setup of devices for quality testing.

\item[2016] Jan$-$Jun: Setup and testing of purification plants at Daya Bay.

Jul$-$Dec: Installation and debugging the purification system.

\item[2017] Determination of the final recipe of LS production based on the experimental data and the purification methods.

Final measurement of light out and attenuation length of the LS.

Signing of contracts with the companies providing the purification plants and start of the design.

Signing of contracts with the providers for PPO and Bis-MSB.

Start of the batch production at the end of the year.

Finalization of the design of storage tanks for LAB. Design and build the LS purificaiton hall on the ground.

\item[2018] Signing of all contracts for LAB storage area.

Setup of the large storage tanks before the middle of the year:\\
3$-$5 of 3000$-$1000\,m$^3$ tanks for LAB, LS and white oil

Installation of the purification system until end of the year.

Setup of two 200-ton underground buffer tank.

Setup and testing of the LS QA/QC systems.

\item[2019] Testing of all purification plants.

Setup and testing of ultrapure water and nitrogen plants.

Design and build the LS mixture system.

Production of 8 ktons of LS of sufficient quality for the central detector.

\end{itemize}

\cleardoublepage
\chapter{Veto Detector}
\label{ch:VetoDetector}
\section{Experiment Requirements}

The main goal of JUNO is to determine the neutrino mass hierarchy, which is one of the most important unsolved problems related to neutrinos.

The neutrinos are detected via the inverse beta decay by measuring the correlated positron and neutron signals.
With a careful design of the detector, the neutrino spectrum can be measured precisely, and we are looking for the distortion of the spectrum for the sign of neutrino mass ordering. Compared to the small number of signal events (60/day), the number of background events is still very high due to the large volume of detector (20~kton). The cosmic ray muon induced backgrounds are the main backgrounds
and they are hard to remove.  The cosmic ray generated backgrounds are:

1. $^9$Li/$^8$He background from muon spallation and muon shower particles.

2. Fast neutron background in the detector from muon induced high energy neutrons.

The cosmic ray induced backgrounds also effect the study of the diffuse supernova neutrino flux.

In order to reduce the experimental backgrounds, the neutrino detector must be placed in deep
underground and a veto system is used to tag muons. The muons should be detected with
high efficiency for the purpose of background reduction. This chapter is
mainly about research and design of the veto system.

\begin{figure}[htb]
\begin{center}
\includegraphics[width=12cm]{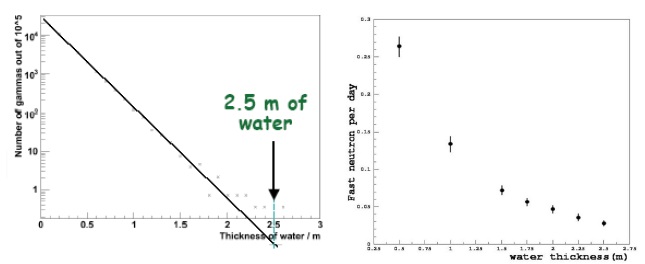}
\caption[Thickness of water vs gamma and neutron attenuation]{Left: Gamma attenuation vs thickness of water shield; Right: Neutron attenuation vs thickness of water shield.}
\label{gamma_neutron_shield}
\end{center}
\end{figure}

Due to the strict requirements  on background suppression,  larger overburden of rocks on top of the detector is needed  to reduce the cosmic ray muon flux.
The experiment is located at a site of about 53~km equal distance to the Yangjiang and Taishan nuclear power plants.
The height of the mountains is 270 meters and the experimental hall is located underground at the depth of 460 meters.
Therefore, there is about 700-meter rock on top of the experimental hall. Muon rate is estimated at about 0.003~Hz/$m^{2}$ and the average muon energy at about 214~GeV from simulation.
The cosmic ray muon flux is reduced by $\sim$60,000 times compared to that at the ground surface. The remaining energetic cosmic ray muons can still produce a large number of neutrons in
the rocks and other detector materials surrounding the central detector.  These neutrons can produce fast neutron background in the central detector which mimics the inverse beta decay signal. This kind of  background
cannot be ignored. In order to shield the neutrons and the natural radioactivities from the surrounding rocks, at least 2 meters of water surrounding the central detector is needed. The water
is effective for shielding against neutrons and gamma (Fig.~\ref{gamma_neutron_shield}). And when being instrumented with photomultiplier tubes (PMTs),  the water pool can serve as a
water Cherenkov detector to tag muons. The Daya Bay experiment shows that the veto system utilizing water Cherenkov detector can be very successful \cite{DYB:muon2014}.
From simulation, muons with relatively long track in the detector can be detected with very high efficiency, and the undetected muons are mainly of short track lengths, which would induce
less background because they are relatively away from central detector. Based on the Daya Bay experimental results (fast neutron background about $\sim$0.2\%), if the water shield thickness is at least 2.5 meters in JUNO, fast neutron background to
signal ratio is $\sim$0.3\%, after taking into consideration the large detector volume and geometry effects.

Another background is produced from $^9$Li/$^8$He resulting from muon spallation in the
scintillator and muon shower particles. The beta-n decay of $^9$Li/$^8$He would mimic inverse beta events. This kind of background can not be reduced from increasing
the thickness of the water shield. The number of $^9$Li/$^8$He background events is estimated to be $\sim$80/day. Their existence would greatly reduce JUNO's capability
to determine the neutrino mass hierarchy.
To reduced the $^9$Li/$^8$He background, we need precise muon track
information. The $^9$Li/$^8$He background is reduced by excluding a
certain cylindrical region along the muon track within a certain period of
time after the muon had passed through the detector. Therefore, the $^9$Li/$^8$He background reduction depends on precise muon track
reconstruction.  We will build a tracking detector on top of the water
Cherenkov detector to help muon tagging as well as track reconstruction. The top tracker can tag muons and  reconstruct muon tracks, independent of the
central detector and water Cherenkov detector. The tracking results from the top tracker can be extrapolated
to the central detector. Therefore, it can provide an independent measurement of the
$^9$Li/$^8$He background. The muons which pass through the water pool but not the central detector can also
produce high energy particles (pions, gamma, neutrons, etc.), which could migrate into
the central detector and produce $^9$Li/$^8$He there. A simple
simulation shows that this kind of background which cannot be detected by the
central detector is $\sim$1.2/day. A top tracker covering a large area can help directly measure such background.
Through optimization of the top detector area and its layout, the detector placed at the top with a long strip
structure provides better results. This configuration can identify up to 1/4 of this kind of
background, which would be helpful for the study of such background.
It can also be used to measure the rock neutron related background.
Also the top tracker and water Cherenkov detector can help identify multi-muon events for the central detector to reduce the $^9$Li/$^8$He background.

\section{Detector Design}

\begin{figure}[htb]
\begin{center}
\includegraphics[width=10cm]{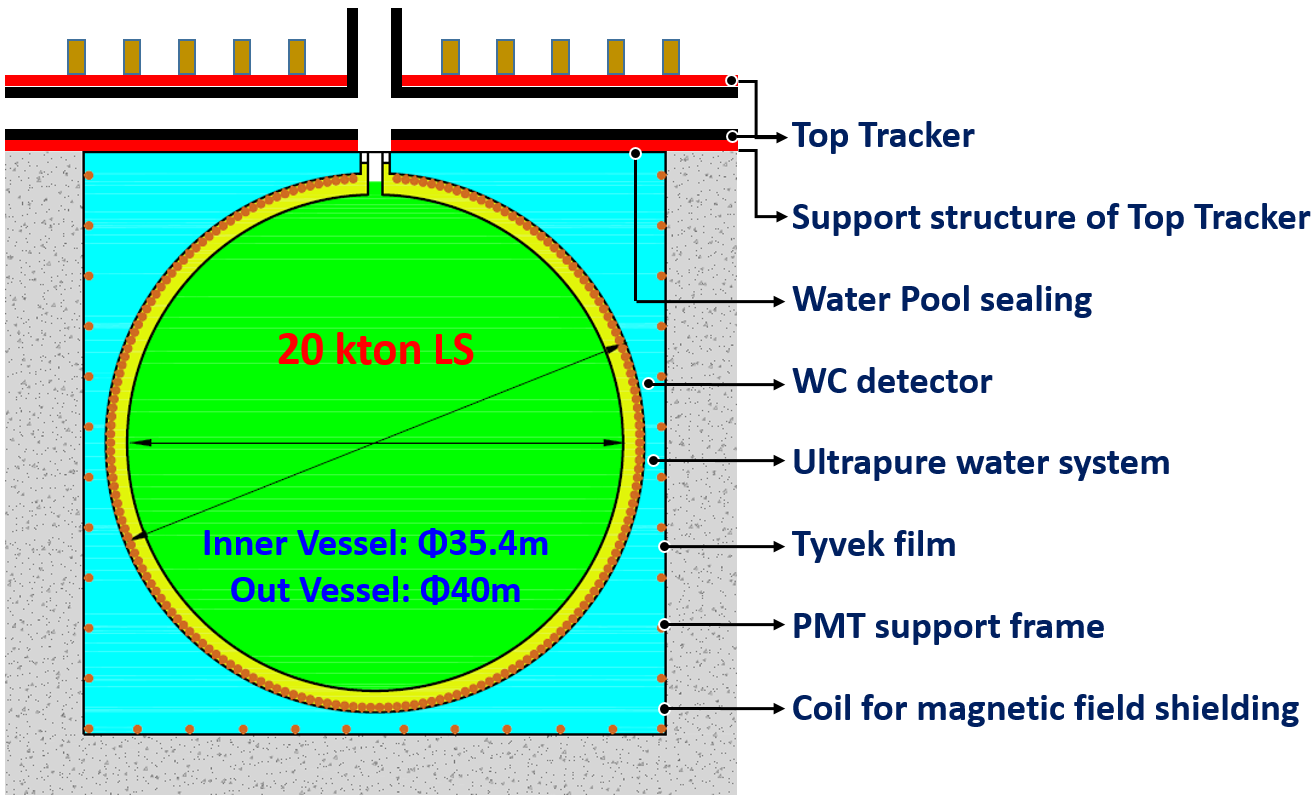}
\caption[Veto systems of JUNO]{Veto systems of JUNO.}
\label{vetodet}
\end{center}
\end{figure}

The JUNO veto system is shown in Fig.~\ref{vetodet}. The system consists of the following components:

1. Water Cherenkov detector system, a pool filled with purified water and instrumented with PMTs.  When energetic muons pass through the water, they can produce Cherenkov light. The Cherenkov
photons can be detected by PMTs. The arrangements of the PMTs as well as the number of PMTs needed in the water pool is currently under study to facilitate the high efficiency detection of muons as
well as good muon track reconstruction. One of the options is to have the water pool surface and central detector outer surface covered with reflective Tyvek to increase the light detection by PMTs without using a large number of PMTs.

2. Water circulation system. There will be 20,000-30,000 tons of water in the pool depending on the different central detector designs. This system will include a water production system on the ground and
a purification/circulation system underground in the experimental hall.

3. Top tracker system.
It can provide independent muon information to help muon tagging and track reconstruction. The OPERA top tracker are going to be transported and installed in JUNO.

4. Geomagnetic field shielding system.
Though small, the Earth's magnetic field can affect the performance of PMTs. Either compensation coils or magnetic shields will be used to reduce the effect on PMTs.

5. Mechanical system.
The system includes top tracker support structure, water pool PMT support structure as well as a light and air tight cover for the water pool.

\subsection{The Central Detector as a Muon Tracker}

The large liquid scintillator central detector itself is one of the most
important muon trackers in the veto system. The muon rate for
the central detector is about 3~Hz due to 700-meter rock overburden above.
A minimum-ionizing muon deposits about 2~MeV cm$^{2}$/g.
On average, a muon will pass through 23~m of LAB based liquid scintillator
(density of about  0.85~g/cm$^{3}$) inside the central detector, resulting in
about 3.9~GeV energy deposition.

A muon generates Cherenkov photons and scintillation photons along its
track in the detector.  The Cherenkov photons are emitted in a cone with an opening angle $\theta = \cos^{-1}(1/n)$ relative to the muon track trajectory,
where $n$ is the index of refraction for liquid scintillator $\approx 1.45$.
Although the scintillation photons are emitted isotropically,
 the path for the fastest photon arriving at any PMT
follows the same angle as that of the Chrenkov light angle $\theta$ when
the velocity of the muon is nearly the speed of light.
Based on the earliest hit time from each PMT, it is straightforward
to reconstruct the muon track. Such a method, called the ``Fastest-Light''
muon reconstruction, has been proved to be very successful in
reconstructing more than 99\% of the non-showering muons in the KamLAND
experiment.
Based a preliminary muon tracking study on the simulated JUNO muon tracks,
the bias of this method on the zenith angle is less than 5$^{\circ}$ and
the distance to the detector center bias is less than 20~cm.
Although the current study has not considered many real detector effects,
such as the resolution of PMT transition time,  given the much better energy resolution and comparable or better PMT transition time resolution
for JUNO detector compared with KamLAND,
the fitting quality is expected to be further improved.

However, the fitting quality of this algorithm becomes worse
for the corner clipping muons, which are close to the edge of the detector.
And more importantly, this reconstruciton method is also not appropriate for
multiple muons or showering muons inside the detector, which is estimated to be about 10-20\% of all muons.  Since the muon rate for the JUNO central detector is about 10 times higher
than that of the KamLAND detector, it is not acceptable to veto the whole detector for a few milliseconds, which will significant reduce the detector live time.
In order to reduce the $^{9}$Li/$^{8}$He background introduced by those muons,  we have to rely on the other veto system, such as the water Cherekov detector and the top tracker to track those muons.

\subsection{Water Cherenkov Detector}

\begin{figure}[htb]
\begin{center}
\includegraphics[width=6cm]{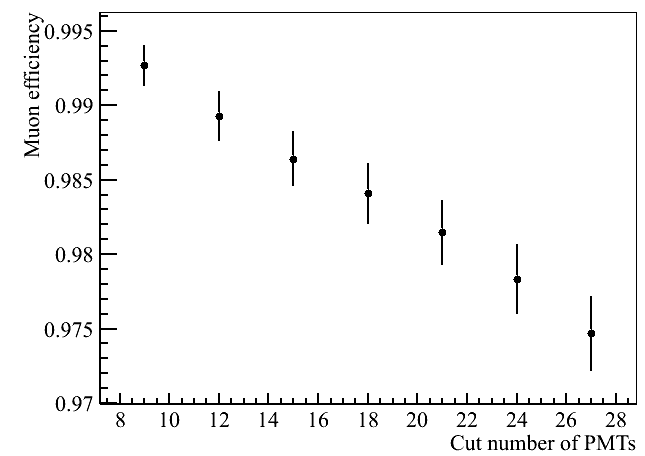}~
\caption[Muon detection efficiency vs Number of PMT threshold]{Muon detection efficiency vs threshold in number of PMT (1,600 PMTs, with a number of PMTs threshold at 20, the efficiency is larger than 97\%).}
\label{junoEffi}
\end{center}
\end{figure}

\begin{figure}[htb]
\begin{center}
\includegraphics[width=8cm]{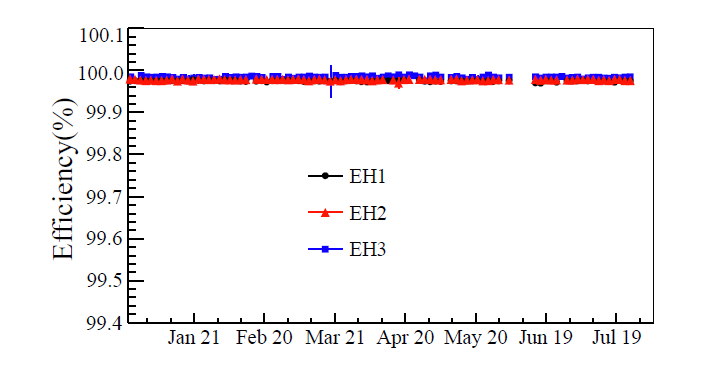}
\caption[Muon efficiency vs time]{Detection efficiency (vs time) for long track muons going through antineutrino detector of Daya Bay. }
\label{iwsEffi}
\end{center}
\end{figure}

PMTs are placed in the water pool to tag cosmic ray muons by detecting water Cherenkov light that is produced. The number of photons produced is proportional to the muon track length in the water.
Due to the limited thickness of water as well as the presence of the central detector, the majority of the muon tracks going through the water pool are not very long. Therefore, a large photo-coverage
in the water pool is desired to detect muons with high efficiency. For the purpose of tagging muon, normally the pool surface would be covered with high reflectivity Tyvek film to help collect the photons
 without using a large number of PMTs.
Based on the experience of the Daya Bay experiment and with Geant4 for  simulation, muon detection efficiency is about 98\% in (Fig.~\ref{junoEffi}) by using 1,600 8-inch PMTs with at least 2.5 meter
thickness of water in the pool, if we can keep the noise level as low as that of the Daya Bay experiment. The long track muons can be detected with extremely high efficiency, as shown in
Fig.~\ref{iwsEffi}. Since the central detector size is larger than that of the Daya Bay detector, the large surface area of stainless steel tank might have a big effect on light transmission. We carried out a
simulation and found that there is no major impact on detector performance if the surface of the steel tank is covered with reflective Tyvek.

The challenge for the JUNO water Cherenkov detector is that we not only need to tag the muon with high efficiency but also want to reconstruct the tracks of those muons.
Ideally full photo-coverage on the water pool surface would be the best choice. Yet the budget constraint limit the number of PMTs available for the water pool.
Considering the directional nature of the Cherenkov light and that all muons are coming from outside, it would be natural to have PMTs installed on the central detector.
Only few clipping muons at certain particular incident angles would not have direct photons projected on the central detector if not utilizing reflective water pool surface.
Some of these clipping muons would be detected by the top tracker. Most of these clipping muons are further away from central detector ($>$ 2 m) which would less likely to
produce cosmogenic backgrounds in the central detector.
The possibilities of different configurations of the central detector as well as the uncertainty  to utilize the outer surface of the central detector, require careful study on PMT arrangements,
placement as well as the number of PMTs in the water pool. Large photo coverage enhances the muon detection but also drives the cost high. Therefore, it needs study of
utilizing larger size PMTs or wavelength shifting plates around the PMTs as done in the Super-Kamiokande experiment \cite{SK:detector2003} and find the optimal solution.

Since the muon track reconstruction is important for $^9$Li/$^8$He background reduction, we would consider to increase the water pool muon track reconstruction capability for the
purpose of background reduction. For JUNO, the shower muons will induce non-negligible  dead time if the central detector can't reconstruct the shower muon tracks due to detector
saturation. If the water Cherenkov detector can help reconstruct those muon tracks, it would reduce the central detector dead time.
Reflective surfaces can enhance the Cherenkov photon detection, and therefore help tag muons. But the reflection would smear out the position information. We can use more PMTs to build
optically segmented water Cherenkov detectors to improve muon reconstruction resolution. But these segmented detectors could normally provide only one point along the muon track, therefore
it might need to have multiple layers of these kind of detectors installed to reconstruct the muon tracks, which would pose engineering as well as installation challenges.
In the mean time, developing better algorithms for track reconstruction utilizing both the PMT charge and timing information would be useful. For this purpose, a better water pool PMT calibration
system is essential to not only calibrate the PMT gains but also the PMT timing.  Now, all these options are under study.

\begin{figure}[htpb]
\centering
\includegraphics[width=0.36\textwidth]{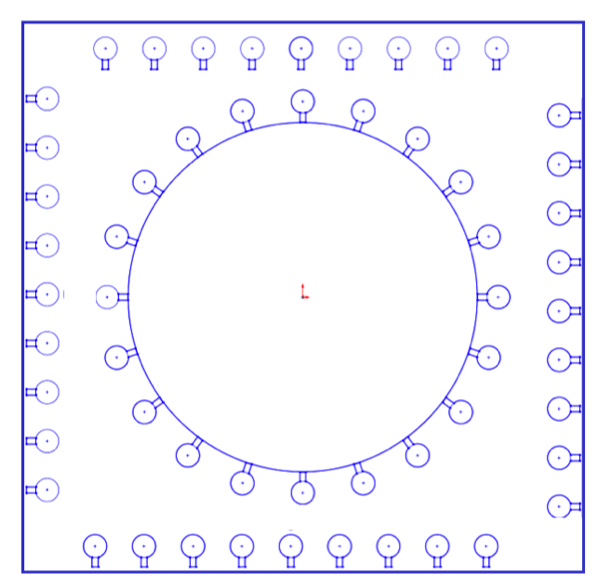}
\caption{Simple water Cherenkov detector option. Uniformed distribution of PMTs.
PMTs at the top of water pool are facing outward. PMTs at the bottom and on the
barrel are facing inward.}
\label{fig:watertank_config1}
\end{figure}

\begin{figure}[htpb]
\centering
\includegraphics[width=0.42\textwidth]{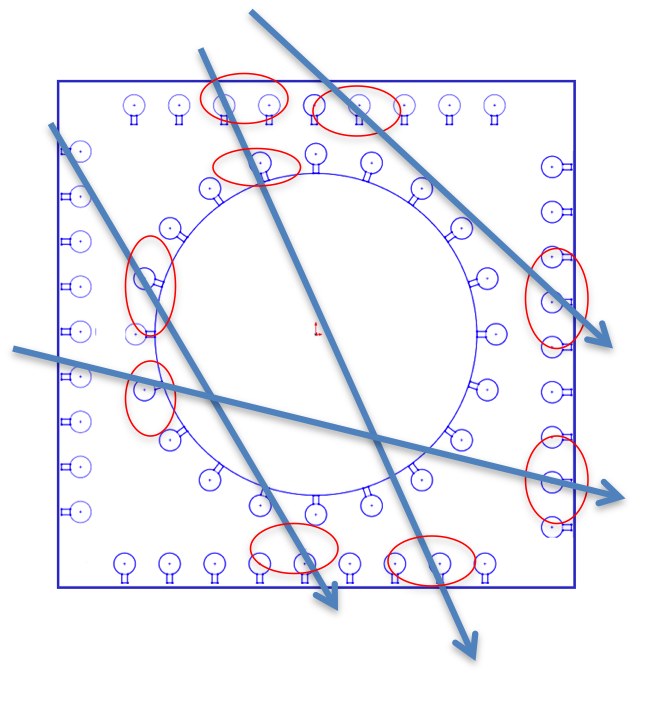}
\caption{Various muon tracks with reconstructed points at pool surface or
central detector. }
\label{fig:watertank_config2}
\end{figure}

The simplest option would be to have a single water Cherenkov detector by
installing PMTs on the surface of the water pool. There are other options, such  as having multiple layers of  PMTs or modular
PMT boxes installed. Since there is not much space in the water pool and there
will be structures for the central detector in the pool, therefore from the
engineering point of view, a simple water Cherenkov
detector is desired.  A simulation was performed to study the muon
reconstruction capability of the simple water Cherenkov detector. PMTs are
uniformly distributed on the surface of the water pool
as well as the outer surface of the central detector
(Figure~\ref{fig:watertank_config1}). Of the 2,000 PMTs, 200 PMTs are on the top
facing toward the outside of the pool, 200 PMTs are at the bottom and
800 PMTs are on the barrel all facing inward. Another 800 PMTs are on the outer
surface of the central detector. This gives roughly 2.5~m distance
between PMTs. All the surfaces are covered by Tyvek with reflectivity of 95\%.
Figure~\ref{fig:watertank_config2} shows various muon tracks with points on
different kinds of surface where the PMTs are installed

\begin{figure}[htpb]
\centering
\includegraphics[height=2.in,
width=0.25\textwidth]{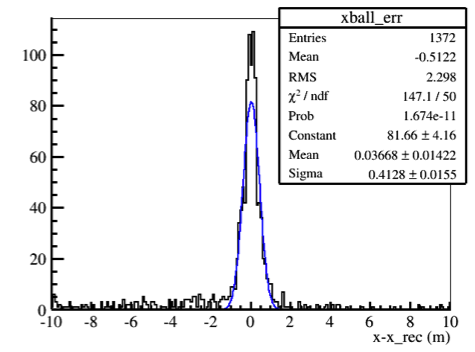}
\includegraphics[height=2.in,
width=0.25\textwidth]{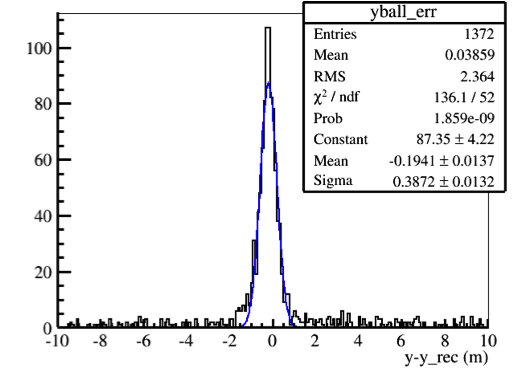}
\includegraphics[height=2.in,
width=0.25\textwidth]{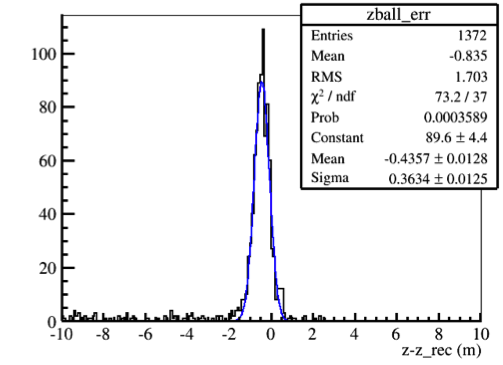}
\caption{Resolution of the reconstructed points on the central detector}
\label{fig:watertank_CDresol}
\end{figure}

\begin{table}[htdp]
\caption{Resolution of reconstructed points on the muon tracks.}
\begin{center}
\begin{tabular}{|c|c|c|c|}
\hline
 & X & Y & Z \\
\hline
\multicolumn{4}{|c|}{Points on the top of the pool}\\
\hline
Resolution & $\sim$1.5 m & $\sim$1.5 m & \\
\hline
\hline
\multicolumn{4}{|c|}{Points on the bottom of the pool}\\
\hline
Resolution & $\sim$0.8 m & $\sim$0.75 m & \\
\hline
\hline
\multicolumn{4}{|c|}{Points on the barrel of the pool}\\
\hline
Resolution & $\sim$0.3 m & $\sim$0.4 m & $\sim$0.8 m \\
\hline
\hline
\multicolumn{4}{|c|}{Points on the surface of CD}\\
\hline
Resolution & $\sim$0.4 m & $\sim$0.4 m & $\sim$0.4 m \\
\hline
\end{tabular}
\end{center}
\label{tab:watertank_muonpoints}
\end{table}

The muon tracks are represented by the interceptive points on either the top,
barrel, bottom of the pool or on the central detector. The points are obtained
using charge center method.
The reconstructed points on the barrel and bottom are the exit points of the
muon track. And the reconstructed points on the top and central detector are the
entry points.
Table \ref{tab:watertank_muonpoints} shows the resolution of the reconstructed
points at various location. The reconstructed points on the top are poor due to
thin water the muons going
through and thus having fewer photons for PMTs. The reconstructed points on the
barrel are good for X and Y and average with a bias of  $\sim$0.5~m for Z, this
could be the results of  all the tracks
are going downward. The reconstructed points are very good on the central detector
for all coordinates. It might be possible to use the reconstructed points from
top tracker for muons passing through the top of the pool.

It is essential to have good reconstructed muon track in central detector. Though the central detector can be excellent muon tracker when the muon path is close
to the center of the central detector, the muon
track reconstruction gets worse when the muon is close to edge of the central
detector. Therefore, the presence of good reconstructed points from the water
Cherenkov detector would definitely help and cross
check with the central detector muon track reconstruction. Optimizing the number
of PMTs, the PMT arrangement as well as taking into account the PMT hit timing
information in reconstruction definitely warrant
further study.

Of all the muons, about 10\% are multiple muons. For double muons, it was
estimated that the central detector could distinguish the muons with distance
greater than 4~m. Water Cherenkov detector with good resolution
of muon entry points on the central detector has the potential to be able to
separate multiple muon tracks by looking for clusters of hit PMTs. This would be
helpful to reduce the dead time caused by multiple
muons.

\subsubsection{Tyvek Reflector Film}

\begin{figure}[htb]
\begin{center}
\includegraphics[width=8cm]{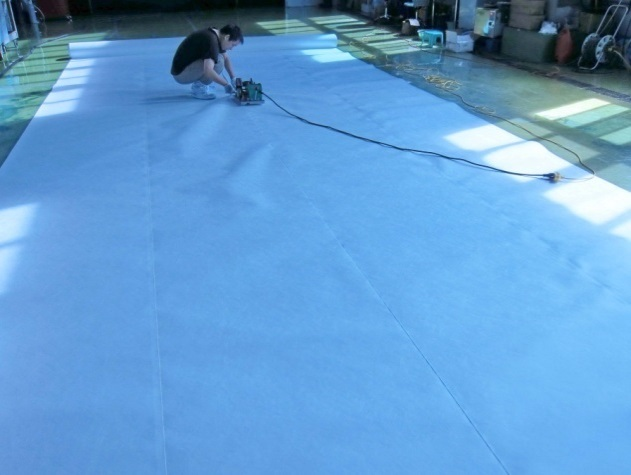}
\caption[Production of the Tyvek reflector]{The process of making the tyvek reflector.}
\label{MakeTyvek}
\end{center}
\end{figure}

\begin{figure}[htb]
\begin{center}
\includegraphics[width=8cm]{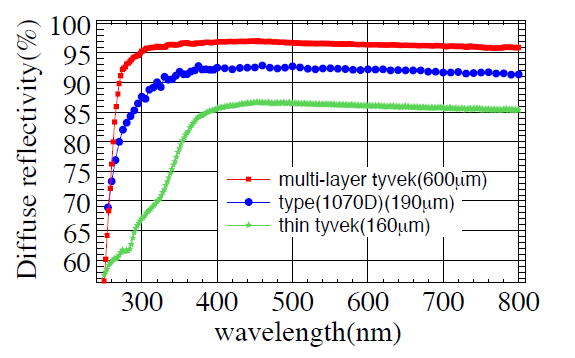}
\caption[Reflectivity of different types of Tyvek]{Reflectivity of different type Tyvek films .}
\label{TvyekReflect}
\end{center}
\end{figure}

As proved in the Daya Bay experiment, reflective surface is effective in tagging muons without larger number of PMTs.
Tyvek film as a reflective material has the merits of  high reflectivity and stable performance, therefore it is widely used in water Cherenkov detectors. The Daya Bay experiment uses Tyvek film as to
make reflective surface in the water Cherenkov detector. In JUNO, the Tyvek reflective film could be used to cover the inner wall of the pool and the outer surface of the central detector.
The Cherenkov photons produced in the pool can be detected by the PMTs after multiple reflections to improve the detection efficiency. Figure~\ref{MakeTyvek} shows the process of making
reflective Tyvek. Tyvek pieces with a width of 1 meter and short lengths can be welded together to form a larger piece of reflector for large detector installation.
The performance of various Tyvek reflective films is shown in Fig.~\ref{TvyekReflect}.

The reflectivity of the Tyvek depends on the thickness of Tyvek. The Daya Bay experiment uses two layers of thick Tyvek (1082 D) with PE between them to form a multi-layer Tyvek.  The reflectivity
of the multilayer Tyvek  is shown as the red points in Fig.~\ref{TvyekReflect}. The reflectivity is >95\% when the wavelength > 300~nm.

\subsubsection{Calibration system}

\begin{figure}[htb]
\begin{center}
\includegraphics[width=8cm]{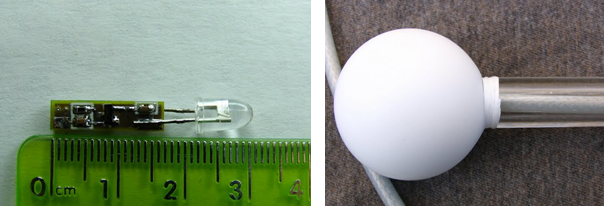}
\caption[LED]{LED and diffuse ball for PMT calibration.}
\label{LED}
\end{center}
\end{figure}

 The gain and the timing calibration of the PMTs will be monitored by a LED system.  No radioactive sources will be required.
   As Fig.~\ref{LED} shows, we will use LED or diffuse ball by LED for the PMT calibration. LED flashing is triggered by pluse generator. The diffuse balls could be put at differnt positions of water pool. So every PMT can receive the photons from diffuse ball. PMTs gain and timing calibration could be done once a week.

\subsection{Top Tracker}

The JUNO cosmic muon tracker will help enormously to evaluate the contamination of the cosmogenic background to the signal.
The OPERA detector has to be dismounted soon and the OPERA TT ~\cite{Adam:2007ex}($\sim4.5$M EUR) will become available by mid--2016.
It will be placed on top of the JUNO water Cherenkov detector to be used as a cosmic muon tracker.

\subsubsection{OPERA Target Tracker}

The TT is a plastic scintillating detector which had several critical roles in OPERA: it was used to trigger the neutrino events, to identify the brick in which the neutrino interaction took place and to reconstruct muons  therefore reducing the Charm background.
Its performances well met the expectations: only limited aging was observed over the 2007 - 2012 data taking period, and the detector understanding was well demonstrated by showing very good data/MC agreement in particular on the muon identification and energy reconstruction~\cite{Agafonova:2011zz}.

\begin{figure}[hbt]
\begin{minipage}[b]{.45\linewidth}
\centering
\includegraphics[width=7cm]{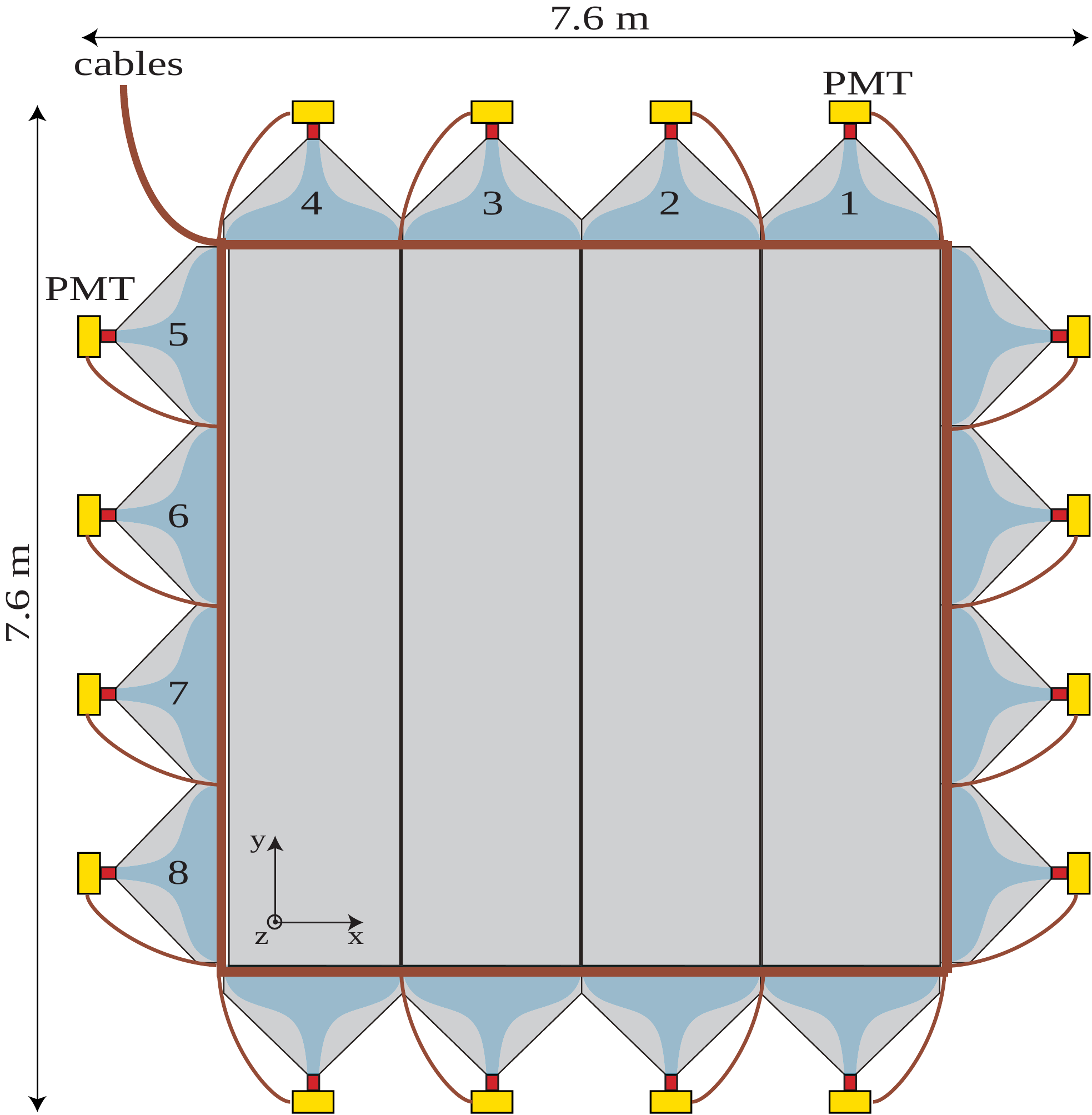}
\caption{\small Schematic view of a plastic scintillator strip wall.} \label{wall_schematic}
\end{minipage} \hspace{1.cm}
\begin{minipage}[b]{.45\linewidth}
\centering
\includegraphics[width=7cm]{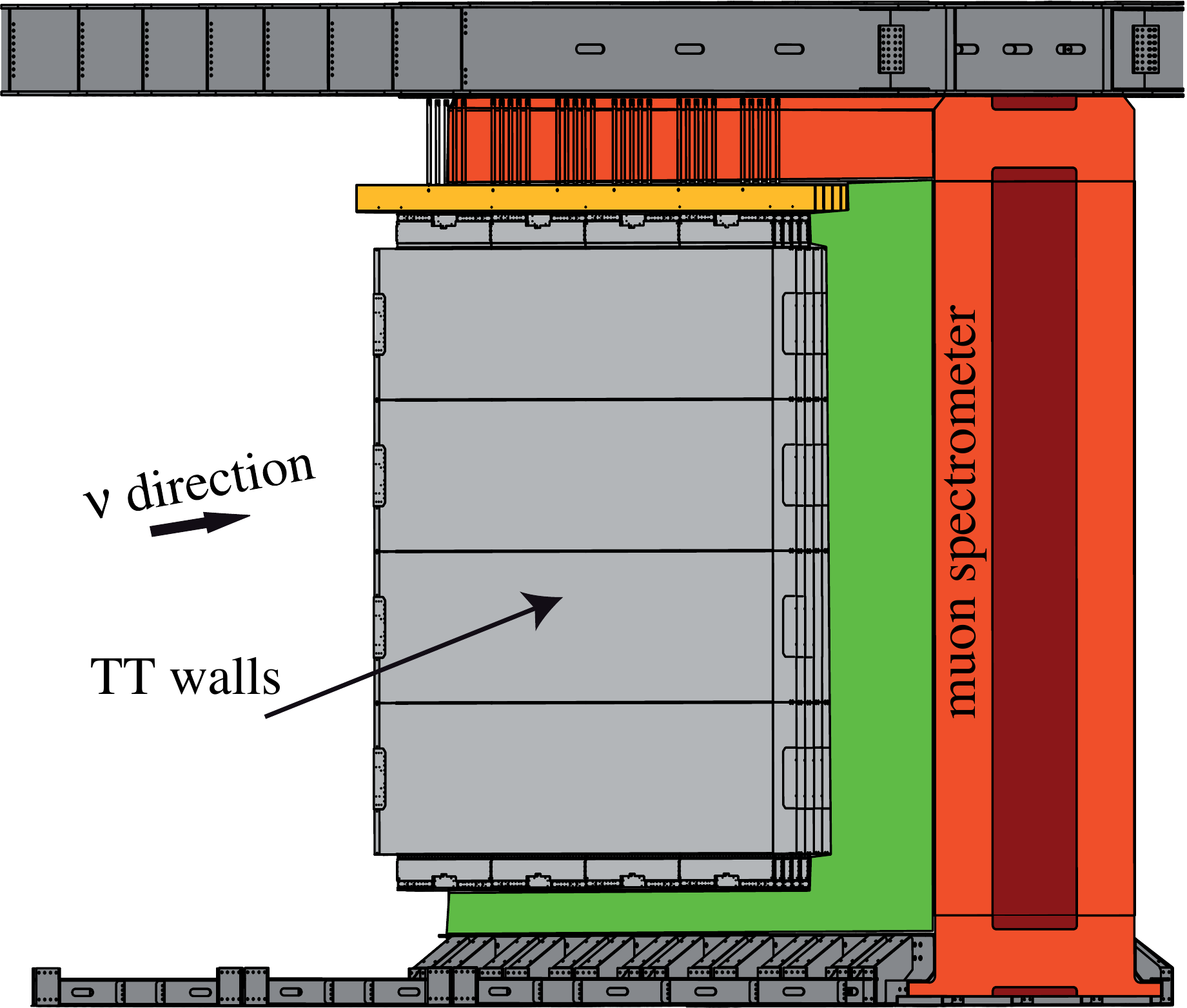}
\caption{\small Target Tracker walls hanging in between two brick walls inside the OPERA detector.} \label{hanging}
\end{minipage}
\end{figure}

The TT is composed of 62 walls each with a sensitive area of 6.7$\times$6.7~m$^2$.
Each wall is formed by four vertical ($x$) and four horizontal ($y$) modules (Fig.~\ref{wall_schematic}).
The TT module is composed of 64 scintillating strips, 6.7~m length and 26.4~mm wide.
Each strip is read on both sides by a Hamamatsu 64-channel multi-anode PMT.
The total surface which could be covered by the 62 x-y walls is 2783~m$^2$.
All TT walls in OPERA are hanged by the top part of the detector and are thus in vertical position (Fig.~\ref{hanging}).
In the case of JUNO, all TT modules have to be placed in horizontal position, in which case more supportive mechanical structure is needed.

\begin{figure}[hbt]
\begin{minipage}{.45\linewidth}
\centering
\includegraphics[width=7cm]{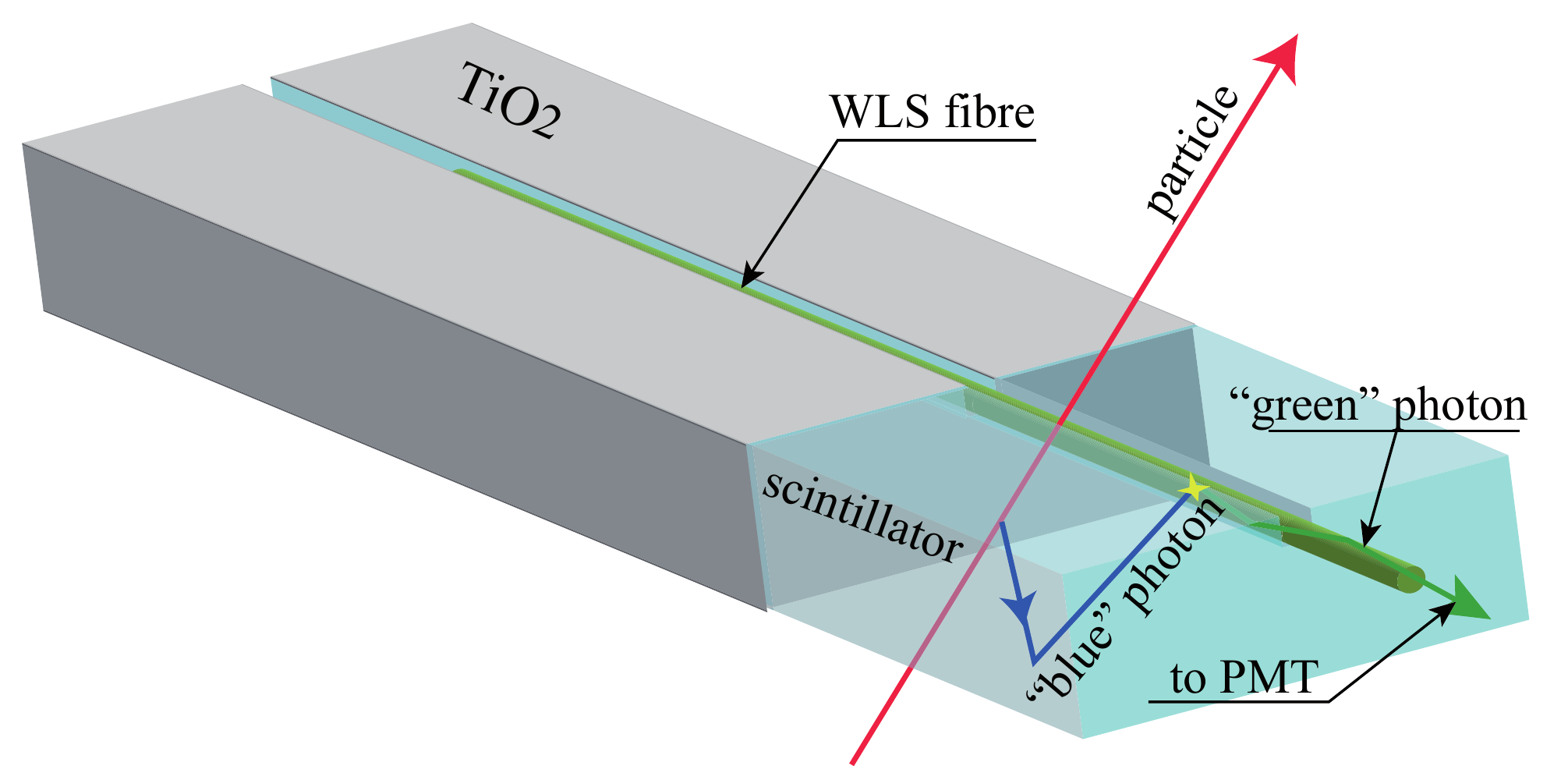}
\caption{\small Particle detection principle in a scintillating strip.}\label{principle}
\end{minipage} \hspace{1.cm}
\begin{minipage}{.45\linewidth}
\centering
\includegraphics[width=7cm]{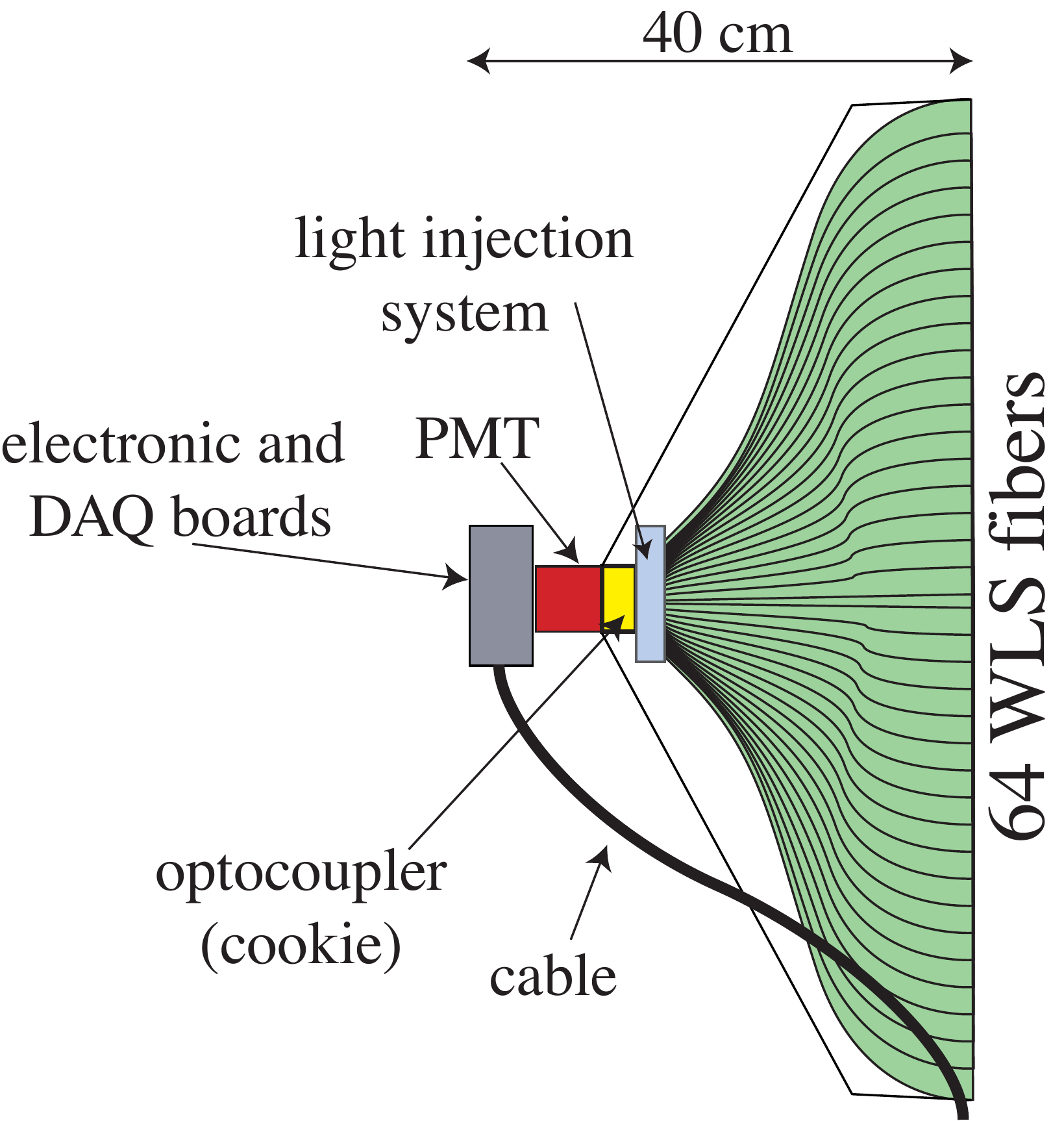}
\caption{\small Schematic view of an end--cap of a scintillator
strip module.}\label{endcap_schematic}
\end{minipage}
\end{figure}

The particle detection principle used by the TT is depicted by Fig.~\ref{principle}.
The scintillator strips have been produced by extrusion, with a $TiO_2$ co-extruded reflective and diffusing coating for better light collection.
A long groove running on the whole length and at the center of the scintillating strips, houses the wavelength shifting (WLS) fiber which is glued inside the groove using a high transparency glue.
This technology is very reliable due to the robustness of its components.
Delicate elements, like electronics and PMTs are located outside the sensitive area where they are accessible (Fig.~\ref{endcap_schematic}).

\begin{figure}[htb]
\centering
\includegraphics[width=12cm]{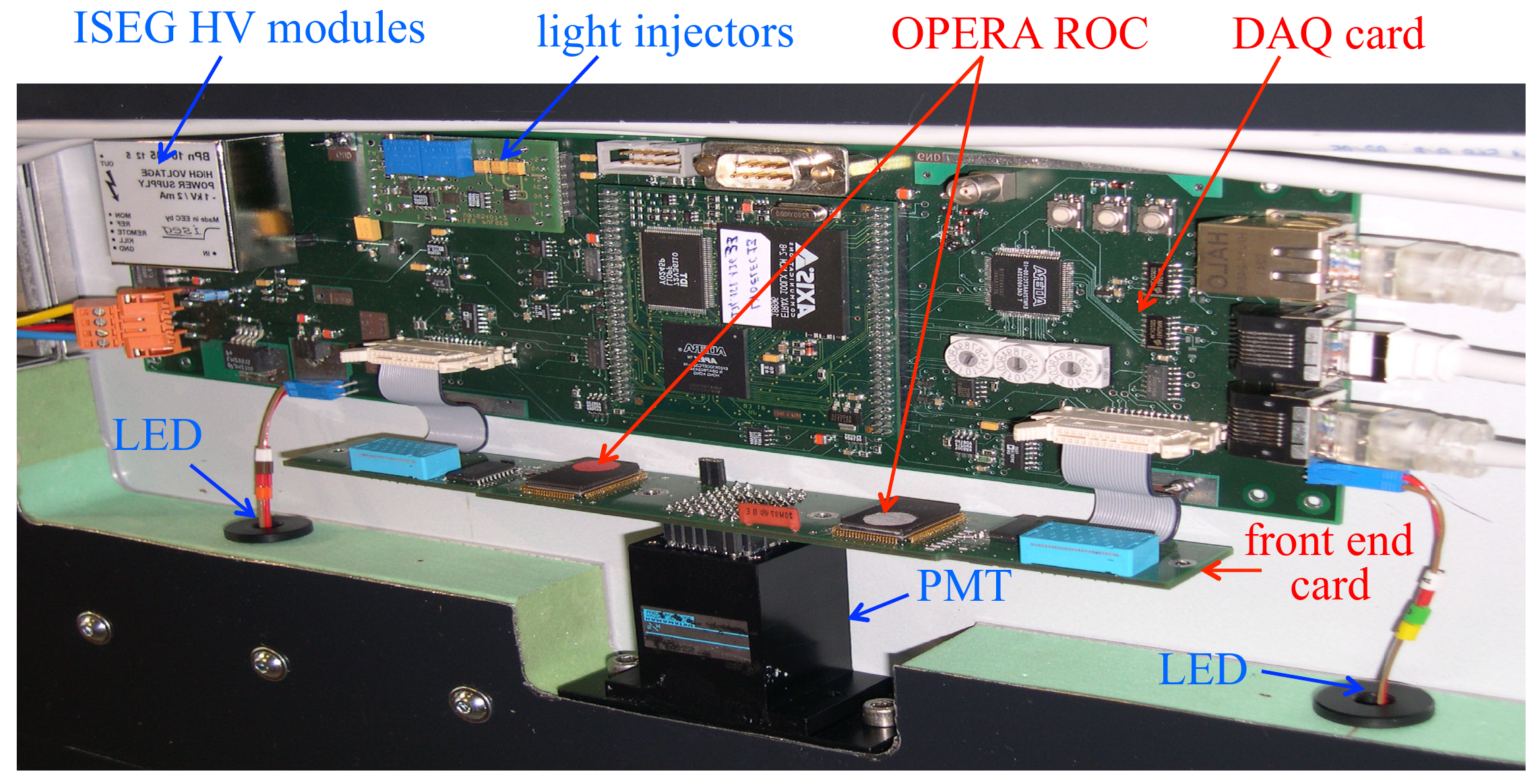}
\caption{TT electronics including DAQ card. All elements in red will be replaced.}
\label{ttelectronics}
\end{figure}

Figure~\ref{ttelectronics} presents details about the end-caps of the TT modules hosting all electronics, front end and acquisition.
These electronics are composed of:
\begin{itemize}
\item a front end card located just behind the multi-anode Hamamatsu PMT, hosting the two OPERA ROC chips (32 channels each, BiCMOS 0.8 microns)~\cite{Lucotte:2004mi},
\item an acquisition card (DAQ)~\cite{Marteau:2009ct}, also hosting an ADC for charge digitization,
\item a light injection card located on the DAQ card,
\item two LEDs able to inject light at the level of the WLS fibers near the PMT and driven by the light injection card, this system is used to regularly calibrate the detector (PMT gain, stability etc.),
\item an ISEG High Voltage module located on the DAQ card.

\end{itemize}

\begin{figure}[htb]
\centering
\includegraphics[width=12cm]{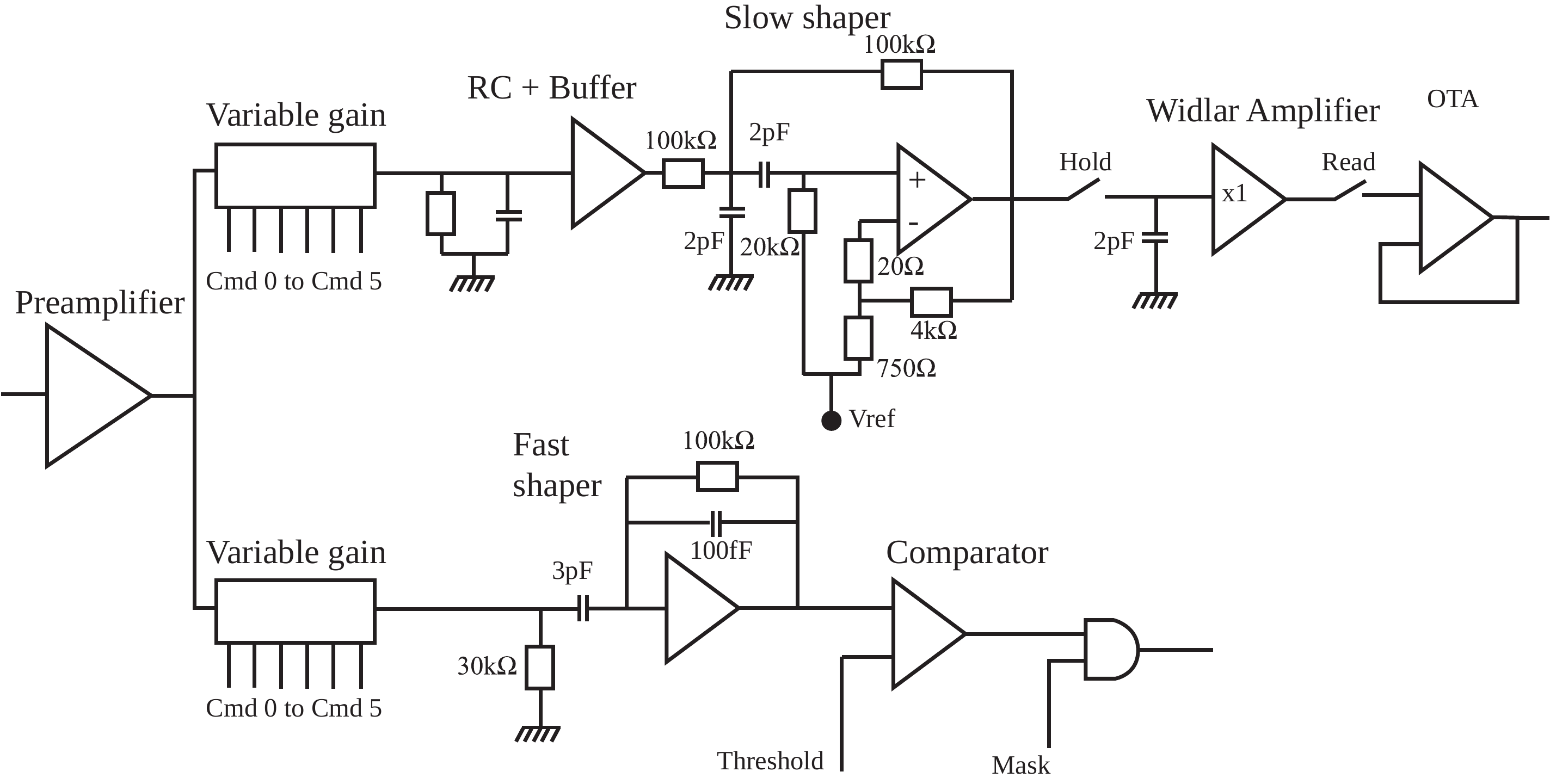}
\caption{Architecture of a single channel of the OPERA--ROC chip.}
\label{operaroc}
\end{figure}

The TT electronics record the triggered channels and their charge, thanks to the OPERA-ROC chip.
A schematic view of this chip is given by Fig.~\ref{operaroc}.
Each channel has a low noise variable gain preamplifier that feeds both a trigger and a charge measurement arms.
The adjustable gain allows an equalization of all PMT channel gains which can vary from channel to channel by a factor 3.
The auto-trigger (lower part) includes a fast shaper followed by a comparator.
The trigger decision is provided by the logical ``OR" of all 32 comparator outputs, with a threshold set externally.
A mask register allows disabling externally any malfunctioning channel.
The charge measurement arm (upper part) consists of a slow shaper followed by a Track \& Hold buffer.
Upon a trigger decision, charges are stored in 2~pF capacitors and the 32 channels outputs are readout sequentially at a 5~MHz frequency, in a period of 6.4~$\mu$s.
All charges are digitized by an external ADC (12--bit AD9220) placed on the DAQ cards.

{\bf TT dismounting in Gran Sasso}\

The TT dismounting in Gran Sasso underground laboratory will start in summer 2015 (first OPERA Super Module) and will end in Spring 2016 (second OPERA Super Module).
The OPERA detector dismounting and cost sharing among the funding agencies are defined in a special MoU.
The cost of the TT dismounting up to its storage area is part of this MoU and thus it is not considered in the IN2P3 JUNO requests.

All TT modules will be stored in Gran Sasso in 10 containers before sending them to China.
The shipping of the containers will be done when storage halls are ready near the JUNO underground laboratory.
This is expected to take place in 2016.
In all the cases the TT will not be mounted on top of JUNO detector before 2019.
This implies that the TT containers will be stored somewhere for about three years.
The best place, in order to avoid big temperature variations and scintillator aging, is the Gran Sasso underground laboratory.
Negotiations are engaged with LNGS on this possibility.
If this is not possible, the TT will be temporarily stored in a hall in the surface LNGS laboratory waiting to be shipped to China.

{\bf TT in JUNO}\

\begin{figure}[htb]
\centering
\includegraphics[width=\textwidth]{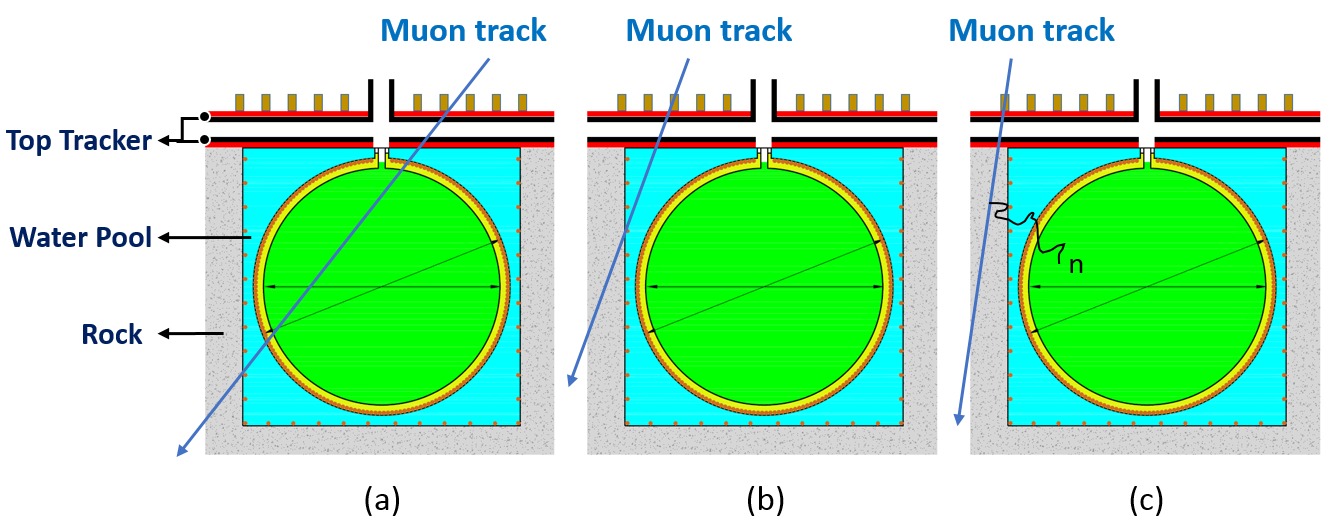}
\caption{Configurations considered for the induced background.}
\label{ttnoise}
\end{figure}

This muon tracker, called now Top Tracker (again TT), will be needed in JUNO in order to well study the cosmogenic background production.
The most dangerous background is induced by cosmic muons generating $^9$Li and $^8$He unstable elements, and fast neutrons, which could fake an IBD interaction inside the central detector.
Figure~\ref{ttnoise} presents schematically the most important noise configurations.
The first two, (a) and (b), mainly concern $^9$Li and $^8$He production directly in the central detector (a) and in the veto water pool (b) while the last one (c) concerns the neutron production in the surrounding rock.

The surface on the top of the JUNO detector is of the order of $40\times 40$~m$^2$.
The total surface that the TT could cover depends on the number of superimposed $x-y$ layers (composed by consecutive TT walls).
In any case it will never be able to cover the entire surface, half of cosmic muons crossing the JUNO detector can pass by the sides.
The number of layers will depend on several parameters:

\begin{itemize}
\item the minimum statistics needed to well measure the cosmogenic background ($^9$Li and $^8$He production),
\item the muon tracking accuracy needed up to the bottom part of the central detector (of the order of the one induced by the multiple scattering),
\item the noise rate reduction using coincidences (affordable by the acquisition system),
\item the rate reduction of fake tracks.
\end{itemize}

First studies show that 3 to 4 layers will be needed that means that the total surface on top of JUNO liquid scintillator detector will be 690~m$^2$ to 920~m$^2$, respectively.

The noise rate in JUNO underground laboratory is expected to be significantly higher than that in Gran Sasso.
During OPERA operation the noise rate for 1/3 p.e. threshold was of the order of 10~Hz.
This low rate was mainly due to the low PMT dark current (2.5~Hz) and the fact that the TT walls were shielded by the OPERA lead/emulsion bricks.
Before the insertion of the bricks the noise rate was of the order of 25~Hz/channel.
In JUNO this rate is expected to be significantly higher for several reasons:

\begin{itemize}

\item The JUNO overburden is lower (2,000~m.w.e.) than for Gran Sasso (4,200~m.w.e.) implying more cosmic muons crossing the detector.

\item There will be no shielding between the TT walls.

\item The radioactivity of the rock, measured by the Daya Bay experiment 200~km away, is expected to be significantly higher than the one observed in the Gran Sasso underground laboratory.

\item There will be no concrete on the walls surrounding the JUNO underground laboratory (contrary to the Gran Sasso underground laboratory), concrete that could absorb part of the $\gamma$'s emitted by the rock.

\end{itemize}

\begin{figure}[htb]
\centering
\includegraphics[width=\textwidth]{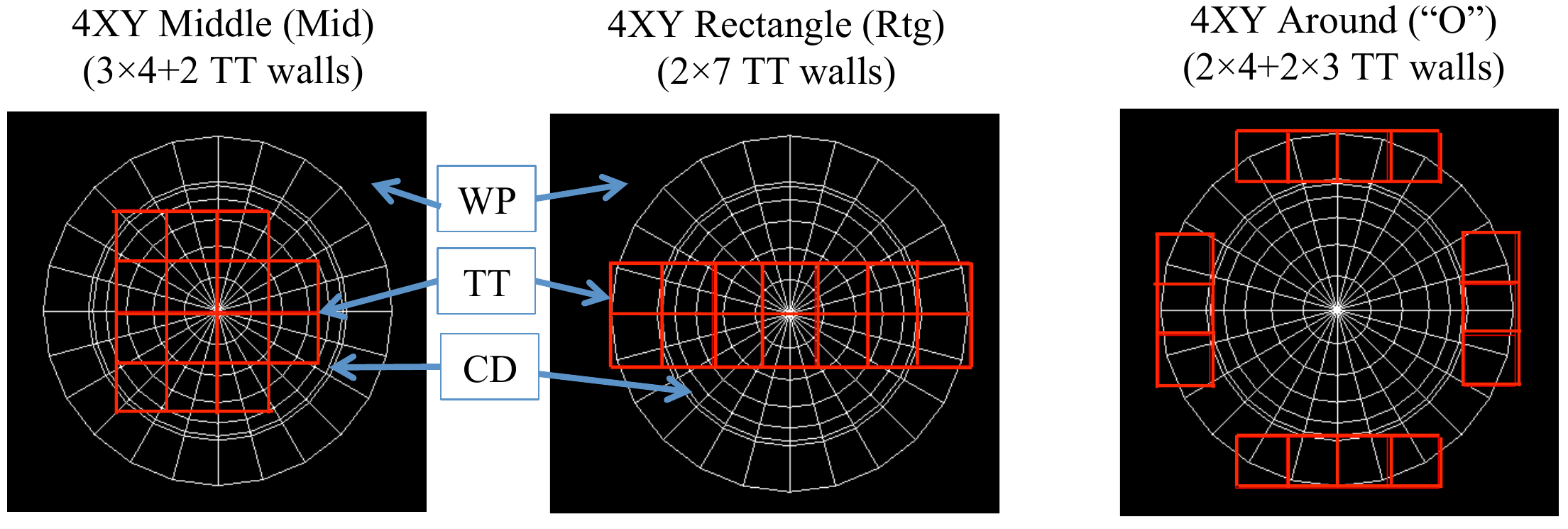}
\caption{Configurations considered for the Top Tracker coverage.}
\label{coverage}
\end{figure}

In the case of 4-layers cover, the configurations of Fig.~\ref{coverage} have been considered.
The first configuration has the advantage compared to the others of covering the maximum surface of the central detector region.
The second one is more representative of all detector configurations and noise measurements could be extrapolated to the total detector top surface.
The third configuration is best for rock muon production estimation but is not very representative of the rest of the detector.
Finally, as baseline it has been defined the second configuration.
In all the cases, the TT has to well cover the chimney region (central region not covered by the veto water Cherenkov) from where radioactive sources will be introduced to calibrate the detector.

{\bf Noise rate estimations}\

Table~\ref{noise} presents the rock noise rate for one $x$ layer on the top, for one $y$ layer just below and for the case of one $x-y$ coincidence of these two layers ($x-y$ correlated coincidence).
For the $x-y$ correlated coincidences the $x$ and $y$ hits must come from the same radioactivity event while for the $x-y$ accidental coincidences come from different events occurring during 200~ns time--window.
The slight noise reduction from the $x$ layer to the $y$ one is due to the fact that the top layer shields a bit the layer above.
A strong noise rate reduction is observed when an $x-y$ coincidence is required.
For all the cases a threshold of one and two photoelectrons is considered.
The results are shown considering no concrete on the JUNO cavern walls and with 10~cm concrete.
The presence of concrete would reduce the radioactivity noise by about a factor of 4.

\begin{table}[htdp]
\caption{Noise rate for several TT conditions. (a) is under 1 p.e. threshold and (b) is under 2 p.e. threshold.
(1)no concrete; (2)10~cm concrete.}
\begin{center}
\begin{tabular}{|c|c|c|c|c|}\hline
 Config. & \multicolumn{2}{c|}{Rate (Hz/m$^2$),(a)} & \multicolumn{2}{c|}{Rate (Hz/m$^2$),(b)} \\
\cline{2-5} & (1) & (2) & (1) & (2) \\
\hline $1-x$ layer (top)  & 6800 & 1780     & 2700  & 710 \\
\hline $1-y$ layer (bottom)  & 6500 & 1780     & 1200  & 310 \\
\hline $x-y$ layer (correlated)  & 540   & 150     & 29   & 7 \\
\hline $x-y$ layer (accidentals)     & 400   & 29 & 29   & 2 \\
\hline
\end{tabular}
\end{center}
\label{noise}
\end{table}

The noise rate per channel is not the only parameter to take care during the design of the new TT configuration.
The number of $x-y$ layers will mainly be determined by the number of ``fake'' muons (virtual tracks due to many hits in a time--window of 200~ns) produced by the environmental radioactivity.
Table~\ref{fake} presents the ``fake'' muons' rate estimation considering two and three TT $x-y$ layers.
For the case of the two layers, the distance between is considered to be 2~m, while for the case of three layers the third layer is inserted at the middle between the two considered layers.
It can be observed that the rate is considerably reduced for the case of three layers compared to the one of only two layers.
From this table the case of two layers can be excluded.
The rate being of the order of few Hz and less for the case of three layers, the case of four layers initially considered probably will not be necessary.
A fourth layer would provide some redundancy in case of inefficiencies due to scintillator aging, dead time or dead zones.
It has also to be noted that each ``fake'' muon will bias the cosmogenic $^9$Li and $^8$He production and thus this rate has to be very low.

\begin{table}[htdp]
\caption{Rate of ``fake'' tracks for several TT conditions.(a) is under 1 p.e. threshold and (b) is under 2 p.e. threshold.
(1)no concrete; (2)10~cm concrete.}
\begin{center}
\begin{tabular}{|c|c|c|c|c|}\hline
 Config. & \multicolumn{2}{c|}{Fake muon rate (Hz), (a)} & \multicolumn{2}{c|}{Fake muon rate (Hz), (b)} \\
\cline{2-5} & (1) & (2) & (1) & (2) \\
\hline 2 TT $x-y$ layers  & 340,000 & 12,500     & 1300  & 31 \\
\hline 3 TT $x-y$ layers  & 2.6        & 0.02        & $6\times 10^{-4}$  & $2\times 10^{-6}$ \\
\hline
\end{tabular}
\end{center}
\label{fake}
\end{table}

In order to study the background induced by $^9$Li, $^8$He and fast neutrons, enough statistics has to be collected using the TT.
To estimate the background rate induced by these three source for the three configurations of Fig.~\ref{ttnoise}, the following parameters have been considered:

\begin{itemize}
\item $E_\mu$: the muon energy,
\item $L_\mu$: the average track length of muons in the liquid scintillator (case a),
\item $R_\mu$: the muon rate in the liquid scintillator (case a),
\item $F_{nCap}$: the neutron capture ratio (1 for JUNO, case a),
\item $N$: the muon number (case b),
\item $L_{att}$: the attenuation length of $^9$Li/$^8$He in water pool (0.5~m, case b),
\item $S_{wp}$: surface of water pool (case c).
\end{itemize}

The following formulas give the background rate of the three considered configurations:

\begin{itemize}
\item ${R_{Li,He}} \propto E_\mu ^{0.74} \cdot {L_\mu } \cdot {R_\mu } \cdot {f_{nCap}}$ for case (a),
\item ${R_{Li,He}} \propto \frac{{\sum\limits_{i = 1}^N {E_{{\mu ^i}}^{0.74}(\sum\limits_{j = 1}^M {L_{{\mu ^i}}^j{e^{ - \frac{{{d_j}}}{{{L_{att}}}}}}} )} {f_{nCap}}}}{N} \cdot {R_{wp}}$ for case (b),
\item $R_n^{rock} \propto E_{{\mu ^i}}^{0.74} \cdot {R_\mu } \cdot {S_{wp}} \cdot {f_{nCap}}$ for case (c)
\end{itemize}

Table~\ref{casea} presents the expected background rate from $^9$Li and $^8$He for all three TT coverage configurations of Fig.~\ref{coverage}.
From these results, the TT configuration (4XY, ``O'') can be excluded because of the poor rate observed of the $^9$Li and $^8$He induced background. It has to be noted that in absence of the TT the $^9$Li and $^8$He rate per day is of the order of 90 to be compared to $\sim$50~IBD interactions induced by nuclear reactors.
Obviously, this background has to be reduced or at least its rate must be well known.

\begin{table}[htdp]
\caption{Noise rate induced by $^9$Li and $^8$He for case (a) of Fig.~\ref{ttnoise}.}
\begin{center}
\begin{tabular}{|c|c|c|c|}\hline
                       & $L_\mu$ (m) & $R_\mu$ (Hz) & $^9$Li/$^8$He rate/day \\
                       \hline
all muons       & 22.5               & 3.5                  & 90  \\
TT (4XY, Mid) & 23.5              & 0.94                & 27   \\
TT (4XY, Rtg) & 23.4              & 0.80                & 23   \\
TT (4XY, ``O") & 21.8              & 0.30                & 9   \\
\hline
\end{tabular}
\end{center}
\label{casea}
\end{table}

Table~\ref{caseb} presents the expected background rate from $^9$Li and $^8$He for all three TT coverage configurations of Fig.~\ref{coverage} where the muons only cross the water pool without passing through the central detector.
As already said, the TT configuration (4XY, Rtg) is the most representative of all parts of the detector and, not seen a significant difference between this configuration and the configuration (4XY, Mid), the configuration (4XY, Rtg) has been considered as baseline.

\begin{table}[htdp]
\caption{Noise rate induced by $^9$Li and $^8$He for case (b) of Fig.~\ref{ttnoise}.}
\begin{center}
\begin{tabular}{|c|c|c|c|}\hline
                       & $^9$Li/$^8$He rate/day  & Back./signal (\%) \\
                       \hline
all muons       & 1.2               & 3.0          \\
TT (4XY, Mid) & 0.35            & 0.9          \\
TT (4XY, Rtg) & 0.3              & 0.7           \\
TT (4XY, ``O") & 0.16           & 0.4           \\
\hline
\end{tabular}
\end{center}
\label{caseb}
\end{table}

Table~\ref{casec} presents the expected background rate from fast neutrons obtained by extrapolations for Daya Bay  observations for case (c) of Fig.~\ref{ttnoise} and for the baseline configuration of TT (4XY, Rtg) of Fig.~\ref{coverage}.
It is expected to have a fast neutron induced background rate of the order of 0.13 per day which represents 0.3\% of the expected signal.

\begin{table}[htdp]
\caption{Expected noise rate from fast neutrons for case (c) of Fig.~\ref{ttnoise} and for the configuration of TT (4XY, Rtg) of Fig.~\ref{coverage}.}
\begin{center}
\begin{tabular}{|c|c|c|c|}\hline
                              & Daya Bay  & Far detector & JUNO \\
                       \hline
$E_\mu$ (GeV)     & 57             & 137               & 215     \\
$R_\mu$ (Hz)        & 1.2            & 0.055            & 0.003  \\
$S_{wp}$ (m$^2$) & 724           & 1032             & 3740   \\
$f_{nCap}$ (\%) & 45                 & 45                 & 100     \\
Rock neutron bkg. rate/day & 1.7 & 0.22 & 0.13  (B/S$\sim$0.3\%)\\
\hline
\end{tabular}
\end{center}
\label{casec}
\end{table}

The noise measured with the help of the TT will be extrapolated to the whole detector and will be introduced in the simulations in order to well estimated the related systematic errors.

{\bf Modifications to the Target Tracker}\

 The present OPERA TT acquisition system can afford up to 20--25~Hz trigger rate per channel.
 For higher rates (as those expected in JUNO) this system would have severe limitations inducing huge dead time and thus significantly reducing the TT efficiency.
Even if the $x-y$ coincidences already reduce the noise rate (see above), the acquisition system specifically developed for OPERA needs will not be able to be used for the TT installed in JUNO cavern.

\begin{figure}[htb]
\centering
\includegraphics[width=\textwidth]{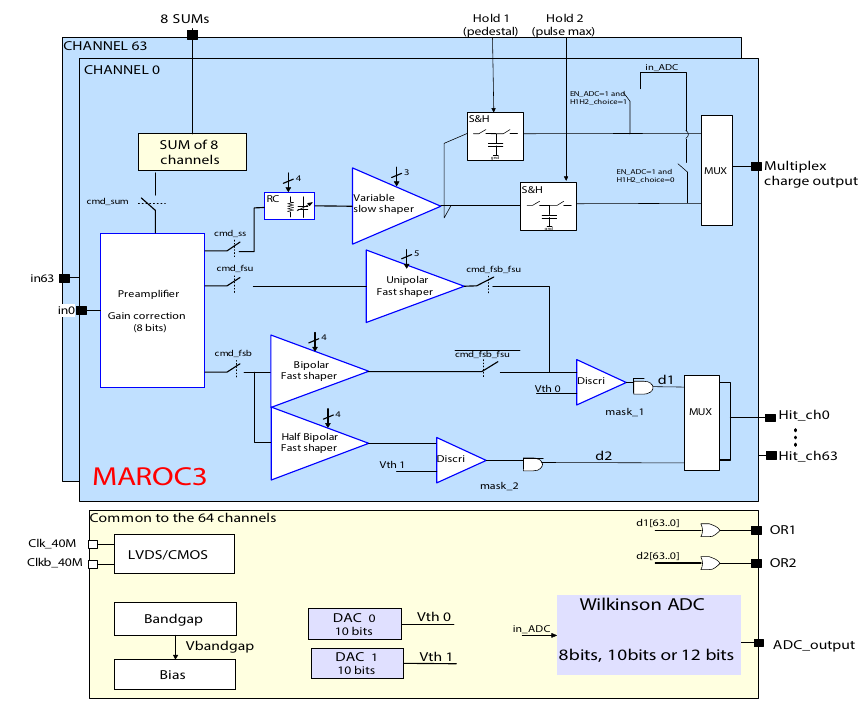}
\caption{Architecture of the MAROC3 chip.}
\label{maroc}
\end{figure}

The high noise rate expected in JUNO
obliges the replacement of the acquisition system (DAQ) by a new one.
By the same way, the front end electronics based on OPERA-ROC~\cite{Lucotte:2004mi} chip will be replaced.
Indeed, this chip is now obsolete and not enough spares are available.
The new generation chip MAROC3~\cite{Blin:2010tsa} (AMS SiGe 0.35~$\mu$m technology, developed by IN2P3 Omega laboratory) disposing of more functionalities can replace the OPERA-ROC chip.
On OPERA-ROC the word registering the triggered channels was not functioning properly while on MAROC3 this has been corrected.
This gives the possibility to only read the triggered channels and not all of them as it was done in OPERA TT, thereby gaining in readout speed.
The MAROC3 chip has a similar architecture (Fig.~\ref{maroc}) than the OPERA--ROC (fast part for triggering and slow one for charge measurement).
The difference is that in MAROC3 the ADC is already integrated in the chip while for OPERA--ROC the ADC was externally implemented on the DAQ cards.
The MAROC3 chip has two ``OR''s, one could be used for the timestamp while the second one could be used for fast coincidences with other sensors.
This would allow to do fast coincidences with at least $x$ and $y$ strips to reduce the DAQ trigger rate.
For all coincidences among TT sensors a dedicate electronics card has to be developed receiving the ``OR'' of all sensors and making coincidences.
The timestamp can be provided by a special clock card to be developed.

OPERA institutes participating now in JUNO have expressed a strong interest to develop the DAQ cards which will be located just behind the multi--anode PMTs, as done in OPERA.
The front end card where the MAROC3 chip will be located could be produced by the same institutes, which would also have the responsibility to check and characterize all MAROC3 chips produced for this application (992 plus spares).
These institutes could also develop the clock card and the coincidence card.

The drawings for the installation of the TT in JUNO will be produced by IN2P3 and IHEP--Beijing.
It is already agreed with IHEP that the TT supporting structure will be financed and produced by Chinese institutes.
This structure, very different from the one used in OPERA due to the fact that in JUNO all modules will be in horizontal position instead of vertical position in OPERA, has to be drawn with care and in a close collaboration between IN2P3 and IHEP.
A special care has also to be taken in order to leave accessibility to all electronics of all TT modules.

\section{Water System}

For the JUNO water Cherenkov detector, it requires ultrapure water
for high muon detection efficiency. There will be different kinds of
materials submerged in the water, including stainless steel, Tyvek,
PMT glass, cables, etc. The complexity of the underground
environment makes it difficult to seal the pool completely and
almost impossible to passively keep the water quality good for a
long period of time.
Therefore, it is necessary to build a reliable
ultrapure water production, purification and circulation system.

Another important function of the water system is to keep the
overall detector temperature stable.  The stability of the
temperature of the central detector is critical for the entire
experiment.
JUNO is located in southern China, and the temperature
of the surrounding rocks can reach 32 degrees Celsius. We intend to
maintain the temperature of the central detector at 20 degrees
Celsius.
Water is a good conductor of heat. Lowering the temperature
of the water flowing into the pool is a natural choice to offset the
heat from the surrounding rocks.

The Daya Bay experiment has 1,300 tons of water in the near
experimental hall water pool and 2,000 tons of water in the far
experimental hall water pool. The water production rate is
8~tons/hour and
the purification rate is 5$\sim$8 tons/hour. It can
circulate one volume of water in two weeks. The water Cherenkov
detector's muon detection efficiency is high to 99.9\% and its
long-term performance is stable.
The designed water purification
system satisfies the experiment requirements~\cite{Wilhelmi:2014rwa}.

\begin{figure}[htb]
\begin{center}
\includegraphics[width=8cm]{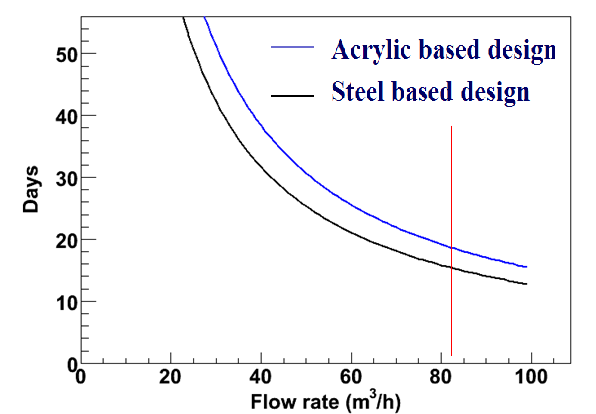}
\caption[Flow rate vs one cycle time]{Flow rate vs one cycle time.}
\label{flowRate}
\end{center}
\end{figure}

The JUNO water Cherenkov detector will have 20,000 - 30,000 tons of
ultrapure water. Based on the experience of Daya Bay experiment,
circulating one volume of water in two weeks requires a
flow rate of
about 80~tons/hour at JUNO. With this flow rate, we can keep good
water quality as well as maintain pool water at stable temperature.
Fig.~\ref{flowRate} shows the number of days to
circulate one volume
of water as function of  the flow rates of the proposed water system
for the two candidate designs of the central detector. The red line
indicates the flow rate at 80~tons/hour.
It would take about 16 to
20 days to circulate one volume of water and this would satisfy the
experiment requirements.

In the proposed design, we divide the water system into two parts.
One part is for the production of ultrapure water and the other part
is for the circulation/purification of water. For the water
production
system,  one part of the system would be on the surface
ground and the other part of the system would be underground. The
reason for this design is that there is a strict requirement of
waste water rate
to be less than 5~tons/hour at 500 meters
underground.  Therefore, we have the part of the water production
system which generates the most waste water on the surface ground.

\subsection{Water Production on Surface}

\begin{figure}[htb]
\begin{center}
\includegraphics[width=12cm]{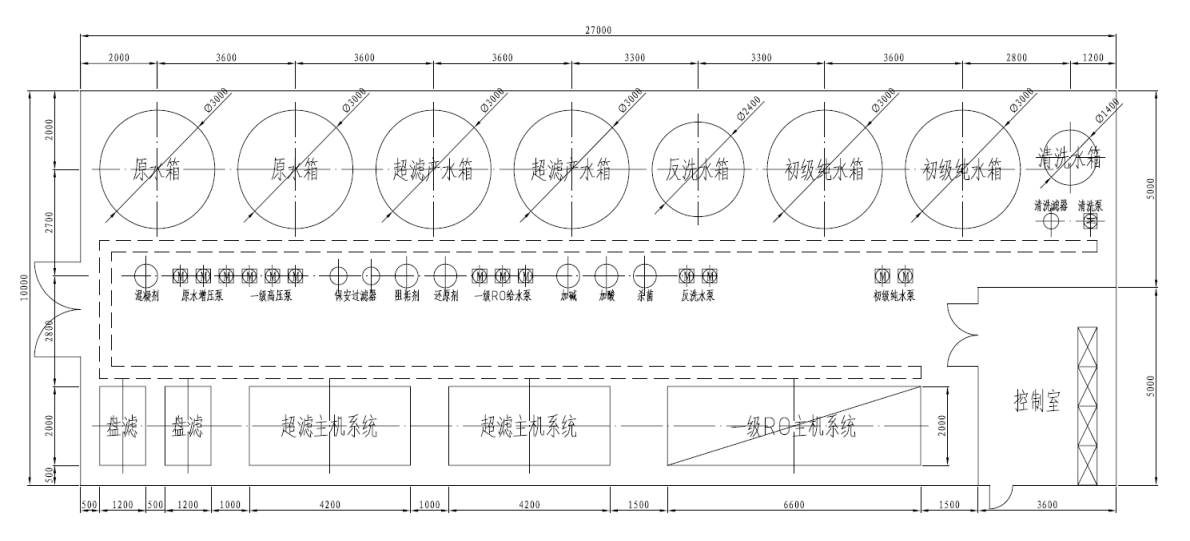}
\caption[General view of water production system on ground]{General view of water production system on surface ground.}
\label{WaterProdGround}
\end{center}
\end{figure}

As Fig.~\ref{WaterProdGround} shows, water flows through the disk
filter first, which can block solid sediments and other large
particles from flowing into the raw tank before goes through the
ultrafiltration hosts.  The ultrafiltration hosts can filter out
small particles of insoluble material in the water before it goes
into ultrafiltration tank. The water then passes through a reverse
osmosis
(RO) system, which can remove most of the ions in the water.
 Finally the water flows into the primary pure water tank. After the
RO  process,  a third of the original water is produced as
waste
water with high ion concentration at a rate of 15~tons/hour. The
waste water is of no harm to the environment and can be directly
drained.

\subsection{Water Production Underground}

\begin{figure}[htb]
\begin{center}
\includegraphics[width=12cm]{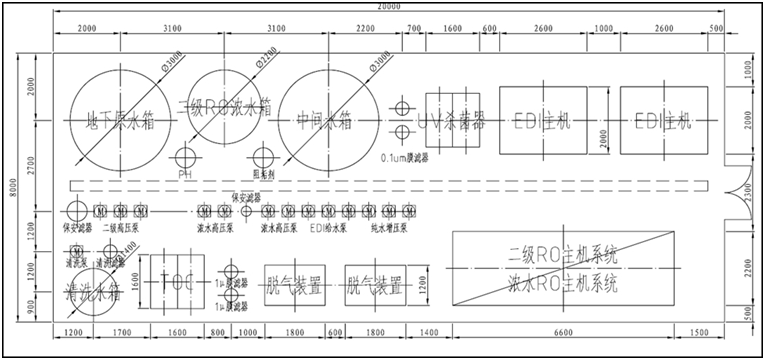}
\caption[General view of water production system underground]{General view of water production system underground.}
\label{WaterProdUnderGround}
\end{center}
\end{figure}

From primary pure water tank on the surface ground, the water is
pumped through pipes at about 85~tons/hour going down a height of
500 meters into the underground raw water tank.
From there the
water will go through another RO system as well as other water
treatment systems, like reduction of Total Organic Carbons (TOC) and
Electrodeionization (EDI) systems.
The resulted ultrapure water with
resistance greater than 16~MOhm flows into experiment hall
at a rate
of 80~tons/hour.

\subsection{Water Circulation/Purification Underground}

\begin{figure}[htb]
\begin{center}
\includegraphics[width=12cm]{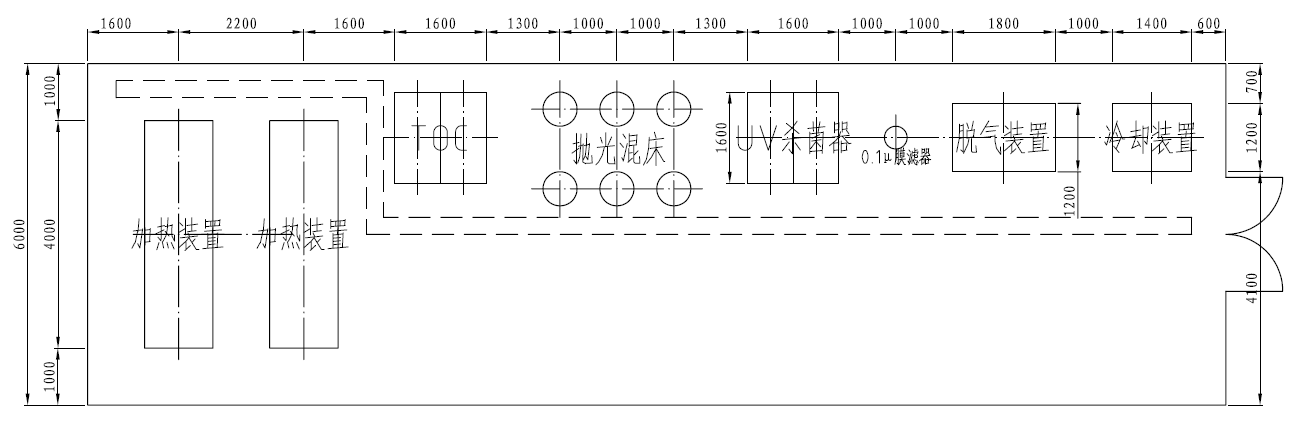}
\caption[General view of water circulation/purification system on ground]{General view of water circulation/purification system on ground.}
\label{WaterCircuUnderGround}
\end{center}
\end{figure}

We would start the circulation process after the pool is fully
filled with water. The water will be drawn drained from the pool at
80~tons/hour. The water will first go through the heat/cooling
equipment . Then the
water will flow through the TOC system (to
remove organic carbon),  the polishing mixed bed (to remove ions in
the water to improve water quality),  UV sterilization system,
membrane filters,
the degassing apparatus (to remove oxygen and
other gases in water), and finally a cooling device (to control the
water temperature). In the end, the purified water will return to
the pool.

For the whole water production, purification and circulation system,
we will use a water control system to automate the running of the
system. The people on duty will only need to replace filters and
other supplies on a  regular basis.

\section{Geomagnetic Field Shielding System}

\begin{figure}[htb]
\begin{center}
\includegraphics[width=12cm]{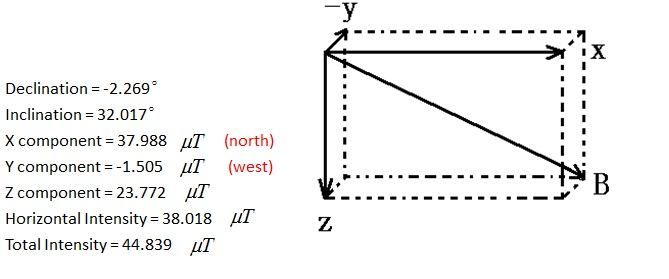}
\caption[Earth magnetic in Jiangmen]{Earth magentic in Jiangmen.}
\label{JUNO_filed}
\end{center}
\end{figure}

\begin{figure}[htb]
\begin{center}
\includegraphics[width=8cm]{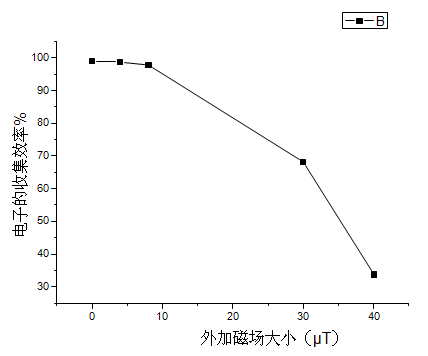}
\caption[PMT collection efficiency vs the external magnetic field]{PMT collection efficiency vs the external magnetic field.}
\label{CEvsField}
\end{center}
\end{figure}

At the experimental site, the horizontal component of the geomagnetic field is about 40~$\mu$T and vertical component is about 24~$\mu$T.  Since JUNO will use 20 inch PMTs as the main device
for central detector, the geomagnetic field will have a big effect on large size PMT. It will deteriorate the whole detector performance. Figure~\ref{CEvsField} shows the relationship between the PMT collection efficiency and the external magnetic field intensity. We need to establish an effective magnetic shielding system to ensure good performance of central detector.

\begin{figure}[htb]
\begin{center}
\includegraphics[width=8cm]{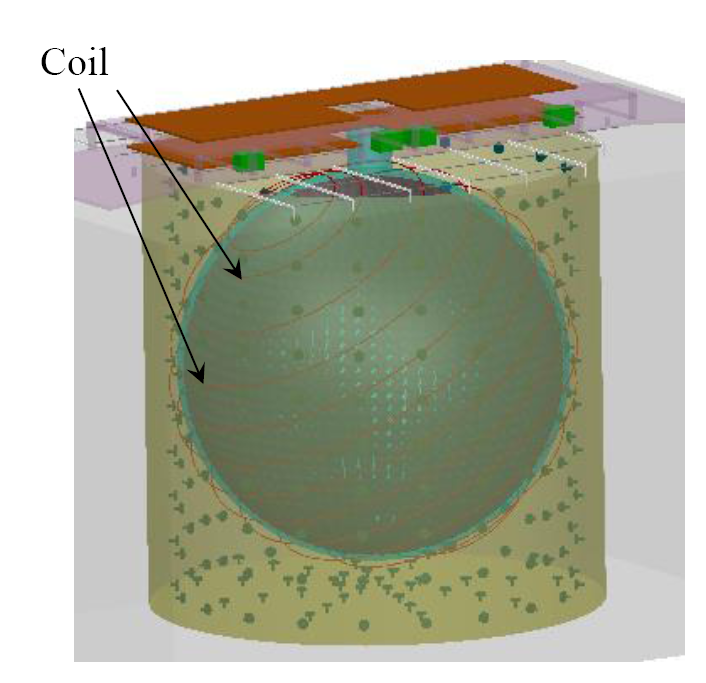}
\caption[One group earth magnetic shield system]{One group earth magnetic shield system.}
\label{OneGroupCoil}
\end{center}
\end{figure}

\begin{figure}[htb]
\begin{center}
\includegraphics[width=12cm]{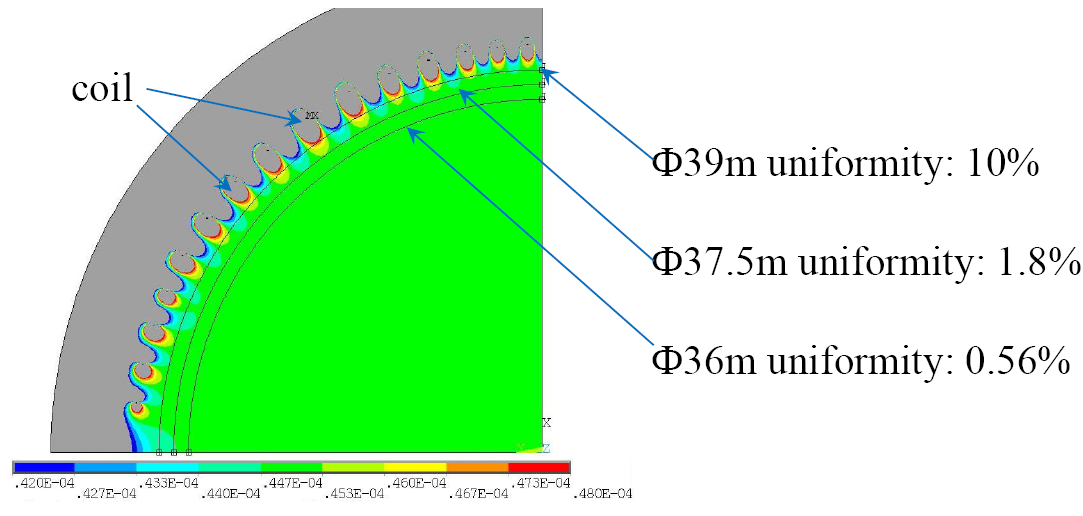}
\caption[Magnetic field distribution with one group system]{Magnetic field distribution with one group system.}
\label{MagDis}
\end{center}
\end{figure}

We intend to ensure that a spherical region with a radius of 37.5~m would be well shielded from geomagnetic field.
We plan to use coils with current flowing through them to compensate the geomagnetic field. The residual field intensity could be reduced to be below 10~$\mu$T.
The baseline design is of one coil system. We can perform accurate measurement of the geomagnetic field to determine its main direction (mapping the geomagnetic field in the pool)
and then do compensation along its main direction. As  Fig.~\ref{OneGroupCoil} shows, this system is easy to control.
Figure~\ref{MagDis} is the magnetic field distribution in the sphere from one coil system. It can have a good shielding. The uniformity within the 37.5~m volume is 1.8\%.
The PMTs in central detector will not be effected by the Earth's magnetic field after compensation.

Careful selection of non-magnetized materials used in the detector for parts such as the supporting structure for PMTs as well as stainless steel for the detector, etc. is also important.
They could cause an additional local magnetic field in additional to the Earth's field, which could affect the performance of the PMTs. The compensation coils would be installed on the central detector which means the PMTs of the water Cherenkov detector would
still be subject to the Earth's magnetic field plus some additional field generated from the coils. Therefore, it needs to have the pool PMTs shielded against magnetic field with passive magnetic shield like the one used in the Daya Bay experiment \cite{DeVore:2014nim}.

\section{Mechanical Structures and Installation}
\subsection{Support Structure over the Pool}

\begin{figure}[htb]
\begin{center}
\includegraphics[width=12cm]{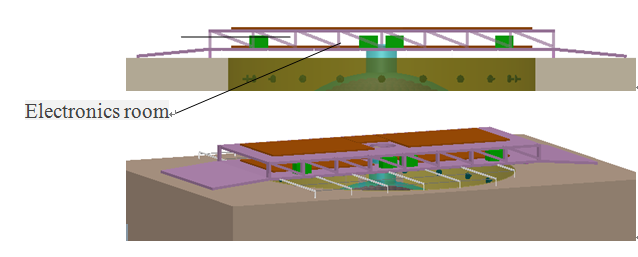}
\caption[The top support structure]{The top support structure.}
\label{SupportStr}
\end{center}
\end{figure}

The tracking detectors over the pool are divided into two groups, each
consisting of several layers of detectors stacked together. One group
is placed on the steel bridge, while the other is put three meters above
the first group. Hence a two-story bridge is needed, as shown in
 Fig.~\ref{SupportStr}.
The parameters of the two-story bridge are as follows: bridge width
is between 17~m and 20~m and the distance between two layers of the
bridge is 3 meters. The support structure has three functions:

 (1) supporting the tracking detectors;

(2) supporting the calibration systems of the central detector;

(3) possibly supporting the electronics rooms.

The diameter of pool is 42 meters and the supports of bridge are located
at both ends of the bridge. Considering the height limit of the underground
space and the structure stiffness, a box structure is adopted for the bridge.
Two columns of tracking detectors, whose width is about 15~m, are placed
on the bridge. Considering the passageway and the width of the electronics
room (about 3 meters), the bridge width is between 17~m and 20~m. The load
of the bridge includes the weight of the central detector cable and
electronics rooms (about 100 tons), weight of the tracking detectors
(about 70 tons), weight of the electronics room (20 tons) and weight
of the supports of the tracking detectors.

 \begin{figure}[htb]
\begin{center}
\includegraphics[width=10cm]{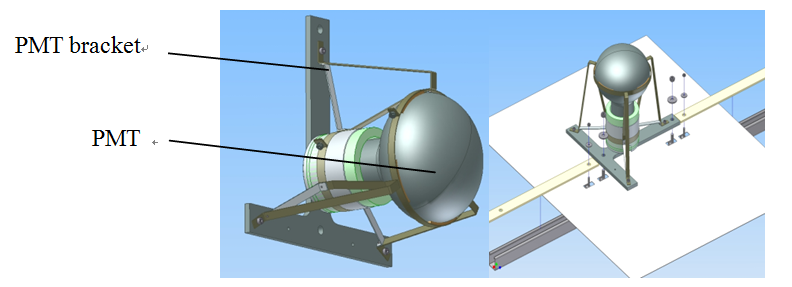}
\caption[PMT support structure of DayaBay]{PMT support structure of DayaBay.}
\label{PMTstr}
\end{center}
\end{figure}

\begin{figure}[htb]
\begin{center}
\includegraphics[width=12cm]{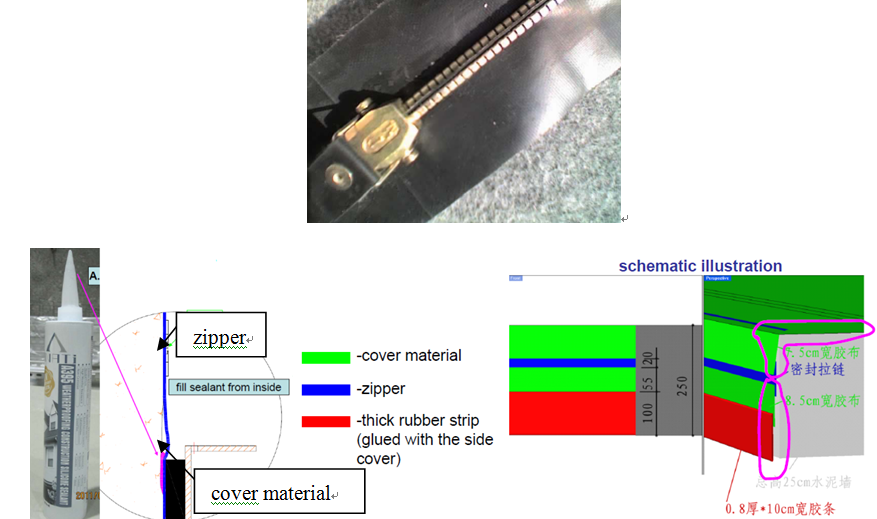}
\caption[The cover design of Daya Bay]{The cover design of Daya Bay. a) gastight zipper  b) Th connection between pool cover and pool dege}
\label{DYBcover}
\end{center}
\end{figure}

\begin{figure}[htb]
\begin{center}
\includegraphics[width=10cm]{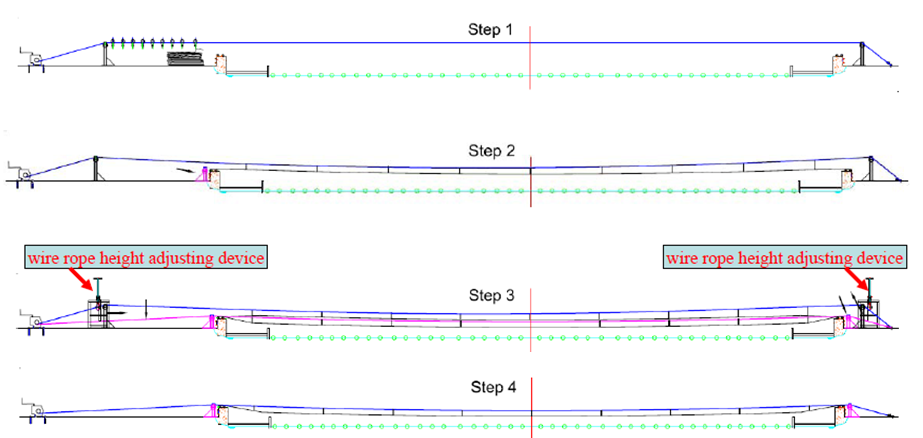}
\caption[The wire design option for cover installation of Daya Bay experiment]{The wire design option for cover installation of Daya Bay experiment.}
\label{cover1}
\end{center}
\end{figure}

\subsection{Support Structure for PMTs}

The support structure of PMT adopts the design for the Daya Bay veto PMTs,
as shown in Fig.~\ref{PMTstr}. The PMT support structure is then attached
to a steel-frame fixed on the waterpool wall. The design of the steel-frame structure will be similar to that for the central detector. The tyvek
film will also be fixed to the steel frame.

\subsection{Sealing of the Pool}
A black rubber cloth served as the pool cover in the Daya Bay experiment.
The pool cover is connected to the pool edge through an air-tight zipper
which shields the pool from the external air and light, as shown in
Fig.~\ref{DYBcover}. With the seal zipper, the pool is easily accessible
for routine maintenance, such as replacing the liquidometer. We will
adopt the same technique for JUNO's pool seal.

As the JUNO pool is very large with a 40-meter span, it is difficult
to unfold the black rubber cloth. Therefore, we propose two solutions:
the first is the tight-wire scheme used in the Daya Bay experiment;
the second is a slide-guide scheme. The two schemes are similiar.
For the slide-guide scheme shown in Fig.~\ref{DYBcover}, the cloth is
unfolded along the guide, instead of the tight-wire in the tight-wire
scheme shown in Fig.\ref{cover1}. For comparison, the tight-wire scheme
is more difficult to operate. In order to prevent the cloth from touching
the ground, we will adjust the height of the wire rope pillar.
The adjustment process is shown in Fig.\ref{cover1}. Firstly, wire is
put on the higher rope pillar and pulled tight; secondly, unfold the
black rubber cloth and fix the edge of the cloth; finally, put the
tight-wire on the lower pillar slowly.

\begin{figure}[htb]
\begin{center}
\includegraphics[width=12cm]{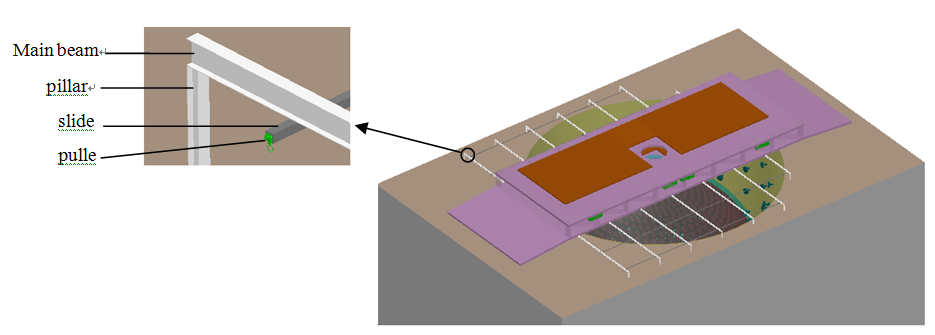}
\caption[Lead rail for rubberized fabric cover option installation]{Lead rail for rubberized fabric cover option installation.}
\label{Detview}
\end{center}
\end{figure}

The slide-guide scheme costs more, but is more easily implemented.
The main structure contains the main beams, columns, slide guides
and pulleys. A couple of slide guides is fixed on the main beams
under bridge. One end of the main beam is supported on the bridge,
and the other end is connected to the pillar fixed on the ground.

Each of the two schemes have advantages and disadvantages, and the
final design choice is still under discussion.

\subsection{Lining and Thermal Insulation Layer of the Pool}

The cylindrical pool with 42-meter diameter and height will be filled
with 20,000-30,000 tons of ultra-pure water. Since the laboratory is located 400 meters
underground, the pressure of underground water is high. A powerful lining
is needed to prevent water from the surrounding rock seeping into the pool.
We propose to adopt high-density polyethylene (HDPE) material as the
waterproof lining between rock and pool. HDPE has good corrosion resistance,
high strength, good ductility, and does not pollute ultrapure water.
The design can be similar to that for landfill, but the lining for the JUNO
water pool contains many through holes and further studies are needed.

The operating temperature of the JUNO detector system is 20 degrees,
while the temperature near the laboratory rock is around 32 degrees,
significantly higher than the operating temperature. The granites
surrounding the pool have poor permeability, which has certain heat
preservation effect. The concrete wall of the pool is a poor heat conductor
with a heat transfer coefficient of only 1.63  W/(mK).

According to an independent heat transfer calculation and the example of calculation for
the Daya Bay experiment, the pool thermal insulation may not require
additional insulating layer. However, to reduce operating costs,
it is proposed to add an insulating layer.

\subsection{Installation}

The detector installation includes the assembling of PMT and bracket, and
the fixation of PMT bracket. The assembling of PMT and bracket is shown
in Fig.~\ref{PMTstr}. To make full use of the embedded parts in concrete
pool wall, the PMT bracket is considered to be fixed on the embedded parts.
In addition, special tools and operation management are needed to ensure
safe and smooth detector installation.

\section{Prototype}

 \subsection{ Geomagnetic Shielding Prototype }

Through numerical calculation, we provide an option for the geomagnetic
shielding system for JUNO. A prototype of coil system is required for
validation and material property study. After the prototype study, we
will have a more accurate estimate for the shielding effect and
can then optimize the design of the geomagnetic shielding system.

\section{Detector Construction and Transportation}

The veto system mainly consists of the water Cherenkov detector and
the tracking detector. For the water Cherenkov detector, the most important
tasks are PMT test and transportation, water purification system construction
and transportation, and support structure manufacturing and transportation.
The veto PMT test and transportation will be done together with the
central detector PMTs. Water system manufacturing and transportation
and support structure construction await the final design decisions before
the bidding and other processes can be started. The option for the
top trackers also needs to be finalized before the detailed plans for the
constructions and transporting can be made.

\section{Risk and Safety Issues}

The large size of the JUNO detector limits the available space
in the experimental hall. For the veto system, it is important to
carefully coordinate the installations of various detectors and
hardwares. If the assembly coordination is not optimal, it could
influence the whole veto systems installation. To avoid this risk,
a careful coordination and strict schedule should be performed
for various assembly tasks. Any potential conflicts in the
installation schedule should be discussed and resolved
among the detector coordinators to ensure a smooth installation process.

Most of the safty issues occur during the detector installation stage.
The safety procedure  must be strictly followed when the large JUNO
detector is installed.

\section{Schedule}

The plan for the period 2014-2020 is as follows:

\begin{itemize}

\item
2014:
Setup the following two kinds of veto detector prototypes and
take data with cosmic muons:

1) liquid scintillator + fiber prototype;

2) a multilayer resistive plate chamber (RPC).

\item
2015: Finalize the top tracker option.

\item
2016: Determine the pool liner design and begin to prepare lining
 installation;

Complete the design of water pool insulating layer design and start
installation preparation;

 Determine geomagnetic shielding design and start the installation preparation.

\item
2017: Start to install water pool insulating layer;

Start to install water pool lining;

Start to install the geomagnetic shielding system.

\item
2018: Complete the production of water Cherenkov detector PMT
support structure and top track detector supporting structure;

Start the installation of the top
track detector and water Cherenkov
detector PMT installation.

\item
2019: Complete water Cherenkov detector installation and top track
detector installation;

Start pump ultra pure water into water pool.

 \item
2020: Complete pump
process and start veto system commissioning.

\end{itemize}

\cleardoublepage
\chapter{PMT}
\label{ch:PMT}

\section{Design Goals and Specifications}

\subsection{Introduction}

Photon detectors measure the scintillation light created by interactions of the
neutrinos with liquid scintillator and are key components for accomplishing the
physics goals of JUNO. Important requirements for photon detectors used in JUNO
include high detection efficiency for scintillation photons, large area, low
cost, low noise, high gain, stable and reliable, long lifetime. JUNO plans to
use approximately 20,000 tons or 23175~m$^{3}$ liquid scintillator hosted in a
spherical container of diameter 35.4~m. Photon detectors will be mounted on a
sphere 40~m in diameter covering at least 75\% of the 4654~m$^{2}$ surface area
of the sphere according to the reference design of the central detector. Large
area vacuum photomultipliers are the only viable choice for the JUNO photon detectors.
Table~\ref{tab:PMTsrequiredbyJUNO} shows the numbers and sizes of PMTs required for JUNO in two scenarios with photocathode coverage ratio of 75\% and 70\%. Assuming 508~mm (20~inch) diameter PMTs with a photocathode diameter 500~mm are used, $\sim$17,000 such PMTs are required by the central detector of JUNO to cover 75\% of the detector surface.

\begin{table}[!hbp]
\centering
\caption{PMTs required by JUNO\label{tab:PMTsrequiredbyJUNO}}
\begin{tabular}{|c|c|c|}
\hline
\hline
	PMT mounting diameter &		photocathode coverage  &     508 mm   	 \\
\hline
	40 m &	70\%       &   	16,000  	 \\
\hline
	40 m &	75\%    	&    17,000 	 \\
\hline
\end{tabular}
\end{table}

\subsection{PMT Spectications}

\begin{figure}[htb]
\begin{center}
\includegraphics[width=0.8\textwidth]{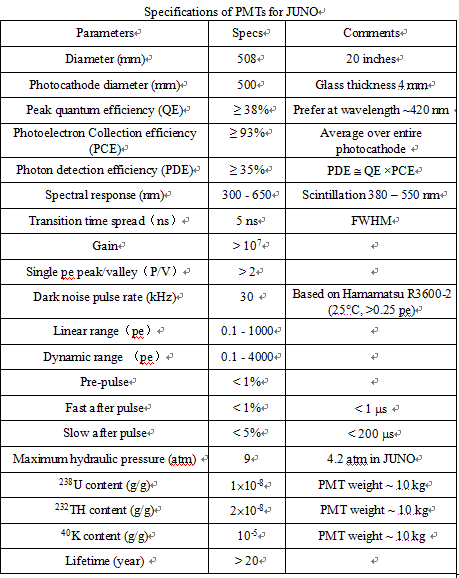}
\caption[ Specifications of PMTs for JUNO]{\label{fig:specificationOfPMTs} Specifications of PMTs for JUNO }
\end{center}
\end{figure}

The main specifications for the 508~mm PMTs in JUNO are given in
Fig.~\ref{fig:specificationOfPMTs}. Bialkali (BA) photocathode will be used. The
spectral response of the BA photocathode matches the light emission spectrum of
the liquid scintillator well. One of the most critical requirements for JUNO
PMTs is high photon detection efficiency (PDE), which is equal to the photocathode
quantum efficiency (QE) multiplied by the efficiency of photoelectron
collection efficiency (PCE) in the PMT vacuum. In order to achieve the
required peak PDE of 35\%, the peak photocathode QE is set to 38\% by assuming
the PCE to be 93\%. We hope the average PDE for scintillation photons,
which have a rather broad spectrum, can be > 30\%.

Note that PMTs are measured by manufacturers in air and with incident light
beams perpendicular to the surface of the glass entrance while PMTs are
submerged either in water or in oil when used in JUNO, and the scintillation
light can penetrate the curved glass window of a PMT at any angle. Small
discrepancies regarding the photocathode QE between the manufacturers,
measurements and those experimentally observed can occur as a result of
differences in light reflections at the medium/glass interface.

JUNO PMTs will work in photon counting mode and the typical gain required is
$10^{7}$. The PMT noise pulse level must be low so that the single
photoelectron peak can be clearly separated from the PMT noise. In JUNO cosmic
events the minimum momentum of cosmic ray muons is 218~GeV and the penetrating
muons' rate is 3.8~Hz. Energy deposition by muons that do not generate showers
would be as high as 7~GeV and a total of $7\times10^{7}$ scintillation photons
would be produced. The required dynamic range of JUNO PMTs is given as 0.1-4000~pe's.
The photocathode noise rate, rate of prepulse and after-pulses are given in
Fig.~\ref{fig:specificationOfPMTs}. The radioactivity levels of the PMTs,
which mostly comes from the glass, are also given.

In addition to the 508~mm diameter PMTs for measuring scintillation light
generated by neutrino reactions in the detector, approximately 2000 of
203~mm (8~inch) diameter PMTs are needed for cosmic ray muon veto by detecting
Cherenkov light generated by cosmic muons. These PMTs will be mounted on the
outside surface of the stainless steel structure that supports the liquid
scintillator vessel. The total area to be covered is approximately
5000~m$^{2}$. The requirements for these PMTs are similar to those used in the
Daya Bay neutrino detector.

\section{MCP-PMT R\&D }

In order to ensure optimum performance of the detector, the PDE of PMTs in the
detector is preferred to be $\ge$~35\%, averaging over the entire photocathode
and scintillation spectrum.  No commercial large PMTs can achieve this. A R\&D
effort intended to accomplish such an ambitious goal was started in early 2009.
This effort was led by IHEP (Beijing) and joined by research institutes,
university groups and related companies in China, and a formal R\&D group was
established in early 2011. Significant progress in all aspects of PMT R\&D has
been made. The goal of the design and manufacturing 508~mm PMTs that meet the
requirements listed in Fig.~\ref{fig:specificationOfPMTs} and in particular to achieve the 35\% PDE is
not easy and cannot be taken for granted. If we are unable to accomplish our
stated goal in time, we will make corresponding adjustments to the design of
the experiment. Depending on the status of our R\&D effort by the time of the
decision on the PMT choice, commercially available PMTs that may or may not
have the desired photon detection efficiency can be considered.

It is worth noting that in early 2009, a team from IHEP lead by Yifang Wang,
then the deputy director of IHEP visited a former PMT manufacturer, the No.
741 factory of the Chinese Electronics Administration in Nanjing, China. The
manager of that factory became interested after hearing of our effort to
develop large PMTs for JUNO (then Daya bay II). He made an effort to line up
financial support from two entrepreneurs in Beijing and successfully negotiated
a deal with PHOTONIS France S.A.S. and bought their stalled PMT production line
and all related technologies. A Chinese PMT company, SCZ PHOTONIS, was
established in 2011 and have now started to make 12 inch PMTs, with a plan in
place to make 20 inch PMT samples soon.

Meanwhile, IHEP representatives have been in contact with Hamamatsu, Japan,
which have been developing high efficiency 20 inch PMTs for the future Hyper
Kamiokande experiment in Japan and made good progress in the last few years.
Hamamatsu has agreed to deliver 20 inch PMT samples to IHEP for testing.

We will discuss the R\&D effort for JUNO PMTs by the collaboration led by IHEP (Beijing) in this section. Options provided by the two commercial companies will be discussed in section 6.4.2.

\subsection{MCP-PMT R\&D at IHEP}

The R\&D effort to develop PMTs for JUNO started in early 2009 when the  Daya
bay II experiment was in the very early conceptual development stage. The main
obstacle for achieving the physics goals of the experiment was the lack of
commercially available PMTs that can reach the photon detection efficiency
(PDE) required. Of the two major PMT manufacturers in China, the aforementioned
No. 741 Factory in Nanjing had stopped PMT production for a number of years and
another PMT manufacturer, the No. 261 Factory in Beijing was bought by
Hamamatsu and the resulted Hamamatsu Photonics (Beijing) produced only low-end
small PMTs. The R\&D work on PMT related technologies such as Bialkali
photocathode had ceased for more than a decade in China.

\begin{figure}[htb]
\begin{center}
\includegraphics[width=0.8\textwidth]{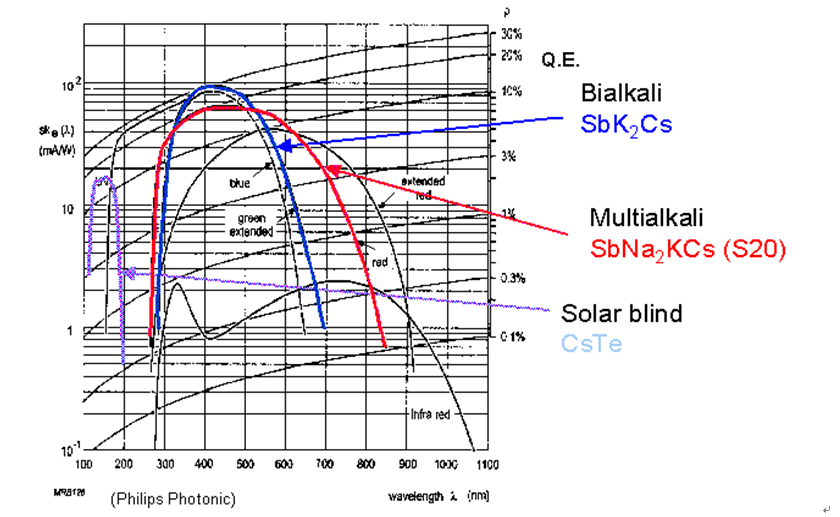}
\caption[ The photocathode sensitivity and quantum efficiencies of various photocathodes]{\label{fig:Fig.5.2} The photocathode sensitivity and quantum efficiencies of various photocathodes}

\end{center}
\end{figure}

The quantum efficiency (QE) of a vacuum PMT is conventionally defined as the
number of photoelectrons emitted from the photocathode layer into vacuum by the
number of incident photons. When the QE of a PMT is measured, the device is
placed in air and the incident light beam is perpendicular to its glass window
action. Typically 4\% of photons are reflected by the glass surface and more
light is reflected at the interface of the glass substrate and the photocathode
layer, which has a high index of refraction. The cathode sensitivity and QE as
functions of wavelength of the incident photons are plotted in Fig.~\ref{fig:Fig.5.2} (taken
from the PMT manual published by Philips Photonics). The blue curve is QE of
conventional bialkali photocathode (SbK2Cs) with a peak value of about 30\%.
The spectral response of the conventional BA  photocathode matches the spectrum
of scintillation light from the liquid scintillators well and the estimated
average QE for the entire scintillation light emission spectrum is about 25\%
based on this plot. However for large PMTs, the QE is normally lower than the
values shown in Fig.~\ref{fig:Fig.5.2}.

For the 20 inch Hamamatsu R3600-02 used in the KamLand experiment, the peak QE
is about 20\% and the peak photon detection efficiency (PDE) is only about 16\%
since the photoelectron collection efficiency is 80\% as stated by Hamamatsu
for the R3600-02. The 16\% peak PDE is far lower than the desired 35\% PDE for
JUNO. Another factor that also needs to be considered is the effective
photocathode area. The R3600-02 PMT has an outside diameter of 508~mm and its
effective photocathode diameter is only 460~mm. This factor puts a limit on the
capability for JUNO to maximize its photocathode coverage. Given these factors,
it became apparent that we need to start our own large PMT R\&D program mainly
to improve the photon detection efficiency. Note that new 20 inch PMT (R12860
Under development) samples with super bialkali photocathode from Hamamatsu have
shown great improvement for the PDE.

\begin{figure}[htb]
\begin{center}
\includegraphics[width=\textwidth]{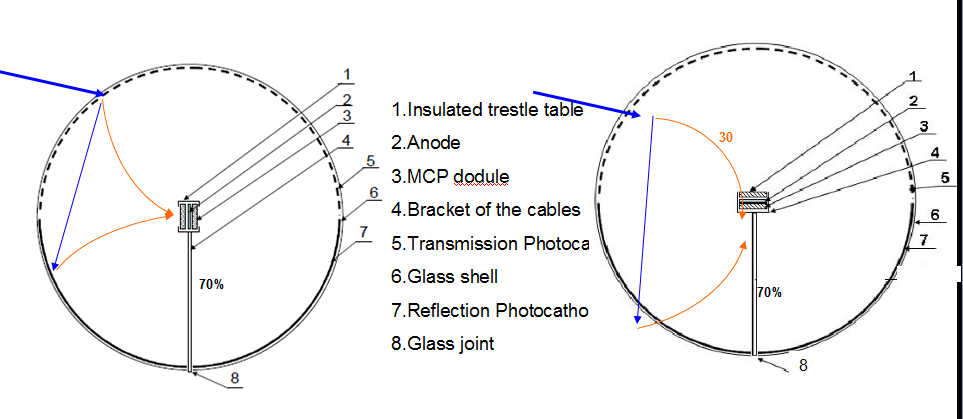}
\caption[The conceptual drawing of the high efficiency spherical PMT with two sets of MCP electron multipliers]{\label{fig:Fig.5.3} The conceptual drawing of the high efficient spherical PMT with two sets of MCP electron multipliers }
\end{center}
\end{figure}

Our PMT R\&D was based on a concept proposed by IHEP (Beijing) in early 2009.
The conceptual schematic is shown in Fig.~\ref{fig:Fig.5.3}. The concept calls
for a spherical PMT with top hemisphere, as shown in the figure, used as
transmission photocathode and its bottom hemisphere used as reflective
photocathode which converts the photons that passed through the transmission
photocathode. It is hoped that by optimizing the thickness of the photocathode
in the top and bottom halves, the QE of such a PMT can be significantly
improved compared to the conventional design that has only the top transmission
photocathode. This new design requires two back-to-back compact electron
multipliers, since conventional dynode chains would be too bulky. The proposed
electron multipliers are two back-to-back pairs of microchannel plates (MCPs),
so that photoelectrons from the top and bottom photocathodes are focused on the
two sets of MCPs facing left and right, as shown in Fig.~\ref{fig:Fig.5.3}.

As a starting point, a contract for a 5 inch MCP-PMT prototype was signed with
the Nanjing Electronic Devices Institute, which was the only place in China
that had made MCP PMTs. The progress was slow at the beginning and after about
two years, working prototypes were successfully made. While working with the
Nanjing Electronic Devices Institute, it became apparent that we needed to
expand our efforts in order to broaden our technology bases to solve many
complex and difficult technical issues. The Large Area MCP-PMT Development
Collaboration was formally established in 2011. Coordinated by IHEP, the
collaboration consists of the CNNC Beijing Nuclear Instrument Factory, the
Xi'An Institute of Optics and Precision Mechanics (XIOPM) of the CAS, Shanxi
Baoguang Vacuum Electric Device Co., Ltd., China North Industries Group
Corporation (NORINCO GROUP), Nanjing University, and the PMT Division of the
CNNC Beijing Nuclear Instrument Institute (formally No. 261 Factory). Great
progress has been made since the start of the collaboration almost three years
ago.

\subsection{Large Area High Efficiency MCP-PMT Project}

Many critical components and technologies need to be developed for the success
of the Large Area High Efficiency MCP-PMT Project. First, the raw materials and
methods to produce large glass bulbs with high strength, high optical quality
and extremely low radioactivity must be identified. Among many requirements,
the glass bulbs must be able to work in pure water under high pressure. The
technology for vacuum sealing between the glass and the Kovar metal leads must
be developed. Technologies required for making large area high QE photocathodes
are also needed. The alkali metals and Antimony sources and thin film
deposition technologies necessary for photocathode fabrication also require
development work. Fabricating MCP pairs as electron multipliers with low noise
and gain exceeding 10$^{7}$ is very complex and requires a lot of development
work. Other key technologies include electro-optic, mechanic and electric
designs for MCP-PMTs and the testing of such PMTs. At the time when the
collaboration was started, most of the required technologies did not exist in
China. After three years of collective work, 8 inch and 20 inch MCP-PMT
prototypes have been successfully made and evaluated.

\subsubsection{Basic Concept of Large Area MCP-PMT}

1) Effect of combining the transmission and reflection photocathodes

Conventional wisdom says that the QE of a reflective photocathode, that can be
made thicker, has much higher QE than that of a transmission photocathode.
Coating the lower half hemisphere of a spherical PMT bulb to convert the light
that has passed through the transmission photocathode on the top hemisphere
into photoelectrons can in principle improve the QE of the device. Scientists
at XIOPM have done theoretical analysis and computer simulation, showing that
by optimizing the thicknesses of transmission and reflection photocathodes, the
QE of the spherical PMT can indeed be improved. Such improvement has also been
seen in prototypes of MCP-PMTs.

2) Electro-optical Simulation

\begin{figure}[htb]
\begin{center}
\includegraphics[width=0.8\textwidth]{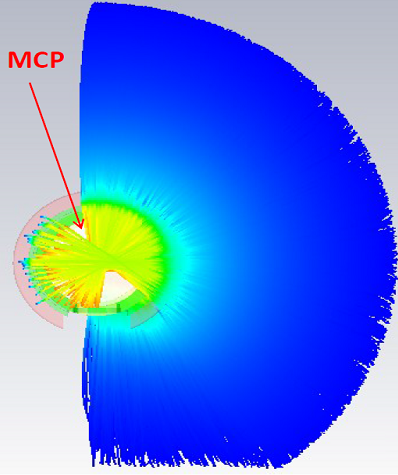}
\caption[MCP-PMT electro-optic simulation with Earth magnetic field]{\label{fig:Fig.5.4} MCP-PMT electro-optic simulation with Earth magnetic field }
\end{center}
\end{figure}

The shapes of photocathodes of the focusing electrodes required for effective
photo-electron collections, and of the other structures inside the PMT bulb,
require careful electro-optical calculations. Commercial computer software
specially written for this purpose are extensively used by XIOPM and NORINCO to
optimize the MCP-PMT design both for 8 inch and 20 inch MCP-PMTs. An example of
the computer simulation results is given in Fig.~\ref{fig:Fig.5.4}. The
simulation includes the photocathode, MCP assembly, its support structure and
metal leads, focusing electrodes and the Earth's magnetic field. Electric
potentials for various electrodes were optimized.

3) MCP computer simulation

\begin{figure}[htb]
\begin{center}
\includegraphics[width=0.8\textwidth]{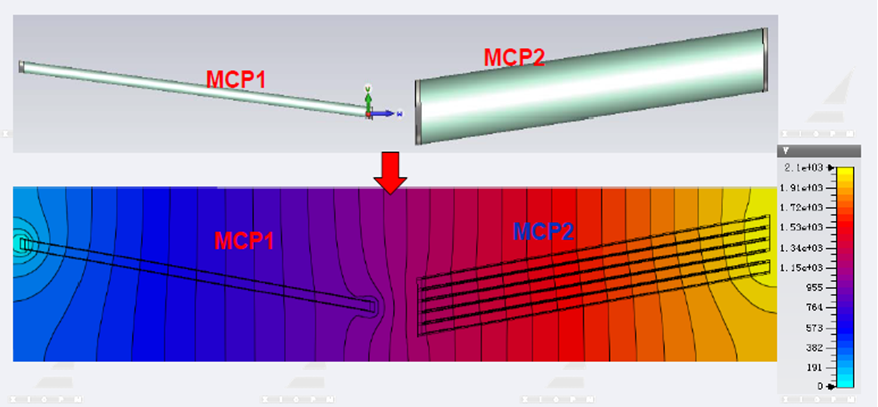}
\caption[Simulation for MCP functionality]{\label{fig:Fig.5.5} Simulation for MCP functionality }
\end{center}
\end{figure}

XIOPM has performed computer simulations for photo-electron collection and
amplification in the microchannels of MCPs. Electrode arrangement, gap size
between two MCP plates, size of microchannels and inclination angles were
studied. A diagram of such simulation is given in Fig.~\ref{fig:Fig.5.5}. The
MCP assembly has several advantages such as compactness insensitivity of the
electron amplification to   the Earth's magnetic field, high dynamic range,
etc. Additional concerns, including the photoelectron collection efficiency and
lifetime, etc. still need to be addressed.

\subsubsection{MCP and MCP Assembly}

NORINCO has decades of extensive experience in MCP production and development.
The company has organized a special group to develop the MCP and MCP assembly
required for the MCP-PMT project. Large capital and manpower investments have
been made. MCP parameters and processing procedures have been adjusted. High
gain MCP assemblies with gain exceeding 10$^{7}$ for a pair of MCPs have been
developed after the design of the MCP assembly was optimized. This level of
gain is critical for single photoelectron detection.

\subsubsection{Glass Tube and Raw Materials}

\begin{figure}[htb]
\begin{center}
\includegraphics[width=0.8\textwidth]{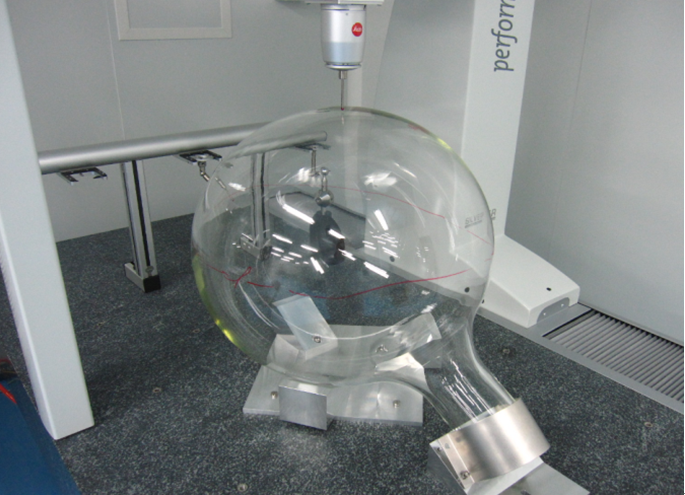}
\caption[Vacuum sealed 20inch glass bulb on a CMM for mechanical accuracy measurements]{\label{fig:Fig.5.6} Vacuum sealed 20inch glass bulb on a CMM for mechanical accuracy measurements }
\end{center}
\end{figure}

Our 8 inch MCP-PMTs use the traditional electro-vacuum glass, whose coefficient
of thermal expansion (CTE) matches the CTE of Kovar, which is used for vacuum
feedthrough pins. The vacuum seal is easy to make. However, the electro-vacuum
glass is difficult to blow into 20 inch glass tubes due to its poor mechanical
properties above the softening point. Also there are other issues that make the
electro-vacuum glass unsuitable for 20 inch PMT tubes even if it can be used to
make 20 inch glass tubes. Its mechanical strength is not good enough to
withstanding the high hydraulic pressure, while oxides of alkaline metal in the
glass can slow leach out and affect the optical transparency of the glass.
Therefore, the Pyrex type of borosilicate glass that has the required
mechanical strength and chemical stability is preferred for making 20 inch
glass tubes. Unfortunately, the CTE of Pyrex glass does not match the CTE of
Kovar so a section of glass tube that joins the Perex to electro-vacuum glass
needs to be made by using glasses with different CTEs. Techniques and glass
materials must be developed for making this multiple transition glass tube
section. The aforementioned technical difficulties have been solved by the
collaborative effort of our collaboration and industrial partners. The
resulting 20 inch glass bulb has a good appearance and excellent quality. Its
mechanical accuracy and thickness uniformity are also excellent. Some completed
glass bulbs, including the Pyrex sphere and neck, multiple transient section
and Kovar seal, as shown in Fig.~\ref{fig:Fig.5.6}, have been tested in a pressure test
set-up with pressure up to 10 bars.

Since the neutrino event rate in JUNO is extremely low, radioactivity of the
PMTs for JUNO must be tightly controlled. The radiative decay of radioactive
materials in PMT can potentially be a major source of background for a low
energy neutrino experiment. The radioactive elements $^{238}$U, $^{232}$TH, and
$^{40}$K in the glass are the major contributors of PMT radiation backgrounds.
The glass of the Hamamatsu 8 inch R5912 used in Daya bay detector has rather
high radioactivity level, which is acceptable for a relatively high rate
experiment like Daya bay, but far exceeds the level that can be tolerated by
JUNO. To control the radioactive level in the PMT glass tubes, care must be
taken to select the raw materials and also to control the glass melting process
preventing the radioactive materials contained in the crucibles to be dissolved into the molten glass.

The main component of Pyrex glass is SiO2. We have identified sources of high
purity quartz sand in China and indeed the measured $^{238}$U, $^{232}$TH, and
$^{40}$K levels are satisfactory for JUNO. Low radioactivity raw materials for
other ingredients of the Pyrex glass have also been identified. Costs of high
purity raw materials are significantly higher than those of materials commonly
used to make Pyrex glassware. However, the cost of raw materials for glass,
even the high grade materials used, are still an insignificant part of the total PMT production cost.

Although we believe we know how to do it, so far low radioactivity glass tubes
have not yet been manufactured. The 20 inch glass tubes for our prototype PMTs
have been made with regular materials and using regular procedures in the
factory. This is because even just to make a test run for low radioactivity
glass requires the glass factory to stop their regular production for an
extended period to purge their ovens, and the test would cost a significant
amount of money. We will do a low radioactive test production when we are ready
to make a small batch of 20 inch PMTs.

\subsubsection{Photocathode}

\begin{figure}[htb]
\begin{center}
\includegraphics[width=0.8\textwidth]{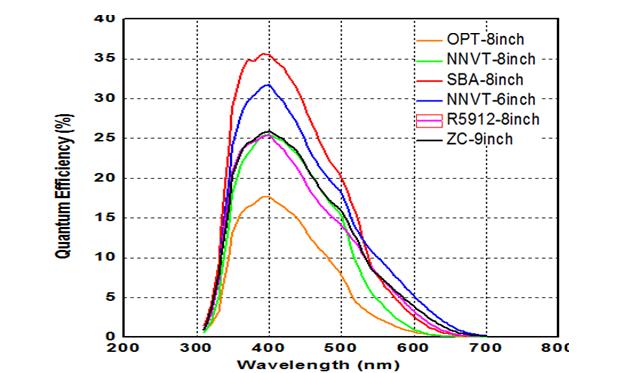}
\caption[QE]{\label{fig:sec2QE} QE }
\end{center}
\end{figure}

The Photocathode is a critical component of our PMT R\&D project. Typically to
make bialkali photocathode, a very thin Antimony layer is first deposited on
the inner surface of an evacuated glass tube, then layers of the alkali metals
Potassium and Cesium are deposited on top. Under high temperature a
semiconductor layer of K$_{2}$CsSb photocathode is formed. Both XIOPM and
NORINCO, on the basis of their previous experiences, have tried to make
bialkali photocathode for 8 inch MCP-PMTs successfully. Recently NORINCO has
also made a 20 inch photocathode with good results as shown in
Fig.~\ref{fig:sec2QE}.

\section{Organization of the MCP-PMT R\&D Project}

\subsection{Introduction}

The required effort to develop a brand new product such as 20 inch MCP-PMT is
very  technically demanding, and the R\&D process is complex and time consuming
both for skilled technicians and scientists. The cost of the R\&D, including
the basic equipment and the investment for developing the various technologies
involved, is quite high. IHEP (Beijing) would be unable to take on this task by
itself. The approach we have taken is to organize a concerted effort by
factories that have experience of producing PMTs and MCPs, as well as by
research institutions and university groups that have strong theoretical and
practical background in opto-electronic devices. This approach led to creating
a collaboration able to exploit effectively the expertise of each individual
member institution toward the final goal of designing and fabricating 20 inch
PMTs that meet the requirements for JUNO. 

\begin{figure}[htb]
\begin{center}
\includegraphics[width=\textwidth]{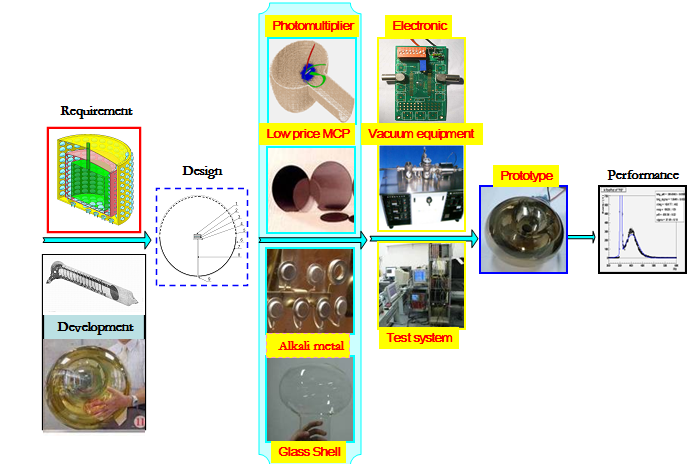} 
\caption[Flow chart of the large area MCP-PMT R\&D process]{\label{fig:Fig.5.7} Flow chats of the large area MCP-PMT R\&D process}
\end{center}
\end{figure}

Unlike other electronics devices, PMTs often cannot be tested at every
production stage; instead meaningful tests can only be made after all the
production steps are completed and the tube is evacuated to high vacuum level.
We must do as much as possible of the testing and Q\&A work before the mass PMT
fabrication is started. Therefore, we divide our R\&D project into the
following categories, as shown in Fig.~\ref{fig:Fig.5.7}.

The development process includes many different aspects that must proceed in parallel as listed below.

1. Alkali sources: Develop and produce standardized alkali sources that ensure high QE and stable photocathode.
  
2. Glass: Develop 20 inch high quality and low radioactivity glass bulbs.

3.  MCP: Develop and produce low noise, high gain MCPs and optimize the performance of MCP assemblies 

4.  Electronics: Optimize the signal propagation from the anode to readout electronics

5. R\&D to fabricate large high vacuum equipment for high throughput PMT production and testing .
  
6. Vacuum seal: Techniques for hot Indium seal and glass seal

7. Electro-optic: Design and performance simulation

8.  Performance tests: Platform for 20 inch PMT testing that can be used in the PMT factory and other places.

\subsection{Capability of the Collaboration}

\subsubsection{IHEP (Beijing)}

\begin{figure}[htb]
\begin{center}
\includegraphics[width=\textwidth]{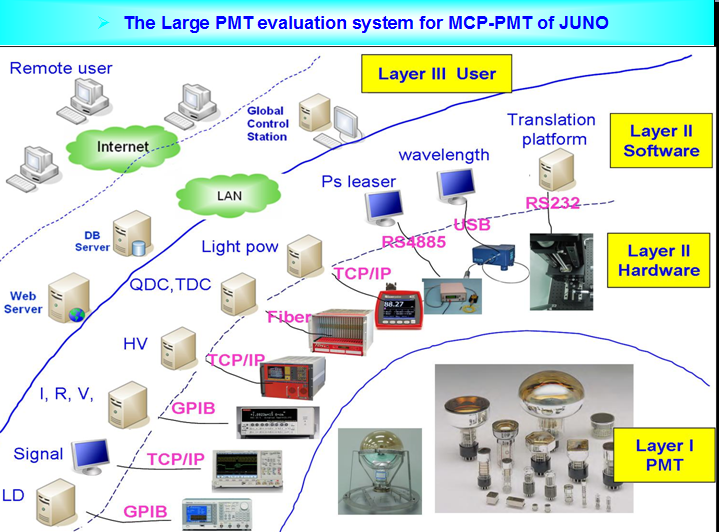} 
\caption[ Organization of the existing PMT test lab at IHEP]{\label{fig:Fig.5.8} Organization of the existing PMT test lab at IHEP}
\end{center}
\end{figure}

The Institute of High Energy Physics, Chinese Academy of Sciences, is the
initiator and coordinator of the Large Area MCP-PMT Project. IHEP has extensive
experience in nuclear and particle instrumentation, vacuum technology, PMT
testing, and PMT applications in large particle physics experiments. The Large
Area MCP-PMT Project is organized within the Time-of-Flight detector group,
which is a subgroup of the State Key Laboratory of Particle Detection and
Electronics at IHEP. Funded partly by the Major Research Instrumentation
Initiatives of CAS, the IHEP group has built a PMT Performance Testing Lab.
Many PMT parameters such as the single photoelectron (SPE) spectrum, absolute
QE and photocathode sensitivity, dark current and dark pulse rate, linearity
and gain curves, etc. can be measured quickly and accurately. Figure~\ref{fig:Fig.5.8} shows graphically the functionality of the PMT testing lab. 

The PMT Performance Lab is equipped with a 10000 class clean room, a dark room
with electromagnetic shielding, an optic testing system based on LEDs, a
variable frequency laser, light flux calibration systems and associated optic
components. Advance testing instrumentation, including pA meters and digital
oscilloscopes has been installed. With a VME based automated PMT input surface
scanning system and testing system for photo gain curves, the absolute
photocathode sensitivity (accurate to 1~$\mu$A/lm), QE (relative uncertainty <
2\%) and single photon transit time spread TTS can be measured in the frequency
range 230~nm - 1000~nm (accurate to 50~ps). The dark current can be measured to
pA level in the range of 1~pA - 10~nA. This test system is completely
automated and the HV, laser wavelength, data recording and data analysis can be
controlled by computers.
 
\begin{figure}[htb]
\begin{center}
\includegraphics[width=\textwidth]{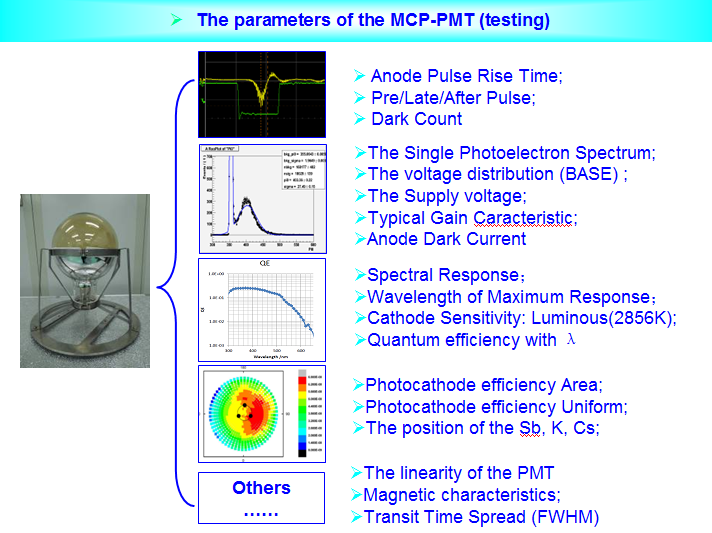} 
\caption[PMT measurement results at IHEP]{\label{fig:Fig.5.9}PMT measurement results at IHEP}
\end{center}
\end{figure}

Moreover, a VME based automated single photoelectron (SPE) spectrum recording
system has been built. The SPE testing usually requires the PMT gain to be set
to gain > 10$^{7}$, but with a preamplifier the SPE spectrum can be acquired 
also at a reduced gain > 10$^5$. Fig.~\ref{fig:Fig.5.9} shows measurement results by such a system.  

In addition, IHEP has recently assembled a small group to perform fundamental
studies on MCP optimization to prepare for the MCP-PMT production and to study
optic thin films. An ultra low radioactivity analysis system based on a
Canberra High-purity Germanium (HPGe) Detector with low background shielding
and BGO veto counter has been built at IHEP. This system is used to study the
radioactivity of PMT glass and raw material samples.

\subsubsection{XIOPM}

The XiAn Institute of Optics and Precision Mechanics (XIOPM) of CAS has
extensive experience in research and fabrication of various opto-electronic
devices and related technologies. With significant government funding, the
State Key Laboratory of Transient Optics, Photonics and Optoelectronics
Laboratory and the Photoelectric Measurement and Control Technology Research
Department at XIOPM have recently upgraded laboratories with new clean rooms
and new equipment, including three research platforms.

1. Platform for opto-electronic devices: Design software for opto-electronic devices, high precision assembly table, plasma surface cleaning, vacuum seal and welding machines, vacuum leak detection systems, photocathode activation station, photocathode testing station, and others. This platform provides basic tools for high sensitivity, high quality opto-electronic devices R\&D and fabrication. 

2. Platform for high speed electronic fabrication and testing: Software for designing radiofrequency circuits, circuit fabrication facility.

3. Platform for opto-electronics testing: static and dynamic opto-electronics testing system, image intensify and fast DAQ system.

\begin{figure}[htb]
\begin{center}
\includegraphics[width=\textwidth]{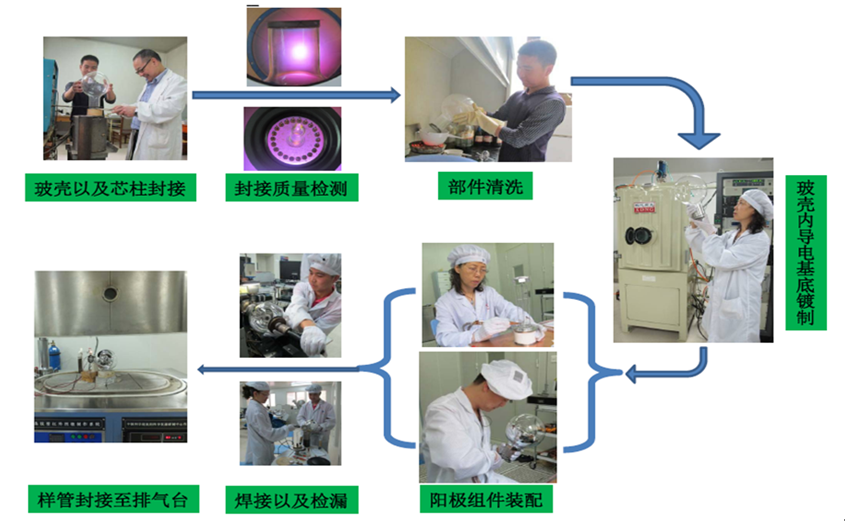} 
\caption[MCP-PMT fabrication process at XIOPM]{\label{fig:Fig.5.10}MCP-PMT fabrication process at XIOPM}
\end{center}
\end{figure}

The 8 inch MCP-PMT design and prototype fabrication were started at XIOPM in
2011, followed by the design and fabrication of the equipment needed for the 20
inch MCP-PMTs after some success in their 8 inch MCP-PMT program. In this
process, they have performed extensive opto-electronics simulation, theoretical
analysis for high QE photocathode, photocathode fabrication and measurements,
MCP functionality studies and optimization. Part of their work is shown in
Fig.~\ref{fig:Fig.5.10}. The 20 inch MCP-PMT fabrication equipment,
both the non-photocathode transfer and photocathode transfer types, is now almost
ready at XIOPM and they are essentially ready to produce some 20 inch MCP-PMT
prototypes.

\subsubsection{NORINCO Group}

China North Industries Group Corporation (NORINCO GROUP) is one of five 
producers of image intensifier tubes based on MCPs. NORINCO has teams of
engineers to design and fabricate special equipment for fabricating and testing
vacuum opto-electronic devices. In particular it has more than 30 years 
experiences in R\&D, design, fabrication and testing for various photocathodes,
MCP and MCP assemblies. Their original technology came from PHOTONIS in the
Netherlands. After many years refinement, they are now able to routinely
realize multialkali photocathodes in small image tubes with sensitivity
exceeding 900~$\mu$A/lm, which is higher than that reached by any other 
manufacturer in the world. For the Large Area MCP-PMT R\&D Project, their
expertise in bialkali photocathode is assisted by the expertise of No. 741 PMT 
factory.

For the R\&D of Large Area MCP-PMT project, NORINCO GROUP has organized a team of experienced engineers, made available workshops, lab spaces and equipment,  and invested a large amount of capital. Three of their platforms, some new and some existing with additional equipment, can be used for various MCP-PMT R\&D tasks. 
 
1. MCP Research Platform: Glass ovens, glass fiber pulling machines, MCP
slicing machines, automated etching machines, electrode deposition stations,
MCP testing station.

2. Platform for low level opto-electronic devices: Software for opto-electronics design and simulation, instrument for high sensitivity photocurrent measurements, photocathode production station, including plasma cleaning, monitoring during photocathode fabrication,  thin film deposition stations for high QE photocathodes, photcathode sensitivity testing station, light transmission and reflection testing station, HV power supply design and fabrication and testing, etc.  
 
\begin{figure}[htb]
\begin{center}
\includegraphics[width=0.8\textwidth]{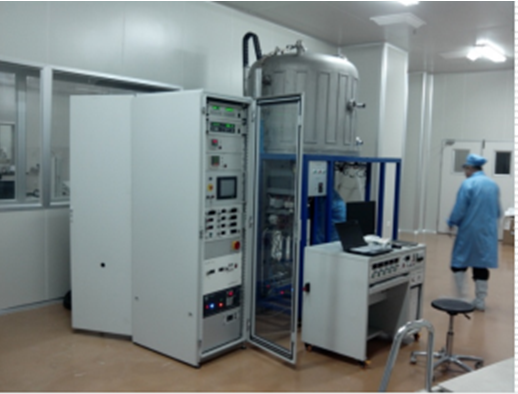} 
\caption[ Equipment for fabricating 20 inch MCP-PMTs developed by NORINCO using photocathode transfer technology]{\label{fig:Fig.5.11} Equipment for fabricating 20 inch MCP-PMTs developed by NORINCO using the photocathode transfer technology}
\end{center}
\end{figure}

3. Platform for Large Area PMT R\&D and prototyping: Glass lathe, large
annealing oven, RF welding machines, large ultrasound cleaning stations, clean
baking oven, point welding machines, thin film deposition stations, 20 inch PMT
fabrication station using non-vacuum transfer technology, 20 inch PMT
fabrication station using vacuum transfer technology (shown in
Fig.~\ref{fig:Fig.5.11}), photocathode activation monitoring station, PMT
dynamic testing system, PMT static testing system, opto-electronics design
software, etc.

These platforms provide a solid foundation for large area MCP-PMT design, R\&D, prototype testing and industrialization.

\section{R\&D Results and Plan}

\subsection{Status of MCP-PMT R\&D}

The main achievements up to June of 2014 made by the Large Area MCP-PMT
Collaboration are summarized in Table~\ref{Achievementsandstatus}.

\begin{table}[!hbp]
\centering
\caption{Achievements and status\label{Achievementsandstatus}}
\begin{tabular}{|c|c|}
\hline
\hline
	Items	 & Status	 \\
\hline
	8 inch MCP-PMT prototypes   &	Completed  	 \\
\hline
	PMT testing system    &	8 inch system completed at IHEP and XIOPM  \\
                           &     20 inch system near completion at NORINCO    \\
\hline
	20 inch  fabrication equipment  &	Near completion at XIOPM; Completion at NORINCO \\
\hline
	20 inch glass bulb  &  	Completed and met expectations  	 \\
\hline
     20 inch  Prototypes & 	Three produced at NORINCO   \\
\hline
\end{tabular}
\end{table}

\subsubsection{8 inch MCP-PMT Prototype R\&D}

\begin{figure}[htb]
\begin{center}
\includegraphics[width=\textwidth]{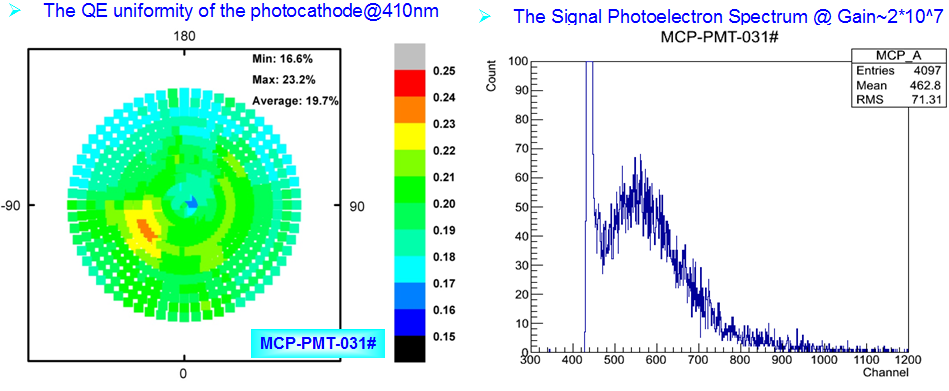}
\caption[QE values and SPE spectra of the 8 inch MCP-PMT]{\label{fig:Fig.5.12}QE values and SPE spectra of the 8 inch MCP-PMT}
\end{center}
\end{figure}

The Large Area MCP-PMT collaboration had its effort focused on the 8 inch
MCP-PMT prototypes in the first two years in order to develop the necessary
technologies for the 20 inch MPC-PMT. The opto-electronic design and
simulation, photocathode fabrication techniques, design of the MCP assembly,
MCP scrubbing, electrode design, etc. have gradually become mature and many
difficult problems have been solved. The highest QE achieved in 8 inch MCP-PMT
slightly exceeds 30\%. The required gain 10$^{7}$ has been achieved and the
dark current can be as low as 10~nA. The Peak/Valley (P/V) ratios of SPE spectra for
prototype tubes have been in the range of 1.6 to 2.5, with one prototype tube
reaching 3.8. QE and SPE spectra of the 8 inch MCP-PMT made at NORINCO and
tested at IHEP are given in Fig.~\ref{fig:Fig.5.12}.

\subsubsection{20 inch Glass Bulb and Transition Section}

As discussed previously, we use the so-called Beijing hard glass (Chinese
standard CG-17), that is equivalent to the Pyrex borosilicate glass, to make
the 20 inch glass bulbs. The CTE of the CG-17 glass is very low at
0.0000033/\SI{}{\degreeCelsius}. It has very high mechanical strength and can
withstand various strong acids, alkalis and pure water. Glass bulbs as large as
508 mm are hand-blown by skilled glass technicians. Very few people have such
skills and experience in the country. The collaboration was able to identify a
glass factory that was very cooperative, which successfully made more than 50
sample bulbs with very good quality. In particular, the achieved mechanical
tolerance (< 0.6 mm) is outstanding.

\begin{figure}[htb]
\begin{center}
\includegraphics[width=\textwidth]{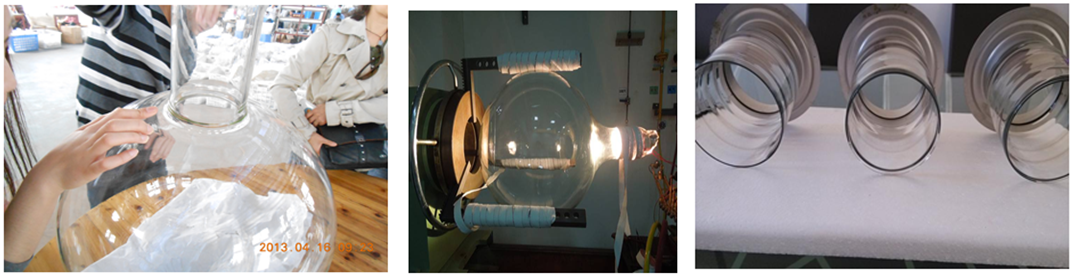}
\caption[20 inch glass bulb, its defect inspection and transition sections]{\label{fig:Fig.5.13}20 inch glass bulb, its defect inspection and transition sections}
\end{center}
\end{figure}

We have tried three methods to make the glass transition section match the low
CTE of the Pyrex glass with the slightly higher CTE of the glass used to make
the vacuum seal with the Kovar pins and the end flanges. After being sealed and
annealed, the evacuated glass bulb was placed in a test container filled with
water at 1~MPa pressure (equivalent to 100 meters of water) and observed for 24
hours without any problem. Photos of a 20 inch glass bulb, its visual defect
inspection and transition sections are shown in Fig.~\ref{fig:Fig.5.13}.

20 inch MCP-PMT prototypes were successfully made at the Nanjing factory of the
NORINCO in June, 2014 by using the photocathode vacuum transfer equipment.
After increasing the pumping capacity of the fabrication station, a QE level of
26\%  has been reached. More detailed testing is in progress and we expect
there will be further  improvements when more prototypes will be made. In
addition, prototypes of 20 inch MCP-PMT will be fabricated by using the
non-vacuum transfer station and comparison will be made of the 20 inch
prototypes made by both technologies.

\begin{figure}[htb]
\begin{center}
\includegraphics[width=\textwidth]{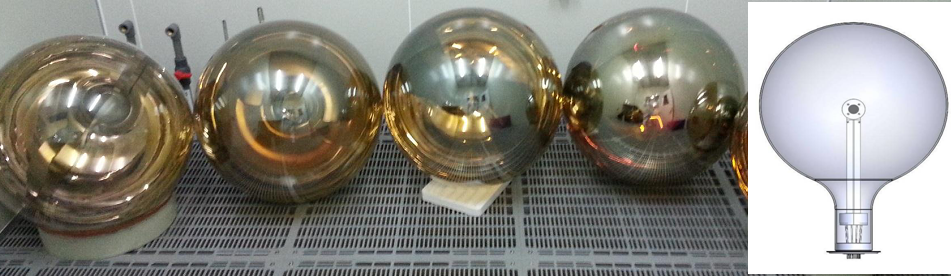}
\caption[ 20 inch MCP-PMT Prototype]{\label{fig:Fig.5.14} 20 inch MCP-PMT Prototype}
\end{center}
\end{figure}

\subsubsection{Improving the Testing System}

\begin{figure}[htb]
\begin{center}
\includegraphics[width=\textwidth]{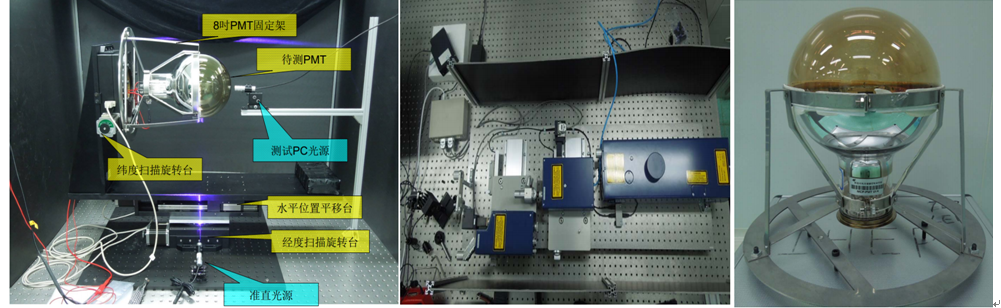}
\caption[ PMT test system at IHEP]{\label{fig:Fig.5.15} PMT test system at IHEP}
\end{center}
\end{figure}

The testing system for large MCP-PMTs with spherical photocathodes are
different in many aspects from the system used for conventional PMTs. IHEP,
XIOPM and NORINCO are all in the process to improve their current testing
system. The existing testing system at IHEP is shown in
Fig.~\ref{fig:Fig.5.15}. Photocathode sensitivity, QE, dark current, SPE
spectra, gain, etc. can be measured. The automated scanning station can
accommodate 8 inch PMTs. Test results using this system agree well with
parameters provided by Hamamatsu and Photonis for their standard PMTs.

\subsection{Status of Commercial PMTs}

\subsubsection{Hamamatsu}
	
In recent years, some PMT manufacturers have been developing so called SBA
photocathode and try to use such  photocathode in large size PMTs.
Hamamatsu has made a new SBA 20 inch R3600-02 with Venetian Blind type dynode.
Five such tubes have been recently tested. \cite{Nishimura2014Jan}. Hamamatsu is also developing
a 20 inch PMT (R12860) with box and line type dynodes, expecting 93\% photoelectron
collection efficiency compared to about 80\% for the R3600-02. The TTS will
also be improved with the box and line type dynodes. The QE of the Hamamatsu
20 inch SBA is expected to be about 30\% (Private communication with Hamamatsu)
at the sensitivity peak. The sensitivity peak of the current SBA photocathode
is located at 390~nm, which matches the requirement of the HyperK water Cherenkov
detector, but slight mismatchs the scintillation light, which is mainly emitted
at 390~nm to 450~nm. Based on what we know now, we can expect that the PDE of
Hamamatsu R12860 may reach 25\% for scintillation photons. Another problem of
using theR12860 for JUNO is that the size of its effective photocathode is
somewhat smaller than the 508 mm outside diameter , which would affect the
global useful coverage. Hamamatsu has promised to deliver samples of the R12860
to IHEP in June 2014.

\subsubsection{HZC Photonics}

HZC Photonics in Hainan Province, China acquired PMT production line and
related technologies, including the SBA technology, from PHOTONIS, France in
2011. HZC Photonics started their PMT production in late 2013 and now have a
12 inch PMT XP1807 available. Its stated photocathode sensitivity 60~\SI{}{\uA} /lm is
rather low. Their preliminary product, the XP53B20 3 inch circular tube, has a
very high peak photocathode sensitivity, listed as 160~\SI{}{\uA} /lm, but with rather
low gain of $ 6.25 \times10^{5}$ . HZC Photonics have promised IHEP to produce
samples of 20 inch SBA PMT for JUNO soon.

\section{Risk Analysis}

\subsection{Risk due to Low Photon Detection Efficiency}

PMTs are critical components for JUNO. The current MCP-PMT R\&D plan and backup
options carry certain risks that need to be considered. In order to achieve the
required energy resolution of less than 3\% at 1 MeV, which is expected to be
dominated by photon statistics, high efficiency liquid scintillator with a long
attenuation length must be used to detect enough photons when neutrinos
interact in the liquid scintillator. In addition, the fraction of the detector
surface that is covered by PMT photocathodes must be as large as possible and
the PDE of PMTs must be as high as possible. For a 508 mm diameter spherical
PMT, assuming glass wall thickness of 4~mm, the maximum diameter of the
photocathode would be 500~mm, or 97\% of the area of the PMT is actually
covered by photocathode. Space needed for anti-implosion covers and gaps among
PMTs futher reduce the maximum possible photocathode coverage.

\begin{figure}[htb]
\begin{center}
\includegraphics[width=0.8\textwidth]{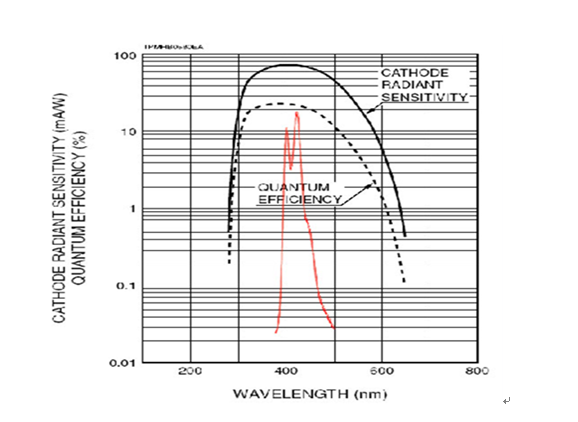}
\caption[Spectral response of common bialkali photocathode and the emission spectrum of liquid scintillator (red curve) ]{\label{fig:Fig.5.16}Spectral response of common bialkali photocathode and the emission spectrum of liquid scintillator (red curve) }
\end{center}
\end{figure}

We also must make the QE of PMTs as large as possible. As stated earlier, the
PDE of PMTs equals the QE of the photocathode multiplied by the photoelectron
collection efficiency (PCE). The goal of the peak photocathode QE for the JUNO
PMTs is 38\%, which is a very demanding goal. In Fig.~\ref{fig:Fig.5.16} (taken
from Hamamatsu PMT handbook), the solid black curve is the typical cathode
radiance sensitivity in unit of mA/W. The typical QE curve can be calculated
from it. The computed QE as a function of wavelength is given as the dashed
curve in Fig.~\ref{fig:Fig.5.16} and the scintillation light spectrum is plotted as the red
solid curve (linear scale) with double peaks at 400~nm and 425~nm.

\begin{figure}[htb]
\begin{center}
\includegraphics[width=0.8\textwidth]{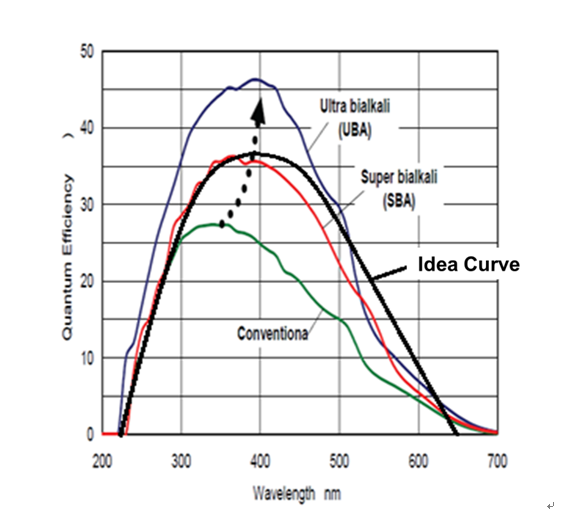}
\caption[QEs vs. wavelengths for conventional bialkali, SBA and UBA photocathodes]{\label{fig:Fig.5.17}QEs vs. wavelengths for conventional bialkali, SBA and UBA photocathodes}
\end{center}
\end{figure}

It can be seen from Fig.~\ref{fig:Fig.5.17} that the typical QE of regular
bialkali photocathode is 20 - 25\% within the wavelength range 350~nm to
450~nm. The value of the peak QE typically is no more than 30\% in commercial
PMTs. In recent years, Hamamatsu has developed SBA and UBA photocathodes with
higher QE. The QEs as functions of wavelengths of these new photocathodes are
given in Fig.~\ref{fig:Fig.5.17}. We will not discuss the UBA photocathode since the UBA is
only available in small flat faced PMTs made by Hamamatsu. The peak QE of the
SBA photocathode shown in Fig.~\ref{fig:Fig.5.16} is about 35\% and the peak
sensitivity occurs at ~390~nm. There are two important issues to be considered.
It can be seen that the spectral response curve of the Hamamatsu SBA is rather
narrow and there is a slight mismatch between the SBA spectrum response peak at
390 nm and the scintillation light emission peak at ~ 425~nm. In Fig.~\ref{fig:Fig.5.17}, the
black curve that is shifted slightly toward longer wavelengths indicates the
desired spectrum response, which matches the spectrum of emitted scintillation
light better. The second issue is that according to our conversation with
Hamamatsu, the typical peak sensitivity they expect for the future 20 inch SBA
PMT is 30\%, not the 35\% given in their published documents. The average QE
for scintillation light is expected to be less than 30\%. With the stated 93\%
photoelectron collection effiiency of the new SBA R12860, we speculate that its
average PDE may be less than 25\% due to the spectral response mismatch
discussed above.

The specified peak QE for JUNO PMTs as listed in
Table~\ref{tab:PMTsrequiredbyJUNO} is 38\% and we hope the average PCE of JUNO
PMTs can be higher than 30\% in the wavelength range 400~nm to 450~nm.
Therefore one of the critical risks is whether or not we can achieve this. Our
approaches to limit this risk are the following.

1. Transmission + Reflection Photocathode
Use the spherical photocathode, combining the power of transmission and
reflection photocathodes. We hope the targeted peak QE 38\% and average QE >
30\% within 400 - 450~nm can be achieved by optimizing the photocathode design
and fabrication process.

2. Transmission photocathode + reflection film
Coat a layer of thin film with high reflectivity in the bottom half of the sphere to reflect photons that have penetrated the transmission photocathode back and convert these photons into photoelectrons.

\subsection{Risk Related to MCP Aging}

 The operational lifetime of JUNO is 20 years and the working lifetime of the
 PMTs must also be greater than 20 years. It is expected that as long as the
 vacuum in the MCP-PMTs can be maintained, their lifetime will mainly be
 determined by the gain reduction due to MCP aging. It is expected that in the
 completely dark environment of JUNO, accumulated charges will mainly be caused
 by the photocathode thermal noise and the dark current of MCPs. At gain 10$^{7}$
 and dark counting rate 10~kHz (equivalent to dark current 16~nA), the
 accumulated charge due to photocathode thermal emissions will be ~10~C for 20
 years operation. The dark counting rate of a PMT is typically measured by
 counting dark pulses with a threshold of 0.25~pe. The true dark current,
 including the MCP dark current and dark current from other sources, can be much
 higher than the 16~nA used in the above estimations. Some of this
 additional dark current should be considered when the MCP aging effect is
 evaluated.

At this level of charge accumulation, MCP aging can become a problem for
MCP-PMT operation. The charge accumulation level of the MCPs developed by
NORINCO can reach 30~C. In addition, PMT anode assembly typically has a gain of
3 - 5 and the charge accumulation is reduced by the same factor if such an
anode design is used. This implies that the accumulated charge that the MCP-PMT
can be operated with can reach the level of 100~C.

In order to control the dark pulse rate and dark current due to MCPs, we are
testing two different methods to fabricate the MCP-PMTs. The photocathode
vacuum transfer method has separate vacuum chambers for photocathode
fabrication and MCP processing and can in principle lower the risk of MCP
contamination by alkali materials, thus achieving lower dark current.

In the past few years, MCP manufacturers and research labs have developed a new
type of MCP that use the ALD technique to deposit a thin secondary electron
emission layer on top of a resistive layer on the wall of the microchannels of
the MCPs. It has been shown that the lifetime of ALD coated MCPs can be an
order of magnitude higher than the conventional MCPs. IHEP has started a R\&D
program to develop MCPs using ALD technology.

\subsection{Risk due to PMT Radiation Background}

\subsubsection{Radiation Background Requirements}

The PMT radiation backgrounds are dominated by the radioactive elements, most
importantly $^{238}$U, $^{232}$Th and $^{40}$K in the glass bulb. In the current 20 inch spherical
MCP-PMT design, the weight of the glass bulb is ~9~kg. According to the current
detector physics simulation, the upper limits for $^{238}$U, $^{232}$Th and $^{40}$K contents
in the PMT glass are 12.1 ppb, 26.1 ppb and 1.5 ppb, respectively. For
comparison, the contents of these elements in Hamamatsu R5912 8 inch PMT glass
are 153 ppb, 335 ppb and 18.5 ppb, while they are 540.3 ppb, 568.3 ppb and 75.2 ppb in the
20 inch R3600-02 PMTs. The radioactivity levels of Hamamatsu PMTs are many times
higher than that allowed by JUNO. In a sample of regular CG-17 borosilicate
glass, the type of glass to be used to fabricate the 20inch bulbs for MCP-PMTs,
$^{238}$U, $^{232}$Th and $^{40}$K contents are measured to be $349.1 \pm 18.7$ ppb,
$851.2 \pm 72.8$ ppb and $26\pm3.$1 ppb, which are 29, 32 and 17.5 times higher
than the limits for JUNO PMTs. Clearly new glass materials with lower
radioactive contents must be developed.

In order to meet the required extra low radiation background requirements, we
must use ultra-pure raw materials for making the glass and also control the
possible contamination occurring in the material handling and glass melting
process. Chemical composition of the GG-17 glass and of raw materials are
listed in Table~\ref{tab:PMTglass}.

\begin{table}[!hbp]
\centering
\caption{Chemical composition of the GG-17 glass and raw materials\label{tab:PMTglass}}
\begin{tabular}{|c|c|c|c|c|c|}
\hline
\hline
	Items	 & Status	 \\
\hline
	Compostion  &	SiO2	&  B2O3	&   Na2O	&  Al2O3	&   NaCl 	 \\
\hline
	Content (\%) &  80.0	&  13.0	&   4.0  &	  2.5  &	0.5 \\
\hline
	Raw materials & Quartz sand	 &   Boric acid  &	Na2CO3  & 	AlOH	&  Salt \\
\hline
\end{tabular}
\end{table}

We have acquired a large number of raw material samples from different
manufacturers in different regions of China and measured contents of the three
radioactive elements in these samples carefully. We have identified suppliers
who can provide the necessary raw materials with $^{238}$U, $^{232}$Th and
 $^{40}$K contents that are low enough to meet JUNO PMT requirements. Obviously
 these high purity materials must be handled properly to prevent any
 contamination. We have also investigated the possibilities for contamination
 during the glass melting process. The most worrisome concern is that at high
 temperature radioactive elements can be introduced from crucibles and other
 debris into the molten glass in the oven. A Platinum crucible can mostly
 eliminate this problem, but it would be a very expensive solution. We believe
 corundum crucibles commonly used in Pyrex glass factories should be good
 enough based on the radiation background test for AlOH. So far, for the 50 or
 so  20 inch glass bulbs that have been made, regular raw materials were used
 and we know their radiation levels are very high. We have not attempted to
 make low radiation background PMT bulbs yet.

\subsection{Risk due to Nonuniformity of MCP PCE}

The photon detection efficiency (PDE) of PMTs can be expressed as $PDE=QE*PCE$,
where QE is the photocathode quantum efficiency and PCE is the photoelectron
collection efficiency of the electron multiplier. In our case the electron
multiplier is the MCP assembly. With a spherical glass bulb and the
photocathode vacuum transfer approach, we believe it is possible to fabricated
completely uniform photocathode since we can place the evaporation alkali
sources at the center of the sphere. To make the photoelectron collection
completely uniform in our current MCP-PMT design is more difficult, since
photoelectrons generated by a spherical photocathode arrive at the flat MCP
input surface from different angles and they may have different efficiencies of
collection and amplification. We are following several possible approaches to
find remedies for this potential problem.

1. Optimize the focusing design to limit the incident angular spread of photoelectrons.

\begin{figure}[htb]
\begin{center}
\includegraphics[width=0.8\textwidth]{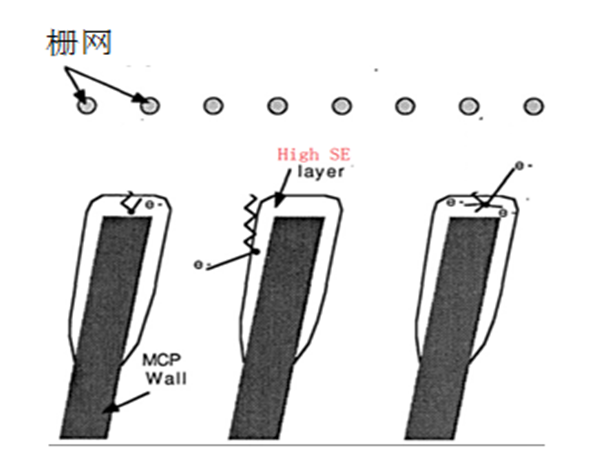}
\caption[The MCP coating scheme]{\label{fig:Fig.5.18} The MCP coating scheme}
\end{center}
\end{figure}

2. Optimize the MCP assembly design to add a metal mesh in front of MCP and
coating the MCP with MgO or Al2O3 thin film, which has a high secondary
electron emission coefficient, or use other schemes to improve the MCP
photoelectron collection. Fig.~\ref{fig:Fig.5.18} shows the metal mesh and MCP
coating schemes.

3. Further increase the gain saturation level of the second MCP such that the
output pulse heights are less dependent on the numbers of secondary electrons
produced by the photoelectron striking the first MCP.

\section{Schedule}
The global schedule from prototyping to mass production  is shown in
Table~\ref{the schedule of the mass production}.

\begin{table}[!hbp]
\centering
\caption{the schedule of the mass production\label{the schedule of the mass production}}
\begin{tabular}{|c|c|c|}
\hline
\hline
	Year	&  Tasks	&  Parameters \\
\hline
     2013 &  8inch  prototype & 	QE $\ge$ 25\%, P/V $\ge$ 1.5      \\
\hline
     2014	&  20 inch prototype & 	QE $\ge$ 30\%, P/V $\ge$ 2.0,\\
     &  &   other parameters meet specs     \\
\hline
     2015  &  Engineering design &  QE $\ge$ 30\%, P/V $\ge$ 2.0,   \\
          &   &  other parameters meet specs and Yield $\ge$ 80\%     \\
\hline
     2016	&  Preproduction   &	1,000 20inch PMTs        \\
\hline
     2017 &  Mass production	&  5,000 20inch  PMTs      \\
\hline
     2018	& Mass production	&  5,000 20inch  PMTs     \\
\hline
     2019	& Mass production	&  5,000 20inch  PMTs        \\
\hline
\end{tabular}
\end{table}

\cleardoublepage
\chapter{The Central Detector Calibration System}
\label{ch:Calibration}
\section{Requirements on the Calibration System}
\label{sec:cal_req}

The neutrino oscillation signal in the JUNO experiment will be
quantified by the measurement  of the positron energy in the inverse
beta decay (IBD) interaction.
The oscillation patterns are driven by the atmospheric mass-squared
splitting $\Delta m^2_{32}$ and the third neutrino mixing angle $\theta_{13}$,
and the solar mass-squared splitting $\Delta m^2_{21}$ and the solar mixing
angle $\theta_{12}$. The interplay of the two
contains the neutrino mass hierarchy (MH) information
through the energy dependence of the
effective atmospheric mass-squared splitting.
Therefore, one needs to determine the positron energy spectrum with high
precision in order to determine the neutrino MH and
to carry out the precision measurements of the oscillation parameters.
As stressed in Chapter~\ref{chap:intro} as well as in
independent studies~\cite{Qian:2013}, the JUNO central detector energy
resolution needs to be better than 3\% at 1~MeV (1.9\% at 2.5 MeV where
the MH signal lies), and the absolute energy
scale uncertainty to be much better than 1\% for the MH determination.
On the other hand, the requirement on the
energy resolution and the energy scale uncertainty is more relaxed for
the sub-percent precision measurements of the oscillation parameters.

There are two main contributors for the energy resolution of the liquid
scintillator (LS) detectors. The first is the number of collected scintillation photons,
which is influenced by the intrinsic light yield and attenuation length
of the LS, the photocathode coverage of the detector,
and the quantum and collection efficiencies of the photon detectors.
The second is the non-uniformity of the detector response, governed by
the detector geometry as well as the optical properties
of the relevant detector components (e.g., the attenuation of the liquid
scintillator and acrylic vessel).  Nevertheless, the non-uniformity can in principle
be reliably evaluated by deploying radioactive sources with known
energies inside the detector during the calibration data taking.

Based on experiences from the Daya Bay experiment,
liquid scintillator detector has two major sources of energy non-linearity.
The first one is the intrinsic energy non-linearity of the liquid scintillator,
which includes the quenching of the scintillation light and the amount
of the Cerenkov light within the energy ranges of interest.
The second one is the energy
non-linearity introduced by the electronics. A precise determination of both
effects to sub-percent level is the primary challenge
that the JUNO calibration system has to confront.
The most reliable handle on the non-linearity effects
is through the \emph{in-situ} calibrations, since the
dependence on particle types as well as the Cerenkov radiation is difficult
to evaluate in a bench setup elsewhere.

To accurately address both the the non-uniformity
and non-linearity in the detector energy response, the calibration system
is required to deploy multiple sources (light or radioactive)
to a wide range of positions inside the detector.
Specifically, a summary of different types of detector response calibration
envisioned
is shown in Table~\ref{table:calib}. In addition to actively
deploying sources, the reconstructed background events can also provide
critical inputs to the detector response.
\begin{table}
\caption{\label{table:calib}Summary of the different types of calibration goals, methods and requirements}
\begin{tabular}{|p{1.25in}|p{2.1in}|p{2.1in}|}
\hline
Calibration Goal & Calibration Method & Requirements\tabularnewline
\hline
\hline
PMT gain & Low intensity light sources at the center & Periodical, low intensity, automatic\tabularnewline
\hline
Light yield (energy scale) & Radioactive sources at the center & Periodical and automatic\tabularnewline
\hline
PMT timing, time walk & Light sources at the center & Periodical and variation in the light intensity\tabularnewline
\hline
Optical property of scintillator & Radioactive sources at various positions & Periodical with optimized positions\tabularnewline
\hline
Boundary effect of energy response & Pre-installed guide tube system with radioactive sources & Scan the
detector boundary and 1-2 times during whole period compare source and cosmogenic data\tabularnewline
\hline
Detector response non-uniformity & Radioactive sources at various positions and full volume cosmogenic data
 & Need based calibration frequency, targeted positions, compare source
and cosmogenic data\tabularnewline
\hline
Capture time non-uniformity & Neutron source at various positions & Need based calibration frequency, targeted positions, compare source and spallation neutrons \tabularnewline
\hline
Energy non-linearity & Various radioactive sources & Full calibration with various sources 1-2 times during the entire period,  compare source and cosmogenic
data, periodical check for 2-3 types of sources at targeted positions, \tabularnewline
\hline
Energy non-linearity of positrons & Mono-energetic positrons to the center and various position along
the central axis & Need based calibration frequency \tabularnewline
\hline
Position-dependent energy non-linearity & Various radioactive sources at fine position coverage &  1-2 times during the whole period, compare source and cosmogenic data, periodic check at targeted positions with 1-2 sources\tabularnewline
\hline
\end{tabular}
\end{table}

Based on the considerations above, we set the following
requirements to the calibration system,
\begin{enumerate}
\item In order to carry out periodical calibrations, the most frequently used
calibration system should be fully automated (similar to the ACU system in
the Daya Bay experiment). It should be simple and extremely reliable.
Such a subsystem should be able to deploy multiple
radioactive and light sources into the detector, likely along the central
axes;
\item The calibration system is required to access the off-center positions in
order to control the position non-uniformity. The source deployment to such
locations is likely to occur, on the other hand, less frequently compared to
that of the central axis;
\item Gamma source energies should cover a significant part of the IBD energy
spectrum (1-7~MeV). Positron and
neutron sources are also needed to calibrate the detection efficiencies;
\item A dedicated deployment system that has nearly $\sim$ 4$\pi$ coverage
to the central detector should be considered;
\item The absolute source positions should be controlled to better than
5~cm (preliminary Monte Carlo simulation has shown that the event
vertex reconstruction accuracy can be $\sim$ 10~cm);
\item An accelerator that can supply mono-energetic $e^{+}$ and $e^-$ beams
is important to measure the positron non-linearity directly;
\item Multiple types of particles such as $\alpha$, $e^-$, $\gamma$,
and $e^{+}$ should be deployed to understand the LS quenching and
Cerenkov light contribution to the energy non-linearity;
\item With the photo-luminescence effect in liquid scintillator,
a UV laser and optical fiber system has superior properties than those of an LED.
Such more stable and versatile laser systems should be considered as
light sources to calibrate the nonlinearity of the photo-sensors and the
electronics.
\end{enumerate}

Considering all these requirements and the mechanical constraints, we
selected the following mutually complementary calibration options
for further investigations.
\begin{enumerate}
\item A central automated calibration unit (ACU) for vertical source deployment;
\item A rope loop system for off-center source deployment;
\item A pelletron system which can provide mono-energetic positron beams;
\item Remotely operated under-liquid-scintillator vehicles (ROV) for
``4$\pi$'' coverage;
\item Pre-installed guide tubes; and
\item A diffuse system that can introduce short-lived radioactive isotopes
into the central detector.
\end{enumerate}

The ACU, the rope loop system and the ROV can all scan the interior of the
central detector but with different ranges and frequencies - we will describe
them together. The pre-installed guide tube system is uniquely
designed to understand the
boundary effects of the central detector. The pelletron calibration system
provides an excellent bench mark of the energy non-linearity for the IBD.
A diffuse source system can validate the energy non-linearity and its
position dependence. We will describe these three systems separately.

\section{Conceptual Designs}

\subsection{Housing and Interfaces}

\begin{figure}[!htbp]
\begin{centering}
\includegraphics[width=0.7\textwidth]{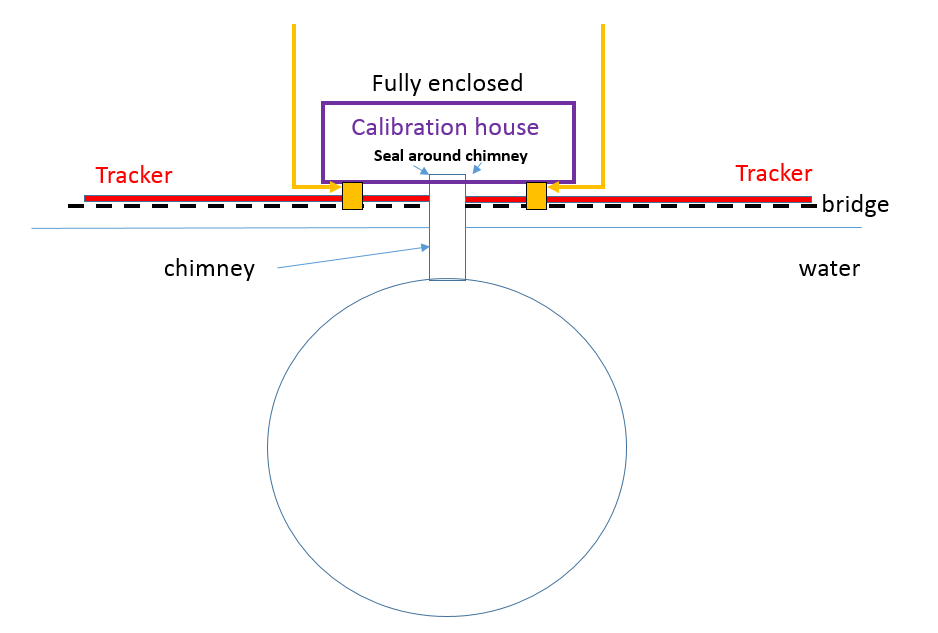}
\par\end{centering}
\caption{\label{fig:calib_house}Illustration of a seal calibration house
on top of the detector.}
\end{figure}

Calibration sources are deployed into the detector through the top
chimney. To avoid Rn contamination, the entire calibration system should be
enclosed in a clean and Rn free volume (``calibration house''),
as shown schematically in Fig.~\ref{fig:calib_house}.

\begin{figure}[!htbp]
\begin{centering}
\includegraphics[width=0.7\textwidth]{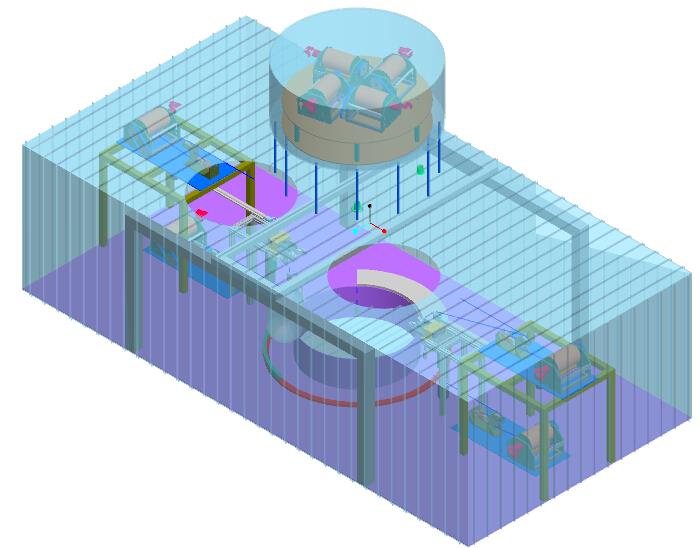}
\par\end{centering}
\caption{\label{fig:cal_overview}Concept of a calibration house.}
\end{figure}

A concept of the calibration house is shown in Fig.~\ref{fig:cal_overview}. The ACU
system is located at the top. Two rope loop systems are placed
inside the house. The source changing area can be accessed via two glove
boxes to allow manual operation. The ROV system is also kept inside the
house and moved along the rails attached to the roof.

\subsection{Sources}
\subsubsection{Routine Sources}
To set the energy scales and to calibrate the positron and neutron detection
efficiencies, radioactive sources for JUNO include
routine gamma and positron sources together with neutron
sources with correlated high energy gamma ray emissions, shown in
Table~\ref{tab:sources}.

\begin{table}
\caption{\label{tab:sources}Radioactive sources under consideration in JUNO.}
\begin{tabular}{ccc}
\hline
Source & Type & Radiation \\\hline
$^{40}$K & $\gamma$ & 1.461 MeV\\
$^{54}$Mn & $\gamma$ & 0.835 MeV \\
$^{60}$Co & $\gamma$ & 1.173 + 1.333 MeV \\
$^{137}$Cs & $\gamma$ & 0.662 MeV\\
$^{22}$Na & e$^{+}$ & annil + 1.275 MeV\\
$^{68}$Ge & e$^{+}$ & annil 0.511 + 0.511 MeV\\
$^{241}$Am-Be & n, $\gamma$  & neutron + 4.43 MeV \\
$^{241}$Am-$^{13}$C or $^{241}$Pu-$^{13}$C & n, $\gamma$ & neutron + 6.13 MeV \\
$^{252}$Cf & multiple n, multiple $\gamma$ & prompt $\gamma$'s, delayed n's\\
\hline
\end{tabular}
\end{table}

\begin{figure}[!htbp]
\begin{centering}
\includegraphics[width=0.5\textwidth]{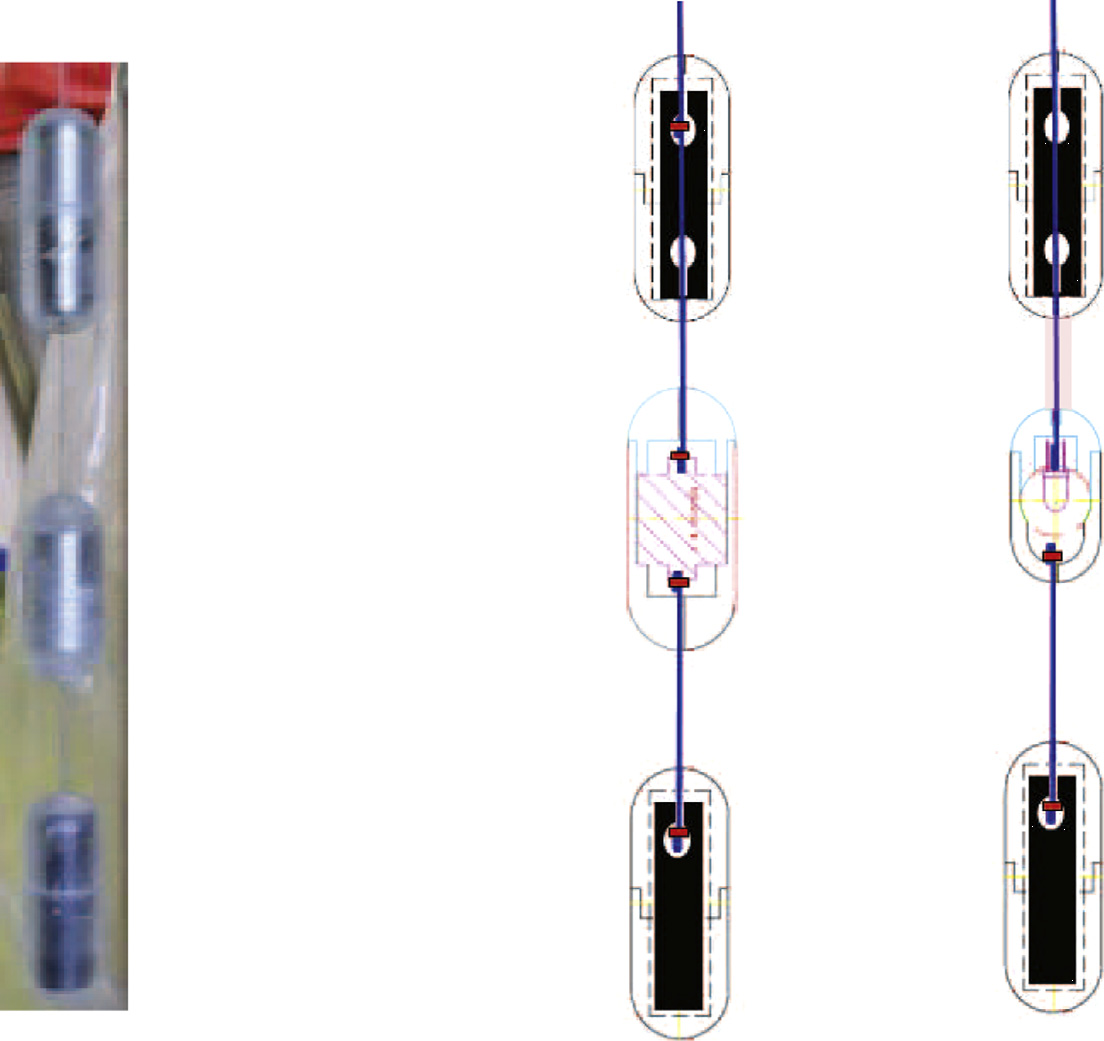}
\par\end{centering}
\caption{\label{fig:source}Illustration of a typical source/weight assembly
used in Daya Bay, envisioned for JUNO as well.}
\end{figure}

To minimize the risk due to contamination of the radioactivities,
the enclosure of these sources should be thin-walled ($\sim$ mm)
stainless steel (SS) capsule, enclosed by a round-headed acrylic
shell to ensure chemical compatibility with the liquid scintillator.
The deployment rope, made with SS, shall be robustly
attached to the SS source enclosures. Two steel weights are attached
above and below the source to maintain a minimum tension in the rope. A
typical source/weight assembly used in Daya Bay is shown in Fig.~\ref{fig:source}.

\begin{figure}[!htbp]
\begin{centering}
\includegraphics[width=0.5\textwidth]{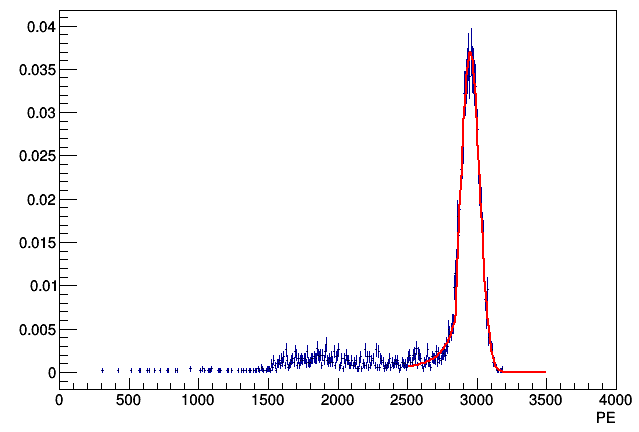}
\par\end{centering}
\caption{\label{fig:Co60}an example JUNO Monte Carlo
spectrum in photoelectrons for the $^{60}$Co source with source geometry.}
\end{figure}

Simulations are underway to optimize the geometry in order to
minimize impacts of the source assembly to the energy scale. In
Fig.~\ref{fig:Co60} a typical $^{60}$Co spectrum is shown. The ``dead volume''
of the source assembly affects the low energy shoulder, but has a much
smaller effect on the full absorption peak. The optical
shadowing of the source assembly, on the other hand, will bias the location
of the full absorption peak.

\subsubsection{Mini-balloon}
\begin{figure}[!htbp]
\begin{centering}
\includegraphics[width=0.5\textwidth]{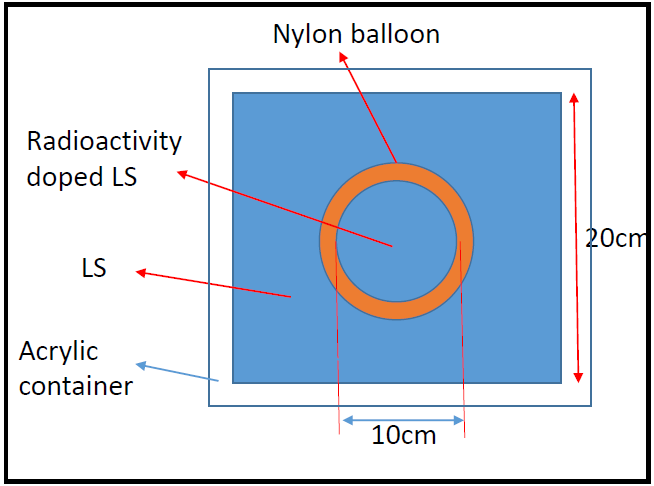}
\par\end{centering}
\caption{\label{fig:balloon}The mini-balloon source concept.}
\end{figure}

\begin{figure}[!htbp]
\begin{centering}
\includegraphics[width=0.5\textwidth]{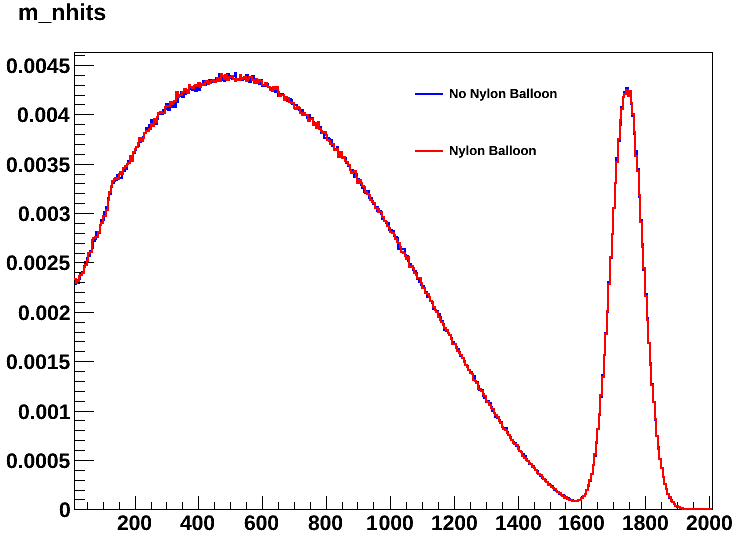}
\par\end{centering}
\caption{\label{fig:K40}Comparison of $^{40}$K spectra with a 10 $\mu$m balloon
and without.}
\end{figure}

In order to study the liquid scintillator response to charged particles
(electrons, positrons, $\alpha$'s) thereby to control the quenching effects,
we are pursuing a ``mini-balloon'' concept (Fig.~\ref{fig:balloon}).
Radioactive isotope is loaded
into the liquid scintillator inside a small, thin-walled,
and transparent balloon ($\sim$ 10 cm OD). The balloon is further
enclosed by an acrylic cylinder ($\sim$ 20 cm OD) filled with undoped
liquid scintillator.
Such a design will minimize the energy loss of the charged particles
across the balloon wall to simulate real events, while maintaining
a double encapsulation of the radioactive isotopes. A comparison of the
$^{40}$K-loaded 10~$\mu$m thick balloon and that without the balloon from
the Monte Carlo is shown
in Fig.~\ref{fig:K40}, in which the bias in beta energy is estimated to be less than
0.3\%.
One important
advantage is that the choice of the dopant in the balloon is very versatile,
for example $^{222}$Rn ($\alpha$'s), $^{40}$K ($\gamma$'s and $\beta$'s), $^{137}$Cs
($\gamma$'s, $\beta$'s and conversion electrons), $^{68}$Ge (positrons), and
even short-lived isotopes. The source deployment system can then take
the source into given locations inside the detector for calibration.

\subsubsection{Pulsed Light Source}
Unlike the radioactive sources, the pulsed light source can generate
photons at a given time with tunable intensity. These features are important
to calibrate the gains of the photomultipliers as well as
the response of the electronics system. Based on the experience in the Daya Bay
calibration system, the stability of the pulses from light emitting diodes or
LEDs ($\sim$5 ns timing) is not as good as pulsed laser
system (which could achieve a ns level timing precision). Furthermore,
a UV laser of $\sim$260~nm
wavelength can excite the LAB molecules thereby producing photons with
similar timing characters as those generated by real particle interactions.
In addition, commercial light sensors, e.g., Si diode can monitor
the intensity of the laser pulses to very high precision.
Based on these considerations, we are developing
a UV laser system coupled to a flexible optical fiber with
a diffuser sphere at the end to achieve a stable, uniform, tunable,
and deployable light source for JUNO.

\subsection{Source Deployment Systems for the Detector Interior}
As mentioned in Sec.~\ref{sec:cal_req}, three complementary subsystems are
under consideration, the ACU system, the rope loop system, as well as
the ROV system. They cover different ranges inside the central detector,
and are envisioned to be used at different frequencies. In addition to
the mechanical means to position a source, independent source locating
systems will be critical to ensure the required high position accuracy.

\subsubsection{Source Locating Systems}
Although the rope loop can position itself via the lengths of different
rope sections, mechanical uncertainty such as deformation of the ropes
would introduce uncertainty to the absolute source position. Similar
uncertainty arises in the ROV system as well.
A dedicated source locating system is required to
determine the source position to better than 5~cm.

\begin{figure}
\begin{centering}
\includegraphics[height=2.5in]{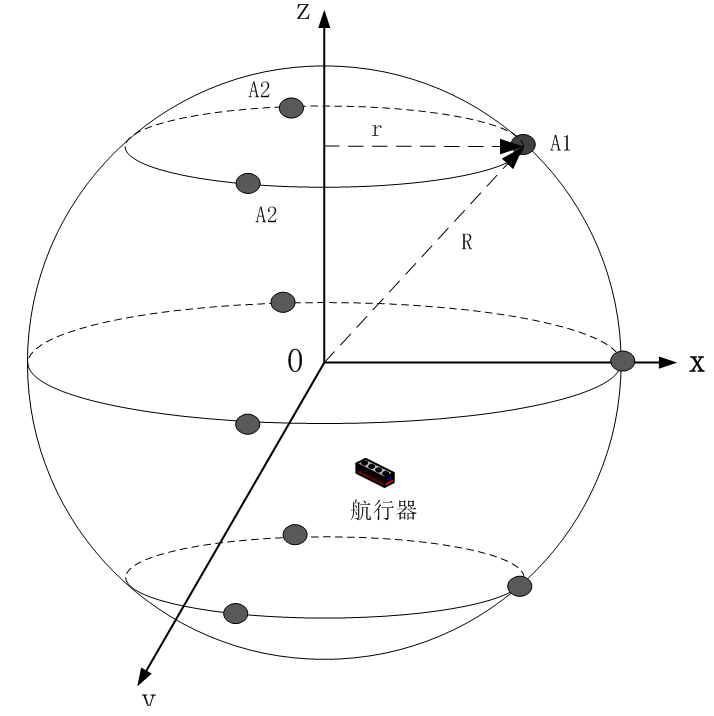}
\par\end{centering}
\begin{centering}
\caption{\label{fig:ultra}A typical placement of the ultrasonic receivers.}
\par\end{centering}
\end{figure}

Two options are
being considered. The first option is an ultrasonic system
with an array of ultrasonic receivers and an ultrasonic emitter attached to the
source. The emitter generates pulsed sonar waves and the receivers
reconstruct the origin via the timing or phase difference. The wavelength of
150~kHz sonar wave is $\sim$ 1~cm, introducing negligible uncertainty.
The dominating systematics is the positioning accuracy of the receivers
as well as the speed of sound in LAB under different temperature/pressure
conditions. Reflections at the interface are also systematics under study.
To allow package of the sound waves, the receivers have to be installed in the
inner wall of the vessel if the central detector housing is acrylic,
and not so critical if the balloon option is adopted.
In principle, the larger the receiver array, the higher the positioning
accuracy. Studies are undergoing to optimize the array. A typical
arrangement is shown in Fig.~\ref{fig:ultra}, with 9 redundant
receivers mounted on the wall.

An alternative technology is to locate the source via optical cameras (CCD)
and one reconstruct source position via image analysis.
Such a system is used by the Borexino experiment. The infrared
lighting of the CCDs poses minimal risk to the PMTs. Alternatively,
a battery driven LED can be attached to the source assembly to ease the
analysis.

\subsubsection{ACU}
\begin{figure}[!htbp]
\begin{centering}
\includegraphics[width=0.6\textwidth]{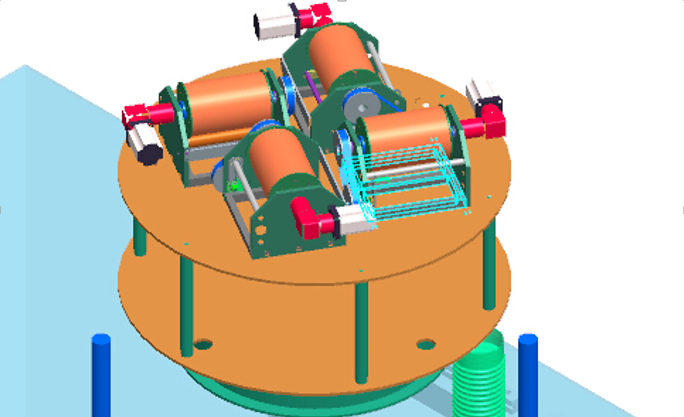}
\par\end{centering}
\caption{\label{fig:ACU}A sketch of the ACU system.}
\end{figure}

The Automated Calibration Unit (ACU) system is a unit very similar to
that used in Daya Bay, capable of deploying a few different
sources along the central axis of the detector. As illustrated in
Fig.~\ref{fig:ACU}, four source deployment
units will be arranged on a turntable, which can select the source
to be deployed. A hole on the bottom plate of the ACU will be aligned to
the center of the chimney to the central detector.
Three of the sources (one light source
and two radioactive sources) will be attached permanently to three deployment
units. The fourth unit will have a changeable source fixture, providing
flexibility for the attachment of special sources.
The operation of the system will be fully automated to ensure routine (weekly)
deployment of calibration sources with high reliability. Based on the
experience from the Daya Bay experiment, a weekly deployment of the light
source allows PMT gain calibration to subpercent precision. A weekly
deployment of a gamma source, e.g., $^{60}$Co, is sufficient to track the
slow drift ($\sim$1\% per year) of the overall energy scale due to changes
in detector properties. Lastly, a routine deployment of the neutron source
can ensure the stability of the detection efficiency.

\subsubsection{Rope Loop System}
\begin{figure}[!htbp]
\begin{centering}
\includegraphics[height=2in]{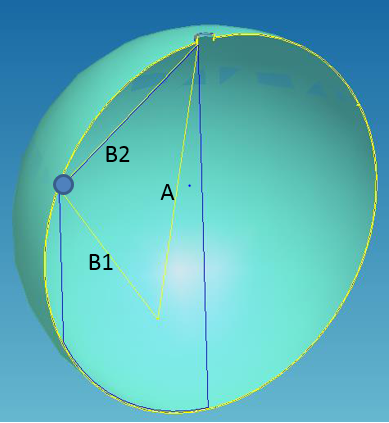}
\includegraphics[height=2in]{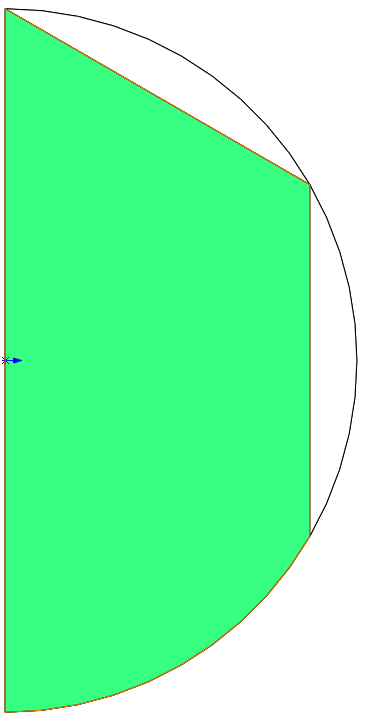}
\par\end{centering}
\caption{\label{fig:Rope-loop-solution}Rope loop solution and its 2-dimensional coverage}
\end{figure}

The concept of a rope loop system is similar to that used in the SNO
experiment, illustrated in Fig.~\ref{fig:Rope-loop-solution}.
The source deployment position can be controlled
by adjusting the lengths of sections A and B of the rope.
In this design, one of the hanging points is through the central
chimney, and the other anchor is located at around 30$\circ$
latitude, allowing a theoretically 90\% 2-dimensional coverage
of a half vertical plane.

\begin{figure}[!htbp]
\begin{centering}
\includegraphics[height=2in]{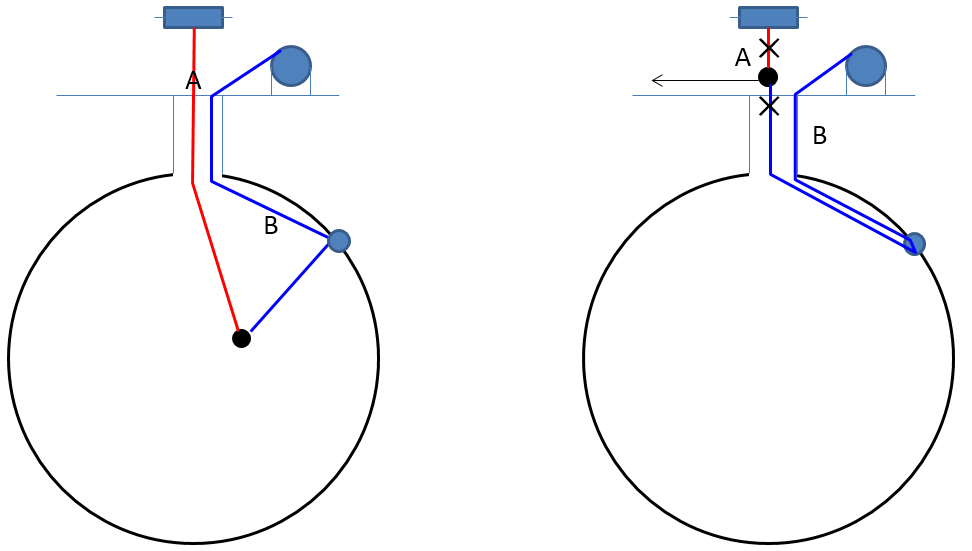}
\par\end{centering}
\caption{\label{fig:dep_seq}3-point hang design and volume coverage}
\end{figure}

A sketch to illustrate a deployment sequence is shown in Fig.~\ref{fig:dep_seq}.
With section A going through the central chimney
sources can be taken out of the detector and get changed. We are developing
a scheme which allows both an automated source change while maintaining
the option of manually changing the source via glove box.

\begin{figure}[!htbp]
\begin{centering}
\includegraphics[width=0.8\textwidth]{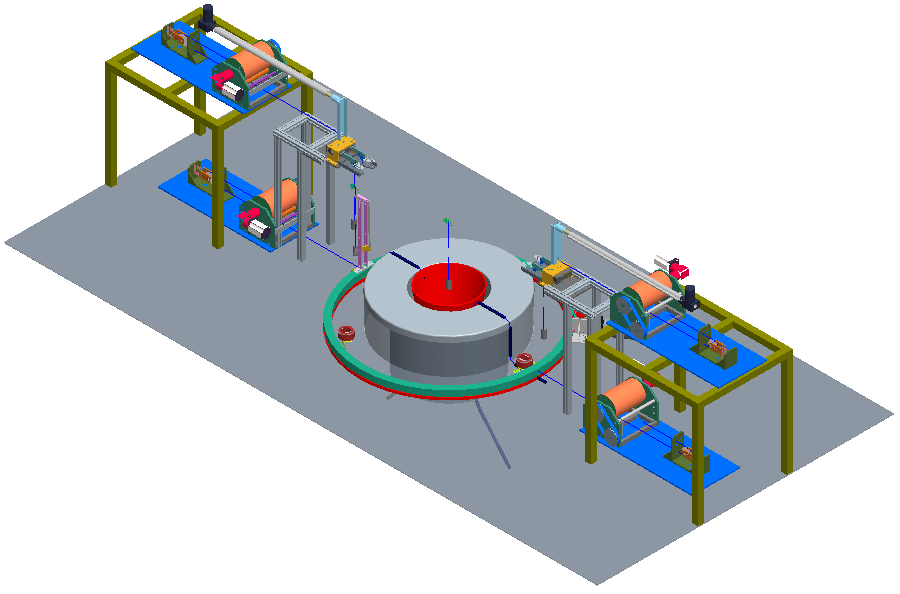}
\par\end{centering}
\caption{\label{fig:rope_layout}An overview sketch of the rope loop
system.}
\end{figure}

The overall layout of the rope loop system on top of the central detector
is shown in Fig.~\ref{fig:rope_layout}. Two independent rope loop systems
are being considered, each covering a half vertical plane to allow
some control of the azimuthal symmetry of the detector response. Each system
has two spool drives A and B to adjust the lengths of the A/B rope in a
synchronous fashion. Rope A goes through a pulley, which is
attached at an end of an extendable lever arm. During the deployment,
the level arm is extended to move the pulley towards the center of the
chimney to lower the source into the detector. Once a deployment is completed,
the source is retrieved by extending B rope and shortening A rope. Once the
source is out, the pulley arm gets retracted to the side so that the source
change operation will be performed away from the chimney.

\begin{figure}[!htbp]
\begin{centering}
\includegraphics[height=3in]{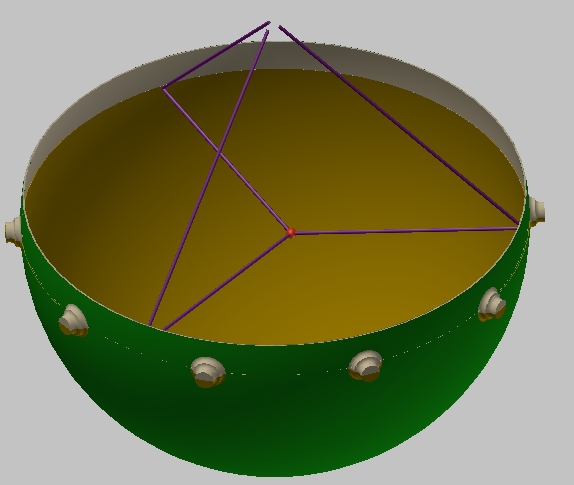}
\par\end{centering}
\caption{\label{fig:alter_rope_system}Illustration of a 4-rope design.}
\end{figure}

Alternative designs allowing three-dimensional coverage are also under
considerations.
For example, if the anchor point is allowed to move on the latitude
circle, a 3-dimensional deployment can be made possible. However, with a
large diameter circle attached to the inner
surface of the detector, maintaining a smooth rail and reliable drive
of the anchor point is a non-trivial engineering challenge. Another
alternative 4-point rope design is shown in
Fig.~\ref{fig:alter_rope_system}. Three hang points, separated by 120$\circ$,
  are placed
around the equator of the central detector. Three ropes, each going
through one hanging point and
the central calibration port, will connect to the source. By adjusting the
length of these three ropes,
one can achieve a volume calibration in the bottom half of the sphere.
With a fourth rope
connecting the source directly from the central calibration port,
the volume calibration on the upper sphere can be performed. In general,
three dimensional rope system would introduce significantly
more mechanical complexity. In such a case, a ROV system described below would become
a viable alternative.

\begin{figure}[!htbp]
\begin{centering}
\includegraphics[width=0.6\textwidth]{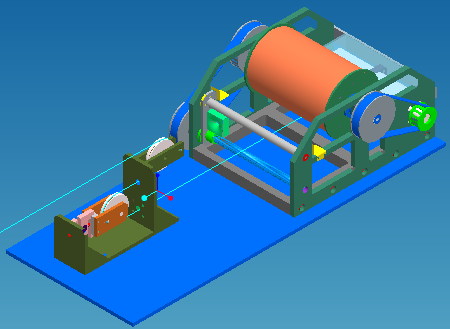}
\par\end{centering}
\caption{\label{fig:spool}An illustration of a spool drive.}
\end{figure}

Details of a single spool drive in the rope loop system
are shown in Fig.~\ref{fig:spool}. The
spool in the ACU will be under an identical design.  The spool will be
made either with acrylic or PTFE. Helical grooves with $\sim$mm
separation will be machined on the spool,
and the deployment cable of less than 1.5~mm diameter will be wound into
the grooves without overlap. The rope capacity of each spool is set to be
50~m. Several additional measures will be implemented to avoid rope slipping
out of the groove, including a spring loaded Teflon press to constrain
the rope, a load cell to constantly monitor the tension in the cable, as
well as a co-moving spooling tracker so that the rope always
gets unwound perpendicular to the spool.

\begin{figure}[!htbp]
\begin{centering}
\includegraphics[width=0.6\textwidth]{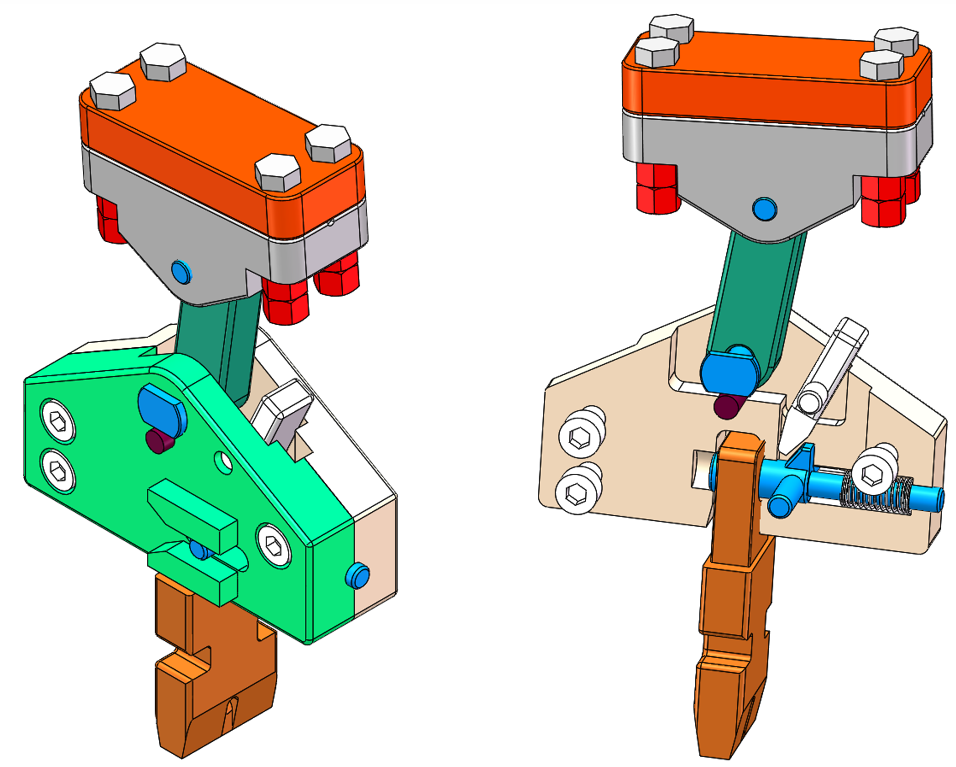}
\par\end{centering}
\caption{\label{fig:quick1}One of the quick connection option.}
\end{figure}

To achieve automated source changing, a key component in the rope
system is a ``quick connection'' connector, which allows an automated
system to attach and detach a source to the rope. One of the options
under considerations is illustrated in Fig.~\ref{fig:quick1}.
The upper piece is attached to the rope loop permanently.
We envision that an ultrasonic emitter can be attached to it for accurate
positioning with the SS rope being the core of a coaxial cable
carrying the electrical signals. The middle piece is the female connector,
which is connected permanently to the upper piece with a rotary joint.
Gravity will keep the middle piece horizontal to ease automated grabbing.
The lower piece is the male connector, permanently attached to the source
and the lower weight.
The lock between the male and female is achieved with a spring-loaded key.
During the attachment (detachment), both the male and female pieces have
to be held by mechanical hands and pushed (pulled) relatively.
One additional degree of freedom is needed to
actuate the key in order to lock or unlock the connector. A manual
key is located on the female connector to easy the manual operation through
a glove box.

\begin{figure}[!htbp]
\begin{centering}
\includegraphics[width=0.4\textwidth]{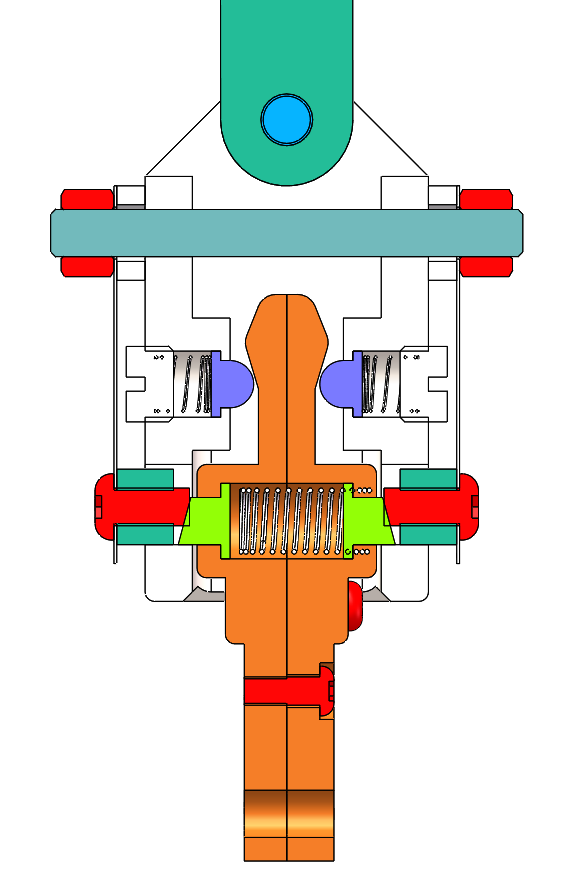}
\par\end{centering}
\caption{\label{fig:quick2}An alternative quick connection option.}
\end{figure}

An alternative option of the connection is illustrated in Fig.~\ref{fig:quick2},
where the upper piece is the same and omitted in the drawing. The connection
is made doubly safe with two spring loaded locks. The head of the male key
has a dumb bell shape, which can be pushed in and pulled out of the upper
lock with force. The lower lock requires a hand to press the side buttons
in order to get unlocked.

\begin{figure}[!htbp]
\begin{centering}
\includegraphics[width=0.7\textwidth]{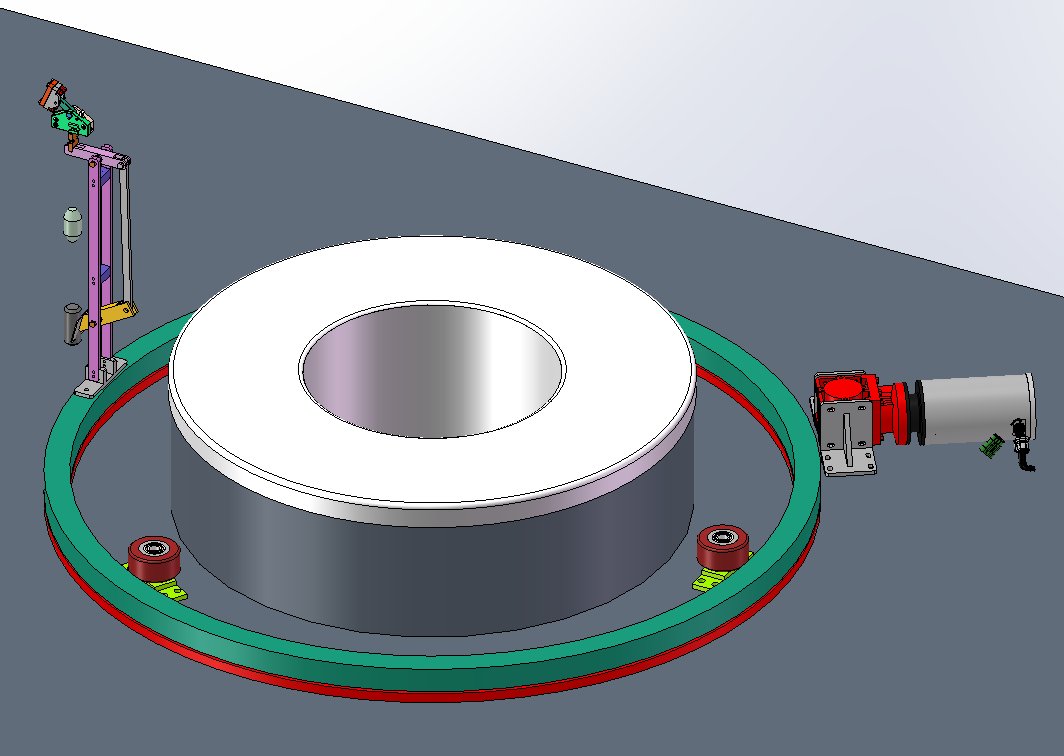}
\par\end{centering}
\caption{\label{fig:source_ring}The source storage ring.}
\end{figure}

All sources will be placed in a circular storage ring surrounding the
chimney (Fig.~\ref{fig:source_ring}), which can be driven automatically
in order to select a source to get attached/detached to the rope loop.
Such a design allows the two rope loops to share the same source storage.
For illustration, a
source assembly placed in the storage is also shown in the figure. For
each rope loop, all sources share the same mechanical hand system to perform
the source changes.

\subsubsection{Remotely Operated Under-liquid-scintillator Vehicles}
\begin{figure}
\begin{centering}
\includegraphics[height=3in]{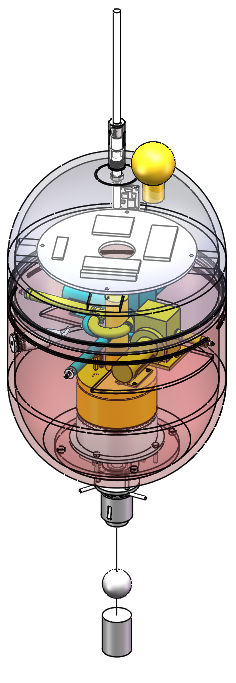}
\par\end{centering}
\caption{\label{fig:ROV}ROV, its umbilical cable and the loaded radioactive
  source}
\end{figure}

A ROV system can deploy the source to nearly everywhere inside the central
detector. Such a system is used in the SNO experiment to deploy the
$^3$He neutron tubes. A conceptual design of the ROV is shown in
Fig.~\ref{fig:ROV}. It can be made into an egg-shaped capsule with
a diameter less than 300~mm and a height less than 500~mm.
The ROV is driven by pump jet propulsion without external propeller
in order to maintain a simple external geometry and a compact size.
The ROV motion speed is set to be about 1 m per min, and the actual
position is feedback via an ultrasonic emitter. The power as well as
signals are transmitted to the ROV via a umbilical cord adjusted to
nearly zero buoyancy. The ROV can also
carry lighting, CCD camera and a magnet to allow emergency rescue or
monitoring the interior of the detector.
Radioactive source assembly is attached below the ROV, with a quick
connection which would allow automated source change.
For material compatibility with LAB, the enclosure of the ROV will be made
with Teflon, which is also highly reflective to
avoid photon losses. Monte Carlo studies are underway to optimize
the ROV geometry in order to minimize impact due to dead materials or
optical shadowing.

Based on its current location and destination location, the ROV software
will automatically select the optimal trajectory. The pump jet propulsion
engine controls the speed and direction of the device. The ROV should
approach the designated locations with a speed less than 1~mm/s or less,
then shuts off the engine. A piston driven buoyancy adjustment
mechanism is used as the depth control. In case of system failure such as
power failure, the control will shut down the pump propeller and pull the
piston of the depth control via a spring to increase the buoyant force to
ensure the ROV float to the surface.

\subsection{Conceptual Design of the Guide Tube System}

\begin{figure}[!htbp]
\begin{centering}
\includegraphics[height=3.0in]{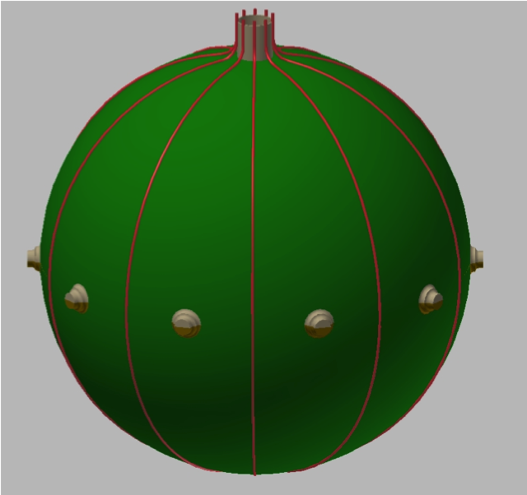}\includegraphics[height=3.0in]{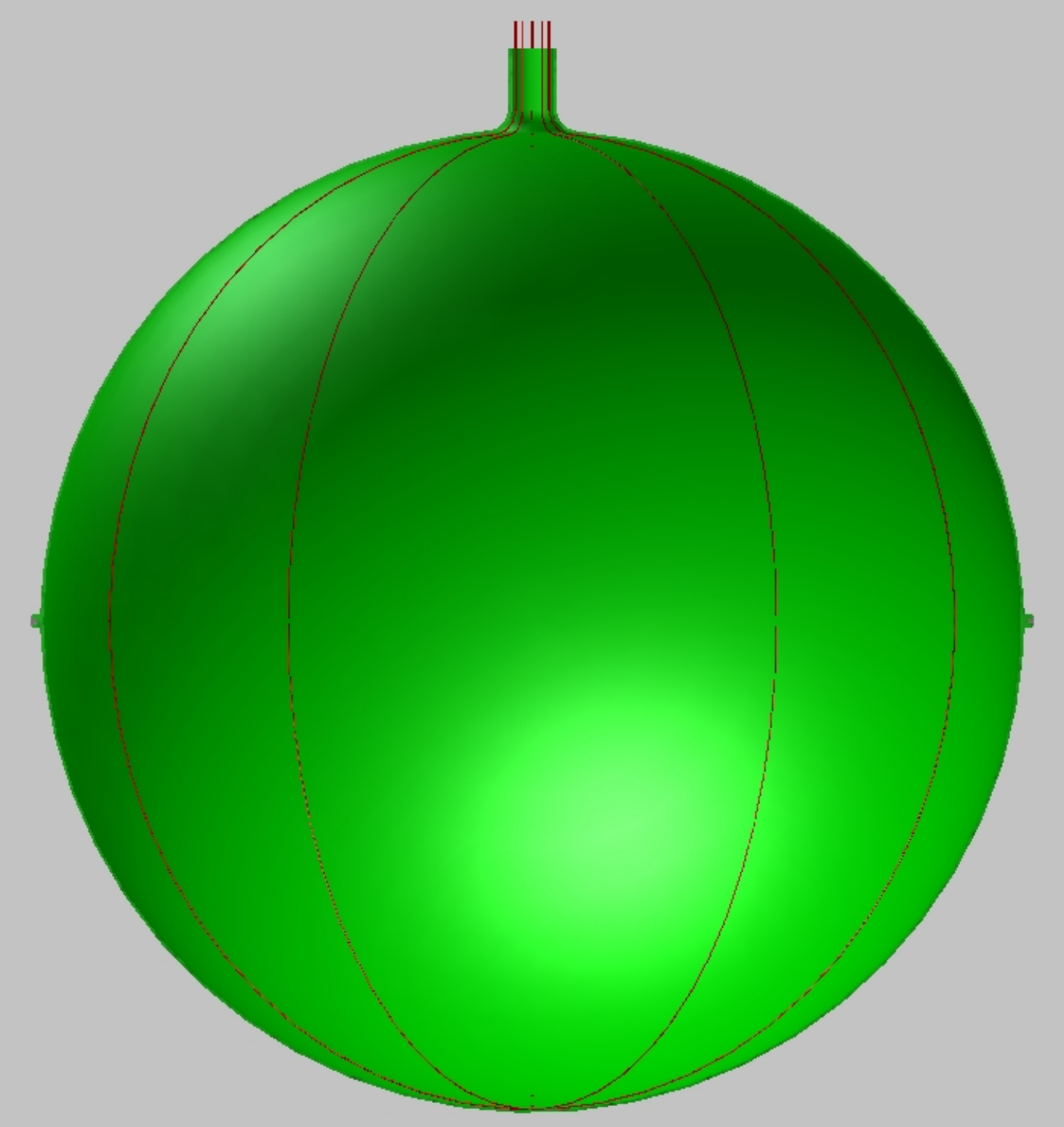}
\par\end{centering}
\centering{}\caption{\label{fig:guide_tube}The guide tube system. }
\end{figure}

\begin{figure}[~!htbp]
\begin{centering}
\includegraphics[height=3.0in]{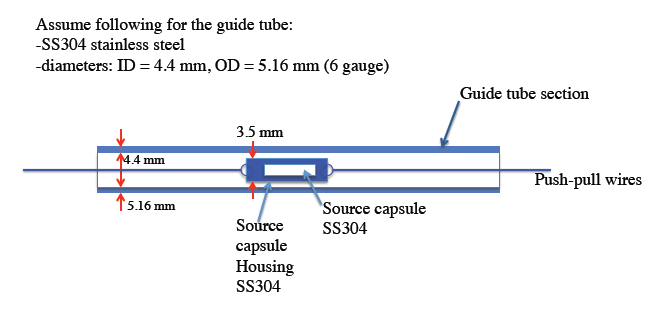}
\par\end{centering}
\centering{}\caption{\label{fig:guide_tube1} The conceptual design
of the source capsule inside the guide tube.}
\end{figure}

Based on positive experiences of the Double Chooz experiment,
a guide tube system
is important to understand detector response from anti-neutrino
interactions occurring at the
boundaries of the central detector volume. Guide tubes will be
positioned along the surface of
the central detector vessel. The tubes could be installed at the
inner and/or outer surfaces of the vessel wall.
Figure~\ref{fig:guide_tube} shows the conceptual design of the guide tube system
with the tubes at the outer surface
(left), and with tubes at the inner surface (right) of the
central detector vessel. The number
of tubes shown is for illustrative purposes only, and the final
number and distribution
of the tubes will be based on detailed simulations of the guide
tube system and studies of detector energy response, currently
under development.
The tube can be made with stainless steel to provide a good
mechanical strength, or it could
be manufactured from an acrylic tube to be compatible with the
central detector vessel materials.
Radioactive sources as small as a few mm in diameter and $\sim$2~cm
in length could be positioned within the tube with use of a stainless
steel wire driven by a stepper motor.
Figure~\ref{fig:guide_tube1} shows one considered design of source capsule inside the
stales steel guide tube, with the source connected on both sides.
Similar designs are considered with the acrylic guide tube solution.
Source positioning accuracy of $\sim$1~cm is anticipated.

\subsection{A Pelletron-based Beam Calibration System}

\begin{figure}
\begin{centering}
\includegraphics[width=0.8\textwidth]{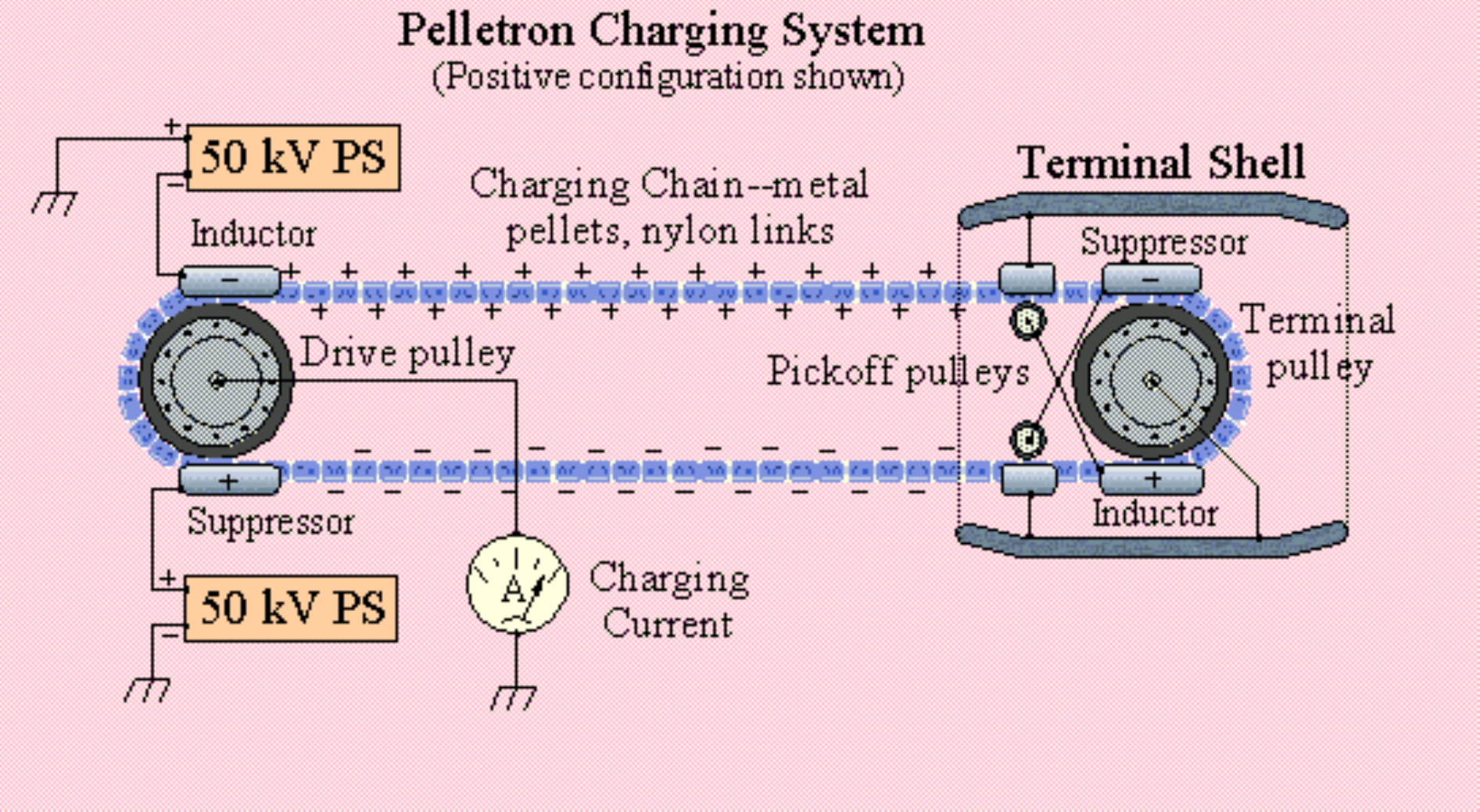}
\par\end{centering}
\caption{\label{fig:Diagram-of-Pelletron}Diagram illustrating the
principle of a pelletron. As the drive pulley and the terminal pulley
rotate clockwise, the chain transports the positive charge to the terminal
shell and builds up a high voltage that could be higher than 25 MV.
By reversing the polarities of the charging voltages, the pelletron can
easily switch to accelerate electrons and negative ions.}
\end{figure}

A pelletron is one type of electrostatic accelerator. The electric charge
is transported mechanically to its high voltage terminals by a chain
of pellets, which are connected with insulating materials (such as nylon).
Figure~\ref{fig:Diagram-of-Pelletron} shows a diagram of a pelletron,
taken from the website of National Electrostatics Corporation
(NEC)~\cite{NEC:2014}. Compared with a LINAC, the electrostatic
accelerator has the advantage of being more stable. This is crucial
for the intended application in JUNO, as the calibration of the Pelletron
energy and the JUNO detector response calibration cannot be
performed at the same time. Furthermore, below
\textasciitilde{}5 MeV in kinetic energy~\cite{Hinterberger:1997ur},
the pelletron solution is
compact and economical enough for an underground installation.


\begin{figure}
\begin{centering}
\includegraphics[width=0.85\textwidth]{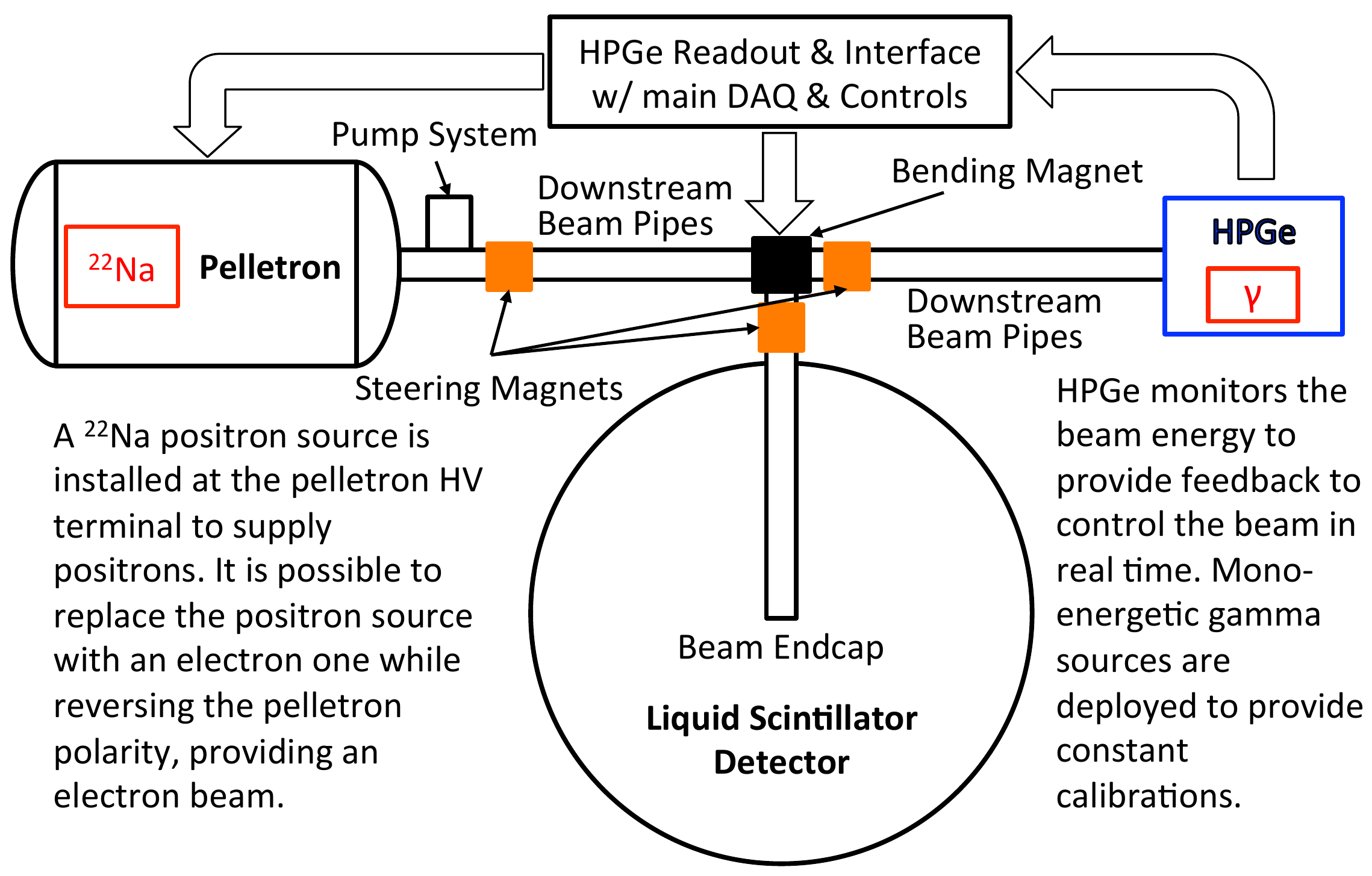}
\par\end{centering}
\caption{\label{fig:sys-diagram} The conceptual design of the pelletron
system}
\end{figure}

Expected positron spectra is shown in Fig.~\ref{fig:Espec}.
We see that most events are between 1 and 6~MeV
prompt energy. Due to finite energy resolution, the mass hierarchy signal
only shows up above 2~MeV prompt energy. Therefore, a pelletron system,
which can provide 1-5~MeV kinetic energy corresponding to 2-6~MeV
prompt energy, can satisfy the requirement of precision energy calibration
for mass hierarchy determination.  Figure~\ref{fig:sys-diagram} shows
the conceptual design of the pelletron system, following the ideas
from Refs.~\cite{Bauer:1990zz,Huomo:1988jw}. The high purity Ge (HPGe)
detector, which is calibrated by mono-energetic gamma sources,
will be used to control the beam energy precisely. A precision
of  $10^{-4}$, which is well below the 0.1\% goal of the JUNO experiment,
has been achieved with existing facilities~\cite{Bauer:1990zz,Berg:1992jr,Huomo:1988jw}.

\begin{figure}
\begin{centering}
\includegraphics[height=4.0in]{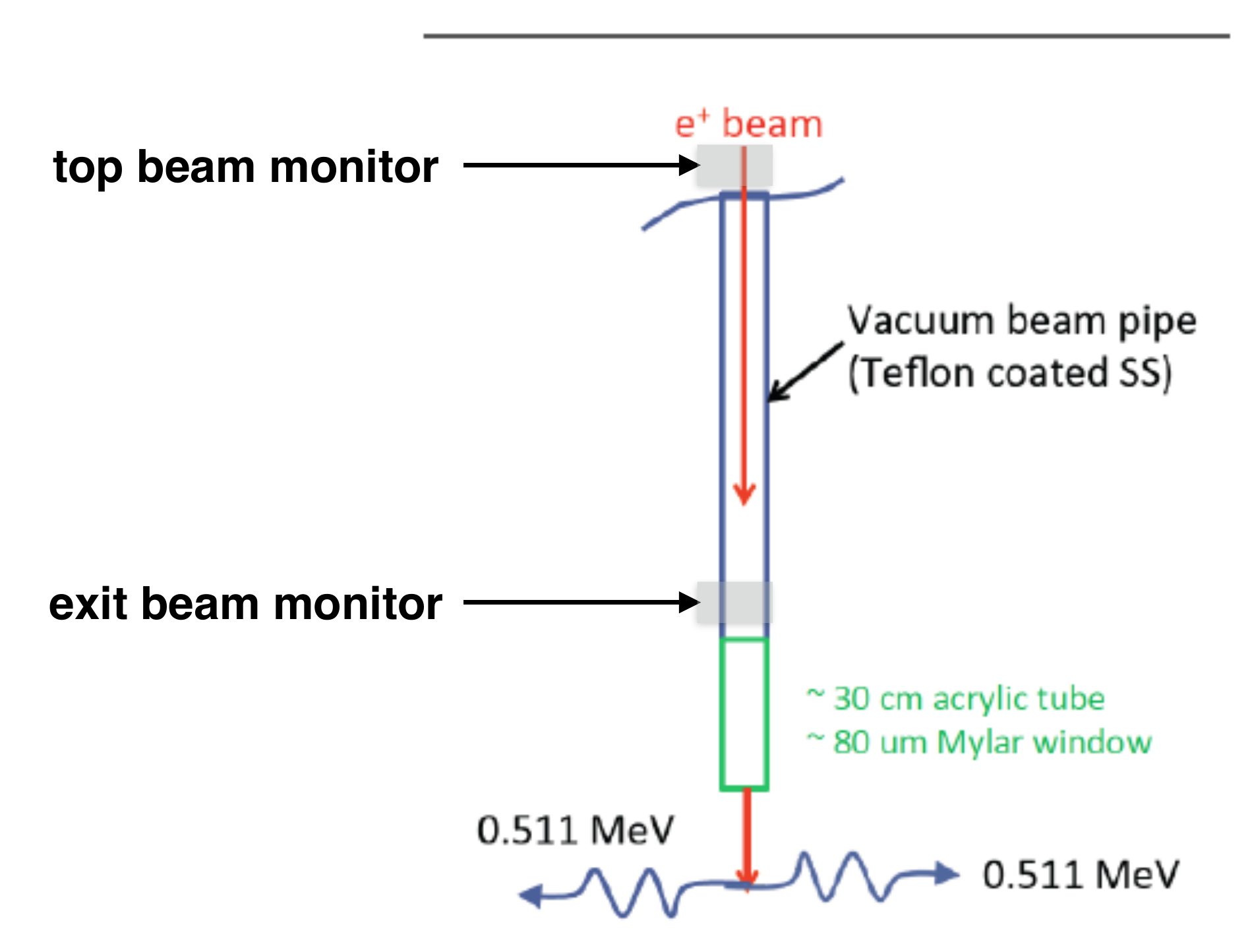}
\par\end{centering}
\caption{\label{fig:endcap-concept}A conceptual design of the beam pipe endcap
inside the LS detector. The transparent acrylic pipe and the Mylar
window make sure the scintillation light can largely pass through.
Two beam monitors near the top and bottom of the beam will allow diagnostics of
beam position and area when the beam first enters the detector vertically downward
and near the exit point of the beam out of the beam.}
\end{figure}


The calibration of the JUNO detector with a pelletron requires the delivery of the
positron beam into the detector center through an evacuated beam pipe. The positron
beam will exit the beam pipe through a window designed to minimize energy losses in
the beam while maintaining the vacuum of the beam pipe at the depth of the detector.
The design and instrumentation of the pelletron beam pipe inside the detector is an
engineering and R\&D challenge.

\underline{Detector beam pipe requirements:} The technical requirements for the
positron beam pipe inside the detector can be summarized as follows:

\begin{itemize}

\item Retractable, telescoping beam pipe that can be deployed into the detector during
the time of calibration and retracted during normal data taking to avoid shadowing
effects in the detector. During normal data taking the beam pipe is stored completely
out of the detector volume.

\item Beam pipe is compatible with the scintillator. To first order, the exposure of
the scintillator to the beam pipe materials is limited to the duration and time
period of the calibration. However, one has to account for possible long-term
interaction of the scintillator with the beam pipe material after the beam pipe
has been retracted from the detector. Excess scintillator may drip back into the
detector or lead to unwanted contamination that may be introduced during the time
of the next calibration. Compatibility issues include possible interaction with the
materials of the beam pipe, damage of the deployment mechanism, leaching out of beam
pipe materials into the scintillator, and degradation of the scintillator itself. The
amount of beam line material to be deployed into the detector poses a challenge for
the purity of the scintillator and to JUNO physics goals that require ultra low backgrounds.

\item Beam pipe and its deployment mechanism have to be leak tight against the
liquid scintillator to maintain the vacuum inside the beam pipe.

\item Beam pipe deployment mechanism is failsafe and can always be retracted out of
the detector. Since the beam pipe does interfere with the normal data taking the
deployment mechanism must allow the retraction of the beam pipe even under unusual
circumstances such as power failure etc. The system has to be tested to be mechanically
reliable for the 10+ year lifetime of the experiment.

\item It is desirable to be able to deploy the beam pipe to different depths. This would
allow calibration of the detector with positrons at different points along the z-axis.
The interaction region of the beam with the scintillator detector is defined by the exit
point of the beam out of the mylar window. A continuous deployment along the z-axis may
not be necessary, discrete steps in z are sufficient.

\item Deployment of the beam pipe is automated as much as possible to allow for
regular calibration and minimize the demands on the on-site personnel.

\end{itemize}

Inside the beam pipe instrumentation to monitor the beam position and profile are
required to help determine the exit point of the beam and the interaction region of
 the calibration beam inside the detector. This monitoring instrumentation will also
 aid in diagnostics of the beam during tuning and setup. One of the challenges of the
proposed calibration scheme is that a positron beam has to be delivered some 17~m into
the detector center without active steering. We expect that from the entrance point of
the beam into the detector region (at the top of the detector) to the point where it
exits the beam pipe through the mylar window no active steering components can be used.
As a result the beam will spread and its final position at the end of the beam pipe will
depend on its direction when it enters the detector and possible divergence. As a result,
we obtain the requirements for the monitoring instrumentation inside the
detector beam pipe (see Fig.~\ref{fig:endcap-concept}):

\begin{itemize}

\item \underline{Top beam monitor:} Very precise beam position monitor at the top of the
detector where the beam enters the detector to determine its direction and profile into the
detector right after the last active magnetic steering and focusing. This beam position monitor
can be fixed along the z-axis. It can be flipped or moved into the beam during tuning and
setup and removed from the beam path during the beam delivery and calibration. The position
resolution requirements for this beam monitor are set by the distance of the beam delivery
into the detector and the width of the beam pipe. Depending on the position of the 90-degree
bending magnet of the beam into the detector this monitoring setup can still be outside
the detector but already in the vertical region of the beam pipe.

\item \underline {Exit beam monitor:} A second beam position monitor near the end of the
beam pipe that determines the position and area of the beam spot before it exits through
the mylar window. Again, this beam monitor may be designed to be movable so that it can
be placed in the beam during monitoring and setup and removed (rotated or flipped) during
calibration. It may be desirable to make this beam monitor movable along the z-axis to
allow for a vertical scan of the beam position. Since this beam monitor will be located
towards the end of the telescoping beam pipe all cabling and supplies have to be routed
through the retractable beam pipe structure. This be may not required for the beam monitor
at the top of the detector. With a beam position monitor near the top and the bottom of
the beam pipe it will be possible to diagnose potential issues in the delivery of the
position beam. More than two position monitors will likely lead to unnecessary and
increased complexity.

\end{itemize}

In the following, we summarize various factors that can induce a bias in the beam energy
calibration. The first one is the energy loss in beam window. At the center
of the detector, the pressure is about 3 atm. A 76~$\mu$m Mylar window is
more than enough to handle this pressure. The resulting energy loss is
about 12~keV and can be calibrated with the HPGe detector.
The variation of the energy loss due to curved window
is well below the 0.1\% of the prompt energy. The second factor
is the shadowing effect of the calibration pipe and endcap.
In order to minimize the impact of these two factors, a transparent
endcap design using acrylic pipe (Fig.~\ref{fig:endcap-concept})
is proposed. The residual shadowing effect is about 1-2\%. The range of
the shadowing effect is due to the kinetic energy range of positrons. A high
energy positron would in general travel longer inside the liquid scintillator,
thus has less shadowing effect due to smaller solid angle. This shadowing
effect can be calibrated by injecting both electron and positron beams into
the central detector and a bench setup. The principle is illustrated in the following:
\begin{itemize}
\item The energy responses of electron and positron are strongly correlated.
The positron energy response is essentially the sum of the
electron energy response and positron annihilation energy. The latter contains two
0.511~MeV gammas and can be calibrated using dedicated $^{68}$Ge radioactive source.
\item The shadowing effect for the positron annihilation energy can be calibrated
with the dedicated $^{68}$Ge source deployed together with the calibration tube.
\item The shadowing effect can be calibrated by comparing a
dedicated bench measurement with the results from the central detector calibration.
\item The light pattern observed by JUNO detector provides additional handle
for the shadowing effect.
\item The obtained shadowing correction for positron ionization
can then be applied to obtain the positron energy response together with
the shadowing of the annihilation gammas.
\item The resulting positron energy non-linearity model can then compared with that
of the electron to further validate the energy model.
\end{itemize}
The following formula summarizes the correction strategy.
\begin{equation}
E_{e+}^{corr} = (E_{e+}^{rec} - E_{1.022}\cdot S_{gamma}^{shadowing})\cdot S_{ionization}^{shadowing} + E_{1.022} + C_{window}^{eloss}.
\end{equation}
Here, $E_{e+}^{rec}$ and $E_{e+}^{corr}$ are the energies before and after
the corrections, respectively. The $E_{1.022}$ is the annihilation
gamma energy for positron at rest, which can be calibrated directly with
dedicated positron source. The $S_{gamma}^{shadowing}$ is the shadowing
correction for annihilation gamma, which can be calibrated by combining
dedicated source with the calibration tube. The $C_{window}^{eloss}$
represents the energy loss correction in the window, it can be calibrated
directly with the HPGe detector. The $S^{shadowing}_{ionization}$ represents
the shadowing effect for the scintillation light for positron ionization energy. This piece
will be calibrated by i) comparing the bench data vs. calibration data vs. MC,
and ii) comparing the observed light pattern vs. MC. The initial simulation
shows that the residual uncertainty can be controlled to about
0.1\% with this strategy.

\begin{figure}
\begin{centering}
\includegraphics[height=2.5in]{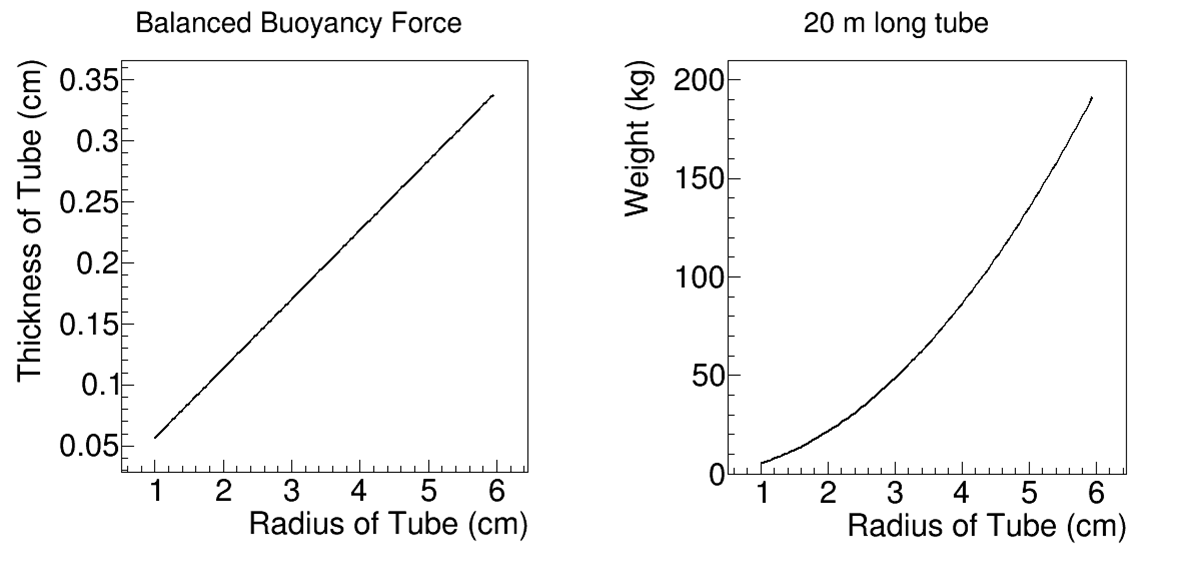}
\par\end{centering}
\caption{\label{fig:ct_force} (Left panel) Calculated required wall thickness
of the stainless steel calibration tube is plotted as a function of the radius
of the calibration tube to balance the buoyancy force. (Right panel) The corresponding
weight of the 20 m long calibration tube is plotted.}
\end{figure}

\begin{figure}
\begin{centering}
\includegraphics[height=1.5in]{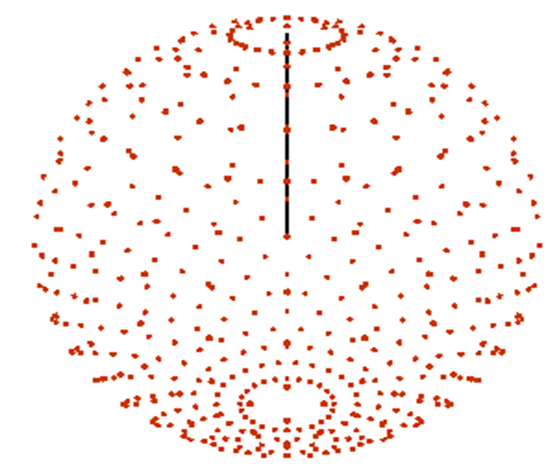}
\includegraphics[height=1.5in]{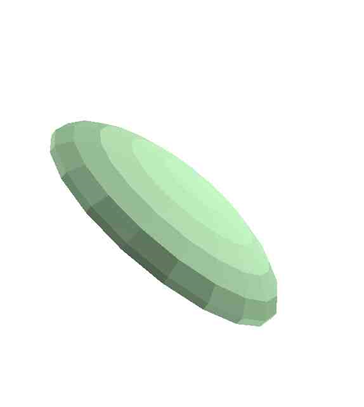}
\includegraphics[height=1.5in]{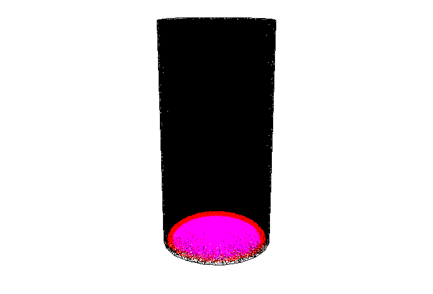}
\par\end{centering}
\caption{\label{fig:tube} (Left) Illustration of calibration tube of Pelletron
inside the JUNO central detector.
(Middle) Curved Mylar window for the calibration tube.
(Right) Endcap of the calibration tube with the curved Mylar window. }
\end{figure}

\begin{figure}
\begin{centering}
\includegraphics[height=3.5in]{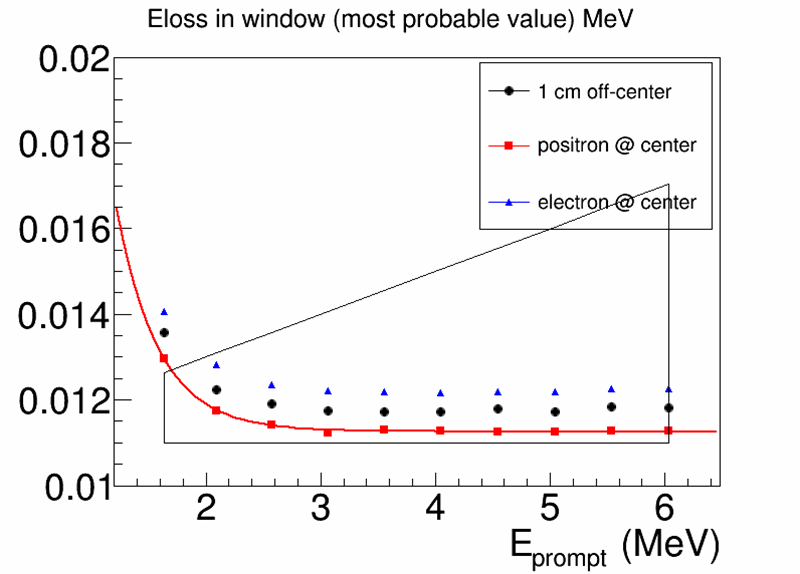}
\par\end{centering}
\caption{\label{fig:ct_eloss} Energy loss of the electrons and positrons
inside the curved Mylar window. For electron, the prompt energy was shifted
to match that of positron. The traipzoid represents 0.1\% band of the energy scale.}
\end{figure}

Simulation of Pelletron calibration has been carried out in the general JUNO
simulation framework to demonstrate the feasibility of reaching 0.1\% targeted
energy scale uncertainty. First, for a 20 m long calibration tube inside the liquid
scintillator, the buoyancy force needs to be taken into account. The left panel
of Fig.~\ref{fig:ct_force} shows the required wall thickness of the stainless steel
calibration tube to balance the buoyancy force. The right panel of
Fig.~\ref{fig:ct_force} shows the weight of the 20~m calibration tube.
Figure~\ref{fig:tube} shows the implemented geometry of calibration tube including
i) $\sim$17.3~m long stainless steel calibration tube, ii) 30~cm long
acrylic endcap, and iii) 7.6 $\mu$m thick curved Mylar window.
The energy loss of the electrons and positrons inside the Mylar window was
simulated and shown in Fig.~\ref{fig:ct_eloss}. For positrons at off-center
locations, the energy loss is slightly higher than those at the center of
calibration tube. This is consistent with the fact that the vertical thickness
of the window at off-center location is bigger than that in the center.
The energy loss of electron is slightly bigger than that of positron
due to additional annihilation process that positrons can go through.
The difference is much smaller than the targeted 0.1\% energy scale
uncertainty band.

\begin{figure}
\begin{centering}
\includegraphics[height=1.6in]{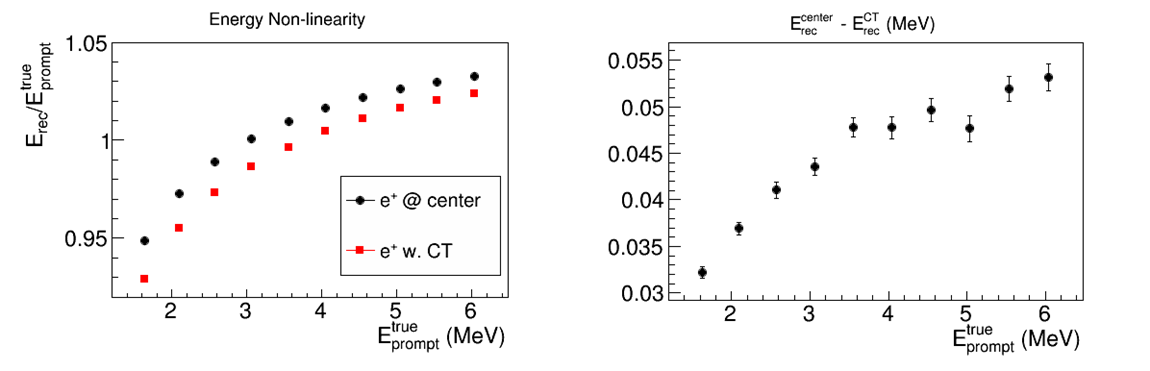}
\par\end{centering}
\caption{\label{fig:ct_result1} (Left) The energy response of positron
injecting through the calibration tube (red) is compared to that
of the positron in the detector without the calibration tube(black).
(Right) The difference in energy is shown with respect of the true
positron prompt energy. The increase of difference reflects the shadowing
effect of the calibration tube}
\end{figure}

\begin{figure}
\begin{centering}
\includegraphics[height=2.5in]{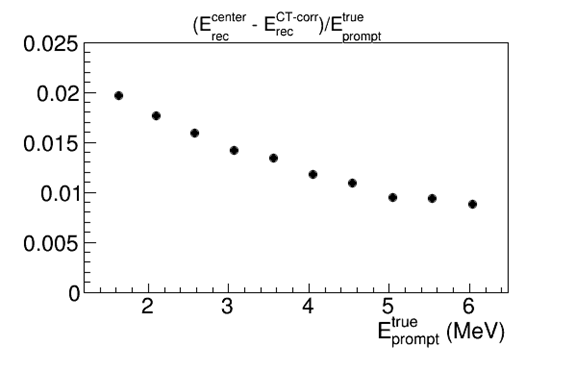}
\includegraphics[height=2.5in]{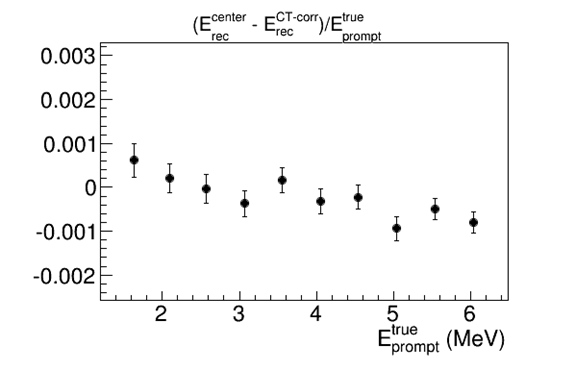}
\par\end{centering}
\caption{\label{fig:ct_result2}(Top) The bias of energy response (1-2\%)
due to the shadowing effect of the calibration tube is shown with respect of
the true prompt energy.
(Bottom) The residual bias of energy response ($<$ 0.1\%) after the
correction of energy loss and the shadowing effect is shown with respect of
the true prompt energy.
}
\end{figure}

A beam of positrons are shooting through the calibration tube to study the
energy response. The positron beam cross section is assumed to be a circle
with radius of 1~cm. The spread in momentum is assumed to be 0.1\%.
The results are compared with those of injecting positrons directly inside
the detector center without any calibration tube (Fig.~\ref{fig:ct_result1}).
The increase of the difference between two cases with respect of the true
prompt energy reflects the shadowing effect of the calibration tube.
The bias without any correction (top panel of Fig.~\ref{fig:ct_result2})
indicate the magnitude of bias is about 1-2\%. The bias becomes smaller at
high energy, because high-energy positrons can penetrate into liquid scintillator
further so that the shadowing effect is smaller. The bottom panel of
Fig.~\ref{fig:ct_result2} shows the residual bias after the energy loss
and shadowing correction. The residual bias can be controlled to below 0.1\%.

\subsection{Diffused Short-lived Isotope Calibration}

Homogeneous energy calibration of a large liquid scintillator detector throughout
the entire volume can only be achieved with distributed sources in
the scintillator. The use of uniformly distributed sources was successfully
demonstrated by the Sudbury Neutrino Observatory using $^{16}$N,
$^{8}$Li, as well as \emph{in-situ} spikes from $^{24}$Na and $^{222}$Rn.
Injection and deployment of such sources in a 20 kt-detector pose challenges
including the production and injection of isotopes over long distances
and their distribution throughout the detector volume. We will survey
suitable short-lived isotopes, their production method and develop
concepts for the injection with a gas or scintillator stream into
the detector. We will pay particular attention to the possible use
of positron sources. We will identify possible isotopes and develop
a technical concept for the injection and distribution of sources.
The principles of the isotope injection will be studied with the test
chamber developed for various R\&D purposes or a re-purposed Daya Bay
detector.


\section{R\&D Status and Plans}
\subsection{Sources}
Source selection and conceptual design of the geometry
need to be determined via Monte Carlo. Mini-balloon source requires
prototype both mechanically as well as test loading of radioactive
isotopes. A robust and simple UV laser system needs to be developed
allowing $\sim$ns level timing and sub-percent intensity control.

\subsection{Rope Loop System}
We have learned much from the Daya Bay calibration experiences. The
challenges of the rope loop approach lie in the automatic source swapping
devices, interfacing with the central detector and the installation
procedure. A 1:1 prototype for the mechanical system is required before
the final design. We shall also find high bay area to performance deployment
tests similar to that in the real detector. Additional key considerations
are:
\begin{itemize}
\item {The rope loop subsystem will be installed after the cleaning of the
  central detector, or even after LS filling, thus installation
  procedures should be simple and they must be carried out quickly to avoid
  contamination;}
\item {LS detector with such a large size is unprecedented. Calibration
system should be designed so it can accommodate potential structural
distortions; and}
\item {The reliability of the source changing devices and life span should be
  understood well with destructive tests.}
\end{itemize}

\subsection{ROV}
To achieve good physics performance, we need
to optimize the shape and the dimension of the ROV via extensive
simulations.

The ROV shall be designed with high cleanliness and be fully compatible with
liquid scintillator. Prototype studies are needed in a realistic deployment
condition.

\subsection{Source Positioning System}
An ultrasonic emitter's diameter is roughly 30~mm and its working frequency
150~kHz with a pulse width of $\sim$1 ms.
The emitter can be installed either on the source or on
the ROV.

The receiver array is mounted on the detector sphere and at least
three receivers can
realize 3D positioning. The actual arrangement of the array need to be
optimized. To increase the reliability, we plan to have $\sim$10
receivers so when some malfunction or break down, the system can still
function. Longer array baselines means better positioning
precision, which implies receivers are better placed near the equator from
the perspective of positioning. However, this also means the cables serving
these receivers are longer and it might cause more potential issues. The
material compatibility between cables and LS needs to be checked and
addressed. Compared with the oceanic environment, JUNO central detector
is an isolated environment. Impact to positioning uncertainty
due to reflections and refraction shall be well studied and understood.

\subsection{Pre-installed Guide Tubes}
How to mount the guide tubes next to the acrylic panels is quite
challenging due to the distortion after the detector is filled. The maximal
distortion of the acrylic panel is expected to be $\sim$5~cm thus if the guide
tubes are fully attached to acrylic panels, this level of distortion is
highly harmful for the smoothness of the tube even the joint does not get
destroyed. We need to allow certain level of distortion thus probably the
two should not be fully coupled. Instead we should allow the relative
movement to some level while still keep the positioning accurate
enough.

In addition to the mechanical concerns, the impact of these additional,
tube materials to the detector performance needs to be
understood better. It should not make much difference they are deployed either inside or
outside of the acrylic sphere if refraction indices match well with LS
and/or mineral oil. Acrylic, stainless steel, and Teflon tubes tubes are considered.

Right now, the system realization and impact to the detector
performance are undergoing discussion. Argonne National Lab has a
high bay area, suitable for performing deployment
tests of the rope loop system, and tests of the guide tube systems.

\subsection{Pelletron System}

The need of space and transportation underground have been checked and no
obvious obstacles were observed. Currently, we need to detail out the beam
line design and the interfacing with the central detector.

\subsection{Diffused Short-lived Isotope Calibration}

Various sources are being studied. The concept of source injection is being
investigated.


\section{System Reliability and Safety Concerns}

It would be difficult to make large scale repair to the
calibration system during the life span of the
experiment. This places very stringent challenges on the reliability and
longevity of the calibration system, especially that there should absolutely
be no source dropping.

For the automatic source swapping component of the rope loop system, we have
considered the following multi-layer safety protection mechanism,
\begin{enumerate}
\item {The position and the clamping force of the robotic hands are
  constantly monitored after each motion;}
\item {Robotic hands and other clamping devices are designed to be
  normally-close to make sure that sources are still secured during power or
  control signal loss;}
\item {To make sure of the reliability of the quick-connects, they are
  pull-tested each time they engage;}
\item {A fall-prevention shutter is installed beneath the source-swapping
  device. During source swapping, the shutter is closed; and}
\item {A piece of annealed iron will be installed inside the source. In case
  of dropping, a ROV with a strong magnet can be deployed to recover it.}
\end{enumerate}

The ROV umbilical cable can take certain load besides functioning as a power
and control signal cable. Thus in case of malfunction, it can be pulled out
of the central detector. In addition, the design of the ROV also has some
built-in safety considerations and the ability of self test.

The effective density of the ROV should be roughly 95\% of LAB. During power
loss or other situations of losing motion, the ROV would float
automatically. The liquid dragging force shall be chosen so that the rising
speed is not too high, avoiding damaging the vessel. The internal pressure
of the ROV should be either positive or negative and should be constantly
monitored. Larger-than-normal fluctuations would indicate a leak might have
developed and an alarm signal shall be sent out.

\section{Schedule}

A major contribution to the calibration system and additional contributions in support
of central detector, liquid scintillator and PMT QA are being investigated for US JUNO
deliverables.






\clearpage
\begin{longtable}{|p{1.5cm}|p{4cm}|p{3cm}|p{5cm}|}
\caption{Yearly targets} \\
\hline
Year & Goals & Deadlines & Milestones \tabularnewline
\endfirsthead
\multicolumn{4}{r}{(See Next Pages.)}
\endfoot
\bottomrule
\endlastfoot
\hline
2013 &Form a working group, start conceptual designs and
simulation &Form a few preliminary mechanical designs &Form 3 alternative
conceptual mechanical designs before November\tabularnewline
\hline
2014 &Continue design and gradually converge; Technique R\&D; Design light
sources and build mock-ups &Finish the simulation of source selection;
finish the design of a stable light source &Finish the stable light source
design before June; Finish the simulation and analyses of the 3 alternative
designs; Select 2 out 3 in September\tabularnewline
\hline
2015 &Build and test a prototype; finalize both the mechanical and the
electronic designs &Finish the R\&D of selecting radioactive sources;
finalize the choice of radioactive sources & Finalize the mechanical design
before June; finalize the construction and testing of the prototype, and the
radioactive source designs before December\tabularnewline
\hline
2016 &Bid, order and start the construction of calibration systems & finish
one complete calibration system and testing & finish bidding before June and
complete one calibration system and finish its testing before
December\tabularnewline
\hline
2017 &Continue the construction of the calibration system & Complete the
remaining half of the project &Complete the remaining half of the project
\tabularnewline
\hline
2018 &Complete the construction of the calibration system; order all
radioactive sources; complete DAQ and the interface with DCS &Compelete the
last half of the project & Complete all remaining constructions, DAQ and the
interface between DAQ and DCS before December\tabularnewline
\hline
2019-2020 &Installation and commissioning &Complete the installation and
commissioning of the system & complete the calibration system before the
completion of the central detector installation\tabularnewline
\hline
\end{longtable}


\cleardoublepage

\chapter{Readout electronics and trigger}
\label{ch:ReadoutElectronicsAndTrigger}

The JUNO readout electronics system will have to cope with the signals of 17,000 PMTs of the central detector as well as 1,500 PMTs installed in the surrounding water pool. The average number of photoelectrons being registered by an individual PMT will reach from below one for low-energy events up to several thousands in case of showering muons or muon bundles. In both extreme cases, the number of photoelectrons has to be counted and their arrival time profiles have to be determined. A dedicated trigger system will be necessary to perform a pre-selection of correlated PMT hits caused by physical events from a sea of random dark noise hits. Based on the selected arrival time and pe pattern, software algorithms will reconstruct energy, type and position of the event vertex. Thus, the main task of the readout electronics is to receive and digitize the analog signals from the PMTs and transmit all the relevant information to storage without a significant loss of data quality. The basic concepts considered in the design of the system as well as a list of specifications are given in Sec.~\ref{sec:roe:specs}.

The PMTs will be submerged in the water pool surrounding the acrylic sphere that contains the LS. Based on this layout, three different approaches for the realization of the read-out electronics are discussed: A mostly "dry scheme" for which individual cables carry the analog signals out of the water pool to an external electronics hut containing front-ends and digitization units. A "wet scheme" for which digitization happens at the PMT bases and the digital signals are bundled in submerged central underwater units before being passed on to an external online computer-farm. And finally, an intermediate readout scheme where a considerable part of the electronics is submerged inside the water pool but still accessible for replacements. The three schemes are presented in Secs.~\ref{sec:roe:dry}$-$\ref{sec:roe:wet}.

Sec.~\ref{sec:roe:dev} will provide an overview of further R\&D efforts to be undertaken. A first time line for the realization of the system and the most important aspects to be dealt with is laid down in Sec.~\ref{sec:roe:rea}. Possible risks are summarized in Sec.~\ref{sec:roe:risk}.

\section{Design considerations and specifications}
\label{sec:roe:specs}

The concept for the design of the electronics is driven by the following goals, which are listed here in order of their priority:

\begin{enumerate}
\item {\bf Energy reconstruction:} Optimization of the energy measurement of $\bar\nu_e$, especially at the lowest energies. A basic limitation on the energy resolution arises from the statistics of detected photoelectrons (pe). This limit must not be worsened significantly by the effect of electronics. At a photoelectron yield of 1100 pe per MeV,  the achievable relative resolution is $\sim$3\,\% at 1\,MeV.
\item {\bf Reconstruction of the photon arrival time pattern:} A precise measurement of the arrival times of the photons is required to reconstruct the position of the events. This is particularly important for the IBD events induced by $\bar\nu_e$'s because (a) the energy response of the detector will be position-dependent and (b) the formation of spatial coincidences between prompt positron and  delayed neutron events will support background suppression. Moreover, track reconstruction for cosmic muons and especially for muon bundles is very sensitive to timing and will be an important tool for the veto of cosmogenic backgrounds.
		
\item {\bf Low dead time and dynamic acquisition rate:} During usual data taking conditions, all neutrino events from reactor, solar or geological origin should be acquired without in-efficiency, i.e. with a minimum or zero dead time. In addition, readout electronics have to be able to cope with short episodes of extremely high trigger rate that might be caused by the neutrino burst from a nearby galactic Supernova. Dependent on the distance, thousands of neutrino events are expected within a period of 10 seconds. Therefore, the system has to be able to bear a dramatical increase in its acquisition rate for short periods without significant data loss or dead time.
\end{enumerate}	

\subsection{Energy reconstruction}

For the low-energy events, the deposited energy can be almost directly inferred from the overall amount of photons or pe detected by the PMTs. However, there are several ways of performing this integral measurement of collected pe (see below). Additional complications arise for high energy events, where the accuracy of determining the number of pe detected for a given PMT will depend on the dynamic range of the readout electronics.
		
\subsubsection{Photoelectron counting}

There are two basic methods to determine the visible energy created by an interaction in the detector. One can either integrate the charge created by the PMTs or one can try to identify individual photons and count them. The reconstruction is complicated by the spread in the arrival time of the photons, which is caused both by the fluorescence time of the LS, photon time-of-flight and photon scattering which all we contribute to delays in the order of several hundred nanoseconds\footnote{Note that once the vertex position is known, the delay from the time-of-flight can be approximately corrected offline. Based on this, most of the light will be concentrated towards the start of the event.}. An additional distortion of the time pattern will be introduced in case of cosmic muons where the light is generated along the extended particle track, causing a delay by the time-of-flight of the muon ($\leq$120\,ns). The trigger system will have to take into account this spread in arrival times when defining a threshold condition.
\medskip\\
\noindent{\bf Charge Integration} is technically the simpler solution. During a certain time window the signal of the PMTs is integrated and the integrals are summed up into a global charge. This global charge is proportional to the visible energy of the event. For an online measurement the time window has to be in the order of 300\,ns from the arrival of the first photon at a given PMT. If time-of-flight effects are taken into account, e.g. for an off-center event close to the surface of the acrylic sphere or a cosmic muon, the acquisition gate has to be extended to 500\,ns. The length of the time window must be optimized between two aspects. With longer time windows more and more of the scintillation light is included, but the contribution from the electronics noise increases. Due to the electronics noise the potential of the baseline will vary around zero. Integrating such a baseline even without a signal will not give exactly zero. The contribution from the baseline noise will scatter around zero with a width of the distribution that increases with the square root of the integration time.

There are two subversions of the charge integration. One can either sum all the PMTs or just those PMTs above a certain threshold (e.g.~0.25\,pe). Due to fluctuations in the amplification process of the PMTs (in the dynode structures of a conventional PMT or in the MCP) the charge produced by a single photoelectron varies. The variation is usually classified by the so-called peak-to-valley ratio of the PMT. Due to these fluctuations it can happen that the signal of a single pe is not visible above the baseline noise level. But the information is not completely lost. In the absence of electronics noise the sum of all PMTs will give a better estimate of the visible energy than just the sum of those PMTs above a certain threshold. For a real system one has to optimize the threshold. It largely depends on the noise of the system and on the average number of pe per PMT for the relevant events . For JUNO it is not yet clear what the best value is. For Daya Bay it turned out to be 0.25\,pe. Note that this charge integration threshold is not necessarily the same as the trigger threshold applied for self-triggering of a channel.
\medskip\\
{\bf Photon Counting} is usually a self-triggered method. One analyzes the wave-form of each PMT individually in an attempt to identify the arrival of individual photons. Then the number of these photons is counted over at time window comparable to the above. The total number is proportional to the visible energy. The advantage of this method is that the number of photons is more directly related to the energy than the total charge. Some photoelectrons produce higher charges and some lower. This introduces an additional fluctuation into the charge measurement which is unrelated to the visible energy. With photon counting this additional fluctuation is eliminated.

With the method of photon counting a threshold for each PMT is unavoidable. Photo electrons with very low charge due to fluctuations in the amplification process will be lost. This degrades the resolution by an amount that strongly depends on the peak-to-valley ratio of the PMTs. Whether photon counting is superior or inferior to charge integration depends on the details of the detector and the electronics. For JUNO the answer is not yet known. A detailed simulation is necessary to produce it. For Borexino photon counting turned out to be the preferred method at low energies and also DoubleChooz is using a method of approximate photon counting. For Daya Bay the electronics does not allow to do photon counting.

To extract the number of photons from the waveform of a PMT, algorithms of different levels of sophistication are possible. In case of a software trigger, a very simple algorithm that counts the number of transitions of the photon pulse(s) across a threshold is probably sufficient. A more sophisticated algorithm will identify the charge of a pulse and its shape. Effectively at very low occupancies any pulse consistent with the charge of a single pe will be counted as one photon. At very high occupancies, the shape of the pulse will not contain any significant information. Therefore the number of photons will be derived from the integrated charge. The two basic methods become identical at very high occupancies. At intermediate occupancies, when the pulses of a few photons partially overlap in time, shape information like satellite peaks in the wave-form combined with the integrated charge will give the best answer

\subsubsection{Dynamic range}
The electronic system must be able to cope with a large range of signal heights (and also durations). For low-energy neutrino events, the average occupancy is below 1\,pe per PMT. Therefore the electronics must have good resolution for single pe's. Cosmic muon events are forming the high-energy end of the the dynamic range. These events can reach hit occupancies of 4000 pe for individual PMTs, and this has been defined as the maximum occupancy the electronics will have to cover. However, even larger signals are expected in case the muons arrive in bundles or create hadronic showers inside the target volume.

\subsection{Photon timing}
\label{sec:roe:timing}

For all event types, the arrival time of the first photons at the individual PMTs will be the primary information available for the reconstruction of the vertex position (in case of neutrino events) or the particle track (in case of muons).
This will be of particular importance for the energy reconstruction of antineutrino events because the effective pe yield will be position-dependent. A precise reconstruction of the vertex position will allow for low systematic uncertainties when applying correction terms.

\subsubsection{Reconstruction of point-like events}

Reconstruction algorithms will start from the hypothesis of a simultaneous emission of all photons from a single event vertex inside the detector. Then, the assumed vertex position will be adjusted to fit the observed photon arrival time pattern when considering the time-of-flight of the photons. The assumption of simultaneous emission is usually well fulfilled for the first photons detected for an event. More sophisticated algorithms will take into account the time delays introduced by the finite fluorescence times, photon scattering in the LS and the transit time spread of the PMTs by using appropriate probability density functions. Moreover, also the density of photon hits distributed over the surface of the detector sphere bears information on the vertex position.

What is more, the precision of the photon timing will be decisive for the practicability of pulse shape discrimination. Heavy particles like $\alpha$'s or neutrons (proton recoils) create more light than electrons in the slow components of the fluorescence profile. Based on the information of the vertex position, individual hit times can be corrected for the time-of-flight of the photons. The resulting sum pulse of all detected pe corresponds to the original scintillation profile of the event (plus impact of light scattering). Profiles can be characterized for their long-lived decay component, allowing to quite clearly distinguish electrons from $\alpha$-particles and proton recoils and with lower efficiency even positrons from electrons.


The quality of the separation will largely depend on the gate length over which the photoelectrons are acquired. The time window should cover a significant portion of the slow fluorescence component which is of the order of several 100\,ns.  In this respect, a decentralized trigger schemes featuring a variable gate start time and gate width for each channel seems preferable. In this case, acquisition will stop only when the acquired pulse becomes indistinguishable from dark noise (see below).


\subsubsection{Reconstruction of cosmic muons}

Reconstruction of cosmic muon tracks is more complex. For single, minimum-ionizing muon tracks, the emitted light front will closely resemble a Cherenkov cone in the forward-running direction\footnote{This shape is due to the relative speeds of muon (speed of light in vacuum) and photons (speed of light in the medium) which results in the same geometrical condition as in case of the Cherenkov effect.}, plus a superposition of spherical back\-ward-running light fronts. Such events are reconstructed in a similar fashion as point-like events, relying on a fit of the muon time-of-flight plus the photon time-of-flight for a given track hypothesis to the arrival time pattern of first photons observed at the surface of PMT photocathodes. Also charge information from the many hundred photons per channel may be included as the collected number of photoelectron at a given PMT will depend on the distance from the muon track.

However, showering muons as well as muon bundles create much more complex event topologies that lead in effect to a superposition of several light fronts arriving at the photocathodes with short time-delays. First attempts at the reconstruction of such events suggest that the full time profile of photon arrival times have to be recorded for each individual PMT in order to be able to discern secondary maxima or at least substructures that are created by the delayed arrival of subsequent light fronts. While a variety of algorithms are currently investigated, none has reached a state from which hard requirements on the necessary photon timing information can be formulated.

Note that especially for these events, the ability for a separate detection of the true Cherenkov cone might provide valuable information for disentangling the contributions of individual particles. However, it seems unlikely that conventional or MCP-PMTs will reach the necessary time resolution.

\subsubsection{Limitations in timing}

While accurate timing information is for sure of great value, there are intrinsic limitations to the accuracy that can be reached in JUNO that are inherent to the LS detection technique. The first one arises from the finite time width of photon emission from the scintillator that is described by the fluorescence time profile. The fast component that is most relevant for timing questions features a decay time in the order of a few ns. This leads to a smearing of the timing information especially in the regime of low-photon statistics, i.e. at low energies. The second and more severe constraint arises from the transient time spread (tts) of the PMTs. The tts characterizes the variation of the time between the arrival of the photon and the creation of of the electrical pulse. As the final PMT is not yet available, the precise tts profile is not known. But the ($1\sigma$) uncertainty introduced is expected to be in the range of 3$-$4\,ns.

Taking both inaccuracies into account, MC simulations of scintillation signals have been performed to obtain realistic PMT signals that were then "digitized" with different sampling rates. These studies indicate that a time resolution of 1\,ns resp.~a sampling rate of 1\,GS/s will be sufficient to record the signals without loss of information.

%

\subsection{Triggering schemes}

A further important consideration is the triggering scheme: It may either be global, i.e. based on the information of all channels, or de-centralized, i.e. based on self-triggering channels that send all their digitized signals (including all dark noise photoelectron) to a computing farm. In the latter case, a (online) software trigger is used to identify physical events. Both options are discussed more closely in the following.
\medskip\\
In case of a {\bf Global Trigger}, the information of all PMTs is combined in order to form a global triggering decision that starts the readout of all PMT signals. The information from a limited local group of PMTs will not be sufficient to form the trigger decision. Information from all of the PMTs is required. Monte Carlo studies have shown that a simple trigger logic based on the total number of active PMTs (PMTs with at least one hit) within a 300\,ns window is sufficient to suppress the background from random coincidences of dark noise and to guarantee 100\,\% efficiency for the detection of $\bar\nu_e$ events. Threshold would be set to 500 hits ($\sim0.5\,$MeV)  which is well above the average number of 225 dark noise hits within the same time span\footnote{This number is based on a trigger coincidence window of 300\,ns and 17,000 PMTs featuring a dark noise rate of 50\,kHz each.}.

Such a trigger can easily be realized in the more conventional scheme of the external readout electronics (sec.~\ref{sec:roe:dry}) that relies on the collection of the analog signals in a physically confined space. In this case, the delay introduced by the decision for triggering reaching the digitization units is fairly small. However, a global trigger is much harder to apply for a de-centralized digitization as it is propagated by the underwater electronics scheme (sec.~\ref{sec:roe:wet}). Hit information from the spread-out PMTs would have to be transmitted either by individual cables/fibers or by combining the information from a group of PMTs underwater into one cable/fiber per group. The trigger decision would then be sent back along the same network to the PMTs to start the readout. During the latency of the trigger (several ms), the wave forms of the PMTs would have to be buffered either on the PMTs or in the underwater units.

An important aspect regarding energy reconstruction that potentially favors a global trigger is the fact that this is the only configuration that will allow the recording of waveforms for channels in which the PMT signals are below an individual triggering threshold, i.e. the detection of low-charge photoelectron. On the other hand, the existence of a global trigger threshold at $\sim$0.5\,MeV poses a special challenge for data acquisition during a Supernova. Besides a large number of $\bar\nu_e$ events, a substantial amount of the resulting neutrino interactions in JUNO will be from elastic scattering of protons. Due to reaction kinematics and quenching, the visible light will be fairly low and necessitate a temporal lowering of the threshold below 500\,pe. While this does not pose a problem from point of view of radioactive background (there might be an interference with dark noise, though), the only information available to make the system aware of the on-going neutrino burst is the sudden increase in event rate. It will take a certain number of triggers to realize that neutrinos from a nearby supernova are arriving. Then the threshold will have to be lowered in retrospective. This is not impossible with an appropriate amount of buffers, but it does impose a significant complication of the trigger scheme.
\medskip\\
\noindent{\bf Decentralized triggering.} In principle this "untriggered" scheme is much simpler. The signal from each PMT is continuously monitored. If it exceeds the acquisition threshold for a single photoelectron (SPE), i.e. probably around 0.25 of the single photoelectron charge, the signal is digitized and transmitted to the online farm. On the online farm the signals from the individual PMTs are combined into events and a decision is formed whether to store the event or to drop it. It is mainly a question of data reduction and storage capacity to decide which fraction of events or even individual PMT waveforms to store.

The advantage is that much more information is available for the decision on storage of the event. Time-of-flight information can be taken into account to associate true photon hits inside a 300\,ns window to the emission by a single vertex. This potent mean for the suppression of background from random dark noise coincidences will allow to lower the threshold below 0.5\,MeV during normal data taking.

For the underwater readout electronics, this scheme bears the considerable advantage of reducing the required cabling. There is no need for a network to send trigger information to the central trigger and to return the trigger decision. The disadvantage is a higher load on the data links to the farm and on the required computing power. In this scheme not only the waveforms of signal but also background events have to be transferred to the farm in order to form a trigger decision. The difference in the data transmission rates between the two trigger schemes is somewhat reduced by the length of the time windows required for the readout. For the global trigger scheme it must cover the full events ($\geq300$\,ns) for all PMTs while in the untriggered scheme the gate width can be adjusted to the number of photoelectrons detected, e.g. 32\,ns per single photoelectron pulse. If for a given event multiple photoelectrons are detected by a single PMT, several internal triggers will occur and several such windows will be transferred.

Also the timing between the PMTs is simpler in the untriggered scheme. There is no need to synchronize the clocks between the PMTs. $t_0$-corrections can be applied in the farm. A White Rabbit system as proposed for the global triggering scheme is not necessary. Only the synchronization of the FADC frequencies has to be ensured, which will be achieved by locking the internal oscillators by a centrally distributed clock.

Last, but not least one should mention that a software algorithm for data reduction on a farm is much more flexible to react to unforeseen complications or opportunities than a hardware-based trigger.

\subsection{List of specifications}

Based on the above discussion, the following list of specifications has been compiled:
\begin{itemize}

\item {\bf Waveform sampling} should be available over the whole energy range with a sampling rate of 1\,GS/s.

\item {\bf Photoelectron resolution:}
In the signal range from 1$-$100\,pe, the charge resolution should increase linearly from 0.1$-$1\,pe.\\
In the background range of 100$-$4000\,pe, the charge resolution should be 1\,\%.\\
Signal range and background range should overlap.

\item {\bf Dynamic range} should reach from 1 to 4000\,pe per channel.

\item {\bf Arrival time resolution}, e.g. by fitting of the signal leading edge, should be $\sigma_t\approx100ps$.

\item {\bf Noise level} should be below 0.1\,pe for single photoelectron detection.

\item {\bf Maximum acquisition rate} should be on the order of $10^3-10^4$ triggers per second.

\item Acquisition should be low in or without any dead time.

\end{itemize}
	
\noindent Beyond these performance requirement, it has to be ensured that the system can be reliably and continuously operated for at least 10 years with no or minimal access to the parts of the readout system submerged inside the water pool. It is also important to facilitate the installation by minimize the cabling between the PMT matrix and the outside electronics.

\section{External readout scheme}
\label{sec:roe:dry}

The philosophy of the external read-out scheme foresees to place as large a part of the electronics as possible in a dedicated electronics area or hut outside the detector. In particular, the analog-to-digital converters are to be located outside the water pool.

\begin{figure}[htb]
\begin{center}
\includegraphics[width=\textwidth]{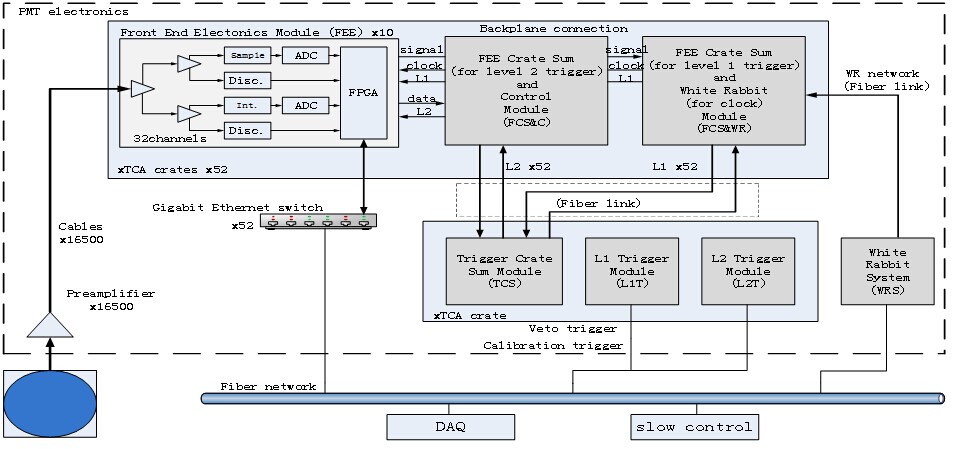}
\caption[Scheme for external readout electronics]{\label{fig:out water scheme}Schematic drawing of the external readout scheme.}
\end{center}
\end{figure}

\subsection{System layout}

The general layout is shown in Fig.~\ref{fig:out water scheme}: The electronics attached directly to the PMTs is kept to a minimum. At the base of the PMT, the signal is decoupled from high voltage and increased by a pre-amplifier. From there, the analog signal travels along a 100\,m long cable to the input at the external front-end electronics (FEE). These will be arrayed in crates of the ATCA chassis specification. Each crate will hold ten FEE modules and two trigger modules. Each FEE module will provide for 32 channels backend amplification, signal distribution and processing, analog-to-digital conversion and data processing, as well as signal processing for the trigger system.

The FEE modules will send the digitized data to the trigger modules via the backplane of the ATCA crate. These will create the trigger information that will be sent to the central trigger crate by fiber links. In case of a positive trigger decision, the trigger signal will be sent back to each signal processing crate and then fanned out to each FEE module. Each FEE module buffers the data and sends the necessary data asynchronously to the DAQ. Trigger and slow control system will follow the White Rabbit protocol to transfer the clock and other service signals to each crate. The clock signal will be distributed together with the DAQ and slow control through an optical fiber network.

This has several advantages: The simple structure of the external scheme allows for an easy implementation. A global trigger can be easily formed. Maybe most importantly, broken components of the FEE and trigger logic are easy to access for repairs or replacement, and an upgrade of the system is easy to realize during the 10 years or more of detector operation time. The main disadvantage is the transmission of the analog signals from each individual PMT over the 100\,m long cables. While signal attenuation can be compensated by placing the preamplifier at the PMT, some shaping of the leading edge of the analog pulses will be unavoidable. In addition, the sheer amount of underwater cables complicates the installation and is a severe cost driver.

\subsection{Signal amplification and digitization}

\begin{figure}[htb]
\begin{center}
\includegraphics[width=10cm]{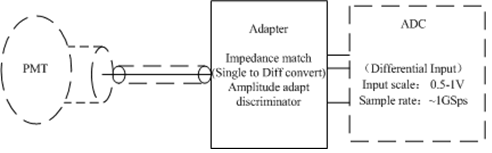}
\caption[Scheme for signal amplification and digitization]{\label{fig:Amplification}Scheme for signal amplification and digitization.}
\end{center}
\end{figure}

In the external electronics scheme, pre-amplification will be performed directly at the PMTs and a second amplification stage is located directly after the input of the FEE modules. The bandwidth of the amplification circuit has to be chosen appropriately to minimize information loss by shaping. From there, as displayed in Fig.~\ref{fig:Amplification}, the signal will be divided into two paths: One is fed to a discriminator in order to provide a trigger flag, the other is sent to an FADC for charge measurement. Due to the large dynamic range of the PMT signals (1\,pe to $\geq$4,000\,pe), the use of a single FADC for the digitization seems hard to achieve while meeting the resolution requirements for low pulses. Therefore, the amplification will be divided into two or three paths with different amplification gains \cite{Qiuju}. In each path, signal amplitude will be adjusted to match the input range of the FADCs. For most of the fast, high-resolution FADCs \cite{AD9234} the maximum range of the input voltage is only 0.5$-$1\,V. Furthermore, minimization of the input electronics noise is quite important to keep up the specified voltage resolution. This can be achieved both by design and by filtering.

The FADCs will be operated in differential mode to increase the performance. The foreseen sampling rate is 1\,GS/s, the required voltage resolution 14 bit. There are two possibilities: Either to purchase commercial chips or to design a customized ASIC. In the first case, the ADS5409 from TI and AD9680 from Analog Device would fulfill the requirements. If relying on a self-designed ASIC, the most practicable approach is a multi-channel time-interleaved architecture which will feature a low power consumption compared to a realization based on a single channel. For each channel, a 12\,bit, 500\,MS/s pipeline quantizer (or a 12\,bit, 250\,MS/s SAR quantizer) will be used for meeting a good balance between power, speed and accuracy. Offsets and gain mismatches between channels can be calibrated offline.  A differential clock input will be used to control all internal conversion cycles.

\subsection{Data processing and transmission}

In a first step the FEE system will process the data and provide some simple information to the trigger system. Only when a trigger decision is formed and a corresponding signal is sent back to the crates, the full data (including waveforms) of all channels will be read out, bundled to a combined output in units of modules and crates and then transferred to the DAQ.

\begin{figure}[htb]
\begin{center}
\includegraphics[width=12cm]{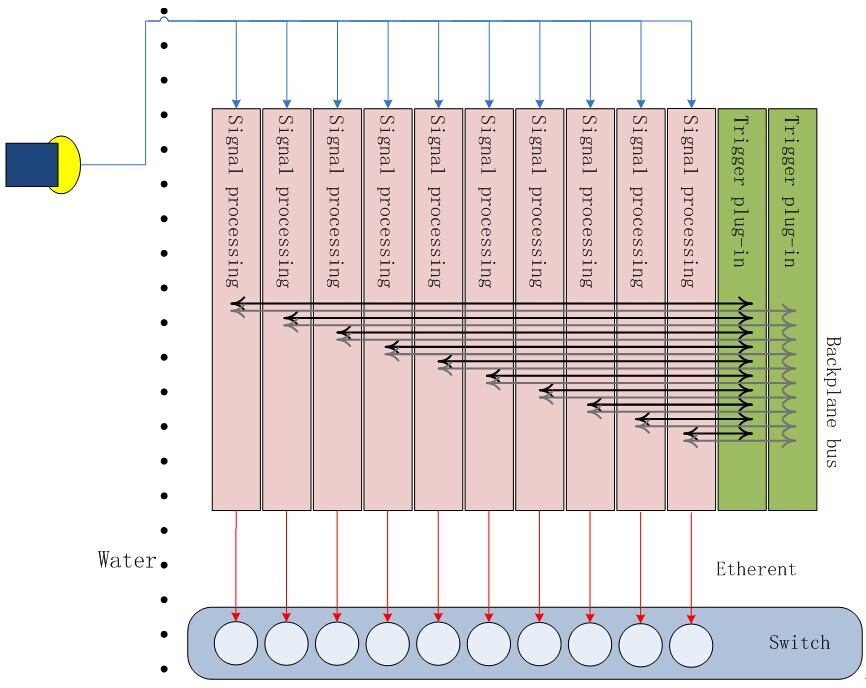}
\caption[Data processing and transmission in the external scheme]{\label{fig:DataProcessingAbove}Block Diagram of the data processing and transmission of the external electronics scheme}
\end{center}
\end{figure}

There are 10 FEE modules plus 2 trigger modules in each crate. Data is transmitted between the modules via the backplane BUS system of the crates. Communication is done by a dual-star structure. Each FEE module can receive the trigger signal from either of the two trigger modules via a pair of high-speed serial connections. The digitized data is transferred via ethernet connection to a switchboard and from there to the DAQ. A block diagram of the system is shown in Fig.~\ref{fig:DataProcessingAbove}.

%

\subsection{Trigger system}

The design goals for the trigger system are the following:
\begin{itemize}
\item The trigger efficiency for IBD events induced by reactor $\bar\nu_e$ should be high, i.e.~reach 99\,\% detection efficiency for an energy deposit of $\sim$0.5\,MeV, so well below the minimum energy deposit of a prompt positron.

\item The system should be able to suppress detector-related backgrounds as PMT dark noise coincidences and surface radioactivity to reduce the overall event rate.

\item  The latency introduced for the processing of 17,000 PMTs should not exceed a few $\mu$s.
\end{itemize}

\subsubsection{Trigger rates and threshold}

The expected trigger rates generated by neutrino signals and backgrounds were studied by simulations. Signal sources considered were IBD signals, residual radioactive background events in the LS, cosmic-ray muons, and the dark noise of PMTs. We assumed that the buffer was thick enough to shield the gamma-ray background from the PMTs and support structures.

The details on signal and background assumed are summarized in Tab.~\ref{tab:trigger}: Radioactive background levels in the LS were estimated conservatively based on the conditions in the Daya Bay detectors \cite{Qiuju}. However, radiopurity requirements for JUNO are much stricter. If i.e.~levels from KamLAND \cite{K.Eguchi} are used, the rates will be lower by several orders of magnitude. Muon background was estimated with 1,500\,mwe of rock overburden, taking into account a coarse image of the three-dimensional surface topology. Muon-induced spallation isotopes (i.e.~$^9$Li etc.) were not considered.

\begin{table}[htbp]
\begin{center}
\begin{tabular}{llc}
\hline
Signal or background & Calculation standard & Event rate before trigger \\
\hline
IBD & 80\,/day & 80\,/day \\
Background in LS & ${}^{238}$U: 2.0$\cdot$10$^{-5}$\,ppb & 69\,Hz \\
 & ${}^{232}$Th: 4.0$\cdot$10$^{-5}$\,ppb & 32\,Hz \\
Muons & 1,500\,mwe & 3\,Hz \\
PMT dark noise & 50\,kHz/PMT & 3.3\,MHz \\		
\hline
\end{tabular}
\caption[Input for trigger rate MC]{\label{tab:trigger}Input for trigger rate simulations}
\end{center}
\end{table}

For the parameters characterizing the detector, we assumed an LS light yield of 11,000 photons per MeV, an attenuation length of 20\,m, a PMT coverage of 78\,\%, and a quantum efficiency of 35\,\% \cite{Dayabay}.

\begin{figure}[htb]
\begin{center}
\includegraphics[width=10cm]{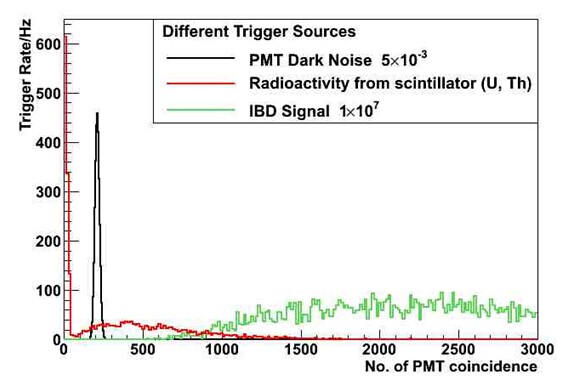}
\caption[Event spectrum for coinciding PMT hits]{\label{fig:trigger coincidence} Event rates as a function of the number of PMT hits coinciding in a 300\,ns trigger window.}
\end{center}
\end{figure}

In the simplest approach, the trigger decision will be based on the hit multiplicity inside a coincidence window of 300\,ns (see above),~i.e. on the number of simultaneously occurring PMT hits. Fig.~\ref{fig:trigger coincidence} shows the event rate as a function of this number of coinciding PMT hits. Based on these spectra, it seems that a simple hit multiplicity trigger will return both sufficiently low trigger rates and high trigger efficiency for IBD events if the trigger threshold is placed at $\sim$500 coincident hits. This coincidence trigger scheme is technically easy to implement and will create only low latencies.

\subsubsection{Trigger setup}

Each FEE will handle 32 PMT signals. The hit flags of ten FEEs inside a crate will be bundled by one Level 1 Trigger Collection Board (L1TCB). In turn the information of ten L1TCB will be fanned to one level-2 board (L2TCB). The Trigger Board (TB) itself will receive data from six L2TCBs. This will be sufficient to read out the trigger flags of 17,000 channels. The L1TCBs and L2TCBs will also work as a fan-out for re-transmitting the trigger decision from the TB to individual FEEs.

\subsection{Clock system}

The clock system for JUNO has two major functions. It will provide:
\begin{itemize}
\item a standard reference frequency to all FEE modules to support the waveform sampling, TDC measurement and process logic;
\item an absolute timestamp for each event to correlate the events of the central and veto detectors and for comparison with other experiments.
\end{itemize}
Based on the experience gained in the Daya Bay experiment and following the latest developments, a distributed clock network based on the "White Rabbit" standard will be applied for JUNO.

The "White Rabbit" (WR) system is a technology originally developed by CERN and GSI for the use at accelerators. It is based on IEEE1588 (PTP) including two further enhancements: The precise knowledge of the link delay and a possibility of clock synchronization over the physical layer with Synchronous Ethernet (SyncE). Applying WR in JUNO will offer the following advantages:
\begin{itemize}
\item The low-jitter recovery clock from SyncE provides a reference frequency for the electronics.

\item Sub-ns clock phase alignment can be achieved among all nodes by phase measurement and phase compensation.

\item The precise time protocol (PTP) synchronizes the timestamp among all nodes.

\item An external Rubidium oscillator and GPS can be used for frequency and UTC reference.
\end{itemize}
The system will synchronize all electronics crates. Each crate will contain a custom-designed clock module to recover the reference frequency and the absolute timestamp. This will then be broadcasted to all FEE and trigger modules in the crate via the dedicated high-speed clock traces on the back-plane for ATCA crates and via front-panel fan-out cables for non-ATCA crates.

\subsection{Power supply and electronics crates}

Conventionally, high voltage for the PMTs would be generated by a dedicated HV supply outside the water pool. However, the preferred solution is the generation of the HV by a Cockroft-Walton type power supply either attached to the PMT base or in underwater units. In this case, these generators have to be fed with low or medium voltage cables from outside the tank. Solutions with voltages of 5\,V or 24\,V are conceivable, the former requiring a larger cable diameter to transfer the larger current without loss. In order to avoid the power loss by using a linear power supply, a DC/DC module should be used for high efficiency. In order to reduce the ripple, the power supply should be independent.
	
\begin{figure}[htb]
\begin{center}
\includegraphics[width=10cm]{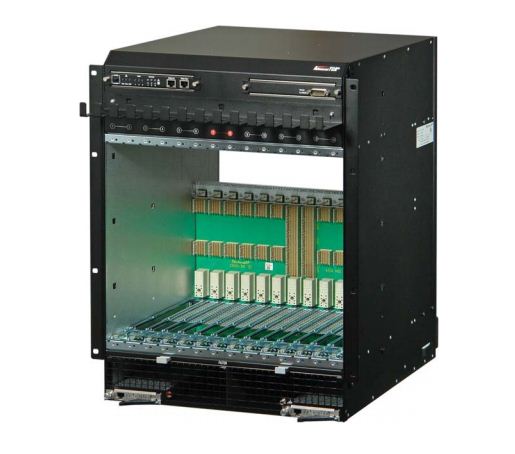}
\caption[ATCA crate]{\label{fig:ATCA Crate}The ATCA crate}
\end{center}
\end{figure}

The FEE and trigger modules will be designed according to the specifications of the ATCA standard. Fig.~\ref{fig:ATCA Crate} shows the ATCA crate that will house the modules. It provides a large bandwidth for digital transmission which allows to pass trigger information between FEE and trigger modules via the bus system of the crate.

\section{Intermediate readout scheme}
\label{sec:roe:int}

\begin{figure}[htb]
\begin{center}
\includegraphics[width=\textwidth]{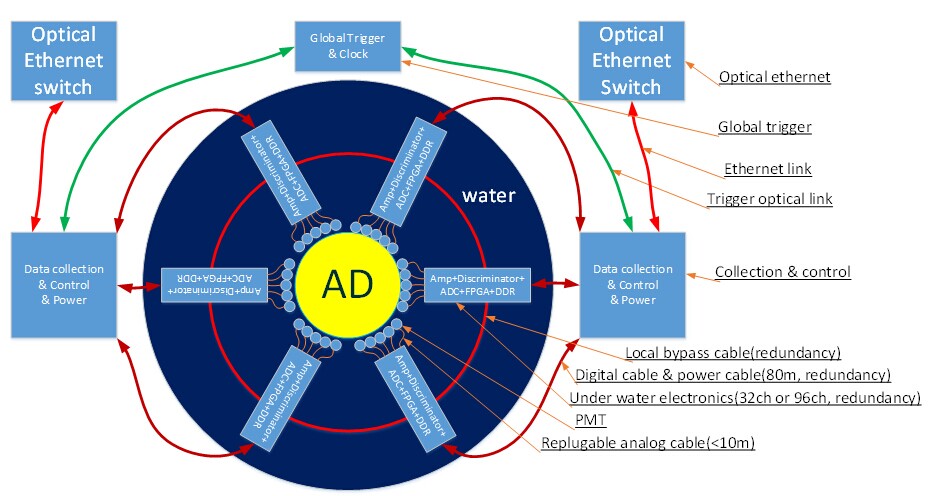}
\caption[Intermediate scheme for readout electronics]{\label{fig:under water scheme}Schematic drawing of the intermediate, partially submerged readout scheme.}
\end{center}
\end{figure}

The primary disadvantage of the external electronics scheme (Sec.~\ref{sec:roe:dry}) is the length of the coaxial cables needed for the transmission of the analog PMT signals. This will lead to a shaping of the leading edge of the signals, deteriorating the time resolution. To avoid this problem, the intermediate scheme shown in Fig.~\ref{fig:under water scheme} places not only the pre-amplifier but most of the FE electronics including digitization in a water-tight box inside the tank, considerably shortening the signal transmission length for the analog signals. Each of these 500 submerged processing units will supply an array of 32 PMTs. From the box, only few cables carrying the digitized signals as well as trigger and clock information will run to the external electronics.

\subsection{Signal processing}

\begin{figure}[htb]
\begin{center}
\includegraphics[width=10cm]{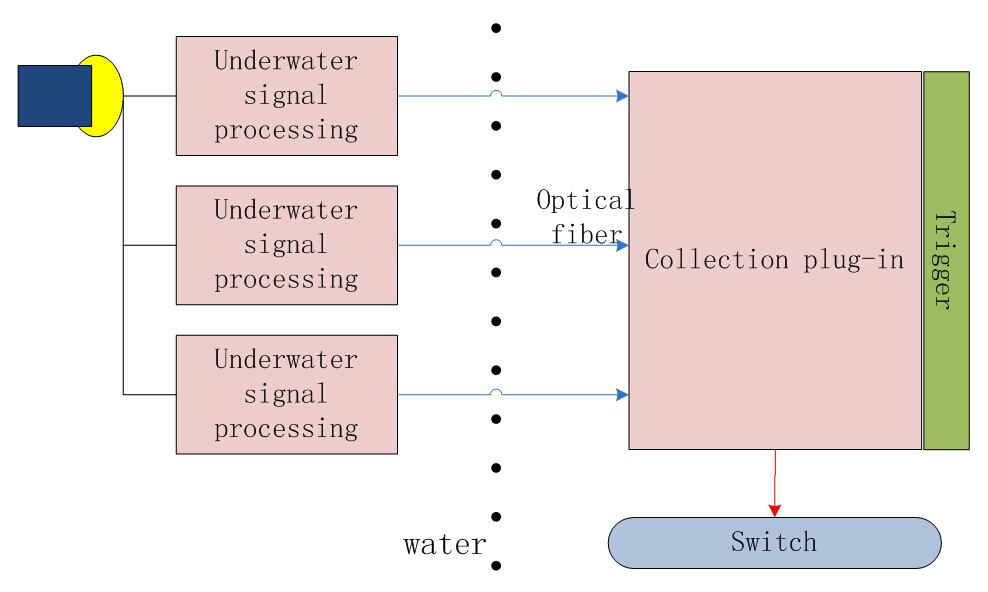}
\caption[Readout Electronics]{\label{fig:DataProcessingUnder} Block Diagram of the data processing and transmission of the under-water scheme}
\end{center}
\end{figure}

In this partially submerged configuration, the analog PMT signal will be transmitted only over a short cable to the amplification stage included in the underwater processing units (cf.~figure\ref{fig:Amplification}). Therefore, the signal shape will not change significantly. Otherwise, the design of the FEE boards inside the processing units will be mostly analogous to the external scheme. The digitized signals will be bundled for all channels connected to one box and passed by an optical fibers to external modules collecting the acquired data. From there, data will be passed on via ethernet. The block diagram is shown in figure\ref{fig:DataProcessingUnder}.

\subsection{Trigger and clock systems}

To simplify the design and maintenance of the trigger system, the trigger modules will be located with the other external electronics. The trigger flags from the FEE discriminators as well as the trigger signals can in principle be transmitted on the same optical fibers as the data. It is possible to transmit the same optical cable  the data. However, separate cables might be used for the trigger system in order to increase the flexibility.

As for the external scheme, a White Rabbit system can be applied for synchronization of the FEEs in the watertight boxes, providing both a reference frequency for FADCs and an absolute timestamp. In case the optical fibers required by White Rabbit are not compatible to oil, a dedicated clock and data recovery (CDR) link could be designed to provide the reference frequency. The physical calibration system could be exploited to determine the absolute timestamps of events relative to calibration pulses fed in from outside for which GPS time would be known.

\subsection{Submerged processing units}

\begin{figure}[htb]
\begin{center}
\includegraphics[width=\textwidth]{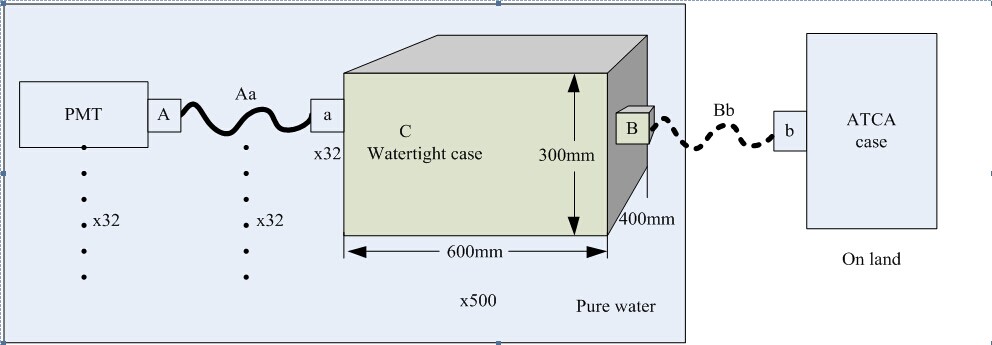}
\caption[Representation of the intermediate scheme]{\label{fig:underwater changeable system}Structural representation of the readout chain in the intermediate scheme.}
\end{center}
\end{figure}

\begin{figure}[htb]
\begin{center}
\includegraphics[width=10cm]{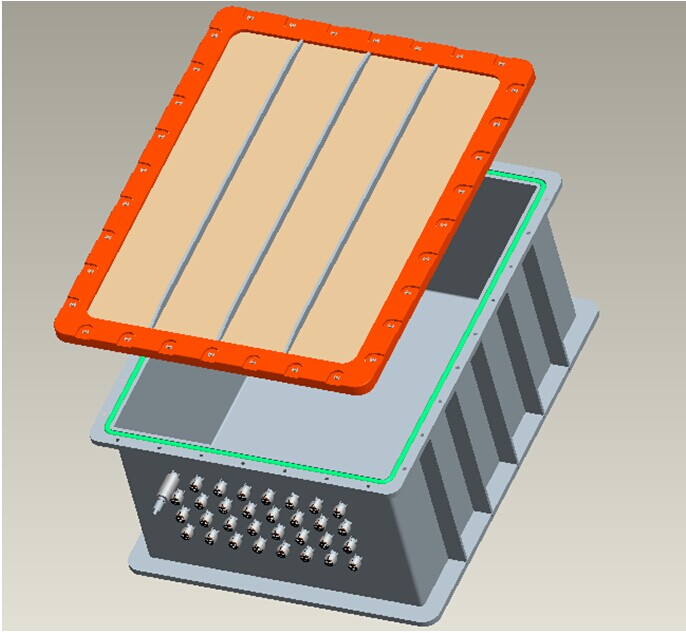}
\caption[Readout Electronics]{\label{fig:watertight case appearance}The diagram of the watertight case appearance}
\end{center}
\end{figure}

\begin{figure}[htb]
\begin{center}
\includegraphics[width=10cm]{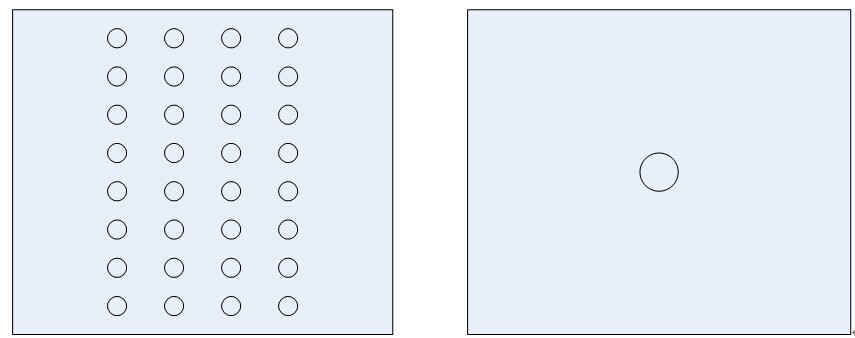}
\caption[Side panels of submerged processing units]{\label{fig:watertight case}Diagrams of the side panels of the submerged processing units}
\end{center}
\end{figure}

An important aspect for the long-term stability of the system is the possibility to replace the part of the electronics submerged in the water, i.e.~the processing units. The structure of a readout strand including one processing unit is shown in Figure\ref{fig:underwater changeable system}. The enclosing box will be 30\,cm high, 40\,cm wide and 60\,cm long. There will be mounting holes and the possibility to remove the top lid. Fig.~\ref{fig:watertight case appearance} shows a three-dimensional representation, Fig.~\ref{fig:watertight case} displays the side panels: The left panel of each watertight case is connected to coaxial-cables by 32 connectors, while the right panel is connected to a composite cable including an optical fibre for signal transmission and a power cable for HV generation inside the underwater box.

The further components shown in Fig.~\ref{fig:underwater changeable system} have to fulfill a number of requirements:
\begin{itemize}
\item Cable $Aa$ should be coaxial, 10$-$20 meters long, and with a characteristic resistance of 50\,$\Omega$.
\item Connector $A$ is watertight.
\item Connector $a$ is watertight and has to be pluggable under water.
\item Cable $Bb$ is a composite cable of 100\,m length containing at least 4 optical fibres and 4 power wires (24-48V, 10-20A) usable for underwater environments.
\item Connector $B$ is watertight and does not have to be pluggable under water.
\item No special requirements are set for connector $b$.
\item Parts $B$, $Bb$ and $b$ can be replaced by non-watertight cables covered by a corrugated pipe.
\end{itemize}

\section{Underwater readout scheme}
\label{sec:roe:wet}

The underwater concept places almost the entire readout electronics inside the water pool, as close as possible to the PMTs themselves. Front-end electronics (FEE) and HV generation are done at the PMT base, all data transfer is done based on digitized data and bundled in central underwater units (CUU). Thus, the underwater scheme minimizes the deteriorating effects of analog signal transmission over long cables and the overall amount of cables needed. On the other hand, it seems unlikely that broken FEE channels can be replaced during operation, while CUUs probably can.

\begin{figure}[htb]
\begin{center}
\includegraphics[width=10cm]{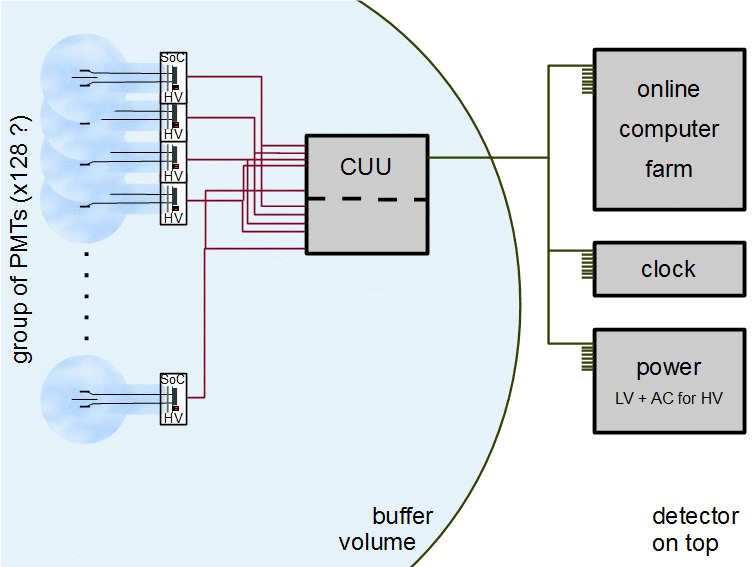}
\caption[Readout Electronics]{\label{fig:Scheme 2}Schematic drawing of the underwater readout scheme.}
\end{center}
\end{figure}

\subsection{Overview}

The layout of the fully-submerged readout scheme is sketched in Fig.~\ref{fig:Scheme 2}. At the end of each PMT there is a small water-tight housing that contains the front-end electronics (FEE) with all the essential electronics of the system. It contains the base of the PMT and a module that generates the HV from a low-voltage AC input. And it contains a PC-board with a system on a chip (SoC) that includes the digitization of the signal, the discriminator to detect photoelectrons, and other functionalities.

The FEE is connected to the online computer farm through data links. It transmits the data to the farm and receives commands from it. It also receives a clock signal from a central clock unit and the power for the PCB and the HV from corresponding power supplies.

In order to reduce the cabling the PMTs are grouped together under water. We estimate that up to 128 PMTs might be grouped together. In this case the full detector would consist of 128 groups. The grouping of the PMTs takes place in central underwater units (CUU) mounted between the backend of the PMTs. To reduce the length of the cables from the PMTs to the CUUs the groups will be formed from nearby PMTs. The cables from the PMTs to the CUUs will have a unique length somewhere between 5 and 10 meters. They contain a by-directional data link, the clock signal, the low voltage power for the PCB (probably 3V), a 24 V AC power for the generation of the HV and a ground potential. On the PMT end the cables will be glued into the housing of the FEE. On the end of the CUU there will be a water-tight connector to allow for the mounting. From the CUU there will be a by-directional data link to the farm (probably two for redundancy) and power cables and the connection to the clock on top of the detector.

\subsubsection{The `intelligent' PMT}

\begin{figure}[htb]
\begin{center}
\includegraphics[width=12cm]{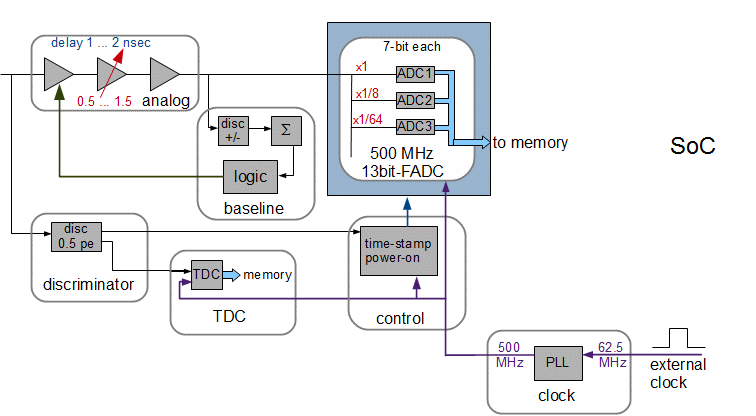}
\caption[Readout Electronics]{\label{fig:SoC}The SoC scheme}
\end{center}
\end{figure}

With the FEE the PMT is turned into an `intelligent' PMT that can be connected directly to a computer. No further electronics is needed. Besides the module that generates the HV and contains also the voltage divider to derive all potentials there is a PCB in the FEE with an ASIC. It contains the SoC with the following functionalities:

\begin{itemize}
\item Amplification of the signal.

\item Shaping of the signal, should it be necessary.

\item Regulation of the baseline to zero.

\item Detection of single photon signals through a leading-edge discriminator.

\item Generation of a time-stamp for each readout window.

\item Digitization of the input signal with an FADC.

\item If necessary, a TDC to measure the precise start of each pulse (relative to the time-stamp of the readout window.

\item Feature extraction, i.e. determination of the number of photons contributing to a pulse and their arrival-time.

\item Self-calibration, i.e.~adjustment of the charge integral of single photons to a predefined value through a regulation of the HV and the amplification of the input stage.

\item Creation of test pulses and injection into the input of the chip and into the PMT.
\end{itemize}
These functionalities are provided for each of the two MCPs of one PMT individually (2 channels). The following additional functionalities are common to both channels:
\begin{itemize}
\item Creation of an optical test pulse through an LED on the PCB that illuminates the photo cathode.

\item Creation of an internal clock signal with the appropriate frequency locked to the input of the external clock.

\item Internal functionality tests.

\item Data exchange with the CUU.

\item Limited data buffer.
\end{itemize}
A sketch of the SoC components is displayed in Fig.~\ref{fig:SoC}. Its functionalities are described in more detail in Sec.~\ref{sec:roe:ipmt}.

\subsubsection{Central Underwater Units}

The CUU functions as the connection between the PMTs of the assorted array to the infrastructure on top of the detector. It receives the data from the PMTs. It combines the data and sends it by a single optical link to the computer farm. It must have a sufficient data buffer to compensate fluctuations in the data rate. At the moment it is envisioned to use a commercial PC-board for the CUU. For the data-links to the PMTs the USB-2 standard might be appropriate. Commercial USB-routers could combine the data from the PMTs into a single data-stream which is then sent to the farm through an optical 10\,GBit/s link. If this works out, no development is necessary for the hardware of the CUU. The CPU power of the CUU could be used for monitoring, pre-sorting of the data, data reduction, or other purposes.

The most critical parts of the CUU are the many connections attached to the CUU in an under-water environment (at least one connector per PMT plus the up-links). The distribution of the power and the clock to the PMTs will also be done inside the housing of the CUU, so that only one combined cable with connector is necessary for each PMT. For the uplink this is less critical. One could use a power cable separate from the fiber(s).

A substantial amount of monitoring will be necessary to control the detector. The CUUs are also the proper place where the data can be injected into the data stream. Sensors for temperatures, voltages, scintillator flows, can be mounted in or near the CUUs and connected to dedicated inputs of the PC. Monitoring data on the PMTs will be collected by the SoC in the FEE and transmitted through the data link to the CUU.

An interesting alternative to the routing of the data as described above is a passive routing of many data streams through one fiber. Such systems transmit each data stream with light of a different color. They are commercially available. They are potentially more reliable than a CUU based on a PC in a water-tight housing. In this case we would need a connector on the PMT to connect its fiber and a separate power cable to connect to a power distribution box under water.

\subsubsection{The Online Computer-Farm}

The online computer farm will receive the data through 128 optical links with 10\,GBit/s capacity. The data will arrive asynchronous. First is has to be sorted into proper time order. Then events are created as time-slices with a minimum activity of the PMTs. These proto-events are subjected to a reconstruction algorithm and from there the decision is derived whether to keep the event or to drop it (or to keep a reduced amount of data). After another sorting in time the accepted events are permanently stored. More details need to be worked out by the DAQ group.

\subsubsection{Expected data rates and buffer sizes}

As PMT pulses are converted early into digitized data, the amount of data to be transferred via fibers and data links as well as local buffers for temporary storage are important characteristics in this scheme. An estimate of the data volume to be processed starts with the amount of digital data caused by the detection of a single photoelectron (SPE). We assume a sampling rate of 500 MHz with one out of three 7-bit FADCs. With a typical rise time of 5\,ns and a fall time not much longer a time window of 32\,ns per spe is quite conservative. With the MCP-PMTs we will always digitize both channels at the same time. The data volume per spe is summarized in the table\ref{tab:Data Volume}:  A single waveform is recorded with 144\,bits. For two channels and with the auxiliary information this adds up to 360 bits per photoelectron. For the potential TDC we have assumed 8\,bits with 125\,ps per unit to cover the full length of the readout window.

\begin{table}[htbp]
\begin{center}
\begin{tabular}{lcl}
\hline
{\bf Data volume of a single waveform}\\
sampling frequency & 500\,MHz  \\
readout window	& 32\,ns \\
number of time-slices per readout & 16 \\
number of bits & 7 \\
range bits	 & 2 \\
number of channels & 2 \\
\hline
Intermediate sum & 288\,bit \\
\hline
{\bf Additional information} \\
address (PMT-ID within group, channel) & 8 \\
time stamp & 40  \\
control flags & 8 \\
TDC (2 x 8 bit)	& 16 \\
\hline
{\bf Total sum} & 360 bit \\
\hline
\end{tabular}
\caption{\label{tab:Data Volume}Data volume needed per single photoelectron.
\label{Data volume per single photoelectron.}}
\end{center}
\end{table}

Based on the spe volume, the data rate to be transmitted between a PMT and the CUU can be derived. It is completely dominated by the dark rate of the PMT consisting of single photons. We assume a conservative 50\,kHz for the dark rate, which corresponds to 18.2\,Mbit/s. Any data volume of physics events is negligible compared to the dark rate, even if the physics events contain much more than single photons. For comparison the maximum data rate of a USB-2 link is 280\,Mbit/s.

In the event of a supernova the data rate will probably exceed the capacity of the link to the CUU. Therefore a buffer is needed that allows a temporary storage of the data on the FEE itself. A nearby ($\sim$3\,kpc) supernova will create 60,000 neutrino events in the detector within $\sim$10 seconds. The event energy ranges up to 50~100\,MeV and will create relatively large pulses in almost all of the PMTs. Therefore, we assume that the average sampling period per neutrino event corresponds to five 32\,ns readout-windows for each PMT. This produces a data volume of $60,000\times5\times360\,{\rm bit} = 13.7$\,MByte for each FEE channel. On top of this there is the data volume from the dark noise which adds up to 22.8\,MByte in 10 seconds. The total sum is 36.4\,MByte. This is the minimum size of the buffer on each FEE.

Further along the readout chain, the uplink from the CUU to the computer farm has to be considered. A maximum of 128 PMTs will be combined into one group read out by a single CUU. For smaller numbers the requirements are less severe. To obtain rates and data volumes, the requirements for of a single PMT have to be multiplied by 128, corresponding to a data rate of 2.3\,Gbit/s for the link. A link capacity of 10\,Gbit/s which is available for standard PCs today will be sufficient. The overall buffer storage needed for a supernova event amounts to 4.7\,GByte. The data can potentially be stored in the RAM of the CUU.

\subsection{The `intelligent' PMT}
\label{sec:roe:ipmt}

This section describes the functionalities of the single elements contributing to the underwater readout scheme.

\subsubsection{High Voltage}

\begin{figure}[htb]
\begin{center}
\includegraphics[width=12cm]{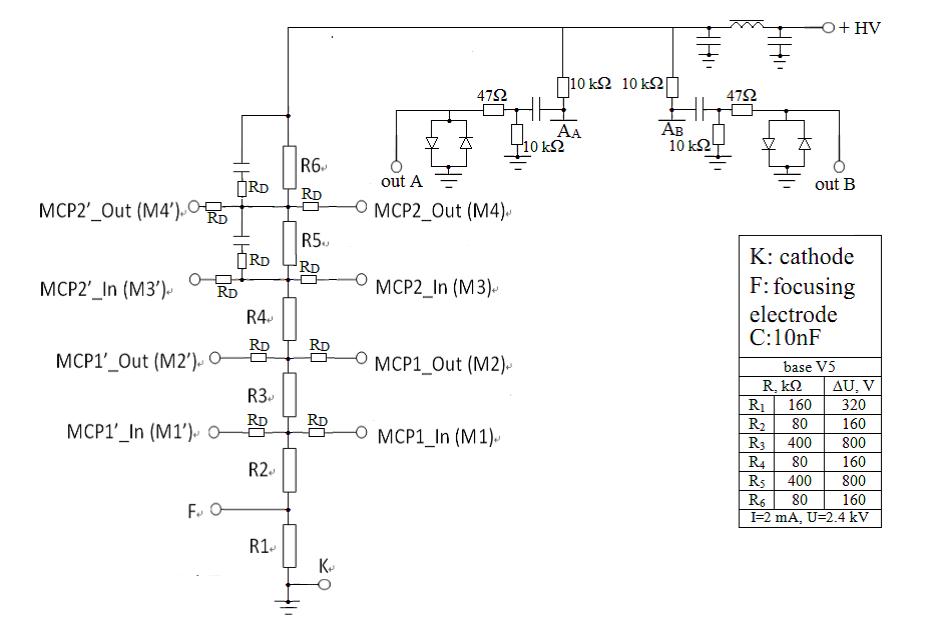}
\caption[Voltage divider for the MCP-PMT base]{\label{fig:MCP base}Circuit for the voltage divider at the MCP-PMT base}
\end{center}
\end{figure}

\begin{figure}[htb]
\begin{center}
\includegraphics[width=10cm]{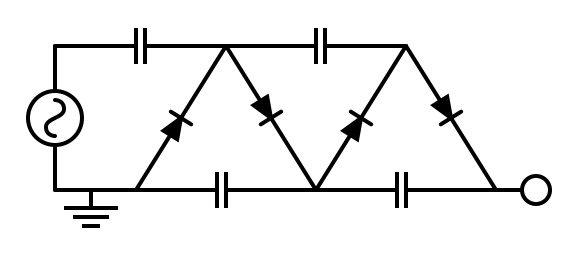}
\caption[Cockroft-Walton multiplier]{\label{fig:Greinacher}The  Cockroft-Walton multiplier for HV generation.}
\end{center}
\end{figure}

The high voltage (HV) will be generated directly on the PMT from a low-voltage input. A single HV potential will be generated for each PMT from which all voltages required for the photo cathode, the field shaping electrodes, and the two MCPs are derived through a voltage divider. The final PMTs will have their cathodes on ground potential. Consequently the output of the signal will be on positive HV. Fig.~\ref{fig:MCP base} shows the corresponding circuit proposed by the Dubna group. The HV will be derived with a Cockcroft-Walton multiplier, also known as a Greinacher cascade, displayed in Fig.~\ref{fig:Greinacher}.

\begin{figure}[htb]
\begin{center}
\includegraphics[width=10cm]{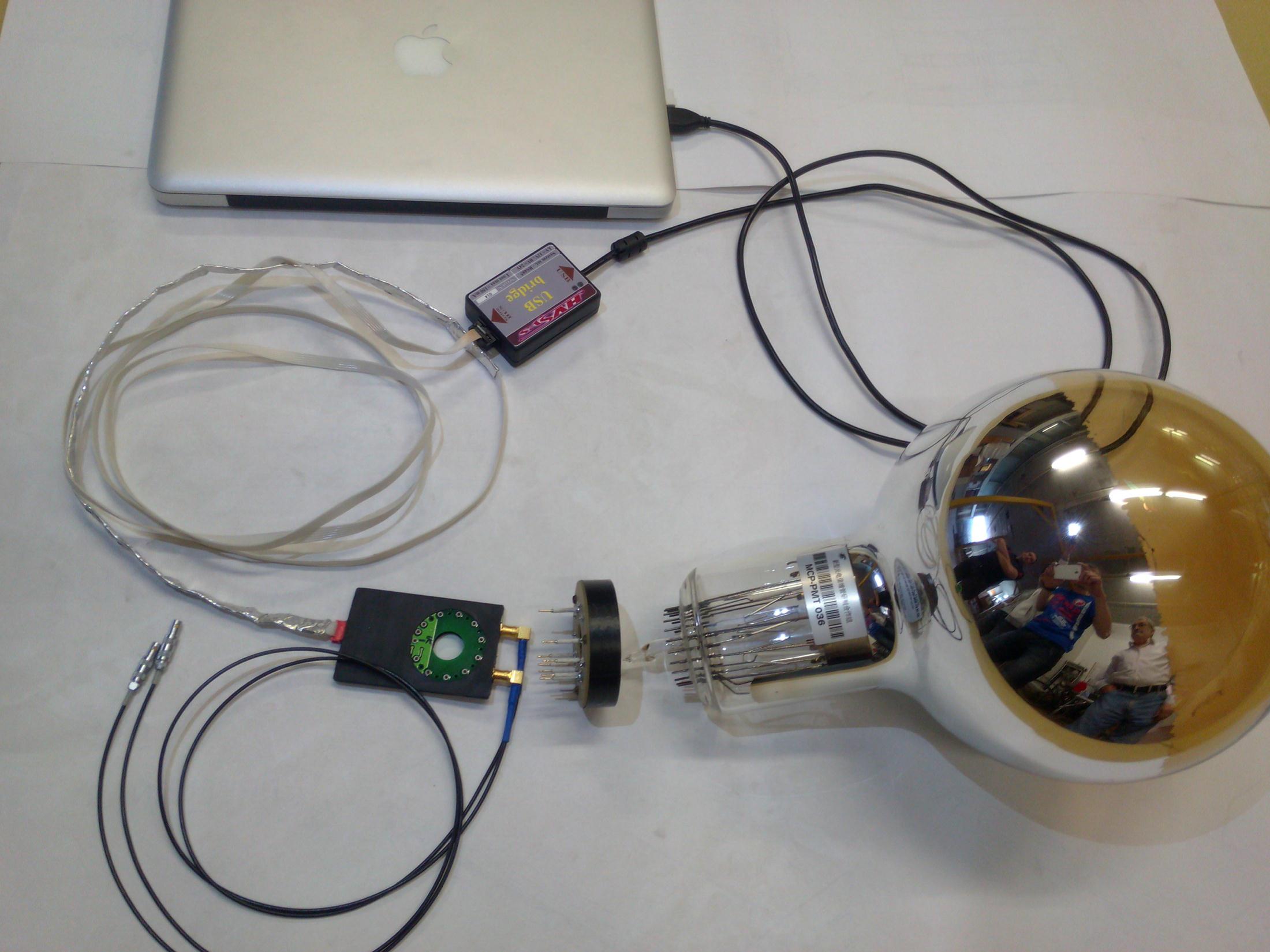}
\caption[Prototype for HV module]{\label{fig:MCP-PMT prototype}Prototype of the HV module for the 8'' MCP-PMT prototype}
\end{center}
\end{figure}

The Dubna group has already developed a working module for the 8'' MCP-PMTs, which is displayed in Fig.~\ref{fig:MCP-PMT prototype}. The input is an AC voltage of 24V. For the final module the input voltage might be higher (up to 200V). The value still needs to be decided. The cascade is embedded in a HV-module that allows regulation of the output voltage. The module already includes the voltage divider and the capacitor necessary to decouple the signal from the HV. It also has the ability to monitor and limit some of the parameters, i.e.~the output current. The current prototype is controlled through a USB interface. The final version will be controlled by the SoC on the PMT.

\subsubsection{Clock System}

It is essential that all PMTs are running with the same frequency. This frequency will be provided by a central clock in the cavern above the detector. The clock signal is distributed to all PMTs. For most physics topics knowledge of the absolute time is not required, with the exception of the signals from a nearby supernova.  A GPS-system that allows absolute timing should be considered for this case. The GPS receiver must be mounted on the surface, probably in one of the entrance areas.

The cables or fibers that connect a PMT to the central clock generate a time delay between the clock in the PMT and the central clock. These are called time-offsets $t_0$. For the reconstruction of events in the detector these offsets must be calibrated on the level of 100\,ps. If the system runs with a global trigger, a rough calibration is necessary already during running (maybe on the level of 1\,ns). For the untriggered system online-calibration is not required. The values of the $t_0$'s can be measured from calibration runs with a light-pulser (LED) in the center of the detector and from the data itself. The electronics must ensure stable $t_0$'s over longer running periods to reduce the calibration runs to a reasonable frequency.

A White Rabbit system has been proposed to calibrate the offsets between the PMTs and the central clock . Such a system seems well suited for a triggered system. For the untriggered system such an effort is not necessary. Offsets ($t_0$'s) from the previous calibration run will be more than sufficient. In the event of a nearby supernova all offsets (including the offset between the GPS receiver and the central clock) must be calibrated in absolute terms (maybe on the level of 1\,ns). The calibration can be done before or after the supernova event. A calibration after the event would save the effort, if no supernova appears, but a plan must be available in any case.

The simplest method to distribute the clock to the PMTs is through individual cables. Distribution of the clock signal could use the same infrastructure as the CUUs, i.e.~a single cable transmitting the signal from the surface to the CUU that than spreads it out to the whole group of PMTs. Alternatively one could add the clock signal to the power cables or distribute it through the data links. In principle a wireless distribution is also possible. The central clock will produce a reference frequency from which the PMTs derive their frequency through a phase lock loop (PLL). Most likely the reference frequency will be lower than the PMTs frequency. To give an example: For a 500 MHz digitization on the PMTs one could distribute a 62.5 MHz reference frequency and divide the clock cycles by a factor 8 on the PMTs.

\subsubsection{Baseline Adjustment}

The output of the PMTs will be on the potential of a positive high voltage. A capacitor will be used to decouple the signal from the HV. This capacitor will be integrated into the HV module. The potential behind the decoupling capacitor will be referenced to ground. For the determination of the charge integral of signal pulses a precise knowledge of this potential (i.e.~the signal baseline) is essential. One possible approach uses the FADC to measure the baseline directly prior to the pulse. The digitization of the input is started already before the arrival of the pulse. It is important not to under-sample the baseline. It must be determined with the same precision as the pulse itself. The baseline must be sampled for the as long as the pulse itself, doubling the data output.

\begin{figure}[htb]
\begin{center}
\includegraphics[width=10cm]{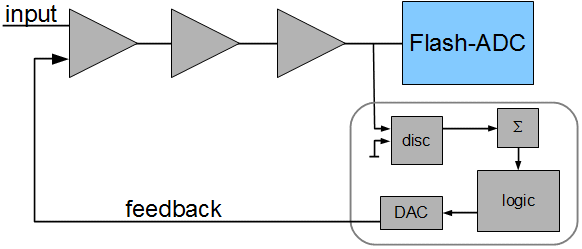}
\caption[Circuit for baseline adjustment]{\label{fig:Baseline adjustment}Circuit for the baseline adjustment}
\end{center}
\end{figure}

Here we propose to use a different approach. We want to measure the baseline continuously and adjust it to ground potential through a feedback loop (Fig.~\ref{fig:Baseline adjustment}). The concept is very simple. A discriminator compares the input against ground potential with the frequency of the clock. Due to noise on the input the input potential will fluctuate between positive and negative potentials. On average, time slices with positive ($+$'s) and negative ($-$'s) voltage bits will be balanced. The discriminator will produce as many $+$'s as $-$'s. If the input potential changes, the balance between $+$'s and $-$'s will be distorted. For example an excess of $+$'s will indicate a deviation of the input potential to positive values. A simple logic will determine the difference of the number of time slices with $+$'s and $-$'s. The result will be used to adjust the potential in one of the input amplifiers until the balance is restored.

Two special situations must be considered to avoid a bias in the baseline:
\begin{itemize}
\item If a true pulse arrives the feedback loop will detect it as a positive deviation of the baseline and counter-correct the deviation. To avoid this bias the feedback loop has to be stopped if an input pulse is detected.
	
\item The return of the signal to the baseline after a pulse will take some time. If a second pulse arrives before the signal has returned to zero, its charge integral will be measured with a too large value. This effect can be corrected offline. All photon pulses will be read out. Time and charge of the previous pulse can be determined independently. An algorithm on the online-farm can either correct for the previous pulse, or store the previous pulse for later correction, or do both.
\end{itemize}	
There is, however, one special case, where this will not work. If the pulseheight of the previous pulse is below detection threshold, the correction cannot be applied. But one should keep in mind, that close pre-pulses will be rare, the undetected fraction of pre-pulses even rarer, and the bias will be very small, especially for pulses below threshold.

A number of parameters of the baseline adjustment need to be optimized. The difference of $+$'s and $-$'s outputs of the discriminator needs to be averaged over a certain amount of time slices. The larger the number, the more precise  the adjustment becomes, but it also gets slower. This parameter should be kept flexible to adapt to the situation in the real experiment. Another important parameter is the step-size with which the baseline is adjusted.

The overall concept of the intelligent PMT will include a random trigger. That is, at adjustable time intervals the electronics will digitize a readout window and store the output. These readout slices should be empty and can be used to check the performance of the baseline adjustment.

\subsubsection{Internal Trigger}

A simple level discriminator will be used to trigger the readout of a single PMT. It will continuously monitor the output of the PMT. The threshold must be programmable. It will be set directly above the noise, as low as acceptable in terms of rate of noise triggers. The threshold can be adjusted individually for each PMT. We envisage a threshold at a level of 0.3 photoelectrons. In the case of the MCP-PMTs there will be separate triggers for each MCP. Most likely the electronics will be configured in such a way that a signal above threshold from one of the MCPs will trigger the readout of both channels.

The readout window will be adjusted to the length of a pulse from a single photoelectron, probably around 32\,ns. In the case of high energy events such as cosmic muons or atmospheric neutrinos several or even many photons will be detected by a PMT. The output signal will extend over more than 32\,ns. If the photons are separated in time, they will be read out as individual triggers. If not, the full signal will be covered by reading several consecutive 32\,ns windows. The trigger logic monitors the threshold discriminators. If at a certain time during the readout (i.e. in the middle of the 32\,ns window) the signal has not returned to below threshold, the logic will continue the readout for another 32\,ns. The readout will be sustained until the signal finally falls below threshold. No information is lost.

\subsubsection{Flash-ADC}

A FADC will digitize the input signal. For the MCP-PMTs two separate FADCs will be needed for the two MCPs. The required dynamic range of amplitude has been estimated to 1000\,pe for a whole 20'' MCP-PMT, or 500\,pe per MCP. The prototypes of the MCP-PMTs show an electronic noise level which is in the order of 10\,\% of the amplitude of a single photo-electron. Therefore a 10\,\% resolution on the amplitude of a single pe will be appropriate. The corresponding resolution on the spe charge integral will be better, on the level of a few percent. If the amplitude of a single pe corresponds to 10 ADC units, a dynamic range of 5,000 to 10,000 ADC units will be needed, corresponding to 13 bits. It is technically not possible to derive 13 bits from a single ADC. Multiple ADCs with staggered ranges are required. We propose to use 3 ADCs with 7 bits each, featuring the following scheme:
\begin{center}
\begin{tabular}{ccc}
\hline
amplitude & range & resolution  \\
\hline
$\times1$	& 0$-$13\,pe & 0.1\,pe \\
$\times1/8$	& 0$-$100\,pe & 0.8\,pe \\
$\times1/64$ & 0$-$800\,pe & 6.4\,pe \\
\hline
\end{tabular}
\end{center}

\subsubsection{Time Measurements}

For the reconstruction of vertices or tracks it is necessary to measure the time of arrival of the photons on the PMTs. A resolution of 1\,ns is required. The arrival time can be extracted from the waveform recorded by the FADC. With the proposed 500-MHz FADC (2\,ns time slices) it is possible to extract the starting time with 1\,ns resolution.

More precise timing information is encoded in the digital pulse of the level discriminator of the internal trigger. The discriminator triggers the first photon of the pulse. This is the most valuable information for the reconstruction. A TDC could be included to measure the time of the trigger relative to the preceding clock pulse. For example to get a 500\,ps resolution a 2-bit TDC would be sufficient.

\subsubsection{Self-Calibration}

\begin{figure}[htb]
\begin{center}
\includegraphics[width=10cm]{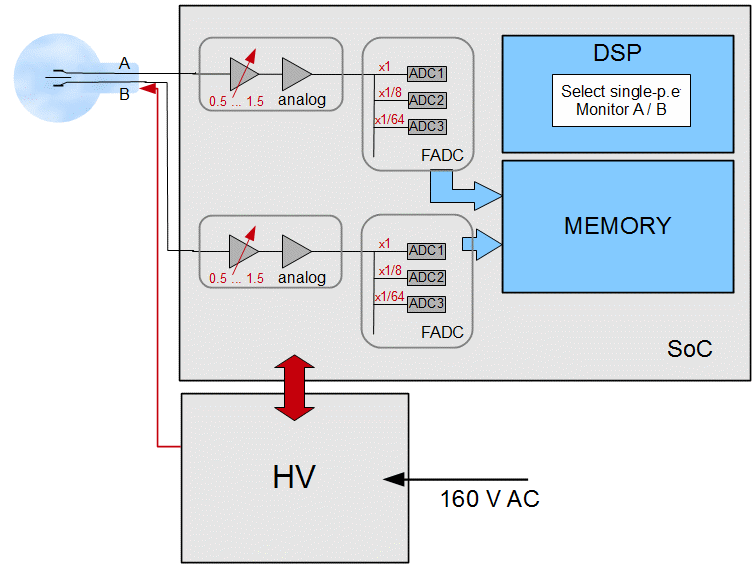}
\caption[Self-calibration scheme]{\label{fig:Self-calibration}Schematic drawing of the self-calibration feedback loop}
\end{center}
\end{figure}

The ASIC will contain a mechanism to calibrate its PMT to a predefined charge integral of a single photoelectron (SPE). The basic scheme is shown in Fig.~\ref{fig:Self-calibration}. Whenever the input signal crosses the trigger threshold for a single photoelectron the flash-ADC will digitize the input and write the data into an internal memory. The logic in the digital part of the chip will analyze the pulses within a single photoelectron window and determine the average charge integral, $\langle q_{\rm pe}\rangle$. The $\langle q_{\rm pe}\rangle$ of each channel is compared to a default charge value, and the HV applied to the PMT adjusted until the predefined value is reached.

The presence of two MCPs with different gains inside one PMT adds a complication because both MCPs are powered from a common high voltage. The algorithm on the chip will compare the $\langle q_{\rm pe}\rangle$ of both MCPs to the predefined value. It will first adjust the gain of the input amplifiers until the $\langle q_{\rm pe}\rangle$ of the two MCPs agree with each other. Then it will adjust the high voltage to match the $\langle q_{\rm pe}\rangle$ to the predefined value.

\subsubsection{Internal test devices}

The circuitry will incorporate extensive possibilities for testing. This is important to guarantee the functionality of the system and to understand how to react in case of problems. The details still need to be worked out. Two test installations that will be integrated for sure are:
\begin{itemize}
\item A test pulser within the SoC: The pulser can be used to inject charge into the input of the chip. It could either be an analogue circuit which produces pulses similar to the real pe's or a digital pulse generator.
\item A light pulser: A LED will be mounted in the electronics in a position where the light illuminates the transmission cathode from the backside (through the glass neck of the PMT).
\end{itemize}

\begin{figure}[htb]
\begin{center}
\includegraphics[width=10cm]{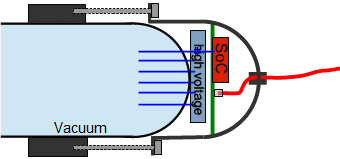}
\caption[Mechanical realization of the intelligent PMT ]{\label{fig:PMT Mechanics}Mechanical realization of the intelligent PMT design.}
\end{center}
\end{figure}

\subsubsection{Mechanics}

The scheme for mounting the FEE directly to the base of the PMT is shown in Fig.~\ref{fig:PMT Mechanics}: The HV module plugs directly onto the pins that connect to the electrodes in the PMT. It includes the wire(s) that carries the output signal. A capacitor is needed to decouple the signal from the high voltage. This capacitor will be included in the high voltage module. On top of the high voltage module there will be a PCB that carries the ASIC with the SoC and some external components (voltage regulator for the low voltage, etc.). The cable to the central underwater units is connected to this PCB. The whole electronics is inside a watertight metal cap that connects to the mechanical fixture of the PMT. The details on how to seal the housing still need to be worked out.

For the default scenario, where data is routed through the central underwater units (CUU) there is a single cable from each PMT to the CUU. It contains power, clock and the data line. It is firmly attached to the electronics housing on the PMT without a connector. The cable length will be between 5 and 10 meters. It has a watertight connector that connects to the CUU. For redundancy there might be two such cables per PMT connecting to two different CUUs.

In case passive data links will be used, an optical fiber will be connected to each PMT. The driver circuit and the laser will be included on the PCB. There must be a watertight optical connector on the fiber where it connects to the PMT. For each PMT there will be a separate cable for the power that connects the PMT to the power distribution box.

\subsubsection{Power}

Each PMT needs a low voltage DC-power for the SoC and a low voltage AC-power for the high voltage module. The exact values have not yet been defined. For the time being we assume a +3V DC-voltage for the SoC and a +24V AC-voltage for the high voltage module. The voltages are generated in power supplies in the cavern on top of the detector. They are accessible and can be replaced in case of failures. From the power supplies, cables will transmit the power to distribution boxes inside the water pool. Each PMT is connected to one of the boxes with a cable and a water tight connector on the power distribution box. The cable contains the DC and the AC voltage, the ground potential and maybe the clock. In case the scheme with active data links is adopted, the power distribution will be integrated into the CUUs. In the case of passive data links there will be distribution boxes just for the power.

One power distribution box or CUU will supply power to a group of 128 PMTs. This system must be protected against failure of the whole group. If a cable to one of the PMTs breaks, a single PMT will be lost which is not catastrophic. However, a short in one PMT could take out the whole group. Therefore a serial resistor RP has to be introduced in the PMT contacting scheme shown in Fig.~\ref{fig:MCP base}. These resistors must be large enough to limit the current through a PMT with a short, at the same time avoiding a too large power consumption in the resistor if the voltage on the PMT changes. For example, if 3\,V is required for the SoC, a 6\,V power supply could be used. We choose a value for RP that creates a 1\,V voltage drop. A voltage regulator located on the PCB at the PMT base will reduce the remaining 5\,V to 3\,V. The protection resistor increases the power consumption by 20\,\%. which will still be acceptable. In case of a short the power of this channel goes up by a factor of 5. This is still negligible compared to the power of the whole group.

Another critical item is the cable from the power supply to the distribution box. It might break or shorten out. We propose to have two identical cables, but only one is connected to the power supply. Inside the distribution box there are switches on the power. Without power the switches are closed. Let's assume cable A is connected to the power supply. If we turn on the power, it will open the switch on cable B and disconnect it completely from the system. If cable A fails, it will be disconnected from the power supply. This closes the switch to cable B. Cable B can now be connected. It will open the switch on cable A and disconnect it from the system, bringing the system is back into operation.

%
%

\section{Development plan}
\label{sec:roe:dev}

\subsection{Characterization of PMT response}

To get a better understanding of the dynamic range and voltage span required, the PMT response and in particular the pulse height and linearity will be studied as a function of the incident number of photo-electrons. For this, a laboratory test stand is currently realized. As the MCP-PMTs are not yet available, characteristics will be first determined for the 20'' PMT produced by Hamamatsu. For covering the whole range specified, waveforms containing 1$-$4000 pe have to be recorded.

\begin{figure}[htb]
\begin{center}
\includegraphics[width=10cm]{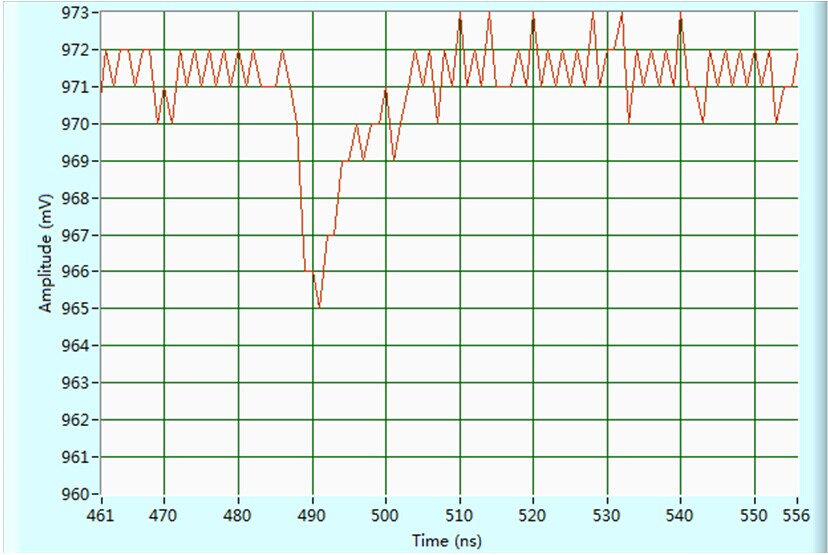}
\caption[Readout Electronics]{\label{fig:waveform for SPE}The waveform for SPE}
\end{center}
\end{figure}

To match this requirement, the tests will be performed following this procedure:
\begin{enumerate}
\item Record {\bf single photoelectron pulses} at a gain level of $~$10$^7$. The right gain is assured by integrating the single photoelectron waveform (cf.~Fig.~\ref{fig:waveform for SPE}), subtracting the baseline pedestal and dividing by the elementary charge. An additional factor of 2 has to be taken into account as a double-end termination is used in the PMT base.

\item {\bf Ramp up of LED intensity} until 4000\,pe are reached, while continuously recording the waveforms for different LED output and thus multiple photoelectron levels. As before, the equivalent charge will be calculated based on waveform integration.
\end{enumerate}
At low hit multiplicities, a discret cascade of broadening peaks will be observed that corresponds to defined numbers of photoelectron. At higher LED intensities, the charge spectrum will become a Gaussian distribution. The number of incident photoelectrons can be calculated based on the gain.

\subsection{Signal amplification}

A careful study of the amplification requirements will be performed for the three readout schemes, including pre-amplification, gain division, and ASIC design. The relevant design specifications are the dynamic range of 1$-$4000\,pe and an equivalent input noise RMS of less than 0.1\,pe. The following aspects should be investigated:
\begin{itemize}
\item {\bf Impedance matching for long cable.} Reduce the reflection between transmission lines and other devices, considering the effect of parasitics and couplings of cables and transmission lines.

\item {\bf Gain division.} The dynamic range is divided into three overlapping FADC acquisition ranges, 1$-$67\,pe, 1$-$529\, and 1$-$4000\,pe. As the original amplification by the PMT dynode chain or MCP will depend on the PMT chosen in the final design, so will the amplification stages.

\item {\bf Amplifier selection.} Commercial amplifiers have to be selected for large dynamic range and low noise. Signal clipping and overload protection will be studied carefully for MCP signal.

\item {\bf Amplifier for FADC driver.} Since the PMT signal amplitude is large, the amplification factor will be small (less than 2), which makes the noise of ADC driver important to the resolution of the total system. Low noise of this amplifier is thus mandatory. Other performance parameters such as settling time have to be studied.

\item  {\bf  Discriminator selection}. A discrimination level of less than 1/4 pe is required.

\item  {\bf  ASIC design.} Amplifiers and discrimators have to be integrated in an ASIC holding 16 or 32 channels.
\end{itemize}

\subsection{Cables and connectors}

\subsubsection{External readout scheme}

The external scheme will employ 20,000 coaxial cables of 80$-$100\, in length and compatible with the underwater environment. Due to the great length, special attention should be given to low signal delay and attenuation to ensure a high quality for the transmission of the analog signals.

A selection of coaxial cables from the have been surveyed Tianjin 609 Cable Co. Ltd.. Samples of 100\,m length have been tested for attenuation. Results are listed in Table\ref{tab:Coaxial Cable}. While the Type III cable clearly shows the best performance in terms of signal dampening, it is also the most expensive.

\begin{table}[htbp]
\begin{center}
\begin{tabular}{lccc}
\hline
Parameter & Type I & Type II & Type III \\
\hline
Product Model & SYVF -50-3-1 & 609C5021A & 609C5019A \\
Outer Diameter & 10.5\,mm & 3.5\,mm & 4.5\,mm \\
outer sleeve & ordinary sleeve & FEP & FEP \\
Attenuation @ 100\,MHz  & 17.7\,dB & 15.9\,dB & 12.3\,dB \\
Attenuation @ 200\,MHz &  24.8\,dB & 23.4\,dB & 17.7\,dB \\
Price (RMB/m) & 10 & 14 & 18 \\
\hline
\end{tabular}
\caption{\label{tab:Coaxial Cable} Survey and attenuation results for three samples of 100\,m long coaxial cables from Tianjin 609 Cable co.}
\end{center}
\end{table}

\begin{figure}[htb]
\begin{center}
\includegraphics[width=0.9\textwidth]{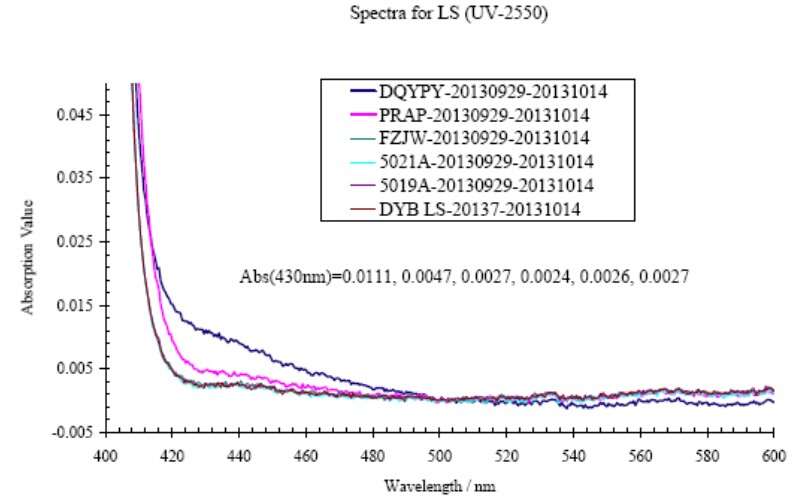}
\caption[Compatibility tests of cables and LS]{\label{fig:compatibility of the sleeves and LS}Effect of the exposure of LS to the cable sleeves on the LS attenuation spectrum.}
\end{center}
\end{figure}

The sleeves have also been tested for compatibility with the LS. The effects on the attenuation spectra of LS samples that have been left for some time in contact with the cables are shown in Fig.~\ref{fig:compatibility of the sleeves and LS}. The resulting changes for cables of Type II and III are less severe than for Type I.

\subsubsection{Intermediate readout scheme}

\begin{figure}[htb]
\begin{center}
\includegraphics[width=10cm]{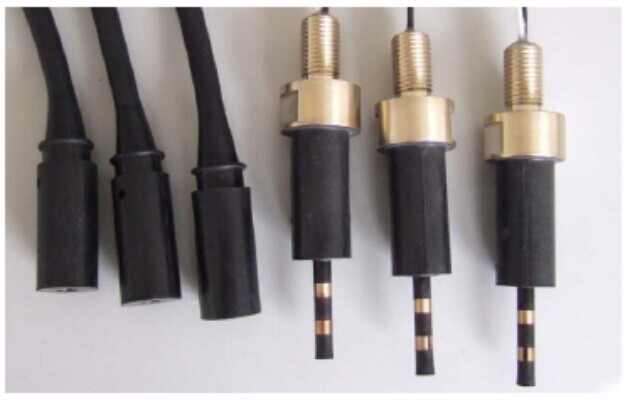}
\caption[Underwater connectors]{\label{fig:pluggable connector}Under-water connectors from Shanghai Rock co.}
\end{center}
\end{figure}

The requirements for the cabling of the intermediate readout scheme are somewhat different. As the FEE are moved into the submerged processing units, watertight boxes as well as underwater connectors are needed. In detail, this scheme needs:
\begin{itemize}
\item About 20,000 coaxial cables of 10$-$20\,m length for the signal transmission from PMT to the processing units. Cables should be covered by a sleeve from fluoroplastics.

\item About 20,000 connectors that can be plugged under water. Fig.~\ref{fig:pluggable connector} displays suitable products by the Shanghai Rock-firm Co.~Ltd.

\item About 500 watertight cases for the central processing units

\item About 500 composite cables for data transmission via optical fibers and low-voltage power supply
\end{itemize}

The feasibility of this scheme is currently evaluated at IHEP. Contacts have been established to a variety of providers for watertight connectors and cases, including Beijing the Great Wall Electronic Equipment Co., Ltd. China aviation optical-electrical technology CO., Ltd. Shanghai Rock-firm CO., Ltd. No.23 Research Institute of China Electronics Technology Group Corporation and others. While watertight cables can in any case be provided, the sleeve material and the fluorination treatment of the rubbers are still under investigation. This is an important aspect as fluorination will protect the rubber against the corrosive strength of pure water, allowing for a life time of 10$-$20 years. Further important questions are the impermeability of the watertight cases, the characteristics of the signal transmission and last but not least the additional cost introduced by this scheme.

\subsection{ADC selection and circuit design}
A further important step is the selection of the FADC chip. Several FADC chips have been included in a preliminary survey. Tab.~\ref{tab:ADC Select} summarizes the main characteristics for a variety of commercial and self-developed chips. Circuit designs are currently studied to evaluate the FADC performance and its compatibility.

\begin{table}[hbtp]
\begin{tabular}{l|lcccc}
\hline
 & & Sampling & No.~of & Resolution & \\
Type & Model & rate [GS/s]  & channels & bits & ENOB \\
\hline
FADC	& ADC10D100D & 2/1 & 2 & 10 & 9.1 \\
     & ADC12D1000 & 2/1 & 2 & 12 & 9.6 \\
	 & EV10AQ190A & 1.25 & 4 & 10 & 8.6 \\
	 & ADS5409 & 0.9 & 2 & 12 & 9.8 \\
	 & ADS5407 & 0.5 & 2 & 12 & 10.3 \\
	 & Stefan's (ASIC) & 1 & 1 & N/A & N/A\\
	 & Fule's (ASIC) & 1 & 1 & 12 & >10\\
\hline
Sampling  & DRS4 + AD9252 & 1-5 & 8 & 11 & 8(?) \\
+ ADC	 & Weiw's (ASIC) & 1 & 1 & N/A & N/A\\
\hline

\end{tabular}
\caption[Overview of FADC chips]{\label{tab:ADC Select}Overview of selected (Flash-)ADC chips}
\end{table}

\subsection{Design of waveform digitization ASIC }

To fulfill the requirements of JUNO concerning timing and pulse height resolution, the acquisition and digitization of PMT waveforms at a sampling rate of 1\,GS/s and an acquisition range of 12\,bit is mandatory. However, commercial high-speed FADCs meeting these specifications are currently banned from import to mainland China. While possibilities exist to obtain suitable chips by transfer via a Sino-American collaboration, the resulting price will be undoubtedly high. In order to reduce the cost per electronics channel, design efforts for implementing a new form of ADC on an ASIC or for a self-made high-speed FADC have been started.

\subsubsection{Design of an ASIC for high-speed waveform sampling}

The bottleneck of the design of the high speed-waveform sampler is the huge data throughput. It is very difficult to store a huge data volume within a very short time as it would be encountered during a supernova. However, as long as the event rate is limited, continuous storage will not be necessary as most of the acquired date is either baseline or background signals.

The waveform-sampling chip will rely on the online selection and caching of the analog data. The chip will first sample the waveform at high speed, and save the resulting waveform information in an analog memory. When there are no physics triggers, the sampling process will keep running, thus refreshing the cached analog data and overwriting the old one; when a trigger is issued, the waveform stored in the analog cache will be digitized immediately. The post-processing logic therefore only has to deal with waveforms of the length of the analog buffer. This method allows for both high-speed sampling and caching. Moreover, the conventional approach of simultaneous analog-to-digital conversion and readout are thus separated. The analog memory first acquires the pulse, then the ADC is converting the stored waveform and sends only a short piece of waveform to the backend system for storage. Therefore, the data processing load for the backend systems is greatly reduced.

\begin{figure}[htb]
\begin{center}
\includegraphics[width=10cm]{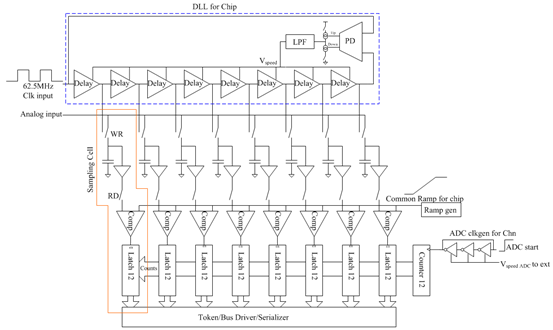}
\caption[Readout Electronics]{\label{fig:Waveform ASIC}The architecture of one channel of the waveform sampling ASIC}
\end{center}
\end{figure}

The architecture of the waveform sampling chip is given in Fig.~\ref{fig:Waveform ASIC}. The high speed sampling and analog memory is realized by a large switched-capacitor array (SCA). When no trigger comes, the SCA will keep sampling in a cycling style, and the voltage values held on the capacitors are consequently refreshed; when a trigger is issued, the held voltage will be compared with the common ramp signal of the chip by an exclusive comparator for each capacitor. Meanwhile, a counter will monitor the time duration from the start of the common ramp signal until stop signals are issued by the comparators. The stored analog information is then converted to digital output in a single step. The individual start-to-stop time counts will be stored in the corresponding latches. When the the ramp signal ends, the latched data will be transmitted by the serializer.

In order to be able to cope with the expected dark rate per channel of 50\,kHZ, each ASIC channel will include 256 storage cells.

\subsubsection{Design of a high-speed FADC}

A further solution is a self-designed high-speed FADC. The idea is based on a more conventional architecture, for which a high-speed ADC will take charge of sampling the waveform and digitization in a single step. All the converted digital data will then be transmitted to the backend system for data buffering, trigger selection processes and so on. Due to the import prohibitions, a self-designed FADC put into mass production will greatly decrease the cost of the system.

The ADC will be based on a hybrid architecture combining a pipeline A/D convertor and flash type. The signal will be sampled through an input buffer and a sample-and-hold circuit, and then be pre-converted by a four-stage pipeline conversion. A flash conversion stage will finally digitize the signal. In this way, the stress of high sampling speed and high voltage resolution are distributed to multiple stages. Also power dissipation per area will be reduced. The chip is expected to achieve asampling rate of 1\,GS/s and a resolution of 12\,bit, with a guaranteed effective number of 10\,bits.

\subsection{Data processing and transmission}

\subsubsection{External readout scheme: ATCA modules}

\begin{figure}[htb]
\begin{center}
\includegraphics[width=0.8\textwidth]{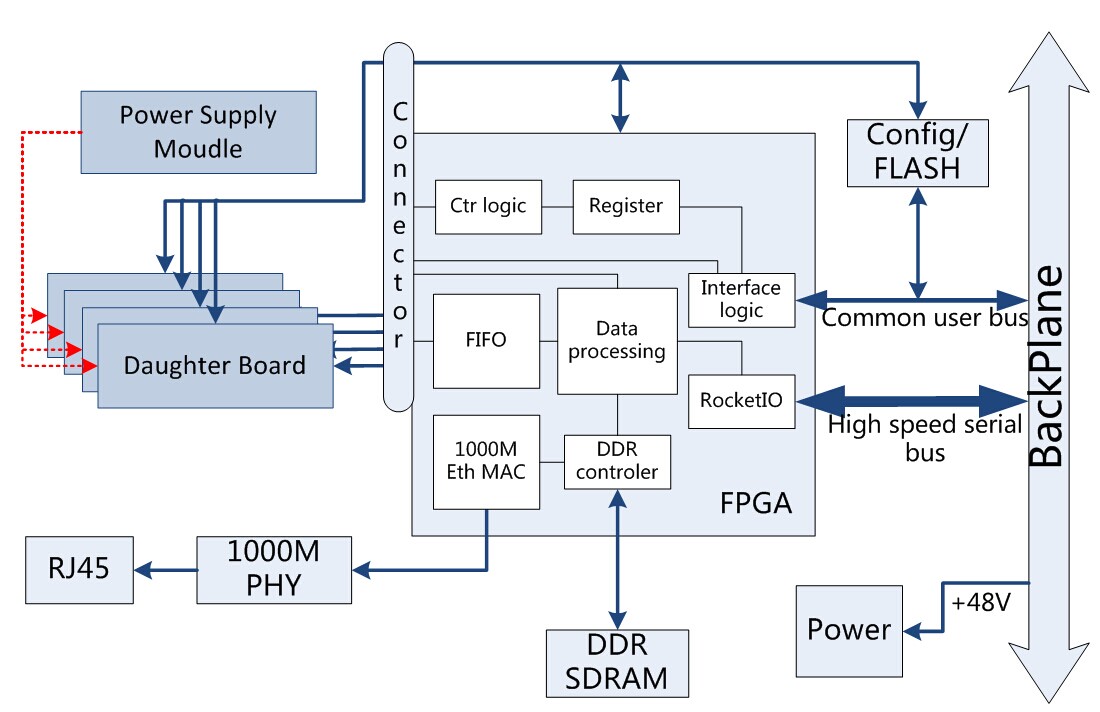}
\caption[FEE block diagram for ATCA modules]{\label{fig:ATCA Structure}Standard ATCA module structure block diagram}
\end{center}
\end{figure}

The FEE modules used in this electronics scheme will be based on the ATCA standard. The structure block diagram is shown in figure\ref{fig:ATCA Structure}: The modules will be based on a daughter-mother board structure. The function of the daughter board is to receive the PMT signals and to perform the digitization. The mother board is a standard ATCA module. As the main unit of data transmission and processing, it will aggregate the data from all daughter boards, transfer the clock and trigger signals via the high-speed backplane bus, transfer the final data to the DAQ via a Gigabit ethernet connection and take care of system control and configuration. The FPGA of the module will perform the digital signal processing and information extraction.

\subsubsection{Intermediate readout scheme}

\begin{figure}[htb]
\begin{center}
\includegraphics[width=10cm]{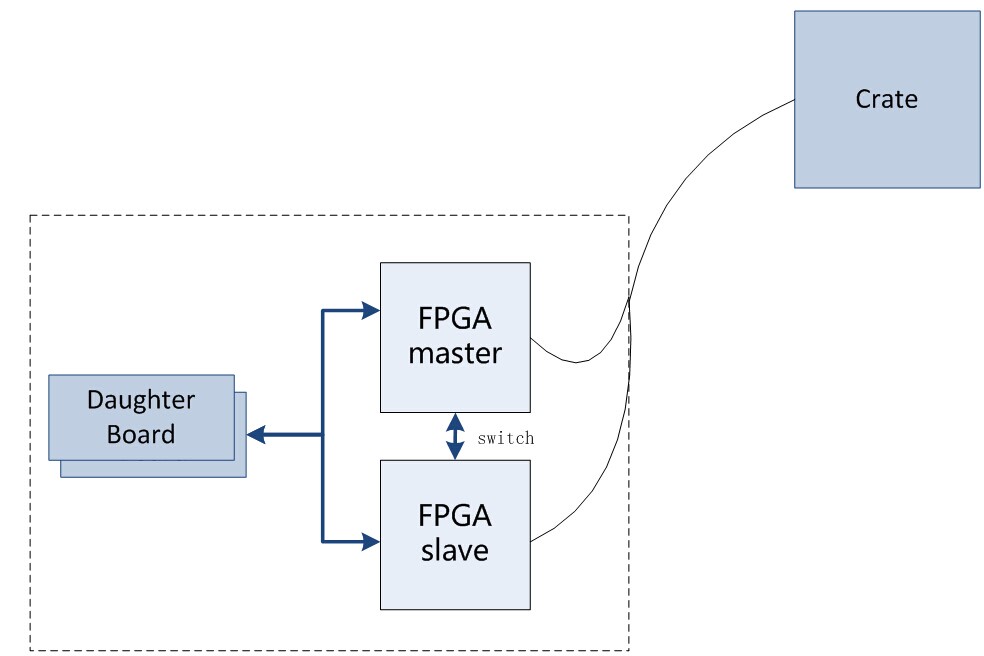}
\caption[Block diagram of processing unit]{\label{fig:UnderWaterSignalProcess} Diagram of the submerged processing unit}
\end{center}
\end{figure}

{\bf Submerged processing unit.} In the partially submerged scheme, the signal processing unit will still be based on a daughter-mother board structure. The FEE daughter board will be mostly identical to the external case. The mother board will package the data and transfer it to the external electronics via an optical fiber. For reliable data transmission, some redudancy will be introduced for the mother board:two FPGAs will serve as backup for each other. When one of the FPGAs is not working properly, the system will switch to the other one. A corresponding diagram is shown in Fig.~\ref{fig:UnderWaterSignalProcess}.
\medskip\\

\noindent {\bf Signal collection unit.} The data sent by the underwater processing units via optical fibers will be received by a collecting ACTA module based on a daughter-mother board structure. The motherboard is an ATCA standard module, while the daughter board completes some simple function like the collection of the optical data and the transmission to the mother board.

\subsection{Design of trigger and clock}
 The White Rabbit technology has been used extensively in the LHAASO experiments. A compact FMC-form White Rabbit node (cute-wr) has been developed which can be easily integrated as a network mezzanine card on other electronics system. It will provide a uniform 125\, MHz clock signal, a Pulse-Per-Second signal and an encoded UTC timestamp signal. First tests with a prototype system show a 100\,ps accuracy and 21\,ps precision among several cute-wr cards.

\begin{figure}[htb]
\begin{center}
\includegraphics[width=\textwidth]{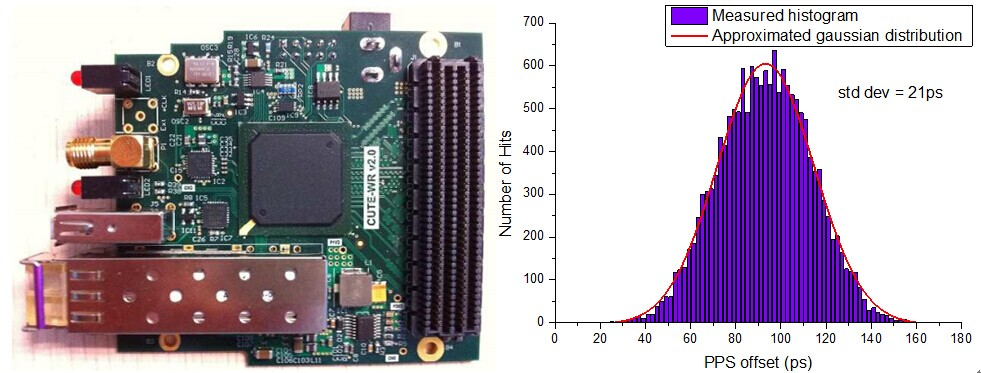}
\caption[White Rabbit system]{\label{fig:WhiteRabbit}Compact FMC-form White Rabbit nodes and the measurement of PPS synchronization}
\end{center}
\end{figure}

\subsection{Design for system reliability}

\subsubsection {Specifications for reliability}

The reliability requirement for the electronic system is a failure rate of less than 0.1\,\% per year of operational time for each of the 16,500 detector channels. The design life time of the system is 20 years. At the desired failure rate, it should be able to run for at least 10 years without need of maintenance action.

In the intermediate and underwater schemes, most of the electronics will be put into sealed boxes at a maximum water depth of 40\,m. The temperature of the water will be in the range of +10 to +20 Centigrade. The ambient radiation levels will be normal.

\subsubsection {Reliability engineering}

As an important aspect of product assurance management, reliability engineering will cover the entire electronics design and development for JUNO. This will include:

\begin{itemize}

\item {\bf Management of components.} The used parts will extremely affect the quality and reliability of the final product. There will be a management plan for the components, and a list of preferred components to provide guidance in their selection and quality control.

\item {\bf Reliability design,} including reliability definition, system reliability analysis, reliability allocation, reliability design and prediction

\item {\bf Analysis of failure modes and effects}, including a fault tree analysis.

\item Analysis of {\bf sneaks}

\item Analysis of  {\bf ASIC reliability}

\item Assurance of {\bf firmware reliability}.
\end{itemize}

For achievement of the high quality and reliability requirements for the electronics, all steps of production and installation of the JUNO electronics (except the reliability design in R\&D procedure) should be controlled by the overall Product Assurance (PA) management plan. This will include
\begin{itemize}
\item Risk management
\item Safety management
\item Control of critical items
\item Procurement, verification, screening, handling and storage of components
\item Control of techniques and production procedure
\item Verification and Environmental tests.
\item Software assurance, etc.
\end{itemize}
There will be a PA organization supervised by PA manager to establish and implement the management plan.

\section{Manufacture and assembly}
\label{sec:roe:rea}

\subsection{Production}
Production of $\sim$20,000 channels will be a demanding task that must be subject to a strict quality management and continuous supervision. The final design will be chosen after a phase of elaboration and prototyping of the three design concepts that have been introduced above and a careful evaluation of their respective merits and disadvantages. Once the design is fixed, production will start in several steps:
\begin{itemize}
\item {\bf Ordering of electronic components}, including cables and crates. Each supplier and batch will be strictly controlled to ensure the reliability.

\item {\bf PCB manufactory and assembly} will be carried out by professional manufacturers. A close personal contact with the companies will be established to allow for quality supervision and sampling at the production site.

\item {\bf Aging tests} with the components that should be performed well in advance of the start of JUNO data taking. Components failing the tests can be repaired and re-introduced after the performance has been tested.
\end{itemize}

\subsection{Aging and long-term tests}
According to typical reliability curves for electronic devices, the failure rate will be relatively high in the beginning while forming a plateau of constant low failure rate after. Therefore, all critical components should be installed and running for some time before the experiment starts, allowing to pick out and repair malfunctioning components.

Aging tests will be performed by exposing electronic parts for a period of the order of one week to an environment of increased temperature. If 1000 channels can be tested simultaneously, the whole process will require about 20 weeks. In order to optimize the conditions of the aging tests, pre-studies risking the destruction of some of the components should be conducted.

Before installation, the system needs to undergo a long-term test run at a dedicated test stand at a laboratory that allows for debugging, identification of failing components and further aging control. This thorough test of at least part of the system should last at least for 3$-$4 months, and should be performed with real PMTs connected.

\subsection{Installation}

The readout system will be installed along with the central detector and PMTs. The foreseen time span for installation is 6$-8$ months. It will start from the bottom and end at the top cap. The external testing stand should be completed before installation to allow to perform the tests during the installation phase. After completing the installation, a debugging phase to test the stability of the whole system is foreseen to ensure optimal working conditions at the start of the experiment.

\subsection{Reliability}

All steps require rigorous quality and reliability management and control, from ordering the components to the finalization of the installation. This will ensure the system stability, reduce losses in the production process, and minimize the consumption.

\section{Risk analysis}
\label{sec:roe:risk}

\subsection{Three read-out schemes}
Some risk obviously exists for such a 20,000-channel high-speed sampling system, in which all of the PMTs are located inside the water pool and therefore not immediately accessible to repairs. The "external readout scheme" (Sec.~\ref{sec:roe:dry}) will place most of the electronics outside the water pool. This will greatly simplify the replacement of broken modules but potentially reduces the measurement quality and increases the costs. In the "underwater readout scheme" (Sec.~\ref{sec:roe:wet}), almost all the critical components of the electronics will be placed underwater and are possible without reach for replacement (especially the FEE). This assures optimum measurement performance, but will increase the risk of failure unless special attention is given to an increased redundancy of vital system components. In case of the "intermediate readout scheme" (Sec.~\ref{sec:roe:int}), a large part fraction of the electronics is located inside the water. All vital electronics are concentrated inside the central processing units that are designed to be replaceable. Compared to the other schemes, both performance and repairability are optimized but costs and design difficulty are increased. The final design will be chosen after all three designs have been properly elaborated and prototypes have been tested.

\subsection{Design risks}

Design risks comprise mainly four aspects:
\begin{itemize}
\item Quality of signal measurement
\item Reliability
\item Cost control
\item Design integrity
\end{itemize}
The biggest risk is the signal measurements, which followed by the risk of reliability and design integrity. The reliability is associated with cost control. Risks in design integrity will be mostly caused by two aspects: Insufficient communication of the groups designing different parts of the system and a lack of understanding of the system itself. These risks can be mitigated by a good collaboration in between the participating groups.

\subsection{Production risks}
Production risks are mainly arising in the procurement of electronic components, PCB manufactory and assembly, aging tests and cost control. A very strict quality control system will be established.

\subsection {Installation risks}

Risk in installation will also be minimized by a strict management and quality control,  including three aspects: installation quality, time and safety.

\section{Schedule}

\begin{itemize}
\item[2013] Start of design of the readout schemes.

\item[2014] R\&D for key components.

\item[2015] Test of a prototype system of 200 channels for the "external readout scheme" at Daya Bay.

Finalization of a 32-channel demonstrator for the "intermediate readout scheme".

Review of all three schemes and choice of the final scheme.

\item[2016] Prototype system for the final scheme, about 200 channels

\item[2017] Mass production

\item[2018] Mass production, testing ans start of installation

\item[2019] Installation and testing
\end{itemize}

\cleardoublepage

\chapter{DAQ and DCS}
\label{ch:DAQAndDCS}


The data acquisition (DAQ) system and detector control system (DCS) are introduced separately as followed.

\section{Data Acquisition System}

The main task of the data acquisition (DAQ) system is to record antineutrino candidate events observed in the antineutrino detectors. In order to understand the backgrounds, other types of events are also recorded, such as cosmic muon events, low energy radioactive backgrounds etc. Therefore, the DAQ must record data from the electronics of antineutrino and muon detectors with precise timing and charge information. DAQ should build event with separated data fragment from all electronics devices. Then DAQ needs to analyze and process data to compress data and monitor data quality. At the last step, DAQ should save most relevant data to disk.

\subsection{System Requirements}
This section presents the DAQ main design requirements.

\subsubsection{Event Rate}
The central detector (CD) of JUNO is composed of about $\sim$17,000~PMTs. According to MC simulation, order of tens antineutrino candidates per day can be acquired by the CD. Most events of the CD come from radioactive backgrounds of PMT glasses, steel and liquid scintillator material. The trigger system could reduce PMT dark noise and radioactive backgrounds event rate to order of 1~kHz. DAQ system will be designed to handle a maximum trigger rate of 1~kHz.

The water shielding detector is composed of about 1500~PMTs and adopt the same electronics design. But the trigger rate could be less than 100~Hz with radioactive backgrounds and muon events. The top muon detector is under design and the bandwidth requirements should be small and ignored.

The calibration system could bring several hundreds to 1~kHz additional event rate. But the trigger rate could be set to a reasonable level. A supernova explosion will generate many neutrino events over a short time scale. For example, about $\sim$10k events will be generated in $\sim$10~seconds from a 10~kpc distance supernova explosion.

\subsubsection{Readout Data Rate Estimation}
The electronics system plans to use 12 bits FADC at 1 GHz sample rate with 2 ranges to acquire signal waveform in maximum 1~$\mu$s time window. In case of 16 bits occupation of one sample, the data size of single PMT signal could be estimated about 2k bytes including bytes of time stamp and header.  Table \ref{DAQRate} shows DAQ readout data estimation, and the total data rate of CD is about $\sim$2.4~GBytes/s which means data rate per PMT is $\sim$140~kBytes/s in Physics mode. The high energy cosmic muon events of CD are estimated to be about 3 Hz and will fire all $\sim$17,000~PMTs of the CD. Neutrino and background low energy events are estimated to be less than 1 kHz with less than 3,000 and 1,000 fired PMTs in average. Water shielding detector is estimated to be with $\sim$100~Hz event rate and $\sim$10\% channels fired per event.

We can set up both high and low energy calibration with similar data rate. The data rate of calibration run modes is 4.6 GB/s and means 273 KByte/s per PMT, which is about 2 times of the data rate of normal physics run mode. According to the table, 3,000~PMTs are assumed to be fired by one supernova event. The additional supernova event data rate is about 3 times larger than that of normal physics running modes. So DAQ was required 11.6 GB/s of total readout bandwidth at least and means 675~KB/s per PMT as 5 times physics mode when supernova happens with calibration together. But it will make remarkable increase for closed supernova explosion.

Supernova events will only continue about 10~seconds, LED based calibration run can be paused to improve the performance. So DAQ only needs to keep about 4.6 GB/s data processing performance for calibration run mode and buffer additional 60 GB supernova data in memory to process latter.

Due to the signal width is very small than 1~$\mu$s data window, we can assume to compress waveform to about 50~ns effective width with about 100~Bytes per signal. As estimated in the table about 95\% low energy event data will be removed. Then only about 200~MB/s data in Physics run mode and 300~MB/s in calibration run mode need to be built to full events and save to disk. The maximum data storage requirements is 600`MB/s when supernova happens during calibration run.

%
%
%
\begin{table}[!hbp]
\caption{JUNO DAQ Data Rate Estimation}
\label{DAQRate}
\newcommand{\tabincell}[2]{\begin{tabular}{@{}#1@{}}#2\end{tabular}}
\begin{tabular}{p{3.5cm}|p{1.5cm}|p{2cm}|p{2cm}|p{2cm}|p{2cm}}
\hline
	&	Channels &		Data size per channel(Bytes) & 	Fired nPMT & 	Trigger rate(Hz) &	Data Rate (MBytes/s) \\
\hline
\tabincell{c}{CD\\(Low energy events)} & 17,000 	&	 100(50ns)/ 2,000(1us) & 1,000 &	1k &	100/2,300	\\
\hline
\tabincell{c}{CD\\(High energy events)}  &	17,000	   &	 2,000(1us) 	     & 17,000 & 	3  &	96	\\
\hline
Water Shelf Detector				&		1,500	& 100(50ns)/ 2,000(1us) & 150 & 	100 & 1.5/30 \\
\hline
\tabincell{c}{Sum\\(Physics mode)} 	&		&				&	 &	  	&	197.5/2426	\\
\hline
\hline
\tabincell{c}{CD\\(Calibration events)} &	17,000 & 100(50ns)/ 2,000(1us) & 1,000-17,000 & 1k &100/2,300\\
\hline
\tabincell{c}{CD\\(Supernova events)}  &	17,000	 &100(50ns)/ 2,000(1us)	 & 3000	    & 1k   & 300/6,900\\
\hline
\tabincell{c}{Sum\\(Maximum)} 	&		 &				&	 &		& 597.5/11,626\\
\hline
\end{tabular}
\end{table}

\subsubsection{Data Process Requirements}
There are several types of read out data links between DAQ and electronic. Ethernet with TCP/IP protocol could be a simple option. DAQ could organize readout branch with groups of PMTs. To deal with tens GB/s readout bandwidth, DAQ could use multiple 10 Gbps link to readout. Then DAQ should assemble completed events with data fragment from different links by trigger number or time stamp.

DAQ needs to finish waveform compression if electronics cannot do that. If we want to reduce data rate further, we need software trigger to remove more background events.

DAQ also needs to merge different detectors' events together and sorts events by time stamp.

\subsubsection{Other Functional Requirements}
DAQ also needs provide common functions like run control, run monitoring, information sharing, distributed process manager, software configure, bookkeeping, Elog, data quality monitoring, remote monitoring and so on.

\subsection{Conceptual Design Schema}
\begin{figure}[htb]
\begin{center}
\includegraphics[width=\textwidth]{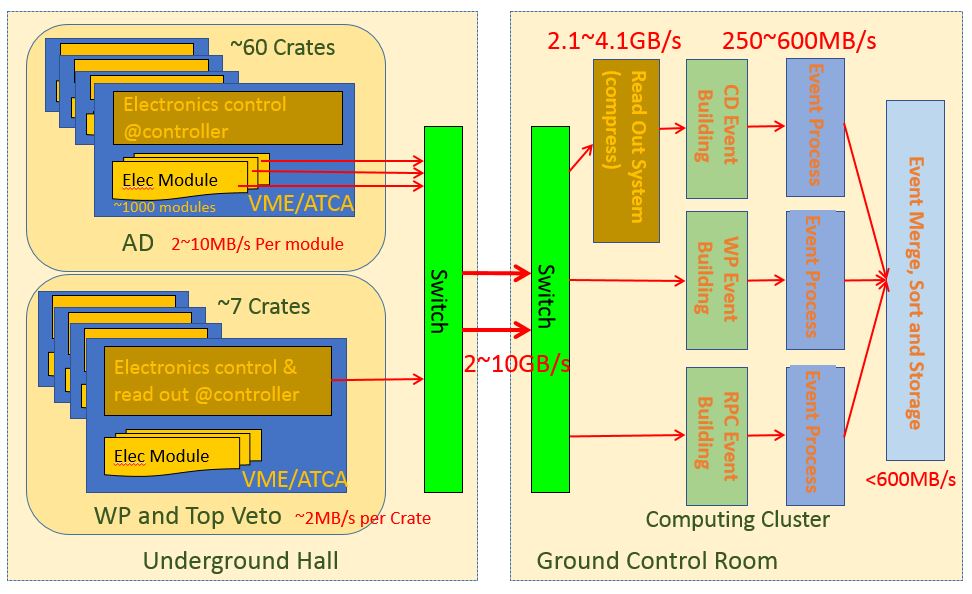}
\caption{DAQ Conceptual Architecture Design Diagram}
\label{daqArc}
\end{center}
\end{figure}

The JUNO experiment has many similar cases with BESIII and the Daya Bay experiments in DAQ part. We can design and develop JUNO DAQ based on BESIII, Daya Bay and ATLAS DAQ\cite{Bes3TDR,DybTDR,DybDaqTDR}. Figure \ref{daqArc} is the conceptual design schema of JUNO DAQ, DAQ readout electronics data through network. Except that network switches are placed in underground experiment hall, all other DAQ computers are deployed at ground computer room. DAQ constructs a blade servers' farm to process data, and the farm is connected with underground switches through multiple 10 gigabit fibers.

Due to electronics systems are not easy to dynamically distribute different DAQ hosts, it is better to design a two level event building at DAQ. The first level (ROS, read out system) reads out front-end electronics (FEE) data and builds event fragments and then the second level finishes full event building. It is better that DAQ compresses waveform data at the first level to reduce unuseful data transferring. The simplest waveform compression algorithm is zero compression with a configurable threshold. Only sample data exceeding the threshold can be reserved.

\begin{figure}[htb]
\begin{center}
\includegraphics[width=\textwidth]{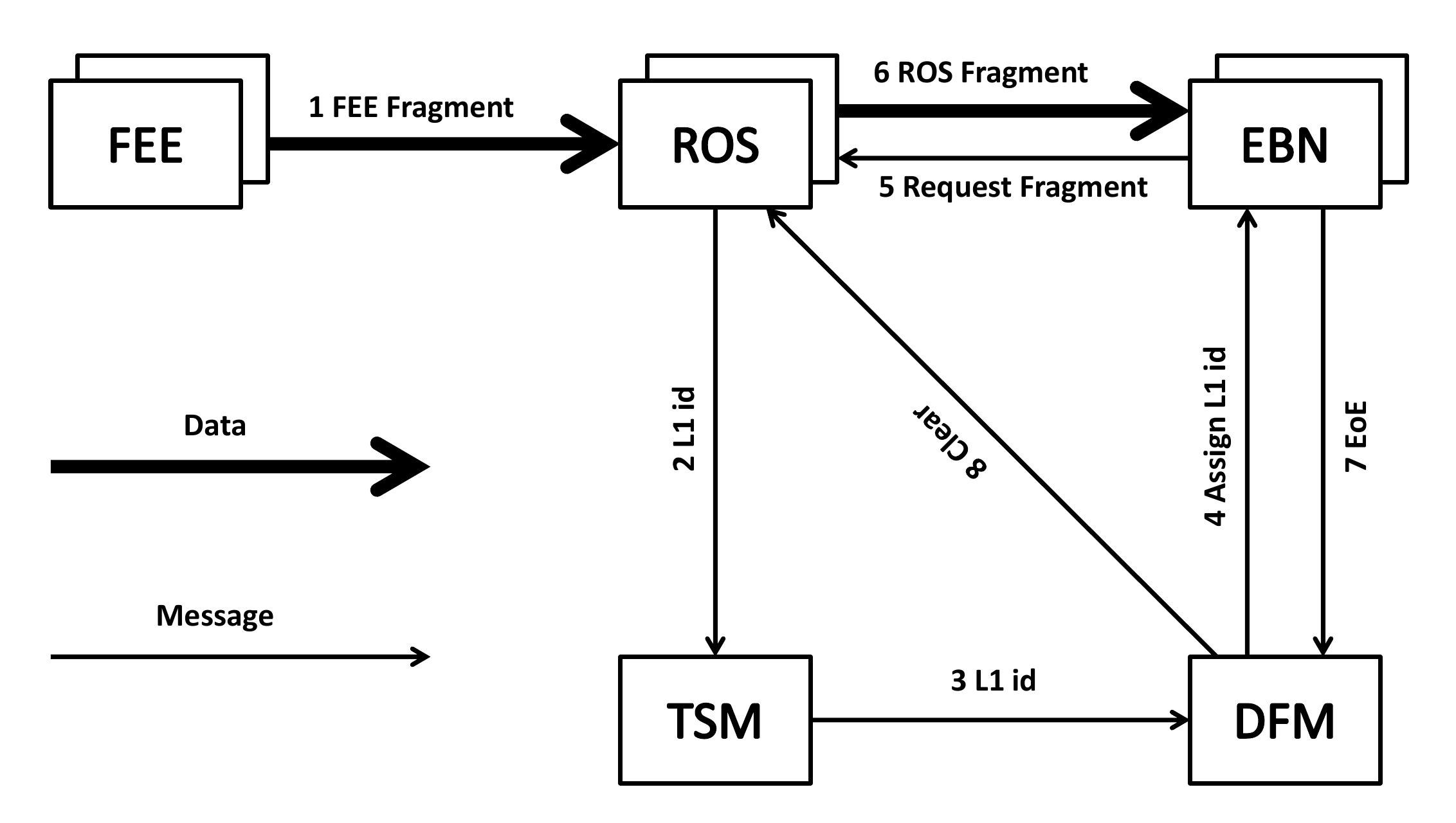}
\caption{DAQ Event Building Collaboration Diagram}
\label{daqEB}
\end{center}
\end{figure}

JUNO DAQ can use the same two level event building data flow scheme refer to BESIII event building as shown in Fig. \ref{daqEB} \cite{LifeiThesis}:
\begin{enumerate}
 \item front-end electronics (FEE) send data to ROS(read out system) through network.
 \item ROS receive all data slice of one event and send event id to TSM(trigger synchronizing manager).
 \item TSM send event id to DFM(data flow manager) when TSM get all same event id from all ROSs.
 \item DFM assign event id to a free EBN(event building node).
 \item EBN send data request to each ROSs.
 \item ROSs send requested data to EBN.
 \item EBN receive all ROSs data fragments of one event and finish full event building, then send event id back to DFM.
 \item DFM send event id to ROSs to clear data buffer. \ldots
\end{enumerate}

\begin{figure}[htb]
\begin{center}
\includegraphics[width=\textwidth]{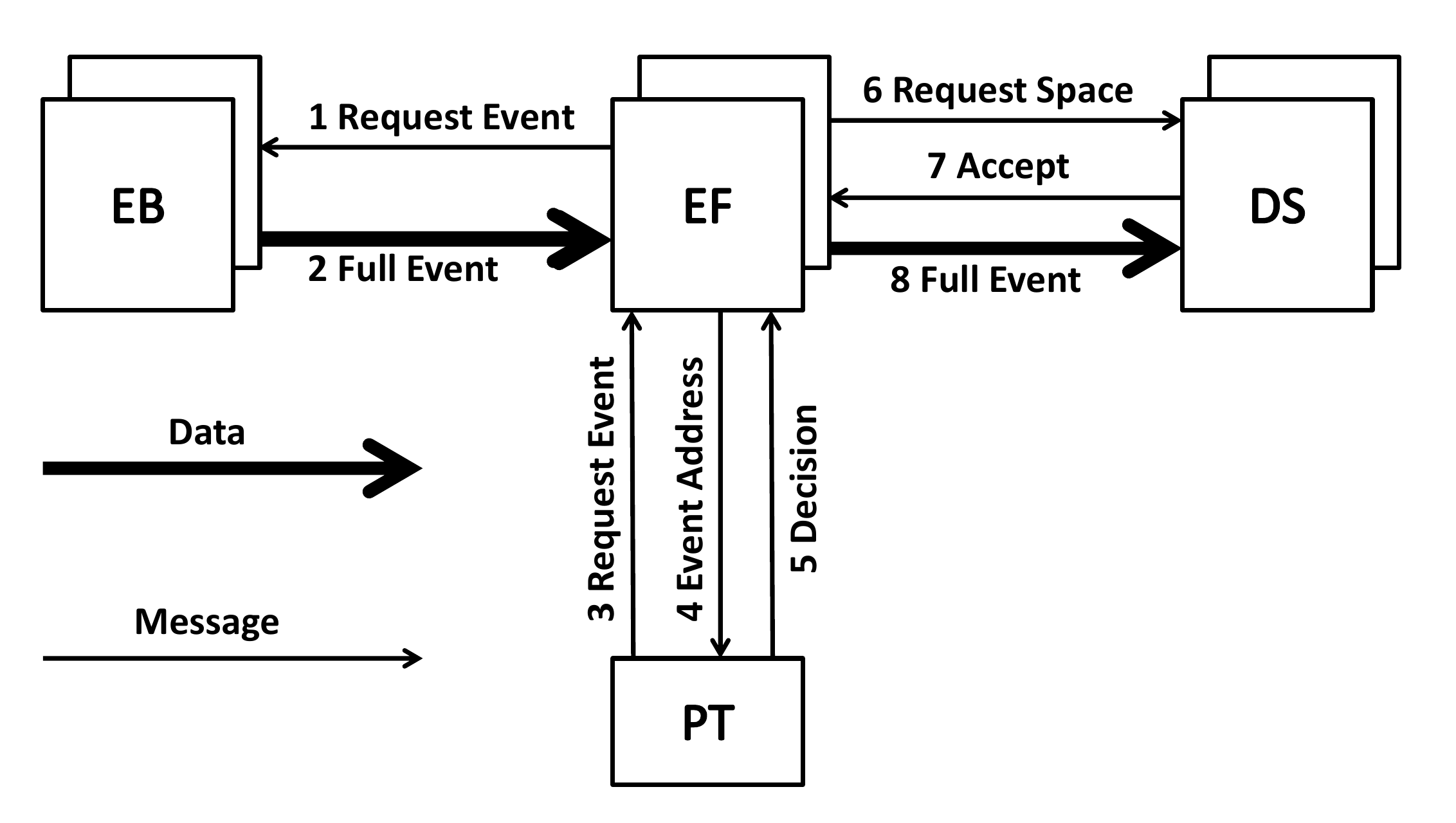}
\caption{DAQ Event Filter Collaboration Diagram}
\label{daqEF}
\end{center}
\end{figure}

Software trigger can be performed as BESIII event filter\cite{LiuyjThesis}. The figure \ref{daqEF} shows event filter collaboration diagram, each EF (event filter) node requests events from EB node, then sends them to PT (process task) to analyze data for software trigger and data quality monitoring, at the end sends triggered event to DS (data storage) node for storage. Another option is performing software trigger before full event building as Atlas Level 2 trigger. It is better if DAQ needs compress waveform.

All three detectors can share one DS node to save data to disks. Then event merging and sorting can be performed at DS node.

\subsection{R$\&$D Plan}
JUNO DAQ needs the following technical research for detail design:
\begin{enumerate}
 \item Waveform compression algorithms and read out host performance.
 \item Software trigger algorithms.
 \item Event building network and performance.
 \item Event sorting and storage.
 \item Big scale computer farm computing and software developing.
 \item Integrating and test with dedicated electronics and detector system.
\end{enumerate}

\subsection{Manufacture, Developing and Installation}
DAQ will not rely on custom made DAQ hardware, so there are no manufacture issues. But we need to investigate and survey different brands and types to integrate design schema. Hardware installation will be outsourced to vendors or system integration company. DAQ software will be designed and developed based on BESIII, Daya Bay and ATLAS DAQ software.

\subsection{Risk Analysis and Safety Evaluation}
The waveform data compression ratio could not reach expectation of detectors and electronics real performance.

There are another two challenges for DAQ up to now. One is that electronics might adopt none hardware trigger schema, the other is that supernova explosion could happen in a closer location.

\subsection{Schedule}
Major technical issues research according to electronics design. Finish DAQ technical design in 2013-2015.

Software developing, test and debug with electronics in 2016-2017.

Installation, deployment and integration in 2018-2020.

\section{Detector Control System}

\subsection{Requirements}

The main task of the Detector Control System is to establish long-term monitoring of the parameters affecting the performance of the experimental equipment. The parameters include pressure, temperature, humidity, liquid level of the scintillation, electronics, gas pressure and the pressure in the lobby and the entire electrical and mechanical environment working status of the devices. Some subsystems need to provide device control such as calibration system, gas system, water cycle system and power system. The real time operation states of the devices will be monitored and recorded into database. When an exception occurs the system can issue timely warnings at the same time through a secure interlock. Meanwhile, the devices can be protected by the safety interlock automatically to prevent equipment damage and personal protection.
\subsubsection{System Requirement}

The system will meet the requirements of an effective mass of 20,000 tons of central detector, which contains about 17,000 photomultiplier tubes. Due to the large scale of the detector there are about $\sim$1000 temperature and humidity monitoring points, about $\sim$20,000 channels of high voltages, and thousands of power supplies etc. There is also pure water system, gas system as well. The collection of the system requirements is based on the design of the detector, electronics and trigger system. The design requirements include the hardware and software design, the test bed of the subsystem design and the system integration.

\subsubsection{Function Requirement}

According to the experimental requirements the general software framework will be designed. The high voltage, the temperature monitoring module, the hardware interface, and the hardware and the software protocols will be defined. Drivers and data acquisition framework of the exchange process will be developed. The function modules of the DCS system are shown below.

\begin{figure}[htb]
\label{dcsFW}
\begin{center}
\includegraphics{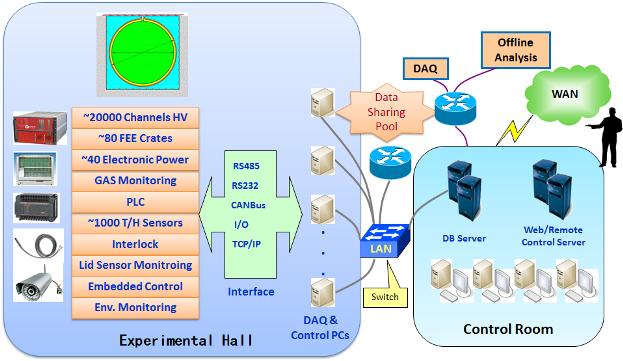}
\caption{Framework Design of Detector Control System}
\end{center}
\end{figure}

\subsubsection{System Schema Design}
\begin{itemize}
\item Content

According to the actual hardware and software requirements of the experiment the system uses a hierarchical design framework. The system will be divided into the global control layer, the control layer and the data acquisition layer. The global control layer will realize the overall experimental equipment information collection, safety interlock, remote control, data storage, network information release and permissions management. Local control layer will realize the local control of the local equipment in the experiment hall. And support the local device monitoring, data recording, data upload and alarm. The data acquisition layer is response for the various hardware interfaces. It can support for embedded systems such as ARM, FPGA and standard industry interface hardware such as PLC, USB and serial RS232 interface and acquisition, which based on the network devices such as the TCP/IP data acquisition interface.

The system will be built on a distributed system development method. According to the experimental equipment distribution characteristics the distributed data exchange platform will be used for the development. Global control systems share the data and the interactively control commands by information sharing pool. Configuration files will use the text format specification which can realize the remote configuration, distributed sharing and management.

The system uses a module based approach of development. This method can achieve rapid integration of complex systems. From a functional view the system will be divided into data acquisition, control module, alarm module, memory module, data sharing and processing, system configuration, privilege management and user interface. Each subsystem can choose module assembly interface based on the actual system requirements. The modules can be divided by subsystem as following:

\begin{enumerate}
 \item High voltage monitoring system, including the central detector PMT high voltage system, RPC high voltage etc.
 \item Detector electronics chassis monitoring system, including the central detector electronics, RPC detector electronics, both inside and outside the pool detector electronics, monitoring content includes electronic temperature, power supply, fan, over current protection, low-voltage power supply.
 \item Temperature and humidity monitoring system, including liquid temperature detector, temperature monitoring room.
 \item Gas monitoring system, including the gas support system and center detector cover gas system
 \item Center detector overflow tank  monitoring, oil monitoring, liquid level clarity, camera monitoring and the calibration system of monitoring center.
 \item Experimental hall of environmental temperature, humidity, pressure monitoring system, the radon monitoring system, video monitoring.
 \item Control room monitoring system, database system and Webpage remote monitoring system.
 \item Water system
\end{enumerate}

\item Key Technology

The key technology of detector control system is the integration framework development of the software. The system will develop a set of integrated management module according to the detector hardware requirements. The system functions will be realized by the conceptual design and the detailed modular design. For the hardware system using commercial framework or special design system the data acquisition module will be designed for the integration. The commercial software will interface with the DCS framework by the integration. The software design is completed after repeated assembly test and performance test before put into use. The user feedback will be fully considered during the software design and the commissioning period. The software will enter the formal running when the stage of training finished.

The traditional design of the JUNO central detector is a stainless-steel tank plus an acrylic sphere, where stainless-steel tank is used to separate the shielding liquid (mineral oil or LAB, to be decided) from the pure water in the water pool, and the acrylic sphere is to hold the 20kt LS. The design principle and key technology of the Daya Bay neutrino experiment could be a reference. The LS separated by an organic glass container is 35.4~meters in diameter. For the quality of the target substance which is one of the most important factors affecting the precision of experiment the quality change of the target material this requires for long-term supervisory.  At the same time, the temperature inside the detector needs to be monitored. Environmental temperature variation not only impact on detector performance, the detector energy scale but also on photomultiplier tube noise, amplification and electronics precision. This requires many point of monitoring from the experimental hall. Therefore, the design of reliable, compatible of monitoring system is needed. Various experimental parameters such as temperature, humidity, voltage, electronics currents, radon density accurately level need to be achieved in the long-term and stable running of the experiment.

Software tools and applications which provide a software infrastructure for use in building distributed control systems will be used to operate devices. Such distributed control systems typically comprise tens or even hundreds of computers, networked together to allow communication between them and to provide control and feedback of the various parts of the device from a central control room, or even remotely over the internet. System will use Client/Server and Publish/Subscribe techniques to communicate between the various computers. Most servers (called Input/Output Controllers or IOCs) perform real-world I/O and local control tasks, and publish this information to clients using the Channel Access (CA) network protocol.

\item Development Process
Top development platform can be used to configure software such as LabVIEW or open source platforms such as EPICS (a widely used software platform for the large-scale control system physical device). The device driver can be developed by hardware I/O controller (IOC) server. Subsystems in the pre-design stage will build a testbed for the development of hardware drivers to test the hardware and software functions. Each detector subsystem model will be tested in testbed platform, as well as data acquisition interface. For the hardware which has no testbed device simulation models should be made with the software development team and a common definition of the interface specification, including data format, the transmission frequency, control flow, the interface distribution, database table and so on.
\end{itemize}

\cleardoublepage
\chapter{Offline Software And Computing}
\label{ch:OfflineSoftwareAndComputing}

\section{Design Goals and Requirements}
\subsection{Requirement Analysis}

The offline software plays an important role in improving physics analysis quality and efficiency. From the point of view of physics analysis, the following requirements need to be applied to the offline software system.
\begin{itemize}
\item Compared to collider experiments, neutrino experiments have two specific characteristics: 1). There is time correlation between events. 2).There is only a very small fraction of signal events among a large number of backgrounds. The software framework should therefore provide a mechanism of flexible data I/O and event buffering to enable high efficiency data access and storage, as well as the capability to retrieve the events within a user-defined time window.

\item An interface should be provided for different algorithms to interchange event data. There should be a unified geometry management. Software and physics analysis parameters should be managed by a conditional database. An interface to retrieve correct parameters according to software versions should also be provided.

\item There are many kinds of events in the JUNO experiment. Apart from reactor neutrino events, they are supernova neutrinos, geo-neutrinos, solar neutrinos,  atmospheric neutrinos, radioactivity etc. The offline software should provide accurate simulations to describe interactions of neutrinos and background events with the detector.

\item The detector simulation software should be able to simulate detector performance and guide detector design. In addition, the calculation of detector efficiencies and systematic errors, which are needed by physics analysis, also relies on detector simulation.

\item The detector performance depends on event reconstruction. In order to get better energy resolution, optical model in liquid scintillator detector should be constructed, using PMT charge and time information to get event vertex and energy. Event vertex can also be used to veto the natural radioactivity background events from outside the liquid scintillator. To reduce the isotope background from muons, accurate muon track reconstruction is needed. For atmospheric neutrinos, the reconstruction of short tracks inside the liquid scintillator and identification of different charged particles are needed.

\item Event display software is needed in order to show the detector structure, physics processes in the detector, and reconstruction performance.
\end{itemize}

The JUNO detector will produce about 2 PB raw data every year, which will be transferred back to the Computing Center at the Institute of High Energy Physics (IHEP) in Beijing through a dedicated network connection. Due to the large data volume, a big scale of offline computing platform is required by Monte Carlo production, raw data processing and physics analysis, as well as data storage and archiving. A rough estimation suggests a level of 10,000 CPU cores, 10 PB disk storage and 30 PB archive in the future. The computing nodes and storage servers will be connected to each other by a 40-Gbps backbone high-speed switching network. In order to improve data processing capacity, the platform will integrate the computing resources contributed by outside members via a distributed computing environment. So the experiment data can be shared among collaboration members and computation tasks can be dispatched to all of the computing sites managed by the platform. Considering the above needs, requirements on data processing, storage, data transfer and sharing, can be summarized as follows:

\paragraph{Data Transfer}
A data transfer and exchange system with high performance, security and reliability is required. To guarantee the timely transfer of raw data from the experiment site to the IHEP data center, the system should monitor and trace the online status of data transfers and it should provide a data transfer visualization interface. This should also guarantee data is dispatched and shared among different data sites within the collaboration smoothly.

\paragraph{Computing Environment}

A computing farm with 10,000 CPU cores will be established at the IHEP Computing Center to facilitate successful processing of the JUNO data. The collaboration members will provide additional computing resources. The computing environment includes not only the local computing farm, but also those remote resources contributed by outside collaboration members.

\paragraph{Data Storage}

JUNO will generate a huge amount of experiment data. Data from the experiment site will be transferred and stored at the IHEP Computing Center. The capacity of storage at IHEP will be 10 PB for storing both real and simulated data. It should support concurrent access by applications running on 10,000 CPU cores. Data migration between tape repository and disk pools should be transparent to users.

\paragraph{Maintenance, Monitoring and Service Support}
To improve the stability of the JUNO computing resources, an intelligent monitoring system will be developed. With this system, the status of both platform and network connection can be monitored through a fine-grained control. The monitor system can not only give the status assessment of the whole platform but also give predictions and prior warnings for various platform services. A set of information management tools should be developed including conference support, document management and online-shift management etc.

\subsection{Introduction}
The offline software and computing system consists of two separate parts: offline software and computing. It not only closely connects data processing and physics analysis, but also it can be regarded as a bridge between detector operations and physics analysis. The primary tasks of this system are to process raw data collected by the detector and produce reconstructed data, to produce Monte Carlo data, to provide software tools for physics analysis, and to provide the networking and computing environment needed by data processing and analysis.

The offline software includes software framework, event generator, detector simulation, event reconstruction and event display etc. The framework is the underlying software supporting the whole system. Based on this framework, experiment related software will be developed such as software for event data model, data I/O, event generation, detector simulation, event reconstruction, physics analysis, geometry service, event display, and database service etc. To facilitate the development, the infrastructure of offline software provides various useful tools such as tools for compiling, debugging, deployment and installation etc.

The offline computing is designed to build a high efficiency and reliable computing environment for the JUNO experiment. It has the following major parts: a data transfer platform to transfer raw data from the experiment site to the Computing Center at IHEP, a large scale of computing farm for data processing and data analysis, a massive data storage subsystem and a monitoring and management platform providing fine-grained monitoring and management of the computing resources.

\section{Offline Software Design}
\subsection{Software Platform}

The offline software system is designed to meet the various requirements of data processing. As JUNO software developers are dispersed all over the world, a unified software platform is required, which provides a formal working environment. This has advantages for resource optimization and manpower integration, which can improve the software development, usage and maintenance.

Based on our experience, Linux OS and GNU compiler are the first choice for the JUNO software platform. However other popular OSes and development environments should be considered for the sake of compatibility. The CMT \cite{CMT} tool, which can calculate package dependencies and generate Makefiles, is used for the software configuration and management. An automated installation tool can ease the deployment of the JUNO software, and is also helpful for daily compilation and testing. Users are able to concentrate on the implementation of software functionality, without suffering from different development environments.

Nowadays, mixed programming with multiple languages is practical, so we can choose different technologies for different parts of our software. The main features of the application are implemented via C++ to guarantee efficiency. User interfaces are provided in Python for additional flexibility. Boost.Python, as a widely used library, is a good choice to integrate C++ and Python. If it is included properly at the beginning of the system design, most users will be able to enjoy the benefits of mixed programming without knowing the details of Boost.Python.

\begin{figure}[htb]
\begin{center}
\includegraphics[width=.8\textwidth]{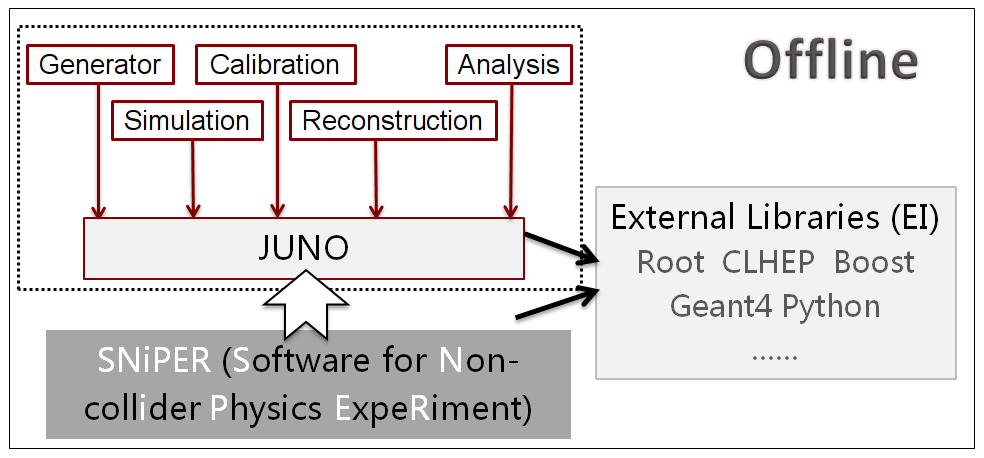}
\caption[Offline Software Platform]{\label{fig:OfflinePlatform}The structure of the offline software system}
\end{center}
\end{figure}

We will implement a software framework, SNiPER, as the core software of the JUNO platform. SNiPER stands for "Software for Non-collider Physics ExpeRiments". As shown in Fig.\ref{fig:OfflinePlatform}, components of the JUNO software, such as simulation and reconstruction, are executed as plug-ins based on SNiPER. The principle of this new framework is simplicity and high efficiency. It depends on a minimal set of external libraries, fully meeting our requirements without over-engineering or loss of efficiency.

The SNiPER kernel is the foundation of the JUNO software system, and determines the vitality of the whole platform. We design and manage SNiPER modularly. Every functional element is implemented as a module, and can be dynamically loaded and configured. Modules are designed as high cohesion units with low couplings between each other. They communicate only through interfaces. We can replace or modify one module without affecting any of the others. Since it's flexible and expandable, users can contribute to it independently.

The model is inspired by other pioneering software frameworks such as Gaudi \cite{Gaudi}. Expandable modules are strategically distinguished as algorithms and services. An algorithm provides a specific procedure for data processing. A sequence of algorithms composes the whole data processing chain. A service provides useful features that can be called by users when necessary. Algorithms and services are plugged and executed dynamically. They can be selected and combined flexibly for different requirements.

Data to be processed may come from different places and be in different types and formats. The results may also be stored in different ways. In the framework we reserve interfaces for multi-stream I/O support, so that the I/O service can communicate conveniently with other modules via the interface. At present, the disk access could be a serious bottleneck in massive data processing. This should be considered in I/O service developing. Many techniques, such as lazy loading, are adopted to optimize the I/O efficiency.

The framework also involves many frequently used functions, such as the logging mechanism, particle property lookup, system resource loading, database access and histogram booking, etc. These contents are wrapped in services in SNiPER, which will ease the development procedure. Users can configure a job interactively with the Python command line or in batch mode with script files. The flexibility of Python will ease the job execution procedure.

The JUNO experiment has a long lifetime and software technology is still evolving rapidly, so the framework has to be very expandable. It should be able to integrate with new tools and libraries in future. This will save the manpower needed for developing, and improve the system robustness.

It is expected that JUNO will have a huge amount of data. In order to make use of all possible computing resources, it is necessary to consider parallel computing techniques. Modern CPUs are generally multi-core. Multi-threaded programming extracts the capacity of multi-core CPUs, and speeds up a single job significantly. In PC clusters, a server/client distributed system can be implemented and deployed, too. The latest computing technology and HEP computing techniques, including GPUs, Grid and Cloud, should be explored and used as appropriate.

\subsection{Event Model}
The event model defines the key data units to be processed in the offline data processing. It not only defines the information included in  one event at different data processing steps such as event generation, detector simulation, calibration, reconstruction and physics analysis, but also provides the ways for events to be correlated between different processes.

Since ROOT \cite{ROOT} has been extensively used as the data storage format in most particle physics experiments, it is decided that the JUNO event model is based on the ROOT TObject. So all the event classes are inherited from TObject. In this way, we can directly take advantage of the powerful functionalities provided by ROOT, such as schema evolution, input/ouput streamer, runtime type information (RTTI), inspection and so on.

In order to quickly view event information and make the decision whether or not a given event is selected, the event information is divided into two levels: Event Header and Event. The Event Header defines the sequential information (such as RunId, EventId and Timestamp etc.) and characteristic information (such as energy of events), while Event defines more detailed information about the event. In this way, the event header can be read in from data files or other storage and its characteristic information can be used for fast event selection without taking extra CPU time or memory required to read in the contents of a full event.

\begin{figure}[htb]
\begin{center}
\includegraphics[width=.8\textwidth]{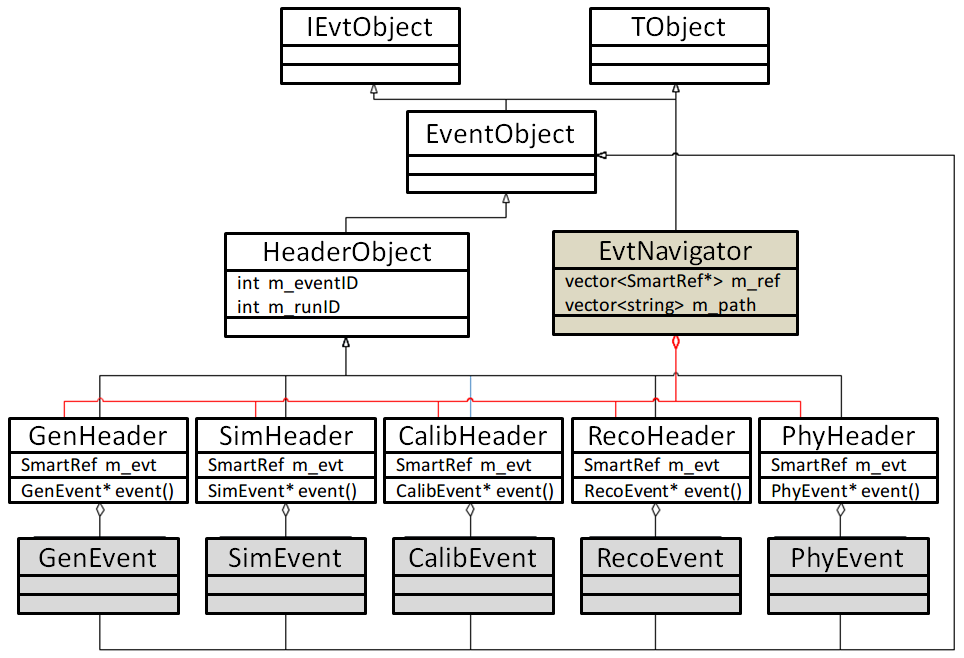}
\caption[Event Model]{\label{fig:EventModel} Design Schema of JUNO Event Model}
\end{center}
\end{figure}

Event correlations are very important for neutrino physics analysis since inverse beta decay (IBD) events will be regarded as two separate event signals: prompt signal and delayed signal. Both of them actually refer to the same physics event. In the JUNO experiment, we design SmartRef to fully meet this requirement and to avoid the duplication in writing these information into ROOT files. SmartRef is based on the TRef class in ROOT but has more powerful functions. It not only provides correlations between events but also provides the information for lazy loading of the data.

For the JUNO event model, the class EvtNavigator is designed to organize all the event information at different processing steps and acts as an index when processing events. In Figure \ref{fig:EventModel}, we can see that EvtNavigator has several SmartRefs which refer to the headers of this event at different processes. Event model classes are usually simple. Only some quality variables and functions need to be defined. These functions can be divided into three types: Set-functions to set new values, Get-functions to return the variable's value, and streamer functions for reading/writing data from/to files. Most of the code is similar event by event, so an XOD (Xml Object Description) tool is developed and used to describe the event information with a more readable XML file. XOD can automatically produce the event header file, the event source file and the dictionary file needed by ROOT.

\subsection{Geometry}

The geometry system describes and manages the detector data and structure in software, providing a consistent detector description for applications like simulation, reconstruction, calibration, alignment, event display and analysis. All subsystems, including the Central Detector, Top Veto Tracker, Water Cherenkov Pool and the Calibration system, will be described in the geometry system. The design of the JUNO detector also requires the geometry system to be flexible and able to handle time-dependent geometry and the co-existence of different designs.

Several detector description tools have been developed and implemented in high energy physics experiments, with some of the most popular languages being XML (eXtensible Markup Language) \cite{XML}, HepRep \cite{HepRep}, VRML(Virtual Reality Modeling Language, updated to X3D) \cite{VRML} \cite{X3D} and GDML (Geometry Description Markup Language) \cite{GDML}. At the stage of conceptual design for the JUNO detector, the detector description geometry is based on GDML, including geometry data, detector structure, materials and optical parameters. In consideration of the co-existence of multiple major designs and frequent minor modifications to the designs, we adopt GDML as the main tool to transfer detector description data between different applications, because GDML requires less human work and has the advantage of automatic translation of detector data between Geant4 \cite{Geant4} and ROOT, the two most popular HEP programs, which are widely used in simulation, reconstruction and analysis. However, it is noteworthy that GDML also has some limitations. For example, its Geant4 and ROOT interfaces are not completely consistent, so some information like optical surface description and matrix elements may be lost in the detector data translation. The GDML interfaces do not support complicated (user defined) shapes and require more work to implement the extensible functions. These small problems need to be solved in future geometry software development.

At the current stage, the detector geometry is described in each individual Geant4 simulation code, which also serves as the unique source of detector description. When a simulation job is running, it constructs the detector structure in Geant4 and exports it in GDML format; at the end of the simulation job, the GDML detector data is read back and converted into ROOT geometry object format, which is bundled with the output event data ROOT file while being written out. When another application such as reconstruction or event display wants to read in the geometry information, it searches the bundled ROOT file for the geometry object to initialize the geometry offline, and uses the geometry service package to retrieve the detector unit information it requires. This mechanism gives every detector designer the maximum flexibility to modify his specific design, uses a single module to handle multiple detector designs at the same time and guarantees the consistency of detector description between different applications.

In the future, when the detector design is fixed, it is preferred to generate the detector description within the geometry system and provide it to all applications through interfaces, rather than the current solution of generating geometry data in Geant4.     

\subsection{Generator}

Generators produce simulated particles with desired distribution of  momentum, direction, time, position and particle ID.

The generators for the JUNO experiment include inverse beta decay (IBD) generator, radioactivity generator, muon generator, neutron generator, and calibration source generator. Some of the generators can be imported from the existing Daya Bay software, while the others need to be developed for JUNO specifically. The generators are grouped into three kinds according to their development platform. One is based on Geant4 ParticleGun for shooting particles with specified particle ID, momentum and position distribution. The second kind, such as ${}^{238}U$ and ${}^{232}Th$ radioactivity generators, is developed with existing Fortran libraries. The third kind of generators, such as IBD and GenDecay, will be developed in C++ language,. The output data of generators should be in the format of HepEvt or HepMC \cite{HepMC} for better communication with other applications. The design of particle vertex position generation need to be flexible and allow the distribution to be geometry related, for example, a random distribution on the surface of a PMT, in liquid scintillator or in acrylic.

The muon generator particularly depends on the muon flux, energy spectrum and angular distribution at the JUNO site. A digitized description of the actual landform at the JUNO site need to be acquired first, and then used as input to MUSIC \cite{MUSIC} to simulate muons passing through the rock, getting the flux, energy spectrum and angular distribution of the muons finally reaching the JUNO detector. The effect of muon bundles also needs to be considered for such a large detector as JUNO.

\subsection{Detector Simulation}

The detector simulation package for JUNO is based on Geant4. Detector simulation includes detector geometry description, physics processes, hit recording, and user interfaces.
\subsubsection{Description of detector geometry}
An accurate description of detector geometry is the foundation of detector simulation. In JUNO simulation, the geometry of the liquid scintillator, acrylic/nylon ball, PMTs, stainless steel tank or trusses need to be described in simulation software. About 17,000 PMTs are placed on a spherical surface. Geant4 is very sensitive to the geometry problem of volume overlaps or extrusions, which in general reduce the simulation speed or waste time in infinite loops. Some built-in or external tools can help to find such problems. At the stage of detector design, the detector geometry keeps changing. It is preferable that the geometry and event data can be exported in the same data file after simulation, so that the reconstruction job can read both information at the same time to ensure geometry consistency between simulation and reconstruction. In addition, the detector description method should allow easy modifications to the geometry with changes made to the detector design.
In simulation of the liquid scintillator detector, the simulation of optical photon processes is a key issue. The optical parameters for materials such as liquid scintillator, acrylic, mineral oil and PMT should be carefully defined. The optical parameters include refractive index, absorption length, light yield factor, quantum efficiency and so on. Each optical parameter varies with the wave-length of optical photons and may be time-dependent in the future. Due to the production technique, the parameters of each individual PMT may be different. How to effectively manage these optical parameters is an important issue in the detector data description. A possible method is to combine Geant4 with GDML/XML and database technology to manage the detector geometry and optical parameters.
In addition, simulation of the optical processes requires the definition of an optical surface, on which optical photons can have different behavior like refraction, reflection and absorption.

\subsubsection{Physics processes}
The physics processes in the JUNO detector simulation include standard electromagnetic, ion, hadronic and optical photon processes and optical model for the PMTs. Optical photon processes include scintillation, Cerenkov effect, absorption and Rayleigh scattering.
When a charged particle passes through liquid scintillator, ionization energy loss is converted to scintillation light. The non-linearity between scintillation light and the ionization energy loss is called the quenching effect. During the propagation of light in the liquid scintillator, the re-emission effect needs to be considered, which is not included in standard Geant4 processes and needs to be implemented in our experiment.
For such a large liquid scintillator detector as JUNO,  the simulation of high energy muons in liquid scintillator can be too CPU-time consuming to be affordable. Reducing light yield or using a parameterized model for fast simulation are possible solutions to speed up the muon simulation.

\subsubsection{Hit recording}
The hit information, including the number of hits in each PMT and the hit time, is the output at the stage of Geant4 tracking. This information will be used as input to electronics simulation in the next step. In addition, particle history information should be saved, including the information of the initial primary particles, neutron capture time and position, energy deposition and position, direction and position of the hit on the PMT, and the relationship between track and hits.

\subsubsection{User interface}
A friendly user interface is necessary to use the simulation package, which should provide easy configuration of simulation parameters such as radius of LS volume, buffer thickness, vertex position, particle type and flags to switch on/off a specific physics process.

\subsection{Digitizer Simulation}

The digitizer simulation consists of electronics simulation, trigger simulation and readout simulation. The main goal is to simulate the real response of the three systems and apply their effects to physics results. The readout simulation transforms Monte Carlo data into the same format as experimental data and simplifies the analysis procedure. The digitizer simulation will be implemented in the JUNO offline framework, SNiPER.

Electronics simulation plays an important role in the digitizer simulation. The waveform from each PMT will be recorded using FADC and analyzed offline to get the final amplitude and timing information for each PMT. In electronics simulation, to transform the input hit information (given by Geant4) into electronic signals which are as real as possible, we need to build a model to exactly describe the single photo-electron response, which includes the effects of signal amplitude, rising time, pre-pulse, after-pulse, dark rate, etc. Electronics simulation should also have the ability to handle background noise, overlapping waveforms from multiple hits on the same PMT, and the effects from waveform distortion and non-linear effects. For very large signals like cosmic muons, however, it is difficult to save all FADC information so only integrated charge will be read out. At the current stage, since the detailed JUNO electronics design has not been finally determined, the electronics simulation imported from the Daya Bay experiment is implemented as a temporary substitute for the JUNO electronics simulation. It will be updated to the real case when the JUNO electronics design is finalized.

Trigger simulation takes the output from the electronics simulation as its input to simulate the real trigger logic and clock system, and decides whether or not to send a trigger signal if the current event passes a trigger. Currently, we have implemented a simple nHit trigger simulation.

Based on the trigger signals provided by the trigger simulation, FADC information in the readout window will be saved for each PMT and form an event. At the same time, the time stamp will be tagged. The final data format will be the same as for real data.

\subsection{Background mixing}
In real data, most events come from background, like natural radioactive events and cosmic muon induced events. In order to simulate the real situation, a background mixing algorithm needs to be developed to mix signal events with background events. It is an essential module if we want to make Monte Carlo data match well with real data. There are two options for background mixing: hit level mixing and readout level mixing. For hit level mixing, the hits from both signal and background are sorted by time first, and then handled by the electronics simulation. Hit level mixing is closer to the real case, but requires a lot of computing resources. Readout level mixing is much easier to implement and requires less computing resources, but it can not accurately model the overlapping between multiple hits. Since both options have advantages and disadvantages, more studies are necessary before an option is finally selected as the official JUNO background mixing algorithm.

\subsection{Event Reconstruction}

\subsubsection{PMT waveform reconstruction}
The total number of photoelectrons and the first hit time of each PMT can be reconstructed based on the waveform sampling from the readout electronics with two algorithms, charge integration or waveform fitting. A fast reconstruction algorithm has been developed using the charge integration with appropriate baseline subtraction. The first hit time is determined as the time above threshold with time walk correction. A waveform fitting algorithm based on a PMT response model is under development, which is expected to give better timing and can reduce the distortion of the charge measurement caused by the PMT and electronics effects such as after-pulse and overshoot. The fast reconstruction is designed for event selection, while the waveform fitting is expected to be used for small subset samples such as IBD candidates or calibration data.

\subsubsection{Vertex reconstruction in the Central Detector}
Since the energy of a reactor antineutrino is usually below 10 MeV, it can be approximated as a point source in the JUNO detector. Then the event vertex can be reconstructed based on the relative arrival time from the point source to different PMTs, by minimizing a likelihood function:
\begin{equation}
F(x_0,y_0,z_0,T_0)=-\sum_i \log(f_t(t_i-T_0-\frac{L(x_0,y_0,z_0,x_i,y_i,z_i}{c_{eff}})),
\end{equation}
where $x_0,y_0,z_0,T_0$ are the event vertex position components and start time to be fitted, $x_i,y_i,z_i,t_i$ are the position components and hit time of the $i_{th}$ PMT, $c_{eff}$ is the speed of light in the detector, and $f_t$ is the probability density function of the PMT hit time with the time-of-flight being subtracted, which contains the decay time of scintillation light and the time resolution of the PMT. If the multiple photoelectrons in the same PMT can be clearly separated, $f_t$ can be expressed analytically, otherwise it relies on MC simulation. A charge center method is used to calculate the event vertex to the first order, which can be used as the initial value of the likelihood fitting.

\subsubsection{Visible energy reconstruction}
The visible energy in the central detector of a point source is reconstructed using the maximum likelihood fitting:
\begin{equation}
\mathcal{L}(E_{vis})=\prod_{no-hit}{P_{no-hit}(E_{vis},i)}\times\prod_{hit}{P_{hit}(E_{vis},i,q_i)},
\end{equation}
where $E_{vis}$ is the visible energy to be fitted, $P_{no-hit}(E_{vis},i)$ and $P_{hit}(E_{vis},i,q_i)$ are the probability that the $i_{th}$ PMT has no hit and has the total charge of $q_i$, respectively. This probability is calculated based on the knowledge of the detector response model, including the light yield of the liquid scintillator, the attenuation length of the liquid scintillator and the buffer, the angular response function of the PMT, and the PMT charge resolution, as well as the reconstructed vertex as input. Another fast reconstruction algorithm has also been developed, which is based on the total number of photoelectrons collected from all PMTs with corrections using the calibration sources.

\subsubsection{Muon tracking}
\begin{figure}[htb!]
\centering
\includegraphics[width=.5\textwidth]{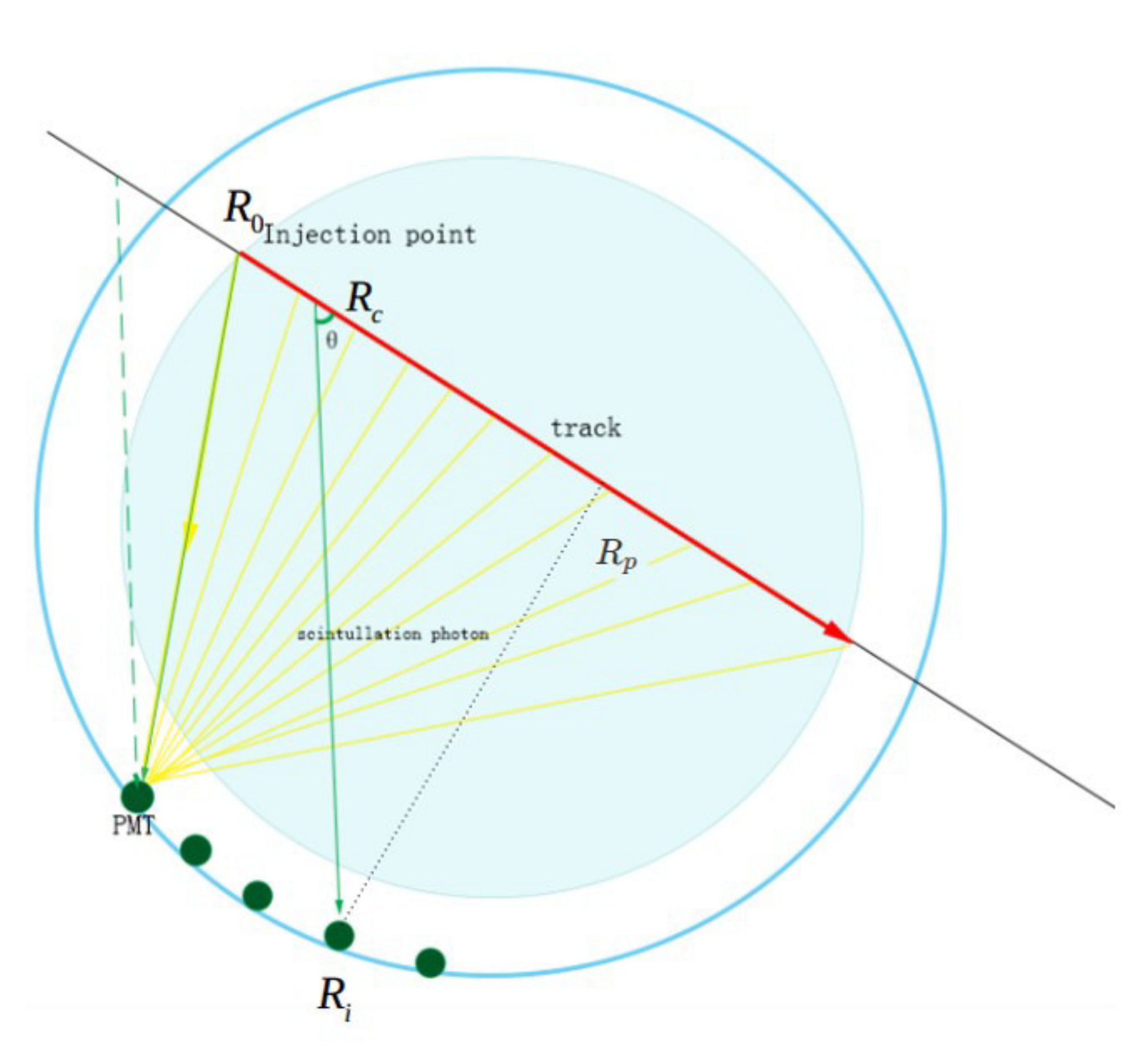}
\vspace{-0.15cm}
\caption{An illustration of the muon tracking. The red line is the muon track to be fitted, $R_c$ is the light source which provides the first hit to the PMT at $R_i$, determined by the intersection angle $\theta$. If $R_c$ is calculated to be outside the detector, it will be set as the injection point $R_0$.}
\label{fig:rec:muon_tracking}
\end{figure}

The muon tracking in the central detector is based on the PMT hit time. Since the energy deposit from cosmic muons is at the level of GeV, the average number of photoelectrons at each PMT will be greater than 100, therefore the first hit time of the PMTs is dominated by the fast component of the scintillation light, which approximately follows a Gaussian distribution. A $\chi^2$ function is defined:
\begin{equation}
\chi^2=\sum_i(\frac{t_i^{exp}-t_i^{mea}}{\sigma})^2,
\end{equation}
where $t_i^{exp}$ and $t_i^{mea}$ are the expected and measured first hit time of the $i_{th}$ PMT, respectively, $\sigma$ is the time resolution of the PMT, and $t_i^{exp}$ is calculated based on the tracking parameters and the optical model in the detector. As shown in Fig.~\ref{fig:rec:muon_tracking}, since the scintillation light is isotropic, the intersection angle $\theta$ can be calculated analytically. For muons passing through the detector, the PMTs around the injection point and the outgoing point see more light and form two clusters, which can be used to estimate the initial tracking parameters. A muon tracking algorithm combined with the veto detectors is still under development.

\subsection{Event Display}

The event display is an indispensable tool in high energy physics experiments. It helps to develop simulation and reconstruction packages in the offline software, to analyze the physics in a recorded interaction and to illustrate the whole situation to general audiences in an imaginable way.

\subsubsection{Detector description and display}
The display of the detector is provided by the geometry service in the offline software. The visual attributes of every detector unit are associated and controlled by identifiers. The event display software will support two display modes -- 3D display and 2D projection display. The 2D projection will be realized by the 2D histograms that are provided by ROOT. For 3D display, several popular 3D graphic engines, like OpenGL \cite{OpenGL}, OpenInventor \cite{OpenInventor} and X3D \cite{X3D}, have been proven as practical choices for event display software for some other HEP experiments. More studies are necessary to determine which infrastructure 3D graphic engine will work best for JUNO and its software environment. Currently a solution based on OpenGL is under study to realize the 3D display mode.

\subsubsection{Event data and display}
The event data in the offline system has various formats, including raw data, simulation data, calibrated data, reconstruction data and data analysis results. Their data formats are uniformly defined by the JUNO offline event model. In processing the event data for display, we need to study how to display the particle trajectories, their interactions with materials and hit response in simulation data, the reconstructed event vertices, energy and display attributes, as well as the relationships between these.

\subsubsection{Graphical User Interface}
A Graphical User Interface (GUI) provides a convenient interface for users to execute complicated commands with simple actions such as pushing buttons, selections, sliding a ruler or pushing hot keys. Frequently used functions, like event control, 3D view rotation and translation, zooming and time controls can be realized through widgets. A ROOT-based GUI framework has been implemented to realize these functions, giving the advantages of seamless binding with the ROOT analysis platform, ROOT-based geometry service and event model.

\subsubsection{Graphic-based analysis and tuning}
Graphic-based analysis will further provide some useful functions such as reconstruction algorithm tuning, event trigger and event type discrimination, which can help users to quickly filter, analyze and understand the events in which they are interested. 

\subsection{Database Service}
Databases play more and more important roles in offline data processing. They are used to store many kinds of information, such as, sizes and positions of complicated detector geometry, optics information, running parameters, calibration constants, schema evolution, bookkeeping and so on. At the same time databases provide access to all kinds of information through their management services, such as creation, query, modification and deletion, which will be frequently used during offline data processing.

Many types of database have been used in particle physics experiments, of which the most popular are Oracle \cite{Oracle}, Mysql \cite{MySQL}, PostgreSQL \cite{postgreSQL} and SQLite \cite{SQLite}. Many of them provide user-friendly APIs interfaces for different programming languages. Recently, NoSQL databases have become more and more popular and powerful in some other fields, so we are also investigating the possibility of using some NoSQL database.

Regardless of which kind of underlying database will be chosen,the general schema of database services will be implemented in three levels. The lowest level is the APIs of the underlying database; the middle level is the new extension of the APIs according to the requirements of the JUNO experiment; the upper level is to provide user-friendly services for different applications such as detector simulation, event reconstruction and physics analysis.

Since Daya Bay already has one good database interface, we will implement it, optimizing and extending the current interface to provide more flexible and more powerful access to the different types of information stored by the databases. We will also set up two types of database servers, master servers,and the slave servers. The former are used to store and manage all information and can only be accessed by the slave servers, while the latter are responsible for retrieving information from the master servers and providing access for all applications.

\section{Computing System Design}

\subsection{Computing Network and Public Network Environment}

A 1 Gbps dedicated link, named IHEP-JUNO-LINK, is planned between the JUNO experiment site and IHEP. It is proposed that this link be used for data transfer between the experiment site and the IHEP data center as well as for connecting the onsite office network and the external network. The 1 Gbps bandwidth is sufficient to fully meet the experiment's requirements for stable data transfer, based on an estimated data volume of 2 PB per year, which should take an estimated average bandwidth of 544 Mbps. The remaining bandwidth can be used for network crash recovery, experiment remote control and office network connectivity for JUNO onsite researchers.

\begin{figure}[htb]
\begin{center}
\includegraphics[width=.8\textwidth]{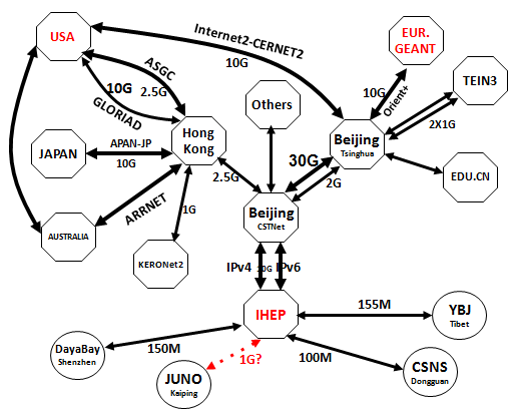}
\caption[WAN topology]{\label{fig:storage}WAN Topology }
\end{center}
\end{figure}

The JUNO network link is planned to be provided by the Chinese Science and Technology Network (CSTNet) \cite{CSTNet}. The data will first be transferred from the experiment site to IHEP through IHEP-JUNO-LINK, then relayed to collaborating sites through CSTNet. At present, IHEP is connected to the CSTNet core network through two 10 Gbps links, one of which supports IPv4 and the other IPv6. The bandwidth from IHEP to the USA is 10 Gbps and from IHEP to Europe is 5 Gbps, both of which are through CSTNet and have good network performance.

\subsubsection{Network Information System Security}
Network security is becoming more and more important in all kinds of information systems. The JUNO application system and even the computing environment also face security risks, so we are planning to apply risk control and safety protection in the following aspects: physical layer, network layer, system layer, transport layer, virus threats layer and management layer.

The safety of the physical layer is the foundation of the whole information system security. In general, the security risks to the physical layer mainly include: system boot failure and database information loss due to power failure, whole systematic destruction caused by natural disasters, probable information loss caused by electromagnetic radiation and so on. In order to ensure the robustness of the physical layer, a powerful UPS should be deployed for the IT system and devices; proper access control policies should also be implemented for some special areas.

The risks to the network layer consist of risks to the network boundaries and risks to safety risk of the data transfer. A powerful network firewall system will be deployed in the JUNO network to avoid risks to the network boundaries. To ensure the safety of data transfers, there are two things we must consider: data integrity and confidentiality. To ensure data integrity, we can add checksums and provide multi-layer data buffering during transfers.To ensure data confidentiality, the information system should use encrypted transmission.

The system layer risk usually refers to the safety of operation systems and applications. In JUNO, a rigorous hierarchical control policy and auditing operation logs will be used to reduce and prevent security threats to the system layer.

For virus threats, while traditional viruses mainly spread through a variety of storage media, modern viruses mainly spread through the network. When a computer in the LAN is infected, the virus will quickly spread through the network to hundreds or thousands of machines in the same network. In JUNO, rigorous access control policy and real-time detection for the network traffic will be implemented for all the network nodes and IT systems to provide a safe network environment.

The risk to the management layer is usually caused by unclear responsibilities and illegal operation of the IT system, so the most feasible solution is to combine management policies and technical solutions together to achieve the integration of technology and policy.

\subsection{Data transfer and sharing}
Data transfer and sharing is the foundation of physical analysis in JUNO. In steady operations mode, raw data will be transferred over the network from the experiment site to the Computing Center for long term data storage, then distributed to the collaboration group members according to their requirements. The raw data from the JUNO experiment is open to all collaboration group members for local or remote processing and analysis.

The raw data from JUNO will be acquired by the online data acquirement system (DAQ) which will be deployed onsite, and then stored in a local disk cache with sufficient storage to keep one month's raw data. After that, the raw data will be transferred to the IHEP Computing Center in Beijing. During the transfers, the checksum of the data will also be transferred to make sure of the data integrity. If the data integrity check is failed, the raw data should be retransferred. After the data transfer is done and the data integrity check is ok, the status of the data in the DAQ local disk cache will be marked as TRANSFERRED. Moreover, a high/low water line deletion algorithm will be used to clear the outdated data in DAQ local disk cache.

Most of the data will be transferred automatically by the data transfer system, so some possible human errors can be reduced. To ensure the stability and robustness of the data transferring system, a monitoring system will be developed and deployed which will provide the data transfer and sharing status in real-time and track the efficiency of the data transfer system. Combined with the status of the IT infrastructure (including network bandwidth), the data transfer system will optimize the transfer path and recover transfer failures automatically to improve the performance and stability of the system. 
\subsection{Computation and Storage}
\subsubsection{Local Cluster}

About 10,000 CPU cores should be used for raw data simulation, reconstruction and data analysis to meet JUNO offline computing requirement. All the resources will be managed by the batch system of the IHEP Computing Center. Several job queues with different priorities will be established to share the resources in the most efficient way.

An optimized scheduling algorithm will be developed based on the features of the the JUNO computing environment and hardware performance. The scheduler should dispatch jobs to the most suitable computing resource so that high overall computing efficiency is obtained. Job types can be defined based on the computing model, characteristics and I/O bandwidth consumption for different jobs. For this, it is necessary to study and test different types of jobs running on different machines with different CPUs. It is also necessary to analyze the effect of the distributed storage system on job running time. New scheduling optimization algorithms based on the results of this study should be implemented as a plug-in which can be integrated with the batch job system. This will be used for resource management and job scheduling to improve the efficiency of single jobs and the overall resource usage.

\subsubsection{Computing Virtualization}

Local clusters are managed by job management systems, with Torque \cite{Torque}, Condor \cite{HTCondor} and LSF \cite{LSF} being popular choices in HEP systems. These are responsible for scheduling jobs to a physical machine. With ever more powerful CPUs available, virtual machine clusters - cloud computing - can be used to relieve the pressure on computing resources at peak times. Cloud computing also reduces the burden of system management in applying upgrades to operating system and offline software, as well as satisfying requests for different software versions. Running jobs on virtual machines (VMs) can also avoid the trouble caused by hardware heterogeneity and give more flexibility in software configuration, without requiring changes in the job itself. Virtual clusters can therefore improve the availability and reliability of the computing resources as well as providing elastic resource expansion and flexible resource allocation. Rather than allocating a job to a node, as for physical cluster resource allocation, jobs are allocated to a CPU core. One CPU core supports one VM running a JUNO job. The lifetime of the VM depends on the job running time. This makes it possible to pre-empt jobs based on priority, and to migrate jobs. The VM can be migrated online and the job running on the VM can be hung up before the migration and released after the migration. These features of VMs will make the computing platform more robust.

\subsubsection{Distributed Storage Management and Data Sharing}

JUNO data processing is a typical massive data computation task. Distributed storage management is necessary for its large scale of data. The following figure shows the architecture of the storage.
\begin{figure}[htb]
\begin{center}
\includegraphics[width=.8\textwidth]{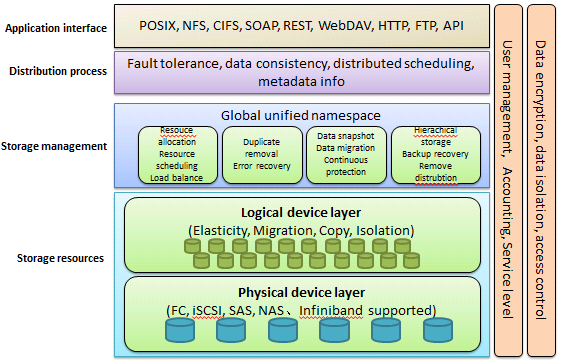}
\caption[Storage architecture]{\label{fig:storage}The storage architecture }
\end{center}
\end{figure}

The logical device layer, which is over the physical layer, provides stable and reliable storage for the upper layers. The logical storage devices are independent of the physical devices and isolated from them. They can be elastically extended or narrowed, copied and be migrated among physical devices. The features of logical devices, including high availability, high performance and high security, make it easy to satisfy the storage demands of JUNO. The logical device layer gives the storage system strong scalability with PB level capacity.

The storage management layer provides the functions of resource allocation, scheduling and load balancing. It provides a global unified name space, called the virtual storage layer. This can shield the heterogeneity of physical devices and integrate a unified virtual storage resource for users.
The distribution process layer is over the storage management, providing fault tolerance, consistency, distributed task scheduling support (such as map-reduce \cite{mapreduce}) and metadata management.

The application interface layer provides rich interfaces compatible with Posix semantics, and is transparent to users.

This virtual distributed storage system should fully satisfy JUNO's requirement of 10 PB storage, providing a high-reliability, high-performance storage service which can be extended elastically as needed.
.
\subsubsection{Distributed computation}
Since JUNO has a large computation scale, it needs cooperative work among the
group members. Distributed computation integrates heterogeneous resources from
different sites, which could be shared by all the sites located in different
region. A unified user interface would be provided to receive user jobs and job
is dispatched to the suitable sites by the system. Job ``pushing'' and
``pulling'' are the two main scheduling methods.

\subsection{Maintenance and service support}
\subsubsection{Network support platform}
Network support platform provides professional and comprehensive solutions for JUNO network and public services. On the one hand, it integrates related resources on the network side, and then transfers them to the expert of  the user service, which can realize quick response to network complaints, and effectively raise user service front-end settlement ratio, reduce complaint processing links, reduce complaint processing duration, and enhance user satisfaction.On the other hand,as for the information related to network complaint received from user sides, the network side will obtain it automatically and pay active attention to it, and realize the comprehensive correlation between user complaint and network execution situation, and the quality degradation warning of terminal-to-terminal network, so as to guide network optimization, enhance network quality, and further realize active care for users and enhance service quality.

\subsubsection{Fine-grained monitoring and maintenance}
Since JUNO data process is a large scale task including many devices, a fine-grained monitoring system is necessary to guarantee smooth running of the platform. It is required that the monitoring system be real-time, easy to use and can recover from some unexpected errors itself. For those errors which can not be recovered, a warning should be sent in time.

\begin{figure}[htb]
\begin{center}
\includegraphics[width=.8\textwidth]{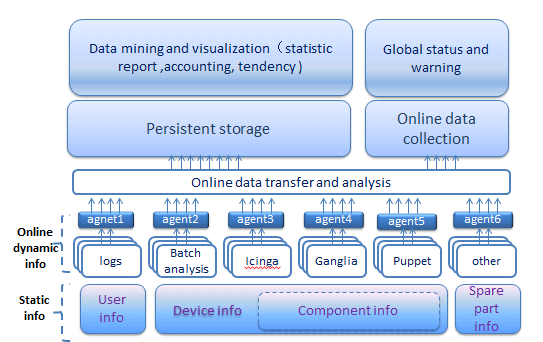}
\caption[Monitoring tool architecture]{\label{fig:monitor}Fine-grained monitoring architecture }
\end{center}
\end{figure}

\subsubsection{  Information Service Platform}
The information Service Platform is an information system which provides services for the research and management activities of the JUNO experiment. The Information Service Platform will provide system services for many aspects of the experimental research and management, such as the JUNO Experiment website, wiki, conference management, document management, collaboration management, video conference, workshop live broadcasting/recording, remote shifts, personal data sharing, science advocacy and so on.

\section{Software Development and Quality Control}
The JUNO offline software system is composed of several sub-systems such as the framework, event generator, detector simulation, calibration, reconstruction, physics analysis and so on.

Each sub-system contains one or more packages,which are divided according to their functionalities. In principle, they are relatively independent in order to allow easy dynamic loading at the run time. The relationship between the necessary dependencies are implemented using the Configuration Management Tool (CMT) which is also used to compile each package.

Developers from different institutes and universities are collaborating together to write code in parallel. Subversion (SVN) \cite{svn} is deployed as the code repository and version control system to track the development of the offline software and tag it for releases. One central SVN server has been set up. All users and developers can easily check out the required release version and commit modifications according to their privileges.

Trac \cite{Trac} is an enhanced wiki and issue tracking system for software development projects. Trac and SVN can be used successfully working together to provide cross searching and references.

Some specific tools, testing algorithms/packages will be created with the development of the offline software in order to test and monitor its functionality and make sure that the whole software is going in the correct direction. Every time a new version is released, a complete test will be performed and a report given showing the results of some characteristic quantities, including CPU consumption and memory usage.
Several kinds of documentation, such as Wiki \cite{Wiki} pages and DocDB \cite{DocDB} will be provided.The Wiki pages will generally record the objectives and progress of each subsystem, while the DocDB will store technical documentation and status reports.

\cleardoublepage
\chapter{Civil Design and Facility}
\label{ch:CivilConstruction}

\section{ Experimental Site Location and Layout}

The location of the JUNO experiment and its supporting facilities should meet the following criteria:

\begin{enumerate}
\item The site should have an equal distance from both the Yangjiang and the Taishan Nuclear Power Station with a largest possible overburden;
\item The site plan and support facilities should be comprehensive for safety, access to the site, and ease in managing the logistics of equipment and personnel above and below the ground for construction and operational considerations of the experiment;
\item The site engineering design and construction plan should be comprehensive in its consideration for energy conservation and emissions reduction, along with minimizing the impact on the local ecological environment;
\item The site should have readily available access to adequate utility water and electric power;
\item The surrounding transportation infrastructure of road should be reasonable for transportation of experimental equipment to the site. The road on site should be comprehensive in its consideration for easy equipment transportation;
\item The site plan for the above ground campus should be comprehensive in its design consideration for seamless and harmonious integration with the local topography, landforms and foliage for civil construction, experimental installation and onsite workforce;
\item All the design should meet applicable national building and construction codes and specifications for civil construction should be followed.
\end{enumerate}

\subsection{ Experiment Site Location}
The site location is driven by the physics requirements to be
optimally equidistant from both nuclear power complexes, along a
central axis determined by the latitudes and longitudes of the (6)
reactors at Yanjiang Station and (4) reactors in Taishan Station.
The site, with surface and underground facilities, and experimental
detector should be located in an area approximate 2,000~m in length
and 200~m in width along this central axis at a distance approximate
53~km from the Yangjiang and Taishan Nuclear Power stations for the
baseline physics requirements.  In addition, the experimental hall
should be at a depth greater than 700~m. The lithology and rock
formation should be carefully considered for optimum stability,
along with mountain topography to maximize rock overburden.

\begin{figure}[htb]
\begin{center}
\includegraphics[width=0.8\textwidth]{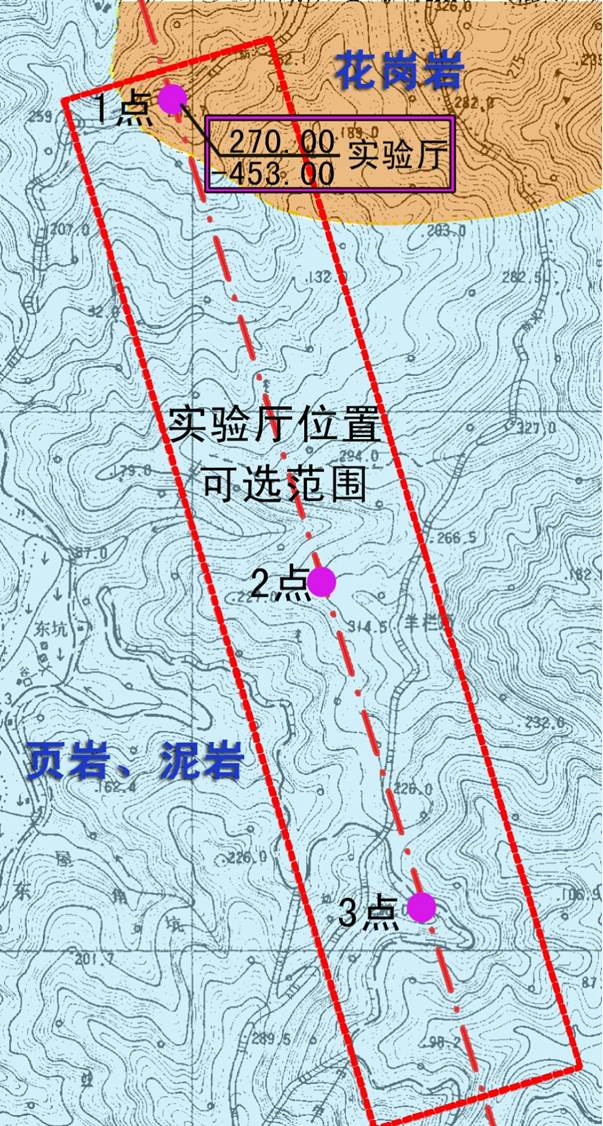}
\caption{ The site region (red dash line) required by Physics,
orange shadow is granite area } \label{Fig10-1}
\end{center}
\end{figure}

\begin{figure}[htb]
\begin{center}
\includegraphics[width=0.8\textwidth]{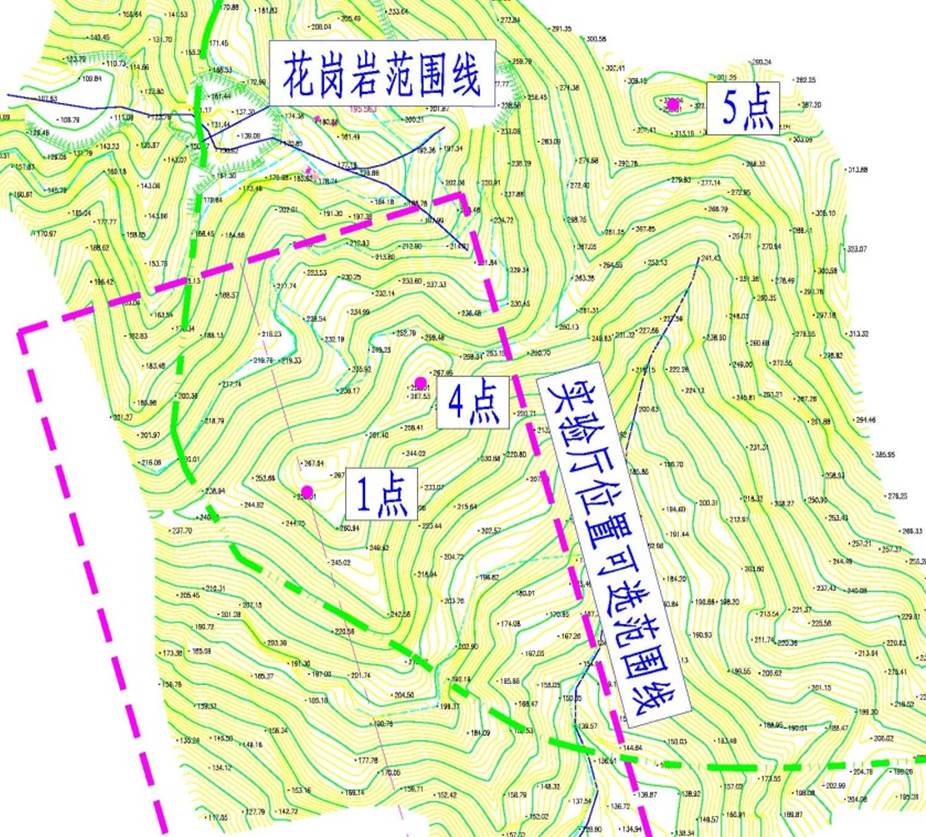}
\caption{ Experimental Hall Location (position 4) in the granite
area with green dash line} \label{Fig10-2}
\end{center}
\end{figure}

Prospective site locations along the central axis are shown in
Fig.~\ref{Fig10-1} and Fig.~\ref{Fig10-2}, at points (1-3) with
elevations of 230~m, 210~m and 205~m respectively. In view of the
changes of boundary condition of the Mesozoic granite stocks
intruding into the Paleozoic rock mass, point (4) at 268.8 m
elevation away from this boundary of granite was selected. This
preliminary location for the experimental hall is far enough away
from the granite boundary to offer adequate safety margin.

\subsection{ Experiment Hall Layout}
With site location determined, access options to the underground
experimental hall must be considered. Life safety and emergency
egress require that at least two independent access paths from the
underground experimental hall to the surface should be available.
With this requirement in mind, the site plan layout for underground
access is either: "vertical shaft and inclined shaft" or "vertical
shaft and normal tunnel". Both options would have the same vertical
shaft with cage-lift, but the normal tunnel with an 8\% grade slope
would be 6~km in length as compared to an inclined shaft at a grade
slope of 42.6\% and 1.34~km in length. The normal tunnel
construction cycle is too long and costly with transportation safety
issues using motorized vehicles for access. The inclined shaft
option was selected for its short construction cycle, lower cost,
and the use of rail cars for equipment transport underground
provides a safer and controlled environment.

\begin{figure}[Fig10-3]
\begin{center}
\includegraphics[width=0.8\textwidth]{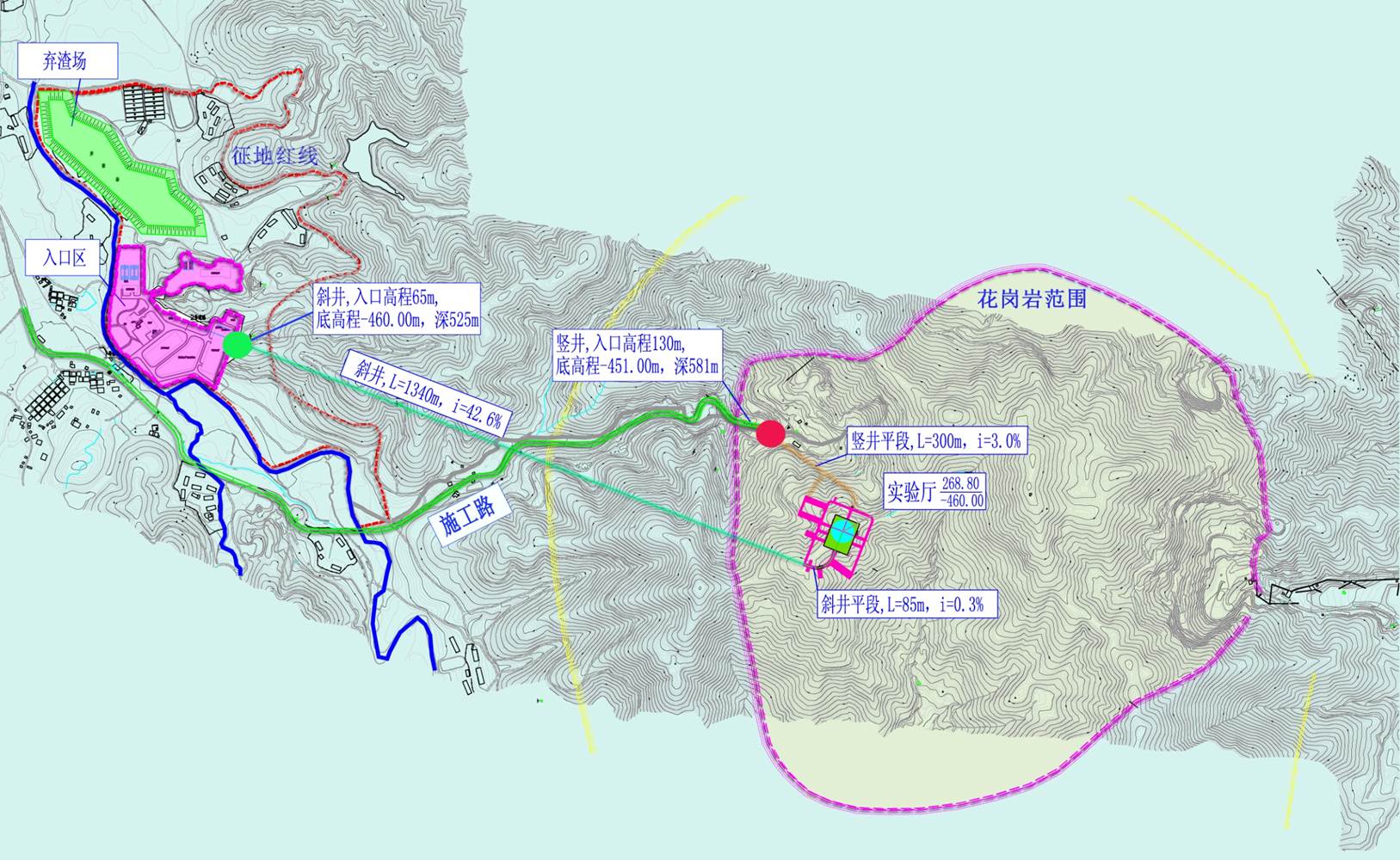}
\caption{ Layout of the Site }
\label{Fig10-3}
\end{center}
\end{figure}

Fig.~\ref{Fig10-3} is a general layout of the site and facilities
with the  "vertical shaft + inclined shaft" scheme. Here the
vertical shaft is 581~m in depth; the inclined shaft is 1,340~m in
length with a slope of 42.6\%; and the overburden at the
experimental hall is 728.8~m. The access entry to the inclined shaft
will be close to the experimental hall, located in the granite
mountain area. All of these factors are favorable for civil
construction and stability of the rock. A site survey was conducted
to determine the optimal location of both access shafts. The
vertical shaft is located at the abandoned quarry. Impact of this
site during the construction and the operation to the surrounding
area, traffic conditions and local residents, have to be studied
well, as well as the construction cycle. To facilitate the
construction and reduce the impact to the local residents, the
construction yard and rock disposal area should be arranged near the
entry of shaft construction site. With full consideration of these
site conditions and other factors, such as difficulties in land
acquisition, etc., the entrance portal of the inclined shaft was
determined to be at Shenghe Village, Jinji Town, Kaiping City.

The entrance portal of the inclined shaft (Fig.~\ref{Fig10-3})
includes permanent buildings, a construction yard area, and a waste
disposal area. All stone debris will be disposed in the waste
disposal area, along with that from the vertical shaft. The waste
transportation route to the waste disposal area does not pass
through neighboring villages or poultry farms, to limit the impact
to the local life and economy.

\subsubsection{ Underground Cavities }
\begin{figure}[Fig10-4]
\begin{center}
\includegraphics[width=\textwidth]{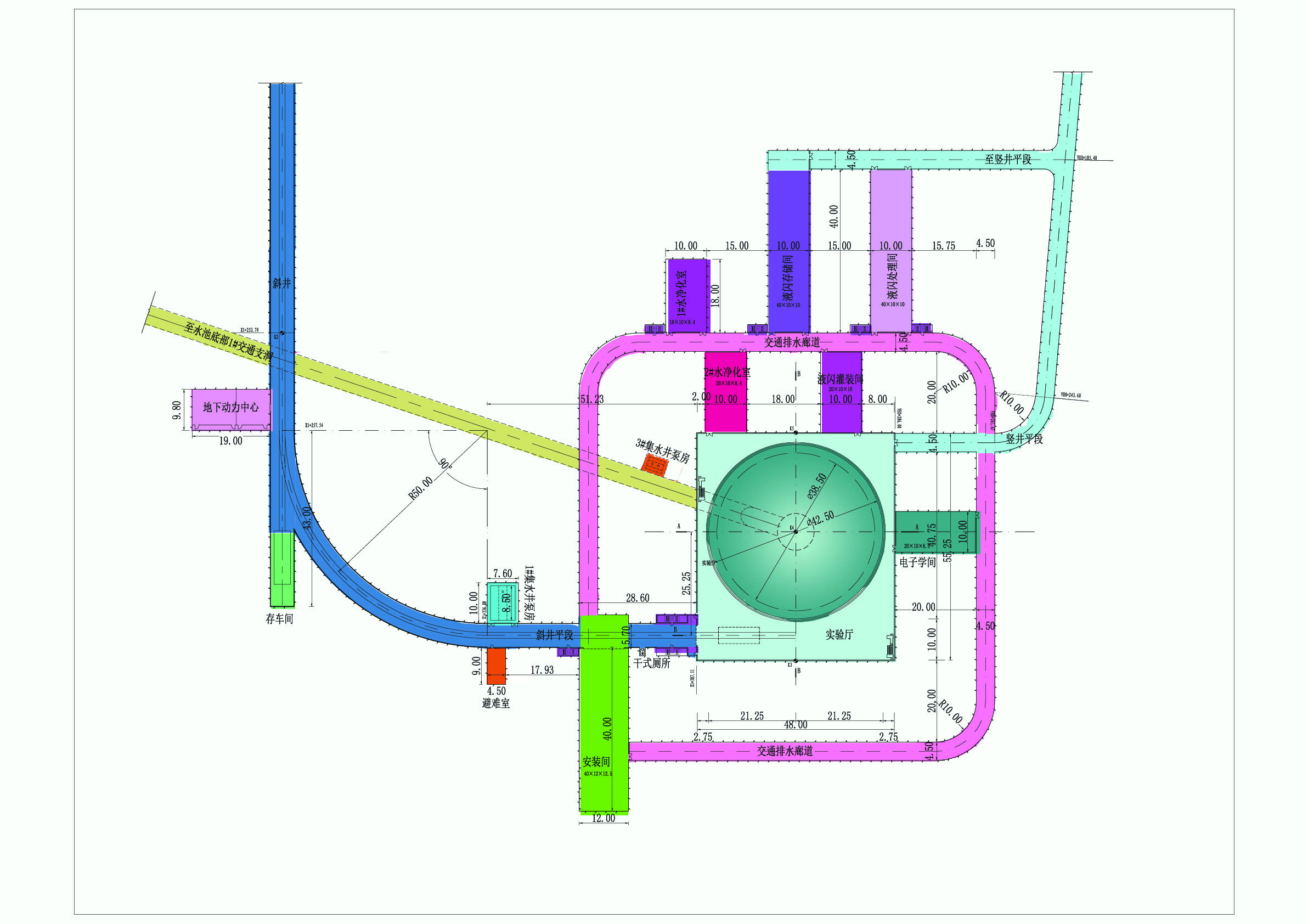}
\caption{ Layout of the Underground Facility }
\label{Fig10-4}
\end{center}
\end{figure}

The layout of the underground facilities (Fig.\ref{Fig10-4}) includes a cavity for the main experimental hall, ancillary rooms with connecting tunnels.

\begin{figure}[Fig10-5]
\begin{center}
\includegraphics[width=\textwidth]{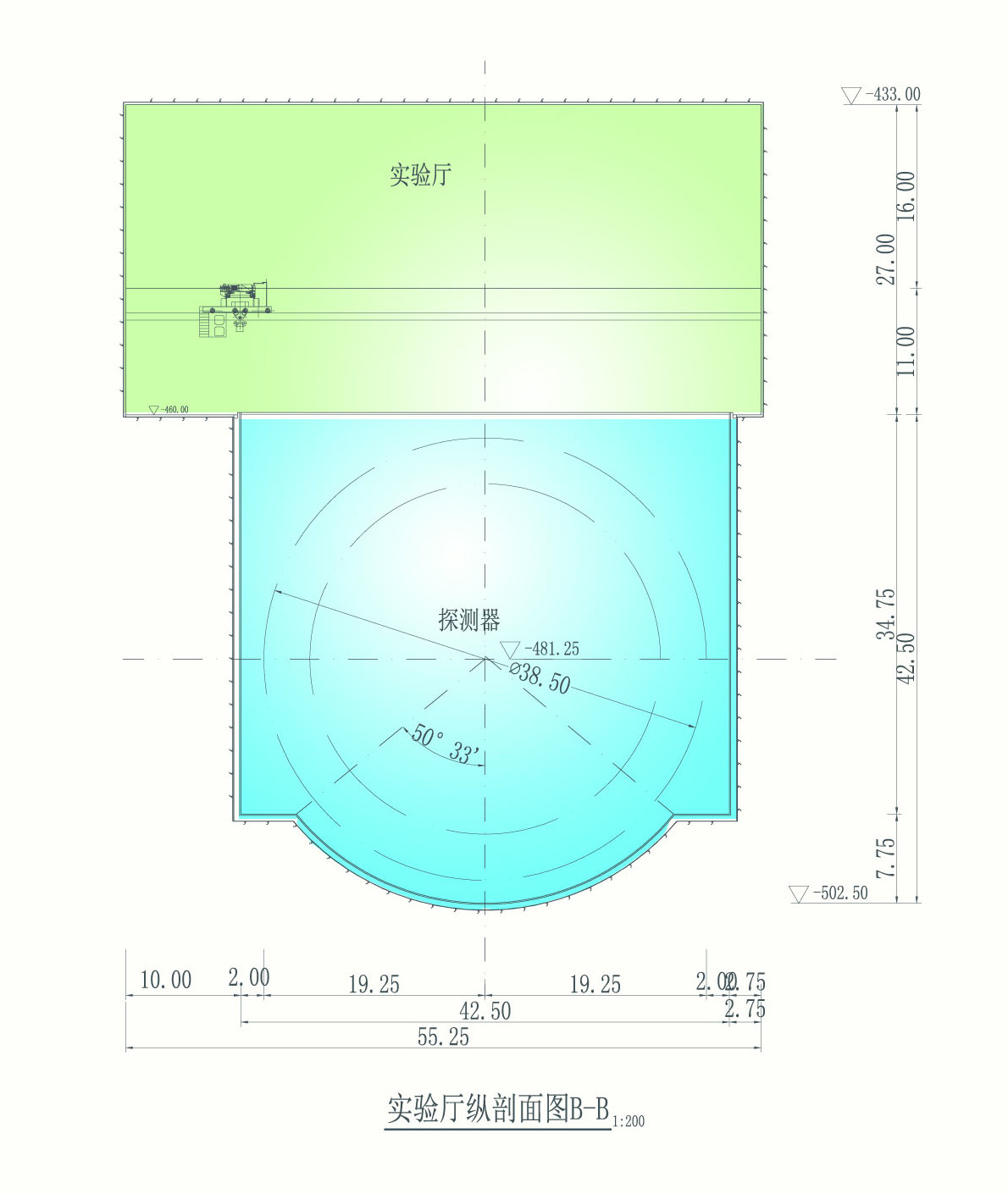}
\caption{ Sideview of the Experimental Hall }
\label{Fig10-5}
\end{center}
\end{figure}

The experimental hall contains a spherical detector tank 38.5~m in diameter located in a cylindrical water pool with inner diameter of 42.5m, as shown in Fig\ref{Fig10-5}. Shape of the experimental hall ceiling, either a dome or a vault scheme, was studied. Criteria such as workflow, construction cost and schedule, rock stability, experimental equipment installation and operation were considered. Vault scheme was selected with two crane bridges on a common rail to span installation hall to water pool. A 2.75~m wide walkway on both side of pool is chosen, and the span of upper vault structure becomes 48~m. A 10~m wide area at the entry of experiment hall will be used for equipment un-loading, resulting in a hall 55.25~m long. The total height of the hall from vault apex to the bottom of the water pool will be 69.5m.

The underground assembly hall, pump room, and electrical equipment room are arranged at the same elevation as the inclined shaft entry into the experimental hall. The assembly hall is right next to the experimental hall. This hall is 40~m in length, 12~m span, and 12.5~m height and connected with the inclined shaft through a tunnel with cross-section of 5.7~m in width and 6.5~m in height. The rail car can transport the equipment directly to the experimental hall or to the door of assembly hall. Two bridge cranes on the same rail, each equipped with two 12.5~ton hooks, are arranged in the experimental hall, for handling of experimental equipment. An additional 10~ton crane is located in the assembly hall, for handling of the cargo and equipment from the inclined shaft rail car.

Liquid scintillator storage room, processing room, and filling room are located at the same elevation and opposite to the assembly hall towards the vertical shaft at a distance far enough away from the experimental hall for fire safety consideration. Electronics room, water purification room, gas room, and refuge room are also arranged around the experimental hall. A single access/drainage tunnel with 4.5~m in width that encompasses the experimental hall interconnects these ancillary rooms.

The inclined shaft being 1,340~m in length, 5.7~m in width, and 5.6~m height, bottoms out at elevation -460~m. From there it proceeds at the same level to the experimental hall by way of an 85~m long interconnecting tunnel with a turning radius of 35~m. The slope of the inclined shaft at 42.6\% ($23.08^{\circ}$) is consistent with requirements for debris transportation during the excavation and rail car vehicle specification, while minimizing the length of the inclined shaft. The transport equipment in the inclined shaft is a single-drum cable winch, which operates at a speed of 4~m/s. A walkway is located on one side of the inclined shaft for the maintenance personnel to carry out patrol check and for underground evacuation in case of an emergency. Utilities such as cable bridges, fire water pipeline, air conditioning water pipeline, and liquid scintillator pipeline, etc. are secured on the side walls of the inclined shaft, while air conditioning conduit and air supply conduit, etc., are located under the upper crown of the shaft.

The access entry to the vertical shaft is at 130~m elevation. The
shaft is 581~m in depth, with the bottom access entry connected to
the experimental hall through an interconnection tunnel of 300~m in
length at 3\% slope. The cross-section of this tunnel is 4.5~m
$\times$ 5.2~m. A lift-cage elevator is used in the vertical shaft
with a maximum lifting speed of 6~m/s, and a load capacity of 8-10
people. In addition, a vertical ladder stairway is installed for
personnel evacuation in case of emergencies and for equipment
maintenance access. Vent duct and cable utilities also pass through
this vertical shaft.

\subsubsection{ Layout of Surface Constructions }
The ground campus comprises inclined shaft entry area and vertical
shaft entry area. The two areas are connected by a 2.5 km road.

Exhaust room, electrical power room, and an access room are arranged near the vertical shaft entry.

The inclined shaft entry area is a comprehensive area for assembly,
office, dorms, and other service. A rock disposal area is also
arranged in this area. The landscape at the rock disposal area will
be recovered after the construction is completed. The living area is
near the entrance of the campus, and is provided with two houses and
one dorm building with 54 rooms. There is a canteen on site to
provide food service. The experimental office building consists of a
control room, a computer room, meeting rooms, and offices. The
assembly area is arranged at the inclined shaft entry, and comprises
one assembly building (3000 $m^{2}$ ) and temporary storage
buildings. The assembly building is air-conditioned and provided
with a crane to facilitate equipment transportation. In addition,
power station, utility buildings for pure water, LN2, et al are all
in the area. The liquid scintillation area is arranged away from the
living area in the valley area at the access entry. It includes
liquid scintillation storage tanks, purification and nitrogen
facilities.

\section{Geological Survey}

\subsection{Geographical Conditions of the Region}
\begin{figure}[htb]
\begin{center}
\includegraphics[width=\textwidth]{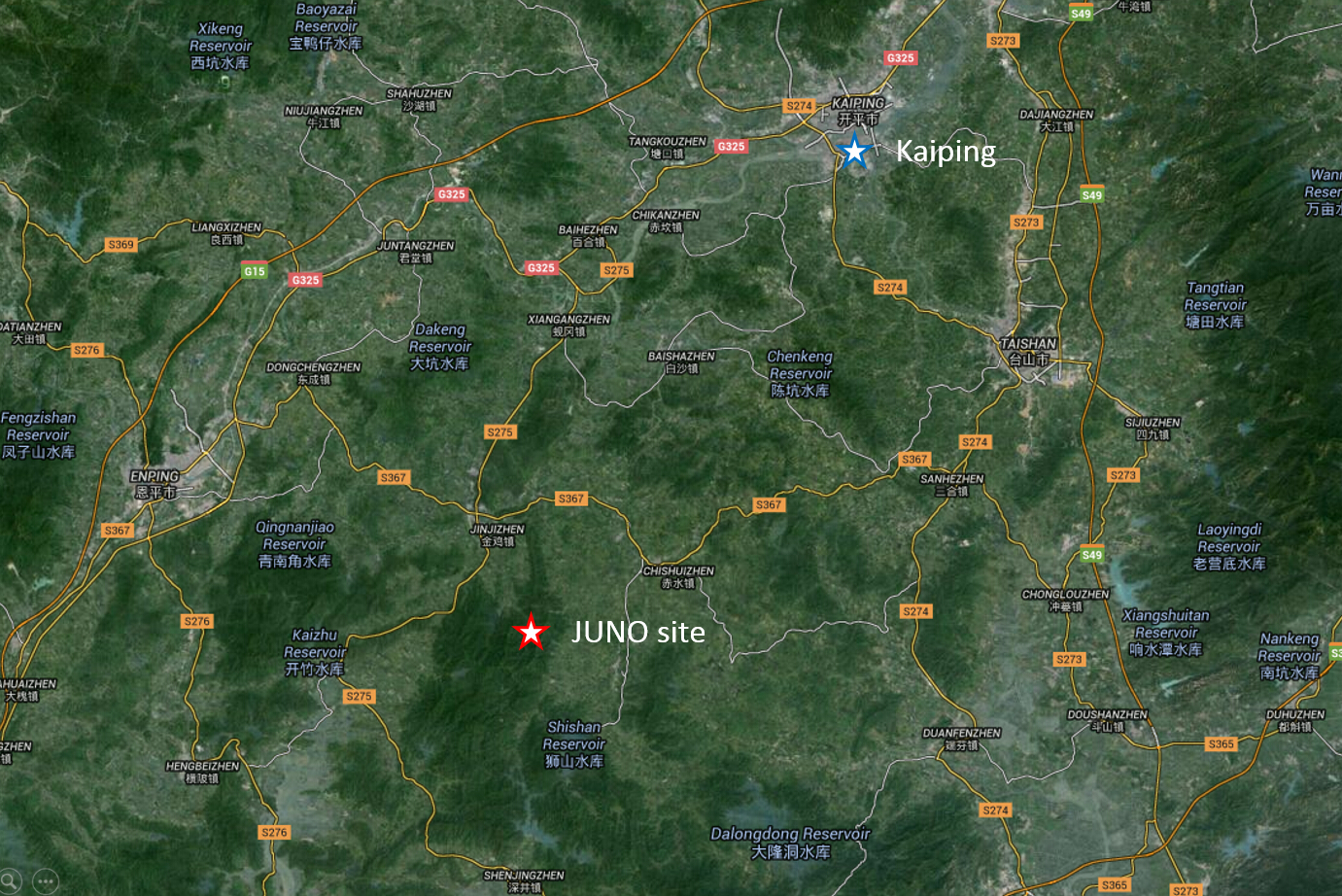}
\caption{ Map of the Project Region }
\label{Fig10-6}
\end{center}
\end{figure}

The site is 5~km away to Jinji Town, which is 40 km away to Kaiping
City (Fig. (\ref{Fig10-6})). There are provincial road S367, S275
and village roads connecting the towns and villages. A new road will
be built to connect the site to the outside.

The underground facility is underneath the Dashi Mountain area in Jinji Town, which has an elevation of 326 meters. The main mountain ridge extending approximately in south-north direction. The elevation of the hill area is about 30m${\sim}$300m while the topographic slope is usually 10$^{\circ}$ to 20$^{\circ}$. The main gulch exists and extends approximately in east-west direction. The depth of gulch cuts is 50 meters to 100 meters, and there is perennial water in the gulch. The largest river in the area, Dongkeng River, originates from Dongkeng Forest Farm and flows across the project area and Shenghe Village. Rivers in the region usually flow approximately in south-north direction. All of rivers are branches of Chishui River. Niuao Water Reservoir (a small reservoir) and several ponds are distributed at the southeast side of the mountain.

The project area is of Southeast Asian tropical maritime monsoon
climate, with frequently strong typhoons in the transition period
from summer to autumn, which bring abundant rainfall. Kaiping City
is surrounded by the plenty of rivers with wide water area. It is
not severely cold in winter and very hot in summer. The climate is
mild and abundant in rain. The annual mean air temperature is
21.7$^{\circ}$C while the annual mean humidity is as high as 82\%.
Annual rainfall varies from 1,700mm to 2,400mm. The rainy season
mainly concentrates from April to September while the dry climate
distributes from October to February. The perennial dominant wind
direction is east. Affected by the subtropical monsoon, the period
from June to October in every year is a strong wind season with a
force of 6-9 from east.

The main active fracture zone in the region is the Enping-Xinfeng
zone, which is at approx. 60 km distance to the project area. A
magnitude-6 earthquake once happened near Mingcheng long fracture
zone. Magnitude-4 or lower earthquakes happened near Kaiping, and a
magnitude-4 earthquake happened in Hecheng area. The rmoluminescence
survey value of the fracture near Gaoming is 245,200 years, and the
the rmoluminescence survey value of the Hecheng-Jinji fracture is
153,300 to 350,000 years. The fractures belong to Mid-Pleistocene
active faults.

According to the earthquake catalogue of China Seismic Network
(CSN), although there were more than 64 earthquakes happened  within
150 km range of project area with magnitude-3.0 or higher since
1970, wherein, only 3 moderate earthquakes higher than magnitude
4.75 happened.  No earthquake higher than 4.0 happened near the
project area (within 25 km range). The earthquake distribution map
in the project region shows that the earthquakes are mainly
distributed in Yangjiang region in southwest. The highest earthquake
is a magnitude-4.9 in Nanhai region on Mar. 26, 1995, which is 100
km away from the project area. Therefore, the impact of earthquakes
on the project area is low.

According to the "Seismic ground motion parameter zonation map of China", the seismic ground peak acceleration is 0.05g in 50 years. The characteristic time of the seismicresponse spectrum is 0.35 s, and the corresponding basic seismic intensity is degree VI.



\subsection{Geological Survey}
A detailed geological survey for the inclined shaft, vertical shaft, and experimental hall areas of the experimental station was carried out in 2013 to ascertain the geological conditions of the main constructions in the project region and provide a geological basis for engineering design and construction and raise proposals for handling main geological problems.

\begin{figure}[htb]
\begin{center}
\includegraphics[width=\textwidth]{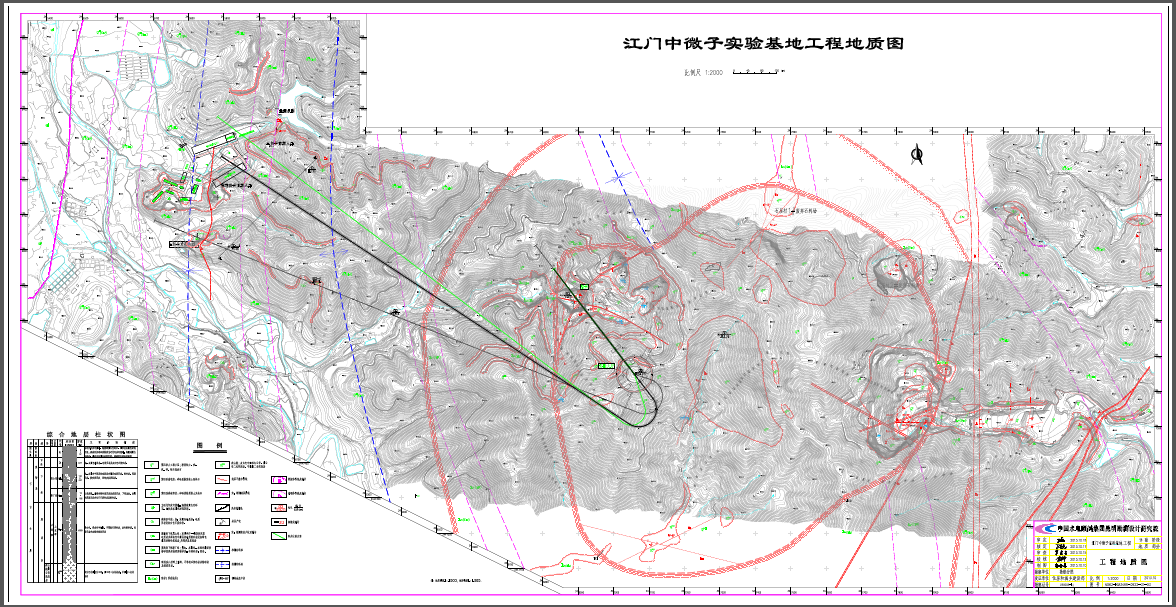}
\caption{ Engineering Geological survey in the Experiment Region }
\label{Fig10-9}
\end{center}
\end{figure}

The survey result is summarized in Fig. {\ref{Fig10-9}}. There are
Yanshanian granite stocks at the location of the experimental hall.
The granite mass intruded into the Palaeozoic sand rocks. The
contact zone between them is a hornfelsic zone. Details are
described in the following.

\subsubsection{Engineering Geological Conditions of the Inclined Shaft}
\begin{figure}[htb]
\begin{center}
\includegraphics[width=\textwidth]{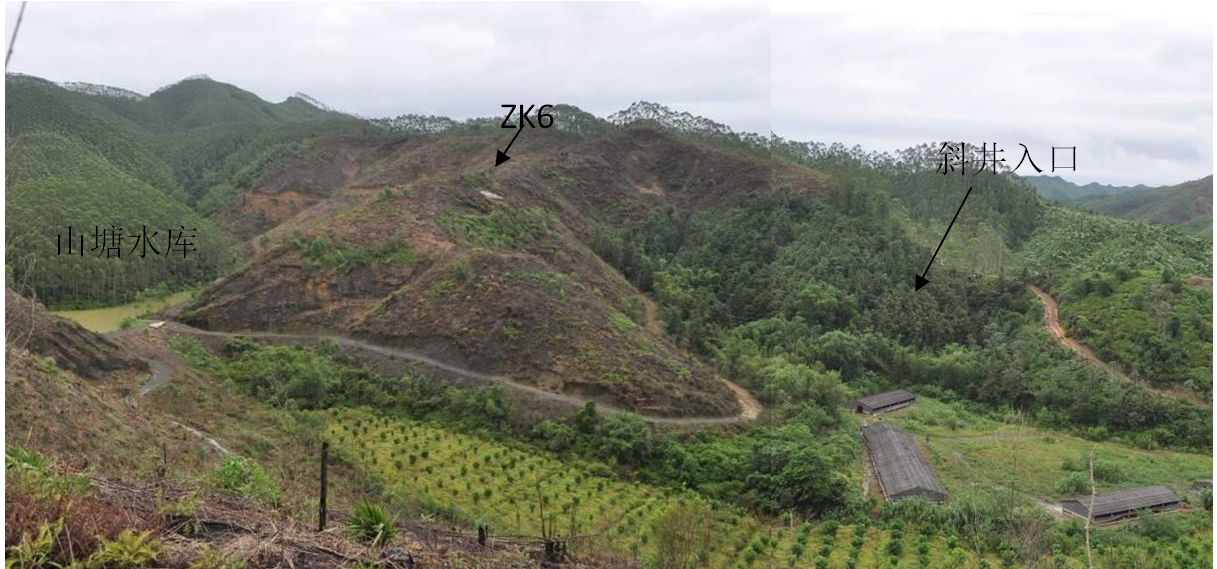}
\caption{ Topography near the Access Entry of Inclined Shaft. The underground facility is at left. The entry of inclined shaft is at right. }
\label{Fig10-10}
\end{center}
\end{figure}

The access entry of the inclined shaft is in a mild gulch with width of 400~m in northeast to Shenghei Village. Vegetation near the access entry is dense. The gulch in front of the access entry is open and favorable for organization of construction and arrangement of surface buildings (see Fig. \ref{Fig10-10}). The formation penetrated by the inclined shaft mainly consists of Ordovician Xinchang ($O_{1x}$), Cambrian Bacun (${\in}bc^c$), hornfels and granitic intrusions.

There is a water reservoir (Shantan Water Reservoir) at the upstream
of the main gulch at the access entry of the inclined shaft (see
Fig. \ref{Fig10-10}). The water reservoir has only $5\times10^4 m^3$
storage capacity, which has no feeding river at its upstream. The
water is mainly supplied by rainfall and ground water discharge. The
catchment area of the water reservoir is approximately $0.3 km^2$.
Therefore, the water inflow into the water reservoir is not
dangerous even in the rainy season, and will not cause severe flood
discharge. There are two floodways behind the dam of the water
reservoir. The minimum distance from the floodways to the access
entry of the inclined shaft is approx. 130 meter. The elevation of
the floodways is lower than that of the access entry by approx. 5
meter. The area at the downstream of the flood discharge area is a
lower and wider plain terrain. In summary, the flood discharge from
Shantang Water Reservoir in the rainy season will have no impact on
the access entry of the inclined shaft.

From the geological survey, the covering layer at the access entry is silty clay gravel soil in $1m\sim3m$ thickness. The bedrocks are mainly Xinchang Fm($O_{1x}$) gray and dark gray siltstone mixed with feldspar-quartz sandstones and clay shale stones in $N6^{\circ}E$, $SE{\angle}58^{\circ}$ occurrence. The bedrocks are completely weathered, locally intensely weathered, and fractured in fractured-loose structure. The condition is not perfect for shaft construction. Slope protection and water drainage should be considered.

\begin{figure}[htb]
\begin{center}
\includegraphics[width=\textwidth]{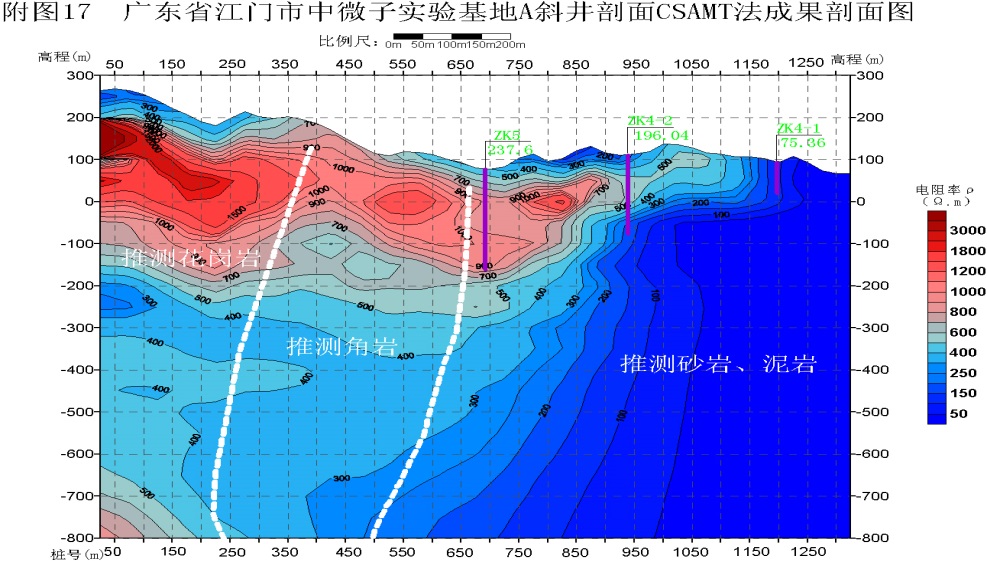}
\caption{ Geophysical Investigation of Inclined Shaft }
\label{Fig10-11}
\end{center}
\end{figure}

The classification of the tunnel enclosing rocks is predicted and evaluated as follows (see Fig. \ref{Fig10-11}):

\begin{enumerate}
\item 1413m${\sim}$1350m section:  The enclosing rocks near the tunnel entrance are mainly silt rocks with locally distributed quartz sandstones and clay shale stones. They are fully or intensely weathered while some are moderately weathered. Joints and fractures are developed and filled with silt. The rocks are mainly in silt filled cataclastic structure while some are in mosaic cataclastic structure. The crown and side walls have water seepage or water dripping. 
The quality rating of the enclosing rocks is grade V. The cavity stability is very poor without self-stabilization time or with very short self-stabilization time.

\item 1350m${\sim}$1283m section: the tunnel enclosing rocks in the region are mainly silt rocks and quartz sandstones with moderately or slightly weathered Joints and fractures. The rocks are mainly in mosaic cataclastic structure while some are in layer structure. The crown and side walls have water seepage or water dripping. 
The quality rating of the enclosing rocks is grade IV. The crown enclosing rocks are unstable, and the self-stabilization time is short. Various large-size deformations and damages may occur.

\item 1283m${\sim}$808m section: The tunnel enclosing rocks are mainly quartz sandstones, with silt rocks distributed locally, and are slightly weathered. Joints and fractures are developed. The rocks are mainly in mosaic cataclastic structure while some are in layer structure. The crown and side walls have water seepage or water dripping. 
The quality rating of the enclosing rocks is grade III. The crown enclosing rocks have poor local stability.

\item 808m${\sim}$418m section: The rocks are hornstones and slightly weathered. Joints are developed or undeveloped. The rocks are mainly in layer structure. The crown and side walls have water seepage or water dripping. 
the quality rating of the enclosing rocks is mainly grade III, and grade II locally. The cavity enclosing rocks have high overall stability.

\item 418m${\sim}$0m section: the rocks are granite rocks and slightly weathered. The joints are undeveloped. The rocks are mainly in block structure. The crown and side walls have water seepage or water dripping. 
The quality rating of the enclosing rocks is grade II. The cavity enclosing rocks have high stability and will not have plastic deformation.
\end{enumerate}

Attention should be paid to water abundance in the sandstones as
indicated on the cross section of geophysical prospecting along the
route (A) near the ditch of Dongkeng River in the original inclined
shaft route (see Fig. \ref{Fig10-11}).

\subsubsection{ Engineering Geological Conditions of the Vertical Shaft}
The vertical shaft is located in the Dongkeng Quarry. It is 589 m in depth and 6 meter in diameter.

\begin{figure}[Fig10-12]
\begin{center}
\includegraphics[width=0.8\textwidth]{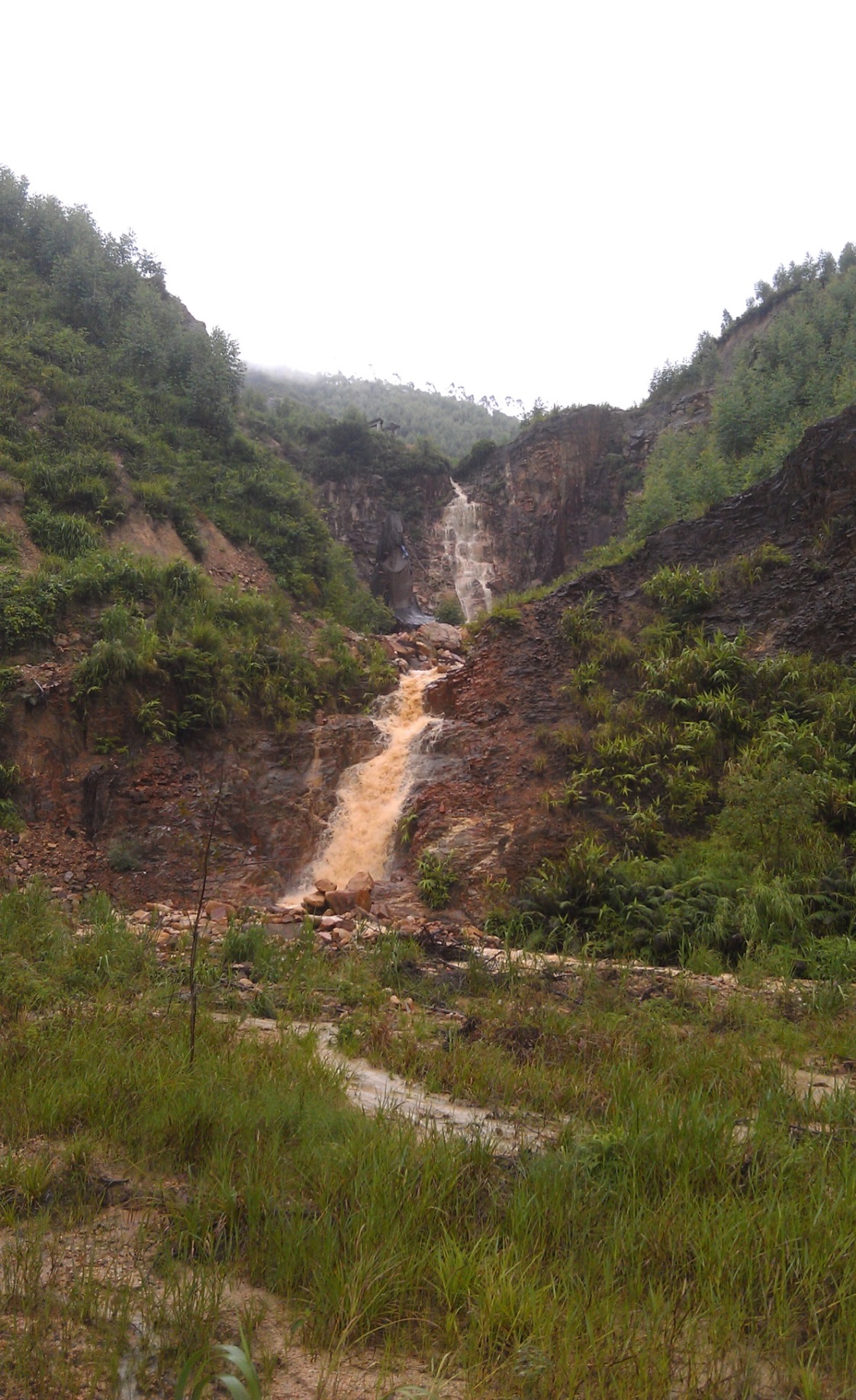}
\caption{ Topography at the Entry of the Vertical Shaft }
\label{Fig10-12}
\end{center}
\end{figure}

The vertical shaft is located in a quarry and at the end of the
gulch (see Fig. \ref{Fig10-12}). The upstream catchment area of the
gulch is approx. $0.1 km^2$. In the rainy season, severe torrential
flood will be produced in the gulch. Water drainage and special care
must be considered during construction and operation.

The vertical shaft is in granite mountain. An exploring borehole ZK1
was drilled at the location of vertical shaft for the geological
survey. The borehole was deeper than bottom of water pool in the
experimental hall. Various borehole tests were carried out,
including terrestrial stress test, high pressure water test,
specific resistance test, natural gamma test, borehole televiewer
test, acoustic wave test, and ground temperature test, etc.

The vertical shaft is near the granite intrusion boundary. In the area of vertical shaft, the joints, fractures, and faults are developed. 5 exposed faults in different sizes are distributed nearby. Eleven or more faults or compresso-crushed zones in different sizes were unveiled in the borehole.

\begin{figure}[Fig10-13]
\begin{center}
\includegraphics[width=\textwidth]{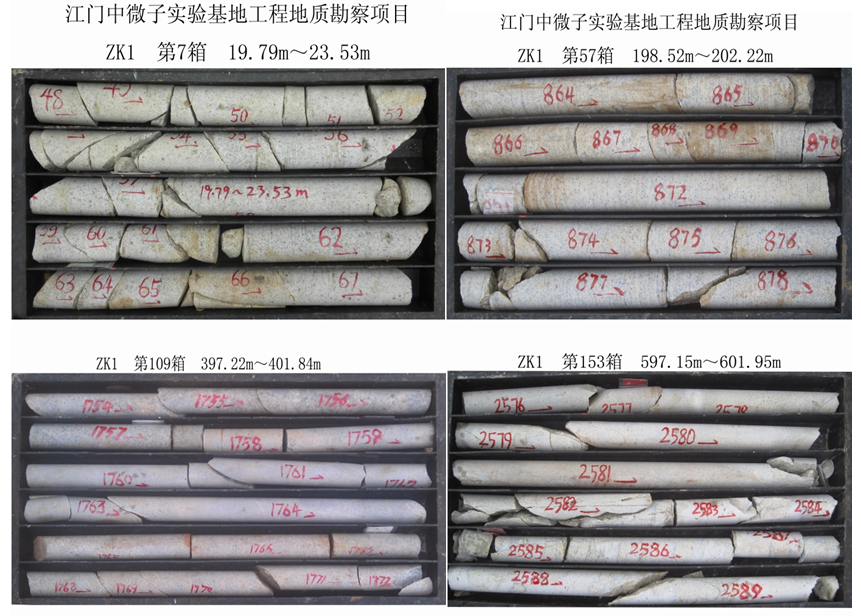}
\caption{ Some Core Samples from the exploring drill well in the
Vertical Shaft } \label{Fig10-13}
\end{center}
\end{figure}

The classification of the vertical shaft enclosing rocks (Fig. \ref{Fig10-13}) is predicted and evaluated as follows:

The rock quality of vertical shaft entry section with depth of 10
meters is grade IV. The granite rocks are moderately weathered.
Joints and fractures are well developed. The section is in mosaic
cataclastic structure or block structure.

The shaft depth to the -118 meter, the rocks are mainly slightly weathered or moderately weathered rocks in some regions. As unveiled in the borehole surveying work, there are several small developed faults and compressive structural planes (grade IV structural planes). The rocks are mainly in sub-block structure or in block structure locally at some places. The quality rating of the enclosing rocks is grade III.

The rocks of shaft bottom section are slightly weathered or fresh. The complete section is in perfect block structure. The quality rating of the enclosing rocks is grade II. The shaft wall enclosing rocks have high overall stability and will not have plastic deformation.

The water inflow in the vertical shaft is predicted with Dupuit formula. According to the water pressure test result, the permeability factor of the rock mass of the vertical shaft is approximately 0.009 m/d. The shaft is approx. 589 meters in depth and 3 meters in radius. The calculated water inflow in the vertical shaft is estimated $1473m^3/d$.

There is only one concern that a fault of F2 may cross the horizontal tunnel between bottom of vertical shaft and experimental hall. The rock in the F2 region is grade IV. According to the survey, the F2 fault is a compressional-shear shifted reversed fault, in $N20^{\circ}{\sim}55^{\circ}E$ and $SE{\angle}59^{\circ}{\sim}90^{\circ}$ occurrence. The occurrence variation is severe. The fault is mainly developed in granite mass, in 0.5 meters${\sim}$3.0 meters width, and extending approximately 400 meters. On the ground surface, the fault zone mainly consists of flake rocks, mylonite, cataclastic rocks, and lens with inclined scratches.  The cementation is poor to moderate. As unveiled in the ZK1 borehole, the deep fault zone is mainly filled with quartz. The cementation is good to very good and is usually in healed form. The buried depth of the F2 fault is approx. 660 meters. It is speculated that the cementation in the deep part of the fault is good to very good. The F2 fault area has no surface water and is close to the dividing ridge in which the rock mass is thin and has weak water permeability. In summary, it is speculated that the F2 fault zone has weak water permeability while the water inflow is low. The fault will not have severe water inflow at the level the tunnel is excavated to the F2 fault.

\subsubsection{ Engineering Geological Conditions of the Experimental Hall }
The experimental hall is located under a small hill at the southeast side of the vertical shaft. The surface elevation is 268 m, and the buried depth is approx. 763 m. The profile of hill is a mountain ridge. The experimental hall is in the granite mass, which consists of gray medium-fine grained adamellite. In the plan view, the experimental hall is relatively close to the central zone of the intruding small stocks.

\begin{figure}[Fig10-14]
\begin{center}
\includegraphics[width=\textwidth]{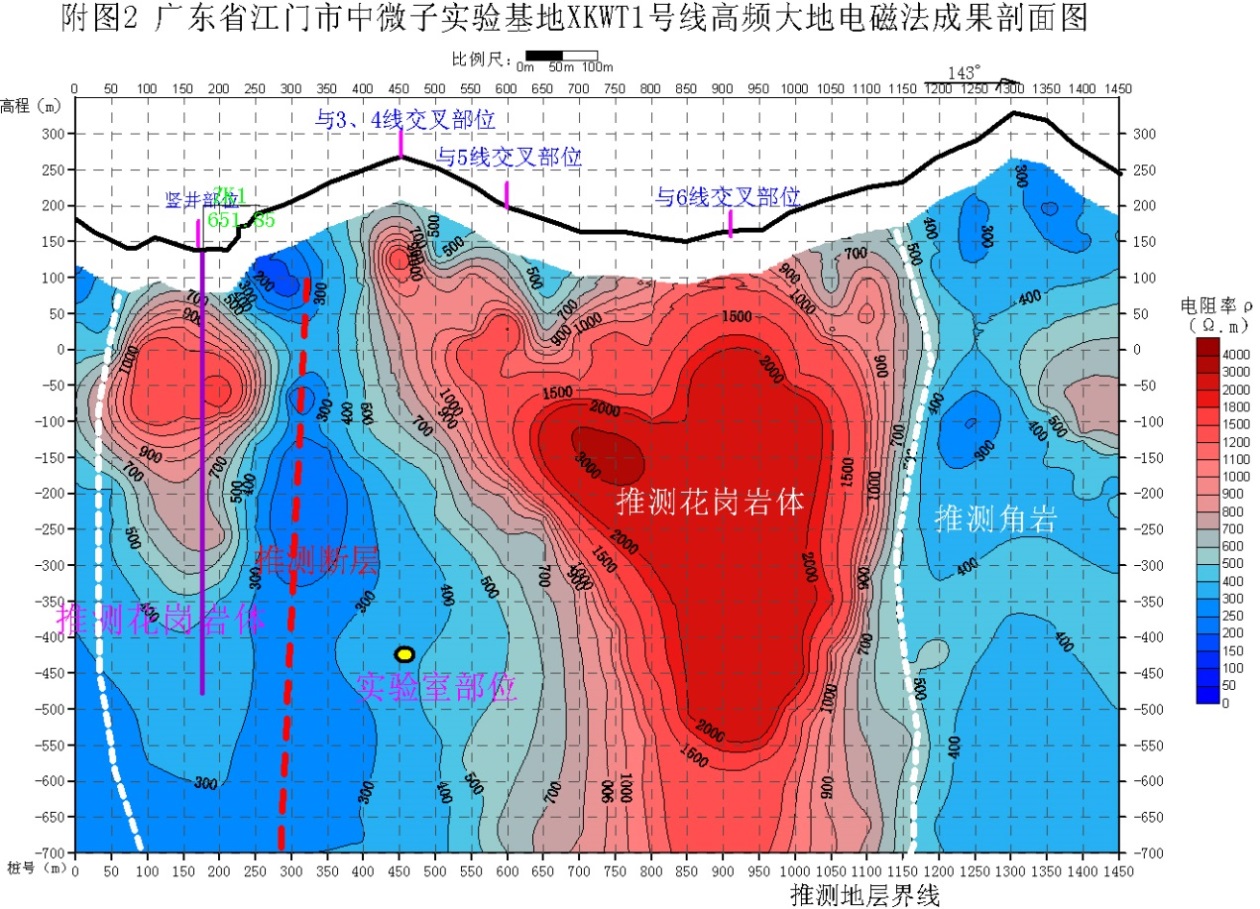}
\caption{ Geologic Diagram Obtained with the High-Frequency
Magnetotelluric Method near the Experimental Hall } \label{Fig10-14}
\end{center}
\end{figure}

A geophysical prospecting diagram of the experimental hall and the vertical shaft area obtained with high-frequency magnetotelluric method is shown in Fig.~\ref{Fig10-14}. Both the vertical shaft area and the experimental hall area consist granite mass. It is speculated that there are faults in the horizonal connecting tunnel between the vertical shaft and the experimental hall.

The major engineering geological issues include:

{\bf Terrestrial Stress and Rock Burst Analysis}
According to the laboratory test, the volumetric weight of slightly weathered (or fresh) granite is approximately $26.6{\times}10^3kN/m^3$, the saturated uniaxial compressive strength is 80 MPa${\sim}$120 MPa as usual. Since the maximum buried depth of the experimental hall is 763 meter, according to the terrestrial stress test, it is estimated that the terrestrial stress at the experimental hall is not higher than 20MPa and the rock strength/stress ratio is 4.85. The deep core obtained in the borehole ZK1 has high integrality in whole. The core near the borehole bottom (651m buried depth) is still in long column shape. Since the experimental hall is relatively close to the center of the intruding rock stocks, it is speculated that the rock mass there has high integrality. Through comprehensive analysis, it is speculated that the stress in the experimental hall is moderated, and will not result in severe rock burst.

{\bf External Water Pressure Analysis}
According to the water pressure test and the high-pressure water test, the permeable rate of the granite rock is usually lower than 1 Lu and considered as weak permeability. The difference between the ground water level and the bottom of experimental hall is 577.79 meter. According to the "Code for Engineering Geological Investigation of Water Resources and Hydropower", the external water pressure reduction factor of weak permeable rocks is 0.1${\sim}$0.2. Thus, the effective external water pressure of the experimental hall is only 58 meters${\sim}$116 meters, corresponding pressure is 0.58MPa${\sim}$1.16MPa. The high-pressure water test shows that most of the joints and fracture structure of the granite did not change under the water pressure and are mainly of turbulence type while some of them are of filling type. Dilatation or erosion phenomenon is not found. In summary, the external water pressure at the experimental hall is relatively low, and its impact on the cavity stability of the experimental hall is trivial.

{\bf Cavity Water Inflow Analysis} The water inflow in the
experimental hall is predicted with underground water dynamics
methods. The maximum water inflow in the early stage and the
long-term stable water inflow are predicted with several methods.

According to the water pressure test, the deep granite is slightly
permeable. The maximum water inflow in early stage calculated with
HirishiOshima empirical equation is approximately $395 m^3/d$. It is
$527 m^3/d$ with the empirical formula method. The long-term stable
water inflow was estimated to be $244 m^3/d$ and $120 m^3/d$
respectively with Ochiai Toshiro formula and empirical formula
methods.

In summary, the maximum water inflow in early stage and the long-term water inflow in the experimental hall are low.

\begin{figure}[Fig10-15]
\begin{center}
\includegraphics[width=\textwidth]{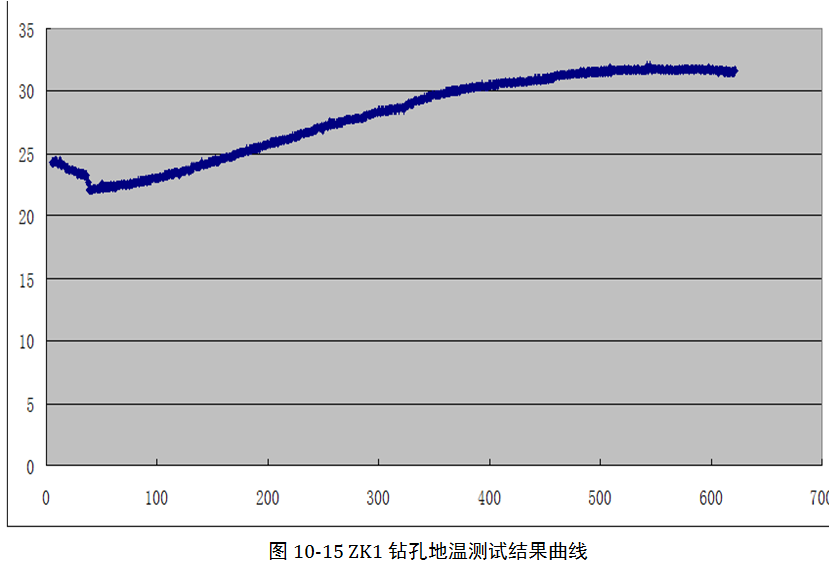}
\caption{ Measured Underground Temperature}
\label{Fig10-15}
\end{center}
\end{figure}

{\bf Underground Temperature} According to the underground
temperature measurement in the borehole, ZK1, the underground
temperature at 450 meter depth (-314 meter elevation) or deeper is
$31{\sim}32 ^{\circ}C$ (see Fig. \ref{Fig10-15}). Therefore,
temperature is not a big issue for the experiment.

{\bf pH value}
The pH value of underground water in borehole ZK1 at the location of the vertical shaft is analyzed to be 7.25${\sim}$7.75 while the pH value of ground water is near 7. It indicates that the deep ground water is medium corrosive.

{\bf Rock Radioactivity}
Table~\ref{tab:rockrad} shows the result of rock radioactivity measurement at different depths in borehole ZK1. All sample results meet the requirements for main building materials (as per "Code for Radionuclides Limits in Building Materials" (GB6566-2010)) except for the external exposure index Ir of three samples, which are slightly greater than 1.0. The rock material meets the criteria for Class B products according to the "Code for Radiation Protection Classification and Control of Natural Stone Material Products".

\begin{table}[H]
\caption{Result of Rock radioactivity Test in Borehole ZK1}
\label{tab:rockrad}
\centering
\scalebox {0.8} {
\begin{tabular}{|c|c|p{1.8cm}|p{1.8cm}|p{1.8cm}|p{1.8cm}|p{1.8cm}|p{1.8cm}|}
    \hline
    Site No. & Sampling  & \multicolumn{6}{|c|}{Radioactivity} \\
    \cline{3-8}
             & Depth               &
      Specific Activity CRa of $^{226}Ra$ (Bq/kg) & 
      Specific Activity CTh of $^{232}Th$ (Bq/kg) & 
      Specific Activity CK of $^{40}K$ (Bq/kg) &   
      Radium Equivalent Concentration C[e]Ra & 
      Internal Irradiation Index IRa &    
      External Irration Index Ir \\   
    \hline

    FSYY-1 & 299.18$\sim$300.28 & 166.6 & 122.1 & 1066.4 & 425.2782 & 0.8 & 1.2 \\ \hline
    FSYY-2 & 403.11$\sim$404.11 & 155.9 & 110.0 & 1106.3 & 401.7544 & 0.8 & 1.1 \\ \hline
    FSYY-3 & 557.28$\sim$558.38 & 88.6  & 83.6  & 1239.7 & 310.5536 & 0.4 & 0.9 \\ \hline
    FSYY-4 & 616.47$\sim$617.12 & 173.1 & 129.5 & 536.6  & 395.1458 & 0.9 & 1.1 \\ \hline
    FSYY-5 & 633.02$\sim$633.77 & 116.4 & 111.2 & 342.2  & 296.6336 & 0.6 & 0.8 \\ \hline
    FSYY-6 & 650.23$\sim$650.93 & 130.1 & 113.0 & 1062.4 & 376.1412 & 0.7 & 1.0 \\ \hline

    \hline
\end{tabular}
}
\end{table}

\subsubsection{ Stability Assessment of the Enclosing Rock }
The experimental hall is  48 meters wide, and can be divided into two parts: the upper cavity is enclosed by the crown and high side walls (vertical section) that connect the crown part, and is mainly used to accommodate lifting appliances and other facilities (e.g., cable channel, etc.), and, the lower cavity is a cylindrical pool, 42.5 meters in diameter and 42.5 meter in depth. Utility halls are arranged around the experimental hall. The experimental hall is large and complex in structure. The  rock types and long term stabilibty of underground facility should be assessed thouroughly.

The length of the experimental hall is 55.25 meters. The axial is directed in  $N37^{\circ}W$. The cavity enclosing rocks are granite, slightly weathered, in block structure. The developed joints are mainly steeply inclined while locally developed joints are slowly inclined and moderately inclined. The joint planes are straight, coarse, and are usually enclosed without filled material. Some joints are slightly open or fully open and filled with rock debris or kaolin film. There exists local water seepage or water dripping at some places.
The quality rating of the enclosing rock is considered as grade II.

Overall, the cavity enclosing rocks are stable. However, local unstable blocks may be produced under the inter-cutting action of the three groups of steeply inclined joints because of the large span of the experimental hall.

According to the terrestrial stress test, the direction of maximum horizontal stress is $N10^{\circ}W{\sim}N54^{\circ}W$. Since the upper cavity is approximately in a square shape, the terrestrial stress has little influence on the selection of the axial direction. The terrestrial stress difference among different directions are small and has little impact on the stability of the side walls. Since the cavity span is large, the terrestrial stress is in static state while the redistributed stress of the cavities is complex. Stability verification must be carried out and appropriate engineering measures must be taken.

The geological conditions (e.g., lithology of enclosing rocks, joint development, and underground water, et al) in the pool are the same as those of the upper cavity. However, because the main joints are steeply inclined, water pool may form an unfavorable combination with the side walls. However, the pool enclosing rocks are overall stable.

The geological conditions of the ancillary cavities are the same as
those of the experimental hall. The cavity enclosing rocks have high
overall stability. Although the utility cavities are small, at
intersections among the experimental hall and ancillary cavities,
care must be taken for the blasting control in order to keep the
integrality and stability of structure.

\section{Utility Systems}
\subsection{\bf Power distribution and grounding system}
To meet the requirements of having redundant, reliable electrical
power to the site, two 10~kV commercial power transmission lines
will be introduced to the power distribution centers at the entrance
to the inclined shaft and the vertical shaft from 110~kV transformer
substations in Jinji Town and in Chishui Town, respectively. They
belong to different public grids. An underground 10~kV power
equipment room is provided to distribute power through the
experimental hall and ancillary rooms. In addition to these
commercial power sources, the surface power distribution center will
be equipped with a self-starting diesel generator, which will supply
power for emergency lighting, ventilation or waste water sump-pumps
during disruptions of the main power supply.

The total expected demand for power distribution is approx. 5,000~kVA from the combined power supply sources, each of which can provide 4,000~kVA of power. Steady-state power loads include ventilation, air-conditioning, and lighting systems along with operational experimental equipment. However, peak power capability is required during the LS distillation process and its transfer to the detector. An isolated transformer will be installed in the underground power equipment room to distribute clean electrical power on a separate grid to experimental equipment to minimize system noise.

The electrical grounding systems include both a clean ground grid for experimental equipment and a safety ground grid for utility systems. The clean ground grid for experimental equipment on clean electrical power will be connected to the secondary winding of the source isolation transformer. The two grounding grids will have no electrical connection between them. The surface safety ground grid is distributed and transferred to underground with power transmission cables, and is connected to the underground safety ground grid. The clean ground for experimental equipment is distributed locally in the underground experimental hall. It is understood that the underground geological formation mainly consists of granite, having a high electrical resistivity, not conductive for electrical grounding. Based on our experience, the electrical grounding systems should include appropriate measures, such as high-efficiency electrolytic ion grounding electrodes to mitigate electrical safety concerns. This grid should also connect to the water pool steel-bar structure, to improve the connectivity with the ground and to increase the overall capacity, hence further stabilize the ground.

\subsection{\bf Ventilation and air-conditioning system} The temperature in the underground experimental hall shall be maintained at a nominal 22$^{\circ}C$ and the relative humidity shall be lower than 70\%, with adequate ventilation to sustain at least 50 people working 24/7 underground. Air quality in the experimental hall can be normal room air that meets cleanliness requirements for personnel safety. The cleanliness in ancillary halls related to
liquid scintillator storage or processing should meet class-100,000-level which is comparable to the requirements for the Experimental Hall-5 at Daya Bay. Special measures (e.g., high efficiency filters) can be taken during liquid scintillator operations to improve the cleanliness to class-10,000-level.

The air exchange rate in the experimental hall is designed for 6~volumes/day, to reduce radon in the air and maintain a comfortable environment for underground workers.

Water chillier units and fresh air handling systems for supplying dehumidified cool air underground are located at the entrances to both the inclined and vertical shafts. This conditioned air supply flows through insulated ducts in each shaft for dispersion in the experimental hall. Air is then drawn from the experimental hall via exhaust ducts to a main duct in the inclined shaft, and then exhausted to the atmosphere using exhaust fans.

\subsection{\bf Water supply and drainage system} The underground utility water supply comes from the fire-fighting water system. Normal water usage is obtained from the fire-fighting water system main pipeline, so both water systems are from a common source.

Ultra-pure water for experimental use either in the detector water pool or in scintillator processing will have its own supply by stainless steel pipes from the surface to underground facilities.

The underground wastewater comes from natural rock seepage, utility
water or fire-fighting water. It is collected in a series of
grate-covered tunnel floor trenches and diverted to a number of
sump-pits for pump-out to the surface. Since most of wastewater
during normal operations comes from rock seepage, five sump-pit
drainage systems have been strategically located through the
underground facilities, based on the distribution of water seepage.
Sump-pit-1, the main sump-pit, is located at the bottom of the
inclined shaft, and is distributed through the curved
interconnecting tunnel to the experimental areas. Sump-pit-2 is
located at elevation of -200~m in a side portal area of the inclined
shaft. Sump-pit-3 is located at the bottom area of the water pool,
and is used during pool cleaning. Sump-pit-4 is located at bottom of
vertical shaft. Sump-pit-5 is located at an elevation of -100~m in a
side portal area of the inclined shaft where the rock formation is
poor and water seepage is severe.

The water from whole underground facilities is pumped up to surface via sump-pit-1 and then sump-pit-2.  This design may choose less power pump for sump-pit-1, which have less impact on the experiment during its operation.

Water pool draining will install additional dedicated pump system to sump-pit-1 temporarily.

\subsection{\bf Fire control system} The fire control system consists of automatic fire alarms and interlocks, fire-fighting water distribution to hydrant-hose units, numerous portable fire extinguishers, emergency smoke exhaust ventilation, emergency lighting, personnel evacuation planning, and refuge-rooms. The automatic fire alarm system covers smoke detection and alarming for all underground cavities, tunnels and shafts, along with surface facilities. All alarm signals will be sent to the surface fire control center to be monitored by dedicated fire-watch personnel around the clock. Relevant zones are interlocked to respond to alarm signals by taking the following actions: cut off main power supply sources and start up emergency power generator, close fire partition doors and relevant air exhaust valves to isolate the ignition source, and start emergency exhaust fans. Fire hydrant-hose units and portable fire extinguishers are the preferred method for fighting fires in the underground complex over that of sprinklers systems. The emergency smoke exhaust ventilation systems can be energized based on the location of the incident and the personnel evacuation planned route. For example, smoke can be exhausted through the vertical shaft while personnel escape via the inclined shaft; alternatively, smoke can be exhausted through the inclined shaft while personnel escape via the vertical shaft. In the case of a fire/smoke emergency the main power feeds will be shut off automatically by interlocks, and the emergency power generator will start, to supply power for smoke exhaust ventilation and lighting to ensure safe personnel evacuation.

\subsection{\bf Communication system} A variety of personnel communication systems will be employed throughout the surface and underground facilities of the experimental complex. All fire emergency communication systems are directly connected to the surface fire control center monitored by dedicated fire-watch personnel around the clock. An intercom communication system that has broadcast, point-to-point calling, and central command functionality will be used for working communications throughout the experimental complex. Local telephone systems are furnished at several key locations at the surface construction site for off-site communication.

High-bandwidth and private optical fiber channels will be acquired from local network providers for the experimental complex, to meet the demand for transmission of experimental data and general office communication.

\section{Risk Analysis and Measures}
The risk assessment for the project depends on the results of the detailed geological survey (section 10.2). Special risk assessment of the civil construction project include: risk assessment of geological hazards, seismic studies, water and soil conservation, and ecological impact on the surrounding environment. Geological survey and recommendations will be fully complied. Only major risks during the construction and operation phase, especially unpredictable risks and mitigation measures, are discussed here.

The underground cavities in this project are deep with large open
spans, and the inclined shaft passes through complex geological
area. Therefore, unpredictable collapse of the cavities, shafts or
tunnels, along with water seepage are the major risks in the
construction phase, and they will have direct impact on the project
schedule and construction cost. The geological survey also indicates
that faults may exist in the interconnecting tunnels at the
experimental hall elevation from the inclined shaft or the vertical
shaft, to the experimental hall. This represents the greatest
uncertainty to the construction and worker safety. Possible
mitigation measures include: perform geological forecast in advance
to make appropriate engineering plans, and prepare structural design
solutions based on geological conditions as the project is
implemented. Another major risk in the construction phase is rock
burst. Although the rock burst intensity in the experimental hall
predicted by the geological survey is not high, plans should be
there to prepare for the worst.

During the operational phase of the project, underground collapse and flooding are unpredictable risks that must be considered. It is recommended that a real-time rock monitoring system, such as a 3-D visual early warning system, should be established. This would monitor the stress and deformation of the cavity enclosing rock in real-time, to provide automatic early warning so appropriate and timely measures can be taken. Additionally, rock-safety consultants can be employed to analyze the data taken from  monitoring system to provide hazard analysis and timely safety warnings. Dual-egress design also provides safety assurance for underground evacuation of personnel. Underground water conditions will be monitored and alarmed for accumulation and daily discharge to the surface. In conjunction with patrol checks of underground cavities and tunnels, any water abnormalities can be acted upon.

\section{Schedule}
The total civil construction period for the entire site complex is
36 months. This work has started on January 2015, and the civil
construction phase of the project should be finished and handed over
to the experimental users by December 2017.

%
%
%
%
%
%
%
%

\cleardoublepage
\chapter{Assembly, Installation and Engineering Issues}
\label{ch:AssemblyAndInstallation}

\section{Introduction}

After the completion of civil construction, through systems integration, assembly, installation and commissioning, all subsystems will be formed as a whole JUNO detector.

In this process, we need to establish various engineering 
regulations and standards, and to coordinate subsystems' assembly, installation, testing and commissioning, especially their onsite work.

\subsection{Main Tasks}

Main tasks of the integration work include:
\begin{itemize}
\item To prepare plans and progress reports of each phase; 
\item To establish a project technology review system; 
\item To standardize the executive technology management system; 
\item To have strictly executive on-site work management system; 
\item To develop and specify security management system on-site; 
\item To prepare common tools and equipment for each system, and to guarantees project progress; 
\item To coordinate the installation progress of each system according to the on-site situation.
\end{itemize}

\subsection{Contents}

The work contents mainly include:
\begin{itemize}
\item To summarize the design, and review progress of each subsystem;
\item To organize preparation work for installation in the experiment region; 
\item To inspect and certify Surface Buildings, underground Tunnels, and Experiment Hall with relevant utilities;
\item To coordinate technology interfaces between key parts;
\item To coordinate the procedure of assembly and installation both on surface and underground;
\end{itemize}

\section{ Design Standards, Guidelines and Reviews}

\subsection{Introduction}

There will be a document to outline the mechanical design standards or guidelines that will be applied to the design work. It also describes the review process of engineering design that will be implemented to ensure that experimental equipment meets all requirements for performance and safety. The following is a brief summary of the guidance that all mechanical engineers/designers should follow in the process. For reasons ranging from local safety requirements to common sense practices, the following information should be understood and implemented in the design process.

\subsection{Institutional Standards}

When specific institutional design standards or guidelines exist, they should be followed. The guidelines outlined are not meant to replace but instead to supplement institutional guidelines. The majority of equipment and components built for the JUNO Experiment will be engineered and designed at home institutions, procured or fabricated at commercial vendors, then eventually delivered, assembled, installed, tested and operated at the JUNO Experimental facilities, Jiangmen, China. The funding agencies in other countries as well as Jiangmen, China will have some guidelines in this area, as would your home institutions. Where more than one set of guidelines exist, use whichever are the more stringent or conservative approaches.

\subsection{Design Loads}

The scope of engineering analysis should take handling, transportation, assembly and operational loads into account as well as thermal expansion considerations caused by potential temperature fluctuations.

For basic stability and a sensible practice in the design of experimental components an appropriate amount of horizontal load (0.10~g) should be applied.

In addition, seismic requirements for experimental equipment is based on a National Standard of People's Republic of China Code of Seismic Design of Buildings GB 50011-2010. At the location of the JUNO Experiment the seismic fortification intensity is grade 7 with a basic seismic acceleration of 0.10~g horizontal applied in evaluating structural design loads. A seismic hazard analysis should be performed and documented based on this local code. The minimum total design lateral seismic base shear force should be determined and applied to the component. The direction of application of seismic forces would be applied at the CG of the component that will produce the most critical load effect, or separately and independently in each of the two orthogonal directions. A qualitative seismic performance goal defines component functionality as: the component will remain anchored, but no assurance it will remain functional or easily repairable. Therefore, a seismic design factor of safety F.S. > 1 based on the Ultimate Strength of component materials would satisfy this goal.

Where an anticipated load path as designed above, allows the material to be subjected to stresses beyond the yield point of the material, redundancy in the support mechanism must be addressed in order to prevent a collapse mechanism of the structure from being formed.

The potential for buckling should also be evaluated. It should be
noted that a rigorous analytical seismic analysis may be performed
in lieu of the empirical design criteria. This work should be
properly documented for review by the Chief Engineer and appropriate
Safety Committee personnel.

\subsection{Materials Data}

All materials selected for component design must have their engineering data sources referenced along with those material properties used in any structural or engineering analysis; this includes fastener certification. There are many sources of data on materials and their properties that aid in the selection of an appropriate component material. There are many national and international societies and associations that compile and publish materials test data and standards. Problems frequently encountered in the purchasing, researching and selection of materials are the cross-referencing of designations and specification and matching equivalent materials from differing countries.

It is recommended that the American association or society standards be used in the
materials selection and specification process, or equivalency to these
standards must be referenced. Excellent engineering materials data have been
provided by the American Society of Metals or Machine Design Handbook Vol. 1-5 and
Mechanical Industry Publishing Co. in PRC, Vol 1-6, 2002, which are worth investigating.

\subsection{Analysis Methods}

The applicable factors of safety depend on the type of load(s) and how well they can be estimated, how boundary conditions have been approximated, as well as how accurate your method of analysis allows you to be.

\paragraph{Bounding Analyses:}
Bounding analysis or rough scoping analyses have proven to be
valuable tools. Even when computer modeling is in your plans, a
bounding analysis is a nice check to avoid gross mistakes. Sometimes
bounding analyses are sufficient. An example of this would be for
the case of an assembly fixture where stiffness is the critical
requirement. In this case where deflection is the over-riding
concern and the component is over-designed in terms of stress by a
factor of 10 or more, then a crude estimation of stress will
suffice.

\paragraph{Closed-Form Analytical Solutions:}
Many times when boundary conditions and applied loads are simple to
approximate, a closed-form or handbook solution can be found or developed. For
the majority of tooling and fixture and some non-critical experimental
components, these types of analyses are sufficient. Often, one of these
formulas can be used to give you a conservative solution very quickly, or a
pair of formulas can be found which represent upper and lower bounds of the
true deflections and stresses. Formulas for Stress and Strain by Roark and
Young is a good reference handbook for these solutions.

\paragraph{Finite Element Analysis:}
When the boundary conditions and loads get complex, or the correctness of the solution is critical, computer modeling is often required. If this is the case, there are several rules to follow, especially if you are not intimately familiar with the particular code or application.
\begin{enumerate}
\item Always bound the problem with an analytical solution or some other approximate means.
\item If the component is critical, check the accuracy of the code and application by modeling a similar problem for which you have an analytical or handbook solution.
\item Find a qualified person to review your results.
\item Document your assumptions and results.
\end{enumerate}

\subsection{Failure Criteria}

The failure criterion depends upon the application. Many factors such as the rate or frequency of load application, the material toughness (degree of ductility), the human or monetary risk of component failure as well as many other complications must be considered.

Brittle materials (under static loads, less than 5\% yield prior to failure), includes ceramics, glass, some plastics and composites at room temperature, some cast metals, and many materials at cryogenic temperatures. The failure criterion chosen depends on many factors so use your engineering judgment. In general, the Coulomb-Mohr or Modified Mohr Theory should be employed.

Ductile materials (under static loads, greater than 5\% yield prior to failure), includes most metals and plastics, especially at or above room temperature. The failure criterion chosen again ultimately rests with the cognizant engineer because of all the adverse factors that may be present. In general, the Distortion- Energy Theory, or von Mises-Hencky Theory (von Mises stresses), is most effective in predicting the onset of yield in materials. Slightly easier to use and a more conservative approach is the Maximum-Shear-Stress Theory.

\subsection{Factor of Safety}

Some institutions may have published guidelines which specifically discuss factors of safety for various applications. For the case where specific guidelines do not exist, the following may be used.

Simplistically, if F is the applied load (or S the applied stress),
and $F_f$ is the load at which failure occurs (or $S_s$ the stress
at which failure occurs), we can then define the factor of safety
(F.S.) as:
\begin{displaymath}
F.S. = F_{f} / F \quad\mathrm{or}\quad S_{s} / S
\end{displaymath}
The word failure, as it applies to engineering elements or systems, can be defined in a number of ways and depends on many factors. Discussion of failure criteria is presented in the previous section, but for the most common cases it will be the load at which yielding begins.

\subsection{Specific Safety Guidelines for JUNO}

Lifting and handling fixtures, shipping equipment, test stands, and fabrication
tooling where weight, size and material thickness do not affect the physical
capabilities of the detector, the appropriate F.S. should be at least 3. When
life safety is a potential concern, then a F.S. of 5 may be more appropriate.
Note that since the vast majority of this type of equipment is designed using
ductile materials, these F.S.'s apply to the material yield point. Experimental hardware that does not present a life safety or significant cost/schedule risk if failure occurs, especially where there is the potential for an increase in physics capabilities, the F.S. may be as low as 1.5. Many factors must be taken into account if a safety factor in this low level is to be employed: a complete analysis of worst case loads must be performed; highly realistic or else conservative boundary conditions must be applied; the method of analysis must yield accurate results; reliable materials data must be used or representative samples must be tested. If F.S.'s this low are utilized, the analysis and assumptions must be highly scrutinized. Guidelines for F.S. for various types of equipment are:

\begin{center}
\begin{tabular}{|p{3.5cm}|c|p{3.5cm}|}
\hline
Type of Equipment & Minimum F.S. & Notes \\
\hline
Lifting and handling & 3 - 5  & Where there is a risk to life
safety or to costly hardware,
choose F.S closer to 5. \\
\hline
Test stands, shipping and assembly fixtures. & 3 & \\
\hline
Experimental hardware & 1.5 - 3  & 1.5 is allowable for physics
capability and analysis where
method is highly refined \\
\hline
\end{tabular}
\end{center}

\subsection{Documentation}

It is not only good engineering practice to document the analysis, but it is an
ESH\&Q requirement for experimental projects. For this reason all major
components of JUNO Experiment will have their engineering analyses documented as Project Controlled Documents. Utilize your institutional documentation formats or use the following guidelines.
Calculations and analyses must
\begin{itemize}
\item Be hard copy documented.
\item Follow an easily understood logic and methodology.
\item Be legible and reproducible by photocopy methods.
\item Contain the following labeling elements.
    \begin{itemize}
    \item Title or subject
    \item Originators signature and date
    \item Reviewers signature and date
    \item Subsystem WBS number.
    \item Introduction, background and purpose.
    \item Applicable requirements, standards and guidelines.
    \item Assumptions (boundary conditions, loads, materials properties, etc.).
    \item Analysis method (bounding, closed form or FEA).
    \item Results (including factors of safety, load path and location of critical areas).
    \item Conclusions (level of conservatism, limitations, cautions or concerns).
    \item References (tech notes, textbooks, handbooks, software code, etc.).
    \item Computer program and version (for computer calculations)
    \item Filename (for computer calculations).
    \end{itemize}
\end{itemize}
\subsection{Design Reviews}

All experimental components and equipment whether engineered or
procured as turn-key systems will require an engineering design
review before procurement, fabrication, assembly or installation can
proceed. The L2 subsystem manager will request the design review
from the project management office, which will appoint a chair for
the review and develop a committee charge. The selected chair should
be knowledgeable in the engineering and technology, and where
practical, should not be directly involved in the engineering design
effort. With advice from the L2, the technical board and project
office chair appoints members to the review committee that have
experience and knowledge in the engineering and technology of the
design requiring review. At least one reviewer must represent ESH\&Q
concerns. The committee should represent such disciplines as:
\begin{itemize}
\item Science requirements
\item Design
\item Manufacturing
\item Purchasing
\item Operational user
\item Maintenance
\item Stress analysis
\item Assembly/installation/test
\item Electrical safety
\item ESH\&Q
\end{itemize}

For the JUNO project there will be a minimum of two engineering design reviews: Preliminary and Final. The preliminary usually takes place towards the end of the conceptual design phase when the subsystem has exhausted alternative designs and has made a selection based on value engineering. The L2 along with the chair ensures that the preliminary design review package contains sufficient information for the review along with:
\begin{itemize}
\item Agenda
\item Requirements documents
\item Review committee members and charge
\item Conceptual layouts
\item Science performance expectations
\item Design specifications
\item Supportive R\&D or test results
\item Summaries of calculations
\item Handouts of slides
\end{itemize}
A final design review will take place before engineering drawings or specifications are released for procurement or fabrication. The L2 along with the chair ensures that the final design review package contains sufficient information for the final review along with:
\begin{itemize}
\item Changes or revisions to preliminary design
\item Completed action items
\item Final assembly and detailed drawings
\item Final design specifications
\item Design calculations and analyses
\end{itemize}
The committee chair will ensure that meeting results are taken and recorded along with further action items. The committee and chair shall prepare a design review report for submittal to the project office in a timely manner which should include:
\begin{itemize}
\item Title of the review
\item Description of system or equipment
\item Copy of agenda
\item Committee members and charge
\item Presentation materials
\item Major comments or concerns of the reviewers
\item Action items
\item Recommendations
\end{itemize}
The project office will file the design review report and distribute copies to all reviewers and affected groups.

\section{On-site management}

According to the experience from the Daya Bay Experiment, an effective technology review system has been well practiced. In JUNO, we will take it as a good reference and carry out for standardized review from the start of all system design, scheme
argument, technology checks, we should also establish management system, to cover engineering drawing control, engineering change control procedure, and mechanical design standards, guidelines and reviews, etc.

To control on-site work process, proper process should be introduced as well as a management structure. Safety training and safety management structure should be also introduced.

\section{Equipment and Tools Preparation on Site}

All of key equipment, devices and tools should be in place and get acceptance
for installation on site, including
\begin{itemize}
\item Cranes:
\begin{itemize}
\item 2 sets of bridge crane with 12T/5T load capacity, and lifting range should
cover any point for components delivery and installation in EH;
\item Several lifting equipment are needed in Surface Assembly areas, Assembly
Chamber underground, Storage area for Liquid Scintillation, and other Chambers
accordingly;
\end{itemize}
\item Removable lifting equipment: Forklifts, manual hydraulic carriages, carts,
etc.
\item Mobile vehicle for smaller items: Pickup van, battery car, and so on;
\item Several work aloft platform: Scissor lifts, Boom lifts, Scaffoldings and
Floating platforms, etc.
\item Common Tooling: complete kits of tool, fixtures, jigs, models, work-bench,
cabinets, and tables, etc.
\item Machine shop, Power equipment rooms, Control rooms are ready to be put into
use.
\item Safety measures and relevant equipment in place.
\end{itemize}

\section{Main Process of Installation}

Since there still are several optional design versions which could
be decided later, and different design could have different
requirements for installation, therefore, the specific installation
procedures will be established and developed later, here a very
rough installation procedure is given.
\begin{itemize}
\item Mounting, debugging, testing of basic utilities, such as, cranes, vertical
elevator, scaffoldings, test devices, etc.

\item Survey, alignment, and adjustment for tolerances of form and position, include,
embedded inserts, anchor plates, positioning holes, etc.

\item Mounting Bottom Structure of Veto Cherenkov detector on the floor of Water Pool.

\item Mounting anchor-plate for support poles of CD on the floor of Water Pool

\item Pre-mounting of Top Structure of Veto system along wall of Water Pool

\item Mount Substrate and Rails for Bridge over the water Pool edges

\item Mount Tyvek with support structure for Water Cherenkov Detector, and with PMT
positioning in the Water Pool

\item Installation for CD with its PMT system and cabling in Water Pool details will be established until design final selection.

\item Mount Calibration System

\item Mount Geomagnetic Shielding system around Central Detector in the Pool

\item Final installation for Tyvek and PMT of Veto system in the Pool

\item Mount a temporary protective cover on top of Pool, to prevent from anything
falling into Pool

\item Mount Bridge on the rails, and drive it closed

\item Mount Top Track Detector on Bridge and establish Electronic Room there

\item Have Cabling completion, and test to get normal signals
\item Check Grounding Electrode

\item Dry Run

\item Final Cleaning for Water Pool, then, drain out

\item Filling Water, LS, LAB (or Oil) evenly into Water Pool and CD

\item Mount Final Pool Cover, closed, and get fixation
\end{itemize}

%
%
%
%
%
%
%
%


\cleardoublepage
\renewcommand{\bibname}{}
\bibliography{juno}

\begin{thebibliography}{99}

\bibitem{Weinberg}
S. Weinberg, Phys. Rev. Lett. {\bf 19}, 1264 (1967).

\bibitem{MNS}
Z. Maki, M. Nakagawa, and S. Sakata, Prog. Theor.
Phys. {\bf 28}, 870 (1962); B. Pontecorvo, Sov. Phys. JETP {\bf 26},
984 (1968).

\bibitem{CKM}
N. Cabibbo, Phys. Rev. Lett. {\bf 10}, 531 (1963);
M. Kobayashi and T. Maskawa, Prog. Theor. Phys. {\bf 49}, 652
(1973).

\bibitem{PDG}
J. Beringer {\it et al.} (Particle Data Group),
Phys. Rev. D {\bf 86}, 010001 (2012).

\bibitem{GF1}
F.~Capozzi, G.~L.~Fogli, E.~Lisi, A.~Marrone, D.~Montanino and A.~Palazzo,
Phys.\ Rev.\ D {\bf 89}, 093018 (2014);
D.~V.~Forero, M.~Tortola and J.~W.~F.~Valle,
Phys.\ Rev.\ D {\bf 90}, 093006 (2014);
M.~C.~Gonzalez-Garcia, M.~Maltoni and T.~Schwetz,
JHEP {\bf 1411}, 052 (2014).

\bibitem{XZ}
Z.Z. Xing and S. Zhou, {\it Neutrinos in Particle Physics, Astronomy
and Cosmology} (Zhejiang University Press and Springer-Verlag,
2011).

\bibitem{zhanl2008}
L. Zhan, Y.F. Wang, J. Cao, and L.J. Wen, Phys. Rev. D {\bf 78},
111103 (2008);
   Phys. Rev. D {\bf 79}, 073007 (2009).

\bibitem{yfwang2008}
Yifang Wang, talk at ICFA seminar, 2008,
\url{http://www-conf.slac.stanford.edu/icfa2008/Yifang\_Wang\_102808.pdf}

\bibitem{caoj2009}
Jun Cao, talk at Neutrino Telescope, 2009,
\url{http://neutrino.pd.infn.it/NEUTEL09/Talks/Cao.pdf}

\bibitem{JUNO}
Y.~F.~Li, J.~Cao, Y.~F.~Wang and L.~Zhan, Phys.\ Rev.\ D {\bf 88},
013008 (2013).

\bibitem{DYB} F.P. An {\it et al.} (Daya Bay Collaboration),
Phys. Rev. Lett. {\bf 108}, 171803 (2012); Chin. Phys. C {\bf 37},
011001 (2013); Phys. Rev. Lett. {\bf 112}, 061801 (2014).

\bibitem{KLloe}
A.~Gando {\it et al.}, [KamLAND Collaboration], Phys.\ Rev.\ D {\bf
83}, 052002 (2011).

\bibitem{MINOS}
P.~Adamson {\it et al.}, [MINOS Collaboration], Phys.\ Rev.\ Lett.\
{\bf 110}, 251801 (2013).

\bibitem{SNO}
B.~Aharmim {\it et al.}, [SNO Collaboration], Phys.\ Rev.\ C {\bf
88}, 025501 (2013).

\bibitem{DBloe}
F.~P.~An {\it et al.}  [Daya Bay Collaboration],
arXiv:1505.03456 [hep-ex];
F.~P.~An {\it et al.}, [Daya Bay Collaboration], Phys.\ Rev.\ Lett.\
{\bf 112}, 061801 (2014).

\bibitem{SKth23}
R.~Wendell {\it et al.}, [Super-Kamiokande Collaboration], Phys.\
Rev.\ D {\bf 81}, 092004 (2010).

\bibitem{T2Kth23}
K.~Abe {\it et al.}, [T2K Collaboration], Phys.\ Rev.\ Lett.\ {\bf
111}, 211803 (2013); for the latest update see arXiv:1403.1532
[hep-ex].

\bibitem{Beacom:2002hs}
J.~F.~Beacom, W.~M.~Farr and P.~Vogel, Phys.\ Rev.\ D {\bf 66},
033001 (2002),  [hep-ph/0205220].

\bibitem{Dasgupta:2011wg}
B.~Dasgupta and J.~F.~Beacom, Phys.\ Rev.\ D {\bf 83}, 113006
(2011), [arXiv:1103.2768 [hep-ph]].

\bibitem{Malek:2002ns}
  M.~Malek {\it et al.} (Super-Kamiokande Collaboration),
  Phys.\ Rev.\ Lett.\  {\bf 90}, 061101 (2003).

\bibitem{Bays:2011si}
  K.~Bays {\it et al.} (Super-Kamiokande Collaboration),
  Phys.\ Rev.\ D {\bf 85}, 052007 (2012).

\bibitem{Rolke:2004mj}
  W.~A.~Rolke, A.~M.~L\'opez and J.~Conrad,
  Nucl.\ Instrum.\ Meth.\ A {\bf 551}, 493 (2005).

\bibitem{KL-recent}
S.~Abe {\it et al.}  [KamLAND Collaboration],
 Phys.\ Rev.\ Lett.\  {\bf 100}, 221803 (2008).

\bibitem{Borexino-fase-I}
G.~Bellini {\it et al.}  [Borexino Collaboration],
Phys.\ Rev.\ D {\bf 89} 112007 (2014).

\bibitem{Villante-Serenelli-2014}
F.~L.~Villante, A.~M.~Serenelli, F.~Delahaye and M.~H.~Pinsonneault,
Astrophys.\ J.\  {\bf 787}, 13 (2014).

%
%
%

  \bibitem{bib3}
L.~M.~Krauss, S.~L.~Glashow, and D.~N.~Schramm,
Nature, {\bf 310}, 5974 (1984).

\bibitem{bib4}
W.~F.~McDonough, J.~G.~Learned, S.~T.~Dye,
Physics Today {\bf 65}, 46 (2012).

\bibitem{bib5}
T.~Araki {\it et al.},
Nature {\bf 436}, 499 (2005).

\bibitem{kearns:isoup}
E.~Kearns, talk presented at the ISOUP Symposium (2013).

\bibitem{Aguilar:2001ty}
  A.~Aguilar-Arevalo {\it et al.}  [LSND Collaboration],
  Phys.\ Rev.\ D {\bf 64}, 112007 (2001)
  [hep-ex/0104049].

\bibitem{Aguilar-Arevalo:2013pmq}
  A.~A.~Aguilar-Arevalo {\it et al.}  [MiniBooNE Collaboration],
  Phys.\ Rev.\ Lett.\  {\bf 110}, 161801 (2013)
  [arXiv:1207.4809 [hep-ex], arXiv:1303.2588 [hep-ex]].

\bibitem{Mention:2011rk}
  G.~Mention, M.~Fechner, T.~Lasserre, T.~A.~Mueller, D.~Lhuillier, M.~Cribier and A.~Letourneau,
  Phys.\ Rev.\ D {\bf 83}, 073006 (2011)
  [arXiv:1101.2755 [hep-ex]].

\bibitem{Giunti:2010zu}
  C.~Giunti and M.~Laveder,
  Phys.\ Rev.\ C {\bf 83}, 065504 (2011)
  [arXiv:1006.3244 [hep-ph]].

\bibitem{Giunti:2013aea}
  C.~Giunti, M.~Laveder, Y.~F.~Li and H.~W.~Long,
  Phys.\ Rev.\ D {\bf 88}, 073008 (2013)
  [arXiv:1308.5288 [hep-ph]].

\bibitem{Kopp:2013vaa}
  J.~Kopp, P.~A.~N.~Machado, M.~Maltoni and T.~Schwetz,
  JHEP {\bf 1305}, 050 (2013)
  [arXiv:1303.3011 [hep-ph]].

\bibitem{Lasserre:2014ita}
  T.~Lasserre,
  Phys.\ Dark Univ.\  {\bf 4}, 81 (2014)
  [arXiv:1404.7352 [hep-ex]].

\bibitem{Aartsen:2012kia}
 M.~G.~Aartsen {\it et al.}  [IceCube Collaboration],
 Phys.\ Rev.\ Lett.\  {\bf 110}, 131302 (2013).

\bibitem{Behnke:2012ys}
  E.~Behnke {\it et al.}  [COUPP Collaboration],
  Phys.\ Rev.\ D {\bf 86}, 052001 (2012)
  [arXiv:1204.3094 [astro-ph.CO]].

\bibitem{Felizardo:2011uw}
  M.~Felizardo {\it et al.} (SIMPLE Collaboration),
  Phys.\ Rev.\ Lett.\  {\bf 108}, 201302 (2012)
  [arXiv:1106.3014 [astro-ph.CO]].

\bibitem{Archambault:2012pm}
  S.~Archambault {\it et al.}  [PICASSO Collaboration],
  Phys.\ Lett.\ B {\bf 711}, 153 (2012)
  [arXiv:1202.1240 [hep-ex]].

\bibitem{ohlsson:14a}
T.~Ohlsson, H.~Zhang, S.~Zhou,
  Phys.\ Lett.\ B.\  {\bf 728}, 148 (2014).

\bibitem{Li:2014mlo}
  Y.~F.~Li and Y.~L.~Zhou,
  Nucl.\ Phys.\ B {\bf 888}, 137 (2014).

\bibitem{Li:2014rya}
  Y.~F.~Li and Z.~H.~Zhao,
  Phys.\ Rev.\ D {\bf 90}, 113014 (2014).

\bibitem{Schwetz:2005fy}
  T.~Schwetz and W.~Winter,
  Phys.\ Lett.\ B {\bf 633}, 557 (2006).

  \bibitem{Ohlsson:2012kf}
  T.~Ohlsson,
  Rept.\ Prog.\ Phys.\  {\bf 76}, 044201 (2013)
  [arXiv:1209.2710 [hep-ph]].

\bibitem{Bolanos:2008km}
  A.~Bolanos, O.~G.~Miranda, A.~Palazzo, M.~A.~Tortola and J.~W.~F.~Valle,
  Phys.\ Rev.\ D {\bf 79}, 113012 (2009).

\bibitem{Joshipura:2003jh}
  A.~S.~Joshipura and S.~Mohanty,
  Phys.\ Lett.\ B {\bf 584}, 103 (2004)
  [hep-ph/0310210].

\bibitem{deHolanda:2010am}
  P.~C.~de Holanda and A.~Y.~Smirnov,
  Phys.\ Rev.\ D {\bf 83}, 113011 (2011).

\bibitem{Arpesella:2008mt}
  C.~Arpesella {\it et al.}  [Borexino Collaboration],
  Phys.\ Rev.\ Lett.\  {\bf 101}, 091302 (2008).

\bibitem{Giunti:2014ixa}
  C.~Giunti and A.~Studenikin,
  arXiv:1403.6344 [hep-ph].

\end{thebibliography}

\begin{thebibliography}{99}

\bibitem{SNO} J. Boger et al. Nuclear Instruments and Methods in Physics Research A 449 (2000) 172-207

\bibitem{Handbook} Jerry D. Stachiw Handbook of acrylics for submersibles hyperrodic chambers and aquaria,2003

\bibitem{Borexino} J. Benziger et al. Nuclear Instruments and Methods in Physics Research A 582 (2007) 509-534

\end{thebibliography}

\newpage



\begin{thebibliography}{99}
    \bibitem{An:2012eh}
        An, F.P. \textit{et al.} (DAYA-BAY),
        Phys.Rev.Lett. \textbf{108}, 171803 (2012),
        arXiv:1203.1669 .
    \bibitem{Wurm:2010ad}
        Michael Wurm \textit{et al.},
        Rev. Sci. Instrum. \textbf{81}, 053301 (2010),
        arXiv:1004.0811 [physics].
    \bibitem{Ford:2011zza}
        R. Ford, M. Chen, O. Chkvorets, D. Hallman, E. Vazquez-Jauregui,
        AIP Conf.Proc. \textbf{1338}, 183 (2011).
    \bibitem{Mark:2008gc}
        http://www.ipp.ca/pdfs/SNOp$_-$chen.pdf
    \bibitem{Mike:2008gc}
        Mike Leung, PhD. Thesis Princeton University
\end{thebibliography}

\begin{thebibliography}{00}


\bibitem{DYB:muon2014} 
 F.~P.~An {\it et al.} [Daya Bay Collaboration], 
 ``The Muon System of the Daya Bay Reactor Antineutrino Experiment",
 Nucl. Instrum. Meth. A {\bf 773} (2015)  8-20. 

\bibitem{SK:detector2003} 
 S.~Fukuda {\it et al.} [Super-Kamiokande Collaboration], 
 ``The Super-Kamiokande Detector",
 Nucl. Instrum. Meth. A {\bf 501} (2003)  418-462.
 
 
\bibitem{Adam:2007ex}
T. Adam \textit{et al.},
"The OPERA experiment Target Tracker",
Nucl.Instrum.Meth. \textbf{A577} (2007) 523,
arXiv:physics/0701153.

\bibitem{Agafonova:2011zz}
  N.~Agafonova {\it et al.}  [OPERA Collaboration],
  ``Study of neutrino interactions with the electronic detectors of the OPERA experiment,''
  New J.\ Phys.\  {\bf 13} (2011) 053051
  [arXiv:1102.1882 [hep-ex]].
  
\bibitem{Lucotte:2004mi}
  A.~Lucotte, S.~Bondil, K.~Borer, J.~E.~Campagne, A.~Cazes, M.~Hess, C.~de La Taille and G.~Martin-Chassard {\it et al.},
  ``A front-end read out chip for the OPERA scintillator tracker,''
  Nucl.\ Instrum.\ Meth.\ A {\bf 521} (2004) 378.


\bibitem{Marteau:2009ct}
  J.~Marteau [OPERA Collaboration],
  ``The OPERA global readout and GPS distribution system,''
  Nucl.\ Instrum.\ Meth.\ A {\bf 617} (2010) 291
  [arXiv:0906.1494 [physics.ins-det]].


\bibitem{Blin:2010tsa}
  S.~Blin, P.~Barrillon and C.~de La Taille,
  ``MAROC, a generic photomultiplier readout chip,''
  IEEE Nucl.\ Sci.\ Symp.\ Conf.\ Rec.\  {\bf 2010} (2010) 1690.
  
\bibitem{Collaboration:2011nsa}
  M.~Ageron {\it et al.}  [ ANTARES Collaboration],
  Nucl.\ Instrum.\ Meth.\ A {\bf 656} (2011) 11
  [arXiv:1104.1607 [astro-ph.IM]].
  
  \bibitem{km3net}
  http://www.km3net.org/TDR/KM3NeT-TDR.pdf (ISBN 978-90-6488-033-9)
\bibitem{NOvA}
Physics 37 (Procedia 2012) 1201 - 1208, 2 NOvA Technical Design Report, Oct 8, 2007
\bibitem{MINOS}
 Nucl.\ Instrum.\ Meth.\ A {\bf 463} (2001) 194-204

\bibitem{DeVore:2014nim}
 P.~DeVore {\it et al.}, ``Light-weight Flexible Magnetic Shield For Large-Aperture Photomultiplier Tubes",
 Nucl. Instrum. Meth. A {\bf 737} (2014)  222-228.
 

\bibitem{Wilhelmi:2014rwa} 
J.~Wilhelmi, R.~Bopp, R.~Brown, J.~Cherwinka, J.~Cummings, E.~Dale, M.~Diwan and J.~Goett {\it et al.},
  arXiv:1408.1302 [physics.ins-det].



\end{thebibliography}

\begin{thebibliography}{99}
\bibitem{Nishimura2014Jan} Y.Nishimura, RCCN, University of
Tokyo, Measurement of Large Aperture Photo-Detectors in a Water Tank, 4th
Hyper-Kamiokande open meeting 28 Jan., 2014.
\end{thebibliography}

\begin{thebibliography}{99}
\bibitem{Qian:2013}
X. Qian et al., Phys. Rev. D87 (2013) 3, 033005
\bibitem{NEC:2014}
    National Electrostatic Corp. \url{http://www.pelletron.com/}
\bibitem{Hinterberger:1997ur}
  F.~Hinterberger,
  Berlin, Germany: Springer (2008) 417 p
\bibitem{Bauer:1990zz}
  W.~Bauer,
  Nucl.\ Instrum.\ Meth.\ B {\bf 50}, 300 (1990).
\bibitem{Huomo:1988jw}
  H.~Huomo, P.~AsokaKumar, S.~D.~Henderson, B.~F.~Phlips, R.~Mayer, J.~McDonough, H.~Hacker and S.~McCorkle {\it et al.},
  BNL-41953.
\bibitem{Berg:1992jr}
  W.~Berg, R.~Fuja, A.~Grelick, G.~Mavrogenes, A.~Nassiri, T.~Russell, W.~Wesolowski and M.~White,
  ``Beam measurements of the ANL-APS linac injector test stand,''
\end{thebibliography}

\begin{thebibliography}{99}

\bibitem{Qiuju}
Li Qiuju {\it et al.}, "The PMT Charge and Time Readout Scheme for Daya Bay Reactor Neutrino Experiment''. The 14th Nuclear Electronics and Detection Technology conference, 2008, 84-88

\bibitem{AD9234}
AD9234 datasheet
\url{http://www.analog.com/static/imported-files/data_sheets/AD9234.pdf}

\bibitem{ASIC1}
ANTARES Collaboration, arXiv:1007.2549; S. Kleinfelder {\it et al.}, "A multi-GHz, multi-channel transient waveform digitization integrated circuit'', Nuclear Science Symposium Conference Record, 2002 IEEE, vol.1, 544 - 548

\bibitem{Dayabay}
"Dayabay Reactor Neutrino Experiment report for feasibility study", Institute of High Energy Physics, Internal documents, IHEP-DB-SB-02-R1, 2006

\bibitem{Z.Wang}
"Design of Prototyping PMT Electronic System for Daya Bay Reactor Neutrino Experiment", Z.Wang {\it et al.}, IEEE NSS/MIC Conference proceeding, NSS-2008-N30-456.

\bibitem{Q.J.Li}
"Front-End Electronics System of PMT Readout for Daya Bay Reactor Neutrino Experiment", Q.J.Li {\it et al.}, IEEE NSS/MIC Conference proceeding, NSS-2009-N25-240.

\bibitem{Q.J.Li-Thesis}
"The Study of PMT Charge and Time Measurement of Daya Bay Reactor Neutrino Experiment", Qiu-ju Li, Master Thesis, 2008.

\bibitem{W.Q.Jiang}
"The Study of Readout electronics system of Daya Bay Reactor Neutrino Experiment", Wen-qi, Jiang, Ph.D. Thesis, 2011.

\bibitem{H.S.Xiang}
"High-speed Digitization Technique Study and Application", Hai-sheng Xiang, Ph.D. Thesis, 2010

\bibitem{Z.Li}
"The Research of Key Technologies of JUNO RPC Readout Electronics",  Lei Zheng, Ph.D. Thesis, 2014.

\bibitem{BESIII}
 BESIII EMC readout electronics report, 2003

\bibitem{F.Reines}
 C.L. Cowan, F. Reines, F.B. Harrison, H.W. Kruseand, A.D. McGuire, Science 124,103 (1956); Phys. Rev. 92, 830 (1953).

\bibitem{G.Danby}
G. Danby, J.M. Gaillard, K. Goulianos, L.M. Lederman, N. Mistry, M.Schwartz and J.Steinberger, Phys. Rev. Lett. 9, 36 (1962).

\bibitem{DONUT}
 DONUT Collaboration, K. Kodama {\it et al.}, Phys. Lett. B 504, 218 (2001), hep-ex/0012035.

\bibitem{Benjamin}
 Benjamin W. Lee {\it et al.}, Phys. Rev. Lett. 39, 165 (1977)

\bibitem{Pontecorvo}
B. Pontecorvo, Sov. Phys. JETP 26, 984; V.N. Gribov and B. Pontecorvo, Phys. Lett. 28B, 493 (1969)

\bibitem{Super-K}
 Super-Kamiokande Collaboration, Phys. Rev. Lett.81, 1158, hep-ex/9805021.

\bibitem{SNO}
 SNO Collaboration, Phys. Rev. Lett. 87, 071301(2001), nucl-ex/0106015

\bibitem{K.Eguchi}
 KamLAND Collaboration, Phys. Rev. Lett. 90, 021802 (2003), hep-ex/0212021.

\bibitem{T.Araki}
KamLAND Collaboration, Phys. Rev. Lett .94, 081801(2005), hep-ex/0406035
\end{thebibliography}

\begin{thebibliography}{9}
\bibitem{Bes3TDR} BESIII Experiment TDR, in chinese
\bibitem{DybTDR} Dayabay Neutrino Experiment TDR
\bibitem{DybDaqTDR} F. Li et al, DAQ Architecture Design of Daya Bay Reactor Neutrino Experiment, IEEE Trans. Nucl. Sci., vol. 58, no. 4, pp. 1723-1727, Aug. 2011
\bibitem{LifeiThesis} F. Li, Research and Implementation of BESIII DAQ Online Distributed Data Monitoring , Ph D dissertation
\bibitem{LiuyjThesis} Y. Liu,  Research and Implementation of BESIII DAQ Online Event Filtor, Ph D dissertation

\end{thebibliography}

\begin{thebibliography}{99}
\bibitem{CMT} http://www.cmtsite.net/
\bibitem{Gaudi} Gaudi project
homepage:http://proj-gaudi.web.cern.ch/proj-gaudi/
\bibitem{ROOT} Root an object-oriented data analysis framework.
               http://root.cern.ch/
\bibitem{XML} Extensible Markup Language (XML), http://www.w3.org/XML/
\bibitem{HepRep} http://www.slac.stanford.edu/~perl/heprep/
\bibitem{VRML} http://www.w3.org/MarkUp/VRML/
\bibitem{X3D} http://www.web3d.org/
\bibitem{GDML} http://gdml.web.cern.ch/GDML/
\bibitem{Geant4} S. Agostinelli et al. (Geant4 Collaboration),
                 Nucl. Instr. And Meth. A506(2003) 250
\bibitem{HepMC} http://lcgapp.cern.ch/project/simu/HepMC
\bibitem{MUSIC} V.A. Kudryavtsev, Computer Physics Communications 180 (2009) 339-346
\bibitem{OpenGL} https://www.opengl.org/
\bibitem{OpenInventor} http://oss.sgi.com/projects/inventor/
\bibitem{Oracle} https://www.oracle.com/database/index.html
\bibitem{MySQL} http://dev.mysql.com/downloads/
\bibitem{postgreSQL} http://www.postgresql.org/
\bibitem{SQLite} http://www.sqlite.org/
\bibitem{CSTNet} http://www.cstnet.net.cn/
\bibitem{Torque} http://www.adaptivecomputing.com/
\bibitem{HTCondor} High Through Out computing http://research.cs.wisc.edu/htcondor/
\bibitem{LSF} Platform Load Sharing Facility http://en.wikipedia.org/wiki/Platform\_LSF
\bibitem{mapreduce} Dean J, Ghemawat S. MapReduce: simplified data processing on large clusters[J]. Communications of the ACM, 2008, 51(1): 107-113. 
\bibitem{svn} https://subversion.apache.org/
\bibitem{Trac} http://trac.edgewall.org
\bibitem{Wiki} https://www.mediawiki.org/
\bibitem{DocDB} http://docdb-v.sourceforge.net/

\end{thebibliography}
\bibliographystyle{unsrt}

\end{document}